\documentclass[10pt, a4paper, oneside]{report}
\pdfoutput=1
\usepackage[utf8]{inputenc}
\usepackage[T1]{fontenc}
\usepackage{lmodern, textcomp}
\usepackage[british]{babel}
\usepackage{csquotes}

\usepackage[heightrounded, top=4cm, bottom=4cm, left=3.5cm, right=3.5cm]{geometry}

\usepackage{eurosym, xspace}
\usepackage[nottoc]{tocbibind}
\usepackage{graphicx, subcaption, float, changepage}
\usepackage{threeparttable, makecell}
\usepackage[multiple]{footmisc}
\usepackage{authblk}
\usepackage{multicol}

\usepackage[usenames, dvipsnames]{xcolor}

\usepackage[backend=bibtex8, citestyle=numeric-comp, maxbibnames=99,
			sorting=nyt, sortcase=false, sortcites=true, giveninits=true]{biblatex}
\input{arxiv.def}

\usepackage{amsmath, amsfonts, amssymb}
\usepackage{mathtools}
\usepackage{mathrsfs}
\usepackage{cancel, siunitx}
\usepackage{simpler-wick}

\usepackage{framed}
\usepackage{comment}
\usepackage{makeidx}

\usepackage{maths_min}

\usepackage[pdftex, breaklinks=true, linktocpage,
	colorlinks=true, urlcolor=blue, linkcolor=blue, citecolor=red,
	pdfauthor={Harold Erbin}
]{hyperref}
\usepackage[all]{hypcap}

\graphicspath{{images/}}

\numberwithin{equation}{chapter}

\DeclareUnicodeCharacter{00A0}{~}
\DeclareUnicodeCharacter{202F}{~}

\newcommand{\email}[1]{\thanks{\href{mailto:#1}{\nolinkurl{#1}}}}

\newcommand{\what}[1]{\widehat{#1}}
\newcommand{\wtilde}[1]{\widetilde{#1}}

\newcommand{\psbrbr}[2]{\langle\!\langle #1 , #2 \rangle\!\rangle}

\newcommand{\eadj}[1]{#1^\ddagger}

\newcommand{\M}[0]{\mathsf{M}}
\renewcommand{\K}[0]{\mathsf{K}}

\newcommand{\scr}[1]{\mathscr{#1}}

\DeclareMathOperator{\Vol}{Vol}

\usepackage[sort&compress, english]{cleveref}
\usepackage[nocomputations]{thmenv}

\makeindex

\hypersetup{
	pdftitle={String theory: a field theory perspective},
}

\title{String theory: a field theory perspective}

\author[*]{Harold Erbin\email{erbin@mit.edu}}

\affil[*]{%
	Center for Theoretical Physics
	\protect\\
	Massachusetts Institute of Technology, Cambridge, MA 02139, \textsc{Usa}
}

\affil[*]{%
	\textsc{Cea}, \textsc{List}, Gif-sur-Yvette, F-91191, France
}

\specialcomment{draft}{\begingroup\color{gray}\small}{\endgroup}
\specialcomment{check}{\begingroup\color{darkgray}\small}{\endgroup}

\newcommand{\introchapter}{}
\newcommand{\refchapter}{\section{References}}
\newcommand{\revname}[0]{review}

\newif\ifbook
\newif\ifspringer
\newif\ifreview

\booktrue
\springertrue
\reviewfalse

\renewcommand{\introchapter}{\paragraph{Abstract}}
\renewcommand{\refchapter}{\section{Suggested readings}}
\DefineBibliographyStrings{english}{references = {References}}
\renewcommand{\revname}[0]{book}
\title{String Field Theory -- A Modern Introduction}
\hypersetup{pdftitle={String Field Theory – A Modern Introduction}}
\excludecomment{check}
\excludecomment{draft}

\begin{document}

\maketitle

\begin{abstract}
	This \revname{} provides an introduction to string field theory (SFT).
	String theory is usually formulated in the worldsheet formalism, which describes a single string (first-quantization).
	While this approach is intuitive and could be pushed far due to the exceptional properties of two-dimensional theories, it becomes cumbersome for some questions or even fails at a more fundamental level.
	These motivations have led to the development of SFT, a description of string theory using the field theory formalism (second-quantization).
	As a field theory, SFT provides a rigorous and constructive formulation of string theory.

	The main objective is to construct the closed bosonic SFT and to explain how to assess the consistency of string theory with it.
	The accent is put on providing the reader with the foundations, conceptual understanding and intuition of what SFT is.
	After reading this book, they should be able to study the applications from the literature.

	The \revname{} is organized in two parts.
	The first part reviews the topics of the worldsheet theory that are necessary to build SFT (worldsheet path integral, CFT and BRST quantization).
	The second part starts by introducing general concepts of SFT from the BRST quantization.
	Then, it introduces off-shell string amplitudes before providing a Feynman diagrams interpretation from which the building blocks of SFT are extracted.
	After constructing the closed SFT, it is used to outline the proofs of several important consistency properties, such as background independence, unitarity and crossing symmetry.
	Finally, the generalization to the superstring is also discussed.

	This book grew up from lecture notes for a course given at the Ludwig-Maximilians-Universität LMU (winter semesters 2017--2018 and 2018--2019).
	The current document is the draft of the manuscript published by Springer.
\end{abstract}

\index{adjoint|see{Euclidean, Hermitian adjoint}}

\index{bc@$bc$ CFT|see{first-order CFT, reparametrization $bc$ ghosts}}
\index{bg@$\beta\gamma$ CFT|see{first-order CFT, superconformal $\beta\gamma$ ghosts}}

\index{BRST SFT|see{covariant SFT}}

\index{closed string states!level-matching|see{level-matching}}

\index{conformal!anomaly|see{Weyl anomaly}}
\index{conformal!factor|see{Liouville field}}
\index{conformal!vacuum|see{$\mathrm{SL}(2, \C)$ vacuum}}

\index{conformal algebra!$2d$|see{Witt algebra, Virasoro algebra}}

\index{conformal isometry group|seealso{conformal Killing vector}}
\index{conformal isometry group!sphere|see{$\mathrm{SL}(2, \C)$}}

\index{covariant SFT|seealso{classical (-), free (-), quantum (-)}}

\index{Faddeev--Popov!gauge fixing|see{path integral}}

\index{factorization|see{string amplitude}}

\index{field space|seealso{path integral}}

\index{first-order CFT!Fock space|see{Hilbert space}}

\index{free covariant SFT!path integral|see{string field path integral}}

\index{fundamental vertex|seealso{closed string (-)}}

\index{G@$G$|see{spacetime ghost number}}%

\index{ghosts|see{first-order CFT, reparametrization $bc$ ghosts, superconformal $\beta\gamma$ ghosts}}

\index{mapping class group (MCG)|see{modular group}}

\index{marked Riemann surface|see{punctured Riemann surface}}

\index{metric!local rescaling|see{Weyl transformation}}

\index{Mobius@Möbius group|see{$\mathrm{SL}(2, \C)$}}

\index{puncture|see{Riemann surface}}

\index{reparametrization $bc$ ghosts|seealso{first-order CFT}}

\index{Riemann surface!g0@$g = 0$|see{Riemann sphere}}

\index{Schwinger parameter|see{propagator}}

\index{Sm@$S$-matrix|see{scattering amplitude}}

\index{spacetime momentum|seealso{scalar field CFT}}

\index{SFT|see{string field theory}}

\index{string amplitude!closed string|see{closed string amplitude}}

\index{string Feynman diagram|seealso{momentum-space SFT}}

\index{string field|seealso{string states}}
\index{string field!gauge fixing|see{Siegel gauge}}

\index{string states|seealso{closed, open (-)}}
\index{string states!on-shell|see{on-shell condition}}
\index{string states!physical|see{BRST cohomology}}

\index{superconformal $\beta\gamma$ ghosts|seealso{first-order CFT}}

\index{ultralocality|see{path integral measure}}

\index{vertex operator|seealso{string states}}
\index{vertex operator|seealso{scalar field CFT}}

\index{winding number|seealso{scalar field CFT}}

\index{worldsheet!ghosts|see{reparametrization ghosts}}
\index{worldsheet!path integral|see{Polyakov path integral}}

\index{BRST cohomology|seealso{string states}}
\index{BV formalism|see{Batalin--Vilkovisky formalism}}
\index{CFT|see{conformal field theory}}
\index{CISO|see{conformal isometry group}}
\index{ISO|see{isometry group}}
\index{OPE|see{operator product expansion}}
\index{PCO|see{picture changing operator}}

\index{$^\ddagger$|see{Euclidean adjoint}}
\index{$^\dagger$|see{Hermitian adjoint}}
\index{$^c$|see{conjugate state}}
\index{$^t$|see{BPZ conjugation}}

\index{[]@$\{ \cdots \}_g$|see{fundamental vertex}}
\index{[]@$[ \cdots ]_g$|see{string product}}

\index{x@$\mc X$|see{picture changing operator}}

\index{$\chi_g, \chi_{g,n}$|see{Euler characteristics}}

\index{$\ket{\uparrow}$|see{first-order system, vacuum}}
\index{$\ket{\downarrow}$|see{first-order system, vacuum}}

\index{$\omega_p^{g,n}$|see{$\mc P_{g,n}$ space, $p$-form}}

\index{Agn@$A_{g,n}$|see{(off-shell) string amplitude}}
\index{Rgn@$\mc R_{g,n}$|see{off-shell string amplitude contribution}}

\index{an@$\alpha_n$|see{scalar field CFT, mode expansion}}

\index{f@$f \circ$|see{CFT, finite transformation}}

\index{g@$\hat g_{ab}$|see{background metric}}
\index{g@$g_{ab}$|see{worldsheet metric}}

\index{Kg@$\mc K_g$|see{conformal Killing vector}}

\index{ker P@$\ker \adj{P_1}$|see{quadratic differential}}
\index{ker P@$\ker P_1$|see{conformal Killing vector}}

\index{I@$I(z), I^\pm(z)$|see{inversion}}

\index{Mg@$\mc M_g$|see{moduli space}}
\index{Ngh@$N_{\text{gh}}$|see{ghost number}}

\index{$\Sigma_0$|see{Riemann sphere}}
\index{$\Sigma_g, \Sigma_{g,n}$|see{Riemann surface}}

\index{Vgn@$\mc V_{g,n}$|see{fundamental vertex}}
\index{Fgn@$\mc F_{g,n}$|see{propagator region}}
\index{Vgn@$\mc V_{g,n}^{\text{1PI}}$|see{1PI vertex}}
\index{Fgn@$\mc F_{g,n}^{\text{1PR}}$|see{1PR region}}

\index{lgn@$\ell_{g,n}$|see{string product}}

\index{r@$r(\Sigma_{g,n})$|see{index}}

\setcounter{page}{2}

\chapter*{Preface}
\addcontentsline{toc}{chapter}{Preface}

This \revname{} grew up from lectures delivered within the Elite Master Program “Theoretical and Mathematical Physics” from the Ludwig-Maximilians-Universität during the winter semesters 2017--2018 and 2018--2019.

The main focus of this \revname{} is the closed bosonic string field theory (SFT).
While there are many resources available for the open bosonic SFT, a single review~\cite{Erler:2020:FourLecturesClosed} has been written since the final construction of the bosonic closed SFT by Zwiebach~\cite{Zwiebach:1993:ClosedStringField}.
For this reason, it makes sense to provide a modern and extensive study.
Moreover, the usual approach to open SFT focuses on the cubic theory, which is so special that it is difficult to generalize the techniques to other SFTs.
Finally, closed strings are arguably more fundamental than open strings because they are always present since they describe gravity, which further motivates my choice.
However, the reader should not take this focus as denying the major achievements and the beauty of the open SFT; reading this book should provide most of the tools needed to feel comfortable also with this theory.

While part of the original goal of SFT is to provide a non-perturbative definition of string theory and to address important questions such as classifying consistent string backgrounds or understanding dualities, no progress on this front has been achieved so far.
Hence, there is still much to understand and the recent surge of developments provide a new chance to deepen our understanding of closed SFT.
For example, several consistency properties of string theory have been proven rigorously using SFT.
Moreover, the recent construction of the open-closed superstring field theory~\cite{Moosavian:2020:SuperstringFieldTheory} together with earlier works~\cite{Zwiebach:1993:ClosedStringField, Zwiebach:1998:OrientedOpenClosedString, Sen:2016:BVMasterAction, deLacroix:2017:ClosedSuperstringField} show that all types of string theories can be recast as a SFT.
This is why, I believe, it is a good time to provide a complete \revname{} on SFT.

The goal of this \revname{} is to offer a self-contained description of SFT and all the tools necessary to build it.
The emphasis is on describing the concepts behind SFT and to make the reader build intuitions on what it means.
For this reason, there are relatively few applications.

The reader is assumed to have some knowledge of QFT, and a basic knowledge of CFT and string theory (classical string, Nambu--Goto action, light-cone and old-covariant quantizations).

\section*{Organization}

The text is organized on three levels: the main content (augmented with examples), computations, and remarks.
The latter two levels can be omitted in a first lecture.
The examples, computations and remarks are clearly separated from the text (respectively, by a half-box on the left and bottom, by a vertical line on the left, and by italics) to help the navigation.

Many computations have been set aside from the main text to avoid breaking the flow and to provide the reader with the opportunity to check by themself first.
In some occasions, computations are postponed well below the corresponding formula to gather similar computations or to avoid breaking an argument.
While the derivations contain more details than usual textbooks and may look pedantic to the expert, I think it is useful for students and newcomers to have complete references where to check each step.
This is even more the case when there are many different conventions in the literature.
The remarks are not directly relevant to the core of the text but they make connections with other parts or topics.
The goal is to broaden the perspectives of the main text.

General references can be found at the end of each chapter to avoid overloading the text.
In-text references are reserved for specific points or explicit quotations (of a formula, a discussion, a proof, etc.).
I did not try to be exhaustive in the citations and I have certainly missed important references: this should be imputed to my lack of familiarity with them and not to their value.

This text is a preprint of the textbook~\cite{Erbin:2021:StringFieldTheory} and is reproduced with permission of Springer.
My plan is to frequently update the draft of this \revname{} with new content.
The last version can be accessed on my professional webpage, currently located at:
\begin{quote}
	\url{http://www.lpthe.jussieu.fr/~erbin/}
\end{quote}

\section*{Acknowledgements}

I have started to learn string field theory at \textsc{Hri} by attending lectures from Ashoke Sen.
Since then, I have benefited from collaboration and many insightful discussions with him.
Following his lectures have been much helpful in building an intuition that cannot be found in papers or reviews on the topic.
Through this \revname{}, I hope being able to make some of these insights more accessible.

I am particularly grateful to Ivo Sachs who proposed me to teach this course and to Michael Haack for continuous support and help with the organization, and to both of them for many interesting discussions during the two years I have spent at \textsc{Lmu}.
Moreover, I have been very lucky to be assigned an excellent tutor for this course, Christoph Chiaffrino.
After providing him with the topic and few references, Christoph has prepared all the tutorials and the corrections autonomously.
His help brought a lot to the course.

I am particularly obliged to all the students who have taken this course at \textsc{Lmu} for many interesting discussions and comments: Enrico Andriolo, Hrólfur Ásmundsson, Daniel Bockisch, Fabrizio Cordonnier, Julian Freigang, Wilfried Kaase, Andriana Makridou, Pouria Mazloumi, Daniel Panea, Martin Rojo.

I am also grateful to all the string theory community for many exchanges.
For discussions related to the topics of this book, I would like to thank more particularly: Costas Bachas, Adel Bilal, Subhroneel Chakrabarti, Atish Dabholkar, Benoit Douçot, Ted Erler, Dileep Jatkar, Carlo Maccaferri, Juan Maldacena, Yuji Okawa, Sylvain Ribault, Raoul Santachiara, Martin Schnabl, Dimitri Skliros, Jakub Vošmera.
I have received a lot of feedback during the different stages of writing this \revname{}, and I am obliged to all the colleagues who sent me feedback.

I am thankful to my colleagues at \textsc{Lmu} for providing a warm and stimulating environment, with special thanks to Livia Ferro for many discussions around coffee.
Moreover, the encouragements and advice from Oleg Andreev and Erik Plauschinn have been strong incentives for publishing this \revname{}.

The editorial process at Springer has been very smooth.
I would also like to thank Christian Caron and Lisa Scalone for their help and efficiency during the publishing process.
I am also indebted to Stefan Theisen for having supported the publication at Springer and for numerous comments and corrections on the draft.

Most of this book was written at the Ludwig--Maximilians--Universität (\textsc{Lmu}, Munich, Germany) where I was supported by a Carl Friedrich von Siemens Research Fellowship of the Alexander von Humboldt Foundation.
The final stage has been completed at the University of Turin (Italy).
My research is currently funded by the European Union's Horizon 2020 research and innovation program under the Marie Skłodowska-Curie grant agreement No 891169.

Finally, writing this \revname{} would have been more difficult without the continuous and loving support from Corinne.

\bigskip
November 2020
\begin{flushright}
Harold Erbin
\end{flushright}

\pdfbookmark[0]{\contentsname}{toc}
\tableofcontents

\chapter{Introduction}
\label{chap:intro}

\introchapter

In this chapter, we introduce the main motivations for studying string theory, and why it is important to design a string field theory.
After describing the central features of string theory, we describe the most important concepts of the worldsheet formulation.
Then, we explain the reasons leading to string field theory (SFT) and outline the ideas which will be discussed in the rest of the \revname{}.

\section{Strings, a distinguished theory}
\label{sec:intro:string:distinguished}

\index{string theory!motivations|(}
\index{p-brane@$p$-brane}

The first and simplest reason for considering theories of fundamental $p$-branes (fundamental objects extended in $p$ spatial dimensions) can be summarized by the following question: “Why would Nature just make use of point-particles?”
There is no a priori reason forbidding the existence of fundamental extended objects and, according to Gell-Mann's totalitarian principle, “Everything not forbidden is compulsory.”
If a consistent theory cannot be built (after a reasonable amount of effort) or if it contradicts current theories (in their domains of validity) and experiments, then one can support the claim that only point-particles exist.
On the other side, if such a theory can be built, it is of primary interest to understand it deeper and to see if it can solve the current problems in high-energy theoretical physics.

The simplest case after the point particle is the string, so it makes sense to start with it.
It happens that a consistent theory of strings can be constructed, and that the latter (in its supersymmetric version) contains all the necessary ingredients for a fully consistent high-energy model:\footnotemark{}
\footnotetext{%
	There are also indications that a theory of membranes ($2$-branes) in $10 + 1$ dimensions, called M-theory, should exist.
	No direct and satisfactory description of the latter has been found and we will thus focus on string theory in this \revname{}.
}%
\begin{itemize}
	\item quantum gravity (quantization of general relativity plus higher-derivative corrections);

	\item grand unification (of matter, interactions and gravity);

	\item no divergences, UV finiteness (finite and renormalizable theory);

	\item fixed number of dimensions ($26 = 25 + 1$ for the bosonic string, $10 = 9 + 1$ for the supersymmetric version);

	\item existence of all possible branes;

	\item no dimensionless parameters and one dimensionful parameter (the string length $\ell_s$).
\end{itemize}
It can be expected that a theory of fundamental strings ($1$-branes) occupies a distinguished place among fundamental $p$-branes for the following reasons.

\paragraph{Interaction non-locality}

\begin{figure}[tp]
	\centering
	\includegraphics[scale=1.2]{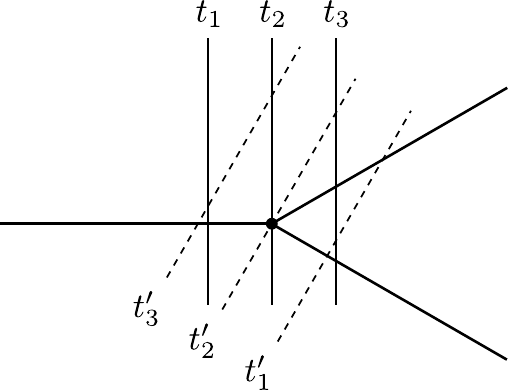}
	\caption{%
		Locality of a particle interaction: two different observers always agree on the interaction point and which parts of the worldline are $1$- and $2$-particle states.
	}%
	\label{intro:fig:particle-locality-interaction}
\end{figure}

\begin{figure}[tp]
	\centering
	\subcaptionbox{Observers at rest and boosted.}{%
		\includegraphics[scale=1.2]{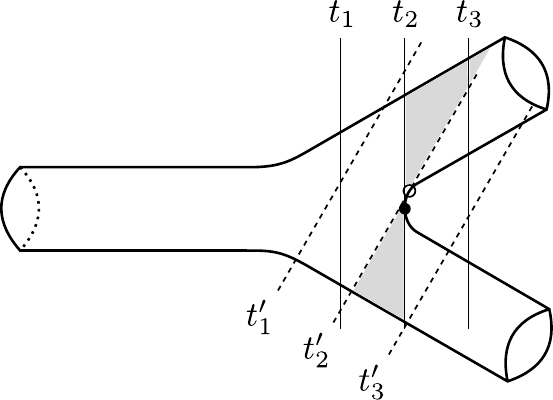}
	}%
	\hspace{1cm}
	\subcaptionbox{Observers close to the speed of light moving in opposite directions. The interactions are widely separated in each case.}{%
		\includegraphics[scale=1.2]{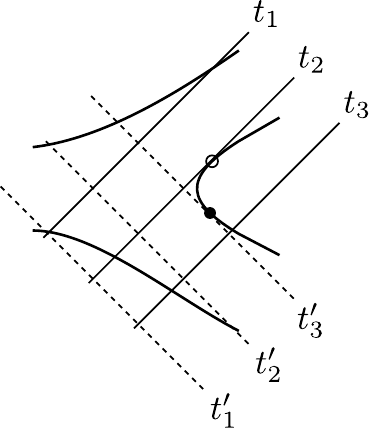}
	}%
	\caption{%
		Non-locality of string interaction: two different observers see the interaction happening at different places (denoted by the filled and empty circles) and they don't agree on which parts of the worldsheet are $1$- and $2$-string states (the litigation is denoted by the grey zone).
	}%
	\label{intro:fig:string-locality-interaction}
\end{figure}

\index{non-locality}
\index{string!interactions}
In a QFT of point particles, UV divergences arise because interactions (defined as the place where the number and/or nature of the objects change) are arbitrarily localized at a spacetime point.
In Feynman graphs, such divergences can be seen when the momentum of a loop becomes infinite (two vertices collide): this happens when trying to concentrate an infinite amount of energy at a single point.
However, these divergences are expected to be reduced or absent in a field theory of extended objects: whereas the interaction between particles is perfectly local in spacetime and agreed upon by all observers (\Cref{intro:fig:particle-locality-interaction}), the spatial extension of branes makes the interactions non-local.
This means that two different observers will neither agree on the place of the interactions (\Cref{intro:fig:string-locality-interaction}), nor on the part of the diagram which describes one or two branes.

The string lies at the boundary between too much local and too much non-local: in any given frame, the interaction is local in space, but not in spacetime.
The reason is that a string is one-dimensional and splits or joins along a point.
For $p > 1$, the brane needs to break/join along an extended spatial section, which looks non-local.

Another consequence of the non-locality is a drastic reduction of the possible interactions.
If an interaction is Lorentz invariant, Lorentz covariant objects can be attached at the vertex (such as momentum or gamma matrices): this gives Lorentz invariants after contracting with indices carried by the field.
But, this is impossible if the interaction itself is non-local (and thus not invariant): inserting a covariant object would break Lorentz invariance.

\paragraph{Brane degrees of freedom}

The higher the number of spatial dimensions of a $p$-brane, the more possibilities it has to fluctuate.
As a consequence, it is expected that new divergences appear as $p$ increases due to the proliferations of the brane degrees of freedom.
From the worldvolume perspective, this is understood from the fact that the worldvolume theory describes a field theory in $(p+1)$ dimensions, and UV divergences become worse as the number of dimensions increase.
The limiting case happens for the string ($p = 1$) since two-dimensional field theories are well-behaved in this respect (for example, any monomial interaction for a scalar field is power-counting renormalizable).
This can be explained by the low-dimensionality of the momentum integration and by the enhancement of symmetries in two dimensions.
Hence, strings should display nice properties and are thus of special interest.

\paragraph{Worldvolume theory}

\index{worldvolume description}%

The point-particle ($0$-brane) and the string ($1$-brane) are also remarkable in another aspect: it is possible to construct a simple worldvolume field theory (and the associated functional integral) in terms of a worldvolume metric.
All components of the latter are fixed by gauge symmetries (diffeomorphisms for the particle, diffeomorphisms and Weyl invariance for the string).
This ensures the reparametrization invariance of the worldvolume without having to use a complicated action.
Oppositely, the worldvolume metric cannot be completely gauge fixed for $p > 1$.

\paragraph{Summary}

As a conclusion, strings achieve an optimal balance between spacetime and worldsheet divergences, as well as having a simple description with reparametrization invariance.

Since the construction of a field theory is difficult, it is natural to start with a worldsheet theory and to study it in the first-quantization formalism, which will provide a guideline for writing the field theory.
In particular, this allows to access the physical states in a simple way and to find other general properties of the theory.
When it comes to the interactions and scattering amplitudes, this approach may be hopeless in general since the topology of the worldvolume needs to be specified by hand (describing the interaction process).
In this respect, the case of the string is again exceptional: because Riemann surfaces have been classified and are well-understood, the arbitrariness is minimal.
Combined with the tools of conformal field theory, many computations can be performed.
Moreover, since the modes of vibrations of the strings provide all the necessary ingredients to describe the Standard model, it is sufficient to consider only one string field (for one type of strings), instead of the plethora found in point-particle field theory (one field for each particle).
Similarly, non-perturbative information (such as branes and dualities) could be found only due to the specific properties of strings.

Coming back to the question which opened this section, higher-dimensional branes of all the allowed dimensions naturally appear in string theory as bound states.
Hence, even if the worldvolume formulation of branes with $p > 1$ looks pathological\footnotemark{}, string theory hints towards another definition of these objects.
\footnotetext{%
	Entering in the details would take us too far away from the main topic of this \revname{}.
	Some of the problems found when dealing with $(p > 2)$-branes are: how to define a Wick rotation for 3-manifolds, the presence of Lorentz anomalies in target spacetime, problems with the spectrum, lack of renormalizability, impossibility to gauge-fix the worldvolume metric~\cite{Polchinski:2005:StringTheory-1, Kikkawa:1986:CanMembraneBe, Bars:1987:MasslessSpectrumCritical, Bars:1988:AnomaliesSuperPbranes, Bars:1988:CentralExtensionsArea, Bars:1988:ThereUniqueConsistent, Bars:1990:MembraneSymmetriesAnomalies, Bars:1988:FirstMassiveLevel, Bars:1990:IssuesTopologySpectrum, deAzcarraga:1991:ClassicalAnomaliesSupersymmetric, deWit:1988:QuantumMechanicsSupermembranes, deWit:1989:SupermembraneUnstable, Duff:1988:SemiclassicalQuantizationSupermembrane, Marquard:1988:ConditionsEmbeddingSpace, Marquard:1989:LorentzAlgebraCriticalBosonic, Marquard:1989:LorentzAlgebraCriticalSuper, Paccanoni:1989:RemarksConsistencyQuantum, Luckock:1989:QuantumGeometryRandom}.

}%

\index{string theory!motivations|)}%

\section{String theory}
\label{sec:intro:string}

\subsection{Properties}
\label{sec:intro:string:properties}

\index{string!properties}%

The goal of this section is to give a general idea of string theory by introducing some concepts and terminology.
The reader not familiar with the points described in this section is advised to follow in parallel some standard worldsheet string theory textbooks.

\subsubsection{Worldsheet CFT}

A string is characterized by its worldsheet field theory (\Cref{chap:bos:ws-int-vac}).\footnotemark{}
\footnotetext{%
	We focus mainly on the bosonic string theory, leaving aside the superstring, except when differences are important.
}%
The worldsheet is parametrized by coordinates $\sigma^a = (\tau, \sigma)$.
The simplest description is obtained by endowing the worldsheet with a metric $g_{ab}(\sigma^a)$ ($a = 0, 1$) and by adding a set of $D$ scalar fields $X^\mu(\sigma^a)$ living on the worldsheet ($\mu = 0, \ldots, D-1$).
The latter represents the position of the string in the $D$-dimensional spacetime.
From the classical equations of motion, the metric $g_{ab}$ is proportional to the metric induced on the worldsheet from its embedding in spacetime.
More generally, one ensures that the worldsheet metric is non-dynamical by imposing that the action is invariant under (worldsheet) diffeomorphisms and under Weyl transformations (local rescalings of the metric).
\index{critical dimension}%
The consistency of these conditions at the quantum level imposes that $D = 26$, and this number is called the \emph{critical dimension}.
\index{worldsheet!CFT}%
Gauge fixing the symmetries, and thus the metric, leads to the conformal invariance of the resulting worldsheet field theory: a conformal field theory (CFT) is a field theory (possibly on a curved background) in which only angles and not distances can be measured (\Cref{chap:cft:general,chap:cft:plane,cft:chap:systems}).
This simplifies greatly the analysis since the two-dimensional conformal algebra (called the Virasoro algebra) is infinite-dimensional.

CFTs more general than $D$ free scalar fields can be considered: fields taking non-compact values are interpreted as non-compact dimensions while compact or Grassmann-odd fields are interpreted as compact dimensions or internal structure, like the spin.

While the light-cone quantization allows to find quickly the states of the theory, the simplest covariant method is the BRST quantization (\Cref{cft:chap:brst}).
It introduces ghosts (and superghosts) associated to the gauge fixing of diffeomorphisms (and local supersymmetry).
These (super)ghosts form a CFT which is universal (independent of the matter CFT).

\begin{figure}[tp]
	\centering
	\subcaptionbox{Open string}[.4\linewidth]{%
		\includegraphics[scale=1]{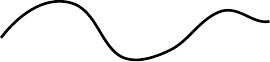}
	}%
	\subcaptionbox{Closed string}[.4\linewidth]{%
		\includegraphics[scale=1]{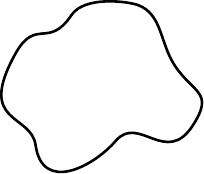}
	}%
	\caption{Open and closed strings.}
	\label{intro:fig:string-open-closed}
\end{figure}

\begin{figure}[tp]
	\centering
	\begin{subfigure}[c]{.25\linewidth}
		\includegraphics[scale=0.8]{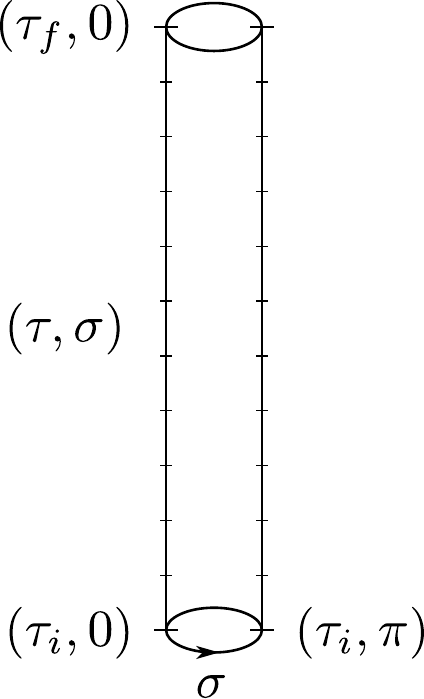}
	\end{subfigure}
	$\xrightarrow{\makebox[1.5cm]{}}$
	\begin{subfigure}[c]{.5\linewidth}
		\includegraphics[scale=0.8]{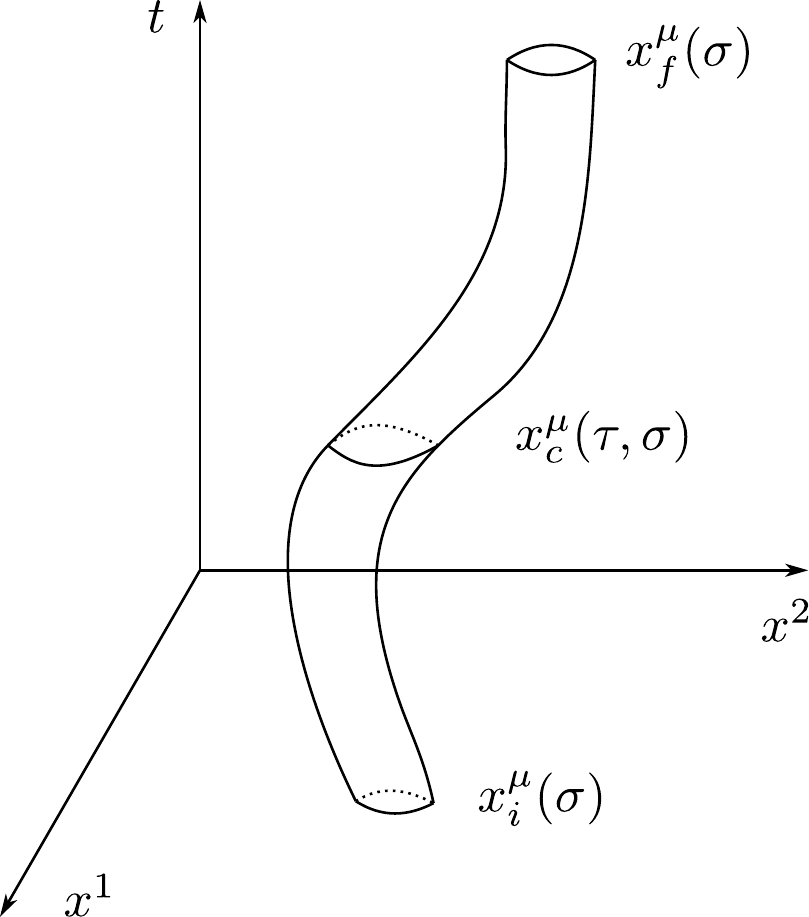}
	\end{subfigure}
	\caption{%
		Trajectory $x^\mu_c(\tau, \sigma)$ of a closed string in spacetime (worldsheet).
		It begins and ends at the circles parametrized by $x_i(\sigma)$ and $x_f(\sigma)$.
		The worldsheet is topologically a cylinder and is parametrized by $(\tau, \sigma) \in [\tau_i, \tau_f] \times [0, 2\pi)$.
	}%
	\label{intro:fig:string-closed-worldvolume}
\end{figure}

\begin{figure}[tp]
	\centering
	\begin{subfigure}[c]{.3\linewidth}
		\centering
		\includegraphics[scale=0.8]{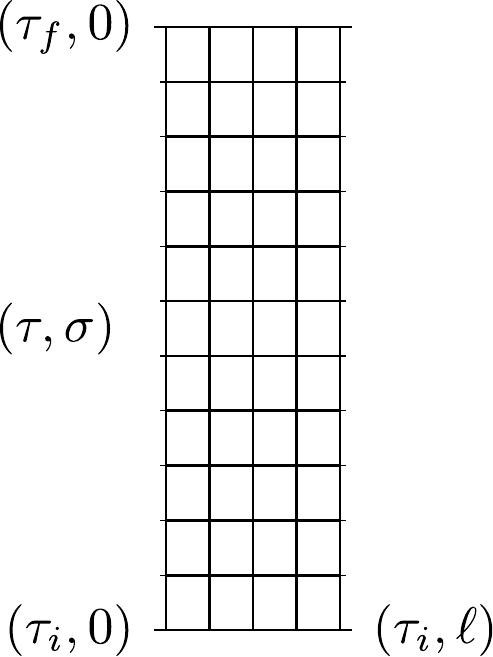}
	\end{subfigure}
	$\xrightarrow{\makebox[1.5cm]{}}$
	\begin{subfigure}[c]{.45\linewidth}
		\centering
		\includegraphics[scale=0.8]{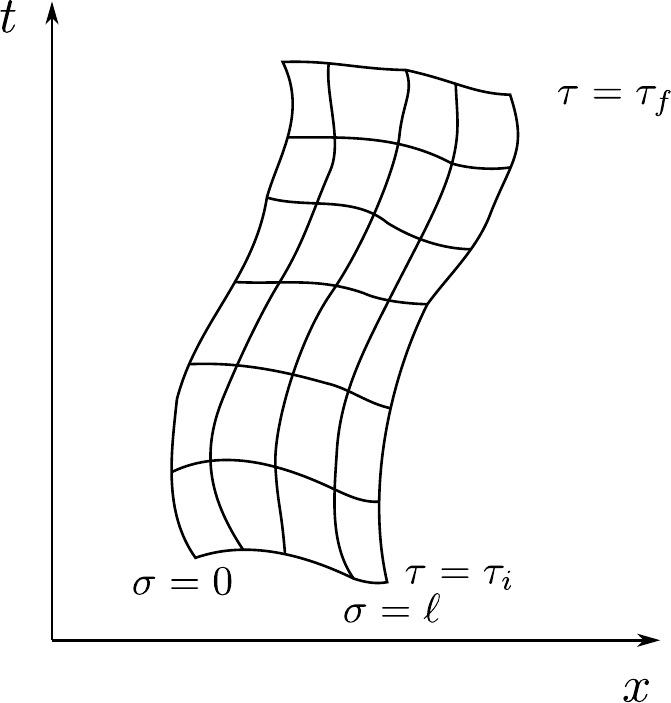}
	\end{subfigure}
	\caption{%
		Trajectory $x^\mu_c(\tau, \sigma)$ of an open string in spacetime (worldsheet).
		It begins and ends at the lines parametrized by $x_i(\sigma)$ and $x_f(\sigma)$.
		The worldsheet is topologically a rectangle and is parametrized by $(\tau, \sigma) \in [\tau_i, \tau_f] \times [0, \ell]$.
	}%
	\label{intro:fig:string-open-worldvolume}
\end{figure}

\index{worldsheet!classification}%
\index{worldsheet!boundary conditions}%
The trajectory of the string is denoted by $x_c(\tau, \sigma)$.
It begins and ends respectively at the geometric shapes parametrized by $x_c(\tau_i, \sigma) = x_i(\sigma)$ and by $x_c(\tau_f, \sigma) = x_f(\sigma)$.
Note that the coordinate system on the worldsheet itself is arbitrary.
The spatial section of a string can be topologically closed (circle) or open (line) (\Cref{intro:fig:string-open-closed}), leading to cylindrical or rectangular worldsheets as illustrated in \Cref{intro:fig:string-closed-worldvolume,intro:fig:string-open-worldvolume}.
To each topology is associated different boundary conditions and types of strings:
\begin{itemize}
	\item closed: periodic and anti-periodic boundary conditions;
	\item open: Dirichlet and Neumann boundary conditions.
\end{itemize}
While a closed string theory is consistent by itself, an open string theory is not and requires closed strings.

\subsubsection{Spectrum}

\index{string states}%
In order to gain some intuition for the states described by a closed string, one can write the Fourier expansion of the fields $X^\mu$ (in the gauge $g_{ab} = \eta_{ab}$ and after imposing the equations of motion)
\index{scalar field CFT!mode expansion}%
\begin{equation}
	X^\mu(\tau, \sigma) \sim x^\mu + p^\mu \tau + \frac{\I}{\sqrt{2}} \sum_{n \in \Z^*} \frac{1}{n} \, \big( \alpha_n^{\mu} \e^{- \I n (\tau - \sigma)} + \bar\alpha_n^{\mu} \e^{- \I n (\tau + \sigma)} \big) \,,
\end{equation}
where $x^\mu$ is the centre-of-mass position of the string and $p^\mu$ its momentum.\footnotemark{}
\footnotetext{%
	In the introduction, we set $\alpha' = 1$.
}%
Canonical quantization leads to the usual commutator:
\begin{equation}
	\com{x^\mu}{p^\nu} = \I \eta^{\mu\nu} \,.
\end{equation}
With respect to a point-particle for which only the first two terms are present, there are an infinite number of oscillators $\alpha_n^{\mu}$ and $\bar\alpha_n^{\mu}$ which satisfy canonical commutation relations for creation $n < 0$ and annihilation operators $n > 0$
\begin{equation}
	\com{\alpha_m^{\mu}}{\alpha_n^{\nu}} = m \, \eta^{\mu\nu} \delta_{m+n,0} \,.
\end{equation}
The non-zero modes are the Fourier modes of the excitations of the embedded string.
The case of the open string is simply obtained by setting $\bar\alpha_n = \alpha_n$ and $p \to 2 p$.
The Hamiltonian for the closed and open strings read respectively
\begin{subequations}
\begin{align}
	H_{\text{closed}} &= - \frac{m^2}{2} + N + \bar N - 2 \,,
	\\
	H_{\text{open}} &= - m^2 + N - 1
\end{align}
\end{subequations}
where $m^2 = - p^\mu p_\mu$ is the mass of the state (in Planck units), $N$ and $\bar N$ (level operators) count the numbers $N_n$ and $\bar N_n$ of oscillators $\alpha_n$ and $\bar\alpha_n$ weighted by their mode index $n$:
\begin{equation}
	\begin{gathered}
		N = \sum_{n \in \N} n N_n \,,
		\qquad
		N_n = \frac{1}{n} \, \alpha_{-n} \cdot \alpha_{n} \,,
		\\
		\bar N = \sum_{n \in \N} n \bar N_n \,,
		\qquad
		\bar N_n = \frac{1}{n} \, \bar \alpha_{-n} \cdot \bar \alpha_{n} \,.
	\end{gathered}
\end{equation}

With these elements, the Hilbert space of the string theory can be constructed.
\index{on-shell condition}%
Invariance under reparametrization leads to the \emph{on-shell condition}, which says that the Hamiltonian vanishes:
\begin{equation}
	H \ket{\psi} = 0
\end{equation}
for any physical state $\ket{\psi}$.
\index{level-matching condition}%
Another constraint for the closed string is the \emph{level-matching condition}
\begin{equation}
	(N - \bar N) \ket{\psi} = 0 \,.
\end{equation}
It can be understood as fixing an origin on the string.

The ground state $\ket{k}$ with momentum $k$ is defined to be the eigenstate of the momentum operator which does not contain any oscillator excitation:
\begin{equation}
	p^\mu \ket{k}
		= k^\mu \ket{k},
	\qquad
	\forall n > 0 :
		\quad
		\alpha_n^{\mu} \ket{k}
			= 0 \,.
\end{equation}
A general state can be built by applying successively creation operators
\begin{equation}
	\ket{\psi} = \prod_{n > 0} \prod_{\mu = 0}^{D - 1} (\alpha^{\mu}_{-n})^{N_{n,\mu}} \ket{k} \,,
\end{equation}
where $N_{n,\mu} \in \N$ counts excitation level of the oscillator $\alpha_{-n}^{\mu}$.
In the rest of this section, we describe the first two levels of states.

\index{closed string states!tachyon $T$}%
\index{open string states!tachyon $T$}%
\index{tachyon}%
The ground state is a tachyon (faster-than-light particle) because the Hamiltonian constraint shows that it has a negative mass (in the units where $\alpha' = 1$):
\begin{equation}
	\text{closed}:
	\quad
		m^2 = - 4 \,,
	\hspace{2cm}
	\text{open}:
	\quad
		m^2 = - 1 \,.
\end{equation}

\index{open string states!gauge field $A_\mu$}%
The first excited state of the open string is found by applying $\alpha_{-1}$ on the vacuum $\ket{k}$:
\begin{equation}
	\alpha_{-1}^{\mu} \ket{k} \,.
\end{equation}
This state is massless:
\begin{equation}
	m^2 = 0
\end{equation}
and since it transforms as a Lorentz vector (spin $1$), it is identified with a $\group{U}(1)$ gauge boson.
\index{open string states!momentum expansion}%
Writing a superposition of such states
\begin{equation}
	\ket{A} = \int \dd^D k \, A_\mu(k) \, \alpha_{-1}^{\mu} \ket{k},
\end{equation}
the coefficient $A_\mu(k)$ of the Fourier expansion is interpreted as the spacetime field for the gauge boson.
Reparametrization invariance is equivalent to the equation of motion
\begin{equation}
	k^2 A_\mu = 0 \,.
\end{equation}
One can prove that the field obeys the Lorentz gauge condition
\begin{equation}
	k^\mu A_\mu = 0 \,,
\end{equation}
\index{open string states!gauge invariance}%
which results from gauge fixing the $\group{U}(1)$ gauge invariance
\begin{equation}
	A_\mu \longrightarrow A_\mu + k_\mu \lambda \,.
\end{equation}
It can also be checked that the low-energy action reproduces the Maxwell action.

\index{closed string states!massless field}%
\index{closed string states!metric $G_{\mu\nu}$}%
\index{closed string states!dilaton $\Phi$}%
\index{closed string states!Kalb--Ramond $B_{\mu\nu}$}%
The first level of the closed string is obtained by applying both $\alpha_{-1}$ and $\bar\alpha_{-1}$ (this is the only way to match $N = \bar N$ at this level)
\begin{equation}
	\alpha_{-1}^{\mu} \bar\alpha_{-1}^{\nu} \ket{k}
\end{equation}
and the corresponding states are massless
\begin{equation}
	m^2 = 0 \,.
\end{equation}
These states can be decomposed into irreducible representations of the Lorentz group
\begin{equation}
	\begin{gathered}
		\left( \alpha_{-1}^{\mu} \bar\alpha_{-1}^{\nu} + \alpha_{-1}^{\nu} \bar\alpha_{-1}^{\mu} - \frac{1}{D} \, \eta^{\mu\nu} \alpha_{-1} \cdot \bar\alpha_{-1} \right) \ket{p} \,,
		\\
		\big( \alpha_{-1}^{\mu} \bar\alpha_{-1}^{\nu} - \alpha_{-1}^{\nu} \bar\alpha_{-1}^{\mu} \big) \ket{p} \,,
		\qquad
		\frac{1}{D} \, \eta_{\mu\nu} \alpha_{-1}^{\mu} \bar\alpha_{-1}^{\nu} \ket{p}
	\end{gathered}
\end{equation}
which are respectively associated to the spacetime fields $G_{\mu\nu}$ (metric, spin $2$), $B_{\mu\nu}$ (Kalb--Ramond $2$-form) and $\Phi$ (dilaton, spin $0$).
The appearance of a massless spin $2$ particle (with low-energy action being the Einstein--Hilbert action) is a key result and originally raised interest for string theory.

\begin{remark}[Reparametrization constraints]
	Reparametrization invariance leads to other constraints than $H = 0$.
	They imply in particular that the massless fields have the correct gauge invariance and hence the correct degrees of freedom.

	Note that, after taking into account these constraints, the remaining modes correspond to excitations of the string in the directions transverse to it.
\end{remark}

Hence, each vibrational mode of the string corresponds to a spacetime field for a point-particle (and linear superpositions of modes can describe several fields).
This is how string theory achieves unification since a single type of string (of each topology) is sufficient for describing all the possible types of fields encountered in the standard model and in gravity.
They correspond to the lowest excitation modes, the higher massive modes being too heavy to be observed at low energy.

\begin{check}
The understanding of the string in terms of spacetime fields follows also from the observation that a fundamental string is very tiny ($\SI{e-33}{cm}$) and as such it appears to be point-like when seen from afar (\Cref{intro:fig:particle-zoom-string}).
The spin and the other properties of the particles are provided by the internal structure of the string (and in particular its vibrational mode).

\begin{figure}[tp]
	\centering
	\begin{subfigure}[c]{.4\linewidth}
		\centering
		\includegraphics[scale=1.2]{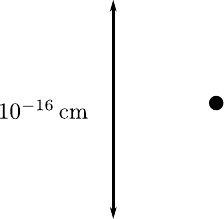}
	\end{subfigure}
	\begin{subfigure}[c]{.4\linewidth}
		\centering
		\includegraphics[scale=1.2]{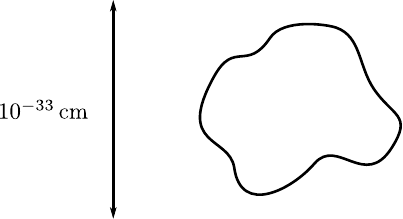}
	\end{subfigure}
	\caption{%
		A string of length $\SI{e-33}{cm}$ looks like a point-like particle at higher scales.
	}%
	\label{intro:fig:particle-zoom-string}
\end{figure}

\end{check}

\index{tachyon!instability}%
Bosonic string theory includes tachyons and is thus unstable.
While the instability of the open string tachyon is well understood and indicates that open strings are unstable and condense to closed strings, the status of the closed string tachyon is more worrisome (literally interpreted, it indicates a decay of spacetime itself).
\index{superstring!motivations}%
In order to solve this problem, one can introduce supersymmetry: in this case, the spectrum does not include the tachyon because it cannot be paired with a supersymmetric partner.

Moreover, as its name indicates, the bosonic string possesses only bosons in its spectrum (perturbatively), which is an important obstacle to reproduce the standard model.
By introducing spacetime fermions, supersymmetry also solves this problem.
The last direct advantage of the superstring is that it reduces the number of dimensions from $26$ to $10$, which makes the compactification easier.

\subsection{Classification of superstring theories}
\label{sec:intro:string:classification}

\index{superstring}%
In this section, we describe the different superstring theories (\Cref{part:superstring}).
In order to proceed, we need to introduce some new elements.

\index{worldsheet!CFT}%
The worldsheet field theory of the closed string is made of two sectors, called the left- and right-moving sectors (the $\alpha_n$ and $\bar\alpha_n$ modes).
While they are treated symmetrically in the simplest models, they are in fact independent (up to the zero-mode) and the corresponding CFT can be chosen to be distinct.

\index{supersymmetry}%
The second ingredient already evoked earlier is supersymmetry.
This symmetry associates a fermion to each boson (and conversely) through the action of a supercharge $Q$
\begin{equation}
	\ket{\text{boson}} = Q \ket{\text{fermion}}.
\end{equation}
More generally, one can consider $N$ supercharges which build up a family of several bosonic and fermionic partners.
Since each supercharge increases the spin by $1/2$ (in $D = 4$), there is an upper limit for the number of supersymmetries -- for interacting theories with a finite number of fields\footnotemark{} -- in order to keep the spin of a family in the range where consistent actions exist:
\footnotetext{%
	These conditions exclude the cases of free theories and higher-spin theories.
}%
\begin{itemize}
	\item $N_{\text{max}} = 4$ without gravity ($-1 \le \text{spin} \le 1$);
	\item $N_{\text{max}} = 8$ with gravity ($-2 \le \text{spin} \le 2$).
\end{itemize}
This counting serves as a basis to determine the maximal number of supersymmetries in other dimensions (by relating them through dimensional reductions).

Let's turn our attention to the case of the two-dimensional worldsheet theory.
The number of supersymmetries of the closed left- and right-moving sectors can be chosen independently, and the number of charges is written as $(N_L, N_R)$ (the index is omitted when statements are made at the level of the CFT).
\index{critical dimension}%
The critical dimension (absence of quantum anomaly for the Weyl invariance) depends on the number of supersymmetry
\begin{equation}
	D(N = 0) = 26,
	\qquad
	D(N = 1) = 10.
\end{equation}
\index{superstring!type II (-)}%
Type II superstrings have $(N_L, N_R) = (1, 1)$ and come in two flavours called IIA and IIB according to the chiraly of the spacetime gravitini chiralities.
\index{superstring!heterotic (-)}%
A theory is called \emph{heterotic} if $N_L > N_R$; we will mostly be interested in the case $N_L = 1$ and $N_R = 0$.\footnotemark{}
\footnotetext{%
	The case $N_L < N_R$ is identical up to exchange of the left- and right-moving sectors.
}%
In such theories, there cannot be open strings since both sectors must be equal in the latter.
Since the critical dimensions of the two sectors do not match, one needs to get rid of the additional dimensions of the right-moving sector; this leads to the next topic -- gauge groups.

\index{string!gauge group}%
Gauge groups associated with spacetime gauge bosons appear in two different places.
In heterotic models, the compactification of the unbalanced dimensions of the left sector leads to the appearance of a gauge symmetry.
The possibilities are scarce due to consistency conditions which ensure a correct gluing with the right-sector.
Another possibility is to add degrees of freedom -- known as Chan--Paton indices -- at the ends of open strings: one end transforms in the fundamental representation of a group $G$, while the other end transforms in the anti-fundamental.
The modes of the open string then reside in the adjoint representation, and the massless spin-$1$ particles become the gauge bosons of the non-Abelian gauge symmetry.

\index{string!orientation}%
Finally, one can consider oriented or unoriented strings.
An oriented string possesses an internal direction, i.e.\ there is a distinction between going from the left to the right (for an open string) or circling in clockwise or anti-clockwise direction (for a closed string).
Such an orientation can be attributed globally to the spacetime history of all strings (interacting or not).
The unoriented string is obtained by quotienting the theory by the $\Z_2$ worldsheet parity symmetry which exchanges the left- and right-moving sectors.
\index{superstring!type I (-)}%
Applying this to the type IIB gives the type I theory.

The tachyon-free superstring theories together with the bosonic string are summarized in \Cref{intro:tab:types-string}.

\begin{table}[ht]
	\begin{adjustwidth}{-2cm}{-2cm}
	\centering
	\begin{threeparttable}
	\begin{tabular}{c|c|cccccc}
		&
			\makecell{worldsheet \\ susy} &
			$D$ &
			\makecell{spacetime \\ susy} &
			gauge group &
			open string &
			oriented &
			tachyon
			\\
		\hline
		bosonic &
			$(0, 0)$ &
			$26$ &
			$0$ &
			any\tnote{*} &
			yes &
			yes / no &
			yes
			\\
		type I &
			$(1, 1)$ &
			$10$ &
			$(1, 0)$ &
			$\group{SO}(32)$ &
			yes &
			no &
			no
			\\
		type IIA &
			$(1, 1)$ &
			$10$ &
			$(1, 1)$ &
			$\group{U}(1)$ &
			(yes)\tnote{\dag} &
			yes &
			no
			\\
		type IIB &
			$(1, 1)$ &
			$10$ &
			$(2, 0)$ &
			none &
			(yes)\tnote{\dag} &
			yes &
			no
			\\
		heterotic $\group{SO}(32)$ &
			$(1, 0)$ &
			$10$ &
			$(1, 0)$ &
			$\group{SO}(32)$ & 
			no &
			yes &
			no
			\\
		heterotic $\group{E}_8$ &
			$(1, 0)$ &
			$10$ &
			$(1, 0)$ &
			$\group{E}_8 \times \group{E}_8$ &
			no &
			yes &
			no
			\\
		heterotic $\group{SO}(16)$ &
			$(1, 0)$ &
			$10$ &
			$(0, 0)$ &
			$\group{SO}(16) \times \group{SO}(16)$ &
			no &
			yes &
			no
	\end{tabular}
	\medskip
	\begin{tablenotes}
		\small
		\item[*] UV divergences beyond the tachyon (interpreted as closed string dilaton tadpoles) cancel only for the unoriented open plus closed strings with gauge group $\group{SO}(2^{13}) = \group{SO}(8192)$.

		\item[\dag] The parenthesis indicates that type II theories don't have open strings in the vacuum: they require a $D$-brane background.
		This is expected since there is no gauge multiplet in $d = 10$ $(1, 1)$ or $(2, 0)$ supergravities (the $D$-brane breaks half of the supersymmetry).
	\end{tablenotes}
	\end{threeparttable}
	\end{adjustwidth}

	\caption{%
		List of the consistent tachyon-free (super)string theories.
		The bosonic theory is added for comparison.
		There are additional heterotic theories without spacetime supersymmetry, but they contain a tachyon and are thus omitted.
	}%
	\label{intro:tab:types-string}
\end{table}

\subsection{Interactions}
\label{sec:intro:string:interactions}

\subsubsection{Worldsheet and Riemann surfaces}

\index{string!interactions}%

After having described the spectrum and the general characteristics of string theory comes the question of interactions.
\index{worldsheet!Riemann surface}%
The worldsheets obtained in this way are Riemann surfaces, i.e.\ $1$-dimensional complex manifolds.
They are classified by the numbers of handles (or holes) $g$ (called the genus) and external tubes $n$.
In the presence of open strings, surfaces have boundaries: in addition to the handles and tubes, they are classified by the numbers of disks $b$ and of strips $m$.\footnotemark{}
\footnotetext{%
	We ignore unoriented strings in this discussion.
	They would lead to an additional object called a cross-cap, which is a place where the surface looses its orientation.
}%
A particularly important number associated to each surface is the Euler characteristics
\begin{equation}
	\chi = 2 - 2 g - b \,,
\end{equation}
which is a topological invariant.
It is remarkable that there is a single topology at every loop order when one considers only closed strings, and just a few more in the presence of open strings.
The analysis is greatly simplified in contrast to QFT, for which the number of Feynman graphs increases very rapidly with the number of loops and external particles.

\begin{figure}[tp]
	\centering
		\includegraphics[scale=1.6]{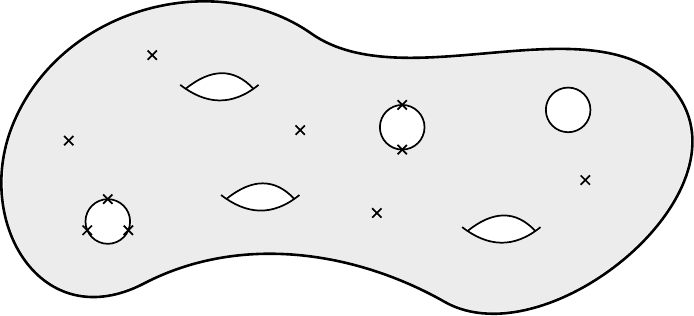}
	\caption{%
		General Riemann surfaces with boundaries and punctures.
	}%
	\label{intro:fig:ws-process-general}
\end{figure}

Due to the topological equivalence between surfaces, a conformal map can be used in order to work with simpler surfaces.
\index{Riemann surface!puncture}%
In particular, the external tubes and strips are collapsed to points called punctures (or marked points) on the corresponding surfaces or boundaries.
A general amplitude then looks like a sphere from which holes and disks have been removed and to which marked points have been pierced (\Cref{intro:fig:ws-process-general}).

\begin{figure}[tp]
	\centering
	\begin{subfigure}[c]{.4\linewidth}
		\centering
		\includegraphics[scale=1.6]{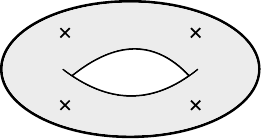}
		\caption{Closed strings}
	\end{subfigure}
	\hspace{1cm}
	\begin{subfigure}[c]{.4\linewidth}
		\centering
		\includegraphics[scale=1.6]{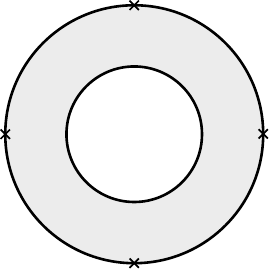}
		\caption{Open strings}
	\end{subfigure}
	\caption{%
		Graphs corresponding to $1$-loop $4$-point scattering after a conformal mapping.
	}%
	\label{intro:fig:ws-process-punctures}
\end{figure}

\subsubsection{Amplitudes}

\index{string amplitude}%
In order to compute an amplitude for the scattering of $n$ strings (\Cref{chap:bos:ws-int-amp,bos:chap:ws-int-complex}), one must sum over all the inequivalent worldsheets through a path integral weighted by the CFT action chosen to describe the theory.\footnotemark{}
\footnotetext{%
	For simplicity we focus on closed string amplitudes in this section.
}%
At fixed $n$, the sum runs over the genus $g$, such that each term is described by a Riemann surface $\Sigma_{g,n}$ of genus $g$ with $n$ punctures.

The interactions between strings follow from the graph topologies: since the latter are not encoded into the action, the dependence in the coupling constant must be added by hand.
For closed strings, there is a unique cubic vertex with coupling $g_s$.
A direct inspection shows that the correct factor is $g_s^{n - 2 + 2 g}$:
\begin{itemize}
	\item for $n = 3$ there is one factor $g_s$, and every additional external string leads to the addition of one vertex with factor $g_s$, since this process can be obtained from the $n-1$ process by splitting one of the external string in two by inserting a vertex;

	\item each loop comes with two vertices, so $g$-loops provide a factor $g_s^{2 g}$.
\end{itemize}

\begin{remark}[Status of $g_s$ as a parameter]
	\index{string!parameters}%
	\index{string coupling constant}%

	It was stated earlier that string theory has no dimensionless parameter, but $g_s$ looks to be one.
	In reality it is determined by the expectation value of the dilaton $g_s = \e^{\mean{\Phi}}$.
	Hence the coupling constant is not a parameter defining the theory but is rather determined by the dynamics of the theory.
\end{remark}

Finally, the external states must be specified: this amounts to prescribe boundary conditions for the path integral or to insert the corresponding wave functions.
Under the conformal mapping which brings the external legs to punctures located at $z_i$, the states are mapped to local operators $V_i(k_i, z_i)$ inserted at the points $z_i$.
The latter are built from the CFT fields and are called vertex operators: they are characterized by a momentum $k^\mu$ which comes from the Fourier transformation of the $X^\mu$ fields representing the non-compact dimensions.
These operators are inserted inside the path integral with integrals over the positions $z_i$ in order to describe all possible conformal mappings.

Ultimately, the amplitude (amputated Green function) is computed as
\begin{equation}
	A_n(k_1, \ldots, k_n) = \sum_{g \ge 0} g_s^{n - 2 + 2 g} A_{g,n}
\end{equation}
where
\begin{equation}
	A_{g,n} = \int \prod_{i=1}^n \dd^2 z_i \int \dd g_{ab} \dd \Psi \, \e^{- S_{\text{cft}}[g_{ab}, \Psi]} \prod_{i=1}^n V_i(k_i, z_i)
\end{equation}
is the $g$-loop $n$-point amplitude (for simplicity we omit the dependence on the states beyond the momentum).
$\Psi$ denotes collectively the CFT fields and $g_{ab}$ is the metric on the surface.

The integration over the metrics and over the puncture locations contain a huge redundancy due to the invariance under reparametrizations, which means that one integrates over many equivalent surfaces.
To avoid this, Faddeev--Popov ghosts must be introduced and the integral is restricted to only finitely many (real) parameters $t_\lambda$.
\index{moduli space}%
They form the \emph{moduli space} $\mc M_{g,n}$ of the Riemann surfaces $\Sigma_{g,n}$ whose dimension is
\begin{equation}
	\dim_{\R} \mc M_{g,n} = 6 g - 6 + 2 n.
\end{equation}

The computation of the amplitude $A_{g,n}$ can be summarized as:
\begin{equation}
	\label{intro:eq:amplitude-Agn}
	A_{g,n}
		= \int_{\mc M_{g,n}} \!\! \prod_{\lambda=1}^{6 g - 6 + 2 n} \! \dd t_\lambda \, F(t).
\end{equation}
The function $F(t)$ is a correlation function in the worldsheet CFT defined on the Riemann surface $\Sigma_{g,n}$
\begin{equation}
	F(t) = \Mean{\prod_{i = 1}^{n} V_i \times \text{ghosts} \times \text{super-ghosts}}_{\mathrlap{\Sigma_{g,n}}}.
\end{equation}
Note that the (super)ghost part is independent of the choice of the matter CFT.

\subsubsection{Divergences and Feynman graphs}

\begin{figure}[tp]
	\centering
	\includegraphics[scale=1.4]{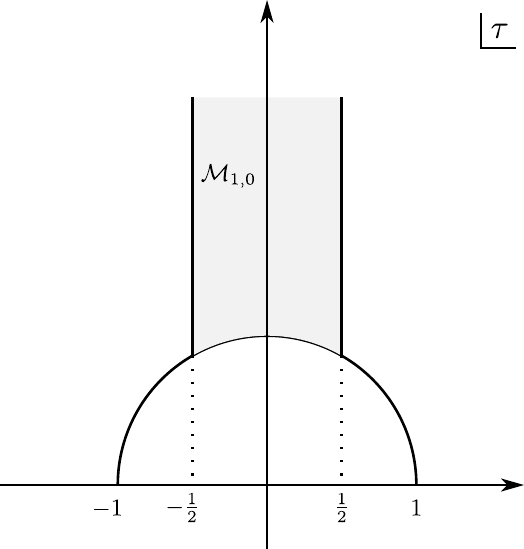}
	\caption{%
		Moduli space of the torus: $\Re \tau \in [- 1/2, 1/2]$, $\Im \tau > 0$ and $\abs{\tau} > 1$.
	}%
	\label{intro:fig:Mgn-torus}
\end{figure}

\index{propagator!Schwinger parametrization}%
Formally the moduli parameters are equivalent to Schwinger (proper-time) parameters $s_i$ in usual QFT: these are introduced in order to rewrite propagators as
\begin{equation}
	\frac{1}{k^2 + m^2} = \int_{0}^{\infty} \dd s \, \e^{- s (k^2 + m^2)},
\end{equation}
such that the integration over the momentum $k$ becomes a Gaussian times a polynomial.
\index{string amplitude!divergence}%
This form of the propagator is useful to display the three types of divergences which can be encountered:
\begin{enumerate}
	\item IR: regions $s_i \to \infty$ (for $k^2 + m^2 \le 0$).
	These divergences are artificial for $k^2 + m^2 < 0$ and means that the parametrization is not appropriate.
	Divergences for $k^2 + m^2 = 0$ are genuine and translates the fact that quantum effects shift the vacuum and the masses.
	Taking these effects into account necessitates a field theory framework in which renormalization can be used.

	\item UV: regions $s_i \to 0$ (after integrating over $k$).
	Such divergences are absent in string theories because these regions are excluded from the moduli space $\mc M_{g,n}$ (see \Cref{intro:fig:Mgn-torus} for the example of the torus).\footnotemark{}
	\footnotetext{%
		There is a caveat to this statement: UV divergences reappear in string field theory in Lorentzian signature due to the way the theory is formulated.
		The solution requires a generalization of the Wick rotation.

		Moreover, this does not hold for open strings whose moduli spaces contains those regions: in this case, the divergences are reinterpreted in terms of closed strings propagating.
	}%

	\item Spurious: regions with finite $s_i$ where the amplitude diverges.
	This happens typically only in the presence of super-ghosts and it translates a breakdown of the gauge fixing condition.\footnotemark{}
	\footnotetext{%
		Such spurious singularities are also found in supergravity.
	}%
	Since these spurious singularities of the amplitudes are not physical, one needs to ensure that they can be removed, which is indeed possible to achieve.
\end{enumerate}

Hence, only IR divergences present a real challenge to string theory.
Dealing with these divergences requires renormalizing the amplitudes, but this is not possible in the standard formulation of worldsheet string theory since the states are on-shell.\footnotemark{}
\footnotetext{%
	The on-shell condition is a consequence of the BRST and conformal invariance.
	While the first will be given up, the second will be maintained to facilitate the computations.
}%

\section{String field theory}
\label{sec:intro:string-field-theory}

\subsection{From the worldsheet to field theory}

\index{local coordinates}%
\index{off-shell string amplitude}%
The first step is to solve the IR divergences problem is to go \emph{off-shell} (\Cref{bos:chap:offshell,bos:chap:offshell-amp}).
This is made possible by introducing \emph{local coordinates} around the punctures of the Riemann surface (\Cref{bos:sec:offshell:geometry}).

\begin{figure}[tp]
	\centering
	\subcaptionbox{Separating.}{%
		\includegraphics[scale=1.8]{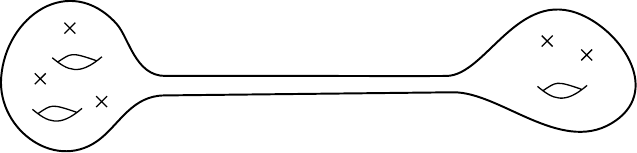}
	}%
	\\
	\bigskip
	\subcaptionbox{Non-separating.}{%
		\includegraphics[scale=1.8]{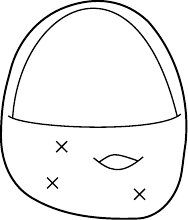}
	}%
	\caption{%
		Degeneration of Riemann surfaces.
	}%
	\label{intro:fig:Sgn-degen}
\end{figure}

\index{Riemann surface!degeneration}%
The IR divergences originate from Riemann surfaces close to \emph{degeneration}, that is, surfaces with long tubes.
The latter can be of \emph{separating} and \emph{non-separating} types, depending on whether the Riemann surface splits in two pieces if the tube is cut (\Cref{intro:fig:Sgn-degen}).
By exploring the form of the amplitudes in this limit (\Cref{bos:chap:feynman}), the expression naturally separates into several pieces, to be interpreted as two amplitudes (of lower $n$ and $g$) connected by a propagator.
The latter can be reinterpreted as a standard $(k^2 + m^2)^{-1}$ term, hence solving the divergence problem for $k^2 + m^2 < 0$.
Taking this decomposition seriously leads to identify each contribution with a Feynman graph.

Decomposing the amplitude recursively, the next step consists in finding the elementary graphs, i.e.\ the interaction vertices from which all other graphs (and amplitudes) can be built.
These graphs are the building blocks of the field theory (\Cref{bsft:chap:bv-sft}), with the kinetic term given by the inverse of the propagator.
Having Feynman diagrams and a field theory allows to use all the standard tools from QFT.

However, this field theory is gauge fixed because on-shell amplitudes are gauge invariant and include only physical states.
For this reason, one needs to find how to re-establish the gauge invariance.
\index{Batalin--Vilkovisky formalism}%
Due to the complicated structure of string theory, the full-fledged \emph{Batalin--Vilkovisky} (BV) formalism must be used (\Cref{bsft:chap:bv-sft}): it basically amounts to introduce ghosts before the gauge fixing.
The final stage is to obtain the 1PI effective action from which the physics is more easily extracted.
But, it is useful to study first the free theory (\Cref{bsft:chap:string-field,bsft:chap:free-brst}) to gain some insights.
\ifbook
The book ends with a discussion of the momentum-space representation and of background independence (\Cref{bsft:chap:background,sft:chap:momentum-sft}).
\fi

\index{string field theory!construction}%
The procedure we will follow is a kind of reverse-engineering: we know what is the final result and we want to study backwards how it is obtained:
\begin{gather*}
	\text{on-shell amplitude}
	\to \text{off-shell amplitude}
	\to \text{Feynman graphs}
	\\
	\to \text{gauge fixed field theory}
	\to \text{BV field theory}
\end{gather*}
In standard QFT, one follows the opposite process.

\begin{remark}
	There are some prescriptions (using for example analytic continuation, the optical theorem, some tricks…) to address the problems mentioned above, but there is no general and universally valid procedure.
	A field theory is much more satisfactory because it provides a unique and complete framework.
\end{remark}

We can now summarise the disadvantages of the worldsheet approach over the spacetime field one:
\begin{itemize}
	\item no natural description of (relativistic) multi-particle states;
	\item on-shell states:
	\begin{itemize}
		\item lack of renormalization,
		\item presence of infrared divergences,
		\item scattering amplitudes only for protected states;
	\end{itemize}
	\item interactions added by hand;
	\item hard to check consistency (unitarity, causality…);
	\item absence of non-perturbative processes.
\end{itemize}
Some of these problems can be addressed with various prescriptions, but it is desirable to dispose of a unified and systematic procedure, which is to be found in the field theory description.

\subsection{String field action}

\index{string field theory}%
A string field theory (SFT) for open and closed strings is based on two fields $\Phi[X(\sigma)]$ (open string field) and $\Psi[X(\sigma)]$ (closed string field) governed by some action $S[\Phi, \Psi]$.
This action is built from a diagonal kinetic term
\begin{equation}
	S_0
		= \frac{1}{2} \, K_\Psi(\Psi, \Psi) + \frac{1}{2} \, K_\Phi(\Phi, \Phi)
\end{equation}
and from an interaction polynomial in the fields
\begin{equation}
	S_{\text{int}}
		= \sum_{m, n} \mc V_{m,n}(\Phi^m, \Psi^n)
\end{equation}
where $\mc V_{m,n}$ is an appropriate product mapping $m$ closed and $n$ open string states to a number (the power is with respect to the tensor product).
In particular, it contains the coupling constant.
Contrary to the worldsheet approach where the cubic interaction looks sufficient, higher-order elementary interactions with $m, n \in \N$ are typically needed.
A second specific feature is that the products also admit a loop (or genus $g$) expansion: a fundamental $n$-point interaction is introduced at every loop order $g$.
These terms are interpreted as (finite) counter-terms needed to restore the gauge invariance of the measure.
These two facts come from the decomposition of the moduli spaces in pieces (\Cref{sec:intro:string:interactions}).

Writing an action for a field $\Psi[X(\sigma)]$ for which reparametrization invariance holds is highly complicated.
The most powerful method is to introduce a functional dependence in ghost fields $\Psi[X(\sigma), c(\sigma)]$ and to extend the BRST formalism to the string field, leading ultimately to the BV formalism.
While the latter formalism is the most complete and ensures that the theory is consistent at the quantum level, it is difficult to characterize the interactions explicitly.
Several constructions which exploit different properties of the theory have been proposed:
\begin{itemize}
	\item direct computation by reverse engineering of worldsheet amplitudes;

	\item specific parametrization of the Riemann surfaces (hyperbolic, minimal area);

	\item analogy with Chern--Simons and Wess--Zumino--Witten (WZW) theories;

	\item exploitation of the $L_\infty$ and $A_\infty$ algebra structures.
\end{itemize}
It can be shown that these constructions are all equivalent.
For the superstring, the simplest strategy is to dress the bosonic interactions with data from the super-ghost sector, which motivates the study of the bosonic SFT by itself.
The main difficulty in working with SFT is that only the first few interactions have been constructed explicitly.
Finally, the advantage of the first formulation is that it provides a general formulation of SFT at the quantum level, from which the general structure can be studied.

\subsection{Expression with spacetime fields}

\index{string field!momentum expansion}%
To obtain a more intuitive picture and to make contact with the spacetime fields, the field is expanded in terms of $1$-particle states in the momentum representation
\begin{equation}
	\ket{\Psi} = \sum_n \int \frac{\dd^D k}{(2\pi)^D} \, \psi_\alpha(k) \ket{k, \alpha},
\end{equation}
where $\alpha$ denotes collectively the discrete labels of the CFT eigenstates.
The coefficients $\psi_\alpha(k)$ of the CFT states $\ket{k, \alpha}$ are spacetime fields, the first ones being the same as those found in the first-quantized picture (\Cref{sec:intro:string:properties})
\begin{equation}
	\psi_\alpha = \{ T, G_{\mu\nu}, B_{\mu\nu}, \Phi, \ldots \}.
\end{equation}
Then, inserting this expansion in the action gives an expression like $S[T, G_{\mu\nu}, \ldots]$.
The exact expression of this action is out of reach and only the lowest terms are explicitly computable for a given CFT background.
Nonetheless, examining the string field action indicates what is the generic form of the action in terms of the spacetime fields.
One can then study the properties of such a general QFT: since it is more general than the SFT (expanded) action, any result derived for it will also be valid for SFT.
This approach is very fruitful for studying properties related to consistency of QFT (unitarity, soft theorems…) and this can provide helpful phenomenological models.

In conclusion, SFT can be seen as a regular QFT with the following properties:
\begin{itemize}
	\item infinite number of fields;

	\item non-local interaction (proportional to $\e^{- k^2 \#}$);

	\item the amplitudes agree with the worldsheet amplitudes (when the latter can be defined);

	\item genuine (IR) divergences agree but can be handled with the usual QFT tools.
\end{itemize}

\subsection{Applications}

\begin{draft}
In \Cref{sec:intro:wv-field}, we have given several more pragmatical reasons to motivate the construction of a string field theory.
\end{draft}
The first aspect is the possibility to use standard QFT techniques (such as renormalization) to study -- and to make sense of -- string amplitudes.
In this sense, SFT can be viewed as providing recipes for computing quantities in the worldsheet theory which are otherwise not defined.
This program has been pushed quite far in the last years.

Another reason to use SFT is gauge invariance: it is always easier to describe a system when its gauge invariance is manifest.
We have explained that string theory contains Yang--Mills and graviton fields with the corresponding (spacetime) gauge invariances (non-Abelian gauge symmetry and diffeomorphisms).
In fact, these symmetries are enhanced to an enormous gauge invariance when taking into account the higher-spin fields.
This invariance is hidden in the standard formulation and cannot be exploited fully.
On the other hand, the full gauge symmetry is manifest in string field theory.

Finally, the worldvolume description of $p$-brane is difficult because there is no analogue of the Polyakov action.
If one could find a first-principle description of SFT which does not rely on CFT and first-quantization, then one may hope to generalize it to build a brane field theory.

\noindent
We can summarize the general motivations for studying SFT:
\begin{itemize}
	\item field theory (second-quantization);

	\item more rigorous and constructive formulation;

	\item make gauge invariance explicit ($L_\infty$ algebras et al.);

	\item use standard QFT techniques (renormalization, analyticity…)
	\\
	→ remove IR divergences, prove consistency (Cutkosky rules, unitarity, soft theorems, background independence…);

	\item worldvolume theory ill-defined for $(p > 1)$-branes.
\end{itemize}
Beyond these general ideas, SFT has been developed in order to address different questions:
\begin{itemize}
	\item worldsheet scattering amplitudes;
	\item effective actions;
	\item map of the consistent backgrounds (classical solutions, marginal deformations, RR fluxes…);
	\item collective, non-perturbative, thermal, dynamical effects;
	\item symmetry breaking effects;
	\item dynamics of compactification;
	\item proof of dualities;
	\item proof of the AdS/CFT correspondence.
\end{itemize}
The last series of points is still out of reach within the current formulation of SFT.
However, the last two decades have seen many important develoments developments:
\begin{itemize}
	\item construction of the open, closed and open-closed superstring field theories:
	\begin{itemize}
		\item 1PI and BV actions and general properties~\cite{Sen:2015:GaugeInvariant1PI-NS, Sen:2015:GaugeInvariant1PI-R, Sen:2015:SupersymmetryRestorationSuperstring, Sen:2016:BVMasterAction, Sen:2016:RealitySuperstringField, Sen:2015:FillingGapsPCOs, Erler:2017:VerticalIntegrationLarge, Sen:2016:WilsonianEffectiveAction, Sen:2017:EquivalenceTwoContour, Sen:2018:BackgroundIndependenceClosed, Moosavian:2020:SuperstringFieldTheory, Moosavian:2019:ExistenceHeteroticStringTypeIISuperstring},

		\item dressing of bosonic products using the WZW construction and homotopy algebra~\cite{Berkovits:1995:SuperPoincareInvariantSuperstring, Okawa:2004:HeteroticStringField, Berkovits:2004:WZWlikeActionHeterotic, Iimori:2014:BerkovitsFormulationWitten, Kroyter:2012:OpenSuperstringField, Erler:2014:ResolvingWittensSuperstring, Erler:2014:NSNSSectorClosed, Erler:2015:RamondEquationsMotion, Erler:2015:RelatingBerkovitsAinftySmall, Erler:2015:RelatingBerkovitsAinftyLarge, Erler:2015:AinftyStructureBerkovits, Erler:2016:CompleteActionOpen, Erler:2017:SuperstringFieldTheory, Erler:2017:SupersymmetryOpenSuperstring, Kunitomo:2014:FirstOrderEquationsMotion, Kunitomo:2014:RamondSectorHeterotic, Kunitomo:2015:SymmetriesFeynmanRules, Kunitomo:2016:CompleteActionOpen, Kunitomo:2017:SpacetimeSupersymmetryWZWlike, Kunitomo:2019:HeteroticStringField, Kunitomo:2020:TypeIISuperstring, Konopka:2016:OpenSuperstringField, Goto:2016:ConstructionActionHeterotic},

		\item light-cone super-SFT~\cite{Ishibashi:2013:MultiloopAmplitudesLightcone, Ishibashi:2016:WorldsheetTheoryLightcone, Ishibashi:2018:MultiloopAmplitudesLightcone-1, Ishibashi:2018:MultiloopAmplitudesLightcone-2},

		\item supermoduli space~\cite{Ohmori:2018:OpenSuperstringField, Takezaki:2019:OpenSuperstringField};
	\end{itemize}

	\item hyperbolic and minimal area constructions~\cite{Moosavian:2019:HyperbolicGeometryClosed-1, Moosavian:2019:HyperbolicGeometryClosed-2, Moosavian:2017:HyperbolicGeometrySuperstring, Headrick:2020:ConvexProgramsMinimalarea, Headrick:2020:MinimalareaMetricsSwiss, Pius:2018:QuantumClosedSuperstring, Costello:2019:HyperbolicStringVertices};

	\item open string analytic solutions~\cite{Erler:2014:StringFieldTheory, Erler:2020:StringFieldTheory};

	\item level-truncation solutions~\cite{Kudrna:2013:GaugeinvariantObservablesMarginal, Kudrna:2016:BCFTModuliSpace, Kudrna:2018:UniversalSolutionsOpen};

	\item field theory properties~\cite{Pius:2016:CutkoskyRulesSuperstring, Sen:2016:UnitaritySuperstringField, Sen:2017:SoftTheoremsSuperstring, Sen:2017:SubleadingSoftGraviton, Laddha:2017:SubsubleadingSoftGraviton, Chakrabarti:2017:SubleadingSoftTheorem, deLacroix:2019:AnalyticityCrossingSymmetry};

	\item spacetime effective actions~\cite{Maccaferri:2018:LocalizationEffectiveActions, Maccaferri:2019:LocalizationEffectiveActions, Vosmera:2019:GeneralizedADHMEquations, Erbin:2020:LocalizationEffectiveActions};

	\item defining worldsheet scattering amplitudes~\cite{Pius:2014:MassRenormalizationStringSpecial, Pius:2014:MassRenormalizationStringGeneral, Pius:2014:StringPerturbationTheory, Sen:2015:OffshellAmplitudesSuperstring, Sen:2015:SupersymmetryRestorationSuperstring, Sen:2016:OneLoopMass, Sen:2020:FixingAmbiguityTwo, Sen:2019:StringFieldTheoryUV, Sen:2020:DinstantonPerturbationTheory};

	\item marginal and RR deformations~\cite{Cho:2020:StringsRamondRamondBackgrounds, Sen:2020:FixingAmbiguityTwo, Vosmera:2019:GeneralizedADHMEquations}.
\end{itemize}
Recent reviews are~\cite{deLacroix:2017:ClosedSuperstringField, Erler:2022:FourLecturesAnalytic, Erler:2020:FourLecturesClosed}.

\begin{draft}
\begin{remark}[Double and exceptional field theories]
	Double (exceptional) field theories are other constructions aiming at encoding more symmetries of the string.
	The idea is to write an action for the lowest modes with a manifest $T$-duality ($U$-duality).
\end{remark}
\end{draft}

\refchapter

For references about different aspects in this chapter:
\begin{itemize}
	\item Differences between the worldvolume and spacetime formalisms -- and of the associated first- and second-quantization -- for the particle and string~\cites[chap.~1]{Kaku:1999:IntroductionSuperstringsMTheory}[chap.~11]{Zwiebach:2009:FirstCourseString}.

	\item General properties of relativistic strings~\cite{Goddard:1973:QuantumDynamicsMassless, Zwiebach:2009:FirstCourseString}.

	\item Divergences in string theory~\cites{Sen:2015:UltravioletInfraredDivergences, deLacroix:2017:ClosedSuperstringField}[sec.~7.2]{Witten:2012:SuperstringPerturbationTheory}.

	\item Motivations for building a string field theory~\cite[sec.~4]{Polchinski:1994:WhatStringTheory}.
\end{itemize}

\part{Worldsheet theory}
\label{part:bosonic-ws}
\label{part:cft}

\chapter{Worldsheet path integral: vacuum amplitudes}
\label{chap:bos:ws-int-vac}

\introchapter

\index{Polyakov path integral}%
In this chapter, we develop the path integral quantization for a generic closed string theory in worldsheet Euclidean signature.
We focus on the vacuum amplitudes, leaving scattering amplitudes for the next chapter.
This allows to focus on the definition and gauge fixing of the path integral measure.

The exposition differs from most traditional textbooks in three ways: 1) we consider a general matter CFT, 2) we consider the most general treatment (for any genus) and 3) we don't use complex coordinates but always a covariant parametrization.

The derivation is technical and the reader is encouraged to not stop at this chapter in case of difficulties and to proceed forward: most concepts will be reintroduced from a different point of view later in other chapters of the \revname{}.

\section{Worldsheet action and symmetries}
\label{bos:sec:ws-int:action}

\index{worldsheet!Riemann surface}%
\index{Riemann surface!genus}%
The string worldsheet is a Riemann surface $\mc W = \Sigma_g$ of genus $g$: the genus counts the number of holes or handles.
Coordinates on the worldsheet are denoted by $\sigma^a = (\tau, \sigma)$. When there is no risk of confusion, $\sigma$ denotes collectively both coordinates.
Since closed strings are considered, the Riemann surface has locally the topology of a cylinder, with the spatial section being circles $S^1$ with radius taken to be $1$, such that
\begin{equation}
	\sigma \in [0, 2\pi),
	\qquad
	\sigma \sim \sigma + 2\pi.
\end{equation}
The string is embedded in the $D$-dimensional spacetime $\mc M$ with metric $G_{\mu\nu}$ through maps $X^\mu(\sigma^a): \mc W \to \mc M$ with $\mu = 0, \ldots, D-1$.

\index{worldsheet!action!Nambu--Goto (-)}%
The Nambu--Goto action is the starting point of the worldsheet description:
\begin{equation}
	\label{bos:eq:action-nambu-goto}
	S_{\text{NG}}[X^\mu]
		= \frac{1}{2\pi \alpha'}
			\int \dd^2 \sigma \,
			\sqrt{\det G_{\mu\nu}(X) \frac{\pd X^\mu}{\pd \sigma^a} \frac{\pd X^\nu}{\pd \sigma^b}},
\end{equation}
where $\alpha'$ is the Regge slope (related to the string tension and string length).
However, quantizing this action is difficult because it is highly non-linear.
To solve this problem, a Lagrange multiplier is introduced to remove the squareroot.
\index{worldsheet metric}%
This auxiliary field corresponds to an intrinsic worldsheet metric $g_{ab}(\sigma)$.
The worldsheet dynamics is described by the Polyakov action:
\index{worldsheet!action!Polyakov (-)}%
\begin{equation}
	\label{bos:eq:action-polyakov}
	S_{\text{P}}[g, X^\mu]
		= \frac{1}{4\pi \alpha'}
			\int \dd^2 \sigma \sqrt{g} \,
			g^{ab} G_{\mu\nu}(X) \frac{\pd X^\mu}{\pd \sigma^a} \frac{\pd X^\nu}{\pd \sigma^b},
\end{equation}
which is classically equivalent to the Nambu--Goto action \eqref{bos:eq:action-nambu-goto}.
In this form, it is clear that the scalar fields $X^\mu(\sigma)$ ($\mu = 0, \ldots D - 1$) characterize the string theory under consideration in two ways.
First, by specifying some properties of the spacetime in which the string propagates (for example, the number of dimensions is determined by the number of fields $X^\mu$), second, by describing the internal degrees of freedom (vibration modes).\footnotemark{}
\footnotetext{%
	Obviously, the vibrational modes are also constrained by the spacetime geometry.
}%

But, nothing prevents to consider a more general matter content in order to describe a different spacetime or different degrees of freedom.
\index{worldsheet!action!sigma model (-)}%
In Polyakov's formalism, the worldsheet geometry is endowed with a metric $g_{ab}(\sigma)$ together with a set of matter fields living on it.
The scalar fields $X^\mu$ can be described by a general sigma model which encodes the embedding of the string in the $D$ non-compact spacetime dimensions, and other fields can be added, for example to describe compactified dimensions or (spacetime) spin.
Different sets of fields (and actions) correspond to different string theories.
However, to describe precisely the different possibilities, we first have to understand the constraints on the worldsheet theories and to introduce conformal field theories (\Cref{part:cft}).
In this chapter (and in most of the book), the precise matter content is not important and we will denote the fields collectively as $\Psi(\sigma)$.

\medskip

\index{Euler characteristics}%
Before discussing the symmetries, let's introduce a topological invariant which will be needed throughout the text: the \emph{Euler characteristics}.
It is computed by integrating the Riemann curvature $R$ of the metric $g_{ab}$ over the surface $\Sigma_g$:
\begin{equation}
	\label{bos:eq:chi-g}
	\chi_g
		:= \chi(\Sigma_g)
		:= 2 - 2 g
		= \frac{1}{4\pi} \int_{\Sigma_g} \dd^2 \sigma \sqrt{g} \, R,
\end{equation}
where $g$ is the genus of the surface.
Oriented Riemann surfaces without boundaries are completely classified (topologically or as complex manifolds) by their Euler characteristics $\chi_g$, or equivalently by their genus $g$.

\medskip

In order to describe a proper string theory, the worldsheet metric $g_{ab}(\sigma)$ should not be dynamical.
This means that the worldsheet has no intrinsic dynamics and that no supplementary degrees of freedom are introduced when parametrizing the worldsheet with a metric.
\index{worldsheet!symmetry}%
A solution to remove these degrees of freedom is to introduce gauge symmetries with as many gauge parameters as there are of degrees of freedom.
\index{worldsheet!symmetry!diffeomorphisms}%
The simplest symmetry is invariance under diffeomorphisms: indeed, the worldsheet theory is effectively a QFT coupled to gravity and it makes sense to require this invariance.
Physically, this corresponds to the fact that the worldsheet spatial coordinate $\sigma$ used along the string and worldsheet time are arbitrary.
However, diffeomorphisms alone are not sufficient to completely fix the metric.
\index{worldsheet!symmetry!Weyl}%
Another natural candidate is \emph{Weyl invariance} (local rescalings of the metric).

\index{diffeomorphism}%
A diffeomorphism $f \in \group{Diff}(\Sigma_g)$ acts on the fields as
\begin{equation}
	\label{bos:eq:sym-diffeo}
	\sigma'^a
		= f^a(\sigma^b),
	\qquad
	g'(\sigma')
		= f^* g(\sigma),
	\qquad
	\Psi'(\sigma')
		= f^* \Psi(\sigma),
\end{equation}
where the star denotes the pullback by $f$: this corresponds simply to the standard coordinate transformation where each tensor index of the field receives a factor $\pd \sigma^a / \pd \sigma'^b$.
In particular, the metric and scalar fields transform explicitly as
\begin{equation}
	g'_{ab}(\sigma') = \frac{\pd \sigma^c}{\pd \sigma'^a} \frac{\pd \sigma^d}{\pd \sigma'^b} \, g_{cd}(\sigma),
	\qquad
	X'^\mu(\sigma') = X^\mu(\sigma).
\end{equation}
The index $\mu$ is inert since it is a target spacetime index: from the worldsheet point of view, it just labels a collection of worldsheet scalar fields.
\index{diffeomorphism!infinitesimal (-)}%
Infinitesimal variations are generated by vector fields on $\Sigma_g$:
\begin{equation}
	\label{bos:eq:sym-diffeo-inf}
	\delta_\xi \sigma^a = \xi^a,
	\qquad
	\delta_\xi \Psi = \dlie_\xi \Psi,
	\qquad
	\delta_\xi g_{ab} = \dlie_\xi g_{ab},
\end{equation}
where $\dlie_\xi$ is the Lie derivative\footnotemark{} with respect to the vector field $\xi \in \alg{diff}(\Sigma_g) \simeq T\Sigma_g$.
The Lie derivative of the metric is
\footnotetext{%
	For our purpose here, it is sufficient to accept the definition of the Lie derivative as corresponding to the infinitesimal variation.
}%
\begin{equation}
	\dlie_\xi g_{ab}
		= \xi^c \pd_c g_{ab} + g_{ac} \pd_b \xi^c + g_{bc} \pd_a \xi^c
		= \grad_a \xi_b + \grad_b \xi_a.
\end{equation}
The Lie algebra generates only transformations in the connected component $\group{Diff}_0(\Sigma_g)$ of the diffeomorphism group which contains the identity.

\index{diffeomorphism!large (-)}%
Transformations not contained in $\group{Diff}_0(\Sigma_g)$ are called large diffeomorphisms: this includes reflections, for example.
\index{modular group}%
The quotient of the two groups is called the modular group $\Gamma_g$ (also mapping class group or MCG):
\begin{equation}
	\label{bos:eq:group-mcg}
	\Gamma_g
		:= \pi_0\big(\group{Diff}(\Sigma_g)\big)
		= \frac{\group{Diff}(\Sigma_g)}{\group{Diff}_0(\Sigma_g)}.
\end{equation}
It depends only on the genus $g$ of the Riemann surface, but not on the metric.
It is an infinite discrete group for genus $g \ge 1$ surfaces; in particular, $\Gamma_1 = \group{SL}(2, \Z)$.

\medskip

\index{Weyl!symmetry}%
A Weyl transformation $\e^{2\omega} \in \group{Weyl}(\Sigma_g)$ corresponds to a local rescaling of the metric and leaves the other fields unaffected\footnotemark{}
\footnotetext{%
	For simplicity, we consider only fields which do not transform under Weyl transformations, which excludes fermions.
}%
\begin{equation}
	\label{bos:eq:sym-weyl}
	g'_{ab}(\sigma) = \e^{2\omega(\sigma)} g_{ab}(\sigma),
	\qquad
	\Psi'(\sigma) = \Psi(\sigma).
\end{equation}
The exponential parametrization is generally more useful, but one should remember that it is $\e^{2\omega}$ and not $\omega$ which is an element of the group.
The infinitesimal variation reads
\begin{equation}
	\label{bos:eq:sym-weyl-inf}
	\delta_\omega g_{ab} = 2 \omega \, g_{ab},
	\qquad
	\delta_\omega \Psi = 0
\end{equation}
where $\omega \in \alg{weyl}(\Sigma) \simeq \mc F(\Sigma_g)$ is a function on the manifold.
Two metrics related in this way are said to be conformally equivalent.
\index{conformal structure}%
The \emph{conformal structure} of the Riemann surface is defined by
\begin{equation}
	\label{bos:eq:conf-structure}
	\group{Conf}(\Sigma_g)
		:= \frac{\group{Met}(\Sigma_g)}{\group{Weyl}(\Sigma_g)},
\end{equation}
where $\group{Met}(\Sigma_g)$ denotes the space of all metrics on $\Sigma_g$.
Each element is a class of conformally equivalent metrics.

Diffeomorphisms have two parameters $\xi^a$ (vector field) and Weyl invariance has one, $\omega$ (function).
Hence, this is sufficient to locally fix the three components of the metric (symmetric matrix) and the total gauge group of the theory is the semi-direct product
\begin{equation}
	\label{bos:eq:gauge-group-worldsheet}
	G
		:= \group{Diff}(\Sigma_g) \ltimes \group{Weyl}(\Sigma_g).
\end{equation}
Similarly, the component connected of the identity is written as
\begin{equation}
	G_0
		:= \group{Diff}_0(\Sigma_g) \ltimes \group{Weyl}(\Sigma_g).
\end{equation}

The semi-direct product arises because the Weyl parameter is not inert under diffeomorphisms.
Indeed, the combination of two transformations is
\begin{equation}
	\label{bos:eq:sym-weyl-diffeo}
	g' = f^* \big(\e^{2 \omega} g\big)
		= \e^{2 f^* \omega} f^* g,
\end{equation}
such that the diffeomorphism acts also on the conformal factor.

\index{metric!gauge fixing}%
The combination of transformations \eqref{bos:eq:sym-weyl-diffeo} can be chosen to fix the metric in a convenient gauge.
\index{metric!gauge!conformal (-)}%
For example, the \emph{conformal gauge} reads
\begin{equation}
	\label{bos:eq:gauge-conformal}
	g_{ab}(\sigma) = \e^{2 \phi(\sigma)} \hat g_{ab}(\sigma),
\end{equation}
\index{background metric}%
\index{Liouville!field}%
where $\hat g_{ab}$ is some (fixed) \emph{background metric} and $\phi(\sigma)$ is the conformal factor, also called the \emph{Liouville field}.
Fixing only diffeomorphisms amount to keep $\phi$ arbitrary: the latter can then be fixed with a Weyl transformation.
\index{metric!gauge!conformally flat (-)}%
For instance, one can adopt the conformally flat gauge
\begin{equation}
	\label{bos:eq:gauge-conf-flat}
	\hat g_{ab} = \delta_{ab},
	\qquad
	\text{$\phi$ arbitrary}
\end{equation}
\index{metric!gauge!flat (-)}%
with a diffeomorphism, and then reach the flat gauge
\begin{equation}
	\label{bos:eq:gauge-flat}
	\hat g_{ab} = \delta_{ab},
	\qquad
	\phi = 0
\end{equation}
with a Weyl transformation.
\index{metric!gauge fixing!uniformization gauge}%
Another common choice is the uniformization gauge where $\hat g$ is taken to be the metric of constant curvature on the sphere ($g = 0$), on the plane ($g = 1$) or on the hyperbolic space ($g > 1$).
All these gauges are covariant (both in spacetime and worldsheet).

\begin{remark}[Active and passive transformations]
	Usually, symmetries are described by \emph{active} transformations, which means that the field is seen to be changed by the transformation.
	On the other hand, gauge fixing is seen as a \emph{passive} transformation, where the field is expressed in terms of other fields (i.e.\ a different parametrization).
	These are mathematically equivalent since both cases correspond to inverse elements, and one can choose the most convenient representation.
	We will use indifferently the same name for the parameters to avoid introducing minus signs and inverse.
\end{remark}

\begin{remark}[Topology and gauge choices]
	While it is always possible to adopt locally the flat gauge \eqref{bos:eq:gauge-flat}, it may not be possible to extend it globally.
	The can be seen intuitively from the fact that the sign of the curvature is given by the one of $1 - g$, but the curvature of the flat metric is zero: curvature must then be localized somewhere and this prevents from using a single coordinate patch.
\end{remark}

The final step is to write an action $S_m[g, \Psi]$ for the matter fields.
According to the previous discussion, it must have the following properties:
\begin{itemize}
	\item local in the fields;
	\item renormalizable;
	\item non-linear sigma models for the scalar fields;
	\item periodicity conditions;
	\item invariant under diffeomorphisms \eqref{bos:eq:sym-diffeo};
	\item invariant under Weyl transformations \eqref{bos:eq:sym-weyl}.
\end{itemize}
The latter two conditions are summarized by
\begin{equation}
	\label{bos:eq:sym-action}
	S_m[f^* g, f^* \Psi] = S_m[g, \Psi],
	\qquad
	S_m[\e^{2\omega} g, \Psi] = S_m[g, \Psi].
\end{equation}
The invariance under diffeomorphisms is straightforward to enforce by using only covariant objects.
Since the scalar fields represent embedding of the string in spacetime, the non-linear sigma model condition means that spacetime is identified with the target space of the sigma model, of which $D$ dimensions are non-compact, and the spacetime metric appears in the matter action as in \eqref{bos:eq:action-polyakov}.
The isometries of the target manifold metric become global symmetries of $S_m$: while they are not needed in this chapter, they will have their importances in other chapters.
Finally, to make the action consistent with the topology of the worldsheet, the fields must satisfy appropriate boundary conditions.
For example, the scalar fields $X^\mu$ must be periodic for the closed string:
\begin{equation}
	\label{bos:eq:field-periodic}
	X^\mu(\tau, \sigma) \sim X^\mu(\tau, \sigma + 2\pi).
\end{equation}

\begin{remark}[$2d$ gravity]
	The setup in two-dimensional gravity is exactly similar, except that the system is, in general, not invariant under Weyl transformations.
	As a consequence, one component of the metric (usually taken to be the Liouville mode) remains unconstrained: in the conformal gauge, \eqref{bos:eq:gauge-conformal} only $\hat g$ is fixed.
\end{remark}

The symmetries \eqref{bos:eq:sym-action} of the action have an important consequence: they imply that the matter action is conformally invariant on flat space $g_{ab} = \delta_{ab}$.
\index{central charge}%
A two-dimensional conformal field theory (CFT) is characterized by a \emph{central charge} $c_m$: roughly, it is a measure of the quantum degrees of freedom.
The central charge is additive for decoupled sectors.
In particular, the scalar fields $X^\mu$ contribute as $D$, and it is useful to define the perpendicular CFT with central charge $c_\perp$ as the matter which does not describe the non-compact dimensions:
\begin{equation}
	\label{bos:eq:c-D-perp}
	c_m = D + c_\perp.
\end{equation}
This will be discussed in length in \Cref{part:cft}.
For this chapter and most of the book, it is sufficient to know that the matter is a CFT of central charge $c_m$ and includes $D$ scalar fields $X^\mu$:
\begin{equation}
	\text{matter CFT parameters: $D$, $c_m$.}
\end{equation}

\begin{draft}
\begin{remark}
	When discussing a curved background, it is customary to also speak about “conformal invariance” as a synonym for Weyl invariance and to say that a field theory is a CFT.
\end{remark}
\end{draft}

\index{worldsheet!energy--momentum tensor}%
The energy--momentum is defined by
\begin{equation}
	T_{m,ab}
		:= - \frac{4\pi}{\sqrt{g}} \, \frac{\delta S_m}{\delta g^{ab}}.
\end{equation}
The variation of the action under the transformations \eqref{bos:eq:sym-diffeo-inf} vanishes on-shell if the energy--momentum tensor is conserved
\begin{equation}
	\grad^a T_{m,ab} = 0
	\qquad
	\text{(on-shell)}.
\end{equation}
\index{worldsheet!energy--momentum tensor!trace}%
On the other hand, the variation under \eqref{bos:eq:sym-weyl-inf} vanishes off-shell (i.e.\ without using the equations of motion) if the energy--momentum tensor is traceless:
\begin{equation}
	g^{ab} T_{m,ab} = 0
	\qquad
	\text{(off-shell)}.
\end{equation}
The conserved charges associated to the energy--momentum tensor generate worldsheet translations
\begin{equation}
	\label{bos:eq:worldsheet-momentum}
	P^a
		:= \int \dd \sigma \, T_m^{0a}.
\end{equation}
The first component is identified with the worldsheet Hamiltonian $P^0 = H$ and generates time translations, the second component generates spatial translations.

\begin{remark}[Tracelessness of the energy--momentum tensor]
	\index{worldsheet!energy--momentum tensor!trace}%
	In fact, the trace can also be proportional to the curvature
	\begin{equation}
		g^{ab} T_{m,ab} \propto R.
	\end{equation}
	Then, the equations of motion are invariant since the integral of $R$ is topological.
	The theory is invariant even if the action is not.
	Importantly, this happens for fields at the quantum level (Weyl anomaly), for the Weyl ghost field (\Cref{bos:sec:ws-int:ghosts}) and for the Liouville theory (two-dimensional gravity coupled to conformal matter).
\end{remark}

\section{Path integral}
\label{bos:sec:ws-int:integral}

\index{string amplitude!gv@$g$-loop vacuum (-)}%
The quantization of the system is achieved by considering the path integral, which yields the genus-$g$ vacuum amplitude (or partition function):
\begin{equation}
	\label{bos:eq:path-int-dg}
	Z_g
		:= \int \frac{\dd_g g_{ab}}{\Omega_{\text{gauge}}[g]} \, Z_m[g],
	\qquad
	Z_m[g]
		:= \int \dd_g \Psi \, \e^{- S_m[g, \Psi]}
\end{equation}
at fixed genus $g$ (not to be confused with the metric).
The integration over $g_{ab}$ is performed over all metrics of the genus-$g$ Riemann surface $\Sigma_g$: $g_{ab} \in \mathrm{Met}(\Sigma_g)$.
The factor $\Omega_{\text{gauge}}[g]$ is a normalization inserted in order to make the integral finite: it depends on the metric (but only through the moduli parameters, as we will show later)~\cite[p.~931]{DHoker:1988:GeometryStringPerturbation}, which explains why it is included after the integral sign.
Its value will be determined in the next section by requiring the cancellation of the infinities due to the integration over the gauge parameters.
This partition function corresponds to the $g$-loop vacuum amplitude: interactions and their associated scattering amplitudes are discussed in \Cref{bos:sec:ws-int:amp}.

	\index{path integral!measure}%
In order to perform the gauge fixing and to manipulate the path integral \eqref{bos:eq:path-int-dg}, it is necessary to define the integration measure over the fields.
Because the space is infinite-dimensional, this is a difficult task.
One possibility is to define the measure implicitly through Gaussian integration over the field tangent space (see also \Cref{sec:form:path-integrals}).
A Gaussian integral involves a quadratic form, that is, an inner-product (or equivalently a metric) on the field space.
The explanation is that a metric also defines a volume form, and thus a measure.
To reduce the freedom in the definition of the inner-product, it is useful to introduce three natural assumptions:
\begin{enumerate}
	\item \emph{ultralocality:} the measure is invariant under reparametrizations and defined point-wise, which implies that it can depend on the fields but not on their derivatives;
	\index{path integral!measure!ultralocality}%

	\item \emph{invariant} measure: the measure for the matter transforms trivially under any symmetry of the matter theory by contracting indices with appropriate tensors;

	\item \emph{free-field} measure: for fields other than the worldsheet metric and matter (like ghosts, Killing vectors, etc.), the measure is the one of a free field.
	\index{path integral!measure!free-field}%

\end{enumerate}
This means that the inner-product is obtained by contracting the worldsheet indices of the fields with a tensor built only from the worldsheet metric, by contracting other indices (like spacetime) with some invariant tensor (like the spacetime metric), and finally by integrating over the worldsheet.

We need to distinguish the matter fields from those appearing in the gauge fixing procedure.
The matter fields live in the representation of some group under which the inner-product is invariant: this means that it is not possible to define each field measure independently if the exponential of inner-products does not factorize.
As an example, on a curved background: $\dd X \neq \prod_\mu \dd X^\mu$.
However, we will not need to write explicitly the partition function for performing the gauge fixing: it is sufficient to know that the matter is a CFT.
In the gauge fixing procedure, different types of fields (including the metric) appear which don't carry indices (beyond the worldsheet indices).
Below, we focus on defining a measure for each of those single fields (and use free-field measures according to the third condition).

\index{field space!norm}%
\index{field space!inner-product}%
Considering the finite elements $\delta \Phi_1$ and $\delta\Phi_2$ of tangent space at the point $\Phi$ of the state of fields, the inner-product $\psp{\cdot}{\cdot}_g$ and its associated norm $\abs{\cdot}_g$ read
\begin{equation}
	(\delta \Phi_1, \delta \Phi_2)_g
		:= \int \dd^2 \sigma \sqrt{g} \, \gamma_g(\delta\Phi_1, \delta\Phi_2),
	\qquad
	\abs{\delta \Phi}_g^2
		:= (\delta \Phi, \delta \Phi)_g,
\end{equation}
where $\gamma_g$ is a metric on the $\delta \Phi$ space.
It is taken to be flat for all fields except the metric itself, that is, independent of $\Phi$.
The dependence in the metric ensures that the inner-product is diffeomorphism invariant, which in turns will lead to a metric-dependent but diffeomorphism invariant measure.
The functional measure is then normalized by a Gaussian integral:
\begin{equation}
	\label{bos:eq:measure-normalisation}
	\int \dd_g \delta \Phi\, \e^{- \frac{1}{2} (\delta \Phi, \delta \Phi)_g}
		= \frac{1}{\sqrt{\det \gamma_g}}.
\end{equation}
This, in turn, induces a measure on the field space itself:
\begin{equation}
	\int \dd \Phi \sqrt{\det \gamma_g}
\end{equation}
The determinant can be absorbed in the measure, such that
\begin{equation}
	\label{bos:eq:measure-normalisation-one}
	\int \dd_g \delta \Phi\, \e^{- \frac{1}{2} (\delta \Phi, \delta \Phi)_g}
		= 1.
\end{equation}
In fact, this normalization and the definition of the inner-product is ambiguous, but the ultralocality condition allows to fix uniquely the final result (\Cref{bos:rem:measure-ambiguities}).
Moreover, such a free-field measure is invariant under field translations
\begin{equation}
	\Phi(\sigma) \longrightarrow \Phi'(\sigma)
		= \Phi(\sigma) + \varepsilon(\sigma).
\end{equation}

The most natural inner-products for single scalar, vector and symmetric tensor fields are
\begin{subequations}
\label{bos:eq:path-int-norm}
\begin{align}
	\label{bos:eq:path-int-norm-scalar}
	\psp{\delta f}{\delta f}_g
		&
		:= \int \dd^2 \sigma \sqrt{g} \, \delta f^2
	\\
	\label{bos:eq:path-int-norm-vector}
	\psp{\delta V^a}{\delta V^a}_g
		&
		:= \int \dd^2 \sigma \sqrt{g} \, g_{ab} \delta V^a \delta V^b,
	\\
	\label{bos:eq:path-int-norm-tensor}
	\psp{\delta T_{ab}}{\delta T_{ab}}_g
		&
		:= \int \dd^2 \sigma \sqrt{g} \, G^{abcd} \delta T_{ab} \delta T_{cd},
\end{align}
\end{subequations}
\index{field space!DeWitt metric}%
where the (DeWitt) metric for the symmetric tensor is
\begin{equation}
	\label{bos:eq:dewitt-metric}
	G^{abcd}
		:= G_\perp^{abcd}
			+ u \, g^{ab} g^{cd},
	\qquad
	G_\perp^{abcd}
		:= g^{ac} g^{bd} + g^{ad} g^{bc} - g^{ab} g^{cd},
\end{equation}
with $u$ a constant.
The first term $G_\perp$ is the projector on the traceless component of the tensor.
Indeed, consider a traceless tensor $g^{ab} T_{ab} = 0$ and a pure trace tensor $\Lambda g_{ab}$, then we have:
\begin{equation}
	G^{abcd} T_{cd}
		= G_\perp^{abcd} T_{cd}
		= 2 T_{ab},
	\qquad
	G^{abcd} (\Lambda g_{cd})
		= 2 u \, (\Lambda g_{ab}).
\end{equation}

\index{Weyl!anomaly}%
While all measures are invariant under diffeomorphisms, only the vector measure is invariant under Weyl transformations.
This implies the existence of a quantum anomaly (the \emph{Weyl} or \emph{conformal anomaly}): the classical symmetry is broken by quantum effects because the path integral measure cannot respect all the classical symmetries.
Hence, one can expect difficulties for imposing it at the quantum level and ensuring that the Liouville mode in \eqref{bos:eq:gauge-conformal} remains without dynamics.

The metric variation (symmetric tensor) is decomposed in its trace and traceless parts
\begin{equation}
	\label{bos:eq:metric-decomposition}
	\delta g_{ab}
		= g_{ab} \, \delta \Lambda + \var g_{ab}^\perp,
	\qquad
	\delta \Lambda
		= \frac{1}{2} \, g^{ab} \delta g_{ab},
	\qquad
	g^{ab} \delta g_{ab}^\perp
		= 0.
\end{equation}
In this decomposition, both terms are decoupled in the inner-product
\begin{equation}
	\label{bos:eq:path-int-norm-metric-decomposition}
	\abs{\delta g_{ab}}_g^2 = 4 u \abs{\delta \Lambda}_g^2 + \abs{\delta g_{\mu\nu}^\perp}_g^2,
\end{equation}
where the norm of $\delta\Lambda$ is the one of a scalar field \eqref{bos:eq:path-int-norm-scalar}.
The norm for $\delta g_{ab}^\perp$ is equivalent to \eqref{bos:eq:path-int-norm-tensor} with $u = 0$ (since it is traceless).
Requiring positivity of the inner-product for a non-traceless tensor imposes the following constraint on $u$:
\begin{equation}
	u > 0.
\end{equation}
One can absorb the coefficient with $u$ in $\delta\Lambda$, which will just contribute as an overall factor: its precise value has no physical meaning.
The simple choice $u = 1/4$ sets the coefficient of $\abs{\delta \Lambda}_g^2$ to $1$ in \eqref{bos:eq:path-int-norm-metric-decomposition} (another common choice is $u = 1/2$).
Ultimately, this implies that the measure factorizes as
\begin{equation}
	\label{bos:eq:measure-factorization}
	\dd_g g_{ab} = \dd_g \Lambda\, \dd_g g_{ab}^\perp.
\end{equation}

\begin{computation}[bos:eq:path-int-norm-metric-decomposition]
	\begin{align*}
		G^{abcd}\, \delta g_{ab} \delta g_{cd}
			&= \big( G_\perp^{abcd} + u \, g^{ab} g^{cd} \big)
				\big(g_{ab}\, \delta \Lambda + \delta g_{ab}^\perp \big)
				\big(g_{cd}\, \delta \Lambda + \delta g_{cd}^\perp \big)
			\\
			&= \big(2 u \, g^{cd}\, \delta \Lambda + G_\perp^{abcd} \delta g_{ab}^\perp \big)
				\big(g_{cd}\, \delta \Lambda + \delta g_{cd}^\perp \big)
			\\
			&= 4 u \, (\delta \Lambda)^2
				+ G_\perp^{abcd} \delta g_{ab}^\perp \delta g_{cd}^\perp
			\\
			&= 4 u \, \delta \Lambda^2 + 2 g^{ac} g^{bd} \delta g_{ab}^\perp \delta g_{cd}^\perp.
	\end{align*}
\end{computation}

\begin{remark}
	Another common parametrization is
	\begin{equation}
		G^{abcd} = g^{ac} g^{bd} + c\, g^{ab} g^{cd}.
	\end{equation}
	It corresponds to \eqref{bos:eq:dewitt-metric} up to a factor $1/2$ and setting $u = 1 + 2c$.
\end{remark}

\begin{remark}[Matter and curved background measures]
	\label{bos:rem:measure-curved}
	As explained previously, matter fields carry a representation and the inner-product must yield an invariant combination.
	In particular, spacetime indices must be contracted with the spacetime metric $G_{\mu\nu}(X)$ (which is the non-linear sigma model metric appearing in front of the kinetic term) for a general curved background.
	For example, the inner-product for the scalar fields $X^\mu$ is
	\begin{equation}
		\psp{\delta X^\mu}{\delta X^\mu}_g
			= \int \dd^2 \sigma \sqrt{g} \, G_{\mu\nu}(X) \delta X^\mu \delta X^\nu.
	\end{equation}
	It is not possible to normalize anymore the measure to set $\det G(X) = 1$ like in \eqref{bos:eq:measure-normalisation-one} since it depends on the fields.
	On the other hand, this factor is not important for the manipulations performed in this chapter.
	Any ambiguity in the measure will again corresponds to a renormalization of the cosmological constant~\cite[p.~923]{DHoker:1988:GeometryStringPerturbation}.
	Moreover, as explained above, it is not necessary to write explicitly the matter partition function as long as it describes a CFT.
\end{remark}

\section{Faddeev--Popov gauge fixing}
\label{bos:sec:ws-int:faddeev-popov}

\begin{figure}[htp]
	\centering
	\includegraphics[scale=1]{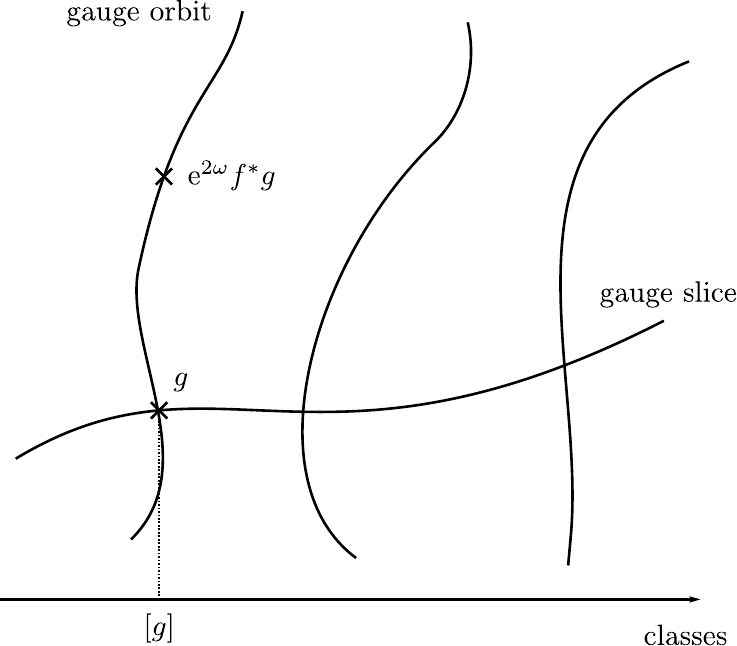}
	\caption{%
		The space of metrics decomposed in gauge orbits.
		Two metrics related by a gauge transformation lie on the same orbit.
		Choosing a gauge slice amounts to pick one metric in each orbit, and the projection gives the space of metric classes.
	}%
	\label{bos:fig-gauge-slice-metrics}
\end{figure}

\index{Polyakov path integral!Faddeev--Popov gauge fixing}%
\index{path integral!Faddeev--Popov gauge fixing}%
The naive integration over the space $\group{Met}(\Sigma_g)$ of all metrics of $\Sigma_g$ (note that the genus is fixed) leads to a divergence of the functional integral since equivalent configurations
\begin{equation}
	(f^* g, f^* \Psi) \sim (g, \Psi),
	\qquad
	(\e^{2\omega} g, \Psi) \sim (g, \Psi)
\end{equation}
gives the same contribution to the integral.
This infinite redundancy causes the integral to diverge, and since the multiple counting is generated by the gauge group, the infinite contribution corresponds to the volume of the latter.
The Faddeev--Popov procedure is a means to extract this volume by separating the integration over the gauge and physical degrees of freedom
\begin{equation}
	\dd(\text{fields})
		= \text{Jacobian} \times \dd(\text{gauge}) \times \dd(\text{physical}).
\end{equation}
The space of fields $(g, \Psi)$ is divided into equivalence classes and one integrates over only one representative of each class (gauge slice), see \Cref{bos:fig-gauge-slice-metrics}.
This change of variables introduces a Jacobian which can be represented by a partition function with ghost fields (fields with a wrong statistics).
This program encounters some complications since $G$ is a semi-direct product and is non-connected.

\begin{example}[Gauge redundancy]
	A finite-dimensional integral which mimics the problem is
	\begin{equation}
		Z = \int_{\R^2} \dd x \, \dd y \, \e^{- (x - y)^2}.
	\end{equation}
	One can perform the change of variables
	\begin{equation}
		r = x - y,
		\qquad
		y = a
	\end{equation}
	such that
	\begin{equation}
		Z = \int_{\R} \dd a \int_0^\infty \e^{- r^2}
			= \frac{\sqrt{\pi}}{2} \, \mathrm{Vol}(\R),
	\end{equation}
	and $\mathrm{Vol}(\R)$ is to be interpreted as the volume of the gauge group (translation by a real number $a$).
\end{example}

\begin{remark}
	Mathematically, the Faddeev--Popov procedure consists in identifying the orbits (class of equivalent metrics) under the gauge group $G$ and to write the integral in terms of $G$-invariant objects (orbits instead of individual metrics).
	This can be done by decomposing the tangent space into variations generated by $G$ and its complement.
	Then, one can define a foliation of the field space which equips it with a fibre bundle structure: the base is the push-forward of the complement and the fibre corresponds to the gauge orbits.
	The integral is then defined by selecting a section of this bundle.
\end{remark}

\subsection{Metrics on Riemann surfaces}

\index{metric!gauge fixing}%
According to the above procedure, each metric $g_{ab} \in \group{Met}(\Sigma_g)$ has to be expressed in terms of gauge parameters ($\xi$ and $\omega$) and of a metric $\hat g_{ab}$ which contains the remaining gauge-independent degrees of freedom.
As there are as many gauge parameters as metric components (\Cref{bos:sec:ws-int:action}), one could expect that there are no remaining physical parameters and then that $\hat g$ is totally fixed.
But, this is not the case and the metric $\hat g$ depends on a finite number of parameters $t_i$ (moduli).
The reason for this is topological: while locally it is always possible to completely fix the metric, topological obstructions may prevent doing it globally.
This means that not all conformal classes in \eqref{bos:eq:conf-structure} can be (globally) related by a diffeomorphism.

\index{moduli space}%
The quotient of the space of metrics by gauge transformations is called the \emph{moduli space}
\begin{equation}
	\label{bos:eq:def-moduli-space}
	\mc M_g
		:= \frac{\group{Met}(\Sigma_g)}{G}.
\end{equation}
Accordingly, its coordinates $t_i$ with $i = 1, \ldots, \dim_\R \mc M_g$ are called moduli parameters.
\index{Teichmüller space}%
The \emph{Teichmüller space} $\mc T_g$ is obtained by taking the quotient of $\group{Met}(\Sigma_g)$ with the component connected to the identity
\begin{equation}
	\label{bos:eq:def-Teich-space}
	\mc T_g
		:= \frac{\group{Met}(\Sigma_g)}{G_0}.
\end{equation}
The space $\mc T_g$ is the covering space of $\mc M_g$:
\begin{equation}
	\mc M_g
		= \frac{\mc T_g}{\Gamma_g},
\end{equation}
where $\Gamma_g$ is the modular group defined in \eqref{bos:eq:group-mcg}.
\index{modular group}%
Both spaces can be endowed with a complex structure and are finite-dimensional~\cite{Nakahara:2003:GeometryTopologyPhysics}:
\begin{equation}
	\label{bos:eq:dim-Mg}
	\M_g
		:= \dim_\R \mc M_g
		= \dim_\R \mc T_g
		=
		\begin{cases}
			0 & g = 0, \\
			2 & g = 1, \\
			6 g - 6 & g \ge 2,
		\end{cases}
\end{equation}
In particular, their volumes are related by
\begin{equation}
	\label{bos:eq:volumes-Teich-Mod}
	\int_{\mc M_g} \!\! \dd^{\M_g} t
		= \frac{1}{\Omega_{\Gamma_g}} \int_{\mc T_g} \!\! \dd^{\M_g} t
\end{equation}
where $\Omega_{\Gamma_g}$ is the volume of $\Gamma_g$.

\medskip

\index{group!volume}%
We will need to extract volumes of different groups, so it is useful to explain how they are defined.
A natural measure on a connected group $G$ is the Haar measure $\dd g$, which is the unique left-invariant measure on $G$.
Integrating the measure gives the volume of the group
\begin{equation}
	\label{bos:eq:volume-group}
	\Omega_G
		:= \int_G \dd g
		= \int_G \dd (h g),
\end{equation}
for any $h \in G$.
Given the Lie algebra $\alg{g}$ of the group, a general element of the algebra is a linear combinations of the generators $T_i$ with coefficients $\alpha^i$
\begin{equation}
	\alpha = \alpha^i T_i.
\end{equation}
Group elements can be parametrized in terms of $\alpha$ through the exponential map.
Moreover, since a Lie group is a manifold, it is locally isomorphic to $\R^n$: this motivates the use of a flat metric for the Lie algebra, such that
\begin{equation}
	\label{bos:eq:volume-group-alg}
	\Omega_G
		= \int \dd\alpha
		:= \int \prod_i \dd \alpha^i.
\end{equation}
Finally, it is possible to perform a change of coordinates from the Lie parameters to coordinates $x$ on the group: the resulting Jacobian is the Haar measure for the coordinates $x$.

\begin{remark}
	While $\mc T_g$ is a manifold, this is not the case of $\mc M_g$ for $g \ge 2$, which is an orbifold: the quotient by the modular group introduces singularities~\cite{Nelson:1987:LecturesStringsModuli}.
\end{remark}

\begin{remark}[Moduli space and fundamental domain]
	\index{group!fundamental domain}%

	Given a group acting on a space, a fundamental domain for a group is a subspace such that the full space is generated by acting with the group on the fundamental domain.
	Hence, one can view the moduli space $\mc M_g$ as a fundamental domain (sometimes denoted by $\mc F_g$) for the group $\Gamma_g$ and the space $\mc T_g$.
\end{remark}

\index{metric!gauge decomposition}%
In the conformal gauge \eqref{bos:eq:gauge-conformal}, the metric $g_{ab}$ can be parametrized by
\begin{equation}
	\label{bos:eq:conformal-gauge-fp}
	g_{ab}
		= \hat g_{ab}^{(f, \phi)}(t)
		:= \e^{2 f^* \phi} f^* \hat g_{ab}(t)
		= f^* \big(\e^{2 \phi} \hat g_{ab}(t) \big)
\end{equation}
where $\phi := \omega$ and $t$ denotes the dependence in the moduli parameters.
To avoid surcharging the notations, we will continue to write $g$ when there is no ambiguity.
In coordinates, this is equivalent to:
\begin{equation}
	g_{ab}(\sigma)
		= \hat g^{(f, \phi)}_{ab}(\sigma; t)
		:= \e^{2 \phi(\sigma)} \hat g'_{ab}(\sigma; t),
	\qquad
	\hat g'_{ab}(\sigma; t) = \frac{\pd \sigma'^c}{\pd \sigma^a} \frac{\pd \sigma'^d}{\pd \sigma^b} \, \hat g_{cd}(\sigma'; t).
\end{equation}

\begin{remark}
	Strictly speaking, the matter fields also transform and one should write $\Psi = \Psi^{(f)} := f^* \hat\Psi$ and include them in the change of integration measures of the following sections.
	But, this does not bring any particular benefits since these changes are trivial because the matter is decoupled from the metric.
\end{remark}

\begin{remark}
	Although the metric cannot be completely gauge fixed, having just a finite-dimensional integral is much simpler than a functional integral.
	In higher dimensions, the gauge fixing does not reduce that much the degrees of freedom and a functional integral over $\hat g$ remains (in similarity with Yang--Mills theories).
\end{remark}

The corresponding infinitesimal transformations are parametrized by $(\phi, \xi, \delta t_i)$.
\index{metric!gauge decomposition}%
The variation of the metric \eqref{bos:eq:conformal-gauge-fp} can be expressed as
\begin{equation}
	\label{bos:eq:metric-variation}
	\delta g_{ab}
		= 2 \phi \, g_{ab}
			+ \grad_a \xi_b + \grad_b \xi_a
			+ \delta t_i \pd_i g_{ab},
\end{equation}
which is decomposed in a reparametrization \eqref{bos:eq:sym-diffeo-inf}, a Weyl rescaling \eqref{bos:eq:sym-weyl-inf}, and a contribution from the variations of the moduli parameters.
\index{Teichmüller deformation}%
The latter are called Teichmüller deformations and describe changes in the metric which cannot be written as a combination of diffeomorphism and Weyl transformation.
Only the last term is written with a delta because the parameters $\xi$ and $\phi$ are already infinitesimal.
There is an implicit sum over $i$ and we have defined
\begin{equation}
	\pd_i
		:= \frac{\pd}{\pd t_i}.
\end{equation}

\index{diffeomorphism!group volume}%
\index{Weyl!group volume}%
According to the formula \eqref{bos:eq:volume-group-alg}, the volumes $\Omega_{\text{Diff}_0}[g]$ and $\Omega_{\text{Weyl}}[g]$ of the diffeomorphisms connected to the identity and Weyl group are
\begin{subequations}
\label{bos:eq:volume-diff-weyl}
\begin{align}
	\label{bos:eq:volume-diff0}
	\Omega_{\text{Diff}_0}[g]
		&
		:= \int \dd_g \xi,
	\\
	\label{bos:eq:volume-weyl}
	\Omega_{\text{Weyl}}[g]
		&
		:= \int \dd_g \phi.
\end{align}
The full diffeomorphism group has one connected component for each element of the modular group $\Gamma_g$, according to \eqref{bos:eq:group-mcg}: the volume $\Omega_{\text{Diff}}[g]$ of the full group is the volume of the component connected to the identity times the volume $\Omega_{\Gamma_g}$
\begin{equation}
	\label{bos:eq:volumes-Diff-Diff0}
	\Omega_{\text{Diff}}[g]
		= \Omega_{\text{Diff}_0}[g] \,
			\Omega_{\Gamma_g}.
\end{equation}
\end{subequations}
We have written that the volume depends on $g$: but, the metric itself is parametrized in terms of the integration variables, and thus the LHS of \eqref{bos:eq:volume-diff-weyl} cannot depend on the variable which is integrated over: $\Omega_{\text{Diff}_0}$ can depend only on $\phi$ and $\Omega_{\text{Weyl}}$ only on $\xi$.
But, all measures \eqref{bos:eq:path-int-norm-vector} are invariant under diffeomorphisms, and thus the result cannot depend on $\xi$.
Moreover, the measure for vector is invariant under Weyl transformation, which means that $\Omega_{\text{Diff}_0}$ does not depend on $\phi$.
This implies that the volumes depend only on the moduli parameters
\begin{subequations}
\label{bos:eq:volume-dep}
\begin{equation}
	\Omega_{\text{Diff}_0}[g]
		:= \Omega_{\text{Diff}_0}[\e^{2\phi} \hat g]
		= \Omega_{\text{Diff}_0}[\hat g],
	\qquad
	\Omega_{\text{Weyl}}[g]
		:= \Omega_{\text{Weyl}}[\dlie_\xi \hat g]
		= \Omega_{\text{Weyl}}[\hat g].
\end{equation}
For this reason, it is also sufficient to take the normalization factor $\Omega_{\text{gauge}}$ to have the same dependence:
\begin{equation}
	\Omega_{\text{gauge}}[g]
		:= \Omega_{\text{gauge}}[\hat g].
\end{equation}
\end{subequations}
These volumes are also discussed in \Cref{bos:rem:measure-ambiguities}.

\begin{computation}[bos:eq:volume-dep]
	\begin{gather*}
		\Omega_{\text{Diff}_0}[\e^{2\phi} \hat g]
			= \int \dd_{\e^{2\phi} \dlie_\xi \hat g} \xi
			= \int \dd_{\e^{2\phi} \hat g} \xi
			= \int \dd_{\hat g} \xi
			= \Omega_{\text{Diff}_0}[\hat g],
		\\
		\Omega_{\text{Weyl}}[\dlie_\xi \hat g]
			= \int \dd_{\e^{2\phi} \dlie_\xi \hat g} \phi
			= \int \dd_{\e^{2\phi} \hat g} \phi
			= \Omega_{\text{Weyl}}[\hat g].
	\end{gather*}
\end{computation}

\begin{remark}[Free-field measure for the Liouville mode]
	\index{Liouville!free-field measure}%

	The explicit measure \eqref{bos:eq:volume-weyl} of the Liouville mode is complicated since the inner-product contains an exponential of the field:
	\begin{equation}
		\abs{\delta\phi}^2
			= \int \dd^2 \sigma \sqrt{g} \, \delta \phi^2
			= \int \dd^2 \sigma \sqrt{\hat g} \, \e^{2\phi} \delta \phi^2.
	\end{equation}
	It has been proposed by David--Distler--Kawai~\cite{David:1988:ConformalFieldTheories, Distler:1989:ConformalFieldTheory}, and later checked explicitly~\cite{Mavromatos:1989:RegularizingFunctionalIntegral, DHoker:1990:2DQuantumGravity, DHoker:1991:EquivalenceLiouvilleTheory}, how to rewrite the measure in terms of a free measure weighted by an effective action.
	The latter is identified with the Liouville action (\Cref{bos:sec:ws-int:faddeev-popov:weyl}).
\end{remark}

In principle, we could follow the standard Faddeev--Popov procedure by inserting a delta function for the gauge fixing condition
\begin{equation}
	\label{bos:eq:gf-cond}
	F_{ab}
		:= g_{ab} - \hat g_{ab}^{(f, \phi)}(t),
\end{equation}
with $\hat g_{ab}^{(f, \phi)}(t)$ defined in \eqref{bos:eq:conformal-gauge-fp}.
However, we will take a detour to take the opportunity to study in details manipulations of path integrals and to understand several aspects of the geometry of Riemann surfaces.
In any case, several points are necessary even when going the short way, but less apparent.

\index{metric!gauge decomposition}%
In order to make use of the factorization \eqref{bos:eq:measure-factorization} of the integration measure, the variation \eqref{bos:eq:metric-variation} is decomposed into its trace (first term) and traceless parts (last two terms) \eqref{bos:eq:metric-decomposition}
\begin{equation}
	\label{bos:eq:metric-variation-fp}
	\delta g_{ab}
		= 2 \tilde\Lambda \, g_{ab}
			+ (P_1 \xi)_{ab}
			+ \delta t_i \, \mu_{iab},
\end{equation}
where\footnotemark{}
\footnotetext{%
	For comparison, Polchinski~\cite{Polchinski:2005:StringTheory-1} defines $P_1$ with an overall factor $1/2$.
}%
\index{P1@$P_1$}%
\begin{subequations}
\begin{gather}
	\label{bos:eq:op-P1}
	(P_1 \xi)_{ab} = \grad_a \xi_b + \grad_b \xi_a - g_{ab} \grad_c \xi^c,
	\\
	\label{bos:eq:def-mu}
	\mu_{iab} = \pd_i g_{ab} - \frac{1}{2} \, g_{ab} \, g^{cd} \pd_i g_{cd},
	\\
	\label{bos:eq:def-Lambda}
	\tilde\Lambda = \Lambda + \frac{1}{2} \, \delta t_i \, g^{ab} \pd_i g_{ab},
	\qquad
	\Lambda = \phi + \frac{1}{2} \, \grad_c \xi^c.
\end{gather}
\end{subequations}
\index{Beltrami differential}%
\index{Teichmüller deformation}%
The objects $\mu_i$ are called \emph{Beltrami differentials} and correspond to traceless Teichmüller deformations (the factor of $1/2$ comes from the symmetrization of the metric indices).
The decomposition emphasizes which variations are independent from each other.
In particular, changes to the trace of the metric due to a diffeomorphism generated by $\xi$ or a modification of the moduli parameters can be compensated by a Weyl rescaling.

One can use \eqref{bos:eq:measure-factorization} to replace the integration over $g_{ab}$ by one over the gauge parameters $\xi$ and $\phi$ and over the moduli $t_i$ since they contain all the information about the metric:
\begin{equation}
	\label{bos:eq:path-int-dL-dPxi}
	Z_g
		= \int \dd^{\M_g} t \, \dd_g \tilde\Lambda \, \dd_g (P_1 \xi) \,
			\Omega_{\text{gauge}}[g]^{-1} \, Z_m[g].
\end{equation}
It is tempting to perform the change of variables
\begin{equation}
	(P_1 \xi, \tilde\Lambda) \longrightarrow (\xi, \phi)
\end{equation}
such that
\begin{equation}
	\dd_g (P_1 \xi) \, \dd_g \tilde\Lambda
		\overset{?}{=} \dd_g \xi \, \dd_g \phi \, \Delta_{\text{FP}}[g]
\end{equation}
where $\Delta_{\text{FP}}[g]$ is the Jacobian of the transformation
\begin{equation}
	\Delta_{\text{FP}}[g]
		= \det \frac{\pd(P_1 \xi, \tilde\Lambda)}{\pd(\xi, \phi)}
		= \det
		\begin{pmatrix}
			P_1 & 0
			\\
			\star & 1
		\end{pmatrix}
		= \det P_1.
\end{equation}
But, one needs to be more careful:
\begin{enumerate}
	\item The variations involving $P_1 \xi$ and $\delta t_i$ are not orthogonal and, as a consequence, the measure does not factorize.

	\item $P_1$ has zero-modes, i.e.\ vectors such that $P_1 \xi = 0$, which causes the determinant to vanish, $\det P_1 = 0$.
\end{enumerate}
A rigorous analysis will be performed in \Cref{bos:sec:ws-int:faddeev-popov:P1} and will lead to additional factors in the path integral.

Next, if the actions and measures were invariant under diffeomorphisms and Weyl transformations (which amounts to replace $g$ by $\hat g$ everywhere), it would be possible to factor out the integrations over the gauge parameters and to cancel the corresponding infinite factors thanks to the normalization $\Omega_{\text{gauge}}[g]$.
A new problem arises because the measures are not Weyl invariant as explained above and one should be careful when replacing the metric (\Cref{bos:sec:ws-int:faddeev-popov:weyl}).

\subsection{Reparametrizations and analysis of \texorpdfstring{$P_1$}{P1}}
\label{bos:sec:ws-int:faddeev-popov:P1}

The properties of the operator $P_1$ are responsible for both problems preventing a direct factorization of the measure; for this reason, it is useful to study it in more details.

\index{P1@$P_1$}%
The operator $P_1$ is an object which takes a vector $v$ to a symmetric traceless $2$-tensor $T$, see \eqref{bos:eq:op-P1}.
Conversely, its adjoint $\adj{P_1}$ can be defined from the scalar product \eqref{bos:eq:path-int-norm-tensor}
\begin{equation}
	\psp{T}{P_1 v}_g
		= \psp{\adj{P_1} T}{v}_g,
\end{equation}
and takes symmetric traceless tensors to vectors.
In components, one finds
\begin{equation}
	\label{bos:eq:op-P1-adj}
	(\adj{P_1} T)_a = - 2 \grad^b T_{ab}.
\end{equation}

The Riemann--Roch theorem relates the dimension of the kernels of both operators~\cite{Nakahara:2003:GeometryTopologyPhysics}:
\begin{equation}
	\label{bos:eq:P1-Riemann-Roch}
	\dim \ker \adj{P_1} - \dim \ker P_1
		= - 3 \chi_g
		= 6 g - 6.
\end{equation}

\subsubsection{Teichmüller deformations}

\index{Teichmüller deformation}%
We first need to characterize Teichmüller deformations, the variations of moduli parameters which lead to transformations of the metric independent from diffeomorphisms and Weyl rescalings.
This means that the different variations must be orthogonal for the inner-product \eqref{bos:eq:path-int-norm}.

First, the deformations must be traceless, otherwise they can be compensated by a Weyl transformation.
The traceless metric variations $\delta g$ which cannot be generated by a vector field $\xi$ are perpendicular to $P_1 \xi$ (otherwise, the former would a linear combination of the latter):
\begin{equation}
	\psp{\delta g}{P_1 \xi}_g = 0
	\quad \Longrightarrow \quad
	\psp{\adj{P_1} \delta g}{\xi}_g = 0.
\end{equation}
Since $\xi$ is arbitrary, this means that the first argument vanishes
\begin{equation}
	\adj{P_1} \delta g = 0.
\end{equation}
Metric variations induced by a change in the moduli $t_i$ are in the kernel of $\adj{P_1}$
\begin{equation}
	\delta g \in \ker \adj{P_1}.
\end{equation}

\index{quadratic differential}%
Elements of $\ker \adj{P_1}$ are called \emph{quadratic differentials} and a basis (not necessarily orthonormal) of $\ker \adj{P_1}$ is denoted as:
\begin{equation}
	\label{bos:eq:basis-adj-P1}
	\ker \adj{P_1} = \Span \{ \phi_i \},
	\qquad
	i = 1, \ldots, \dim \ker \adj{P_1}
\end{equation}
(these should not be confused with the Liouville field).
The dimension of $\ker \adj{P_1}$ is in fact equal to the dimension of the moduli space \eqref{bos:eq:dim-Mg}:
\begin{equation}
	\dim_\R \ker \adj{P_1}
		= \M_{g}
		=
		\begin{cases}
			0 & g = 0,
			\\
			2 & g = 1,
			\\
			6 g - 6 & g > 1.
		\end{cases}
\end{equation}

The last two terms in the variation \eqref{bos:eq:metric-variation-fp} of $\delta g_{ab}$ are not orthogonal.
Let's introduce the projector on the complement space of $\ker \adj{P_1}$
\begin{equation}
	\Pi
		:= P_1 \, \frac{1}{\adj{P_1} P_1} \, \adj{P_1}.
\end{equation}
The moduli variations can then be rewritten as
\begin{equation}
	\delta t_i \, \mu_i
		= \delta t_i \, (1 - \Pi) \mu_i + \delta t_i \, \Pi \mu_i
		= \delta t_i \, (1 - \Pi) \mu_i + \delta t_i \, P_1 \zeta_i.
\end{equation}
The $\zeta_i$ exist because $\Pi \mu_i \in \Im P_1$, and they read
\begin{equation}
	\zeta_i
		:= \frac{1}{\adj{P_1} P_1} \, \adj{P_1} \mu_i.
\end{equation}
The first term can be decomposed on the quadratic differential basis \eqref{bos:eq:basis-adj-P1}
\begin{equation}
	(1 - \Pi) \mu_i = \phi_j (M^{-1})_{jk} \psp{\phi_k}{\mu_i}_g
\end{equation}
where
\begin{equation}
	M_{ij}
		:= \psp{\phi_i}{\phi_j}_g.
\end{equation}

\index{metric!gauge decomposition}%
Ultimately, the variation \eqref{bos:eq:metric-variation-fp} becomes
\begin{equation}
	\label{bos:eq:metric-variation-fp-orth}
	\delta g_{ab}
		= (P_1 \tilde\xi)_{ab}
			+ 2 \tilde\Lambda \, g_{ab}
			+ Q_{iab} \, \delta t_i.
\end{equation}
where
\begin{equation}
	\tilde\xi = \xi + \zeta_i \delta t_i,
	\qquad
	Q_{iab} = \phi_{jab} \, (M^{-1})_{jk} \psp{\phi_k}{\mu_i}_g.
\end{equation}
Correspondingly, the norm of the variation splits in three terms since each variation is orthogonal to the others:
\begin{equation}
	\label{bos:eq:norm-var-g-Lambda-xi-zeta-t}
	\abs{\delta g}_g^2 = \abs{\delta \tilde\Lambda}_g^2
		+ \abs{P_1 \tilde\xi}_g^2
		+ \abs{Q_i \delta t_i}_g^2.
\end{equation}
Since the norm is decomposed as a sum, the measure factorizes:
\begin{equation}
	\dd_g g_{ab} = \dd_g \tilde\Lambda \, \dd_g (P_1 \tilde\xi) \, \dd_g (Q_i \delta t_i).
\end{equation}

One can then perform a change of coordinates
\begin{equation}
	(\tilde\xi, \tilde\Lambda, Q_i \delta t_i)
	\longrightarrow
	(\xi, \Lambda, \delta t_i),
\end{equation}
where $\Lambda$ was defined in \eqref{bos:eq:def-Lambda}.
The goal of this transformation is to remove the dependence in the moduli from the measures on the Weyl factor and vector fields, and to recover a finite-dimensional integral over the moduli:
\begin{equation}
	\label{bos:eq:measure-Lambda-xi-t}
	\dd_g \tilde\Lambda \, \dd_g (P_1 \tilde\xi) \, \dd_g (Q_i \delta t_i)
		= \dd^{\M_g} t \, \dd_g \Lambda \, \dd_g (P_1 \xi) \,
			\frac{\det \psp{\phi_i}{\mu_j}_g}{\sqrt{\det \psp{\phi_i}{\phi_j}_g}},
\end{equation}
where the determinants correspond to the Jacobian.
The role of the determinant in the denominator is to ensure a correct normalization when the basis is not orthonormal (in particular, it ensures that the Jacobian is independent of the basis).
Plugging this result in \eqref{bos:eq:path-int-dg} gives the partition function as
\begin{equation}
	\label{bos:eq:path-int-dPxi-dL-dt}
	Z_g
		= \int_{\mc T_g} \!\! \dd^{\M_g} t \,
			\frac{1}{\Omega_{\text{gauge}}[\hat g]}
			\int \dd_g \Lambda \, \dd_g (P_1 \xi) \,
			\frac{\det \psp{\phi_i}{\mu_j}_g}{\sqrt{\det \psp{\phi_i}{\phi_j}_g}} \,
			Z_m[g].
\end{equation}
The $t_i$ are integrated over the Teichmüller space $\mc T_g$ defined by \eqref{bos:eq:def-Teich-space} because the vectors $\xi$ generate only reparametrizations connected to the identity, and thus the remaining freedom lies in $\group{Met}(\Sigma_g) / G_0$.
Next, we study how to perform the changes of variables to remove $P_1$ from the measure.

\subsubsection{Conformal Killing vectors}

In this section, we focus on the $\dd_g \Lambda \, \dd_g (P_1 \xi)$ part of the measure and we make contact with the rest at the end.

Infinitesimal reparametrizations generated by a vector field $\xi^a$ produce only transformations close to the identity.
For this reason, integrating over all possible vector fields yields the volume \eqref{bos:eq:volume-diff0} of the component of the diffeomorphism group connected to the identity:
\index{diffeomorphism!group volume}%
\begin{equation}
	\int \dd_g \xi = \Omega_{\text{Diff}_0}[\hat g].
\end{equation}
Remember that the volume depends only on the moduli, but obviously not on $\xi$ (integrated over) nor $\phi$ (the inner-product \eqref{bos:eq:path-int-norm-vector} is invariant).
But, due to the existence of zero-modes, one gets an integration over a subset of all vector fields, and this complicates the program, as we discuss now.

\index{CKV|see{conformal Killing vector}}%
\index{conformal Killing!vector}%
Zero-modes $\xi^{(0)}$ of $P_1$ are called \emph{conformal Killing vectors} (CKV)
\begin{equation}
	\label{bos:eq:def-group-Kg}
	\xi^{(0)}
		\in \mc K_g
		:= \ker P_1
\end{equation}
\index{conformal Killing!equation}%
and satisfy the conformal Killing equation (see also \Cref{sec:cft:general:conf-group}):
\begin{equation}
	(P_1 \xi^{(0)})_{ab}
		= \grad_a \xi_b^{(0)} + \grad_b \xi_a^{(0)} - g_{ab} \grad_c \xi^{(0)c}
		= 0.
\end{equation}
CKVs correspond to reparametrizations which can be absorbed by a change of the conformal factor.
They should be removed from the $\xi$ integration in order to not double-count the corresponding metrics.
The dimension of the zero-modes CKV space depends on the genus~\cite{Nakahara:2003:GeometryTopologyPhysics}:
\begin{equation}
	\label{bos:eq:dim-Kg}
	\K_{g}
		:= \dim_\R \mc K_g
		= \dim_\R \ker P_1
		=
		\begin{cases}
			6 & g = 0,
			\\
			2 & g = 1,
			\\
			0 & g > 1.
		\end{cases}
\end{equation}
The associated transformations will be interpreted later (\Cref{chap:cft:general}).
The groups generated by the CKVs are
\begin{equation}
	g = 0:
		\quad
		\mc K_0
			= \group{SL}(2, \C),
	\qquad
	g = 1:
		\quad
		\mc K_1
			= \group{U}(1) \times \group{U}(1).
\end{equation}
Note that the first group is non-compact while the second is compact.

A general vector $\xi$ can be separated into a zero-mode part and its orthogonal complement $\xi'$:
\begin{equation}
	\label{bos:eq:split-xi}
	\xi = \xi^{(0)} + \xi',
\end{equation}
such that
\begin{equation}
	\psp{\xi^{(0)}}{\xi'}_g = 0
\end{equation}
for the inner-product \eqref{bos:eq:path-int-norm-vector}.
Because zero-modes are annihilated by $P_1$, the correct change of variables in the partition function \eqref{bos:eq:path-int-dL-dPxi} maps to $\xi'$ only:
\begin{equation}
	(P_1 \xi, \Lambda) \longrightarrow (\xi', \phi).
\end{equation}
Integrating over $\xi$ at this stage would double count the CKV (since they are already described by the $\phi$ integration).
The appropriate Jacobian reads
\begin{equation}
	\label{bos:eq:metric-change-var-fp}
	\dd_g \Lambda \, \dd_g (P_1 \xi) = \dd_g \phi \, \dd_g \xi' \, \Delta_{\text{FP}}[g],
\end{equation}
\index{Faddeev--Popov!determinant}%
where the Faddeev--Popov determinant is
\begin{equation}
	\label{bos:eq:fp-det}
	\Delta_{\text{FP}}[g]
		= \det' \frac{\pd(P_1 \xi, \Lambda)}{\pd(\xi', \phi)}
		= \det' P_1
		= \sqrt{\det' P_1 \adj{P_1}},
\end{equation}
the prime on the determinant indicating that the zero-modes are excluded.
This brings the partition function \eqref{bos:eq:path-int-dPxi-dL-dt} to the form
\begin{equation}
	\label{bos:eq:path-int-dphi-dxi-dt}
	Z_g
		= \int_{\mc T_g} \!\! \dd^{\M_g} t \,
			\Omega_{\text{gauge}}[\hat g]^{-1}
			\int \dd_g \phi \, \dd_g \xi' \,
			\frac{\det \psp{\phi_i}{\mu_j}_g}{\sqrt{\det \psp{\phi_i}{\phi_j}_g}} \,
			\Delta_{\text{FP}}[g] Z_m[g].
\end{equation}

\begin{computation}[bos:eq:metric-change-var-fp]
	The Jacobian can be evaluated directly:
	\begin{equation}
		\Delta_{\text{FP}}[g]
			= \det' \frac{\pd(P_1 \xi, \Lambda)}{\pd(\xi', \phi)}
			= \det'
			\begin{pmatrix}
				P_1 & 0
				\\
				\frac{1}{2} \grad & 1
			\end{pmatrix}
			= \det' P_1.
	\end{equation}
	As a consequence of $\det' \adj{P_1} = \det' P_1$, the Jacobian can be rewritten as:
	\begin{equation}
		\sqrt{\det' \adj{P_1} P_1} = \det' P_1.
	\end{equation}

	It is instructive to derive this result also by manipulating the path integral.
	Considering small variations of the fields, one has:
	\begin{align*}
		1
			&= \int \dd_g \delta\Lambda \, \dd_g (P_1 \delta\xi) \,
				\e^{- \abs{\delta\Lambda}_g^2 - \abs{P_1 \delta\xi'}_g^2}
			\\
			&= \Delta_{\text{FP}}[g]
				\int \dd_g \delta\phi \, \dd_g \delta\xi' \,
				\e^{- \abs{\delta\phi + \frac{1}{2} \grad_c \delta\xi^c}_g^2 - \abs{P_1 \delta\xi'}_g^2}
			\\
			&= \Delta_{\text{FP}}[g]
				\int \dd_g \delta\phi \, \dd_g \delta\xi' \,
				\e^{- \abs{\delta\phi}_g^2 - \psp{\delta\xi'}{\adj{P_1} P_1 \delta\xi'}_g}
			\\
			&= \Delta_{\text{FP}}[g] \,
				\left( \det' \adj{P_1} P_1 \right)^{-1/2}.
	\end{align*}
	That the expression is equal to $1$ follows from the normalization of symmetric tensors and scalars \eqref{bos:eq:path-int-norm} (the measures appearing in the path integral \eqref{bos:eq:path-int-dPxi-dL-dt} arises without any factor).
	The third equality holds because the measure is invariant under translations of the fields, and we used the definition of the adjoint.
\end{computation}

The volume of the group generated by the vectors orthogonal to the CKV is denoted as
\begin{equation}
	\Omega_{\text{Diff}_0}'[g]
		:= \Omega_{\text{Diff}_0}'[\hat g]
		= \int \dd_g \xi'.
\end{equation}
As explained in the beginning of this section, one should extract the volume of the full $\text{Diff}_0$ group, not only the volume $\Omega_{\text{Diff}_0}'[g]$.
Since the two sets of vectors are orthogonal, we can expect the measures, and thus the volumes, to factorize.
However, a Jacobian can and does arise: its role it to take into account the normalization of the zero-modes.
Denoting by $\psi_i$ a basis (not necessarily orthonormal) for the zero-modes
\begin{equation}
	\label{bos:eq:basis-P1}
	\ker P_1 = \Span \{ \psi_i \},
	\qquad
	i = 1, \ldots, \K_g,
\end{equation}
the change of variables
\begin{equation}
	\xi' \longrightarrow \xi
\end{equation}
reads
\begin{equation}
	\label{bos:eq:change-var-xi}
	\dd_g \xi'
		= \frac{1}{\sqrt{\det \psp{\psi_i}{\psi_j}_g}} \, \frac{\dd_g \xi}{\Omega_{\text{ckv}}[g]},
\end{equation}
where $\Omega_{\text{ckv}}[g]$ is the volume of the CKV group.
The determinant is necessary when the basis is not orthonormal.
The relation between the gauge volumes is then
\index{diffeomorphism!group volume}%
\begin{equation}
	\label{bos:eq:volume-diff-0-perp}
	\Omega_{\text{Diff}_0}[g]
		= \sqrt{\det \psp{\psi_i}{\psi_j}_g} \;
			\Omega_{\text{ckv}}[g] \,
			\Omega_{\text{Diff}_0}'[g].
\end{equation}
Note that the CKV volume is given in \eqref{bos:eq:ckv-vol} and depends only on the topology but not on the metric.
By using arguments similar to the ones which lead to \eqref{bos:eq:volume-dep}, one can expect that each term is independently invariant under Weyl rescaling: this is indeed true (\Cref{bos:sec:ws-int:faddeev-popov:weyl}).

\begin{computation}[bos:eq:change-var-xi]
	Let's expand $\xi^{(0)}$ on the zero-mode basis
	\begin{equation}
		\xi^{(0)} = \alpha_i \psi_i,
	\end{equation}
	where the $\alpha_i$ are real numbers, such that one can write the changes of variables
	\begin{equation}
		\xi \longrightarrow (\xi', \alpha_i).
	\end{equation}
	The Jacobian is computed from
	\begin{align*}
		1
			&= \int \dd \xi \, \e^{- \abs{\xi}_g^2}
			= J \int \dd \xi^{(0)} \, \dd \xi' \, \e^{- \abs{\xi'}_g^2 - \abs{\xi^{(0)}}_g^2}
			\\
			&= J \int \prod_i \dd \alpha_i \, \e^{- \alpha_i \alpha_j \Psp{\psi_i}{\psi_j}_g}
				\int \dd \xi' \, \e^{- \abs{\xi'}_g^2}
			\\
			&= J \left( \det \psp{\psi_i}{\psi_j}_g \right)^{- 1/2}.
	\end{align*}
	Note that the integration over the $\alpha_i$ is a standard finite-dimensional integral.
	This gives
	\begin{equation}
		\dd \xi = \sqrt{\det \psp{\psi_i}{\psi_j}_g} \, \dd \xi' \, \prod_i \dd \alpha_i.
	\end{equation}
	\index{conformal Killing!group volume}%
	Since nothing depends on the $\alpha_i$, they can be integrated over as in \eqref{bos:eq:volume-group}, giving the volume of the CKV group
	\begin{equation}
		\label{bos:eq:ckv-vol}
		\Omega_{\text{ckv}}[g] = \int \prod_i \dd \alpha_i.
	\end{equation}
\end{computation}

Replacing the integration over $\xi'$ thanks to \eqref{bos:eq:change-var-xi}, the path integral becomes
\begin{equation}
	Z_g
		= \int_{\mc T_g} \!\! \dd^{\M_g} t \,
			\Omega_{\text{gauge}}[\hat g]^{-1}
			\int \dd_g \phi \, \dd_g \xi \,
			\frac{\det \psp{\phi_i}{\mu_j}_g}{\sqrt{\det \psp{\phi_i}{\phi_j}_g}} \,
			\frac{\Omega_{\text{ckv}}[g]^{-1}}{\sqrt{\det \psp{\psi_i}{\psi_j}_g}} \,
			\Delta_{\text{FP}}[g] \, Z_m[g].
\end{equation}

Since the matter action and measure, and the Liouville measure are invariant under reparametrizations, one can perform a change of variables
\begin{equation}
	(f^* \hat g, f^* \phi, f^* \Psi)
	\longrightarrow
	(\hat g, \phi, \Psi)
\end{equation}
such that everything becomes independent of $f$ (or equivalently $\xi$).
Since the measure for $\xi$ is Weyl invariant, it is possible to separate it from the rest of the expression, which yields an overall factor of $\Omega_{\text{Diff}_0}[g]$.
This brings the partition function to the form
\begin{equation}
	Z_g
		= \int_{\mc T_g} \!\! \dd^{\M_g} t \,
			\frac{\Omega_{\text{Diff}_0}[\hat g]}{\Omega_{\text{gauge}}[\hat g]}
			\int \dd_g \phi \,
			\frac{\det \psp{\phi_i}{\mu_j}_g}{\sqrt{\det \psp{\phi_i}{\phi_j}_g}} \,
			\frac{\Omega_{\text{ckv}}[g]^{-1}}{\sqrt{\det \psp{\psi_i}{\psi_j}_g}} \,
			\Delta_{\text{FP}}[g] \, Z_m[g]
\end{equation}
where the same symbol is used for the metric
\begin{equation}
	g_{ab} := g^{(\phi)}_{ab}
		= \e^{2\phi} \hat g_{ab}.
\end{equation}

Since the expression is invariant under the full diffeomorphism group $\group{Diff}(\Sigma_g)$ and not just under its component $\group{Diff}_0(\Sigma_g)$, one needs to extract the volume of the full diffeomorphism group before cancelling it with the normalization factor.
Otherwise, there is still an over-counting the configurations.
Using the relation \eqref{bos:eq:volumes-Diff-Diff0} leads to:
\begin{equation}
	Z_g
		= \frac{1}{\Omega_{\Gamma_g}}
			\int_{\mc T_g} \!\! \dd^{\M_g} t \,
			\frac{\Omega_{\text{Diff}}[\hat g]}{\Omega_{\text{gauge}}[\hat g]}
			\int \dd_g \phi \,
			\frac{\det \psp{\phi_i}{\mu_j}_g}{\sqrt{\det \psp{\phi_i}{\phi_j}_g}} \,
			\frac{\Omega_{\text{ckv}}[g]^{-1}}{\sqrt{\det \psp{\psi_i}{\psi_j}_g}} \,
			\Delta_{\text{FP}}[g] \, Z_m[g].
\end{equation}
The volume $\Omega_{\Gamma_g}$ can be factorized outside the integral because it depends only on the genus and not on the metric.
Finally, using the relation \eqref{bos:eq:volumes-Teich-Mod}, one can replace the integration over the Teichmüller space by an integration over the moduli space
\begin{equation}
	\label{bos:eq:path-int-dphi-dt}
	Z_g
		= \int_{\mc M_g} \!\! \dd^{\M_g} t \,
			\frac{\Omega_{\text{Diff}}[\hat g]}{\Omega_{\text{gauge}}[\hat g]}
			\int \dd_g \phi \,
			\frac{\det \psp{\phi_i}{\mu_j}_g}{\sqrt{\det \psp{\phi_i}{\phi_j}_g}} \,
			\frac{\Omega_{\text{ckv}}[g]^{-1}}{\sqrt{\det \psp{\psi_i}{\psi_j}_g}} \,
			\Delta_{\text{FP}}[g] \, Z_m[g].
\end{equation}

\subsection{Weyl transformations and quantum anomalies}
\label{bos:sec:ws-int:faddeev-popov:weyl}

\index{Liouville!field}%
The next question is whether the integrand depends on the Liouville mode $\phi$ such that the Weyl volume can be factorized out.
While the matter action has been chosen to be Weyl invariant -- see the condition \eqref{bos:eq:sym-action} -- the measures cannot be defined to be Weyl invariant.
\index{Weyl!anomaly}%
This means that there is a \emph{Weyl} (or conformal) \emph{anomaly}, i.e.\ a violation of the Weyl invariance due to quantum effects.
Since the techniques needed to derive the results of this section are outside the scope of this \revname{}, we simply state the results.

It is possible to show that the Weyl anomaly reads~\cite[p.~929]{DHoker:1988:GeometryStringPerturbation}\footnotemark{}
\footnotetext{%
	The relation is written for $Z_m$ since the action is invariant and is not affected by the anomaly.
}%
\begin{subequations}
\label{bos:eq:measure-anomaly}
\begin{gather}
	\frac{\Delta_{\text{FP}}[\e^{2\phi} \hat g]}{\sqrt{\det \psp{\phi_i}{\phi_j}_{\e^{2\phi} \hat g}}}
		= \e^{\frac{c_{\text{gh}}}{6} S_L[\hat g, \phi]} \,
			\frac{\Delta_{\text{FP}}[\hat g]}{\sqrt{\det \psp{\hat\phi_i}{\hat\phi_j}_{\hat g}}}
	\\
	Z_m[\e^{2\phi} \hat g]
		= \e^{\frac{c_m}{6} S_L[\hat g, \phi]} Z_m[\hat g],
\end{gather}
\end{subequations}
\index{Liouville!action}%
where $S_L$ is the Liouville action
\begin{equation}
	S_L[\hat g, \phi]
		:= \frac{1}{4\pi} \int \dd^2 \sigma \sqrt{\hat g} \big( \hat g^{ab} \pd_a \phi \pd_b \phi + \hat R \phi \big),
\end{equation}
where $\hat R$ is the Ricci scalar of the metric $\hat g_{ab}$.
These relations require to introduce counter-terms, discussed further in \Cref{bos:rem:measure-ambiguities}.
\index{reparametrization $bc$ ghosts!central charge}%
The coefficients $c_m$ and $c_{\text{gh}}$ are the central charges respectively of the matter and ghost systems, with:
\begin{equation}
	c_{\text{gh}} = - 26.
\end{equation}
This value will be derived in \Cref{cft:sec:systems:ghosts}.

The inner-products between $\phi_i$ and $\mu_j$, and between the $\psi_i$, and the CKV volume are independent of $\phi$~\cites[sec.~14.2.2]{Nakahara:2003:GeometryTopologyPhysics}[p.~931]{DHoker:1988:GeometryStringPerturbation}
\begin{equation}
	\begin{gathered}
		\det \psp{\phi_i}{\mu_j}_{\e^{2\phi} \hat g}
			= \det \psp{\hat\phi_i}{\hat\mu_j}_{\hat g},
		\qquad
		\det \psp{\psi_i}{\psi_j}_{\e^{2\phi} \hat g}
			= \det \psp{\psi_i}{\psi_j}_{\hat g},
		\\
		\Omega_{\text{ckv}}[\e^{2\phi} \hat g]
			= \Omega_{\text{ckv}}[\hat g].
	\end{gathered}
\end{equation}

\begin{remark}[Weyl and gravitational anomalies]
	The Weyl anomaly translates into a non-zero trace of the quantum energy--momentum tensor
	\begin{equation}
		\mean{g^{\mu\nu} T_{\mu\nu}} = \frac{c}{12} \, R,
	\end{equation}
	where $c$ is the central charge of the theory.
	The Weyl anomaly can be traded for a gravitational anomaly, which means that diffeomorphisms are broken at the quantum level~\cite{Jackiw:1995:AnotherViewMassless}.
\end{remark}

Inserting \eqref{bos:eq:measure-anomaly} in \eqref{bos:eq:path-int-dphi-dt} yields
\begin{equation}
	Z_g
		= \int_{\mc M_g} \!\! \dd^{\M_g} t \,
				\frac{\Omega_{\text{Diff}}[\hat g]}{\Omega_{\text{gauge}}[\hat g]} \,
				\frac{\det \psp{\phi_i}{\hat\mu_j}_{\hat g}}{\sqrt{\det \psp{\phi_i}{\phi_j}_{\hat g}}} \,
				\frac{\Omega_{\text{ckv}}[\hat g]^{-1}}{\sqrt{\det \psp{\psi_i}{\psi_j}_{\hat g}}} \,
				\Delta_{\text{FP}}[\hat g] \, Z_m[\hat g]
			\int \dd_g \phi \,
				\e^{- \frac{c_L}{6} S_L[\hat g, \phi]},
\end{equation}
\index{Liouville!central charge}%
with the Liouville central charge
\begin{equation}
	c_L
		:= 26 - c_m.
\end{equation}
\index{critical dimension}%
The critical “dimension” is defined to be the value of the matter central charge $c_m$ such that the Liouville central charge cancels
\begin{equation}
	\label{bos:eq:crit-dim}
	c_L = 0
	\quad \Longrightarrow \quad
	c_m = 26.
\end{equation}
If the number of non-compact dimensions is $D$, it means that the central charge \eqref{bos:eq:c-D-perp} of the transverse CFT satisfies
\begin{equation}
	c_\perp = 26 - D.
\end{equation}

In this case, the integrand does not depend on the Liouville mode (because $\Omega_{\text{Diff}}$ is invariant under Weyl transformations) and the integration over $\phi$ can be factored out and yields the volume of the Weyl group \eqref{bos:eq:volume-weyl}
\begin{equation}
	\int \dd_g \phi
		= \Omega_{\text{Weyl}}[\hat g].
\end{equation}
Then, taking
\begin{equation}
	\Omega_{\text{gauge}}[\hat g]
		= \Omega_{\text{Diff}}[\hat g] \times \Omega_{\text{Weyl}}[\hat g]
\end{equation}
removes the infinite gauge contributions and gives the partition function
\begin{equation}
	Z_g
		= \int_{\mc M_g} \!\! \dd^{\M_g} t \,
			\frac{\det \psp{\phi_i}{\hat\mu_j}_{\hat g}}{\sqrt{\det \psp{\phi_i}{\phi_j}_{\hat g}}} \,
			\frac{\Omega_{\text{ckv}}[\hat g]^{-1}}{\sqrt{\det \psp{\psi_i}{\psi_j}_{\hat g}}} \,
			\Delta_{\text{FP}}[\hat g] \, Z_m[\hat g].
\end{equation}

\begin{draft}

\begin{exercise}[Faddeev--Popov in particular cases]
	In order to better understand the Faddeev--Popov procedures, redo the computation of the full section by focusing on the following specific cases:
	\begin{enumerate}
		\item ignore the CKV and the Teichmüller parameters
		\item ignore the Teichmüller parameters (case of the sphere, $g = 0$)
		\item ignore the CKV (case of a surface with $g > 1$)
	\end{enumerate}
	To simplify even more the problem, you can consider $S_m$ to be the Polyakov action.
\end{exercise}

\end{draft}

\subsection{Ambiguities, ultralocality and cosmological constant}
\label{bos:rem:measure-ambiguities}

Different ambiguities remain in the previous computations, starting with the definitions of the measures \eqref{bos:eq:measure-normalisation-one} and \eqref{bos:eq:path-int-norm}, then in obtaining the volume of the diffeomorphism \eqref{bos:eq:volume-diff0} and Weyl \eqref{bos:eq:volume-weyl} groups, and finally in deriving the conformal anomaly \eqref{bos:eq:measure-anomaly}.

\index{worldsheet!cosmological constant}%
These different ambiguities can be removed by renormalizing the worldsheet cosmological constant.
This implies that the action
\begin{equation}
	S_\mu[g] = \int \dd^2 \sigma \sqrt{g}
\end{equation}
must be added to the classical Lagrangian, where $\mu_0$ is the bare cosmological constant.
This means that Weyl invariance is explicitly broken at the classical level.
After performing all the manipulations, $\mu_0$ is determined by removing all ambiguities and enforcing invariance under the Weyl symmetry at the quantum level.
This amounts to set the renormalized cosmological constant to zero (since it breaks the Weyl symmetry).
The possibility to introduce a counter-term violating a classical symmetry arises because the symmetry itself is broken by a quantum anomaly, so there is no reason to enforce it in the classical action.

We now review each issue separately.
First, consider the inner-product of a single tensor \eqref{bos:eq:measure-normalisation-one}: the determinant $\det \gamma_g$ depends on the metric and one should be more careful when fixing the gauge or integrating over all metrics.
\index{path integral!measure!ultralocality}%
However, ultralocality implies that the determinant can only be of the form~\cite[pp.~923]{DHoker:1988:GeometryStringPerturbation}
\begin{equation}
	\sqrt{\det \gamma_g}
		= \e^{- \mu_\gamma \, S_\mu[g]},
\end{equation}
for some $\mu_\gamma \in \R$, since $S_\mu$ is the only renormalizable covariant functional depending on the metric but not on its derivatives.
The effect is just to redefine the cosmological constant.

Second, the volume of the field space can be defined as the limit $\lambda \to 0$ of a Gaussian integral~\cite[pp.~931]{DHoker:1988:GeometryStringPerturbation}:
\begin{equation}
	\Omega_{\Phi}
		= \lim_{\lambda \to 0} \int \dd_g \Phi \, \e^{- \lambda \, \psp{\Phi}{\Phi}_g}.
\end{equation}
Due to ultralocality, the Gaussian integral should again be of the form
\begin{equation}
	\int \dd_g \Phi \, \e^{- \lambda \, \psp{\Phi}{\Phi}_g}
		= \e^{- \mu(\lambda) \, S_\mu[g]},
\end{equation}
for some constant $\mu(\lambda)$.
Hence, the limit $\lambda \to 0$ gives
\begin{equation}
	\Omega_{\Phi}
		= \int \dd_g \Phi
		= \e^{- \mu(0) \, S_\mu[g]},
\end{equation}
which can be absorbed in the cosmological constant.
However, the situation is more complicated if $\Phi = \xi, \phi$ since the integration variables also appear in the measure, as it was also discussed before \eqref{bos:eq:volume-dep}.
But, in that case, it cannot appear in the expression of the volume in the LHS.
Moreover, invariances under diffeomorphisms for both measures, and under Weyl rescalings for the vector measure, imply that the LHS can only depend on the moduli through the background metric $\hat g$.
The diffeomorphism and Weyl volumes can be written in terms of $\e^{- \hat\mu \, S_\mu[\hat g]}$: since there is no counter-term left (the cosmological constant counter-term is already fixed to cancel the coefficient of $S_\mu[g]$), it is necessary to divide by $\Omega_{\text{gauge}}$ to cancel the volumes.

\index{Weyl!anomaly}%
Finally, the computation of the Weyl anomaly \eqref{bos:eq:measure-anomaly} yields divergent terms of the form
\begin{equation}
	\lim_{\epsilon \to 0} \frac{1}{\epsilon} \int \dd^2 \sigma \sqrt{g}.
\end{equation}
These divergences are canceled by the cosmological constant counter-term, see~\cite[app.~5.A]{DiFrancesco:1999:ConformalFieldTheory} for more details.

\subsection{Gauge-fixed path integral}
\label{bos:sec:ws-int:faddeev-popov:conclusion}

\index{string amplitude!gv@$g$-loop vacuum (-)}%
As a conclusion of this section, we found that the partition function \eqref{bos:eq:path-int-dg} can be written as
\begin{subequations}
\label{bos:eq:path-int-fp}
\begin{align}
	Z_g
		&= \int_{\mc M_g} \!\! \dd^{\M_g} t \,
			\frac{\det \psp{\phi_i}{\hat\mu_j}_{\hat g}}{\sqrt{\det \psp{\phi_i}{\phi_j}_{\hat g}}} \,
			\frac{\Omega_{\text{ckv}}[\hat g]^{-1}}{\sqrt{\det \psp{\psi_i}{\psi_j}_{\hat g}}} \,
			\Delta_{\text{FP}}[\hat g] \, Z_m[\hat g],
		\\
		&= \int_{\mc M_g} \!\! \dd^{\M_g} t \,
			\sqrt{
				\frac{\det \psp{\phi_i}{\hat\mu_j}_{\hat g}^2}{\det \psp{\phi_i}{\phi_j}_{\hat g}} \,
				\frac{\det' \adj{\hat P_1} \hat P_1}{\det \psp{\psi_i}{\psi_j}_{\hat g}}} \;
			\frac{Z_m[\hat g]}{\Omega_{\text{ckv}}[\hat g]}.
\end{align}
\end{subequations}
after gauge fixing of the worldsheet diffeomorphisms and Weyl rescalings.
It is implicit that the factors for the CKV and moduli are respectively absent for $g > 1$ and $g < 1$.
For $g = 0$ the CKV group is non-compact and its volume is infinite.
It looks like the partition vanishes, but there are subtleties which will be discussed in \Cref{bos:sec:ws-int:amp:gauge-fixing-2pt}.

\begin{remark}[Weil--Petersson metric]
	When the metric is chosen to be of constant curvature $\hat R = - 1$, the moduli measure together with the determinants form the Weil--Petersson measure
	\begin{equation}
		\dd(\mathrm{WP})
			= \int_{\mc M_g} \!\! \dd^{\M_g} t \,
				\frac{\det \psp{\phi_i}{\hat\mu_j}_{\hat g}}{\sqrt{\det \psp{\phi_i}{\phi_j}_{\hat g}}}.
	\end{equation}
\end{remark}

\medskip

\index{worldsheet!symmetry!background diffeomorphisms}%
In \eqref{bos:eq:path-int-fp}, the background metric $\hat g_{ab}$ is fixed.
However, the derivation holds for any choice of $\hat g_{ab}$: as a consequence, it makes sense to relax the gauge fixing and allow it to vary while adding gauge symmetries.
The first symmetry is background diffeomorphisms:
\begin{equation}
	\label{bos:eq:sym-bg-diffeo}
	\sigma'^a
		= \hat f^a(\sigma^b),
	\qquad
	\hat g'(\sigma')
		= f^* \hat g(\sigma),
	\qquad
	\phi'(\sigma')
		= f^* \phi(\sigma),
	\qquad
	\Psi'(\sigma')
		= f^* \Psi(\sigma).
\end{equation}
This symmetry is automatic for $S_m[\hat g, \Psi]$ since $S_m[g, \Psi]$ was invariant under \eqref{bos:eq:sym-diffeo}.
Similarly, the integration measures are also invariant.
\index{worldsheet!symmetry!background Weyl}%
\index{metric!gauge decomposition}%
A second symmetry is found by inspecting the decomposition \eqref{bos:eq:conformal-gauge-fp}
\begin{equation}
	g_{ab}
		= f^* \big(\e^{2 \phi} \hat g_{ab}(t) \big),
\end{equation}
which is left invariant under a \emph{background Weyl symmetry} (also called emergent):
\begin{equation}
	\label{bos:eq:sym-bg-weyl}
	g'_{ab}(\sigma)
		= \e^{2\omega(\sigma)} g_{ab}(\sigma),
	\qquad
	\phi'(\sigma)
		= \phi(\sigma) - \omega(\sigma),
	\qquad
	\Psi'(\sigma)
		= \Psi(\sigma).
\end{equation}
Let us stress that it is not related to the Weyl rescaling \eqref{bos:eq:sym-weyl} of the metric $g_{ab}$.
The background Weyl rescaling \eqref{bos:eq:sym-bg-weyl} is a symmetry even when the physical Weyl rescaling \eqref{bos:eq:sym-weyl} is not.
Together, the background diffeomorphisms and Weyl symmetry have three gauge parameters, which is sufficient to completely fix the background metric $\hat g$ up to moduli.

In fact, the combination of both symmetries is equivalent to invariance under the physical diffeomorphisms.
To prove this statement, consider two metrics $g$ and $g'$ related by a diffeomorphism $F$ and both gauge fixed to pairs $(f, \phi, \hat g)$ and $(f', \phi', \hat g')$:
\begin{equation}
	\label{bos:eq:sym-phys-bg-relations}
	g_{ab}'
		= F^* g_{ab},
	\qquad
	g_{ab}'
		= f'^* \big(\e^{2 \phi'} \hat g_{ab}' \big),
	\qquad
	g_{ab}
		= f^* \big(\e^{2 \phi} \hat g_{ab} \big).
\end{equation}
Then, the gauge fixing parametrizations are related by background symmetries $(\hat F, \omega)$ as
\begin{equation}
	\label{bos:eq:sym-phys-bg}
	\hat F
		= f'^{-1} \circ F \circ f,
	\qquad
	\phi'
		= \hat F^* (\phi - \omega),
	\qquad
	\hat g_{ab}'
		= \hat F^* (\e^{2\omega} \hat g_{ab}).
\end{equation}
Moreover, this also implies that there is a diffeomorphism $\tilde f = F \circ f$ such that $g'$ is gauge fixed in terms of $(\phi, \hat g)$:
\begin{equation}
	g_{ab}'
		= \tilde f^* \big(\e^{2 \phi} \hat g_{ab} \big).
\end{equation}

\begin{computation}[bos:eq:sym-phys-bg]
	The functions $F$, $f$, $f'$, $\phi$, $\phi'$ and the metrics $g_{ab}$, $g'_{ab}$, $\hat g_{ab}$ and $\hat g'_{ab}$ are all fixed and one must find $\hat F$ and $\omega$ such that the relations \eqref{bos:eq:sym-phys-bg-relations} are compatible.
	First, one rewrites $g'_{ab}$ in terms of $\hat g_{ab}$ and compare with the expression with $\hat g'_{ab}$:
	\begin{align*}
		g'_{ab}
			&
			= F^* g_{ab}
			= F^* \big( f^* \big(\e^{2 \phi} \hat g_{ab} \big) \big)
			= F^* \big( f^* \big(\e^{2 (\phi - \omega)} \e^{2\omega} \hat g_{ab} \big) \big)
			\\ &
			= f'^* \big(\e^{2 \phi'} \hat g'_{ab} \big).
	\end{align*}
	In the third equality, we have introduced $\omega$ because $\hat g'_{ab} = \hat F^* \hat g_{ab}$ is not true in general since there are $3$ independent components but $\hat F$ has only $2$ parameters, so we cannot just define $f' = F \circ f$ and $\phi' = \phi$.
	This explains the importance of the emergent Weyl symmetry.
\end{computation}

\begin{remark}[Gauge fixing and field redefinition]
	Although it looks like we are undoing the gauge fixing, this is not exactly the case since the original metric is not used anymore.
	One can understand the procedure of this section as a field redefinition: the degrees of freedom in $g_{ab}$ are repackaged into two fields $(\phi, \hat g_{ab})$ adapted to make some properties of the system more salient.
	A new gauge symmetry is introduced to maintain the number of degrees of freedom.
	The latter helps to understand the structure of the action on the background.
	Finally, in this context, the Liouville action is understood as a Wess--Zumino action, which is defined as the difference between the effective actions evaluated in each metric.
	Another typical use of this point of view is to rewrite a massive vector field as a massless gauge field together with an axion~\cite{Preskill:1991:GaugeAnomaliesEffective}.
\end{remark}

\index{2d@$2d$ gravity}%
\begin{remark}[Two-dimensional gravity]
	In $2d$ gravity, one does not work in the critical dimension \eqref{bos:eq:crit-dim} and $c_L \neq 0$.
	Thus, the Liouville mode does not decouple: the conformal anomaly breaks the Weyl symmetry at the quantum level which gives dynamics to gravity, even if it has no degree of freedom classically.
	As a consequence, one chooses $\Omega_{\mathrm{gauge}} = \Omega_{\mathrm{Diff}}$.

	Since the role of the classical Weyl symmetry is not as important as for string theory, it is even not necessary to impose it classically.
	This leads to consider non-conformal matter~\cite{Ferrari:2012:GravitationalActionsTwo, Ferrari:2014:FQHECurvedBackgrounds, Can:2015:GeometryQuantumHall, Bilal:2017:2DQuantumGravity, Bilal:2017:2DGravitationalMabuchi}.
	Following the arguments from \Cref{bos:sec:ws-int:action}, the existence of the emergent Weyl symmetry \eqref{bos:eq:sym-bg-weyl} implies that the total action $S_{\text{grav}}[\hat g, \phi] + S_m[\hat g, \Psi]$ must be a CFT for a flat background $\hat g = \delta$, even if the two actions are not independently CFTs.
\end{remark}

\section{Ghost action}
\label{bos:sec:ws-int:ghosts}

\subsection{Actions and equations of motion}

\index{reparametrization $bc$ ghosts}%
It is well-known that a determinant can be represented with two anticommuting fields, called ghosts.
The fields carry indices dictated by the map induced by the operator of the Faddeev--Popov determinant: one needs a symmetric and traceless anti-ghost $b_{ab}$ and a vector ghost $c^a$ fields:
\begin{equation}
	\label{bos:eq:fp-ghost}
	\Delta_{\text{FP}}[g]
		= \int \dd'_g b \, \dd'_g c \;
			\e^{- S_{\text{gh}}[g, b, c]},
\end{equation}
where the prime indicates that the ghost zero-modes are omitted.
\index{reparametrization $bc$ ghosts!action}%
The ghost action is
\begin{subequations}
\label{bos:eq:action-ghost}
\begin{align}
	S_{\text{gh}}[g, b, c]
		&
		:= \frac{1}{4\pi} \int \dd^2 \sigma \sqrt{g} \,
			g^{ab} g^{cd} b_{ac} (P_1 c)_{bd}
		\\
		&= \frac{1}{4\pi} \int \dd^2 \sigma \sqrt{g} \,
			g^{ab} \big(b_{ac} \grad_b c^c + b_{bc} \grad_a c^c - b_{ab} \grad_c c^c \big).
\end{align}
\end{subequations}
The ghosts $c^a$ and anti-ghosts $b_{ab}$ are associated respectively to the variations due to the diffeomorphisms $\xi^a$ and to the variations perpendicular to the gauge slice.
The normalization of $1/4\pi$ is conventional.
In Minkowski signature, the action is multiplied by a factor $\I$.

Since $b_{ab}$ is traceless, the last term of the action vanishes and could be removed.
However, this implies to consider traceless variations of the $b_{ab}$ when varying the action (to compute the equations of motion, the energy--momentum tensor, etc.).
On the other hand, one can keep the term and consider unconstrained variation of $b_{ab}$ (since the structure of the action will force the variation to have the correct symmetry), which is simpler.
A last possibility is to introduce a Lagrange multiplier.
These aspects are related to the question of introducing a ghost for the Weyl symmetry, which is described in \Cref{bos:sec:ws-int:ghosts:weyl}.

\index{reparametrization $bc$ ghosts!equation of motion}%
The equations of motion are
\begin{equation}
	\label{bos:eq:eom-ghost}
	(P_1 c)_{ab} = \grad_a c_b + \grad_b c_a - g_{ab} \grad_c c^c
		= 0,
	\qquad
	(\adj{P_1} b)_a = - 2 \grad^b b_{ab}
		= 0.
\end{equation}
Hence, the classical solutions of $b$ and $c$ are respectively mapped to the zero-modes of the operators $\adj{P_1}$ and $P_1$, and they are thus associated to the CKV and Teichmüller parameters.

The energy--momentum tensor is
\begin{equation}
	\label{bos:eq:T-ghost}
	T^{\text{gh}}_{ab}
		= - b_{ac} \grad_b c^c - b_{bc} \grad_a c^c
			+ c^c \grad_c b_{ab}
			+ g_{ab} b_{cd} \grad^c c^d.
\end{equation}
Its trace vanishes off-shell (i.e.\ without using the $b$ and $c$ equations of motion)
\begin{equation}
	\label{bos:eq:T-ghost-trace}
	g^{ab} T^{\text{gh}}_{ab} = 0,
\end{equation}
which shows that the action \eqref{bos:eq:action-ghost} is invariant under Weyl transformations
\begin{equation}
	S_{\text{gh}}[\e^{2\omega} g, b, c] = S_{\text{gh}}[g, b, c].
\end{equation}

\begin{draft}

\begin{exercise}
	Derive \eqref{bos:eq:eom-ghost} and \eqref{bos:eq:T-ghost} by varying the action \eqref{bos:eq:action-ghost}.
\end{exercise}

\end{draft}

The action \eqref{bos:eq:action-ghost} also has a $\group{U}(1)$ global symmetry.
\index{ghost number}%
The associated conserved charge is called the \emph{ghost number} and counts the number of $c$ ghosts minus the number of $b$ ghosts, i.e.\
\begin{subequations}
\label{bos:eq:Ngh-cov}
\begin{equation}
	N_{\text{gh}}(b) = - 1,
	\qquad N_{\text{gh}}(c) = 1.
\end{equation}
The matter fields are inert under this symmetry:
\begin{equation}
	N_{\text{gh}}(\Psi) = 0.
\end{equation}
\end{subequations}

\index{string amplitude!gv@$g$-loop vacuum (-)}%
In terms of actions, the path integral \eqref{bos:eq:path-int-fp} can be rewritten as
\begin{equation}
	\label{bos:eq:path-int-ghosts}
	Z_g
		= \int_{\mc M_g} \!\! \dd^{\M_g} t \,
				\frac{\det \psp{\phi_i}{\hat\mu_j}_{\hat g}}{\sqrt{\det \psp{\phi_i}{\phi_j}_{\hat g}}} \,
				\frac{\Omega_{\text{ckv}}[\hat g]^{-1}}{\sqrt{\det \psp{\psi_i}{\psi_j}_{\hat g}}}
			\int \dd_{\hat g} \Psi \, \dd'_{\hat g} b \, \dd'_{\hat g} c \,
				\e^{- S_m[\hat g, \Psi] - S_{\text{gh}}[\hat g, b, c]}.
\end{equation}
One can use \eqref{bos:eq:path-int-fp} or \eqref{bos:eq:path-int-ghosts} indifferently: the first is more appropriate when using spectral analysis to compute the determinant explicitly, while the second is more natural in the context of CFTs.

\subsection{Weyl ghost}
\label{bos:sec:ws-int:ghosts:weyl}

\index{Weyl!ghost}%
Ghosts have been introduced for the reparametrizations (generated by $\xi^a$) and the traceless part of the metric (the gauge field associated to the transformation): one may wonder why there is not a ghost $c_w$ associated to the Weyl symmetry along with an antighost for the trace of the metric (i.e.\ the conformal factor).
This can be understood from several viewpoints.
First, the relation between a metric and its transformation -- and the corresponding gauge fixing condition -- does not involve any derivative: as such, the Jacobian is trivial.
Second, one could choose
\begin{equation}
	\label{bos:eq:gf-cond-traceless}
	F_{ab}^\perp
		= \sqrt{g} g_{ab} - \sqrt{\hat g} \hat g_{ab}
		= 0
\end{equation}
as a gauge fixing condition instead of \eqref{bos:eq:gf-cond}, and the trace component does not appear anywhere.
Finally, a local Weyl symmetry is not independent from the diffeomorphisms.

\begin{remark}[Local Weyl symmetry]
   \label{bos:rem:gauged-weyl}
   \index{Weyl!symmetry}%

   The topic of obtaining a local Weyl symmetry by gauging a global Weyl symmetry (dilatation) is very interesting~\cites[chap.~15]{Freedman:2012:Supergravity}{Iorio:1997:WeylGaugingConformalInvariance}.
   Under general conditions, one can express the new action in terms of the Ricci tensor (or of the curvature): this means that the Weyl gauge field and its curvature are composite fields.

   Moreover, one finds that local Weyl invariance leads to an off-shell condition while diffeomorphisms give on-shell conditions.
   This explains why one imposes only Virasoro constraints (associated to reparametrizations) and no constraints for the Weyl symmetry in the covariant quantization.
\end{remark}

However, it can be useful to introduce a ghost field $c_w$ for the Weyl symmetry nonetheless.
In view of the previous discussion, this field should appear as a Lagrange multiplier which ensures that $b_{ab}$ is traceless.
Starting from the action \eqref{bos:eq:action-ghost}, one finds
\begin{equation}
	\label{bos:eq:action-ghost-weyl}
	S'_{\text{gh}}[g, b, c, c_w]
		= \frac{1}{4\pi} \int \dd^2 \sigma \sqrt{g} \,
			g^{ab} \big(b_{ac} \grad_b c^c + b_{bc} \grad_a c^c
			+ 2 b_{ab} c_w \big),
\end{equation}
where $b_{ab}$ is not traceless anymore.
The ghost $c_w$ is not dynamical since the action does not contain derivatives of it, and it can be integrated out of the path integral to recover \eqref{bos:eq:action-ghost}.

The equations of motion for this modified action are
\begin{equation}
	\label{bos:eq:eom-ghost-weyl}
	\grad_a c_b + \grad_b c_a + 2 g_{ab} c_w = 0,
	\qquad
	\grad^a b_{ab} = 0,
	\qquad
	g^{ab} b_{ab} = 0.
\end{equation}
Contracting the first equation with the metric gives
\begin{equation}
	c_w = - \frac{1}{2} \, \grad_a c^a,
\end{equation}
and thus $c_w$ is nothing else than the divergence of the $c^a$ field: the Weyl ghost is a composite field (this makes connection with \Cref{bos:rem:gauged-weyl}) – see also \eqref{bos:eq:def-Lambda}.
The energy--momentum tensor of the ghosts with action \eqref{bos:eq:action-ghost-weyl} is
\begin{equation}
	\label{bos:eq:T-ghost-weyl}
	 \begin{aligned}
		T'^{\text{gh}}_{ab}
			= &- \big(b_{ac} \grad_b c^c + b_{bc} \grad_a c^c
					+ 2 b_{ab} c_w \big)
				- \grad_c ( b_{ab} c^c)
				\\
				&+ \frac{1}{2} \, g_{ab}
				g^{cd} \big(b_{ce} \grad_d c^e + b_{de} \grad_c c^e
					+ 2 b_{cd} c_w \big).
	\end{aligned}
\end{equation}
The trace of this tensor
\begin{equation}
	\label{bos:eq:T-ghost-weyl-trace}
	g^{ab} T'^{\text{gh}}_{ab}
		= - g^{ab} \grad_c ( b_{ab} c^c)
\end{equation}
does not vanish off-shell, but it does on-shell since $g^{ab} b_{ab} = 0$.
This implies that the theory is Weyl invariant even if the action is not.
It is interesting to contrast this with the trace \eqref{bos:eq:T-ghost-trace} when the Weyl ghost has been integrated out.

The equations of motion \eqref{bos:eq:eom-ghost} and energy--momentum tensor \eqref{bos:eq:T-ghost} for the action \eqref{bos:eq:action-ghost} can be easily derived by replacing $c_w$ by its solution in the previous formulas.

\begin{computation}[bos:eq:T-ghost-weyl]
	The first parenthesis comes from varying $g^{ab}$, the second from the covariant derivatives, the last from the $\sqrt{g}$.
	The second term comes from
	\begin{align*}
		g^{ab} \big(b_{ac} \delta\grad_b c^c + b_{bc} \delta\grad_a c^c \big)
			&= 2 g^{ab} b_{ac} \delta\grad_b c^c
			= 2 g^{ab} b_{ac} \delta\tensor{\Gamma}{^c_{bd}} c^d
			\\
			&= g^{ab} b_{ac} g^{ce} \big( \grad_b \delta g_{de} + \grad_d \delta g_{be} - \grad_e \delta g_{bd} \big) c^d
			\\
			&= b^{ab} \big( \grad_a \delta g_{bc} + \grad_c \delta g_{ab} - \grad_b \delta g_{ac} \big) c^c
			\\
			&= b^{ab} \grad_c \delta g_{ab} c^c,
	\end{align*}
	where two terms have cancelled due to the symmetry of $b^{ab}$.
	Integrating by part gives the term in the previous equation.
\end{computation}

Note that the integration on the Weyl ghost yields a delta function
\begin{equation}
	\int \dd_g c_w \, \e^{- \psp{c_w}{g^{ab} b_{ab}}_g}
		= \delta\big( g^{ab} b_{ab} \big).
\end{equation}

\subsection{Zero-modes}

\index{reparametrization $bc$ ghosts!zero-mode}%
The path integral \eqref{bos:eq:path-int-ghosts} excludes the zero-modes of the ghosts.
One can expect them to be related to the determinants of elements of $\ker P_1$ and $\ker \adj{P_1}$ with Grassmann coefficients.
They can be included after few simple manipulations (see also \Cref{sec:form:path-integrals:zero-mode}).

It is simpler to first focus on the $b$ ghost (to avoid the problems related to the CKV).
\index{string amplitude!gv@$g$-loop vacuum (-)}%
The path integral \eqref{bos:eq:path-int-ghosts} can be rewritten as
\begin{equation}
	\label{bos:eq:path-int-ghosts-zero-b}
	Z_g = \int_{\mc M_g} \!\! \dd^{\M_g} t \,
		\frac{\Omega_{\text{ckv}}[\hat g]^{-1}}{\sqrt{\det \psp{\psi_i}{\psi_j}_{\hat g}}}
		\int \dd_{\hat g} \Psi \, \dd_{\hat g} b \, \dd'_{\hat g} c \,
		\prod_{i=1}^{\M_g} \psp{b}{\hat\mu_i}_{\hat g} \,
		\e^{- S_m[\hat g, \Psi] - S_{\text{gh}}[\hat g, b, c]}.
\end{equation}
In this expression, $c$ zero-modes are not integrated over, only the $b$ zero-modes are.
This is the standard starting point on Riemann surfaces with genus $g \ge 1$.
The inner-product reads explicitly
\begin{equation}
	\psp{b}{\hat\mu_i}_{\hat g}
		= \int \dd^2 \sigma \sqrt{\hat g} \,
			G_\perp^{abcd} b_{ab} \hat\mu_{i,cd}
		= \int \dd^2 \sigma \sqrt{\hat g} \,
			g^{ac} g^{bd} b_{ab} \hat\mu_{i,cd}.
\end{equation}

\begin{computation}[bos:eq:path-int-ghosts-zero-b]
	Since the zero-modes of $b$ are in the kernel of $\adj{P_1}$, it means that the quadratic differentials \eqref{bos:eq:basis-adj-P1} also provide a suitable basis:
	\[
		b = b_0 + b',
		\qquad
		b_0 = b_{0i} \phi_i,
	\]
	where the $b_{0i}$ are Grassmann-odd coefficients.
	The first step is to find the Jacobian for the changes of variables $b \to (b', b_{0i})$:
	\[
		1
			= \int \dd_{\hat g} b \, \e^{- \abs{b}_{\hat g}^2}
			= J \int \dd_{\hat g} b' \prod_i \dd b_{0i} \, \e^{- \abs{b'}_{\hat g}^2 - \abs{b_{0i} \phi_i}^2}
			= J \sqrt{\det \psp{\phi_i}{\phi_j}}.
	\]
	Next, \eqref{bos:eq:path-int-ghosts} has no zero-modes, so one must insert $\M_g$ of them at arbitrary positions $\sigma_j^0$ to get a non-vanishing result when integrating over $\dd^{\M_g} b_{0i}$.
	The result of the integral is:
	\[
		\int \dd^{\M_g} b_{0i} \,
				\prod_{j} b_0(\sigma_j^0)
			= \int \dd^{\M_g} b_{0i} \,
				\prod_{j} \big[ b_{0i} \phi_i(\sigma_j^0) \big]
			= \det \phi_i(\sigma_j^0).
	\]
	The only combination of the $\phi_i$ which does not vanish is the determinant due to the anti-symmetry of the Grassmann numbers.
	Combining both results leads to:
	\begin{equation}
		\frac{\dd_{\hat g} b'}{\sqrt{\det \psp{\phi_i}{\phi_j}_{\hat g}}}
			= \frac{\dd_{\hat g} b}{\det \phi_i(\sigma_j^0)}
				\prod_{j=1}^{\M_g} b(\sigma_j^0).
	\end{equation}
	The locations positions $\sigma_j^0$ are arbitrary (in particular, the RHS does not depend on them since the LHS does not either).
	Note that more details are provided in \Cref{sec:form:path-integrals:zero-mode}.

	An even simpler result can be obtained by combining the previous formula with the factor $\det \psp{\phi_i}{\hat\mu_j}_{\hat g}$:
	\begin{equation}
		\dd_{\hat g} b' \, \frac{\det \psp{\phi_i}{\hat\mu_j}_{\hat g}}{\sqrt{\det \psp{\phi_i}{\phi_j}_{\hat g}}}
			= \dd_{\hat g} b \,
				\prod_{j=1}^{\M_g} \psp{b}{\hat\mu_j}_{\hat g}.
	\end{equation}
	This follows from
	\begin{gather*}
		\prod_{j=1}^{\M_g} b(\sigma_j^0)
			= \prod_{j=1}^{\M_g} \big[ b_{0i} \phi_i(\sigma_j^0) \big]
			= \det \phi_i(\sigma_j^0)
				\prod_{j=1}^{\M_g} b_{0i},
		\\
		\det \psp{\phi_i}{\hat\mu_j}_{\hat g}
				\prod_{j=1}^{\M_g} b_{0i}
			= \prod_{j=1}^{\M_g} \big[ b_{0i} \psp{\phi_i}{\hat\mu_j}_{\hat g} \big]
			= \prod_{j=1}^{\M_g} \psp{b_{0i} \phi_i}{\hat\mu_j}_{\hat g}
			= \prod_{j=1}^{\M_g} \psp{b}{\hat\mu_j}_{\hat g}.
	\end{gather*}
	Note that the previous manipulations are slightly formal: the symmetric traceless fields $b_{ab}$ and $\phi_{i,ab}$ carry indices and there should be a product over the (two) independent components.
	This is a trivial extension and would just make the notations heavier.
\end{computation}

\index{string amplitude!gv@$g$-loop vacuum (-)}%
Similar manipulations lead to a new expression which includes also the $c$ zero-mode (but which is not very illuminating):
\begin{equation}
	\label{bos:eq:path-int-ghosts-zero-bc}
	\begin{multlined}
	Z_g = \int_{\mc M_g} \!\! \dd^{\M_g} t \,
		\frac{\Omega_{\text{ckv}}[\hat g]^{-1}}{\det \psi_i(\sigma_j^0)}
		\int \dd_{\hat g} \Psi \, \dd_{\hat g} b \, \dd_{\hat g} c \,
		\prod_{j=1}^{\K_g^c} \frac{\epsilon_{ab}}{2} c^a(\sigma_j^0) c^b(\sigma_j^0)
		\\
		\times
		\prod_{i=1}^{\M_g} \psp{\hat\mu_i}{b}_{\hat g} \,
		\e^{- S_m[\hat g, \Psi] - S_{\text{gh}}[\hat g, b, c]}.
	\end{multlined}
\end{equation}
The $\sigma^{0a}_j$ are $\K_g^c = \K_g / 2$ fixed positions and the integral does not depend on their values.
Note that only $\K_g^c$ positions are needed because the coordinate is $2$-dimensional: fixing $3$ points with $2$ components correctly gives $6$ constraints.
Then, $\psi_i(\sigma^{0a}_j)$ is a $6$-dimensional matrix, with the rows indexed by $i$ and the columns by the pair $(a, j)$.

The expression cannot be simplified further because the CKV factor is infinite for $g = 0$.
This is connected to a fact mentioned previously: there is a remaining gauge symmetry which is not taken into account
\begin{equation}
	c \longrightarrow c + c_0,
	\qquad
	P_1 c_0 = 0.
\end{equation}
A proper account requires to gauge fix this symmetry: the simplest possibility is to insert three or more vertex operators -- this topic is discussed in \Cref{bos:sec:ws-int:amp}.

Finally, note that the same question arises for the $b$-ghost since one has the symmetry
\begin{equation}
	b \longrightarrow b + b_0,
	\qquad
	\adj{P_1} b_0 = 0.
\end{equation}
That there is no problem in this case is related to the presence of the moduli.

\section{Normalization}

\begin{figure}[ht]
	\centering
	\includegraphics[scale=1.4]{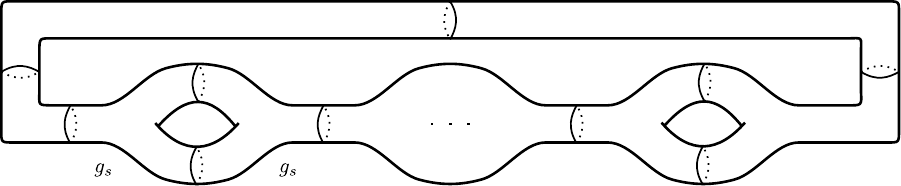}
	\caption{$g$-loop partition function.}
	\label{bos:fig:Sg0-partition}
\end{figure}

\index{string amplitude!normalization}%
In the previous sections, the closed string coupling constant $g_s$ did not appear in the expressions.
Loops in vacuum amplitudes are generated by splitting of closed strings.
By inspecting the amplitudes, it seems that there are $2 g$ such splittings (\Cref{bos:fig:Sg0-partition}), which would lead to a factor $g_s^{2g}$.
However, this is not quite correct: this result holds for a $2$-point function.
Gluing the two external legs to get a partition function (that is, taking the trace) leads to an additional factor $g_s^{-2}$ (to be determined later), such that the overall factor is $g_s^{2g - 2}$.
The fact that it is the appropriate power of the coupling constant can be more easily understood by considering $n$-point amplitudes (\Cref{bos:sec:ws-int:amp}).
The normalization of the path integral can be completely fixed by unitarity~\cite{Polchinski:2005:StringTheory-1}.

The above factor has a nice geometrical interpretation.
Defining
\begin{equation}
	\label{bos:eq:gs-dilaton}
	\Phi_0 = \ln g_s
\end{equation}
and remembering the expression \eqref{bos:eq:chi-g} of the Euler characteristics $\chi_g = 2 - 2 g$, the coupling factor can be rewritten as
\index{worldsheet!action!Einstein--Hilbert (-)}%
\begin{equation}
	g_s^{2 g - 2}
		= \e^{- \Phi_0 \chi_g}
		= \exp \left( - \frac{\Phi_0}{4\pi} \int \dd^2 \sigma \sqrt{g} R \right)
		= \e^{- \Phi_0 S_{\text{EH}}[g]},
\end{equation}
where $S_{\text{EH}}$ is the Einstein--Hilbert action.
This action is topological in two dimensions.
Hence, the coupling constant can be inserted in the path integral simply by shifting the action by the above term.
This shows that string theory on a flat target spacetime is completely equivalent to matter minimally coupled to Einstein--Hilbert gravity with a cosmological constant (tuned to impose Weyl invariance at the quantum level).
The advantage of describing the coupling power in this fashion is that it directly generalizes to scattering amplitudes and to open strings.
The parameter $\Phi_0$ is interpreted as the expectation value of the dilaton.
Replacing it by a general field $\Phi(X^\mu)$ is a generalization of the matter non-linear sigma model, but this topic is beyond the scope of this \revname{}.

\section{Summary}

In this chapter, we started with a fairly general matter CFT -- containing at least $D$ scalar fields $X^\mu$ -- and explained under which condition it describes a string theory.
The most important consequence is that the matter $2d$ QFT must in fact be a $2d$ CFT.
We then continued by describing how to gauge fix the integration over the surfaces and we identified the remaining degrees of freedom -- the moduli space $\mc M_g$ -- up to some residual redundancy -- the conformal Killing vector (CKV).
Then, we showed how to rewrite the result in terms of ghosts and proved that they are also a CFT.
This means that a string theory can be completely described by two decoupled CFTs: a universal ghost CFT and a theory-dependent matter CFT describing the string spacetime embedding and the internal structure.
The advantage is that one can forget the path integral formalism altogether and employ only CFT techniques to perform the computations.
This point of view will be developed for off-shell amplitudes (\Cref{bos:chap:offshell}) in order to provide an alternative description of how to build amplitudes.
It is particularly fruitful because one can also consider matter CFTs which do not have a Lagrangian description.
In the next chapter, we describe scattering amplitudes.

\refchapter

\ifbook
Numerous books have been published on the worldsheet string theory.
Useful (but not required) complements to this chapter and subsequent ones are~\cite{Zwiebach:2009:FirstCourseString, Lawrie:2012:UnifiedGrandTour} for introductory texts and~\cite{Blumenhagen:2014:BasicConceptsString, Polchinski:2005:StringTheory-1, Kiritsis:2007:StringTheoryNutshell, Deligne:1999:QuantumFieldsStrings1, Deligne:1999:QuantumFieldsStrings2} for more advanced aspects.
\fi

\begin{itemize}
	\item The definition of a field measure from a Gaussian integral and manipulations thereof can be found in~\cites[sec.~15.1, 22.1]{Hatfield:1998:QuantumFieldTheory}[chap.~14]{Nakahara:2003:GeometryTopologyPhysics}{Polchinski:1986:EvaluationOneLoop}{DHoker:1988:GeometryStringPerturbation}.

	\item The most complete explanations of the gauge fixing procedure are~\cites[sec.~15.1, 22.1]{Hatfield:1998:QuantumFieldTheory}[sec.~3.4, 6.2]{Blumenhagen:2014:BasicConceptsString}[chap.~5]{Polchinski:2005:StringTheory-1}{Deligne:1999:QuantumFieldsStrings2}[chap.~5]{Kaku:1999:IntroductionSuperstringsMTheory}.
	The original derivation can be found in~\cite{Moore:1986:MeasureModuliPolyakov, DHoker:1986:MultiloopAmplitudesBosonic}.

	\item For the geometry of the moduli space, see~\cite{Nelson:1987:LecturesStringsModuli, Nakahara:2003:GeometryTopologyPhysics}.

	\item Ultralocality and its consequences are described in~\cite{Polchinski:1986:EvaluationOneLoop, DHoker:1988:GeometryStringPerturbation} (see also~\cite[sec.~2.4]{Hamber:2009:QuantumGravitationFeynman}).

	\item The use of a Weyl ghost is shown in~\cites[sec.~8]{tHooft:2004:IntroductionStringTheory}[sec.~9.2]{Wray:2011:IntroductionStringTheory}.

\end{itemize}

\chapter{Worldsheet path integral: scattering amplitudes}
\label{chap:bos:ws-int-amp}

\introchapter

In this chapter, we generalize the worldsheet path integral to compute scattering amplitudes, which corresponds to insert vertex operators.
The gauge fixing from the previous chapter is generalized to this case.
In particular, we discuss the $2$-point amplitude on the sphere.
Finally, we introduce the BRST symmetry and motivate some properties of the BRST quantization, which will be performed in details later.
The formulas in this chapter are all covariant: they will be rewritten in complex coordinates in the next chapter.

\section{Scattering amplitudes on moduli space}
\label{bos:sec:ws-int:amp}

In this section, we describe the scattering of $n$ strings.
The momentum representation is more natural for describing interactions, especially in string theory.
Therefore, each string is characterized by a state $V_{\alpha_i}(k_i)$ with momentum $k_i$ and some additional quantum numbers $\alpha_i$ ($i = 1, \ldots, n)$.
We start from the worldsheet path integral \eqref{bos:eq:path-int-dg}
before gauge fixing:
\begin{equation}
	Z_g
		= \int \frac{\dd_g g_{ab}}{\Omega_{\text{gauge}}[g]} \, Z_m[g],
	\qquad
	Z_m[g]
		= \int \dd_g \Psi \, \e^{- S_m[g, \Psi]}.
\end{equation}

\subsection{Vertex operators and path integral}
\label{bos:sec:ws-int-amp:Mg:vertex}

The external states are represented by infinite semi-tubes attached to the surfaces.
\index{Riemann surface!puncture}%
Under a conformal mapping, the tubes can be mapped to points called \emph{punctures} on the worldsheet.
At $g$ loops, the resulting space is a Riemann surface $\Sigma_{g,n}$ of genus $g$ with $n$ punctures (or marked points).
\index{vertex operator!integrated}%
The external states are represented by \emph{integrated vertex operators}
\begin{equation}
	\label{bos:eq:vertex-integrated}
	V_{\alpha}(k_i)
		:= \int \dd^2 \sigma \sqrt{g(\sigma)} \, V_{\alpha}(k; \sigma).
\end{equation}
The vertex operators $V_{\alpha}(k; \sigma)$ are built from the matter CFT operators and from the worldsheet metric $g_{ab}$.
The functional dependence is omitted to not overload the notation, but one should read $V_{\alpha}(k; \sigma) := V_{\alpha}[g, \Psi](k; \sigma)$.
The integration over the state positions is necessary because the mapping of the tube to a point is arbitrary.
Another viewpoint is that it is needed to obtain an expression invariant under worldsheet diffeomorphisms.
The vertex operators described general states which not necessarily on-shell: this restriction will be found later when discussing the BRST invariance of scattering amplitudes (\Cref{bos:sec:ws-int:brst:states}).

\index{string amplitude!normalization}%
Following \Cref{bos:sec:ws-int:faddeev-popov:conclusion}, the Einstein--Hilbert action with boundary term
\begin{equation}
	S_{\text{EH}}[g]
		:= \frac{1}{4\pi} \int \dd^2 \sigma \sqrt{g}\, R
			+ \frac{1}{2\pi} \oint \dd s \, k
		= \chi_{g,n}.
\end{equation}
is inserted in the path integral equals the Euler characteristics $\chi_{g,n}$ (the $g$ in $\chi_{g,n}$ denotes the genus).
\index{Euler characteristics}%
On a surface with punctures, the latter is shifted by the number of punctures (which are equivalent to boundaries or disks) with respect to \eqref{bos:eq:chi-g}:
\begin{equation}
	\label{bos:eq:chi-gn}
	\chi_{g,n}
		:= \chi(\Sigma_{g,n})
		= 2 - 2 g - n.
\end{equation}
This gives the normalization factor:
\begin{equation}
	g_s^{- \chi_{g,n}}
		= \e^{- \Phi_0 S_{\text{EH}}[g]},
	\qquad
	\Phi_0
		:= \ln g_s.
\end{equation}
The correctness factor can be verified by inspection of the Riemann surface for the scattering of $n$ string at $g$ loops.
In particular, the string coupling constant is by definition the interaction strength for the scattering of $3$ strings at tree-level.
Moreover, the tree-level $2$-point amplitude contains no interaction and should have no power of $g_s$.
This factor can also be obtained by unitarity~\cite{Polchinski:2005:StringTheory-1}.

\index{string amplitudegn@!$g$-loop $n$-point (-)}%
By inserting these factors in \eqref{bos:eq:path-int-dg}, the $g$-loop $n$-point scattering amplitude is described by:
\begin{equation}
	\label{bos:eq:amp-dg}
	A_{g,n}(\{ k_i \})_{\{ \alpha_i \}}
		:= \int \frac{\dd_g g_{ab}}{\Omega_{\text{gauge}}[g]} \,
			\dd_g \Psi \, \e^{- S_m[g, \Psi] - \Phi_0 S_{\text{EH}}[g]}
			\prod_{i=1}^n \left(
				\int \dd^2 \sigma_i \sqrt{g(\sigma_i)} \,
				V_{\alpha_i}(k_i; \sigma_i)
				\right).
\end{equation}
The $\sigma_i$ dependence of each $\sqrt{g}$ will be omitted from now on since no confusion is possible.
The following equivalent notations will be used:
\begin{equation}
	A_{g,n}(\{ k_i \})_{\{ \alpha_i \}}
		:= A_{g,n}(k_1, \ldots, k_n)_{\alpha_1, \ldots, \alpha_n}
		:= A_{g,n}\big(V_{\alpha_1}(k_1), \ldots, V_{\alpha_n}(k_n) \big).
\end{equation}
The complete (perturbative) amplitude is found by summing over all genus:
\begin{equation}
	A_{n}(k_1, \ldots, k_n)_{\alpha_1, \ldots, \alpha_n}
		= \sum_{g=0}^\infty
			A_{g,n}(k_1, \ldots, k_n)_{\alpha_1, \ldots, \alpha_n}.
\end{equation}
We omit a genus-dependent normalization which can be determined from unitarity~\cite{Polchinski:2005:StringTheory-1}.
Sometimes, it is convenient to extract the factor $\e^{- \Phi_0 \chi_{g,n}}$ of the amplitude $A_{g,n}$ to display explicitly the genus expansion, but we will not follow this convention here.
Since each term of the sum scales as $A_{g,n} \propto g_s^{2 g + n - 2}$, this expression clearly shows that worldsheet amplitudes are perturbative by definition: this motivates the construction of a string field theory from which the full non-perturbative $S$-matrix can theoretically be computed.

Finally, the amplitude \eqref{bos:eq:amp-dg} can be rewritten in terms of correlation functions of the matter QFT integrated over worldsheet metrics:
\begin{equation}
	\label{bos:eq:amp-dg-corr}
	A_{g,n}(\{ k_i \})_{\{ \alpha_i \}}
		= \int \frac{\dd_g g_{ab}}{\Omega_{\text{gauge}}[g]} \,
			\e^{- \Phi_0 S_{\text{EH}}[g]}
			\int \prod_{i=1}^n \dd^2 \sigma_i \sqrt{g}
			\Mean{\prod_{i=1}^n V_{\alpha_i}(k_i; \sigma_i)}_{\mathrlap{m, g}}.
\end{equation}
The correlation function plays the same role as the partition function in \eqref{bos:eq:path-int-dg}.
This shows that string expressions are integrals of CFT expressions over the space of worldsheet metrics (to be reduced to the moduli space).

\index{string amplitude!matching QFT|(}%
\index{scattering amplitude|(}%
\index{Green function|(}%
We address a last question before performing the gauge fixing: what does \eqref{bos:eq:amp-dg} computes exactly: on-shell or off-shell? Green functions or amplitudes? if amplitudes, the $S$-matrix or just the interacting part $T$ (amputated Green functions)?
The first point is that a path integral over connected worldsheets will compute connected processes.
We will prove later, when discussing the BRST quantization, that string states must be on-shell (\Cref{bos:sec:ws-int:brst:states,bos:sec:ws-int:brst}) and that it corresponds to setting the Hamiltonian \eqref{bos:eq:worldsheet-momentum} to zero:
\index{on-shell condition}%
\begin{equation}
	H = 0.
\end{equation}
From this fact, it follows that \eqref{bos:eq:path-int-dg} must compute amplitudes since non-amputated Green functions diverge on-shell (due to external propagators).
Finally, the question of whether it computes the $S$-matrix $S = 1 + \I T$, or just the interacting part $T$ is subtler.
At tree-level, they agree for $n \ge 3$, while $T = 0$ for $n = 2$ and $S$ reduces to the identity.
This difficulty (discussed further in \Cref{bos:sec:ws-int:amp:gauge-fixing}) is thus related to the question of gauge-fixing tree-level $2$-point amplitude (\Cref{bos:sec:ws-int:amp:gauge-fixing-2pt}).
It has long been believed that \eqref{bos:eq:path-int-dg} computes only the interacting part (amputated Green functions), but it has been understood recently that this is not correct and that \eqref{bos:eq:path-int-dg} computes the $S$-matrix.

\index{string amplitude!matching QFT|)}%

\begin{remark}[Scattering amplitudes in QFT]
	\label{bos:rk:summary-qft}
	Remember that the $S$-matrix is separated as:
	\begin{equation}
		\label{eq:S-1+T}
		S
			= 1 + \I T,
	\end{equation}
	where $1$ denotes the contribution where all particles propagate without interaction.
	The connected components of $S$ and $T$ are denoted by $S^c$ and $T^c$.
	The $n$-point (connected) scattering amplitudes $A_n$ for $n \ge 3$ can be computed from the Green functions $G_n$ through the LSZ prescription (amputation of the external propagators):
	\begin{equation}
		A_n(k_1, \ldots, k_n)
			= G_n(k_1, \ldots, k_n) \prod_{i=1}^n (k_i^2 + m_i^2).
	\end{equation}
	The path integral computes the Green functions $G_n$; perturbatively, they are obtained from the Feynman rules.
	They include a $D$-dimensional delta function
	\begin{equation}
		G_n(k_1, \ldots, k_n)
			\propto \delta^{(D)}(k_1 + \cdots + k_n).
	\end{equation}

	\index{scattering amplitude!tree-level 2-point (-)}%
	The $2$-point amputated Green function $T_2$ computed from the LSZ prescription vanishes on-shell.
	For example, considering a scalar field at tree-level, one finds:
	\begin{equation}
		T_2
			= G_2(k, k') \, (k^2 + m^2)^2
			\sim (k^2 + m^2) \, \delta^{(D)}(k + k')
			\xrightarrow[k^2 \to - m^2]{} 0
	\end{equation}
	since
	\begin{equation}
		G_2(k, k')
			= \frac{\delta^{(D)}(k + k')}{k^2 + m^2}.
	\end{equation}
	Hence, $T_2 = 0$ and the $S$-matrix \eqref{eq:S-1+T} reduces to the identity component $S^c_2 = 1_2$ (which is a connected process).
	There are several way to understand this result:
	\begin{enumerate}
		\item The recursive definition of the connected $S$-matrix $S^c$ from the cluster decomposition principle requires a non-vanishing $2$-point amplitude~\cites[sec.~5.1.5]{Itzykson:2006:QuantumFieldTheory}[sec.~4.3]{Weinberg:2005:QuantumTheoryFields-2}[sec.~6.1]{Duncan:2012:ConceptualFrameworkQuantum}.

		\item The $2$-point amplitude corresponds to the normalization of the $1$-particle states (overlap of a particle state with itself, which is non-trivial)~\cites[eq.~4.1.4]{Weinberg:2005:QuantumTheoryFields-1}[chap.~5]{Srednicki:2007:QuantumFieldTheory}.

		\item A single particle in the far past propagating to the far future without interacting is a connected and physical process~\cite[p.~133]{Duncan:2012:ConceptualFrameworkQuantum}.

		\item It is required by the unitarity of the $2$-point amplitude~\cite{Erbin:2019:TwoPointStringAmplitudes}.
	\end{enumerate}
	These points indicate that the $2$-point amplitude is proportional to the identity in the momentum representation~\cites[p.~212]{Itzykson:2006:QuantumFieldTheory}[eq.~4.3.3 and 4.1.5]{Weinberg:2005:QuantumTheoryFields-1}
	\begin{equation}
		\label{qft:2pt-amp}
		A_2(k, k')
			= 2 k^0 \, (2\pi)^{D-1} \delta^{(D-1)}(\vec k - \vec k').
	\end{equation}
	The absence of interactions implies that the spatial momentum does not change (the on-shell condition implies that the energy is also conserved).
	This relation is consistent with the commutation relation of the operators with the Lorentz invariant measure\footnotemark{}
	\footnotetext{%
		If the modes are defined as $\tilde a(k) = a(k) / \sqrt{2 k^0}$ such that $\com{\tilde a(\vec k)}{\adj{\tilde a}(\vec k')} = (2\pi)^{D-1} \delta^{(D-1)}(\vec k - \vec k')$, then one finds $\tilde A_2(k, k') = (2\pi)^{D-1} \delta^{(D-1)}(\vec k - \vec k')$.
	}%
	\begin{equation}
		\com{a(\vec k)}{\adj{a}(\vec k')}
			= 2 k^0 \, (2\pi)^{D-1} \delta^{(D-1)}(\vec k - \vec k').
	\end{equation}
	That this holds for all particles at all loops can be proven using the Källen--Lehman representation~\cite[p.~212]{Itzykson:2006:QuantumFieldTheory}.

	On the other hand, the identity part in \eqref{eq:S-1+T} is absent for $n \ge 3$ for connected amplitudes: $S^c_n = T^c_n$ for $n \ge 3$.
	This shows that the Feynman rules and the LSZ prescription compute only the interacting part $T$ of the on-shell scattering amplitudes.
	The reason is that the derivation of the LSZ formula assumes that the incoming and outgoing states have no overlap, which is not the case for the $2$-point function.
	A complete derivation of the $S$-matrix from the path integral is more involved~\cites[sec.~5.1.5]{Itzykson:2006:QuantumFieldTheory}[sec.~6.7]{ZinnJustin:2002:QuantumFieldTheory}{Faddeev:1975:IntroductionFunctionalMethods} (see also~\cite{Collins:2019:NewApproachLSZ}).
	The main idea is to consider a superposition of momentum states (here, in the holomorphic representation~\cite[sec.~5.1, 6.4]{ZinnJustin:2002:QuantumFieldTheory})
	\begin{equation}
		\phi(\alpha)
			= \int \dd^{D-1} \vec k \, \conj{\alpha(\vec k)} \adj a(\vec k).
	\end{equation}
	They contribute a quadratic piece to the connected $S$-matrix and, setting them to delta functions, one recovers the above result.
\end{remark}

\index{scattering amplitude|)}%
\index{Green function|)}%

\subsection{Gauge fixing: general case}
\label{bos:sec:ws-int:amp:gauge-fixing}

\index{Polyakov path integral!Faddeev--Popov gauge fixing}%
\index{vertex operator!Weyl invariance}%
The Faddeev--Popov gauge fixing of the worldsheet diffeomorphisms and Weyl rescaling \eqref{bos:eq:sym-weyl-diffeo} goes through also in this case if the integrated vertex operators are diffeomorphism and Weyl invariant:
\begin{subequations}
\begin{align}
	\label{bos:eq:vertex-op-diffeo-inv}
	\delta_\xi V_{\alpha_i}(k_i)
		&= \delta_\xi \int \dd^2 \sigma \sqrt{g} \, V_{\alpha_i}(k_i; \sigma)
		= 0,
	\\
	\label{bos:eq:vertex-op-weyl-inv}
	\delta_\omega V_{\alpha_i}(k_i)
		&= \delta_\omega \int \dd^2 \sigma \sqrt{g} \, V_{\alpha_i}(k_i; \sigma)
		= 0,
\end{align}
\end{subequations}
with the variations defined in \eqref{bos:eq:sym-diffeo-inf} and \eqref{bos:eq:sym-weyl-inf}.
Diffeomorphism invariance is straightforward if the states are integrated worldsheet scalars.
However, if the states are classically Weyl invariant, they are not necessary so at the quantum level: vertex operators are composite operators, which need to be renormalized to be well-defined at the quantum level.
Renormalization introduces a scale which breaks Weyl invariance.
Enforcing it to be a symmetry of the vertex operators leads to constraints on the latter.
We will not enter in the details since it depends on the matter CFT and we will assume that the operators $V_{\alpha_i}(k_i)$ are indeed Weyl invariant (see~\cite[sec.~3.6]{Polchinski:2005:StringTheory-1} for more details).
In the rest of this book, we will use CFT techniques developed in \Cref{chap:cft:plane}.
The Einstein--Hilbert action is clearly invariant under both symmetries since it is a topological quantity.

Following the computations from \Cref{bos:sec:ws-int:faddeev-popov} leads to a generalization of \eqref{bos:eq:path-int-fp} with the vertex operators inserted for the amplitude \eqref{bos:eq:amp-dg}:
\begin{equation}
	\label{bos:eq:amp-fp}
	\begin{aligned}
	A_{g,n}(\{ k_i \})_{\{ \alpha_i \}}
		= g_s^{- \chi_{g,n}}
			\int_{\mc M_g} \!\! \dd^{\M_{g}} t \,
			\frac{\det \psp{\phi_i}{\hat\mu_j}_{\hat g}}{\sqrt{\det \psp{\phi_i}{\phi_j}_{\hat g}}} \,
			&
			\frac{\Omega_{\text{ckv}}[\hat g]^{-1}}{\sqrt{\det \psp{\psi_i}{\psi_j}_{\hat g}}}
			\\
			&
			\times
			\int \prod_{i=1}^n \dd^2 \sigma_i \sqrt{\hat g}
			\Mean{\prod_{i=1}^n \hat V_{\alpha_i}(k_i; \sigma_i)}_{\mathrlap{m, \hat g}}.
	\end{aligned}
\end{equation}
The hat on the vertex operators indicates that they are evaluated in the background metric $\hat g$.

The next step is to introduce the ghosts: following \Cref{bos:sec:ws-int:ghosts}, the generalization of \eqref{bos:eq:path-int-ghosts-zero-b} is
\begin{equation}
	\label{bos:eq:amp-ghosts-b}
	\begin{aligned}
	A_{g,n}(\{ k_i \})_{\{ \alpha_i \}}
		= g_s^{- \chi_{g,n}}
			\int_{\mc M_g} \!\! \dd^{\M_{g}} t \,
			&
			\frac{\Omega_{\text{ckv}}[\hat g]^{-1}}{\sqrt{\det \psp{\psi_i}{\psi_j}_{\hat g}}}
			\int \dd_{\hat g} b \, \dd'_{\hat g} c \,
			\prod_{i=1}^{\M_{g}} \psp{b}{\hat\mu_i}_{\hat g} \,
			\e^{- S_{\text{gh}}[\hat g, b, c]}
			\\
			&
			\times
			\int \prod_{i=1}^n \dd^2 \sigma_i \sqrt{\hat g}
			\Mean{\prod_{i=1}^n \hat V_{\alpha_i}(k_i; \sigma_i)}_{\mathrlap{m, \hat g}}.
	\end{aligned}
\end{equation}
For the moment, only the $b$ ghosts come with zero-modes.
Then, $c$ zero-modes can be introduced in \eqref{bos:eq:amp-ghosts-b}
\begin{equation}
	\label{bos:eq:amp-ghosts-bc-gen}
	\begin{multlined}
	A_{g,n}
		= g_s^{- \chi_{g,n}}
			\int_{\mc M_g} \!\! \dd^{\M_{g}} t \,
			\frac{\Omega_{\text{ckv}}[\hat g]^{-1}}{\det \psi_i(\sigma_j^0)}
			\int \dd_{\hat g} b \, \dd_{\hat g} c \,
			\prod_{j=1}^{\K_g^c} \frac{\epsilon_{ab}}{2} c^a(\sigma_j^0) c^b(\sigma_j^0)
			\prod_{i=1}^{\M_{g}} \psp{\hat\mu_i}{b}_{\hat g} \,
			\e^{- S_{\text{gh}}[\hat g, b, c]}
			\\
			\times
			\int \; \prod_{i=1}^n \, \dd^2 \sigma_i \sqrt{\hat g}
			\Mean{
				\prod_{i=1}^n \hat V_{\alpha_i}(k_i; \sigma_i)
				}_{\mathrlap{m, \hat g}},
	\end{multlined}
\end{equation}
by following the same derivation as \eqref{bos:eq:path-int-ghosts-zero-bc}.
The formulas \eqref{bos:eq:amp-ghosts-b} and \eqref{bos:eq:amp-ghosts-bc-gen} are the correct starting point for all $g$ and $n$.
In particular, the $c$ ghosts are not paired with any vertex (a condition often assumed or presented as mandatory).
This fact will help resolve some difficulties for the $2$-point function on the sphere.

\index{conformal Killing!vector}%
\index{string amplitude!CKV gauge fixing}%
Remember that there is no CKV and no $c$ zero-mode for $g \ge 2$.
For the sphere $g = 0$ and the torus $g = 1$, there are CKVs, indicating that there is a residual symmetry in \eqref{bos:eq:amp-ghosts-b} and \eqref{bos:eq:amp-ghosts-bc-gen}, which is the global conformal group of the worldsheet.
It can be gauge fixed by imposing conditions on the vertex operators.\footnotemark{}
\footnotetext{%
	In fact, it is only important to gauge fix for the sphere because the volume of the group is infinite.
	On the other hand, the volume of the CKV group for the torus is finite-dimensional such that dividing by $\Omega_{\text{ckv}}$ is not ambiguous.
}%
The simplest gauge fixing condition amounts to fix the positions of $\K_g^c$ vertex operators through the Faddeev--Popov trick:
\begin{equation}
	\label{bos:eq:fp-positions}
	1
		= \Delta(\sigma_j^0) \int \dd \xi \,
			\prod_{j=1}^{\K_g^c}
			\, \delta^{(2)}( \sigma_j - \sigma^{0(\xi)}_j ),
	\qquad
	\sigma^{0(\xi)}_j
		= \sigma_j^0 + \delta_\xi \sigma_j^0,
	\qquad
	\delta_\xi \sigma_j^0 = \xi(\sigma_j^0),
\end{equation}
where $\xi$ is a conformal Killing vector, and the variation of $\sigma$ was given in \eqref{bos:eq:sym-diffeo-inf}.
We find that
\begin{equation}
	\label{bos:eq:fp-delta}
	\Delta(\sigma_j^0) = \det \psi_i(\sigma_j^0).
\end{equation}
A priori, the positions $\sigma_j^0$ are not the same as the one appearing in \eqref{bos:eq:path-int-ghosts-zero-bc} (since both sets are arbitrary): however, considering the same positions allows to cancel the factor \eqref{bos:eq:fp-delta} with the same one in \eqref{bos:eq:path-int-ghosts-zero-bc}.

\begin{computation}[bos:eq:fp-delta]
	The first step is to compute $\Delta$ in \eqref{bos:eq:fp-positions}.
	For this, we decompose the CKV $\xi$ on the basis \eqref{bos:eq:basis-P1}
	\[
		\xi(\sigma_j^0) = \alpha_i \psi_i(\sigma_j^0)
	\]
	and write the Gaussian integral:
	\begin{align*}
		1
			&= \int \prod_{j=1}^{\K_g^c} \dd^2 \delta\sigma_j \,
				\e^{- \sum_j \psp{\delta\sigma_j}{\delta\sigma_j}}
			= \Delta \int \prod_{j=1}^{\K_g} \dd \alpha_i \,
				\e^{- \sum_{j,i,i'} \psp{\alpha_i \psi_i(\sigma_j)}{\alpha_{i'} \psi_{i'}(\sigma_j)}}
			\\
			&= \Delta \big(\det \psi_i(\sigma_j) \big)^{-1}.
	\end{align*}
	Again, we have reduced rigour in order to simplify the manipulations.
\end{computation}

After inserting the identity \eqref{bos:eq:fp-positions} into \eqref{bos:eq:amp-ghosts-bc-gen}, one can integrate over $\K_g^c$ vertex operator positions to remove the delta functions -- at the condition that there are at least $\K_g^c$ operators.
As a consequence, we learn that the proposed gauge fixing works only for $n \ge 1$ if $g = 1$ or $n \ge 3$ if $g = 0$.
This condition is equivalent to
\begin{equation}
	\chi_{g,n}
		= 2 - 2g - n
		< 0.
\end{equation}
\index{string amplitudegn@!$g$-loop $n$-point (-)}%
In this case, the factors $\det \psi_i(\sigma_j^0)$ cancel and \eqref{bos:eq:amp-ghosts-b} becomes
\begin{equation}
	\label{bos:eq:amp-ghosts-bc}
	\begin{multlined}
	A_{g,n}(\{ k_i \})_{\{ \alpha_i \}}
		= g_s^{- \chi_{g,n}}
			\int_{\mc M_g} \!\! \dd^{\M_{g}} t \,
		\int \dd_{\hat g} b \, \dd_{\hat g} c \,
		\prod_{j=1}^{\K_g^c} \frac{\epsilon_{ab}}{2} c^a(\sigma_j^0) c^b(\sigma_j^0)
		\prod_{i=1}^{\M_{g}} \psp{\hat\mu_i}{b}_{\hat g} \,
		\e^{- S_{\text{gh}}[\hat g, b, c]}
		\\
		\times
		\int \; \prod_{\mathclap{i=\K_g^c+1}}^n \, \dd^2 \sigma_i \sqrt{\hat g}
		\Mean{
			\prod_{j=1}^{\K_g^c} \hat V_{\alpha_j}(k_j; \sigma_j^0)
			\prod_{i=\K_g^c+1}^n \hat V_{\alpha_i}(k_i; \sigma_i)
			}_{\mathrlap{m, \hat g}}.
	\end{multlined}
\end{equation}
The result may be divided by a symmetry factor if the delta functions have solutions for several points~\cite[sec.~5.3]{Polchinski:2005:StringTheory-1}.
Performing the gauge fixing for the other cases (in particular, $g = 0, n = 2$ and $g = 1, n = 0$) is more subtle (\Cref{bos:sec:ws-int:amp:gauge-fixing-2pt} and~\cite{Polchinski:2005:StringTheory-1}).

\index{string amplitudegn@!$g$-loop $n$-point (-)}%
The amplitude can be rewritten in two different ways.
First, the ghost insertions can be rewritten in terms of a ghost correlation functions
\begin{equation}
	\label{bos:eq:amp-ghosts-bc-corr}
	\begin{aligned}
	A_{g,n}(\{ k_i \})_{\{ \alpha_i \}}
		= g_s^{- \chi_{g,n}}
			\int_{\mc M_g} \!\! \dd^{\M_{g}} t \,
			\int \; \prod_{\mathclap{i=\K_g^c+1}}^n \, \dd^2 \sigma_i \sqrt{\hat g}
			&
			\Mean{
				\prod_{j=1}^{\K_g^c} \frac{\epsilon_{ab}}{2} c^a(\sigma_j^0) c^b(\sigma_j^0)
				\prod_{i=1}^{\M_{g}} \psp{\hat\mu_i}{b}_{\hat g}
				}_{\mathrlap{\text{gh}, \hat g}}
			\\
			&
			\times
			\Mean{
				\prod_{j=1}^{\K_g^c} \hat V_{\alpha_j}(k_j; \sigma_j^0) \,
				\prod_{\mathclap{i=\K_g^c+1}}^n \hat V_{\alpha_i}(k_i; \sigma_i)
				}_{\mathrlap{m, \hat g}}.
		\end{aligned}
\end{equation}
This form is particularly interesting because it shows that, before integration over the moduli, the amplitudes factorize.
This is one of the main advantage of the conformal gauge, since the original complicated amplitude \eqref{bos:eq:amp-dg} for a QFT on a dynamical spacetime reduces to the product of two correlation functions of QFTs on a fixed curved background.
In fact, the situation is even simpler when taking a flat background $\hat g = \delta$ since both the ghost and matter sectors are CFTs and one can employ all the tools from two-dimensional CFT (\Cref{part:cft}) to perform the computations and mostly forget about the path integral origin of these formulas.
This approach is particularly fruitful for off-shell (\Cref{bos:chap:offshell}) and superstring amplitudes (\Cref{part:superstring}).

\begin{remark}[Amplitudes in $2d$ gravity]
	The derivation of amplitudes for $2d$ gravity follows the same procedure, up to two differences: 1) there is an additional decoupled (before moduli and position integrations) gravitational sector described by the Liouville field, 2) the matter and gravitational action are not CFTs if the original matter was not.
\end{remark}

\index{string amplitudegn@!$g$-loop $n$-point (-)}%
A second formula can be obtained by bringing the $c$-ghost on top of the matter vertex operators which are at the same positions
\begin{equation}
	\label{bos:eq:amp-ghosts-bc-common}
	A_{g,n}(\{ k_i \})_{\{ \alpha_i \}}
		= g_s^{- \chi_{g,n}}
			\int_{\mc M_g} \!\! \dd^{\M_{g}} t \,
			\int \; \prod_{\mathclap{i=\K_g^c+1}}^n \, \dd^2 \sigma_i \sqrt{\hat g}
			\Mean{
				\prod_{i=1}^{\M_{g}} \hat B_i
				\prod_{j=1}^{\K_g^c} \hat{\scr V}_{\alpha_j}(k_j; \sigma_j^0) \,
				\prod_{\mathclap{i=\K_g^c+1}}^n \hat V_{\alpha_i}(k_i; \sigma_i)
				}_{\mathrlap{\hat g}},
\end{equation}
and where
\begin{equation}
	\label{bos:eq:vertex-unintegrated}
	\hat{\scr V}_{\alpha_j}(k_j; \sigma_j^0)
		:= \frac{\epsilon_{ab}}{2} c^a(\sigma_j^0) c^b(\sigma_j^0) \,
			\hat{V}_{\alpha_j}(k_j; \sigma_j^0),
	\qquad
	\hat B_i
		:= \psp{\hat\mu_i}{b}_{\hat g}.
\end{equation}
\index{vertex operator!unintegrated}%
The operators $\scr V_{\alpha_i}(k_i; \sigma_j^0)$ (a priori off-shell) are called \emph{unintegrated operators}, by opposition to the integrated operators $V_{\alpha_i}(k_i)$.
We will see that both are natural elements of the BRST cohomology.

\index{string amplitudegn@!$g$-loop $n$-point (-)}%
To stress that the $\hat B_i$ insertions are really an element of the measure, it is finally possible to rewrite the previous expression as
\begin{equation}
	A_{g,n}(\{ k_i \})_{\{ \alpha_i \}}
		= g_s^{- \chi_{g,n}}
			\int_{\mc M_g \times \C^{n - \K^c_g}}
			\Mean{
				\bigwedge_{i=1}^{\M_{g}} \hat B_i \, \dd t_i
				\prod_{j=1}^{\K_g^c} \hat{\scr V}_{\alpha_i}(k_i; \sigma_j^0) \,
				\prod_{\mathclap{i=\K_g^c+1}}^n \hat V_{\alpha_i}(k_i; \sigma_i) \, \dd^2 \sigma_i \sqrt{\hat g}
				}_{\mathrlap{\hat g}}.
\end{equation}

The result \eqref{bos:eq:amp-ghosts-bc-common} suggests a last possibility for improving the expression of the amplitude.
Indeed, the different vertex operators don't appear symmetrically: some are integrated over and other come with $c$ ghosts.
Similarly, the two types of integrals have different roles: the moduli are related to geometry while the positions look like external data (vertex operators).
However, punctures can obviously be interpreted as part of the geometry, and one may wonder if it is possible to unify the moduli and positions integrals.
It is, in fact, possible to put all vertex operators and integrals on the same footing by considering the amplitude to be defined on the moduli space $\mc M_{g,n}$ of genus-$g$ Riemann surfaces with $n$ punctures instead of just $\mc M_g$~\cite{Polchinski:2005:StringTheory-1} (see also \Cref{bos:sec:offshell:amp-intro:marked-moduli}).

\subsection{Gauge fixing: 2-point amplitude}
\label{bos:sec:ws-int:amp:gauge-fixing-2pt}

\index{string theory!consistency}%
\index{string amplitude!matching QFT}%
\index{string amplitude!tree-level 2-point (-)|(}%
As discussed at the end of \Cref{bos:sec:ws-int-amp:Mg:vertex}, it has long been believed that the tree-level $2$-point amplitude vanishes.
There were two main arguments: there are not sufficiently many vertex operators 1) to fix completely the $\group{SL}(2, \C)$ invariance or 2) to saturate the number of $c$-ghost zero-modes.
Let's review both points and then explain why they are incorrect.
We will provide the simplest arguments, referring the reader to the literature~\cite{Erbin:2019:TwoPointStringAmplitudes, Seki:2020:TwopointStringAmplitudes} for more general approaches.

For simplicity, we consider the flat metric $\hat g = \delta$ and an orthonormal basis of CKV.
The two weight-$(1,1)$ matter vertex operators are denoted as $V_k(z, \bar z)$ and $V_{k'}(z', \bar z')$ such that the $2$-point correlation function on the sphere reads (see \Cref{chap:cft:plane,cft:chap:systems} for more details):
\begin{equation}
	\label{eq:2-point-corr-V}
	\Mean{V_k(z, \bar z) V_{k'}(z', \bar z')}_{S^2}
		= \frac{\I \, (2\pi)^D \delta^{(D)}(k + k')}{\abs{z - z'}^4}.
\end{equation}
The numerator comes from the zero-modes $\e^{\I (k + k') \cdot x}$ for a target spacetime with a Lorentzian signature~\cites[p.~866]{Deligne:1999:QuantumFieldsStrings2}{Polchinski:2005:StringTheory-1} (required to make use of the on-shell condition).

\subsubsection{Review of the problem}

The tree-level amplitude \eqref{bos:eq:amp-fp} for $n = 2$ reads:
\begin{equation}
	\label{eq:2-point-amp}
	A_{0,2}(k, k')
		= \frac{C_{S_2}}{\Vol \mc K_{0,0}}
			\int \dd^2 z \dd^2 z' \,
			\Mean{V_k(z, \bar z) V_{k'}(z', \bar z')}_{S^2},
\end{equation}
where $\mc K_{0,n}$ is the CKV group of $\Sigma_{0,n}$, the sphere with $n$ punctures.
In particular, the group of the sphere without puncture is $\mc K_{0,0} = \group{PSL}(2, \C)$.
The normalization of the amplitude is $C_{S_2} = 8\pi \alpha'^{-1}$ for $g_s = 1$~\cite{Weigand:2012:IntroductionStringTheory, Polchinski:2005:StringTheory-1}.
Since there are two insertions, the symmetry can be partially gauge fixed by fixing the positions of the two punctures to $z = 0$ and $z' = \infty$.
In this case, the amplitude \eqref{eq:2-point-amp} becomes:
\begin{equation}
	\label{eq:2-point-amp-fixed}
	A_{0,2}(k, k')
		= \frac{C_{S^2}}{\Vol \mc K_{0,2}} \,
			\Mean{V_k(\infty, \infty) V_{k'}(0, 0)}_{S^2},
\end{equation}
where $\mc K_{0,2} = \R_+^* \times \group{U}(1)$ is the CKV group of the $2$-punctured sphere -- containing dilatations and rotations.\footnotemark{}
\footnotetext{%
	The subgroup and the associated measure depend on the locations of the two punctures.
}%
Since the volume of this group is infinite $\Vol \mc K_{0,2} = \infty$, it looks like $A_{0,2} = 0$.
However, this forgets that the $2$-point correlation function \eqref{eq:2-point-corr-V} contains a $D$-dimensional delta function.
The on-shell condition implies that the conservation of the momentum $k + k' = 0$ is automatic for one component, such that the numerator in \eqref{eq:2-point-amp-fixed} contains a divergent factor $\delta(0)$:
\begin{equation}
	\label{eq:2-point-amp-fixed-delta}
	A_{0,2}(k, k')
		= (2\pi)^{D-1} \delta^{(D-1)}(\vec k + \vec k') \, \frac{C_{S_2} \, 2\pi \I \, \delta(0)}{\Vol \mc K_{0,2}}.
\end{equation}
Hence, \eqref{eq:2-point-amp-fixed} is of the form $A_{0,2} = \infty / \infty$ and one should be careful when evaluating it.

The second argument relies on a loophole in the understanding of the gauge fixed amplitude \eqref{bos:eq:amp-ghosts-bc-common}.
The result \eqref{bos:eq:amp-ghosts-bc-common} is often summarized by saying that one can go from \eqref{bos:eq:amp-fp} to \eqref{bos:eq:amp-ghosts-bc-common} by replacing $\K_{g}^c$ integrated vertices $\int V$ by unintegrated vertices $c \bar c V$ in order to saturate the ghost zero-modes and to obtain a non-zero result.
For $g = 0$, this requires $3$ unintegrated vertices.
But, since there are only two operators in \eqref{eq:2-point-amp}, this is impossible and the result must be zero.
However, this is also incorrect because it is always possible to insert $6$ $c$ zero-modes, as show the formulas \eqref{bos:eq:path-int-ghosts-zero-bc} and \eqref{bos:eq:amp-ghosts-bc-corr}.
Indeed, they are part of how the path integral measure is defined and do not care of the matter operators.
The question is whether they can be attached to vertex operators (for aesthetic reasons or more pragmatically to get natural states of the BRST cohomology).
To find the correct result with ghosts requires to start with \eqref{bos:eq:amp-ghosts-bc-corr} and to see how this can be simplified when there are only two operators.

\subsubsection{Computation of the amplitude}

In this section, we compute the $2$-point amplitude from \eqref{eq:2-point-amp-fixed}:
\begin{equation}
	A_{0,2}(k, k')
		= \frac{C_{S^2}}{\Vol \mc K_{0,2}} \,
			\Mean{V_k(\infty, \infty) V_{k'}(0, 0)}_{S^2}.
\end{equation}

The volume of $\mc K_{0,2}$ reads (by writing a measure invariant under rotations and dilatations, but not translations nor special conformal transformations)~\cite{DHoker:1988:GeometryStringPerturbation, Dorn:1994:TwoThreepointFunctions}:
\begin{equation}
	\Vol \mc K_{0,2}
		= \int \frac{\dd^2 z}{\abs{z}^2}
		= 2 \int_0^{2\pi} \dd \sigma \int_0^{\infty} \frac{\dd r}{r},
\end{equation}
by doing the change of variables $z = r \e^{\I \sigma}$.
Since the volume is infinite, it must be regularized.
A first possibility is to cut-off a small circle of radius $\epsilon$ around $r = 0$ and $r = \infty$ (corresponding to removing the two punctures at $z = 0, \infty$).
A second possibility consists in performing the change of variables $r = \e^{\tau}$ and to add an imaginary exponential:
\begin{equation}
	\label{eq:vol-dil-delta}
	\Vol \mc K_{0,2}
		= 4\pi \int_0^{\infty} \frac{\dd r}{r}
		= 4\pi \int_{-\infty}^{\infty} \dd \tau
		= 4\pi \lim_{\varepsilon \to 0} \int_{-\infty}^{\infty} \dd \tau \, \e^{\I \varepsilon \tau}
		= 4\pi \times 2\pi \, \lim_{\varepsilon \to 0} \delta(\varepsilon),
\end{equation}
such that the regularized volume reads
\begin{equation}
	\Vol_\varepsilon \mc K_{0,2}
		= 8 \pi^2 \, \delta(\varepsilon).
\end{equation}
In fact, $\tau$ can be interpreted as the Euclidean worldsheet time on the cylinder since $r$ corresponds to the radial direction of the complex plane.

Since the worldsheet is an embedding into the target spacetime, both must have the same signature.
As a consequence, for the worldsheet to be also Lorentzian, the formula \eqref{eq:vol-dil-delta} must be analytically continued as $\varepsilon = - \I E$ and $\tau = \I t$ such that
\begin{equation}
	\label{eq:vol-dil-delta-lorentz}
	\Vol_{M,E} \mc K_{0,2}
		= 8\pi^2 \I \, \delta(E),
\end{equation}
where the subscript $M$ reminds that one considers the Lorentzian signature.
Inserting this expression in \eqref{eq:2-point-amp-fixed-delta} and taking the limit $E \to 0$, it looks like the two $\delta(0)$ will cancel.
However, we need to be careful about the dimensions.
Indeed, the worldsheet time $\tau$ and energy $E$ are dimensionless, while the spacetime time and energy are not.
Thus, it is not quite correct to cancel directly both $\delta(0)$ since they don't have the same dimensions.
In order to find the correct relation between the integrals in \eqref{eq:vol-dil-delta} and of the zero-mode in \eqref{eq:2-point-corr-V}, we can look at the mode expansion for the scalar field (removing the useless oscillators):
\begin{equation}
	X^0(z, \bar z)
		= x^0 + \frac{\I}{2} \, \alpha' k^0 \, \ln \abs{z}^2
		= x^0 + \I \alpha' k^0 \tau,
\end{equation}
where the second equality follows by setting $z = \e^{\tau}$.
After analytic continuation $k^0 = - \I k^0_M$, $X^0 = \I X^0_M$, $x^0 = \I x^0_M$ and $\tau = \I t$, we find~\cite[p.~186]{Zwiebach:2009:FirstCourseString}:
\begin{equation}
	X^0_M
		= x^0_M + \alpha' k^0_M t.
\end{equation}
This indicates that the measure of the worldsheet time in \eqref{eq:vol-dil-delta-lorentz} must be rescaled by $1 / \alpha' k^0_M$
such that:
\begin{equation}
	\Vol_M \mc K_{0,2}
		\longrightarrow \frac{8 \pi^2 \I \, \delta(0)}{\alpha' k^0_M}
		= \frac{C_{S_2} \, 2 \pi \I \, \delta(0)}{2 k^0_M}.
\end{equation}
This is equivalent to rescale $E$ by $\alpha' k^0$ and to use $\delta(a x) = a^{-1} \delta(x)$.

Ultimately, the $2$-point amplitude becomes (removing the subscript on $k^0$):
\begin{equation}
	A_{0,2}(k, k')
		= 2 k^0 (2\pi)^{D-1} \delta^{(D-1)}(\vec k + \vec k')
\end{equation}
and matches the QFT formula \eqref{qft:2pt-amp}.
We see that taking into account the scale of the coordinates is important to reproduce this result.

The computation displayed here presents some ambiguities because of the regularization.
However, this ambiguity can be fixed from unitarity of the scattering amplitudes.
A more general version of the Faddeev--Popov gauge fixing has been introduced in~\cite{Erbin:2019:TwoPointStringAmplitudes} to avoid dealing altogether with infinities.
It is an interesting question whether these techniques can be extended to the compute the tree-level $1$- and $0$-point amplitudes on the sphere.
In most cases, the $1$-point amplitude is expected to vanish since $1$-point correlation functions of primary operators other than the identity vanish in unitary CFTs.\footnotemark{}
\footnotetext{%
	The integral over the zero-mode gives a factor $\delta^{(D)}(k)$ which implies $k = 0$.
	At zero momentum, the time scalar $X^0$ is effectively described by unitary CFT.
	However, there can be some subtleties when considering marginal operator.
}%
The $0$-point function corresponds to the sphere partition function: the saddle point approximation to leading order allows to relate it to the spacetime action evaluated on the classical solution $\phi_0$, $Z_0 \sim \e^{- S[\phi_0] / \hbar}$.
Since the normalization is not known and because $S[\phi_0]$ is expected to be infinite, only comparison between two spacetimes should be meaningful (\emph{à la} Gibbons--Hawking--York~\cite[sec.~4.1]{Poisson:2007:RelativistsToolkitMathematics}).
In particular, for Minkowski spacetime we find naively
\begin{equation}
	Z_0
		\sim \frac{\delta^{(D)}(0)}{\Vol \mc K_0},
\end{equation}
which is not well-defined.
This question has no yet been investigated.

\subsubsection{Expression with ghosts}

There are different ways to rewrite the $2$-point amplitude in terms of ghosts.
In all cases, one correctly finds the $6$ insertions necessary to get a non-vanishing result since, by definition, it is always possible to rewrite the Faddeev--Popov determinant in terms of ghosts.
A first approach is to insert $1 = \int \dd^2 z \, \delta^{(2)}(z)$ inside \eqref{eq:2-point-amp} to mimic the presence of a third operator.
This is equivalent to use the identity
\begin{equation}
	 \bra{0} c_{-1} \bar c_{-1} c_0 \bar c_0 c_1 \bar c_1 \ket{0}
		= 1
\end{equation}
inside \eqref{eq:2-point-amp-fixed}, leading to:
\begin{equation}
	A_{0,2}(k, k')
		= \frac{C_{S^2}}{\Vol \mc K_{0,2}} \,
			\Mean{\scr V_k(\infty, \infty) c_0 \bar c_0 \, \scr V_{k'}(0, 0)}_{S^2},
\end{equation}
where $\scr V_k(z, \bar z) = c \bar c V_k(z, \bar z)$.
This shows that \eqref{qft:2pt-amp} can also be recovered using the correct insertions of ghosts.
The presence of $c_0 \bar c_0$ can be expected from string field theory since they appear in the kinetic term \eqref{bsft:eq:closed-action-siegel}.

The disadvantage of this formula is to still contain the infinite volume of the dilatation group.
It is also possible to introduce ghosts for the more general gauge fixing presented in~\cite{Erbin:2019:TwoPointStringAmplitudes}.
An alternative approach has been proposed in~\cite{Seki:2020:TwopointStringAmplitudes}.

\index{string amplitude!tree-level 2-point (-)|)}%

\section{BRST quantization}
\label{bos:sec:ws-int:brst}

The symmetries of a Lagrangian dictate the possible terms which can be considered.
This continues to hold at the quantum level and the counter-terms introduced by renormalization are constrained by the symmetries.
However, if the path integral is gauge fixed, the original symmetry is no more available for this purpose.
Fortunately, one can show that there is a global symmetry (with anticommuting parameters) remnant of the local symmetry: the BRST symmetry.
It ensures consistency of the quantum theory.
It also provides a direct access to the physical spectrum.

The goal of this section is to provide a general idea of the BRST quantization for the worldsheet path integral.
A more detailed CFT analysis and the consequence for string theory are given in \Cref{cft:chap:brst}.
The reader is assumed to have some familiarity with the BRST quantization in field theory -- a summary is given in \Cref{app:sec:qft:brst}.

\subsection{BRST symmetry}

The partition function \eqref{bos:eq:path-int-ghosts-zero-b} is not the most suitable to display the BRST symmetry.
The first step is to restore the dependence in the original metric $g_{ab}$ by introducing a delta function
\begin{equation}
	Z_g
		= \int_{\mc M_g}
				\frac{\dd^{\M_g} t}{\Omega_{\text{ckv}}[g]}
			\int \dd_{g} g_{ab} \, \dd_{g} \Psi \, \dd_{g} b \, \dd'_{g} c \,
				\delta\big( \sqrt{g} g_{ab} - \sqrt{\hat g} \hat g_{ab} \big)
				\prod_{i=1}^{\M_{g}} \psp{\phi_i}{b}_{g} \,
				\e^{- S_m[g, \Psi] - S_{\text{gh}}[g, b, c]}.
\end{equation}
Note that it is necessary to use the traceless gauge fixing condition \eqref{bos:eq:gf-cond-traceless} as it will become clear.
The delta function is Fourier transformed in an exponential thanks to an auxiliary bosonic field:
\begin{equation}
	\label{bos:eq:path-int-ghosts-aux}
	Z_g
		= \int_{\mc M_g}
				\frac{\dd^{\M_g} t}{\Omega_{\text{ckv}}[g]}
			\int \dd_{g} g_{ab} \, \dd_g B^{ab}\, \dd_{g} \Psi \, \dd_{g} b \, \dd'_{g} c \,
				\prod_{i=1}^{\M_{g}} \psp{\phi_i}{b}_{g} \,
				\e^{- S_m[g, \Psi] - S_{\text{gf}}[g, \hat g, B] - S_{\text{gh}}[g, b, c]}
\end{equation}
\index{worldsheet!action!gauge-fixing (-)}%
where the gauge-fixing action reads:
\begin{equation}
	S_{\text{gf}}[g, \hat g, B]
		= - \frac{\I}{4\pi} \int \dd^2 \sigma \,
			B^{ab} \big( \sqrt{g} g_{ab} - \sqrt{\hat g} \hat g_{ab} \big).
\end{equation}
\index{worldsheet!Nakanishi--Lautrup auxiliary field}%
Varying the action with respect to the auxiliary field $B_{ab}$, called the Nakanish--Lautrup field, produces the gauge-fixing condition.

\index{worldsheet!symmetry!BRST}%
The BRST transformations are
\begin{equation}
	\label{bos:eq:sym-brst-B}
	\begin{gathered}
		\delta_{\epsilon} g_{ab}
			= \I \epsilon \, \mc L_c g_{ab},
		\qquad
		\delta_{\epsilon} \Psi
			= \I \epsilon \, \mc L_c \Psi,
		\\
		\delta_{\epsilon} c^a
			= \I \epsilon \, \mc L_c c^a,
		\qquad
		\delta_{\epsilon} b_{ab}
			= \epsilon \, B_{ab},
		\qquad
		\delta_{\epsilon} B_{ab}
			= 0,
	\end{gathered}
\end{equation}
where $\epsilon$ is a Grassmann parameter (anticommuting number) independent of the position.
If the traceless gauge fixing \eqref{bos:eq:gf-cond-traceless} is not used, then $B_{ab}$ is not traceless: in that case, the variation $\delta_{\epsilon} b_{ab}$ will generate a trace, which is not consistent.
Since the transformations act on the matter action $S_m$ as a diffeomorphism with vector $\epsilon c^a$, it is obvious that it is invariant by itself.
It is easy to show that the transformations \eqref{bos:eq:sym-brst-B} leave the total action invariant in \eqref{bos:eq:path-int-ghosts-aux}.
The invariance of the measure is given in~\cite{Polchinski:2005:StringTheory-1}.

\begin{remark}[BRST transformations with Weyl ghost]
	\index{Weyl!ghost}%

	One can also consider the action \eqref{bos:eq:action-ghost-weyl} with the Weyl ghost.
	In this case, the transformation law of the metric is modified and the Weyl ghost transforms as a scalar:
	\begin{equation}
		\delta_{\epsilon} g_{ab}
			= \I \epsilon \, \mc L_c g_{ab} + \I \epsilon \, g_{ab} c_w,
		\qquad
		\delta_{\epsilon} c_w
			= \I \epsilon \, \mc L_c c_w.
	\end{equation}
	The second term in $\delta_\epsilon g_{ab}$ is a Weyl transformation with parameter $\epsilon c_w$.
	Moreover, $b_{ab}$ and $B_{ab}$ are not symmetric traceless.
\end{remark}

The equation of motion for the auxiliary field is
\begin{equation}
	B_{ab}
		= \I \, T_{ab}
		:= \I \big( T^{m}_{ab} + T^{\text{gh}}_{ab} \big),
\end{equation}
where the RHS is the total energy--momentum tensor (matter plus ghosts).
Integrating it out imposes the gauge condition $g_{ab} = \hat g_{ab}$ and yields the modified BRST transformations
\begin{equation}
	\label{bos:eq:transf-inf-brst-psi-c-b}
	\delta_\epsilon \Psi
		= \I \epsilon \, \mc L_c \Psi,
	\qquad
	\delta_\epsilon c^a
		= \I \epsilon \, \mc L_c c^a,
	\qquad
	\delta_\epsilon b_{ab}
		= \I \epsilon \, T_{ab}.
\end{equation}
Without starting with the path integral \eqref{bos:eq:path-int-ghosts-aux} with auxiliary field, it would have been difficult to guess the transformation of the $b$ ghost.
Since $c^a$ is a vector, one can also write
\begin{equation}
	\delta_\epsilon c^a = \epsilon \, c^b \pd_b c^a.
\end{equation}

Associated to this symmetry is the BRST current $j^a_B$ and the associated conserved BRST charge $Q_B$
\begin{equation}
	Q_B = \int \dd \sigma \, j_B^0.
\end{equation}
The charge is nilpotent
\begin{equation}
	Q_B^2 = 0,
\end{equation}
and, through the presence of the $c$-ghost in the BRST transformation, the BRST charge has ghost number one
\begin{equation}
	N_{\text{gh}}(Q_B) = 1.
\end{equation}
Variations of the matter fields can be written as
\begin{equation}
	\delta_{\epsilon} \Psi = \I \, \com{\epsilon Q_B}{\Psi}_\pm.
\end{equation}
Note that the energy--momentum tensor is BRST exact
\begin{equation}
	\label{bos:eq:T-brst-exact}
	T_{ab} = \com{Q_B}{b_{ab}}.
\end{equation}

\subsection{BRST cohomology and physical states}
\label{bos:sec:ws-int:brst:states}

\index{BRST cohomology}%
\index{BRST cohomology!absolute}%
Physical state $\ket{\psi}$ are elements of the \emph{absolute} cohomology of the BRST operator:
\begin{equation}
	\ket{\psi} \in \mc H(Q_B)
		:= \frac{\ker Q_B}{\Im Q_B},
\end{equation}
or, more explicitly, closed but non-exact states:
\begin{equation}
	Q_B \ket{\psi} = 0,
	\qquad
	\nexists \ket{\chi}: \ket{\psi} = Q_B \ket{\chi}.
\end{equation}
The adjective “absolute” is used to distinguish it from two other cohomologies (relative and semi-relative) defined below.
Two states of the cohomology differing by an exact state represent identical physical states:
\begin{equation}
	\ket{\psi}
		\sim \ket{\psi} + Q_B \ket{\Lambda}.
\end{equation}
This equivalence relation, translated in terms of spacetime fields, correspond to spacetime gauge transformations.
In particular, it contains the (linearized) reparametrization invariance of the spacetime metric in the closed string sector, and, for the open string sector, it contains Yang--Mills symmetries.
We will find that it corresponds to the gauge invariance of free string field theory (\Cref{bsft:chap:free-brst}).

However, physical states satisfy two additional constraints (remember that $b_{ab}$ is traceless symmetric):
\begin{equation}
	\label{bos:eq:phys-state-b-cond}
	\int \dd \sigma \, b_{ab} \ket{\psi}
		= 0.
\end{equation}
These conditions are central to string (field) theory, so they will appear regularly in this \revname{}.
For this reason, it is useful to provide first some general motivations, and to refine the analysis later since the CFT language will be more appropriate.
Moreover, these two conditions will naturally emerge in string field theory.

\index{reparametrization $bc$ ghost!zero-mode}%
In order to introduce some additional terminology, let's define the following quantities:\footnotemark{}
\footnotetext{%
	The objects $b^\pm$ are zero-modes of the $b$ ghost fields.
	They correspond (up to a possible irrelevant factor) to the modes $b_0^\pm$ in the CFT formulation of the ghost system \eqref{cft:eq:bn-cn-pm}.
}%
\begin{equation}
	b^+
		:= \int \dd \sigma \, b_{00},
	\qquad
	b^-
		:= \int \dd \sigma \, b_{01}.
\end{equation}
\index{BRST cohomology!semi-relative}%
\index{BRST cohomology!relative}%
The \emph{semi-relative} and \emph{relative} cohomologies $\mc H^-(Q_B)$ and $\mc H^0(Q_B)$ are defined as\footnotemark{}
\footnotetext{%
	The BRST cohomologies described in this section are slightly different from the ones used in the rest of this \revname{}.
	To distinguish them, indices are written as superscripts in this section, and as subscripts otherwise.
}%
\begin{equation}
	\mc H^-(Q_B)
		= \mc H(Q_B) \cap \ker b^-,
	\qquad
	\mc H^0(Q_B)
		= \mc H^-(Q_B) \cap \ker b^+.
\end{equation}

The first constraint arises as a consequence of the topology of the closed string worldsheet: the spatial direction is a circle, which implies that the theory must be invariant under translations along the $\sigma$ direction (the circle is invariant under rotation).
However, choosing a parametrization implies to fix an origin for the spatial direction: this is equivalent to a gauge fixing condition.
As usual, this implies that the corresponding generator $P_\sigma$ of worldsheet spatial translations \eqref{bos:eq:worldsheet-momentum} must annihilate the states:
\begin{equation}
	P_\sigma \ket{\psi} = 0.
\end{equation}
\index{level-matching condition}%
This is called the \emph{level-matching condition}.
Using \eqref{bos:eq:T-brst-exact}, this can be rewritten as
\begin{equation}
	P_\sigma \ket{\psi}
		= \int \dd \sigma \, T_{01} \ket{\psi}
		= \int \dd \sigma \, \anticom{Q_B}{b_{01}} \ket{\psi}
		= Q_B \int \dd \sigma \, b_{01} \ket{\psi},
\end{equation}
since $Q_B \ket{\psi} = 0$ for a state $\ket{\psi}$ in the cohomology.
The simplest way to enforce this condition is to set the state on which $Q_B$ acts to zero:\footnotemark{}
\footnotetext{%
	The reverse is not true.
	We will see in \Cref{bos:sec:ws-int:brst:states} the relation between the two conditions in more details.
}%
\begin{equation}
	b^- \ket{\psi} = 0,
\end{equation}
which is equivalent to one of the conditions in \eqref{bos:eq:phys-state-b-cond}.

The second condition does not follow as simply.
The Hilbert space can be decomposed according to $b^+$ as
\begin{equation}
	\mc H^-
		:= \mc H_{\downarrow} \oplus \mc H_{\uparrow},
	\qquad
	\mc H_{\downarrow}
		:= \mc H^0
		:= \mc H^- \cap \ker b^+.
\end{equation}
Indeed, $b^+$ is a Grassmann variable and generates a $2$-state system.
In the ghost sector, the two Hilbert spaces are generated from the ghost vacua $\ket{\downarrow}$ and $\ket{\uparrow}$ obeying
\begin{equation}
	b^+ \ket{\downarrow}
		= 0,
	\qquad
	b^+ \ket{\uparrow}
		= \ket{\downarrow}.
\end{equation}
The action of the BRST charge on states $\ket{\psi_\downarrow} \in \mc H_{\downarrow}$ and $\ket{\psi_\uparrow} \in \mc H_{\uparrow}$ follow from these relations and from the commutation relation \eqref{bos:eq:T-brst-exact}:
\begin{equation}
	Q_B \ket{\psi_\downarrow}
		= H \ket{\psi_\uparrow},
	\qquad
	Q_B \ket{\psi_\uparrow}
		= 0,
\end{equation}
where $H$ is the worldsheet Hamiltonian defined in \eqref{bos:eq:worldsheet-momentum}.
To prove this relation, start first with $H \ket{\psi_\uparrow}$, then use \eqref{bos:eq:T-brst-exact}) to get the LHS of the first condition; then apply $Q_B$ to get the second condition (using that $Q_B$ commutes with $H$, and $b^+$ with any other operators building the states).
For $H \neq 0$, the state $\ket{\psi_\downarrow}$ is not in the cohomology and $\ket{\psi_\uparrow}$ is exact.
Thus, the exact and closed states are
\begin{subequations}
\begin{align}
	\Im Q_B
		&
		= \big\{ \ket{\psi_\uparrow} \in \mc H_{\uparrow}
			\mid H \ket{\psi_\uparrow} \neq 0 \big\},
		\\
	\ker Q_B
		&
		= \big\{ \ket{\psi_\uparrow} \in \mc H_{\uparrow} \big\}
			\cup
			\big\{ \ket{\psi_\downarrow} \in \mc H_{\downarrow}
				\mid H \ket{\psi_\downarrow} = 0 \big\}.
\end{align}
\end{subequations}
\index{on-shell condition}%
This implies that eigenstates of $H$ in the cohomology satisfy the on-shell condition:
\begin{equation}
	\label{bos:eq:brst-on-shell-init}
	H \ket{\psi} = 0.
\end{equation}
This is consistent with the fact that scattering amplitudes involve on-shell states.
In this case, $\ket{\psi_\uparrow}$ is not exact and is thus a member of the cohomology $\mc H(Q_B)$, as well as $\ket{\psi_\downarrow}$ since it becomes close.
But, the Hilbert space $\mc H_{\uparrow}$ must be rejected for two reasons: there would be an apparent doubling of states and scattering amplitudes would behave badly.
The first problem arises because one can show that the cohomological subspaces of each space are isomorphic: $\mc H_{\downarrow}(Q_B) \simeq \mc H_{\uparrow}(Q_B)$.
Hence, keeping both subspaces would lead to a doubling of the physical states.
For the second problem, consider an amplitude where one of the external state is built from $\ket{\psi_\uparrow}$: the amplitude vanishes if the states are off-shell since the state $\ket{\psi_\uparrow}$ is exact, but it does not vanish on-shell~\cite[ch.~4]{Polchinski:2005:StringTheory-1}.
This means that it must be proportional to $\delta(H)$.
But, general properties in QFT forbid such dependence in the amplitude (only poles and cuts are allowed, except if $D = 2$).
Projecting out the states in $\mc H_{\uparrow}$ is equivalent to require
\begin{equation}
	\label{bos:eq:brst-siegel-init}
	b^+ \ket{\psi} = 0
\end{equation}
for physical states, which is the second condition in \eqref{bos:eq:phys-state-b-cond}.

In fact, this condition can be obtained very similarly as the $b^- = 0$ condition: using the expression of $H$ \eqref{bos:eq:worldsheet-momentum} and the commutation relation \eqref{bos:eq:T-brst-exact}, \eqref{bos:eq:brst-on-shell-init} is equivalent to
\begin{equation}
	Q_B \int \dd \sigma \, b_{00} \ket{\psi}
		= 0.
\end{equation}
Hence, imposing \eqref{bos:eq:brst-siegel-init} allows to automatically ensure that \eqref{bos:eq:brst-on-shell-init} holds.

Since the on-shell character \eqref{bos:eq:brst-on-shell-init} of the BRST states and of the BRST symmetry are intimately related to the construction of the worldsheet integral, one can expect difficulty for going off-shell.

\begin{draft}

\section{Alternative formulations}
\label{chap:bos:ws-int:bv}
\label{chap:bos:ws-int:ext-brst}

\subsection{Delta function}

An alternative derivation of the Faddeev--Popov gauge fixing is to rewrite the gauge fixing delta function in terms of a functional Fourier transformation.

\subsection{BV quantization}

As explained previously the gauge fixing is not complete due to the $P_1$ and $\adj{P_1}$ zero-modes, leading to a residual gauge invariance.
In this case, the Faddeev--Popov procedure is not sufficient and but one needs to employ the BV formalism (introducing ghosts for the ghosts)~\cite{Craps:2005:CommentsBRSTQuantization}.
This is particularly salient for the $c$ ghost since one divides by an infinite volume.
One does not find the same problem for the $b$ ghosts because the moduli acts as ghosts for the ghosts.

\subsection{Extended BRST quantization}

A last possible formulation is to extend the BRST formalism~\cite[sec.~2]{Witten:2012:SuperstringPerturbationTheory}.

\end{draft}

\section{Summary}

In this chapter, we derived general formulas for string scattering amplitudes.
The general BRST formalism has been summarized.
Moreover, we gave general motivations for restricting the absolute cohomology to the smaller relative cohomology.
In \Cref{cft:chap:brst}, a more precise derivation of the BRST cohomology is worked out.
It includes also a proof of the no-ghost theorem: the ghosts and the negative norm states (in Minkowski signature) are unphysical particles and should not be part of the physical states.
This theorem asserts that it is indeed the case.
It will also be the occasion to recover the details of the spectrum in various cases.

\refchapter

\begin{itemize}
	\item The delta function approach to the gauge fixing is described in~\cites[sec.~3.3]{Polchinski:2005:StringTheory-1}[sec.~15.3.2]{Lawrie:2012:UnifiedGrandTour}, with a more direct computation is in~\cite{Kiritsis:2007:StringTheoryNutshell}.

	\item The most complete references for scattering amplitudes in the path integral formalism are~\cite{DHoker:1988:GeometryStringPerturbation, Polchinski:2005:StringTheory-1}.

	\item Computation of the tree-level $2$-point amplitude~\cite{Erbin:2019:TwoPointStringAmplitudes, Seki:2020:TwopointStringAmplitudes} (for discussions of $2$-point function, see~\cites[p.~936--7]{DHoker:1988:GeometryStringPerturbation}{Seiberg:1990:NotesQuantumLiouville}{Dorn:1994:TwoThreepointFunctions}{Dorn:1995:ConclusionsNoncriticalString}[p.~863--4]{Deligne:1999:QuantumFieldsStrings2}).

	\item The BRST quantization of string theory is discussed in~\cites{Mansfield:1987:NilpotentBRSTInvariance}{Craps:2005:CommentsBRSTQuantization}[chap.~4]{Polchinski:2005:StringTheory-1}.
	For a general discussion see~\cite{Henneaux:1994:QuantizationGaugeSystems, vanHolten:2004:AspectsBRSTQuantization, Weinberg:2005:QuantumTheoryFields-2}.
	The use of an auxiliary field is considered in~\cite[sec.~3.2]{West:2012:IntroductionStringsBranes}.
\end{itemize}

\chapter{Worldsheet path integral: complex coordinates}
\label{bos:chap:ws-int-complex}

\introchapter

In the two previous chapters, the amplitudes computed from the worldsheet path integrals have been written covariantly for a generic curved background metric.
In this chapter, we start to use complex coordinates and finally take the background metric to be flat.
This is the usual starting point for computing amplitudes since it allows to make contact with CFTs and to employ tools from complex analysis.
We first recall few facts on $2d$ complex manifolds before briefly describing how to rewrite the scattering amplitudes in complex coordinates.

\section{Geometry of complex manifolds}
\label{bos:sec:ws-int-complex:geom-C}

Choosing a flat background metric simplifies the computations.
However, we have seen in \Cref{bos:sec:ws-int:faddeev-popov} that there is a topological obstruction to get a globally flat metric.
\index{metric!gauge!conformally flat (-)}%
The solution is to work with coordinate patches $(\sigma^0, \sigma^1) = (\tau, \sigma)$ such that the background metric $\hat g_{ab}$ is flat in each patch (conformally flat gauge):
\begin{equation}
	\dd s^2
		= g_{ab} \dd \sigma^a \dd \sigma^b
		= \e^{2\phi(\tau, \sigma)} \big( \dd \tau^2 + \dd \sigma^2 \big),
\end{equation}
or
\begin{equation}
	g_{ab} = \e^{2\phi} \delta_{ab},
	\qquad
	\hat g_{ab} = \delta_{ab}.
\end{equation}
To simplify the notations, we remove the dependence in the flat metric and the hat for quantities (like the vertex operators) expressed in the background metric when no confusion is possible.

\index{complex coordinates}%
Introducing complex coordinates
\begin{subequations}
\begin{gather}
	z = \tau + \I \sigma,
	\qquad
	\bar z = \tau - \I \sigma,
	\\
	\tau = \frac{z + \bar z}{2},
	\qquad
	\sigma = \frac{z - \bar z}{2 \I},
\end{gather}
\end{subequations}
the metric reads\footnotemark{}
\footnotetext{%
	In \Cref{cft:sec:plane:sphere}, we provide more details on the relation between the worldsheet (viewed as a cylinder or a sphere) and the complex plane.
}%
\begin{equation}
	\label{bos:eq:metric-plane}
	\dd s^2
		= 2 g_{z \bar z} \dd z \dd \bar z
		= \e^{2\phi(z, \bar z)} \abs{\dd z}^2.
\end{equation}
The metric and its inverse can also be written in components:
\begin{subequations}
\begin{gather}
	g_{z \bar z}
		= \frac{\e^{2\phi}}{2},
	\qquad
	g_{zz}
		= g_{\bar z \bar z}
		= 0,
	\\
	g^{z \bar z} = 2 \e^{- 2\phi},
	\qquad
	g^{zz}
		= g^{\bar z \bar z}
		= 0.
\end{gather}
\end{subequations}
Equivalently, the non-zero components of the background metric are
\begin{equation}
	\hat g_{z \bar z} = \frac{1}{2},
	\qquad
	\hat g^{z \bar z} = 2.
\end{equation}
An oriented two-dimensional manifold is a complex manifold: this means that there exists a complex structure, such that the transition functions and changes of coordinates between different patches are holomorphic at the intersection of the two patches:
\begin{equation}
	\label{bos:eq:change-coord-holom}
	w = w(z),
	\qquad
	\bar w = \bar w(\bar z).
\end{equation}
For such a transformation, the Liouville mode transforms as
\begin{equation}
	\e^{2\phi(z, \bar z)}
		= \Abs{\frac{\pd w}{\pd z}}^2 \e^{2\phi(w, \bar w)}
\end{equation}
such that
\begin{equation}
	\dd s^2
		= \e^{2\phi(w, \bar w)} \abs{\dd w}^2.
\end{equation}
This shows also that a conformal structure \eqref{bos:eq:conf-structure} induces a complex structure since the transformation law of $\phi$ is equivalent to a Weyl rescaling.

The integration measures are related as
\begin{equation}
	\dd^2 \sigma
		:= \dd \tau \dd \sigma
		= \frac{1}{2} \, \dd^2 z,
	\qquad
	\dd^2 z := \dd z \dd \bar{z}.
\end{equation}
Due to the factor of $2$ in the expression, the delta function $\delta^{(2)}(z)$ also gets a factor of $2$ with respect to $\delta^{(2)}(\sigma)$
\begin{equation}
	\delta^{(2)}(z) = \frac{1}{2} \, \delta^{(2)}(\sigma).
\end{equation}
Then, one can check that
\begin{equation}
	\int \dd^2 z \, \delta^{(2)}(z)
		= \int \dd^2 \sigma \, \delta^{(2)}(\sigma)
		= 1.
\end{equation}

The basis vectors (derivatives) and one-forms can be found using the chain rule:
\begin{subequations}
\begin{gather}
	\pd_{z}
		= \frac{1}{2} \, (\pd_\tau - \I \pd_\sigma),
	\qquad
	\pd_{\bar z}
		= \frac{1}{2} \, (\pd_\tau + \I \pd_\sigma),
	\\
	\dd z
		= \dd \tau + \I \dd \sigma,
	\qquad
	\dd \bar z
		= \dd \tau - \I \dd \sigma.
\end{gather}
\end{subequations}

The Levi--Civita (completely antisymmetric) tensor is normalized by
\begin{subequations}
\begin{gather}
	\epsilon_{01}
		= \epsilon^{01}
		= 1.
	\\
	\epsilon_{z \bar z}
		= \frac{\I}{2},
	\qquad
	\epsilon^{z \bar z}
		= - 2 \I,
\end{gather}
\end{subequations}
remembering that it transforms as a density.
Integer indices run over local frame coordinates.

The different tensors can be found from the tensor transformation law.
For example, the components of a vector $V^a$ in both systems are related by
\begin{equation}
	V^{z} = V^0 + \I V^1,
	\qquad
	V^{\bar z} = V^0 - \I V^1
\end{equation}
such that
\begin{equation}
	V
		= V^0 \pd_0 + V^1 \pd_1
		= V^z \pd_z + V^{\bar z} \pd_{\bar z}.
\end{equation}
For holomorphic coordinate transformations \eqref{bos:eq:change-coord-holom}, the components of the vector do not mix:
\begin{equation}
	V^w = \frac{\pd w}{\pd z} \, V^z,
	\qquad
	V^{\bar w} = \frac{\pd \bar w}{\pd \bar z} \, V^{\bar z}.
\end{equation}
This implies that the tangent space of the Riemann surface is decomposed into holomorphic and anti-holomorphic vectors:\footnotemark{}
\footnotetext{%
	However, at this stage, each component can still depend on both $z$ and $\bar z$: $V^{z} = V^{z}(z, \bar z)$ and $V^{\bar z} = V^{\bar z}(z, \bar z)$.
}%
\begin{subequations}
\begin{gather}
	T \Sigma_g
		\simeq T \Sigma_g^+ \oplus T \Sigma_g^-,
	\\
	V^{z} \pd_{z} \in T \Sigma_g^+,
	\qquad
	V^{\bar z} \pd_{\bar z} \in T \Sigma_g^-,
\end{gather}
\end{subequations}
as a consequence of the existence of a complex structure.
Similarly, the components of a $1$-form $\omega$ -- which is the only non-trivial form on $\Sigma_g$ -- can be written in terms of the real coordinates as:
\begin{equation}
	\omega_{z}
		= \frac{1}{2} \, (\omega_0 - \I \omega_1),
	\qquad
	\omega_{\bar z}
		= \frac{1}{2} \, (\omega_0 + \I \omega_1)
\end{equation}
such that
\begin{equation}
	\omega
		= \omega_0 \dd \sigma^0 + \omega_1 \dd \sigma^1
		= \omega_{z} \dd z + \omega_{\bar z} \dd \bar z.
\end{equation}
Hence, a $1$-form is decomposed into complex $(1, 0)$- and $(0, 1)$-forms:
\begin{subequations}
\begin{gather}
	T^* \Sigma_g
		\simeq \Omega^{1, 0}(\Sigma_g) \oplus \Omega^{0, 1}(\Sigma_g),
	\\
	\omega_{z} \dd z \in \Omega^{1, 0}(\Sigma_g),
	\qquad
	\omega_{\bar z} \dd \bar z \in \Omega^{0, 1}(\Sigma_g),
\end{gather}
\end{subequations}
since both components will not mixed under holomorphic changes of coordinates \eqref{bos:eq:change-coord-holom}.
Finally, the metric provides an isomorphism between $T \Sigma_g^+$ and $\Omega^{0, 1}(\Sigma_g)$, and between $T \Sigma_g^-$ and $\Omega^{1, 0}(\Sigma_g)$, since it can be used to lower/raise an index while converting it from holomorphic to anti-holomorphic, or conversely:
\begin{equation}
	V_{z} = g_{z \bar z} V^{\bar z},
	\qquad
	V_{\bar z} = g_{z \bar z} V^{z}.
\end{equation}

This can be generalized further by considering components with more indices: all anti-holomorphic indices can be converted to holomorphic indices thanks to the metric:
\begin{equation}
	\tensor{T}
			{^{\overbrace{\scriptstyle z \cdots z}^{q_+ + p_-}}}
			{_{\underbrace{\scriptstyle z \cdots z}_{p_+ + q_-}}}
		=
			(g^{z \bar z})^{p_-} (g_{z \bar z})^{q_-}
			\tensor{T}
			{^{\overbrace{\scriptstyle z \cdots z}^{q_+} \overbrace{\scriptstyle \bar z \cdots \bar z}^{q_-}}}
			{_{\underbrace{\scriptstyle z \cdots z}_{p_+} \underbrace{\scriptstyle \bar z \cdots \bar z}_{p_-}}}.
\end{equation}
Hence, it is sufficient to study $(p, q)$-tensors with $p$ upper and $q$ lower holomorphic indices.
In this case, the transformation rule under \eqref{bos:eq:change-coord-holom} reads
\begin{equation}
	\tensor{T}
			{^{\overbrace{\scriptstyle w \cdots w}^{q}}}
			{_{\underbrace{\scriptstyle w \cdots w}_{p}}}
		= \left( \frac{\pd w}{\pd z} \right)^n
		\tensor{T}
			{^{\overbrace{\scriptstyle z \cdots z}^{q}}}
			{_{\underbrace{\scriptstyle z \cdots z}_{p}}},
	\qquad
	n := q - p.
\end{equation}
The number $n \in \Z$ is called the helicity or rank.\footnotemark{}
\footnotetext{%
	In fact, it is even possible to consider $n \in \Z +1/2$ to describe spinors.
}%
The set of helicity-$n$ tensors is denoted by $\mc T^{n}$.

The first example is vectors (or equivalently $1$-forms): $V^z \in \mc T^{1}$, $V_z \in \mc T^{-1}$.
The second most useful case is traceless symmetric tensors, which are elements of $\mc T^{\pm 2}$.
Consider a traceless symmetric tensor $T^{ab} = T^{ba}$ and $g_{ab} T^{ab} = 0$: this implies $T^{01} = T^{10}$ and $T^{00} = - T^{11}$ in real coordinates.
The components in complex coordinates are:
\begin{equation}
	\label{bos:eq:sym-T-complex}
	T^{z z}
		= 2 (T^{00} + \I T^{01})
		\in \mc T^{2},
	\qquad
	T^{\bar z \bar z}
		= 2 (T^{00} - \I T^{01})
		\in \mc T^{-2},
	\qquad
	\tensor{T}{^z_z} = 0.
\end{equation}
Note that
\begin{equation}
	T_{z z}
		= g_{z \bar z} g_{z \bar z} T^{\bar z \bar z}
		= \frac{1}{2} (T^{00} - \I T^{01}),
\end{equation}
and $\tensor{T}{^z_z} = g_{z \bar z} T^{z \bar z} \in \mc T^0$ corresponds to the trace.

\begin{computation}[bos:eq:sym-T-complex]
	\[
		T^{z z}
			= \left( \frac{\pd z}{\pd \tau} \right)^2 T^{00}
				+ \left( \frac{\pd z}{\pd \sigma} \right)^2 T^{11}
				+ 2 \, \frac{\pd z}{\pd \tau} \frac{\pd z}{\pd \sigma} \, T^{01}
			= T^{00}
				- T^{11}
				+ 2 \I \, T^{01}.
	\]
\end{computation}

\index{Stokes' theorem}%
Stokes' theorem in complex coordinates follows directly from \eqref{app:eq:stokes-thm-2d}:
\begin{equation}
	\label{bos:eq:stokes-complex}
	\int \dd^2 z \, (\pd_z v^z + \pd_{\bar z} v^{\bar z})
		= - \I \oint \big( \dd z \, v^{\bar z} - \dd \bar z v^{z} \big)
		= - 2 \I \oint_{\pd R} (v_z \dd z - v_{\bar z} \dd \bar z),
\end{equation}
where the integration contour is anti-clockwise.
To obtain this formula, note that $\dd^2 x = \frac{1}{2} \dd^2 z$ and $\epsilon_{z\bar z} = \I / 2$, such that the factor $1/2$ cancels between both sides.

\section{Complex representation of path integral}

\index{Polyakov path integral!complex representation}%
In the previous section, we have found that tensors of a given rank are naturally decomposed into different subspaces thanks to the complex structure of the manifold.
Accordingly, complex coordinates are natural and one can expect most objects in string theory to split similarly into holomorphic and anti-holomorphic sectors (or left- and right-moving).
This will be particularly clear using the CFT language (\Cref{chap:cft:plane}).
The main difficulty for this program is due to the matter zero-modes.
In this section, we focus on the path integral measure and expression of the ghosts.

There is, however, a subtlety in displaying explicitly the factorization: the notion of “holomorphicity” depends on the metric (because the complex structure must be compatible with the metric for an Hermitian manifold).
Since the metric depends on the moduli which are integrated over in the path integral, it is not clear that there is a consistent holomorphic factorization.
We will not push the question of achieving a global factorization further (but see \Cref{bos:rem:holom-fact}) to focus instead on the integrand.
The latter is local (in moduli space) and there is no ambiguity.

The results of the previous section indicate that the basis of Killing vectors \eqref{bos:eq:basis-P1} and quadratic differentials \eqref{bos:eq:basis-adj-P1} split into holomorphic and anti-holomorphic components:
\begin{equation}
	\psi_i(z, \bar z)
		= \psi_i^z \pd_z + \psi_i^{\bar z} \pd_{\bar z},
	\qquad
	\phi_i(z, \bar z)
		= \phi_{i,zz} (\dd z)^2 + \phi_{i,\bar z \bar z} (\dd \bar z)^2.
\end{equation}
Similarly, the operators $P_1$ \eqref{bos:eq:op-P1} and $\adj{P_1}$ \eqref{bos:eq:op-P1-adj} also split:
\begin{subequations}
\begin{gather}
	(P_1 \xi)_{zz}
		= 2 \grad_z \xi_z
		= \pd_z \xi^{\bar z},
	\qquad
	(P_1 \xi)_{\bar z \bar z}
		= 2 \grad_{\bar z} \xi_{\bar z}
		= \pd_{\bar z} \xi^z,
	\\
	(\adj{P_1} T)_z
		= - 2 \grad^z T_{zz}
		= - 4 \, \pd_{\bar z} T_{zz},
	\qquad
	(\adj{P_1} T)_{\bar z}
		= - 2 \grad^{\bar z} T_{\bar z \bar z}
		= - 4 \, \pd_z T_{\bar z \bar z}
\end{gather}
\end{subequations}
for arbitrary vector $\xi$ and traceless symmetric tensor $T$ (in the background metric).
As a consequence, the components of Killing vectors and quadratic differentials are holomorphic or anti-holomorphic as a function of $z$:
\begin{equation}
	\psi^{z} = \psi^{z}(z),
	\qquad
	\psi^{\bar z} = \psi^{\bar z}(\bar z),
	\qquad
	\phi_{zz} = \phi_{zz}(z),
	\qquad
	\phi_{\bar z \bar z} = \phi_{\bar z \bar z}(\bar z),
\end{equation}
such that it makes sense to consider a complex basis instead of the previous real basis:
\begin{subequations}
\begin{gather}
	\ker P_1
		= \Span \{ \psi_K(z) \}
			\oplus \Span \{ \bar\psi_K(\bar z) \},
	\qquad
	K = 1, \ldots, \mathsf{K}_g^c,
	\\
	\ker \adj{P_1}
		= \Span \{ \phi_I(z) \}
			\oplus \Span \{ \bar\phi_I(\bar z) \},
	\qquad
	I = 1, \ldots, \M_g^c.
\end{gather}
\end{subequations}

\index{moduli space!complex coordinates}%
The last equation can inspire to search for a similar rewriting of the moduli parameters.
In fact, the moduli space itself is a complex manifold and can be endowed with complex coordinates~\cite{Nelson:1987:LecturesStringsModuli, Polchinski:2005:StringTheory-1}:
\begin{equation}
	m_I = t_{2 I - 1} + \I t_{2 I},
	\qquad
	\bar m_I = t_{2 I - 1} - \I t_{2 I},
	\qquad
	I = 1, \ldots, \M_g^c
\end{equation}
with the integration measure
\begin{equation}
	\dd^{\M_g} t
		= \dd^{2 \M_g^c} m.
\end{equation}

The last ingredient to rewrite the vacuum amplitudes \eqref{bos:eq:path-int-fp} is to obtain the determinants.
The inner-products of vector and traceless symmetric fields also factorize:
\begin{subequations}
\begin{gather}
	\psp{T_1}{T_2}
		= 2 \int \dd^2 \sigma \sqrt{\hat g} \, \hat g^{ac} g^{bd} T_{1,ab} T_{2,cd}
		= 4 \int \dd^2 z \,
			\big(
				T_{1,z z} T_{2,\bar z \bar z}
				+ T_{1,\bar z \bar z} T_{2,zz}
				\big),
	\\
	\psp{\xi_1}{\xi_2}
		= \int \dd^2 \sigma \sqrt{\hat g} \, \hat g_{ab} \xi^a \xi^b
		= \frac{1}{4} \int \dd^2 z \,
			\big(
				\xi_1^{z} \xi_2^{\bar z}
				+ \xi_1^{\bar z} \xi_2^{z}
				\big).
\end{gather}
\end{subequations}
All inner-products are evaluated in the flat background metric.
For (anti-)holomorphic fields, only one term survives in each integral: since each field appears twice in the determinants $\psp{\phi_i}{\phi_j}$ and $\psp{\phi_i}{\phi_j}$, the final expression is a square, which cancels against the squareroot in \eqref{bos:eq:path-int-fp}.
The remaining determinant involves the Beltrami differential \eqref{bos:eq:def-mu}:
\begin{equation}
	\mu_{i z z}
		= \pd_i \bar g_{zz},
	\qquad
	\mu_{i \bar z \bar z}
		= \pd_i \bar g_{\bar z \bar z}
\end{equation}
($\bar g_{zz} = 0$ in our coordinates system, but its variation under a shift of moduli is not zero).
The basis can be changed to a complex basis such that the determinant of inner-products between Beltrami and quadratic differentials is a modulus squared.
All together, the different formulas lead to the following rewriting of the vacuum amplitude :
\index{string amplitude!gv@$g$-loop vacuum (-)}%
\begin{equation}
	Z_g = \int_{\mc M_g} \!\! \dd^{2 \M_g^c} m \,
			\frac{\abs{\det \psp{\phi_I}{\mu_J}}^2}{\abs{\det \psp{\phi_I}{\bar\phi_J}}} \,
				\frac{\det' \adj{P_1} P_1}{\abs{\det \psp{\psi_I}{\bar\psi_J}}} \;
			\frac{Z_m[\delta]}{\Omega_{\text{ckv}}[\delta]},
\end{equation}
where the absolute values are to be understood with respect to the basis of $P_1$ and $\adj{P_1}$, for example $\abs{f(m_I)}^2 := f(m_I) f(\bar m_I)$.

The same reasoning can be applied to the ghosts.
The $c$ and $b$ ghosts are respectively a vector and a symmetric traceless tensor, both with two independent components: it is customary to define
\begin{equation}
	c := c^z,
	\qquad
	\bar c := c^{\bar z},
	\qquad
	b := b_{zz},
	\qquad
	\bar b := b_{\bar z \bar z}.
\end{equation}
\index{reparametrization $bc$ ghosts!action}%
In that case, the action \eqref{bos:eq:action-ghost} reads
\begin{equation}
	\label{bos:eq:action-ghost-complex}
	S_{\text{gh}}[g, b, c]
		= \frac{1}{2\pi} \int \dd^2 z \,
			\big( b \pd_{\bar z} c
			+ \bar b \pd_{z} \bar c \big).
\end{equation}
The action is the sum of two holomorphic and anti-holomorphic contributions and it is independent of $\phi(z, \bar z)$ as expected.
In fact, the equations of motion are
\begin{equation}
	\pd_z c = 0,
	\qquad
	\pd_z b = 0,
	\qquad
	\pd_{\bar z} \bar c = 0,
	\qquad
	\pd_{\bar z} \bar b = 0,
\end{equation}
such that $b$ and $c$ (resp.\ $\bar b$ and $\bar c$) are holomorphic (anti-holomorphic) functions.
Then, the integration measure is simply
\begin{equation}
	\bigwedge_{i=1}^{\M_g} B_i \, \dd t_i
		= \bigwedge_{I=1}^{\M_g^c} B_I \bar B_I \, \dd m_I \wedge \bar m_I,
	\qquad
	B_I := \psp{\mu_I}{b}.
\end{equation}
Note that $B_I$ does not contain $\bar b(\bar z)$, it is built only from $b(z)$.

Finally, the vacuum amplitude \eqref{bos:eq:path-int-ghosts-zero-bc} reads
\begin{equation}
	Z_g = \int_{\mc M_g} \!\! \dd^{2 \M_g^c} m \,
		\frac{\Omega_{\text{ckv}}[\delta]^{-1}}{\abs{\det \psi_I(z_j^0)}^2}
		\int \dd(b, \bar b) \, \dd(c, \bar c) \,
		\prod_{j=1}^{\mathsf{K}_g^c} c(z_j^0) \bar c(\bar z_j^0)
		\prod_{I=1}^{\M_g^c} \abs{\psp{\mu_I}{b}}^2 \,
		\e^{- S_{\text{gh}}[b, c]} \,
		Z_m[\delta].
\end{equation}
The $c$ insertions are separated in holomorphic and anti-holomorphic components because, at the end, only the zero-modes contribute.
The measures are written as $\dd(b, \bar b)$ and $\dd(c, \bar c)$ because proving that they factorize is difficult (\Cref{bos:rem:holom-fact}).

\begin{remark}[Holomorphic factorization]
	\label{bos:rem:holom-fact}

	\index{critical dimension}%
	\index{holomorphic factorization}%

	It was proven in~\cite{Belavin:1986:AlgebraicGeometryGeometry, Bost:1986:HolomorphyPropertyCritical, Catenacci:1986:AlgebraicGeometryPath} (see~\cites[sec.~9]{Nelson:1987:LecturesStringsModuli}[sec.~VII]{DHoker:1988:GeometryStringPerturbation}[sec.~3]{Skliros:2016:HighlyExcitedStrings} for reviews) that the ghost and matter path integrals can be globally factorized, up to a factor due to zero-modes.
	Such a result is suggested by the factorization of the inner-products, which imply a factorization of the measures: the caveat is due to the zero-mode determinants and matter measure.
	Interestingly, the factorization is possible only in the critical dimension \eqref{bos:eq:crit-dim}.
\end{remark}

\begin{draft}

\section{Holomorphic factorization}

The procedure follows directly for the matter if it factorizes like the ghost sector (in particular, if the action splits as the sum of left- and right-moving sectors).
We will see that it is not the case due to the zero-mode of the non-compact scalars $X^\mu$.
The simplest possibility is to split the scalar fields into the zero-mode and its orthogonal part
\begin{equation}
	X^\mu(z, \bar z) = x^\mu + X'^\mu(z, \bar z),
\end{equation}
where $x^\mu$ is constant.
Removing the zero-mode from the vertex operator, the remaining part generically factorizes holomorphically:
\begin{equation}
	V_\alpha(k; z, \bar z)
		= \hat V_\alpha(k; z, \bar z) \e^{\I k \cdot x},
	\qquad
	\hat V_\alpha(k; z, \bar z)
		= \Abs{v_\alpha(z) \e^{\I k \cdot X'(z)}}^2
\end{equation}
(the hat should not be confused with the hat indicating that $V$ is in the background metric).
As a consequence, the matter partition function also factorizes assuming that the action does:
\begin{equation}
	\Mean{\prod_{i=1}^n V_{\alpha_i}(k_i; z, \bar z)}
		= \int \frac{\dd x^\mu}{A^{D/2}} \, \e^{\I (k_1 + \cdots + k_n) \cdot x} \,
			\Abs{\Mean{\prod_{i=1}^n v_{\alpha_i}(z_i) \e^{\I k_i \cdot X'(z_i)}}'}^2,
\end{equation}
where the prime on the correlation function indicates that the zero-modes $x^\mu$ are excluded.
The factor $A^{D/2}$, with $A$ the area of the surface, arises by separating the zero-mode integration from the functional measure..
Finally, the unintegrated vertices can be written as
\begin{equation}
	\scr V_\alpha(k; z, \bar z) = c(z) \bar c(\bar z) V_\alpha(k; z, \bar z).
\end{equation}

\section{Computations of determinants}

\section{Amplitudes on marked moduli space}
\label{bos:sec:ws-int:amp:Mgn}

\end{draft}

\section{Summary}

In this chapter, we have introduced complex notations for the fields, path integral and moduli space.

\refchapter

\begin{itemize}
	\item Good references for this chapter are~\cite{Nelson:1987:LecturesStringsModuli, Nakahara:2003:GeometryTopologyPhysics, Polchinski:2005:StringTheory-1, Blumenhagen:2014:BasicConceptsString, DHoker:1988:GeometryStringPerturbation}.

	\item Geometry of complex manifolds is discussed in~\cites[sec.~6.2]{Blumenhagen:2014:BasicConceptsString}[chap.~14]{Nakahara:2003:GeometryTopologyPhysics}{DHoker:1988:GeometryStringPerturbation}.

\end{itemize}

\ifbook
\chapter{Conformal symmetry in \texorpdfstring{$D$}{D} dimensions}
\fi
\label{chap:cft:general}

\introchapter

Starting with this chapter, we discuss general properties of conformal field theories (CFT).
The goal is not to be exhaustive, but to provide a short introduction and to gather the concepts and formulas that are needed for string theory.
However, the subject is presented as a standalone topic such that it can be of interest for a more general public.

The conformal group in any dimension is introduced in this chapter.
The specific case $D = 2$, which is the most relevant for the current \revname{}, is developed in the following chapters.

\section{CFT on a general manifold}
\label{sec:cft:general:gen}

In this chapter and in the next one, we discuss CFTs as QFTs living on a spacetime $\mc M$, independently from string theory (there is no reference to a target spacetime).
As such, we will use spacetime notations together with some simplifications: coordinates are written as $x^\mu$ with $\mu = 0, \ldots, D - 1$ and time is written as $x^0 = t$ ($x^0 = \tau$) in Lorentzian (Euclidean) signature.

\begin{check}
\subsection{Conformal group}
\end{check}
\label{sec:cft:general:conf-group}

\index{conformal isometry group}%
Given a metric $g_{\mu\nu}$ on a $D$-dimensional manifold $\mc M$, the conformal group $\group{CISO}(\mc M)$ is the set of coordinate transformations (called conformal symmetries or isometries)
\begin{equation}
	x^\mu \longrightarrow x'^\mu = x'^\mu(x)
\end{equation}
which leaves the metric invariant up to an overall scaling factor:
\begin{equation}
	\label{cft:eq:sym-diff}
	g_{\mu\nu}(x)
	\longrightarrow
	g'_{\mu\nu}(x')
		= \frac{\pd x^\rho}{\pd x'^\mu} \frac{\pd x^\sigma}{\pd x'^\nu} \, g_{\rho\sigma}(x)
		= \Omega(x')^2 g_{\mu\nu}(x').
\end{equation}
This means that angles between two vectors $u$ and $v$ are left invariant under the transformation:
\begin{equation}
	\frac{u \cdot v}{\abs{u} \, \abs{v}}
		= \frac{u' \cdot v'}{\abs{u'} \, \abs{v'}}.
\end{equation}
It is often convenient to parametrize the scale factor by an exponential
\begin{equation}
	\Omega := \e^{\omega}.
\end{equation}

Considering an infinitesimal transformation
\begin{equation}
	\delta x^\mu = \xi^\mu,
\end{equation}
\index{conformal Killing!equation}%
the condition \eqref{cft:eq:sym-diff} becomes the conformal Killing equation
\begin{equation}
	\label{cft:eq:ck-eq-D}
	\delta g_{\mu\nu}
		= \mc L_\xi g_{\mu\nu}
		= \grad_\mu \xi_\nu + \grad_\nu \xi_\mu
		= \frac{2}{d} \, g_{\mu\nu} \grad_\rho \xi^\rho,
\end{equation}
such that the scale factor is
\begin{equation}
	\Omega^2 = 1 + \frac{2}{d} \, \grad_\rho \xi^\rho.
\end{equation}
\index{conformal Killing!vector}%
The vector fields $\xi$ satisfying this equation are called conformal Killing vectors (CKV).
Conformal transformations form a global subgroup of the diffeomorphism group: the generators of the transformations do depend on the coordinates, but the parameters do not (for an internal global symmetry, both the generators and the parameters don't depend on the coordinates).

\index{isometry group}%
The conformal group contains the isometry group $\group{ISO}(\mc M)$ of $\mc M$ as a subgroup, corresponding to the case $\Omega = 1$:
\begin{equation}
	\group{ISO}(\mc M) \subset \group{CISO}(\mc M).
\end{equation}
These transformations also preserve distances between points.
\index{Killing vector}%
The corresponding generators of infinitesimal transformations are called Killing vectors and satisfies the Killing equation
\begin{equation}
	\delta g_{\mu\nu}
		= \mc L_\xi g_{\mu\nu}
		= \grad_\mu \xi_\nu + \grad_\nu \xi_\mu
		= 0.
\end{equation}
They form a subalgebra of the CKV algebra.

\index{conformal algebra}%
An important point is to be made for the relation between infinitesimal and finite transformations: with spacetime symmetries it often happens that the first cannot be exponentiated into the second.
The reason is that the (conformal) Killing vectors may be defined only locally, i.e.\ they are well-defined in a given domain but have singularities outside.
When this happens, they do not lead to an invertible transformation, which cannot be an element of the group.
These notions are sometimes confused in physics and the term of “group” is used instead of “algebra”.
We shall be careful in distinguishing both concepts.

\begin{remark}[Isometries of $\mc M \subset \R^{p,q}$]
	\label{cft:rem:isometrie-subset-Rpq}

	In order to find the conformal isometries of a manifold $\mc M$ which is a subset of $\R^{p,q}$ defined in \eqref{cft:eq:def-Rpq}, it is sufficient to restrict the transformations of $\R^{p,q}$ to the subset $\mc M$~\cite{Schottenloher:2008:MathematicalIntroductionConformal}.
	In the process, not all global transformations generically survive.
	On the other hand, the algebra of local (infinitesimal) transformations for $\mc M$ and $\R^{p,q}$ are identical since $\mc M$ is locally like $\R^{p,q}$.
\end{remark}

\begin{check}

\subsection{Conformal field theory}

How to build a conformal field theory (CFT) for some matter fields $\Psi$, i.e.\ a QFT on the curved background $(M, \hat g)$ which is invariant under the conformal group $\group{CISO}(M)$?

To answer this question, we explain first how a background theory can be built from a more general theory.
A background $\hat B$ is a fixed field configuration which couples to the other fields but which does not have any intrinsic dynamics.
Typically, the background is a solution to the equations of motion derived from an action $S_B[B]$
\begin{equation}
	\frac{\delta S_B}{\delta B}(\hat B) = 0,
\end{equation}
but this is not necessary.\footnotemark{}
\footnotetext{%
	For example, the Einstein--Hilbert action with a cosmological constant can be expanded around the Minkowski spacetime, even if it is not a solution to the equation of motion.
}%
One can then consider probe fields $\Psi$ with action $S_p[\hat B, \Psi]$, i.e.\ fields which live on the background without disturbing it (this means that the backreaction is neglected).
If the background theory possesses a gauge symmetry, then any residual symmetry of the background generically becomes a global symmetry of the probe action (if the action $S_p[B, \Psi]$ is invariant under the gauge symmetry).
Note that in this case only the fields $\Psi$ transform under the symmetry.

Before considering the conformal transformations, consider first the invariance under the isometry of the background metric $\hat g$ (often a solution of Einstein equations).
Then, an action $S[\hat g, \Psi]$ which admits $\group{ISO}(M)$ as a global symmetry can be constructed by writing a diffeomorphism invariant action $S[g, \Psi]$ and by freezing the metric $g = \hat g$.
Indeed, isometries does not change the background metric and can be used as symmetries of $S$, where only $\Psi$ transforms.
The rest of the diffeomorphisms are not a symmetry of the action.

The extension to $\group{CISO}(M)$ requires more work because the conformal transformations modify the background metric and does not directly give global symmetries of $S$.
\index{symmetry!Weyl}%
The solution is to introduce an additional gauge symmetry
\begin{equation}
	g'_{\mu\nu}(x) = \e^{2 \omega(x)} g_{\mu\nu}(x),
	\qquad
	\Psi'(x) = \e^{d_\Psi \omega(x)} \Psi(x),
\end{equation}
called the (local) Weyl symmetry, where $d_\Psi$ is the dimension of the field $\Psi$.
The group of Weyl transformations is denoted by $\group{Weyl}(M)$.
If this is a symmetry of the original theory, then the scaling factor $\Omega(x)$ in front of the metric in \eqref{cft:eq:sym-diff} can be compensated with a Weyl transformation.
As a consequence, an action $S[\hat g, \Psi]$ invariant under the conformal group $\group{CISO}(M)$ can be obtained from an action $S[g, \Psi]$ invariant under diffeomorphisms and Weyl transformations.
Then, the conformal group can be understood as the subgroup of the diffeomorphism which transforms the metric like a Weyl transformation.
This fact has been encountered in \Cref{chap:bos:ws-int-vac} from a different perspective.

\begin{remark}
	One may want to reverse the argument by starting with $S[\hat g, \Psi]$ to derive actions $S[g, \Psi]$ which are invariant under diffeomorphisms and (local) Weyl transformations.

	The standard procedure to construct a diffeomorphism invariant theory from $S[\eta, \phi]$ is to use the minimal coupling of the field $\Psi$ by replacing derivatives with covariant derivatives.
	But, not all actions $S[g, \Psi]$ can be found in this way: indeed any term in $S[g, \Psi]$ which vanishes upon fixing the background (for example, a term proportional to the equations of motion if the background is a solution) cannot be recovered from the minimal coupling.

	For Weyl transformations, the subject is more complicated: a necessary condition is that $S[\eta, \Psi]$ be invariant under conformal transformations in flat space, but this condition is sufficient only if the action is at most quadratic in the first derivatives.
	On the other hand, global Weyl transformations require only invariance under global scale transformations.
	These remarks can be important in the construction of string worldsheet theories and they are related to \Cref{bos:rem:gauged-weyl} \cpageref{bos:rem:gauged-weyl}.
	Selected references on this topic are~\cite{Coleman:1971:WhyDilatationGenerators, Polchinski:1988:ScaleConformalInvariance, Iorio:1997:WeylGaugingConformalInvariance, Karananas:2016:WeylVsConformal, Karananas:2016:WeylRicciGauging, Farnsworth:2017:WeylConformalInvariance}.
\end{remark}

\begin{remark}[Axiomatic formulations]
	More axiomatic formulations are given in~\cite{Ribault:2014:ConformalFieldTheory, Schottenloher:2008:MathematicalIntroductionConformal}.
\end{remark}

\end{check}

\section{CFT on Minkowski space}

In this section, we consider the case where $\mc M = \R^{p,q}$ ($D = p + q$) and where $g = \eta$ is the flat metric with signature $(p, q)$:
\begin{equation}
\label{cft:eq:def-Rpq}
\eta = \diag(\underbrace{- 1, \ldots, - 1}_{q}, \underbrace{1, \ldots, 1}_{p}).
\end{equation}
The conformal Killing equation becomes
\begin{equation}
	\big( \eta_{\mu\nu} \lap + (D - 2) \pd_\mu \pd_\nu \big) \pd \cdot \epsilon = 0,
\end{equation}
where $\Delta$ is the $D$-dimensional Beltrami--Laplace operator for the metric $\eta_{\mu\nu}$.
The case $D = 2$ is relegated to the next chapter.
For $D > 2$, one finds the following transformations:
\begin{subequations}
\label{cft:eq:sym-conf-Rpq-inf}
\begin{align}
	\text{translation:}&
		\qquad
		\xi^\mu = a^\mu,
	\\
	\text{rotation \& boost:}&
		\qquad
		\xi^\mu = \tensor{\omega}{^\mu_\nu} x^\nu,
	\\
	\text{dilatation:}&
		\qquad
		\xi^\mu = \lambda\, x^\mu,
	\\
	\text{SCT:}&
		\qquad
		\xi^\mu = b^\mu x^2 - 2 b \cdot x \, x^\mu,
\end{align}
\end{subequations}
where $\omega_{\mu\nu}$ is antisymmetric.
The rotations include Lorentz transformations and SCT means “special conformal transformation”.

All parameters $\{ a^\mu, \omega_{\mu\nu}, \lambda, b^\mu \}$ are constant.
The generators are respectively denoted by $\{ P_\mu, J_{\mu\nu}, D, K_\mu \}$.
The finite translations and rotations form the Poincaré group $\group{SO}(p, q)$, while the conformal group can be shown to be $\group{SO}(p+1, q+1)$:
\begin{equation}
	\group{ISO}(\R^{p,q}) = \group{SO}(p, q),
	\qquad
	\group{CISO}(\R^{p,q}) = \group{SO}(p+1, q+1).
\end{equation}
The dimension of this group is
\begin{equation}
	\dim \group{SO}(p+1, q+1) = \frac{1}{2}\, (p + q + 2) (p + q + 1).
\end{equation}

\begin{draft}

\begin{exercise}
	Prove \eqref{cft:eq:sym-conf-Rpq-inf}, then write the finite transformations.
\end{exercise}

\begin{exercise}
	Compute the algebra of the generators and show the equivalence with $\alg{so}(p+1,q+1)$.
\end{exercise}

\end{draft}

\refchapter

\begin{itemize}
	\item References on higher-dimensional CFTs are~\cite{DiFrancesco:1999:ConformalFieldTheory, Rychkov:2016:EPFLLecturesConformal, SimmonsDuffin:2016:TASILecturesConformal, Qualls:2015:LecturesConformalField, Schottenloher:2008:MathematicalIntroductionConformal}.
\end{itemize}

\chapter{Conformal field theory on the plane}
\label{chap:cft:plane}

\introchapter

Starting with this chapter, we focus on two-dimensional Euclidean CFTs on the complex plane (or equivalently the sphere).
We start by describing the geometry of the sphere and the relation to the complex plane and to the cylinder, in order to make contact with the string worldsheet.
Then, we discuss classical CFTs and the Witt algebra obtained by classifying the conformal isometries of the complex plane.
Then, we describe quantum CFTs and introduce the operator formalism.
This last section is the most important for this \revname{} as it includes information on the operator product expansion, Hilbert space, Hermitian and BPZ conjugations.

As described at the beginning of \Cref{chap:cft:general}, we use spacetime notations for the coordinates, but follow otherwise the normalization for the worldsheet.
In particular, integrals are normalized by $2\pi$.
However, the spatial coordinate on the cylinder is still written as $\sigma$ to avoid confusions: $x^\mu = (\tau, \sigma)$.

\section{The Riemann sphere}
\label{cft:sec:plane:sphere}

\subsection{Map to the complex plane}

\index{Riemann sphere $S^2$}%
The Riemann sphere $\Sigma_0$, which is diffeomorphic to the unit sphere $S^2$, has genus $g = 0$ and is thus the simplest Riemann surface.
\index{Riemann sphere $S^2$!complex plane map}%
\index{extended complex plane $\bar \C$}%
Its most straightforward description is obtained by mapping it to the extended\footnotemark{} complex plane $\bar \C$ (also denoted $\hat \C$), which is the complex plane $z \in \C$ to which the point at infinity $z = \infty$ is added:
\footnotetext{%
	This qualification will often be omitted.
}%
\begin{equation}
	\bar \C = \C \cup \{ \infty \}.
\end{equation}
One speaks about “\emph{the} point at infinity” because all the points at infinity (i.e.\ the points $z$ such that $\abs{z} \to \infty$)
\begin{equation}
	\lim_{r \to \infty} r \, \e^{\I \theta}
		:= \infty
\end{equation}
are identified (the limit is independent of $\theta$).

\begin{figure}[tp]
	\begin{adjustwidth}{-2cm}{-2cm}
	\centering
	\begin{subfigure}[c]{.3\linewidth}
		\centering
		\includegraphics[scale=0.8]{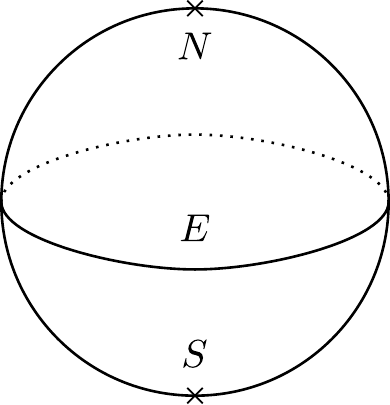}
	\end{subfigure}
	\qquad
	$\xrightarrow{\makebox[1.5cm]{}}$
	\qquad
	\begin{subfigure}[c]{.4\linewidth}
		\centering
		\includegraphics[scale=0.8]{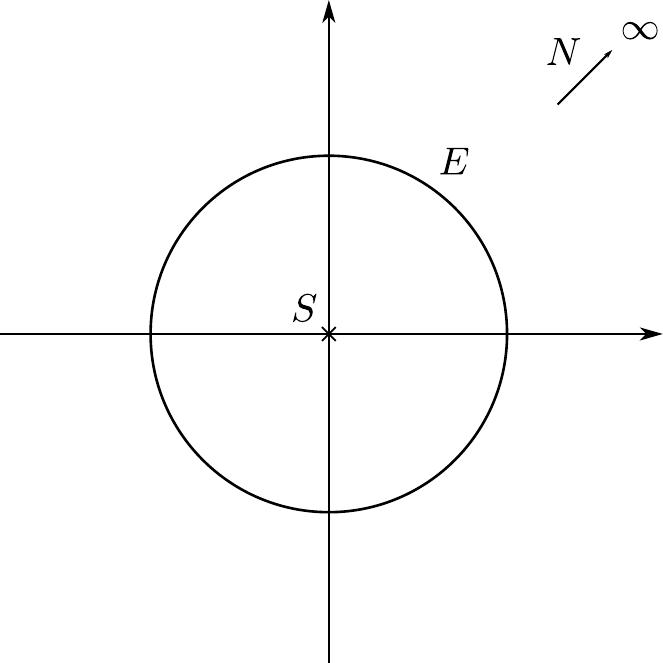}
	\end{subfigure}
	\end{adjustwidth}
	\caption{%
		Map from the Riemann sphere to the complex plane.
		The south and north poles are denoted by the letter $S$ and $N$, and the equatorial circle by $E$.
	}%
	\label{cft:fig:map-sphere-plane}
\end{figure}

The identification can be understood by mapping (say) the south pole to the origin of the plane and the north pole to infinity\footnotemark{} (\Cref{cft:fig:map-sphere-plane}) through the stereographic projection
\footnotetext{%
	Note that the points are distinguished in order to write the map, but they have nothing special by themselves (i.e.\ they are not punctures).
}%
\begin{equation}
	\label{cft:eq:var-stereo-z-theta-phi}
	z = \e^{\I \phi} \cot \frac{\theta}{2},
\end{equation}
where $(\theta, \phi)$ are angles on the sphere.
Any circle on the sphere is mapped to a circle in the complex plane.
Conversely, the Riemann sphere can be viewed as a compactification of the complex plane.

\begin{draft}

\begin{exercise}[Stereographic projection in Cartesian coordinates]
	Find \eqref{cft:eq:var-stereo-z-theta-phi} in Cartesian coordinates.
\end{exercise}

\end{draft}

\index{complex coordinates}%
Introducing Cartesian coordinates $(x, y)$ related to the complex coordinates by\footnotemark{}
\footnotetext{%
	General formulas can be found in \Cref{bos:sec:ws-int-complex:geom-C} by replacing $(\tau, \sigma)$ with $(x, y)$.
	In most cases, the conformal factor is set to zero ($\phi = 0$) in this chapter.
}%
\begin{subequations}
\begin{gather}
	z = x + \I y,
	\qquad
	\bar z = x - \I y,
	\\
	x = \frac{z + \bar z}{2},
	\qquad
	y = \frac{z - \bar z}{2 \I},
\end{gather}
\end{subequations}
the metric reads
\begin{equation}
	\label{cft:eq:metric-plane}
	\dd s^2
		= \dd x^2 + \dd y^2
		= \dd z \dd \bar{z}.
\end{equation}

The relations between the derivatives in the two coordinate systems are easily found:
\begin{equation}
	\pd
		:= \pd_{z}
		= \frac{1}{2} \, (\pd_x - \I \pd_y),
	\qquad
	\bar\pd
		:= \pd_{\bar z}
		= \frac{1}{2} \, (\pd_x + \I \pd_y).
\end{equation}
The indexed form will be used when there is a risk of confusion.
If the index is omitted then the derivative acts directly to the field next to it, for example
\begin{equation}
	\pd \phi(z_1) \pd \phi(z_2)
		:= \pd_{z_1} \pd_{z_2} \phi(z_1) \phi(z_2).
\end{equation}
Generically, the meromorphic and anti-meromorphic parts of a object will be denoted without and with a bar, see \eqref{cft:eq:def-T-Tb} for an example.

The extended complex plane $\bar \C$ can be covered by two coordinate patches $z \in \C$ and $w \in \C$.
In the first, the point at infinity (north pole) is removed, in the second, the origin (south pole) is removed.
On the overlap, the transition function is
\begin{equation}
	w = \frac{1}{z}.
\end{equation}
This description avoids to work with the infinity: studying the behaviour of $f(z)$ at $z = \infty$ is equivalent to study $f(1/w)$ at $w = 0$.

Since any two-dimensional metric is locally conformally equivalent to the flat metric, it is sufficient to work with this metric in each patch.
This is particularly convenient for the Riemann sphere since one patch covers it completely except for one point.

\subsection{Relation to the cylinder -- string theory}
\label{cft:sec:plane:sphere:cylinder}

\index{worldsheet!cylinder}%
The worldsheet of a closed string propagating in spacetime is locally topologically a cylinder $\R \times S^1$ of circumference $L$.
In this section, we show that the cylinder can also be mapped to the complex plane -- and thus to the Riemann sphere -- after removing two points.
Since the cylinder has a clear physical interpretation in string theory, it is useful to know how to translate the results from the plane to the cylinder.

It makes also sense to define two-dimensional models on the cylinder independently of a string theory interpretation since the compactification of the spatial direction from $\R$ to $S^1$ regulates the infrared divergences.
Moreover, it leads to a natural definition of a “time” and of an Hamiltonian on the Euclidean plane.

Denoting the worldsheet coordinates in Lorentzian signature by $(t, \sigma)$ with\footnotemark{}
\footnotetext{%
	Consistently with the comments at the beginning of \Cref{chap:cft:general}, the Lorentzian worldsheet time is denoted by $t$ instead of $\tau_M$.
}%
\begin{equation}
	\label{cft:eq:def-coord-cyl}
	t \in \R,
	\qquad
	\sigma \in [0, L),
	\qquad
	\sigma \sim \sigma + L,
\end{equation}
the metric reads
\begin{equation}
	\dd s^2 = - \dd t^2 + \dd \sigma^2
		= - \dd \sigma^+ \dd \sigma^-,
\end{equation}
where the light-cone coordinates
\begin{equation}
	\dd \sigma^\pm = \dd t \pm \dd \sigma
\end{equation}
have been introduced.
It is natural to perform a Wick rotation from the Lorentzian time $t$ to the Euclidean time
\begin{equation}
	\tau = \I t,
\end{equation}
and the metric becomes
\begin{equation}
	\dd s^2 = \dd \tau^2 + \dd \sigma^2.
\end{equation}

\index{complex coordinates!cylinder}%
It is convenient to introduce the complex coordinates
\begin{equation}
	w = \tau + \I \sigma,
	\qquad
	\bar w = \tau - \I \sigma
\end{equation}
for which the metric is
\begin{equation}
	\dd s^2 = \dd w \dd \bar{w}.
\end{equation}
\index{light-cone coordinates}%
Note that the relation to Lorentzian light-cone coordinates are
\begin{equation}
	\label{cft:eq:coord-cyl-wts}
	w = \I (t + \sigma)
		= \I \sigma^+,
	\qquad
	\bar w = \I (t - \sigma)
		= \I \sigma^-.
\end{equation}
\index{left/right-moving sectors}%
\index{holomorphic/anti-holomorphic sectors}%
Hence, an (anti-)holomorphic function of $w$ ($\bar w$) depends only on $\sigma^+$ ($\sigma^-$) before the Wick rotation: this leads to the identification of the left- and right-moving sectors with the holomorphic and anti-holomorphic sectors of the theory.

\index{Riemann sphere $S^2$!cylinder map}%
The cylinder can be mapped to the complex plane through
\begin{equation}
	\label{cft:eq:coord-cyl-wz}
	z = \e^{2\pi w / L},
	\qquad
	\bar z = \e^{2\pi \bar{w} / L},
\end{equation}
and the corresponding metric is
\begin{equation}
	\label{cft:eq:metric-plane-z2}
	\dd s^2 = \left( \frac{L}{2\pi} \right)^2 \, \frac{\dd z \dd \bar{z}}{\abs{z}^2}.
\end{equation}
A conformal transformation brings this metric to the flat metric \eqref{cft:eq:metric-plane}.
The conventions for the various coordinates and maps vary in the different textbooks.
We have gathered in \Cref{cft:tab:conventions-coord} the three main conventions and which references use which.

The map from the cylinder to the plane is found by sending the bottom end (corresponding to the infinite past $t \to - \infty$) to the origin of the plane, and the top end (infinite future $t \to \infty$) to the infinity.
Since the cylinder has two boundaries (its two ends) the map excludes the point $z = 0$ and $z = \infty$ and one really obtains the space $\bar \C - \{0, \infty \} = \C^*$.
This space can, in turn, be mapped to the $2$-punctured Riemann sphere $\Sigma_{0,2}$.

\begin{figure}[tp]
	\centering
	\begin{subfigure}[c]{.1\linewidth}
		\centering
		\rotatebox{90}{%
			\includegraphics[scale=0.8]{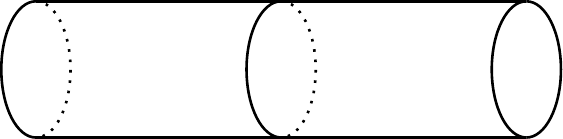}
		}
	\end{subfigure}
	\quad
	$\xrightarrow{\makebox[1cm]{}}$
	\quad
	\begin{subfigure}[c]{.25\linewidth}
		\centering
		\rotatebox{90}{%
			\includegraphics[scale=0.8]{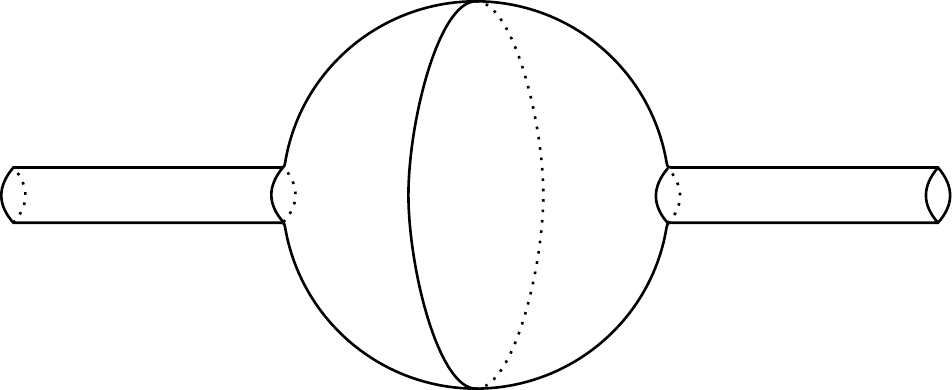}
		}
	\end{subfigure}
	\quad
	$\xrightarrow{\makebox[1cm]{}}$
	\quad
	\begin{subfigure}[c]{.25\linewidth}
		\centering
		\rotatebox{90}{%
			\includegraphics[scale=0.8]{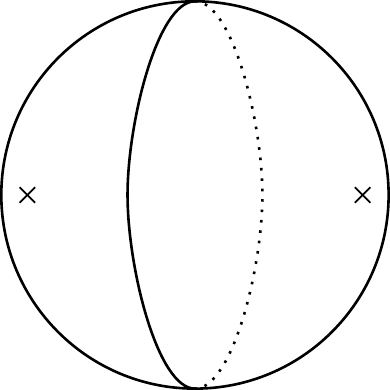}
		}
	\end{subfigure}
	\caption{%
		Map from the cylinder to the sphere with two tubes, to the $2$-punctured sphere $\Sigma_{0,2}$.
	}%
	\label{cft:fig:map-cylinder-S02}
\end{figure}

\begin{figure}[tp]
	\centering
	\begin{subfigure}[c]{.3\linewidth}
		\centering
		\includegraphics[scale=0.8]{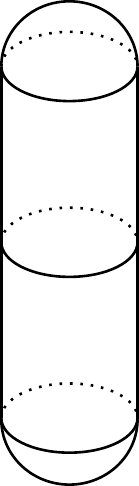}
	\end{subfigure}
	\qquad
	$\xrightarrow{\makebox[1.5cm]{}}$
	\qquad
	\begin{subfigure}[c]{.4\linewidth}
		\centering
		\includegraphics[scale=0.8]{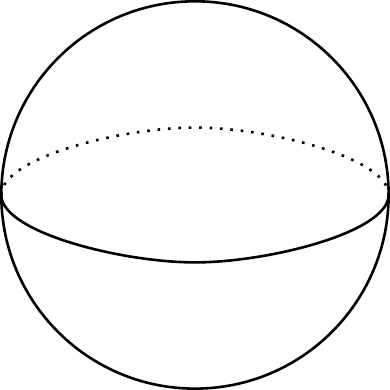}
	\end{subfigure}
	\caption{%
		Map from the cylinder with two caps (half-spheres) to the Riemann sphere $\Sigma_0$.
	}%
	\label{cft:fig:map-cylinder-sphere}
\end{figure}

The physical interpretation for the difference between $\Sigma_0$ and $\Sigma_{0,2}$ is simple: since one considers the propagation of a string, it means that the worldsheet corresponds to an amplitude with two external states, which are the mapped to the sphere as punctures (\Cref{cft:fig:map-cylinder-S02}, \Cref{bos:sec:ws-int-amp:Mg:vertex}).
Removing the external states (yielding the tree-level vacuum amplitude) corresponds to gluing half-sphere (caps) at each end of the cylinder (\Cref{cft:fig:map-cylinder-sphere}).
Then, it can be mapped to the Riemann sphere without punctures.
As a consequence, the properties of tree-level string theory are found by studying the matter and ghost CFTs on the Riemann sphere.
Scattering amplitudes are computed through correlation functions of appropriate operators on the sphere.
This picture generalizes to higher-genus Riemann surfaces.
Moreover, since local properties of the CFT (e.g.\ the spectrum of operators) are determined by the conformal algebra, they will be common to all surfaces.

Mathematically, a difference between $\Sigma_0$ and $\Sigma_{0,2}$ had to be expected since the sphere has a positive curvature (and $\chi = - 2$) but the cylinder is flat (with $\chi = 0$).
Punctures contribute negatively to the curvature (and thus positively to the Euler characteristics).

\begin{remark}
	The coordinate $z$ is always used as a coordinate on the complex plane, but the corresponding metric may be different -- compare \eqref{cft:eq:metric-plane} and \eqref{cft:eq:metric-plane-z2}.
	As explained previously, this does not matter since the theory is insensitive to the conformal factor.
\end{remark}

\section{Classical CFTs}
\label{cft:sec:plane:classical-cft}

\index{conformal field theory!classical}%
In this section, we consider an action $S[\Psi]$ which is conformally invariant.
We first identify and discuss the properties of the conformal algebra and group, before explaining how a CFT is defined.

\subsection{Witt conformal algebra}
\label{cft:sec:plane:class-cft:witt}

\index{Witt algebra}%
Since the Riemann sphere is identified with the complex plane, they share the same conformal group and algebra.
Consider the metric \eqref{cft:eq:metric-plane}
\begin{equation}
	\dd s^2 = \dd z \dd \bar{z},
\end{equation}
then, any meromorphic change of coordinates
\begin{equation}
	z \longrightarrow
	z' = f(z),
	\qquad
	\bar z \longrightarrow
	\bar z' = \bar f(\bar z)
\end{equation}
is a conformal transformation since the metric becomes
\begin{equation}
	\dd s^2
		= \dd z' \dd \bar z'
		= \Abs{\frac{\dd f}{\dd z}}^{2} \dd z \dd \bar{z}.
\end{equation}

However, only holomorphic functions which are globally defined on $\bar \C$ are elements of the group.
At the algebra level, any holomorphic function $f(z)$ regular in a domain $D$ gives a well-defined transformation in this domain $D$.
Hence, the algebra is infinite-dimensional.
On the other hand, $f(z)$ is only meromorphic on $\C$ generically: it cannot be exponentiated to a group element.
We first characterize the algebra and then obtain the conditions to promote the local transformations to global ones.

Since the transformations are defined only locally, it is sufficient to consider an infinitesimal transformation
\begin{equation}
	\label{cft:eq:sym-inf-coord-z}
	\delta z = v(z),
	\qquad
	\delta \bar{z} = \bar v(\bar z),
\end{equation}
where $v(z)$ is a meromorphic vector field on the Riemann sphere.
Indeed, the conformal Killing equation \eqref{cft:eq:ck-eq-D} in $D = 2$ is equivalent to the Cauchy--Riemann equations:
	\begin{equation}
		\bar\pd v = 0,
		\qquad
		\pd \bar v = 0.
	\end{equation}
The vector field admits a Laurent series
\begin{equation}
	v(z)
		= \sum_{n \in \Z} v_n z^{n+1},
	\qquad
	\bar v(\bar z)
		= \sum_{n \in \Z} \bar v_n \bar z^{n+1},
\end{equation}
and the $v_n$ and $\bar v_n$ are to be interpreted as the parameters of the transformation.
A basis of vectors (generators) is:
\begin{equation}
	\ell_n = - z^{n+1} \pd_{z},
	\qquad
	\bar\ell_n = - \bar z^{n+1} \pd_{\bar z},
	\qquad
	n \in \Z.
\end{equation}
One can check that each set of generators satisfies the Witt algebra
\begin{equation}
	\label{cft:eq:algebra-witt}
	\com{\ell_m}{\ell_n} = (m - n) \ell_{m+n},
	\qquad
	\com{\bar\ell_m}{\bar\ell_n} = (m - n) \bar\ell_{m+n},
	\qquad
	\com{\ell_m}{\bar\ell_n} = 0.
\end{equation}

Since there are two commuting copies of the Witt algebra, it is natural to extend the ranges of the coordinates from $\C$ to $\C^2$ and to consider $z$ and $\bar{z}$ as independent variables.
In particular, this gives a natural action of the product algebra over $\C^2$.
This procedure will be further motivated when studying CFTs since the holomorphic and anti-holomorphic parts will generally split, and it makes sense to study them separately.
Ultimately, physical quantities can be extracted by imposing the condition $\bar z = \conj{z}$ at the end (the star is always reserved for the complex conjugation, the bar will generically denote an independent variable).
In that case, the two algebras are also related by complex conjugation.

Note that the variation of the metric \eqref{app:eq:var-inf-diffeo-metric} under a meromorphic change of coordinates \eqref{cft:eq:sym-inf-coord-z} becomes
\begin{equation}
	\label{cft:eq:var-inf-diffeo-metric-z}
	\delta g_{z \bar z} = \pd v + \bar\pd \bar v,
	\qquad
	\delta g_{zz} = \delta g_{\bar z \bar z}
		= 0.
\end{equation}

\subsection{\texorpdfstring{$\group{PSL}(2, \C)$}{PSL(2, C)} conformal group}

The next step is to determine the globally defined vectors and to study the associated group.

First, the conditions for a vector $v(z)$ to be well-defined at $z = 0$ are
\begin{equation}
	\lim_{\abs{z} \to 0} v(z) < \infty
	\quad \Longrightarrow \quad
	\forall n < - 1:
	\quad
	v_n = 0.
\end{equation}
The behaviour at $z = \infty$ can be investigated thanks to the map $z = 1 / w$
\begin{equation}
	v(1/w) = \frac{\dd z}{\dd w} \sum_n v_n w^{-n - 1},
\end{equation}
where the additional derivative arises because $v$ is a vector.
Then, the regularity conditions at $z = \infty$ are
\begin{equation}
	\lim_{\abs{z} \to \infty} v(z)
		= \lim_{\abs{w} \to 0} \frac{\dd z}{\dd w} \, v(1/w)
		= - \lim_{\abs{w} \to 0} \frac{v(1/w)}{w^2}
		< \infty
	\quad \Longrightarrow \quad
	\forall n > 1:
	\quad
	v_n = 0.
\end{equation}
As a result, the globally defined generators are
\begin{equation}
	\{ \ell_{-1}, \ell_0, \ell_1 \}
		\cup \{ \bar\ell_{-1}, \bar\ell_0, \bar\ell_1 \}
\end{equation}
where
\begin{equation}
	\ell_{-1}
		= - \pd_z,
	\qquad
	\ell_0
		= - z \pd_z,
	\qquad
	\ell_1
		= - z^2 \pd_z.
\end{equation}
\index{SL2C@$\mathrm{SL}(2, \C)$ group}%
It is straightforward to check that they form two copies of the $\alg{sl}(2, \C)$ algebra
\begin{equation}
	\label{cft:eq:algebra-sl2C}
	\com{\ell_0}{\ell_{\pm 1}}
		= \mp \ell_{\pm 1},
	\qquad
	\com{\ell_1}{\ell_{-1}}
		= 2 \ell_0.
\end{equation}
The global conformal group is sometimes called Möbius group:
\begin{equation}
	\group{PSL}(2, \C)
		:= \group{SL}(2, \C) / \Z_2
		\sim \group{SO}(3, 1),
\end{equation}
where the additional division by $\Z_2$ is clearer when studying an explicit representation.
It corresponds with $\ker P_1$ defined in \eqref{bos:eq:def-group-Kg}:
\begin{equation}\label{key}
	\mc K_0
		= \group{PSL}(2, \C).
\end{equation}

A matrix representation of $\group{SL}(2, \C)$ is
\begin{equation}
	\label{cft:eq:sl2C-elem-g}
	g =
	\begin{pmatrix}
		a & b \\
		c & d
	\end{pmatrix},
	\qquad
	a, b, c, d \in \C,
	\qquad
	\det g
		= a d - b c
		= 1,
\end{equation}
which shows that this group has six real parameters
\begin{equation}
	\K_0
		:= \dim \group{SL}(2, \C)
		= 6.
\end{equation}
The associated transformation on the complex plane reads
\begin{equation}
	\label{cft:eq:sl2C-transf-fg}
	f_g(z)
		= \frac{a z + b}{c z + d}.
\end{equation}
The quotient by $\Z_2$ is required since changing the sign of all parameters does not change the transformation.
These transformations have received different names: Möbius, projective, homographic, linear fractional transformations…

Holomorphic vector fields are then of the form
\begin{equation}
	\label{cft:eq:vector-SL2C}
	v(z) = \beta + 2 \alpha z + \gamma z^2,
	\qquad
	\bar v(\bar z) = \bar \beta + 2 \bar \alpha \bar z + \bar \gamma \bar z^2,
\end{equation}
where
\begin{equation}
	a = 1 + \alpha,
	\qquad
	b = \beta,
	\qquad
	c = - \gamma,
	\qquad
	d = 1 - \alpha.
\end{equation}

The finite transformations associated to \eqref{cft:eq:sym-conf-Rpq-inf} are:
\begin{subequations}
\label{cft:eq:sym-conf-SL2C-simple}
\begin{align}
	\text{translation:}&
		&
		f_g(z) &= z + a,
		&
		a &\in \C,
	\\
	\text{rotation:}&
		&
		f_g(z) &= \zeta \, z,
		&
		\abs{\zeta} &= 1,
	\\
	\text{dilatation:}&
		&
		f_g(z) &= \lambda \, z,
		&
		\lambda &\in \R,
	\\
	\text{SCT:}&
		&
		f_g(z) &= \frac{z}{c z + 1},
		&
		c &\in \C.
\end{align}
\end{subequations}
Investigation leads to the following association between the generators and transformations:
\begin{itemize}
	\item translation: $\ell_{-1}$ and $\bar\ell_{-1}$;
	\item dilatation (or radial translation): $(\ell_0 + \bar\ell_0)$;
	\item rotation (or angular translation): $\I (\ell_0 - \bar\ell_0)$;
	\item special conformal transformation: $\ell_1$ and $\bar\ell_1$.
\end{itemize}

\index{inversion map}%
The inversion defined by
\begin{subequations}
\label{cft:eq:inversion}
\begin{equation}
	\text{inversion:}
		\qquad
		I^+(z)
			:= I(z)
			:= \frac{1}{z}
\end{equation}
is not an element of $\group{SL}(2, \C)$.
However, the inversion with a minus sign
\begin{equation}
	I^-(z)
		:= - I(z)
		= I(-z)
		= - \frac{1}{z}
\end{equation}
\end{subequations}
is a $\group{SL}(2, \C)$ transformation.

A useful transformation is the circular permutation of $(0, 1, \infty)$:
\begin{equation}
	\label{cft:eq:conf-transf-circ-perm}
	g_{\infty,0,1}(z) = \frac{1}{1 - z}.
\end{equation}

\subsection{Definition of a CFT}

\index{conformal field theory!definition}%
A CFT is characterized by its set of (composite) fields (also called operators) $\mc O(z, \bar z)$ which correspond to any local expression constructed from the fields $\Psi$ appearing in the Lagrangian and of their derivatives.\footnotemark{}
\footnotetext{%
	Not all CFTs admit a Lagrangian description.
	But, since we are mostly interested in string theories defined from Polyakov's path integral, it is sufficient to study CFTs with a Lagrangian.
}%
For example, in a scalar field theory, the simplest operators are of the form $\pd^m \phi^n$.

\index{primary operator/state}%
\index{quasi-primary operator/state}%
Among the operators, two particular categories are distinguished according to their transformation laws:
\begin{itemize}
	\item primary operator:
	\begin{equation}
		\label{cft:eq:transf-primary}
		\text{$\forall f$ meromorphic}:
			\qquad
			\mc O(z, \bar z)
				= \left( \frac{\dd f}{\dd z} \right)^{h} \left( \frac{\dd \bar f}{\dd \bar z} \right)^{\bar h} \mc O'\big( f(z), \bar f(\bar z) \big),
	\end{equation}

	\item quasi-primary (or $\group{SL}(2, \C)$ primary) operator:
	\begin{equation}
		\label{cft:eq:transf-quasi-primary}
		\forall f \in \group{PSL}(2, \C):
			\qquad
			\mc O(z, \bar z)
				= \left( \frac{\dd f}{\dd z} \right)^{h} \left( \frac{\dd \bar f}{\dd \bar z} \right)^{\bar h} \mc O'\big( f(z), \bar f(\bar z) \big).
	\end{equation}
\end{itemize}
\index{conformal!weight}%
The parameters $(h, \bar h)$ are the conformal weights of the operator $\mc O$ (both are independent from each other), and combinations of them give the conformal dimension $\Delta$ and spin $s$:
\index{conformal!dimension}%
\index{conformal!spin}%
\begin{equation}
	\Delta
		:= h + \bar h,
	\qquad
	s
		:= h - \bar h.
\end{equation}
The conformal weights correspond to the charges of the operator under $\ell_0$ and $\bar \ell_0$.
We will use “$(h, \bar h)$ (quasi-)primary” as a synonym of “(quasi-)primary field with conformal weight $(h, \bar h)$”.

\begin{remark}[Complex conformal weights]
	\label{cft:rem:complex-weights}
	\index{Liouville!theory}%

	While we consider $h, \bar h \in \R$, and more specifically $h, \bar h \ge 0$ for a unitary theory (which is the case of string theory except for the reparametrization ghosts), theories with $h, \bar h \in \C$ make perfectly sense.
	One example is the Liouville theory with complex central charge $c \in \C$~\cite{Ribault:2014:ConformalFieldTheory, Ribault:2015:LiouvilleTheoryCentral} (central charges are defined below, see \eqref{cft:eq:algebra-virasoro}).
\end{remark}

Primaries and quasi-primaries are hence operators which have nice transformations respectively under the algebra and group.
Obviously, a primary is also a quasi-primary.
These transformations are similar to those of a tensor with $h$ holomorphic and $\bar h$ anti-holomorphic indices (\Cref{bos:sec:ws-int-complex:geom-C}).
Another point of view is that the object
\begin{equation}
	\mc O(z, \bar z) \, \dd z^{h} \dd \bar z^{\bar h}
\end{equation}
is invariant under local / global conformal transformations.

\index{conformal field theory!finite transformation}%
The notation $f \circ \mc O$ indicates the complete change of coordinates, including the tensor transformation law and the possible corrections if the operator is not primary.\footnotemark{}
\footnotetext{%
	In fact, one has $f \circ \mc O := f^* \mc O$ in the notations of \Cref{chap:bos:ws-int-vac}.
}%
\index{primary operator/state!finite transformation}%
For a primary field, we have:
\begin{equation}
	\label{cft:eq:transf-coord-circ}
	f \circ \mc O(z, \bar z)
		:= f'(z)^{h} \bar f'(\bar z)^{\bar h} \, \mc O'\big( f(z), \bar f(\bar z) \big).
\end{equation}
We stress that it does not correspond to function composition.

Under an infinitesimal transformations
\begin{equation}
	\delta z = v(z),
	\qquad
	\delta \bar z = \bar v(\bar z),
\end{equation}
a primary operator changes as
\begin{equation}
	\label{cft:eq:transf-inf-primary}
	\delta \mc O(z, \bar z)
		= (h \, \pd v + v \, \pd) \mc O(z, \bar z)
			+ (\bar h \, \bar \pd \bar v + \bar v \, \bar \pd) \mc O(z, \bar z).
\end{equation}
The transformation of a non-primary field contains additional terms, see for example \eqref{cft:eq:transf-inf-T}.

\begin{check}
\begin{remark}[Group versus algebra]
	The question is whether one requires the theory to be invariant under the global transformations or rather under local transformations.
	Theories arising from gauge fixing a local Weyl invariance (which is the case for string theory) leads naturally to a local invariance.
	More generally, one expects that a local field theory is sensitive only to local properties.
	On the other hand, this assumption may be too strong (e.g.\ in statistical physics or in systems without a Lagrangian formulation).
	But, making the assumption that only the local properties matter can be useful for a preliminary study.
	This is really because the two-dimensional algebra is infinite-dimensional that so many models can be solved exactly in two dimensions.
	Useful discussions can be found in~\cites{Nakayama:2019:ConformalEquationsThat}[sec.~1.3]{Ribault:2014:ConformalFieldTheory}{Schottenloher:2008:MathematicalIntroductionConformal}.
\end{remark}
\end{check}

\begin{remark}[Higher-genus Riemann surfaces]
	\index{higher-genus Riemann surface!conformal group}%
	According to \Cref{cft:rem:isometrie-subset-Rpq}, all Rie\-mann surfaces $\Sigma_g$ share the same conformal algebra since locally they are all subsets of $\R^2$.
	On the other hand, one finds that no global transformations are defined for $g > 1$, and only the subgroup $\group{U}(1) \times \group{U}(1)$ survives for the torus.
\end{remark}

The most important operator in a CFT is the energy--momentum tensor $T_{\mu\nu}$, if it exists as a local operator.
According to \Cref{bos:sec:ws-int:action}, this tensor is conserved and traceless
\begin{equation}
	\grad^\nu T_{\mu\nu} = 0,
	\qquad
	g^{\mu\nu} T_{\mu\nu} = 0.
\end{equation}
The traceless equation in components reads
\begin{equation}
	g^{\mu\nu} T_{\mu\nu}
		= 4 \, T_{z \bar z}
		= T_{xx} + T_{yy}
		= 0
\end{equation}
which implies that the off-diagonal component vanishes in complex coordinates
\begin{equation}
	T_{z \bar z} = 0.
\end{equation}
Then, the conservation equation yields
\begin{equation}
	\pd_{z} T_{\bar z \bar z} = 0,
	\qquad
	\pd_{\bar z} T_{zz} = 0,
\end{equation}
such that the non-vanishing components $T_{zz}$ and $T_{\bar z \bar z}$ are respectively holomorphic and anti-holomorphic.
This motivates the introduction of the notations:
\begin{equation}
	\label{cft:eq:def-T-Tb}
	T(z)
		:= T_{zz}(z),
	\qquad
	\bar T(\bar z)
		:= T_{\bar z \bar z}(\bar z).
\end{equation}
This is an example of the factorization between the holomorphic and anti-holomorphic sectors.

Currents are local objects and thus one expects to be able to write an infinite number of such currents associated to the Witt algebra.
Applying the Noether procedure gives
\begin{equation}
	J_v(z)
		:= J_v^{\bar z}(z)
		= - T(z) v(z),
	\qquad
	\bar J_v(\bar z)
		:= J_v^{z}(\bar z)
		= - \bar T(\bar z) \bar v(\bar z).
\end{equation}

\section{Quantum CFTs}
\label{cft:sec:plane:quantum-cft}

The previous section was purely classical.
The quantum theory is first defined through the path integral
\begin{equation}
	Z = \int \dd \Psi \, \e^{- S[\Psi]}.
\end{equation}
We will also develop an operator formalism.
The latter is more general than the path integral and allows to work without reference to path integrals and Lagrangians.
This is particularly fruitful as it extends the class of theories and parameter ranges (e.g.\ \Cref{cft:rem:complex-weights}) which can be studied.

\subsection{Virasoro algebra}

\index{Virasoro algebra}%
As discussed in \Cref{bos:sec:ws-int:faddeev-popov:weyl}, field measures in path integrals display a conformal anomaly, meaning that they cannot be defined without introducing a scale.
This anomaly can be traded for a gravitational anomaly by introducing counter-terms in the action~\cites{Fujikawa:1988:RegularizedBRSTcoordinateinvariantMeasure}[sec.~3.2]{Green:1988:SuperstringTheory-1}{Hatsuda:1990:RegularizedPhaseSpace, Guadagnini:1988:CentralChargeTrace, Knecht:1990:ShiftingWeylAnomaly, Jackiw:1995:AnotherViewMassless}.
As a consequence, the Witt algebra \eqref{cft:eq:algebra-witt} is modified to its central extension, the Virasoro algebra.\footnotemark{}
\footnotetext{%
	That the central charge in the Virasoro algebra indicates a diffeomorphism anomaly can be understood from the fact that
}%
\index{Virasoro operators}%
The generators in both sectors are denoted by $\{ L_n \}$ and $\{ \bar L_n \}$ and are called Virasoro operators (or modes).
The algebra is given by:
\index{commutator!$[L_m, L_n]$}%
\index{central charge}%
\begin{subequations}
\label{cft:eq:algebra-virasoro}
\begin{gather}
	\com{L_m}{L_n}
		= (m - n) L_{m+n} + \frac{c}{12} \, m (m - 1) (m + 1) \delta_{m+n},
	\\
	\com{\bar L_m}{\bar L_n}
		= (m - n) \bar L_{m+n} + \frac{\bar c}{12} \, m (m - 1) (m + 1) \delta_{m+n},
	\\
	\com{L_m}{\bar L_n} = 0,
	\qquad
	\com{c}{L_m} = 0,
	\qquad
	\com{\bar c}{\bar L_m} = 0,
\end{gather}
\end{subequations}
where $c, \bar c \in \C$ are the holomorphic and anti-holomorphic central charges.
Consistency of the theory on a curved space implies $\bar c = c$, but there is otherwise no constraint on the plane~\cite{Tong:2009:LecturesStringTheory, Green:1988:SuperstringTheory-1}.

The $\alg{sl}(2, \C)$ subalgebra is not modified by the central extension.
This means that states are still classified by eigenvalues of $(h, \bar h)$ of $(L_0, \bar L_0)$.

\begin{remark}
	\index{Liouville!theory}%

	In most models relevant for string theory, one finds that the central charges are real, $c, \bar c \in \R$.
	Moreover, unitarity requires them to be positive $c, \bar c > 0$, and only reparametrization ghosts do not satisfy this condition.
	On the other hand, it makes perfect sense to discuss general CFTs for $c, \bar c \in \C$ (the Liouville theory is such an example~\cite{Ribault:2014:ConformalFieldTheory, Ribault:2015:LiouvilleTheoryCentral}).
\end{remark}

\subsection{Correlation functions}

\index{correlation function}%
A $n$-point correlation function is defined by
\begin{equation}
	\Mean{ \prod_{i=1}^n \mc O_i(z_i, \bar z_i) }
		= \int \dd \Psi \, \e^{- S[\Psi]} \prod_{i=1}^n \mc O_i(z_i, \bar z_i),
\end{equation}
choosing a normalization such that $\mean{1} = 1$.
The path integral defines the time-ordered product (on the cylinder) of the corresponding operators.

Invariance under global transformations leads to strong constraints on the correlation functions.
\index{correlation function!quasi-primary operator}%
For quasi-primary fields, they transform under $\group{SL}(2, \C)$ as
\begin{equation}
	\Mean{ \prod_{i=1}^n \mc O_i(z_i, \bar z_i) }
		= \prod_{i=1}^n \left( \frac{\dd f}{\dd z}(z_i) \right)^{h_i} \left( \frac{\dd f}{\dd \bar z}(\bar z_i) \right)^{\bar h_i}
			\times \Mean{ \prod_{i=1}^n \mc O_i\big(f(z_i), \bar f(\bar z_i) \big) }.
\end{equation}
Considering an infinitesimal variation \eqref{cft:eq:transf-inf-primary} yields a differential equation for the $n$-point function
\begin{equation}
	\delta \Mean{ \prod_{i=1}^n \mc O_i(z_i, \bar z_i) }
		= \sum_{i=1}^n \big( h_i \pd_i v(z_i) + v(z_i) \pd_i + \cc \big)
			\Mean{ \prod_{i=1}^n \mc O_i(z_i, \bar z_i) }
		= 0,
\end{equation}
where $\pd_i := \pd_{z_i}$ and $v$ is a vector \eqref{cft:eq:vector-SL2C} of $\alg{sl}(2, \C)$.
These equations are sufficient to determine completely the forms of the $1$-, $2$- and $3$-point functions of quasi-primaries:
\index{correlation function!sphere 1-point (-)}%
\index{correlation function!sphere 2-point (-)}%
\index{correlation function!sphere 3-point (-)}%
\begin{subequations}
\label{cft:eq:corr-func}
\begin{gather}
	\label{cft:eq:corr-func-1pt}
	\Mean{\mc O_i(z_i, \bar z_i)}
		= \delta_{h_i, 0} \delta_{\bar h_i, 0},
	\\
	\label{cft:eq:corr-func-2pt}
	\Mean{\mc O_i(z_i, \bar z_i) \mc O_j(z_j, \bar z_j)}
		= \delta_{h_i, h_j} \delta_{\bar h_i, \bar h_j} \,
			\frac{g_{ij}}{z_{ij}^{2 h_i} \bar z_{ij}^{2 \bar h_i}},
	\\
	\label{cft:eq:corr-func-3pt}
	\begin{multlined}
		\Mean{\mc O_i(z_i, \bar z_i) \mc O_j(z_j, \bar z_j) \mc O_k(z_k, \bar z_k)}
			= \frac{C_{ijk}}{z_{ij}^{h_i + h_j - h_k} z_{jk}^{h_j + h_k - h_i} z_{ki}^{h_i + h_k - h_j}}
			\\
				\times \frac{1}{\bar z_{ij}^{\bar h_i + \bar h_j - \bar h_k} \bar z_{jk}^{\bar h_j + \bar h_k - \bar h_i} \bar z_{ki}^{\bar h_i + \bar h_k - \bar h_j}},
	\end{multlined}
\end{gather}
\end{subequations}
where we have defined
\begin{equation}
	z_{ij} = z_i - z_j.
\end{equation}
\index{structure constant (CFT)}%
\index{Zamolodchikov metric}%
The coefficients $C_{ijk}$ are called structure constants and the matrix $g_{ij}$ defines a metric (Zamolodchikov metric) on the space of fields.
The metric is often taken to be diagonal $g_{ij} = \delta_{ij}$, which amounts to use an orthonormal eigenbasis of $L_0$ and $\bar L_0$.
The vanishing of the $1$-point function of a non-primary quasi-primary holds only on the plane: for example the value on the cylinder can be non-zero since the map is not globally defined -- see in particular \eqref{cft:eq:L0-cyl}.

\begin{remark}[Logarithmic CFTs]
	\index{logarithmic CFT}%

	Logarithmic CFTs display a set of unusual properties~\cite{Gurarie:1993:LogarithmicOperatorsConformal, Kogan:1996:WorldSheetLogarithmicOperators, Gaberdiel:2003:AlgebraicApproachLogarithmic, Flohr:2003:BitsPiecesLogarithmic, Flohr:1996:ModularInvariantPartition}.
	In particular, the correlation functions are not of the form displayed above.
	The most striking feature of those theories is that the $L_0$ operator is non-diagonalisable (but it can be set in the Jordan normal form).
\end{remark}

\begin{remark}[Fake identity]
	\label{cft:rk:fake-identity}
	\index{primary operator/state!weight-0}%

	Usually, the only primary operator with $h = \bar h = 0$ is the identity $1$.
	While this is always true for unitary theories, there are non-unitary theories ($c \le 1$ Liouville theory, SLE, loop models) where there is another field (called the indicator, marking operator, or also fake identity) with $h = \bar h = 0$~\cite{Harlow:2011:AnalyticContinuationLiouville, Delfino:2011:ThreepointConnectivityTwodimensional, Picco:2013:ConnectivitiesPottsFortuinKasteleyn, Ribault:2014:ConformalFieldTheory, Ribault:2015:LiouvilleTheoryCentral, Ikhlef:2016:ThreepointFunctionsC, Bautista:2019:QuantumGravityTimelike}.
	The main difference between both fields is that the identity is a degenerate field (it has a null descendant), whereas the other operator with $h = \bar h = 0$ is not.
	Such theories will not be considered in this \revname{}.
	Operators with $h = \hbar = 0$ can also be built by comining several CFTs, and they play a very important role in string theory since they describe on-shell states.
\end{remark}

Finally, the $4$-point function is determined up to a function of a single variable $x$ and its complex conjugate:
\begin{equation}
	\label{cft:eq:corr-func-4pt}
	\Mean{\prod_{i=1}^4 \mc O_i(z_i, \bar z_i)}
		= f(x, \bar x) \prod_{i < j} \frac{1}{z_{ij}^{(h_i + h_j) - h / 3}} \times \cc
\end{equation}
where
\begin{equation}
	h
		:= \sum_{i=1}^4 h_i,
	\qquad
	\bar h
		:= \sum_{i=1}^4 \bar h_i.
\end{equation}
\index{cross-ratio}%
The cross-ratio $x$ is $\group{SL}(2, \C)$ invariant and reads
\begin{equation}
	x
		:= \frac{z_{12} z_{34}}{z_{13} z_{24}}.
\end{equation}
The interpretation is that the $\group{SL}(2, \C)$ invariance allows to fix $3$ of the points to an arbitrary value, and the final result does not depend on this choice.

\begin{draft}

\begin{exercise}
	Prove these formulas.
\end{exercise}

\end{draft}

\section{Operator formalism and radial quantization}
\label{cft:sec:plane:operator}

\index{radial quantization}%
\emph{Radial quantization} is a convenient description of a CFT on the plane in terms of operators.
It relies on the maps given in \Cref{cft:sec:plane:sphere:cylinder}:
\begin{equation}
	z = \e^{\tau + \I \sigma}
		= x + \I y.
\end{equation}
Taking the physical spacetime to be the cylinder, every question is rephrased on the complex plane in order to exploit the powerful tools from complex analysis.
The term “radial quantization” comes from the fact that time translation of the cylinder
\begin{equation}
	\tau \longrightarrow \tau + T
\end{equation}
corresponds to dilatation on the plane
\begin{equation}
	z \longrightarrow \e^{T} z.
\end{equation}
Thus, time evolution on the cylinder and radial evolution (from the origin to the complex infinity) are identified.
In particular, the Hamiltonian of the system of the plane is
\begin{equation}
	H = \frac{2\pi}{L} (L_0 + \bar L_0),
\end{equation}
since the RHS is the dilatation operator.
The cylinder length $L$ was defined in \eqref{cft:eq:def-coord-cyl}.
The theory is quantized according to this Hamiltonian.
In the string theory language, a state with $H = 0$ is said to be on-shell:
\index{on-shell condition}%
\begin{equation}
	\text{on-shell state:}
	\qquad
	h + \bar h = 0.
\end{equation}

\subsection{Radial ordering and commutators}

Time-ordering in $\tau$ becomes radial ordering in the plane:
\begin{equation}
	R\big(A(z) B(w)\big)
		=
		\begin{cases}
			A(z) B(w) & \abs{z} > \abs{w},
			\\
			(-1)^F \, B(w) A(z) & \abs{w} > \abs{z},
		\end{cases}
\end{equation}
where $F = 0$ ($F = 1$) for bosonic (fermionic) operators.
Radial ordering will often be kept implicit.

\begin{figure}[ht]
	\centering
	\includegraphics[scale=1.2]{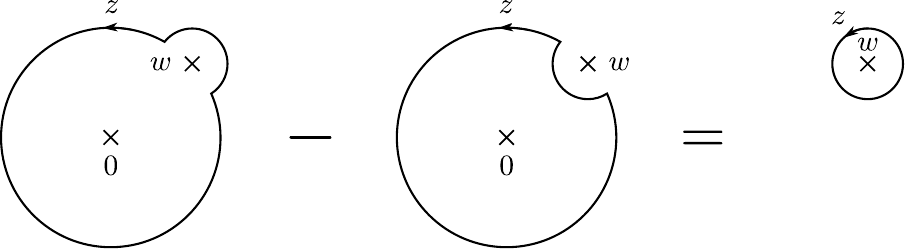}
	\caption{Graphical proof of \eqref{cft:eq:com-int-A}.}
	\label{cft:fig:variation-com-field-contour}
\end{figure}

\index{commutator!CFT|(}%
The equal-time (anti-)commutator becomes an equal radius commutator defined by point-splitting:
\begin{equation}
	\com{A(z)}{B(w)}_{\pm,\abs{z} = \abs{w}}
		= \lim_{\delta \to 0} \big( A(z) B(w)|_{\abs{z} = \abs{w} + \delta} \pm B(w) A(z)|_{\abs{z} = \abs{w} - \delta} \big).
\end{equation}
If $A$ and $B$ are two operators which can be written as the contour integrals of $a(z)$ and $b(z)$ (corresponding to integral over closed curves on the cylinder)
\begin{equation}
	A = \oint_{C_0} \frac{\dd z}{2\pi \I}\, a(z),
	\qquad
	B = \oint_{C_0} \frac{\dd z}{2\pi \I}\, b(z),
\end{equation}
then one finds the following commutators:
\begin{subequations}
\label{cft:eq:com-int-A}
\begin{gather}
	\label{cft:eq:com-int-A-B}
	\com{A}{B}_\pm
		= \oint_{C_0} \frac{\dd w}{2\pi \I} \oint_{C_w} \frac{\dd z}{2\pi \I} \, a(z) b(w),
	\\
	\label{cft:eq:com-int-A-b}
	\com{A}{b(w)}_\pm
		= \oint_{C_w} \frac{\dd z}{2\pi \I} \, a(z) b(w).
\end{gather}
\end{subequations}
The contours $C_0$ and $C_w$ are respectively centered around the points $0$ and $w$.
For a proof, see \Cref{cft:fig:variation-com-field-contour}.
Since these are contour integrals in the complex plane, the Cauchy--Riemann formula \eqref{app:eq:cauchy-riemann} can be used to write the result as soon as one knows the poles of the above expression (ultimately, this amounts to pick the sum of residues).
In CFTs, the poles of such expressions are given by operator product expansions (OPE), defined below (\Cref{cft:sec:plane:operator:ope}).

\index{commutator!CFT|)}%

\index{conserved current!CFT}%
Given a conserved current $j^\mu$
\begin{equation}
	\pd_\mu j^\mu
		= \pd j^{z} + \bar\pd j^{\bar z}
		= 2 (\pd j_{\bar z} + \bar\pd j_{z})
		= 0,
\end{equation}
\index{conserved charge!CFT}%
the associated conserved charge is defined by
\begin{equation}
	\label{cft:eq:charge-contour}
	Q = \frac{1}{2\pi \I} \oint_{C_0} (j_{z} \dd z - j_{\bar z} \dd \bar z),
\end{equation}
where $C_0$ denotes the anti-clockwise contour around $z = 0$ (equivalently the interior of the contour is located to the left).
The difference of sign in the second term follows directly from Stokes' theorem \eqref{app:eq:stokes-thm-2d-complex} (and can be understood as a conjugation of the contour).
The additional factor of $1/2\pi$ is consistent with the normalization of spatial integrals in two dimensions.
The current components are not necessarily holomorphic and anti-holomorphic at this level, but in practice this will often be the case (and each component is independently conserved), and one writes
\begin{equation}
	j(z)
		:= j_{z}(z),
	\qquad
	\bar \jmath(\bar z)
		:= j_{\bar z}(\bar z).
\end{equation}
In this case, the charge also splits into a holomorphic and an anti-holomorphic (left- and right-moving\footnotemark{}) contributions
\footnotetext{%
	For charges, we use subscript $L$ and $R$ to distinguish both sectors to avoid introducing a new symbol for the total charge.
	However, since $Q = Q_L$ in the holomorphic sector, it is often not necessary to distinguish between the two symbols when acting on an operator or a state (however, this is useful for writing mode expansions).
	We do not write a bar on $Q_R$ because the charges don't depend on the position.
}%
\begin{equation}
	Q = Q_L + Q_R,
	\qquad
	Q_L
		:= \frac{1}{2\pi \I} \oint_{C_0} j(z) \dd z,
	\qquad
	Q_R
		:= - \frac{1}{2\pi \I} \oint_{C_0} \bar\jmath(\bar z) \dd \bar z.
\end{equation}
The infinitesimal variation of a field under the symmetry generated by $Q$ reads
\begin{equation}
	\label{cft:eq:variation-field-com}
	\delta_{\epsilon} \mc O(z, \bar z)
		= - \com{\epsilon Q}{\mc O(z, \bar z)}
		= - \epsilon \oint_{C_z} \frac{\dd w}{2\pi \I} \, j(w) \mc O(z, \bar z)
			+ \epsilon \oint_{C_z} \frac{\dd \bar w}{2\pi \I} \, \bar\jmath(\bar w) \mc O(z, \bar z).
\end{equation}
The contour integrals are easily evaluated once the OPE between the current and the operator is known.
This formula gives the infinitesimal variation under the transformation for any field, not only for primaries.

\begin{computation}[cft:eq:charge-contour]
	In real coordinates, the charge is defined by integrating the time component of the current $j^\mu$ over space for fixed time \eqref{app:eq:charge}:
	\[
		Q
			= \frac{1}{2\pi} \int \dd \sigma \, j^0.
	\]
	The first step is to rewrite this formula covariantly.
	Since the time is fixed on the slice, $\dd\tau = 0$ and one can write
	\[
		Q
			= \frac{1}{2\pi} \int (\dd \sigma \, j^0 - \dd \tau \, j^1)
			= - \frac{1}{2\pi} \int \epsilon_{\mu\nu} j^\mu \, \dd x^\nu.
	\]
	The last formula is valid for any contour.
	Moreover, it can be evaluated for complex coordinates:
	\[
		Q
			= - \frac{1}{2\pi} \oint \epsilon_{z \bar z} \big( j^z \, \dd \bar z - j^{\bar z} \, \dd z \big)
			= - \frac{\I}{4\pi} \oint \big( j^z \, \dd \bar z - j^{\bar z} \, \dd z \big)
			= - \frac{1}{2\pi \I} \oint \big( j_z \, \dd z - j_{\bar z} \, \dd \bar z \big).
	\]
	One finds a contour integral because $\tau = \text{cst}$ circles of the cylinder are mapped to $\abs{z} = \text{cst}$ contours.
\end{computation}

\subsection{Operator product expansions}
\label{cft:sec:plane:operator:ope}

\index{operator product expansion}%
The operator product expansion (OPE) is a tool used frequently in CFT: it means that when two local operators come close to each other, it is possible to replace their product by a sum of local operators
\begin{equation}
	\label{cft:eq:ope}
	\mc O_i(z_i, \bar z_i) \mc O_j(z_j, \bar z_j)
		= \sum_k \frac{c_{ij}^k}{z_{ij}^{h_i + h_j - h_k} \bar z_{ij}^{\bar h_i + \bar h_j - \bar h_k}} \,
			\mc O_k(z_j, \bar z_j),
\end{equation}
where the OPE coefficients $c_{ij}^k$ are some constants and the sum runs over all operators.
When $\mc O_k$ is primary, the coefficients $c_{ij}^k$ are related to the structure constants and the field metric by
\begin{equation}
	C_{ijk}
		= g_{k\ell} c_{ij}^\ell.
\end{equation}
The radius of convergence for the OPE is given by the distance to the nearest operators in the correlation function.
The OPE defines an associative algebra (commutative for bosonic operators), and the holomorphic sector forms a subalgebra (called the chiral algebra).

\begin{example}[OPE with the identity]
	\index{operator product expansion!identity-primary}%

	The OPE of a field $\phi(z)$ with the identity $1$ is found by a direct series expansion
	\begin{equation}
		\phi(z) 1 = \sum_{n \in \N} \frac{(z - w)^n}{n!} \, \pd^n \phi(w).
	\end{equation}
	Obviously there are no singular terms.
\end{example}

\medskip

Starting from this point we consider only the holomorphic sector except when stated otherwise.
The formula for the OPE \eqref{cft:eq:ope} can be rewritten as
\begin{equation}
	\label{cft:eq:ope-AB}
	A(z) B(w)
		:= \sum_{n=-\infty}^N \frac{\{A B\}_n(z)}{(z - w)^n}
\end{equation}
to simplify the manipulations.
$N$ is an integer and there are singular terms if $N > 0$.
\index{$\sim$}%
Generally, only the terms singular as $w \to z$ are necessary in the computations (for example, to use the Cauchy--Riemann formula \eqref{app:eq:cauchy-riemann}): equality up to non-singular terms is denoted by a tilde
\begin{equation}
	\label{cft:eq:contraction-AB}
	A(z) B(w)
		\sim \sum_{n=1}^N \frac{\{A B\}_n(z)}{(z - w)^n}
		=: \wick{\c A(z) \c B(w)}.
\end{equation}
\index{contraction}%
The RHS of this expression defines the contraction of the operators $A$ and $B$.

While, most of the time, only singular terms are kept
\begin{equation}
	\phi_i(z_i) \phi_j(z_j)
		\sim \sum_k \theta(h_i + h_j - h_k) \,
			\frac{c_{ij}^k}{(z - w)^{h_i + h_j - h_k}} \, \phi_k(w)
\end{equation}
(with $\theta(x)$ the Heaviside step function), it can happen that one keeps also non-singular terms (the product of two OPE have singular terms coming from non-singular terms multiplying singular terms).
Explicit contractions of operators through the OPE is also denoted by a bracket when there are other operators.

\index{operator product expansion!$T$-primary}%
For a primary field $\phi(z)$, one finds the OPE with the energy--momentum tensor to be
\begin{equation}
	\label{cft:eq:ope-T-phi}
	T(z) \phi(w) \sim \frac{h \, \phi(w)}{(z - w)^2} + \frac{\pd \phi(w)}{z - w},
\end{equation}
where $h$ is the conformal weight of the field.
This OPE together with \eqref{cft:eq:variation-field-com} for $j(z) = - v(z) T(z)$ correctly reproduces \eqref{cft:eq:transf-inf-primary}.

\begin{computation}[cft:eq:transf-inf-primary]
	\begin{align*}
		\delta \phi(z)
			&= \oint_{C_z} \frac{\dd w}{2\pi \I} \, v(w) T(w) \phi(z)
			\sim \oint_{C_z} \frac{\dd w}{2\pi \I} \, v(w) \left( \frac{h \, \phi(z)}{(w - z)^2} + \frac{\pd \phi(z)}{w - z} \right)
			\\
			&= h \, \pd v(z) \, \phi(z) + v(z) \pd \phi(z).
	\end{align*}
\end{computation}

For a non-primary operator, the OPE becomes more complicated (as it is reflected by the transformation property), but the conformal weight can still be identified at the term in $z^{-2}$.
\index{central charge}%
The most important example is the energy--momentum tensor: the central charge is found as the coefficient of the $z^{-4}$ term its OPE with itself:
\index{operator product expansion!$TT$}%
\begin{equation}
	\label{cft:eq:ope-T-T}
	T(z) T(w) \sim \frac{c / 2}{(z - w)^4} + \frac{2 T(w)}{(z - w)^2} + \frac{\pd T(w)}{z - w}.
\end{equation}
The OPE indicates that the conformal weight of $T$ is $h = 2$.
Using \eqref{cft:eq:variation-field-com} for $j(z) = - v(z) T(z)$, one finds the infinitesimal variation
\begin{equation}
	\label{cft:eq:transf-inf-T}
	\delta T = 2 \, \pd v \, T + v \, \pd T + \frac{c}{12} \, \pd^3 v,
\end{equation}
The last term vanishes for global transformations: this translates the fact that $T$ is only a quasi-primary.
\index{energy--momentum tensor!finite transformation}%
The finite form of this transformation is
\begin{equation}
	\label{cft:eq:transf-T}
	T'(w) = \left( \frac{\dd z}{\dd w} \right)^{-2} \left( T(z) - \frac{c}{12} \, S(w, z) \right)
		= \left( \frac{\dd z}{\dd w} \right)^{-2} T(z) + \frac{c}{12} \, S(z, w)
\end{equation}
\index{Schwarzian derivative}%
where $S(w, z)$ is the Schwarzian derivative
\begin{equation}
	\label{cft:eq:Schwarzian}
	S(w, z) = \frac{w^{(3)}}{w'} - \frac{3}{2} \, \left( \frac{w''}{w'} \right)^{2},
\end{equation}
where the derivatives of $w$ are with respect to $z$.
This vanishes if the transformation is in $\group{SL}(2, \C)$, and it transforms as
\begin{equation}
	S(u, z) = S(w, z) + \left( \frac{\dd w}{\dd z} \right)^{2} S(u, w)
\end{equation}
under successive changes of coordinates.

\begin{computation}[cft:eq:transf-inf-T]
	\begin{align*}
		\delta T(z)
			&= \oint_{C_z} \frac{\dd w}{2\pi \I} \, v(w) T(w) T(z)
			\sim \oint_{C_z} \frac{\dd w}{2\pi \I} \, v(w) \left( \frac{c / 2}{(z - w)^4} + \frac{2 T(w)}{(z - w)^2} + \frac{\pd T(w)}{z - w} \right)
			\\
			&= \frac{c}{2 \times 3!} \, \pd^3 v(z)
				+ 2 \pd v(z) \, T(z) + v(z) \pd T(z).
	\end{align*}
\end{computation}

\subsection{Hermitian and BPZ conjugation}

In this section, we introduce two different notions of conjugations: one is adapted for amplitudes because it defines a unitary Euclidean time evolution, while the second is more natural as an inner product of CFT states.
Both can be interpreted as providing a map from in-states to out-states on the cylinder.

\index{Euclidean adjoint|(}%
\index{Hermitian adjoint}%
Given an operator $\mc O$, we need to define an operation $\eadj{\mc O}$ -- called \emph{Euclidean adjoint} (or simply adjoint) -- which, after Wick rotation from Euclidean to Lorentzian signature, can be interpreted as the Hermitian adjoint.\footnotemark{}
\footnotetext{%
	In~\cite{Polchinski:2005:StringTheory-1}, it is denoted by a bar on top of the operator: we avoid this notation since the bar already denotes the anti-holomorphic sector.
	In~\cite{Zwiebach:1993:ClosedStringField}, it is indicated by a subscript $hc$.
	Otherwise, in most of the literature, it has no specific symbol since one directly works with the modes.
}%
This is necessary in order to define a Hermitian inner-product and to impose reality conditions.

To motivate the definition, consider first the cylinder in Lorentzian signature.
Since Hermitian conjugation does not affect the Lorentzian coordinates, the Euclidean time must reverse its sign:
\begin{equation}
	\adj{t}
		= - \I \adj{\tau}
		= t
	\quad \Longrightarrow \quad
	\adj{\tau}
		= - \tau.
\end{equation}
Hence, an appropriate definition of the Euclidean adjoint is an Hermitian conjugation together with time reversal.\footnotemark{}
\footnotetext{%
	The Euclidean adjoint can be used to define an inner product: positive-definiteness of the latter is called \emph{reflection positivity} or OS-positive and is a central axiom of constructive QFT.
}%
Another point of view is that the time evolution operator $U(\tau) := \e^{- \tau H}$ is not unitary when $H$ is Hermitian $\adj{H} = H$: the solution is to define a new Euclidean adjoint $\eadj{U(\tau)} := \adj{U(-\tau)}$ such that $U(\tau)$ is unitary for it.

Time reversal on the cylinder corresponds to inversion and complex conjugation on the complex plane:
\begin{equation}
	z
		\xrightarrow{\tau \to - \tau} \e^{- \tau + \I \sigma}
		= \frac{1}{\conj{z}}
		= I(\bar z),
\end{equation}
where $I(z) = 1/z$ is the inversion \eqref{cft:eq:inversion}.\footnotemark{}
\footnotetext{%
	We do not write “$\adj{z}$” because this notation is confusing as one should not complex conjugate the factor of $\I$ in the exponential (\Cref{cft:sec:plane:class-cft:witt}).
}%
On the real surface\footnotemark{} $\bar z = \conj{z}$, which leads to the definition of the Euclidean adjoint as follows:
\footnotetext{%
	Remember that $\bar z$ is not the complex conjugate of $z$ but an independent variable.
}%
\begin{equation}
	\label{cft:eq:conj-euc}
	\eadj{\mc O(z, \bar z)}
		:= \adj{\big(\bar I \circ \mc O(z, \bar z) \big)},
\end{equation}
where $\bar I(z) := 1 / \bar z$.
If $\mc O$ is quasi-primary, we have:
\begin{equation}
	\label{cft:eq:conj-euc-qprim}
	\eadj{\mc O(z, \bar z)}
		= \adj{\left[\frac{1}{\bar z^{2 h} z^{2 \bar h}} \,
			\mc O\left(\frac{1}{\bar z}, \frac{1}{z} \right) \right]}
		= \frac{1}{z^{2 h} \bar z^{2 \bar h}} \,
			\adj{\mc O}\left(\frac{1}{z}, \frac{1}{\bar z} \right),
\end{equation}
The last equality shows that Euclidean conjugation is equivalent to take the conjugate of all factors of $\I$ but otherwise leaves $z$ and $\bar z$ unaffected.
The Euclidean adjoint acts by complex conjugation of any $c$-number and reverses the order of the operators (acting as a transpose):
\begin{equation}
	\eadj{(\lambda \, \mc O_1 \cdots \mc O_n)}
		= \conj{\lambda} \, \eadj{\mc O_n} \cdots \eadj{\mc O_1},
	\qquad
	\lambda \in \C,
\end{equation}
without any sign.

\index{Euclidean adjoint|)}%

\index{BPZ conjugation}%
A second operation, called the \emph{BPZ conjugation}, is useful.
It can be defined in two different ways:
\begin{equation}
	\label{cft:eq:conj-bpz}
	\mc O(z, \bar z)^t
		:= I^\pm \circ \mc O(z, \bar z)
		= \frac{(\mp 1)^{h + \bar h}}{z^{2 h} \bar z^{2 \bar h}} \, \mc O\left(\pm \frac{1}{z}, \pm \frac{1}{\bar z} \right),
\end{equation}
\index{inversion map}%
where $I^\pm(z) = \pm 1 / z$ is the inversion \eqref{cft:eq:inversion}.
The minus and plus signs are respectively more convenient when working with the open and closed strings.\footnotemark{}
\footnotetext{%
	The index $t$ should not be confused with the matrix transpose: it is used in opposition with $\ddagger$ and $\dagger$ to indicate that no complex conjugation is involved.
}%
The BPZ conjugation does not complex conjugate $c$-number nor changes the order of the operators:\footnotemark{}
\footnotetext{%
	However, the fields become anti-radially ordered after a BPZ conjugation since it sends $z$ to $1/z$.
	The radial ordering can be restored by (anti-)commuting the fields, which can introduce additional signs~\cite{Sen:2016:RealitySuperstringField}.
	This problem does not arise when working in terms of the modes.
}%
\begin{equation}
	(\lambda \, \mc O_1 \cdots \mc O_n)^t
		= \lambda \, \mc O_1^t \cdots \mc O_n^t,
	\qquad
	\lambda \in \C.
\end{equation}

The identity is invariant under both conjugation
\begin{equation}
	\eadj{1} = 1^t = 1.
\end{equation}

\subsection{Mode expansion}
\label{cft:chap:plane:radial:modes}

\index{mode expansion}%
Any field of weight $(h, \bar h)$ can be expanded in terms of modes $\mc O_{m,n}$
\begin{equation}
	\mc O(z, \bar z)
		= \sum_{m,n}
			\frac{\mc O_{m,n}}{z^{m + h} \bar z^{n + \bar h}}.
\end{equation}
Note that the modes $\mc O_{m,n}$ themselves are operators.
\index{mode expansion!mode range}%
\index{boundary condition}%
The ranges of the two indices are such that
\begin{equation}
	\label{cft:eq:mode-range}
	m + h \in \Z + \nu,
	\qquad
	n + \bar h \in \Z + \bar\nu,
	\qquad
	\nu, \bar\nu
		=
		\begin{cases}
			0 & \text{periodic}, \\
			1/2 & \text{anti-periodic}.
		\end{cases}
\end{equation}
The values of $\nu$ and $\bar\nu$ depend on whether the fields satisfies periodic or anti-periodic boundary conditions on the plane (for half-integer weights, the periodicity is reversed on the cylinder):
\begin{equation}
	\mc O(\e^{2\pi\I} z, \bar z)
		= \e^{2\pi\I \nu} \mc O(z, \bar z),
	\qquad
	\mc O(z, \e^{2\pi\I} \bar z)
		= \e^{2\pi\I \bar\nu} \mc O(z, \bar z).
\end{equation}
Depending on whether the weights are integers or half-integers, additional terminology is introduced:
\begin{itemize}
	\item If $h \in \Z + 1/2$, then one can choose anti-periodic (\emph{Neveu--Schwarz} or NS) or periodic (\emph{Ramond} or R) boundary conditions on the cylinder (reversed for the plane):
	\begin{equation}
		\nu, \bar\nu =
			\begin{cases}
				0 & \text{NS}
				\\
				1/2 & \text{R}
			\end{cases}
	\end{equation}
	\index{boundary condition!Neveu--Schwarz (NS)}%
	\index{boundary condition!Ramond (R)}%
	The indices are half-integers (resp.\ integers) for the NS (R) sector.

	\item If $h \in \Z$, periodic (or untwisted) boundary conditions are more natural, but anti-periodic boundary conditions may also be considered:
	\begin{equation}
		\nu, \bar\nu =
			\begin{cases}
				0 & \text{untwisted}
				\\
				1/2 & \text{twisted}
			\end{cases}
	\end{equation}
	\index{boundary condition!untwisted}%
	\index{boundary condition!twisted}%
	The modes of untwisted (resp.\ twisted) fields have integer (half-integers) indices.
\end{itemize}
The mode expansions have no branch cut (fractional power of $z$ or $\bar z$) for periodic fields (bosonic untwisted or fermionic twisted).
We will see explicit examples of such operators later in this \revname{}.

\index{Euclidean adjoint!mode}%
Under Euclidean conjugation \eqref{cft:eq:conj-euc}, the modes are related by
\begin{equation}
	(\eadj{\mc O})_{-m,-n}
		= \adj{(\mc O_{m,n})}.
\end{equation}
In particular, if the operator is Hermitian (under the Euclidean adjoint), the reality condition on the modes relates the negative modes with the conjugated positive modes
\begin{equation}
	\eadj{\mc O}
		= \mc O
	\quad \Longrightarrow \quad
	\adj{(\mc O_{m,n})}
		= \mc O_{-m,-n}.
\end{equation}
When no confusion is possible (for Hermitian operators), we will write $\adj{\mc O_{m,n}}$ instead of $\adj{(\mc O_{m,n})}$.

For a holomorphic field $\phi(z)$, the above expansion becomes
\begin{equation}
	\label{cft:eq:ghost-modes-phi}
	\phi(z) = \sum_{n \in \Z + h + \nu} \frac{\phi_n}{z^{n + h}}.
\end{equation}
Conversely, the modes are recovered from the field through
\begin{equation}
	\phi_n = \oint_{C_0} \frac{\dd z}{2\pi \I} \, z^{n + h - 1} \phi(z),
\end{equation}
where the integration is counter-clockwise around the origin.

\index{mode expansion!Hermiticity}%
If the field is Hermitian, then
\begin{equation}
	\label{cft:eq:modes-adj}
	\eadj{\phi} = \phi
	\quad \Longrightarrow \quad
	\adj{(\phi_n)} = \phi_{-n}.
\end{equation}
The operators $\phi_n$ have a conformal weight of $- n$ (since the weight of $z$ is $- 1$).
\index{BPZ conjugation!mode}%
The BPZ conjugate of the modes is
\begin{equation}
	\label{cft:eq:modes-bpz}
	\phi_n^t = (I^\pm \circ \phi)_n
		= (- 1)^{h} (\pm 1)^{n} \phi_{-n}.
\end{equation}

\begin{computation}[cft:eq:modes-bpz]
	\begin{align*}
		\phi_n^t
			&= (I^\pm \circ \phi)_n
			= \oint \frac{\dd z}{2\pi \I} \, z^{n + h - 1} I^\pm \circ \phi(z)
			\\
			&= \oint \frac{\dd z}{2\pi \I} \, z^{n + h - 1} \left(\mp \frac{1}{z^2} \right)^h \phi\left(\pm \frac{1}{z} \right)
			\\
			&= (\mp 1)^h \oint \frac{\dd z}{2\pi \I} \, z^{n - h - 1} \phi\left(\pm \frac{1}{z} \right)
			\\
			&= (\mp 1)^h \oint \frac{\dd w}{2\pi \I} \, \left(\pm \frac{1}{w}\right)^{n - h} w^{- 1} \phi(w)
			\\
			&= (\mp 1)^h (\pm 1)^{n - h} \oint \frac{\dd w}{2\pi \I} \, w^{- n + h - 1} \phi(w),
	\end{align*}
	where we have set $w = \pm 1 / z$ such that
	\begin{equation}
		\frac{\dd z}{z} = \mp \frac{\dd w}{w^2 z}
			= - \frac{\dd w}{w},
	\end{equation}
	and the minus sign disappears upon reversing the contour orientation.
\end{computation}

\medskip

\index{energy--momentum tensor!mode expansion}%
\index{Virasoro operators}%
The mode expansion of the energy--momentum tensor is
\begin{equation}
	T(z) = \sum_{n \in \Z} \frac{L_n}{z^{n + 2}},
	\qquad
	L_n = \oint \frac{\dd z}{2\pi \I} \, T(z) z^{n + 1},
\end{equation}
where one recognizes the Virasoro operators as the modes.
\index{Virasoro operators!Hermiticity}%
In most situations, the Virasoro operators are Hermitian
\begin{equation}
	\adj{L_n} = L_{-n}.
\end{equation}
The OPE \eqref{cft:eq:ope-T-T} and \eqref{cft:eq:ope-T-phi} together with \eqref{cft:eq:com-int-A-B} help to reconstruct the Virasoro algebra \eqref{cft:eq:algebra-virasoro} and the commutation relations between the $L_m$ and the modes $\phi_n$ of a weight $h$ primary:
\index{commutator!$[L_m, \phi_n]$}%
\begin{equation}
	\label{cft:eq:com-Ln-phin}
	\com{L_m}{\phi_n} = \big( m (h - 1) - n \big) \phi_{m+n}.
\end{equation}
This easily gives the commutation relation for the complete field:
\begin{equation}
	\label{cft:eq:com-Ln-phi}
	\com{L_m}{\phi(z)} = z^m \big(z \pd + (n + 1) h \big) \phi(z).
\end{equation}
We will often use \eqref{cft:eq:algebra-virasoro} and \eqref{cft:eq:com-Ln-phin} for $m = 0$:
\begin{equation}
	\label{cft:eq:com-L0}
	\com{L_0}{L_{-n}} = n L_{-n},
	\qquad
	\com{L_0}{\phi_{-n}} = n \phi_{-n}.
\end{equation}
This means that both $\phi_{n}$ and $L_{n}$ act as raising operators for $L_0$ if $n < 0$, and as lowering operators if $n > 0$ (remember that $L_0$ is the Hamiltonian in the holomorphic sector).
When both the holomorphic and anti-holomorphic sectors enter, it is convenient to introduce the combinations
\index{L@$L^\pm$}%
\begin{equation}
	\label{cft:eq:Ln-pm}
	L_n^\pm = L_n \pm \bar L_n,
\end{equation}
such that $L_0^+$ is the Hamiltonian.

\index{conserved current!mode expansion}%
Finally, every holomorphic current $j(z)$ has a conformal weight $h = 1$ and can be expanded as
\begin{equation}
	\label{cft:eq:modes-j}
	j(z) = \sum_n \frac{j_n}{z^{n+1}}.
\end{equation}
\index{conserved charge}%
By definition, the zero-mode is equal to the holomorphic charge
\begin{equation}
	Q_L = j_0.
\end{equation}

\subsection{Hilbert space}
\label{cft:sec:plane:operator:states}

\index{Hilbert space!CFT}%
The Hilbert space of the CFT is denoted by $\mc H$.
\index{SL2Cv@$\mathrm{SL}(2, \C)$ vacuum}%
The $\group{SL}(2, \C)$ (or \emph{conformal}) \emph{vacuum}\footnotemark{} $\ket{0}$ is defined by the state which is invariant under the global conformal transformations:
\footnotetext{%
	There are different notions of “vacuum”, see \eqref{cft:eq:vacuum-energy}.
	However, the $\group{SL}(2, \C)$ vacuum is unique.
	Indeed, it is mapped to the unique identity operator under the state--operator correspondence (however, there can be other states of weight $0$, see \Cref{cft:rk:fake-identity}).
}%
\begin{equation}
	\label{cft:eq:vacuum-conformal}
	L_0 \ket{0} = 0,
	\qquad
	L_{\pm 1} \ket{0} = 0.
\end{equation}
Expectation value of an operator $\mc O$ in the $\group{SL}(2, \C)$ vacuum is denoted as:
\index{$\mean{\cdot}$}%
\begin{equation}
	\mean{\mc O}
		:= \bra{0} \mc O \ket{0}.
\end{equation}

If the fields are expressed in terms of creation and annihilation operators (which happens e.g.\ for free scalars, free fermions and ghosts), then the Hilbert space has the structure of a Fock space.

\subsubsection{State--operator correspondence}

\index{state--operator correspondence}%
The state--operator correspondence identifies every state $\ket{\mc O}$ of the CFT Hilbert space with an operator $\mc O(z, \bar z)$ through
\begin{equation}
	\ket{\mc O} = \lim_{z, \bar z \to 0} \mc O(z, \bar z) \ket{0}
		= \mc O(0, 0) \ket{0}.
\end{equation}
Such a state can be interpreted as an “in” state since it is located at $\tau \to - \infty$ on the cylinder.
\index{state (CFT)!ket}%
Focusing now on a holomorphic field $\phi(z)$, the state is defined as
\begin{equation}
	\ket{\phi} = \lim_{z \to 0} \phi(z) \ket{0}
		= \phi(0) \ket{0}.
\end{equation}
For this to make sense, the modes which diverge as $z \to 0$ must annihilate the vacuum.
In particular, for a weight $h$ field $\phi(z)$, one finds:
\begin{equation}
	\label{cft:eq:modes-vacuum}
	\forall n \ge - h + 1:
	\quad
	\phi_n \ket{0} = 0.
\end{equation}
Thus, the $\phi_{n}$ for $n \ge - h + 1$ are annihilation operators for the vacuum $\ket{0}$, and conversely the states $\phi_{n}$ with $n < - h + 1$ are creation operators.
As a consequence, the state $\ket{\phi}$ is found by applying the mode $n = - h$ to the vacuum:
\begin{equation}
	\ket{\phi}
		= \phi_{-h} \ket{0}
		= \oint \frac{\dd z}{2\pi\I} \, \frac{\phi(z)}{z} \ket{0}.
\end{equation}
Since $L_{-1}$ is the generator of translations on the plane, one finds
\begin{equation}
	\phi(z) \ket{0}
		= \e^{z L_{-1}} \phi(0) \e^{- z L_{-1}} \ket{0}
		= \e^{z L_{-1}} \ket{\phi}.
\end{equation}
The vacuum $\ket{0}$ is the state associated to the identity $1$.
Translating the conditions \eqref{cft:eq:modes-vacuum} to the energy--momentum tensor gives
\begin{equation}
	\forall n \ge - 1:
	\quad
	L_n \ket{0} = 0.
\end{equation}
This is consistent with the definition \eqref{cft:eq:vacuum-conformal} since it includes the $\group{sl}(2, \C)$ subalgebra.

\index{energy vacuum}%
If $h < 0$, some of the modes with $n > 0$ do not annihilate the vacuum: \eqref{cft:eq:com-L0} implies that some states have an energy lower than the one of $\ket{0}$.
The state $\ket{\Omega}$ (possibly degenerate) with the lowest energy is called the \emph{energy vacuum}
\begin{equation}
	\label{cft:eq:vacuum-energy}
	\forall \ket{\phi} \in \mc H:
		\qquad
		\bra{\Omega} L_0 \ket{\Omega} \le \bra{\phi} L_0 \ket{\phi}.
\end{equation}
It is obtained by acting repetitively with the modes $\phi_{n > 0}$.
This vacuum defines a new partition of the non-zero-modes operators into annihilation and creation operators.
If there are zero-modes, i.e.\ $n = 0$ modes, then the vacuum is degenerate since they commute with the Hamiltonian, $\com{L_0}{\phi_0} = 0$ according to \eqref{cft:eq:com-L0}.
The partition of the zero-modes into creation and annihilation operators depends on the specific state chosen among the degenerate vacua.

\index{zero-point energy}%
The energy $a_\Omega$ of $\ket{\Omega}$, which is also its $L_0$ eigenvalue
\begin{equation}
	\label{cft:eq:0pt-energy}
	L_0 \ket{\Omega}
		:= a_\Omega \ket{\Omega},
\end{equation}
is called \emph{zero-point energy}.
Bosonic operators with negative $h$ are dangerous because they lead to an infinite negative energy together with an infinite degeneracy (from the zero-mode).

The conjugate vacuum is defined by BPZ or Hermitian conjugation
\begin{equation}
	\bra{0}
		= \eadj{\ket{0}}
		= \ket{0}^t
\end{equation}
since both leave the identity invariant.
It is also annihilated by the $\alg{sl}(2, \C)$ subalgebra:
\begin{equation}
	\bra{0} L_0 = 0,
	\qquad
	\bra{0} L_{\pm 1} = 0.
\end{equation}
Since there are two kinds of conjugation, two different conjugated states can be defined.
They are also called “out” states since they are located at $\tau \to \infty$ on the cylinder (\Cref{cft:fig:map-cylinder-S02}).

\subsubsection{Euclidean and BPZ conjugations and inner products}

\index{Euclidean adjoint!state}%
\index{state (CFT)!bra}%
The Euclidean adjoint $\bra{\eadj{\mc O}}$ of the state $\ket{\mc O}$ is defined as
\begin{subequations}
\label{cft:eq:state-herm-O}
\begin{align}
	\bra{\eadj{\mc O}}
		&= \lim_{w, \bar w \to 0}
			\bra{0} \eadj{\mc O(w, \bar w)}
		= \lim_{w, \bar w \to 0}
			\frac{1}{w^{2 h} \bar w^{2 \bar h}}
			\bra{0} \adj{\mc O\left(\frac{1}{\bar w}, \frac{1}{w}\right)}
		\\
		&= \lim_{z, \bar z \to \infty} z^{2 h} \bar z^{2 \bar h}
			\bra{0} \adj{\mc O}(z, \bar z)
		\\
		&= \bra{0} I \circ \adj{\mc O}(0, 0),
\end{align}
\end{subequations}
where the two coordinate systems are related by $w = 1 / \bar z$.
From this formula, the definition of the adjoint of a holomorphic operator $\phi$ follows
\begin{subequations}
\label{cft:eq:state-herm-phi}
\begin{align}
	\bra{\eadj{\phi}}
		&= \lim_{\bar w \to 0}
			\bra{0} \eadj{\phi(w)}
		= \lim_{\bar w \to 0}
			\frac{1}{w^{2 h}}
			\bra{0} \adj{\phi}\left(\frac{1}{w}\right)
		\\
		&= \lim_{z \to \infty} z^{2 h}
			\bra{0} \adj{\phi}(z)
		\\
		&= \bra{0} I \circ \adj{\phi}(0).
\end{align}
\end{subequations}
Then, expanding the field in terms of the modes gives
\begin{equation}
	\bra{\eadj{\phi}} = \bra{0} (\adj{\phi})_{h}.
\end{equation}

\index{state (CFT)!bra}%
\index{BPZ conjugation!state}%
The BPZ conjugated state is
\begin{subequations}
\label{cft:eq:state-bpz-phi}
\begin{align}
	\bra{\phi}
		&
		:= \lim_{w \to 0} \bra{0} \phi(w)^t
		\\
		&= (\pm 1)^{h} \lim_{z \to \infty} z^{2 h} \bra{0} \phi(z)
		\\
		&= \bra{0} I^\pm \circ \phi(0).
\end{align}
\end{subequations}
In terms of the modes, one has
\begin{equation}
	\bra{\phi} = (\pm 1)^{h} \bra{0} \phi_{h}.
\end{equation}
If $\phi$ is Hermitian, then the relation between both conjugated states corresponds to a reality condition:
\begin{equation}
	\label{cft:eq:state-reality}
	\bra{\eadj{\phi}}
		= (\pm 1)^{h} \bra{\phi}.
\end{equation}

Taking the BPZ conjugation of the conditions \eqref{cft:eq:modes-vacuum} tells which modes must annihilate the conjugate vacuum:
\begin{equation}
	\forall n \le h - 1:
		\quad
		\bra{0} \phi_n = 0,
\end{equation}
and one finds more particularly for the Virasoro operators
\begin{equation}
	\forall n \le 1:
		\quad
		\bra{0} L_n = 0.
\end{equation}
This can also be derived directly from \eqref{cft:eq:state-bpz-phi} by requiring that applying an operator on the conjugate vacuum $\bra{0}$ is well-defined.

All conditions taken together mean that the expectation value of the energy--momentum tensor in the conformal vacuum vanishes:
\begin{equation}
	\bra{0} T(z) \ket{0} = 0.
\end{equation}
In particular, this means that the energy vacuum $\ket{\Omega}$, if different from $\ket{0}$, has a negative energy.

The Hermitian\footnotemark{} and BPZ inner products are respectively defined by:
\footnotetext{%
	Depending on the normalization, it can also be anti-Hermitian.
}%
\begin{subequations}
\begin{gather}
	\bracket{\eadj{\phi_i}}{\phi_j}
		= \bra{0} \bar I \circ \phi_j(0) \phi_i(0) \ket{0}
		= \lim_{\substack{z \to \infty \\ w \to 0}} z^{2 h_i} \bra{0} \adj{\phi_i}(z) \phi_j(w) \ket{0},
	\\
	\bracket{\phi_i}{\phi_j}
		= \bra{0} I \circ \phi_j(0) \phi_i(0) \ket{0}
		= (\pm 1)^{h_i} \lim_{\substack{z \to \infty \\ w \to 0}} z^{2 h_i} \bra{0} \phi_i(z) \phi_j(w) \ket{0}.
\end{gather}
\end{subequations}
These products can be recast as $2$-point correlation functions \eqref{cft:eq:corr-func-2pt} on the sphere:
\begin{equation}
	\bracket{\phi_i}{\phi_j}
		= \mean{I \circ \phi_i(0) \phi_j(0)},
	\qquad
	\bracket{\eadj{\phi_i}}{\phi_j}
		= \mean{I \circ \adj{\phi_i}(0) \phi_j(0)}.
\end{equation}
From the state--operator correspondence, the action of one operator on the in-state can be reinterpreted as the matrix element of this operator using the two external states, or also as a $3$-point function:
\begin{equation}
	\bra{\phi_i} \phi_j(z) \ket{\phi_k}
		= (\pm 1)^{h_i} \lim_{w \to \infty} w^{2 h_i} \mean{\phi_i(w) \phi_j(z) \phi_k(0)}.
\end{equation}

\index{conjugate state}%
\index{dual state|see{conjugate state}}%
Given a basis of states $\{ \phi_i \}$ ($i$ can run over both discrete and continuous indices), the \textit{conjugate} or \textit{dual states} $\{ \phi_i^c \}$ are defined by:
\begin{equation}
	\label{cft:eq:bpz-conj-state}
	\bracket{\phi_i^c}{\phi_j}
		= \delta_{ij}
\end{equation}
(the delta function is discrete and/or continuous according to the indices).

\subsubsection{Verma modules}

If $\phi(z)$ is a weight $h$ primary, then the associated state $\ket{\phi}$ satisfies:
\begin{equation}
	L_0 \ket{\phi} = h \ket{\phi},
	\qquad
	\forall n \ge 1:
		\quad
		L_n \ket{\phi} = 0.
\end{equation}
Such a state is also called a highest-weight state.
The descendant states are defined by all possible states of the form
\begin{equation}
	\ket{\phi_{\{n_i\}}}
:		= \prod_i L_{- n_i} \ket{\phi},
\end{equation}
where the same $L_{-n_i}$ can appear multiple times and $n_i > 0$.
\index{Verma module}%
The set of states $\phi_{\{n_i\}}$ is called a \emph{Verma module} $V(h, c)$.
One finds that the $L_0$ eigenvalues of this state is
\begin{equation}
	L_0
		= h + \sum_i n_i.
\end{equation}

\subsubsection{Normal ordering}

\index{normal ordering}%
The \emph{normal ordering} of an operator with respect to a vacuum corresponds to placing all creation (resp.\ annihilation) operators of this vacuum on the left (resp.\ right).
From this definition, the expectation value of a normal ordered operator in the vacuum vanishes identically.
The main reason for normal ordering is to remove singularities in expectation values.

Given an operator $\mc \phi(z)$, we define two normal orderings:
\begin{itemize}
	\index{normal ordering!conformal (-)}%
	\item The conformal normal order (CNO) $\norder{\mc O}$ is defined with respect to the conformal vacuum \eqref{cft:eq:vacuum-conformal}:
	\begin{equation}
		\label{cft:eq:conformal-normal-order}
		\bra{0} \norder{\mc O} \ket{0} = 0.
	\end{equation}

	\index{normal ordering!energy (-)}%
	\item The energy normal order (ENO) $\norderv{\mc O}$ is defined with respect to the energy vacuum \eqref{cft:eq:vacuum-energy}:
	\begin{equation}
		\label{cft:eq:energy-normal-order}
		\bra{\Omega} \norderv{\mc O} \ket{\Omega} = 0.
	\end{equation}
\end{itemize}
We first discuss the conformal normal ordering before explaining how to relate it to the energy normal ordering.

Given two operators $A$ and $B$, the simplest normal ordering amounts to subtract the expectation value:
\begin{equation}
	\label{cft:eq:normal-order-free}
	\norder{A(z) B(w)}
		\overset{?}{=} A(z) B(w) - \mean{A(z) B(w)}.
\end{equation}
This is equivalent to defining the products of two operators at coincident points via point-splitting:
\begin{equation}
	\norder{A(z) B(z)}
		\overset{?}{=} \lim_{w \to z} \Big( A(z) B(w) - \mean{A(z) B(w)} \Big).
\end{equation}
While this works well for free fields, this does not generalize for composite or interacting fields.

\index{Wick theorem}%
The reason is that this procedure removes only the highest singularity in the product: it does not work if the OPE has more than one singular term.
An appropriate definition is
\begin{equation}
	\label{cft:eq:normal-order}
	\norder{A(z) B(w)} \,
		:= A(z) B(w) - \wick{\c A(z) \c B(w)}
		= \sum_{n \in \N} (z - w)^n \{ A B \}_{-n}(z),
\end{equation}
where the contraction between $A$ and $B$ is defined in \eqref{cft:eq:contraction-AB}, and the second equality comes from \eqref{cft:eq:ope-AB}.

Then, the product evaluated at coincident points is found by taking the limit (in this case the argument is often indicated only at the end of the product)
\begin{equation}
	\label{cft:eq:normal-order-z}
	\norder{A B(z)} \,
		:= \norder{A(z) B(z)} \,
		:= \lim_{w \to z} \norder{A(z) B(w)}
		= \{ A B \}_0(z).
\end{equation}
Indeed, since all powers of $(z - w)$ are positive in the RHS of \eqref{cft:eq:normal-order}, all terms but the first one disappear.
The form of \eqref{cft:eq:normal-order-z} shows that the normal order can also be computed with the contour integral
\begin{equation}
	\label{cft:eq:normal-order-z-int}
	\norder{A B(z)}
		= \oint_{C_z} \frac{\dd w}{2\pi \I} \, \frac{A(z) B(w)}{z - w}.
\end{equation}
It is common to remove the colons of normal ordering when there is no ambiguity and, in particular, to write:
\begin{equation}
	A B(z)
		:= \norder{A B(z)}.
\end{equation}

In terms of modes, one has
\begin{subequations}
\label{cft:eq:modes-normal-order}
\begin{gather}
	\norder{A B(z)}
		= \sum_{m} \frac{\norder{A B}_m}{z^{m + h_A + h_B}},
	\\
	\norder{A B}_m = \sum_{n \le - h_A} A_n B_{m-n}
		+ \sum_{n > - h_A} B_{m-n} A_n.
\end{gather}
\end{subequations}
This expression makes explicit that normal ordering is non-commutative and non-associative:
\begin{equation}
	\norder{A B(z)} \neq \norder{B A(z)},
	\qquad
	\norder{A (B C)(z)} \neq \norder{(A B) C(z)}.
\end{equation}

The product of normal ordered operators can then be computed using Wick theorem.
In fact, one is more interested in the contraction of two such operators in order to recover the OPE between these operators: the product is then derived with \eqref{cft:eq:normal-order}.

If $A_i$ ($i = 1, 2, 3$) are \emph{free fields}, one has
\begin{equation}
	\label{cft:eq:contraction-product-free}
	\begin{gathered}
		A_1(z) \, \norder{A_2 A_3(w)}
			= \norder{A_1(z) A_2 A_3(w)}
				+ \wick{ \c A_1(z) \, \norder{A_2 \c{\vphantom{A}} A_3(w)}},
		\\
		\wick{ \c A_1(z) \, \norder{A_2 \c{\vphantom{A}} A_3(w)}}
			= \wick{ \c A_1(z) \c A_2(w) } \, \norder{A_3(w)}
				+ \wick{ \c A_1(z) \c A_3(w) } \, \norder{A_2(w)}.
	\end{gathered}
\end{equation}
If the fields are not free, then the contraction cannot be extracted from the normal ordering.
Similarly if there are more fields, then one needs to perform all the possible contractions.

Given two free fields $A$ and $B$, one has the following identities:
\begin{subequations}
\begin{gather}
	\label{cft:eq:ope-A-Bn}
	A(z) \, \norder{B(w)^n}
		= n \, \wick{ \c A(z) \c B(w)} \, \norder{B(w)^{n-1}},
	\\
	\label{cft:eq:ope-A-expB}
	A(z) \, \norder{\e^{B(w)}}
		= \wick{ \c A(z) \c B(w)} \, \norder{\e^{B(w)}},
	\\
	\label{cft:eq:ope-expA-expB}
	\norder{\e^{A(z)}} \, \norder{\e^{B(w)}}
		= \exp\big(\wick{ \c A(z) \c B(w)} \big) \, \norder{\e^{A(z)} \e^{B(w)}}.
\end{gather}
\end{subequations}
The last relation generalizes for a set of $n$ fields $A_i$:
\begin{subequations}
\begin{gather}
	\prod_{i=1}^n \norder{\e^{A_i}}
		= \norder{\, \exp \left( \sum_{i=1}^n A_i \right)} \, \exp \sum_{i<j} \mean{A_i A_j},
	\\
	\Mean{ \prod_{i=1}^n \norder{\e^{A_i}} }
		= \exp \sum_{i<j} \mean{A_i A_j}.
\end{gather}
\end{subequations}

\begin{computation}[cft:eq:ope-A-expB]
	\[
		A(z) \, \norder{\e^{B(w)}}
			= A(z) \sum_n \frac{1}{n!} \, \norder{B(w)^n}
			= \wick{ \c A(z) \c B(w)} \sum_n \frac{1}{(n - 1)!} \, \norder{B(w)^{n-1}}.
	\]
\end{computation}

\begin{computation}[cft:eq:ope-expA-expB]
	\begin{align*}
		\norder{\e^{A(z)}} \, \norder{\e^{B(w)}}
			&= \sum_{m, n} \frac{1}{m! n!} \, \norder{A(z)^m} \, \norder{B(w)^n}
			\\
			&= \sum_{m, n, k} \frac{k!}{m! n!} \binom{m}{k} \binom{n}{k} \big( \wick{ \c A(z) \c B(w)} \big)^k \norder{A(z)^{m-k}} \, \norder{B(w)^{n-k}}
			\\
			&= \sum_{m, n, k} \frac{1}{k! (m - k)! (n - k)!} \, \big( \wick{ \c A(z) \c B(w)} \big)^k \norder{A(z)^{m-k}} \, \norder{B(w)^{n-k}}.
	\end{align*}
	The factorial $k!$ counts the number of possible ways to contract the two operators.
\end{computation}

\medskip

The general properties of normal ordered expressions are identical for both vacua: what differs is the precise computation in terms of the operators (or modes).
Hence, the energy normal ordering can be defined in parallel with \eqref{cft:eq:modes-normal-order}, but changing the definitions of creation and annihilation operators:
\begin{subequations}
\label{cft:eq:modes-eno}
\begin{gather}
	\norderv{A B(z)}
		= \sum_{m} \frac{\norderv{A B(z)}_n}{z^{m + h_A + h_B}},
	\\
	\norderv{A B}_m = \sum_{n \le 0} A_n B_{m-n}
		+ \sum_{n > 0} B_{m-n} A_n.
\end{gather}
\end{subequations}
To simplify the definition we assume that $A_0$ is a creation operator and it is thus included in the first sum (this must be adapted in function of which vacuum state is chosen if the latter is degenerate).

\index{normal ordering!mode relation}%
The relation between the normal ordered modes is
\begin{equation}
	\label{cft:eq:modes-relations-cno-eno}
	\norder{A B}_m
		= \norderv{A B}_m
			+ \sum_{n=0}^{h_A - 1} \com{B_{m+n}}{A_{-n}}.
\end{equation}

\begin{computation}[cft:eq:modes-relations-cno-eno]
	\begin{align*}
		\norder{A B}_m
			&= \sum_{n \le - h_A} A_n B_{m-n}
				+ \sum_{n > - h_A} B_{m-n} A_n
			\\
			&= \sum_{n \ge h_A} A_{-n} B_{m+n}
				+ \sum_{n > 0} B_{m-n} A_n
				+ \sum_{n=0}^{h_A - 1} B_{m+n} A_{-n}
			\\
			&= \sum_{n \ge 0} A_{-n} B_{m+n}
				+ \sum_{n > 0} B_{m-n} A_n
				+ \sum_{n=0}^{h_A - 1} \com{B_{m+n}}{A_{-n}}
			\\
			&= \norderv{A B}_m
				+ \sum_{n=0}^{h_A - 1} \com{B_{m+n}}{A_{-n}}.
	\end{align*}
\end{computation}

The choice of the normal ordering for the operators is related to the ordering ambiguity when quantizing the system: when the product of two non-commuting modes appears in the classical composite field, the corresponding quantum operator is ambiguous (generally up to a constant).
In practice, one starts with the conformal ordering since it is invariant under conformal transformations and because one can compute with contour integrals.
Then, the expression can be translated in the energy ordering using \eqref{cft:eq:modes-relations-cno-eno}.
But, knowing how the conformal and energy vacua are related, it is often simpler to find the difference between the two orderings by applying the operator on the vacua.

\subsection{CFT on the cylinder}

\index{conformal field theory!cylinder}%
According to \eqref{cft:eq:transf-primary}, the relation between the field on the cylinder and on the plane is
\begin{equation}
	\phi(z) = \left( \frac{L}{2\pi} \right)^{h} z^{-h} \phi_{\text{cyl}}(w)
\end{equation}
(quantities without indices are on the plane by definition).
The mode expansion on the cylinder is
\begin{equation}
	\phi_{\text{cyl}}
		= \left( \frac{2\pi}{L} \right)^{h} \sum_{n \in \Z} \phi_n \e^{- \frac{2\pi}{L} \, w}
		= \left( \frac{2\pi}{L} \right)^{h} \sum_{n \in \Z} \frac{\phi_n}{z^n}.
\end{equation}

Using the finite transformation \eqref{cft:eq:transf-T} for the energy--momentum tensor $T$, one finds the relation
\begin{equation}
	T_{\text{cyl}}(w) = \left( \frac{2\pi}{L} \right)^{2} \left( T(z) z^2 - \frac{c}{24} \right).
\end{equation}
\begin{check}
\index{Casimir energy}%
The vacuum expectation value (Casimir energy) is then proportional to the central charge:
\begin{equation}
	\mean{T_{\text{cyl}}}
		= - \frac{c \pi^2}{6 L^2}.
\end{equation}
This energy is provided by the curvature of the cylinder.
\end{check}
For the $L_0$ mode, one finds
\begin{equation}
	\label{cft:eq:L0-cyl}
	(L_0)_{\text{cyl}}
		= L_0 - \frac{c}{24},
\end{equation}
and thus the Hamiltonian is
\begin{equation}
	H = (L_0)_{\text{cyl}} + (\bar L_0)_{\text{cyl}}
		= L_0 + \bar L_0 - \frac{c + \bar c}{24}.
\end{equation}

\refchapter

\begin{itemize}
	\item The most complete reference on CFTs is~\cite{DiFrancesco:1999:ConformalFieldTheory} but it lacks some recent developments.
	Two excellent complementary books are~\cite{Schottenloher:2008:MathematicalIntroductionConformal, Blumenhagen:2009:IntroductionConformalField}.

	String theory books generally dedicate a fair amount of pages to CFTs: particularly good summaries can be found in~\cite{Kiritsis:2007:StringTheoryNutshell, Blumenhagen:2014:BasicConceptsString, Polchinski:2005:StringTheory-1, Polchinski:2005:StringTheory-2}.

	Finally, a modern and fully algebraic approach can be found in~\cite{Ribault:2014:ConformalFieldTheory, Ribault:2018:MinimalLecturesTwodimensional}.
	Other good reviews are~\cite{Qualls:2015:LecturesConformalField, Yin:2018:AspectsTwoDimensionalConformal}.

	\item There are various other books~\cite{Ketov:1995:ConformalFieldTheory, Itzykson:2000:TheorieStatistiqueChamps-2, Mussardo:2009:StatisticalFieldTheory, Henkel:2010:ConformalInvarianceCritical} and reviews~\cite{Ginsparg:1988:AppliedConformalField, Schellekens:1996:IntroductionConformalField, Gaberdiel:1999:IntroductionConformalField, Cardy:2008:ConformalFieldTheory, Teschner:2017:GuideTwodimensionalConformal}.

	\item The maps from the sphere and the cylinder to the complex plane are discussed in~\cite[sec.~2.6, 6.1]{Polchinski:2005:StringTheory-1}.

	\item Normal ordering is discussed in details in~\cite[chap.~6]{DiFrancesco:1999:ConformalFieldTheory} (see also~\cites[sec.~4.2]{Blumenhagen:2014:BasicConceptsString}[sec.~2.2]{Polchinski:2005:StringTheory-1}).

	\item Euclidean conjugation is discussed in~\cites[sec.~6.1.1]{DiFrancesco:1999:ConformalFieldTheory}[p.~202--3]{Polchinski:2005:StringTheory-1}.
	For a comparison of Euclidean and BPZ conjugations, see~\cites[sec.~2.2]{Zwiebach:1993:ClosedStringField}[p.~11]{Sen:2016:RealitySuperstringField}.

	\item Normal ordering and difference between the different definitions are described in~\cites[chap.~2]{Polchinski:2005:StringTheory-1}[sec.~6.5]{DiFrancesco:1999:ConformalFieldTheory}.
\end{itemize}

\chapter{CFT systems}
\label{cft:chap:systems}

\introchapter

This chapter summarizes the properties of some CFT systems.
We focus on the free scalar field and on the first-order $bc$ system (which generalizes the reparametrization ghosts).
For the different systems, we first provide an analysis on a general curved background before focusing on the complex plane.
This is sufficient to describe the local properties on all Riemann surfaces $g \ge 0$.

\section{Free scalar}
\label{cft:sec:systems:free-scalar}

\subsection{Covariant action}

\index{scalar field CFT!action}%
The Euclidean action of a free scalar $X$ on a curved background $g_{\mu\nu}$ is
\begin{equation}
	\label{cft:eq:scalar-action}
	S
		= \frac{\epsilon}{4\pi \ell^2}
			\int \dd^2 x \sqrt{g} \,
			g^{\mu\nu} \pd_\mu X \pd_\nu X,
\end{equation}
\index{e@$\epsilon$ (scalar action sign)}%
where $\ell$ is a length scale\footnotemark{} and
\footnotetext{%
	To be identified with the string scale, such that $\alpha' = \ell^2$.
}%
\begin{equation}
	\epsilon
		:=
		\begin{cases}
			+ 1 & \text{spacelike}
			\\
			- 1 & \text{timelike}
		\end{cases},
	\qquad
	\sqrt{\epsilon}
		:=
		\begin{cases}
			+ 1 & \text{spacelike}
			\\
			\I & \text{timelike}
		\end{cases}
\end{equation}
denotes the signature of the kinetic term.
\index{scalar field CFT!periodic boundary condition}%
The field is periodic along $\sigma$
\begin{equation}
	\label{cft:eq:scalar-periodic}
	X(\tau, \sigma) \sim X(\tau, \sigma + 2\pi).
\end{equation}

\index{scalar field CFT!energy--momentum tensor}%
The energy--momentum tensor reads
\begin{equation}
	T_{\mu\nu} = - \frac{\epsilon}{\ell^2} \left[ \pd_\mu X \pd_\nu X - \frac{1}{2} \, g_{\mu\nu} (\pd X)^2 \right],
\end{equation}
and it is traceless
\begin{equation}
	T_\mu^\mu = 0.
\end{equation}
\index{scalar field CFT!equation of motion}%
The equation of motion is
\begin{equation}
	\lap X = 0,
\end{equation}
where $\lap$ is the Laplacian \eqref{app:eq:laplacian}.

\index{scalar field CFT!propagator}%
The simplest method for finding the propagator in flat space is by using the identity (assuming that there is no boundary term)
\begin{equation}
	0
		= \int \dd X \, \frac{\delta}{\delta X(\sigma)}
			\left( \e^{- S[X]} X(\sigma') \right),
\end{equation}
which yields a differential equation for the propagator:
\begin{equation}
	\label{cft:eq:scalar-diff-eq}
	\mean{\pd^2 X(\sigma) X(\sigma')}
		= - 2\pi \epsilon \ell^2 \, \cdirac[2](\sigma - \sigma').
\end{equation}
This is easily integrated to
\begin{equation}
	\label{cft:eq:scalar-propagator}
	\mean{X(\sigma) X(\sigma')} = - \frac{\epsilon \ell^2}{2} \, \ln \abs{\sigma - \sigma'}^2.
\end{equation}

\begin{computation}[cft:eq:scalar-propagator]
	By translation and rotation invariance, one has
	\begin{equation}
		\mean{X(\sigma) X(\sigma')} = G(r),
		\qquad
		r = \abs{\sigma - \sigma'}.
	\end{equation}
	In polar coordinates, the Laplacian reads
	\begin{equation}
		\lap G(r) = \frac{1}{r} \, \pd_r(r G'(r)).
	\end{equation}
	Integrating the differential equation \eqref{cft:eq:scalar-diff-eq} over $\dd^2 \sigma = r \dd r \dd \theta$ yields
	\begin{equation}
		- 2\pi \epsilon \ell^2 = 2\pi \int_0^r \dd r' \, r' \times \frac{1}{r'} \, \pd_{r'}(r' G'(r'))
			= 2\pi r G'(r).
	\end{equation}
	The solution is
	\begin{equation}
		G'(r) = - \epsilon \ell^2 \ln r
	\end{equation}
	and the form \eqref{cft:eq:scalar-propagator} follows by writing
	\begin{equation}
		\ln r = \frac{1}{2} \ln r^2
			= \frac{1}{2} \ln \abs{\sigma - \sigma'}^2.
	\end{equation}

\end{computation}

\index{scalar field CFT!$\mathrm{U}(1)$ symmetry}%
The action \eqref{cft:eq:scalar-action} is obviously invariant under constant translations of $X$:
\begin{equation}
	X \longrightarrow X + a,
	\qquad
	a \in \R.
\end{equation}
\index{scalar field CFT!$\mathrm{U}(1)$ current}%
The associated $\group{U}(1)$ current\footnotemark{} is conserved and reads
\footnotetext{%
	The group is $\R$ but the algebra is $\alg{u}(1)$ (since locally there is no difference between the real line and the circle).
}%
\begin{equation}
	\label{cft:eq:scalar-current-U1}
	J^\mu
		:= 2\pi \I \epsilon \, \frac{\partial \mc L}{\partial(\pd_\mu X)}
		= \frac{\I}{\ell^2} \, g^{\mu\nu} \pd_\nu X,
	\qquad
	\grad_\mu J^\mu = 0.
\end{equation}
\index{scalar field CFT!momentum}%
On flat space, the charge follows from \eqref{app:eq:charge}:
\begin{equation}
	\label{cft:eq:scalar-p}
	p
		= \frac{1}{2\pi} \int \dd \sigma \, J^0
		= \frac{\I}{2\pi \ell^2} \int \dd \sigma \, \pd^0 X.
\end{equation}
This charge is called \emph{momentum} because it corresponds to the spacetime momentum in string theory.

\index{scalar field CFT!topological current}%
Moreover, there is a another \emph{topological current}
\begin{equation}
	\label{cft:eq:scalar-current-top}
	\wtilde J^\mu
		:= - \I \, \epsilon^{\mu\nu} J_\nu
		= \frac{1}{\ell^2} \, \epsilon^{\mu\nu} \pd_\nu X,
\end{equation}
which is identically conserved:
\begin{equation}
	\grad_\mu \wtilde J^\mu
		\propto \epsilon^{\mu\nu} \com{\grad_\mu}{\grad_\nu} X
		= 0
\end{equation}
since $\com{\grad_\mu}{\grad_\nu} = 0$ when acting on a scalar field.
Note that $\tilde J^\mu$ is the Hodge dual of $J^\mu$.
\index{scalar field CFT!winding number}%
The conserved charge is called the \emph{winding number} and reads on flat space:
\begin{equation}
	\label{cft:eq:scalar-w}
	w
		= \frac{1}{2\pi} \int \dd \sigma \, \wtilde J^0
		= \frac{1}{2\pi \ell^2} \int_0^{2\pi} \dd \sigma \, \pd_1 X
		= \frac{1}{2\pi \ell^2} \big( X(\tau, 2\pi) - X(\tau, 0) \big).
\end{equation}

\begin{remark}[Normalization of the current]
	\index{conventions!spacetime momentum current}%
	\index{spacetime momentum}%

	The definition of the current \eqref{cft:eq:scalar-current-U1} may look confusing.
	The factor of $\I$ is due to the Euclidean signature, see \eqref{app:eq:current-euc}, and the factor of $2\pi$ comes from the normalization of the spatial integral.
	We have inserted $\epsilon$ in order to interpret the conserved charge $p$ as a component of the momentum contravariant vector in string theory.

	To make contact with string theory, consider $D$ scalar fields $X^a(x^\mu)$.
	Then, the current becomes
	\begin{equation}
		J^\mu_a
			= \frac{\I}{2\pi \ell^2} \, \eta_{ab} \pd^\mu X^b,
	\end{equation}
	where the position of the indices is in agreement with the standard form of Noether's formula \eqref{app:eq:current-euc} (a current has indices in opposite locations as the parameters and fields).
	Since we have $\eta_{00} = - 1 = \epsilon_{X_0}$, we find that $J^{0 \mu} = \epsilon_{X_0} J^\mu_0$ has no epsilon after replacing the expression \eqref{cft:eq:scalar-p} of $J^\mu_0$.

	The transformation $X^a \to X^a + c^a$ is a global translation in target spacetime: the charge $p^a$ is identified with the spacetime momentum.
	The factor of $\I$ indicates that $p^a$ is the Euclidean contravariant momentum vector by comparison with \eqref{app:eq:wick-rot-vector}.

	The convention of this section is to always work with quantities which will become contravariant vector to avoid ambiguity.
\end{remark}

\subsection{Action on the complex plane}

\index{scalar field CFT!complex plane}%
\index{scalar field CFT!action}%
In complex coordinates, the action on flat space reads
\begin{equation}
	\label{cft:eq:scalar-action-cx}
	S
		= \frac{\epsilon}{2\pi \ell^2}
			\int \dd z \dd \bar{z} \,
			\pd_z X \pd_{\bar z} X,
\end{equation}
\index{scalar field CFT!equation of motion}%
giving the equation of motion:
\begin{equation}
	\pd_z \pd_{\bar z} X = 0.
\end{equation}
\index{scalar field CFT!complex components}%
This indicates that $\pd_z X$ and $\pd_{\bar z} X$ are respectively holomorphic and anti-holomorphic such that
\begin{equation}
	X(z, \bar z) = X_L(z) + X_R(\bar z),
\end{equation}
and we will remove the subscripts when there is no ambiguity (for example, when the position dependence is written):
\begin{equation}
	X(z)
		:= X_L(z),
	\qquad
	X(\bar z)
		:= X_R(\bar z).
\end{equation}
It looks like $X_L(z)$ and $X_R(\bar z)$ are unrelated, but this is not the case because of the zero-mode, as we will see below.

\index{scalar field CFT!$\mathrm{U}(1)$ current}%
The $\group{U}(1)$ current is written as
\begin{equation}
	J
		:= J_z
		= \frac{\I}{\ell^2} \, \pd_z X,
	\qquad
	\bar J
		:= J_{\bar z}
		= \frac{\I}{\ell^2} \, \pd_{\bar z} X,
\end{equation}
where we used the relations $J_z = J^{\bar z} / 2$ and $J_{\bar z} = J^z / 2$.
The equation of motion implies that the current $J$ is holomorphic, and $\bar J$ is anti-holomorphic:
\begin{equation}
	\bar\pd J = 0,
	\qquad
	\pd \bar J = 0.
\end{equation}
\index{scalar field CFT!momentum}%
The momentum splits into left- and right-moving parts:
\begin{equation}
	\label{cft:eq:scalar-p-pLR}
	p
		= p_L + p_R,
	\qquad
	p_L
		= \frac{1}{2\pi \I} \oint \dd z \, J,
	\qquad
	p_R
		= - \frac{1}{2\pi \I} \oint \dd \bar z \, \bar J.
\end{equation}

\index{scalar field CFT!topological current}%
The components of the topological current \eqref{cft:eq:scalar-current-top} are related to the ones of the $\group{U}(1)$ current:
\begin{equation}
	\wtilde J_{z}
		= \frac{\I}{\ell^2} \, \pd_{z} X
		= J,
	\qquad
	\wtilde J_{\bar z}
		= - \frac{\I}{\ell^2} \, \pd_{\bar z} X
		= - \bar J.
\end{equation}
\index{scalar field CFT!winding number}%
As a consequence, the winding number is
\begin{equation}
	\label{cft:eq:scalar-w-pLR}
	w
		= p_L - p_R.
\end{equation}
\index{scalar field CFT!momentum}%
Note that we have the relations
\begin{subequations}
\begin{gather}
	p_L
		= \frac{p + w}{2},
	\qquad
	p_R
		= \frac{p - w}{2},
	\\
	\label{cft:eq:scalar-pL2-pR2}
	p^2 + w^2
		= p_L^2 + p_R^2,
	\qquad
	2 p w
		= p_L^2 - p_R^2.
\end{gather}
\end{subequations}

\index{scalar field CFT!energy--momentum tensor}%
The energy--momentum tensor is
\begin{equation}
	T
		:= T_{z z}
		= - \frac{\epsilon}{\ell^2} \, \pd_z X \pd_z X,
	\qquad
	\bar T
		:= T_{\bar z \bar z}
		= - \frac{\epsilon}{\ell^2} \, \pd_{\bar z} X \pd_{\bar z} X,
	\qquad
	T_{z \bar z} = 0.
\end{equation}
Since the $\pd_z X$ ($\pd_{\bar z} X$) is (anti-)holomorphic, so is $T(z)$ ($\bar T(\bar z)$).
Since the energy--momentum tensor, the current and the field itself (up to zero-modes) split in holomorphic and anti-holomorphic components in a symmetric way, it is sufficient to focus on one of the sectors, say the holomorphic one.

\index{scalar field CFT!vertex operator}%
The other primary operators of the theory are given by the vertex operators $V_k(z)$:\footnotemark{}
\footnotetext{%
	The $\epsilon$ in the exponential is consistent with interpreting $X$ and $k$ as a contravariant vector.
}%
\begin{equation}
	\label{cft:eq:scalar-Vk}
	V_k(z, \bar z)
		:= \norder{\e^{\I \epsilon k X(z, \bar z)}}.
\end{equation}

\begin{remark}
	In fact, it is possible to introduce more general vertex operators
	\begin{equation}
		V_{k_L, k_R}(z, \bar z)
			:= \norder{\e^{2 \I \epsilon \big( k_L X(z) + k_R X(\bar z) \big)}},
	\end{equation}
	but we will not consider them in this \revname{}.
\end{remark}

\begin{remark}[Plane and cylinder coordinates]
	\index{scalar field CFT!cylinder}%

	The action in $w$-coordinate (cylinder) takes the same form as a result of the conformal invariance of the scalar field, which in practice results from the cancellation between the determinant and inverse metric.
	As a consequence, every quantity derived from the classical action (equation of motion, energy--momentum tensor…) will have the same form in both coordinate systems: we will focus on the $z$-coordinate, writing the $w$-coordinate expression when it is insightful to compare.
	This is not anymore the case at the quantum level: anomalies may translate into differences between quantities: to differentiate between the plane and cylinder quantities an index “cyl” will be added when necessary (by convention, all quantities without qualification are on the plane).
\end{remark}

\subsection{OPE}

\index{scalar field CFT!OPE|(}%

The OPE between $X$ and itself is directly found from the propagator:
\begin{equation}
	\label{cft:eq:scalar-ope-X-X}
	X(z) X(w)
		\sim - \frac{\epsilon \ell^2}{2} \, \ln (z - w).
\end{equation}
By successive derivations, one finds the OPE between $X$ and $\pd X$
\begin{equation}
	\label{cft:eq:scalar-ope-dX-X}
	\pd X(z) X(w)
		\sim - \frac{\epsilon \ell^2}{2} \, \frac{1}{z - w},
\end{equation}
and between $\pd X$ with itself
\begin{equation}
	\label{cft:eq:scalar-ope-dX-dX}
	\pd X(z) \pd X(w)
		\sim - \frac{\epsilon \ell^2}{2} \, \frac{1}{(z - w)^2}.
\end{equation}
The invariance under the permutation of $z$ and $w$ reflects the fact that $X$ is bosonic and that both operators in \eqref{cft:eq:scalar-ope-dX-dX} are identical.

The OPE between $\pd X$ and $T$ allows to verify that the field $\pd X$ is primary with $h = 1$:
\begin{equation}
	\label{cft:eq:scalar-ope-T-dX}
	T(z) \pd X(w) \sim \frac{\pd X(w)}{(z - w)^2} + \frac{\pd\big(\pd X(w)\big)}{z - w}.
\end{equation}
The OPE of $T$ with itself gives
\begin{equation}
	\label{cft:eq:scalar-ope-T-T}
	T(z) T(w) \sim \frac{1}{2} \, \frac{1}{(z - w)^4} + \frac{2 T(w)}{(z - w)^2} + \frac{\pd T(w)}{z - w}
\end{equation}
\index{scalar field CFT!central charge}%
which shows that the central charge is
\begin{equation}
	c = 1.
\end{equation}

One finds that the operator $\pd^n X$ has conformal weight
\begin{equation}
	h = n
\end{equation}
since the OPE with $T$ is
\begin{equation}
	\label{cft:eq:scalar-ope-T-dnX}
	T(z) \pd^n X(w)
		\sim \cdots + \frac{n \, \pd^n X(w)}{(z - w)^2}
			+ \frac{\pd (\pd^n X(w))}{z - w}
\end{equation}
where the dots indicate higher negative powers of $(z - w)$.
These states are not primary for $n \ge 2$.
Explicitly, for $n = 2$, one finds
\begin{equation}
	T(z) \pd^2 X(w)
		\sim \frac{2 \, \pd X(w)}{(z - w)^3} + \frac{2 \, \pd^2 X}{(z - w)^2}
			+ \frac{\pd (\pd^2 X(w))}{z - w}.
\end{equation}

The OPE of a vertex operator with the current $J$ is
\begin{equation}
	\label{cft:eq:scalar-ope-dX-V}
	J(z) V_k(w, \bar w)
		\sim \frac{\ell^2 k}{2} \, \frac{V_k(w, \bar w)}{z - w}.
\end{equation}
This shows that the vertex operators $V_k$ are eigenstates of the $\group{U}(1)$ holomorphic current with the eigenvalue given by the momentum (with a normalization of $\ell^2$).
Then, the OPE with $T$:
\begin{equation}
	\label{cft:eq:scalar-ope-T-V}
	T(z) V_k(w, \bar w)
		\sim \frac{h_k \, V_k(w, \bar w)}{(z - w)^2}
			+ \frac{\pd V_k(w, \bar w)}{z - w}
\end{equation}
together with its anti-holomorphic counterpart show that the $V_k$ are primary operators with weight
\begin{equation}
	\label{cft:eq:scalar-dim-Vk}
	(h_k, \bar h_k)
		= \left( \frac{\epsilon \ell^2 k^2}{4}, \frac{\epsilon \ell^2 k^2}{4} \right),
	\qquad
	\Delta_k
		= \frac{\epsilon \ell^2 k^2}{2},
	\qquad
	s_k = 0.
\end{equation}
Note that classically $h_k = 0$ since $\ell \sim \hbar$~\cite[p.~81]{Tong:2009:LecturesStringTheory}.
The weight is invariant under $k \to - k$.
Finally, the OPE between two vertex operators is
\begin{equation}
	\label{cft:eq:scalar-ope-V-V}
	V_k(z, \bar z) V_{k'}(w,, \bar w)
		\sim \frac{V_{k + k'}(w, \bar w)}{(z - w)^{- \epsilon k k' \ell^2 / 2}},
\end{equation}
where only the leading term (non-necessarily singular) is displayed.
In particular, correlation functions should be computed for $\epsilon k k' < 0$ in order to avoid exponential growth.

\begin{computation}[cft:eq:scalar-ope-T-dX]
	\begin{align*}
		T(z) \pd X(w)
			= - \frac{\epsilon}{\ell^2} \, \norder{\pd X(z) \pd X(z)} \, \pd X(w)
			\sim - \frac{2 \epsilon}{\ell^2} \, \wick{\norder{\pd X(z) \c1{\pd X}(z)} \, \c1{\pd X}(w)}
			\sim \frac{\pd X(z)}{(z - w)^2}.
	\end{align*}
	The result \eqref{cft:eq:scalar-ope-T-dX} follows by Taylor expanding the numerator.
\end{computation}

\begin{computation}[cft:eq:scalar-ope-T-T]
	\begin{align*}
		T(z) \pd X(w)
			&= \frac{1}{\ell^4} \, \norder{\pd X(z) \pd X(z)} \, \norder{\pd X(w) \pd X(w)}
			\\ &
			\begin{aligned}
				\sim \frac{1}{\ell^4} \, \bigg[
					&
					\wick{\norder{\c1{\pd X}(z) \c2{\pd X}(z)} \, \norder{\c1{\pd X}(w) \c2{\pd X}(w)}}
					+ \wick{\norder{\c2{\pd X}(z) \c1{\pd X}(z)} \, \norder{\c1 {\pd X}(w) \c2{\pd X}(w)}}
					\\
					&
					+ \wick{\norder{\c1{\pd X}(z) \pd X(z)} \, \norder{\c1{\pd X}(w) \pd X(w)}}
					+ \text{perms}
					\bigg]
			\end{aligned}
			\\ &
			\sim 2 \times \frac{1}{4} \, \frac{1}{(z - w)^4}
				- 4 \times \frac{1}{2 \ell^2} \frac{1}{(z - w)^2} \, \norder{\pd X(z) \pd X(w)}
			\\ &
			\sim \frac{1}{2} \, \frac{1}{(z - w)^4}
				- \frac{2}{\ell^2} \frac{1}{(z - w)^2} \, \Big(
					\norder{\pd X(w) \pd X(w)}
					+ (z - w) \, \norder{\pd^2 X(w) \pd X(w)}
					\Big).
	\end{align*}
\end{computation}

\begin{computation}[cft:eq:scalar-ope-T-dnX]
	\begin{align*}
		T(z) \pd^n X(w)
			&
			\sim \pd_w^{n-1} \frac{\pd X(z)}{(z - w)^2}
			\\ &
			\sim n! \, \frac{\pd X(z)}{(z - w)^{n+1}}
			\\ &
			\begin{aligned}
				\sim \frac{n!}{(z - w)^{n+1}} \, \bigg( \cdots
					&+ \frac{1}{(n - 1)!}\, (z - w)^{n-1} \pd^{n-1}(\pd X(w))
					\\
					&+ \frac{1}{n!}\, (z - w)^{n} \pd^{n}(\pd X(w))
					\bigg).
			\end{aligned}
	\end{align*}
\end{computation}

\begin{computation}[cft:eq:scalar-ope-dX-V]
	Using \eqref{cft:eq:ope-A-expB}, one has:
	\[
		\pd X(z) V_k(w, \bar w)
			\sim \I \epsilon k \, \wick{\c{\pd X}(z) \c X(w)} \, V_k(w, \bar w)
			\sim \I \epsilon k \, \left( - \frac{\epsilon \ell^2}{2} \frac{1}{z - w} \right) V_k(w, \bar w).
	\]
\end{computation}

\begin{computation}[cft:eq:scalar-ope-T-V]

	\begin{align*}
		T(z) V_k(w, \bar w)
			&
			\sim - \frac{\epsilon}{\ell^2} \, \wick[offset=1.5em]{ \norder{\pd X(z) \c{\pd X}(z)} \, \norder{\c{\e^{\I \epsilon k X(w, \bar w)}}}}
			\\ &
			\sim \frac{\I \epsilon k}{2} \frac{1}{z - w}\, \pd X(z) \, \norder{\e^{\I \epsilon k X(w, \bar w)}}
				- \frac{\epsilon}{\ell^2} \, \wick[offset=1.5em]{\c{\pd X}(z) \, \norder{\pd X(z) \c{\e^{\I \epsilon k X(w, \bar w)}}}}
			\\ &
			\sim \frac{\I \epsilon k}{2} \frac{1}{z - w} \,
				\left(
					\norder{\pd X(z) \, \e^{\I \epsilon k X(w, \bar w)}}
					+ \wick{\c{\pd X}(z) \, \norder{\c{\e^{\I \epsilon k X(w, \bar w)}}}}
					\right)
				\\ & \hspace{1.5cm}
				+ \frac{\I \epsilon k}{2} \, \frac{\norder{\pd X(z) \e^{\I \epsilon k X(w, \bar w)}}}{z - w}
			\\ &
			\sim \frac{\epsilon k^2 \ell^2}{4} \, \frac{V_k(w, \bar w)}{(z - w)^2}
				+ \I \epsilon k\, \frac{\norder{\pd X(w) \e^{\I \epsilon k X(w, \bar w)}}}{z - w}.
	\end{align*}
	In the first line, we consider a single contraction (hence, there is no factor of $2$): the reason is that considering the contractions symmetrically and not successively counts twice the first term of the last line.
	Indeed, there is only one way to generate this term.
	It is also possible to achieve the same result by expanding the exponential.
\end{computation}

\begin{computation}[cft:eq:scalar-ope-V-V]
	Using \eqref{cft:eq:ope-expA-expB} and keeping only the leading term, one has:
	\begin{align*}
		V_k(z, \bar z) V_{k'}(w, \bar w)
			&
			\sim \exp\big(- k k' \, \wick{\c X(z, \bar z) \c X(w, \bar w)} \big) \,
				\norder{\e^{\I \epsilon k X(z, \bar z)} \e^{\I \epsilon k' X(w, \bar w)}}
			\\ &
			\sim (z - w)^{\epsilon k k' \ell^2 / 2} \, V_{k + k'}(w, \bar w).
	\end{align*}
\end{computation}

\index{scalar field CFT!OPE|)}%

\subsection{Mode expansions}

\index{scalar field CFT!normal ordering}%

\index{scalar field CFT!mode expansion|(}%

Since $\pd X$ is holomorphic and of weight $h = 1$, it can be expanded as:\footnotemark{}
\footnotetext{%
	The Fourier expansion is taken to be identical for $\epsilon = \pm 1$ fields since $\pd X$ is contravariant in target space.
	The difference between the two cases will appear in the commutators.
}%
\begin{equation}
	\label{cft:eq:scalar-exp-dX}
	\pd X
		= - \I \sqrt{\frac{\ell^2}{2}} \sum_{n \in \Z} \alpha_n\, z^{- n - 1},
	\qquad
	\bar\pd X
		= - \I \sqrt{\frac{\ell^2}{2}} \sum_{n \in \Z} \bar\alpha_n\, \bar z^{- n - 1},
\end{equation}
where an individual mode can be extracted with a contour integral:
\begin{equation}
	\label{cft:eq:scalar-modes-int}
	\alpha_n
		= \I \oint \frac{\dd z}{2\pi \I} \, z^{n - 1} \pd X(z),
	\qquad
	\bar\alpha_n
		= \I \oint \frac{\dd z}{2\pi \I} \, z^{n - 1} \bar\pd X(z).
\end{equation}
Integrating this formula gives:
\begin{equation}
	\label{cft:eq:scalar-exp-XLR}
	\begin{aligned}
	X(z)
		&= \frac{x_L}{2}
			- \I \sqrt{\frac{\ell^2}{2}}\, \alpha_0 \ln z
			+ \I \sqrt{\frac{\ell^2}{2}} \sum_{n \neq 0} \frac{\alpha_n}{n}\, z^{-n},
	\\
	X(\bar z)
		&= \frac{x_R}{2}
			- \I \sqrt{\frac{\ell^2}{2}}\, \bar \alpha_0 \ln \bar z
			+ \I \sqrt{\frac{\ell^2}{2}} \sum_{n \neq 0} \frac{\bar \alpha_n}{n}\, \bar z^{-n}.
	\end{aligned}
\end{equation}

\index{scalar field CFT!zero-mode|(}%
The zero-modes are respectively $\alpha_0$ and $\bar\alpha_0$ for $\pd X$ and $\bar\pd X$, and $x_L$ and $x_R$ for $X_L$ and $X_R$.
The meaning of the modes will become clearer in \Cref{cft:sec:systems:scalar:com} where we study the commutation relations.

First, we relate the zero-modes $\alpha_0$ and $\bar\alpha_0$ to the conserved charges $p_L$ and $p_R$ \eqref{cft:eq:scalar-p-pLR} of the $\group{U}(1)$ current:
\begin{equation}
	\label{cft:eq:scalar-pLR-a0}
	p_L
		= \frac{\alpha_0}{\sqrt{2 \ell^2}},
	\qquad
	p_R
		= \frac{\bar\alpha_0}{\sqrt{2 \ell^2}}
\end{equation}
such that
\begin{equation}
	X(z)
		= \frac{x_L}{2}
			- \I \ell^2 \, p_L \ln z
			+ \I \sqrt{\frac{\ell^2}{2}} \sum_{n \neq 0} \frac{\alpha_n}{n}\, z^{-n},
\end{equation}
Then, the relations \eqref{cft:eq:scalar-p-pLR} and \eqref{cft:eq:scalar-w-pLR} allow to rewrite this result in terms of the momentum $p$ and winding $w$:
\index{scalar field CFT!momentum}%
\index{scalar field CFT!winding number}%
\begin{equation}
	\label{cft:eq:scalar-pw-a0}
	p
		= \frac{1}{\sqrt{2 \ell^2}} \,
			\big( \alpha_0 + \bar\alpha_0 \big),
	\qquad
	w
		= \frac{1}{\sqrt{2 \ell^2}} \,
			\big( \alpha_0 - \bar\alpha_0 \big).
\end{equation}
These relations can be inverted as
\begin{equation}
	\label{cft:eq:scalar-a0-pw}
	\alpha_0
		= \sqrt{\frac{\ell^2}{2}} \, (p + w),
	\qquad
	\bar\alpha_0
		= \sqrt{\frac{\ell^2}{2}} \, (p - w).
\end{equation}

\index{scalar field CFT!position (center of mass)}%
\index{scalar field CFT!dual position}%
In the same sense that there are two momenta $p_L$ and $p_R$ conjugated to $x_L$ and $x_R$, it makes sense to introduce two coordinates $x$ and $q$ conjugated to $p$ and $w$.
From string theory, the operator $x$ is called the center of mass.
The expression \eqref{cft:eq:scalar-a0-pw} suggests to write:
\begin{equation}
	\label{cft:eq:scalar-xLR-xq}
	x_L
		= x + q,
	\qquad
	x_R
		= x - q,
\end{equation}
and conversely:
\begin{equation}
	\label{cft:eq:scalar-xq-xLR}
	x = \frac{1}{2} \, (x_L + x_R),
	\qquad
	q = \frac{1}{2} \, (x_L - x_R).
\end{equation}
\index{scalar field CFT!zero-mode|)}%
In terms of these new variables, the expansion of the full $X(z, \bar z)$ reads:
\begin{equation}
	\label{cft:eq:scalar-exp-X}
	X(z, \bar z)
		= x
			- \I \, \frac{\ell^2}{2}\, \left(
				p \ln \abs{z}^2
				+ w \ln \frac{z}{\bar z}
				\right)
			+ \I \sqrt{\frac{\ell^2}{2}}
				\sum_{n \neq 0} \frac{1}{n} \big(\alpha_n \, z^{-n} + \bar \alpha_n \, \bar z^{-n} \big).
\end{equation}
In terms of the coordinates on the cylinder, the part without oscillations becomes:
\begin{equation}
	X(\tau, \sigma)
		= x
			- \I \, \ell^2 \, p \tau
			+ \ell^2 \, w \sigma
			+ \cdots
\end{equation}
Note how the presence of $\ell^2$ gives the correct scale to the second term.
The mode $q$ does not appear at all, and $x$ is the zero-mode of the complete field $X(z, \bar z)$.
As it is well-known, the physical interpretation of $x$ and $p$ is as the position and momentum of the centre-of-mass of the string.\footnotemark{}
\footnotetext{%
	In worldsheet Lorentzian signature, this becomes $X(\tau, \sigma) = x + \ell^2 \, p t + \ell^2 \, w \sigma$ as expected.
}%
If there is a compact dimension, then $w$ counts the number of times the string winds around it, and $q$ can be understood as the position of the centre-of-mass after a $T$-duality.\footnotemark{}
\footnotetext{%
	$T$-duality and compact bosons fall outside the scope of this \revname{} and we refer the reader to~\cites[chap.~17]{Zwiebach:2009:FirstCourseString}[chap.~8]{Polchinski:2005:StringTheory-1} for more details.
}%
\index{T-duality}%

\index{scalar field CFT!mode expansion|)}%

\begin{computation}[cft:eq:scalar-pLR-a0]
	\begin{align*}
		p_L
			&
			= \frac{1}{2\pi \I} \oint \dd z \, J
			= \frac{\I}{\ell^2} \frac{1}{2\pi \I}
				\oint \dd z \, \pd X
			= \frac{\I}{\ell^2} \, \frac{1}{2\pi \I}
				\oint \dd z \, \pd X
			\\
			&
			= \frac{1}{\sqrt{2 \ell^2}} \, \frac{1}{2\pi \I}
				\oint \dd z \, \sum_n \alpha_n\, z^{- n - 1}
			= \frac{1}{\sqrt{2 \ell^2}} \, \alpha_0.
	\end{align*}
	The computation gives $p_R$ after replacing $\alpha_0$ by $\bar\alpha_0$.
\end{computation}

\index{scalar field CFT!boundary condition!periodic}%
If the scalar field is non-compact but periodic on the cylinder, the periodicity condition
\begin{equation}
	X(\tau, \sigma + 2\pi)
		\sim X(\tau, \sigma)
\end{equation}
translates as
\begin{equation}
	X(\e^{2\pi \I} z, \e^{- 2\pi \I} \bar z)
		\sim X(z, \bar z).
\end{equation}
Evaluating the LHS from \eqref{cft:eq:scalar-exp-XLR} gives a constraint on the zero-modes:
\begin{equation}
	X(\e^{2\pi \I} z, \e^{- 2\pi \I} \bar z)
		= X(z, \bar z)
			- \I \sqrt{\frac{\ell^2}{2}}\, (\alpha_0 - \bar\alpha_0),
\end{equation}
which implies
\index{scalar field CFT!momentum}%
\index{scalar field CFT!winding number}%
\begin{equation}
	\label{cft:eq:scalar-nc-periodic-pLR}
	\alpha_0
		= \bar \alpha_0
	\quad \Longrightarrow \quad
	p_L
		= p_R
		= \frac{p}{2},
	\qquad
	w
		= 0.
\end{equation}
The other cases will not be discussed in this \revname{}, but we still use the general notation to make the contact with the literature easier.
This also implies that $X_L$ and $X_R$ cannot be periodic independently.
Hence, the zero-mode couples the holomorphic and anti-holomorphic sectors together.

\index{scalar field CFT!mode expansion|)}%

\index{scalar field CFT!number operator}%
The \emph{number operators} $N_n$ $\bar N_n$ at level $n > 0$ are defined by:
\begin{equation}
	\label{cft:eq:scalar-Nn}
	N_n
		= \frac{\epsilon}{n}\, \alpha_{-n} \alpha_n,
	\qquad
	\bar N_n
		= \frac{\epsilon}{n}\, \bar \alpha_{-n} \bar\alpha_n.
\end{equation}
The modes have been normal ordered.
They count the number of excitations at the level $n$: the factor $n^{-1}$ is necessary because the modes are not canonically normalized.
\index{scalar field CFT!level operator}%
Then, one can build the \emph{level operators}
\begin{equation}
	\label{cft:eq:scalar-N}
	N
		= \sum_{n > 0} n \, N_n.
\end{equation}
They count the number of excitations at level $n$ weighted by the level itself.
This corresponds to the total energy due to the oscillations (the higher the level, the more energy it needs to be excited).

\index{scalar field CFT!Virasoro operators}%
The Virasoro operators are
\begin{equation}
	\label{cft:eq:scalar-Ln}
	L_m
		= \frac{\epsilon}{2} \sum_n \norder{\alpha_n \alpha_{m-n}}
\end{equation}
For $m \neq 0$, we have
\begin{equation}
	m \neq 0:
	\qquad
	L_m
		= \frac{\epsilon}{2} \sum_{n \neq 0, m} \norder{\alpha_n \alpha_{m-n}}
			+ \epsilon \, \alpha_0 \alpha_m,
\end{equation}
there is no ordering ambiguity and the normal order can be removed.
\index{scalar field CFT!$L_0$|(}%
In the case of the zero-mode, one finds
\begin{equation}
	\label{cft:eq:scalar-L0}
	L_0
		= \frac{\epsilon}{2} \sum_{n} \norder{\alpha_n \alpha_{-n}}
		= N + \frac{\epsilon}{2} \, \alpha_0^2
		= N + \epsilon \ell^2 \, p_L^2,
\end{equation}
using \eqref{cft:eq:scalar-N} and \eqref{cft:eq:scalar-pLR-a0}.
It is also useful to define $\what L_0$ which corresponds to $L_0$ stripped from the zero-mode contribution:
\begin{equation}
	\label{cft:eq:scalar-L0-hat}
	\what L_0
		:= N.
\end{equation}
Similarly, the anti-holomorphic zero-mode is
\begin{equation}
	\label{cft:eq:scalar-L0bar}
	\bar L_0
		= \bar N + \epsilon \ell^2 \, p_R^2,
	\qquad
	\what{\bar L}_0
		:= \bar N,
\end{equation}
such that
\begin{subequations}
\label{cft:eq:scalar-L0pm}
\begin{align}
	\label{cft:eq:scalar-L0p}
	L_0^+
		&
		= N + \bar N
			+ \epsilon \ell^2 \, (p_L^2 + p_R^2)
		= N + \bar N
			+ \frac{\epsilon \ell^2}{2} \, (p^2 + w^2),
	\\
	\label{cft:eq:scalar-L0m}
	L_0^-
		&
		= N - \bar N
			+ \epsilon \ell^2 \, (p_L^2 - p_R^2)
		= N - \bar N
			+ \epsilon \ell^2 \, w p,
\end{align}
\end{subequations}
where $L_0^\pm := L_0 \pm \bar L_0$ as defined in \eqref{cft:eq:Ln-pm}.
The last equality of each line follows from \eqref{cft:eq:scalar-pL2-pR2}.
The expression of $L_0^+$ for $N = \bar N = 0$ matches the weights \eqref{cft:eq:scalar-dim-Vk} of the vertex operators for $p_L = p_R = p/2$ (no winding), which will be interpreted below.
It is a good place to stress that $p_L$, $p_R$, $p$ and $w$ are operators, while $k$ is a number.
\index{scalar field CFT!$L_0$|)}%

\subsection{Commutators}
\label{cft:sec:systems:scalar:com}

\index{scalar field CFT!commutator|(}%
The commutators can be computed from \eqref{cft:eq:com-int-A-B} knowing the OPE \eqref{cft:eq:scalar-ope-dX-dX}.
The modes of $\pd X$ and $\bar \pd X$ satisfy
\begin{equation}
	\label{cft:eq:scalar-com-a}
	\com{\alpha_m}{\alpha_n}
		= \epsilon \, m \, \delta_{m+n,0},
	\qquad
	\com{\bar\alpha_m}{\bar\alpha_n}
		= \epsilon \, m \, \delta_{m+n,0},
	\qquad
	\com{\alpha_m}{\bar\alpha_n}
		= 0
\end{equation}
for all $m, n \in \Z$ (including the zero-modes).
The appearance of the factor $m$ in the RHS explains the normalization of the number operator \eqref{cft:eq:scalar-Nn}.

From the commutators of the zero-modes, we directly find the ones for the momentum and winding:
\begin{equation}
	\label{cft:eq:scalar-com-pw}
	\com{p}{w}
		= \com{p}{p}
		= \com{w}{w}
		= 0,
	\qquad
	\com{p}{\alpha_n}
		= \com{p}{\bar\alpha_n}
		= \com{w}{\alpha_n}
		= \com{w}{\bar\alpha_n}
		= 0.
\end{equation}
The OPE \eqref{cft:eq:scalar-ope-dX-X} yields
\begin{equation}
	\label{cft:eq:scalar-com-xpLR}
	\com{x_L}{p_L}
		= \I \epsilon,
		\qquad
	\com{x_R}{p_R}
		= \I \epsilon,
\end{equation}
which can be used to determine the commutators of $x$ and $q$:
\begin{equation}
	\label{cft:eq:scalar-com-xpqw}
	\com{x}{p}
		= \com{q}{w}
		= \I \epsilon,
	\qquad
	\com{x}{w}
		= \com{q}{p}
		= 0.
\end{equation}
This shows that $(x, p)$ and $(q, w)$ are pairs of conjugate variables.
Interestingly, the winding number $w$ commutes will all other modes except $q$, but the latter disappears from the description.
Hence, it can be interpreted as a number which labels different representations: if no other principle (like periodicity) forbids $w \neq 0$, then one can except to have states with all possible $w$ in the spectrum, each value of $w$ forming a different sector.
There are other interpretations from the point of view of $T$-duality and double field theory~\cite{Zwiebach:2009:FirstCourseString, Plauschinn:2018:NongeometricBackgroundsString, Hull:2009:DoubleFieldTheory, Hohm:2013:SpacetimeDoubleField}.

The commutator of the modes with the Virasoro operators is
\begin{equation}
	\label{cft:eq:scalar-com-L-a}
	\com{L_m}{\alpha_n}
		= - n \, \alpha_{m+n}.
\end{equation}
as expected from \eqref{cft:eq:com-Ln-phin}.
For $m = 0$, this reduces to
\begin{equation}
	\label{cft:eq:scalar-com-L0-a}
	\com{L_0}{\alpha_{-n}} = n\, \alpha_{-n},
\end{equation}
which shows that negative modes increase the energy.
The commutator of the creation modes $\alpha_{-n}$ with the number operators is
\begin{equation}
	\label{cft:eq:scalar-com-N-a}
	\com{N_m}{\alpha_{-n}} = \alpha_{-m} \delta_{m,n}.
\end{equation}
\index{scalar field CFT!commutator|)}%

\subsection{Hilbert space}

The Hilbert space of the free scalar has the structure of a Fock space.

From \eqref{cft:eq:scalar-com-L0-a}, the momentum $p$ commutes with the Hamiltonian $L_0^+$ such that it is a good quantum number to label the states:\footnotemark{} this translates the fact the action \eqref{cft:eq:scalar-action} does not depend on the conjugate variable $x$.
\footnotetext{%
	To simplify the discussion, we do not consider winding but only vertex operators of the form \eqref{cft:eq:scalar-Vk}.
}%
As a consequence, there exists a family of vacua $\ket{k}$.

\index{scalar field CFT!vacuum|(}%
The vacua $\ket{k}$ are the states related to the vertex operators \eqref{cft:eq:scalar-Vk} through the state-operator correspondence:
\begin{equation}
	\label{cft:eq:scalar-vac-k}
	\ket{k}
		:= \lim_{z, \bar z \to 0} V_k(z, \bar z) \ket{0}
		= \e^{\I \epsilon k x} \ket{0},
\end{equation}
where $\ket{0}$ is the $\group{SL}(2, \C)$ vacuum and $x$ is the zero-mode of $X(z, \bar z)$.
That this identification is correct follows by applying the operator $p$:
\begin{equation}
	\label{cft:eq:scalar-p-eigenval}
	p \ket{k}
		= k \ket{k}.
\end{equation}
The notation is consistent with the one of the $\group{SL}(2, \C)$ vacuum since $p \ket{0} = 0$.

The vacuum is annihilated by the action of the positive-frequency modes:
\begin{equation}
	\label{cft:eq:scalar-vac-k-annihilation}
	\forall n > 0:
	\qquad
	\alpha_n \ket{k}
		= 0,
\end{equation}
which is equivalent to
\begin{equation}
	N_n \ket{k}
		= 0.
\end{equation}
The different vacua are each ground state of a Fock space (they are all equivalent), but they are not ground states of the Hamiltonian since they have different energies:
\begin{equation}
	L_0^+ \ket{k}
		= 2 \epsilon \ell^2 \, k^2 \ket{k},
	\qquad
	L_0^- \ket{k}
		= 0,
\end{equation}
using \eqref{cft:eq:scalar-L0pm}.
The $\group{SL}(2, \C)$ vacuum is the lowest (highest) energy state if $\epsilon = 1$ ($\epsilon = -1$).
\index{scalar field CFT!vacuum|)}%

\index{scalar field CFT!Fock space}%
The Fock space $\mc F(k)$ built from the vacuum at momentum $k$ is found by acting repetitively with the negative-frequency modes.
A convenient basis, the oscillator basis, is given by the states:
\begin{subequations}
\begin{gather}
	\label{cft:eq:scalar-fock}
	\mc F(k)
		= \Span \Big\{ \Ket{k ; \{ N_n \}} \Big\},
	\\
	\Ket{k ; \{ N_n \}}
		:= \prod_{n \ge 1}
			\frac{(\alpha_{-n})^{N_n}}{\sqrt{n^{N_n} N_n!}} \ket{k},
	\qquad
	N_n \in \N^*
\end{gather}
\end{subequations}
(we don't distinguish the notations between the number operators and their eigenvalues).
\index{scalar field CFT!Hilbert space}%
The full Hilbert space is given by:
\begin{equation}
	\mc H
		= \int_\R \dd k \, \mc F(k).
\end{equation}

\begin{computation}[cft:eq:scalar-vac-k]
	We provide a quick argument to justify the second form of \eqref{cft:eq:scalar-vac-k}.
	Take the limit of \eqref{cft:eq:scalar-exp-X} with $w = 0$:
	\begin{align*}
	\lim_{z, \bar z \to 0} \e^{\I \epsilon k X(z, \bar z)} \ket{0}
		&
		= \lim_{z, \bar z \to 0}
			\exp \I \epsilon k \left[
				x
				- \I \, \frac{\ell^2}{2} \, p \ln \abs{z}^2
				+ \I \sqrt{\frac{\ell^2}{2}}
					\sum_{n \neq 0} \frac{1}{n} \big(\alpha_n \, z^{-n} + \bar \alpha_n \, \bar z^{-n} \big)
				\right] \ket{0}
		\\ &
		= \lim_{z, \bar z \to 0}
			\exp \left[
				\I \epsilon k x
				- \epsilon k \sqrt{\frac{\ell^2}{2}}
					\sum_{n \neq 0} \frac{1}{n} \big(\alpha_n \, z^{-n} + \bar \alpha_n \, \bar z^{-n} \big)
				\right] \ket{0}.
	\end{align*}
	The second term from the first line disappears because $p \ket{0} = 0$.
	For $\epsilon k > 0$, as $z, \bar z \to 0$, the terms with $\alpha_{n}$ and $\bar\alpha_{n}$ for $n < 0$ disappear since they are accompanied with a positive power of $z^n$ and $\bar z^n$.
	The modes with $n > 0$ diverge but the minus sign makes the exponential to vanish.
	A more rigorous argument requires to normal order the exponential and then to use \eqref{cft:eq:scalar-vac-k-annihilation}.
\end{computation}

\begin{computation}[cft:eq:scalar-p-eigenval]
	\begin{align*}
		p \ket{k}
			&
			= \frac{1}{\ell^2} \,
				\frac{1}{2\pi\I}
				\oint \left( \dd z \, \I \pd X(z) + \dd \bar z \, \I \bar\pd X(\bar z) \right)
				V_k(0, 0) \ket{0}
			\\ &
			= \frac{1}{\ell^2} \,
				\frac{1}{2\pi\I}
				\oint \left(
					\frac{\dd z}{z} \, \frac{\ell^2 k}{2}
					+ \frac{\dd \bar z}{\bar z} \, \frac{\ell^2 k}{2}
					\right)
				V_k(0, 0) \ket{0}
			\\ &
			= k \, V_k(0, 0) \ket{0}
	\end{align*}
	using \eqref{cft:eq:scalar-ope-dX-V}.
\end{computation}

\begin{remark}[Fock space and Verma module isomorphism]
	Note that, in the absence of the so-called null states, there is a one-to-one map between states in the $\alpha_{-n}$ oscillator basis and in the $L_{-n}$ Virasoro basis.
	This translates an isomorphism between the Fock space and the Verma module of $V_k$.
	One hint for this relation is that applying $\alpha_{-n}$ and $L_{-n}$ changes the weight (eigenvalue of $L_0$) by the same amount, and there are as many operators in both basis.
\end{remark}

\subsection{Euclidean and BPZ conjugates}

\index{scalar field CFT!Euclidean adjoint!modes}%
Since $X$ is a real scalar field, it is self-adjoint \eqref{cft:eq:conj-euc} such that
\begin{equation}
	\label{cft:eq:scalar-modes-adj}
	\adj{x} = x
	\qquad
	\adj{p} = p,
	\qquad
	\adj{\alpha_n} = \alpha_{-n}.
\end{equation}
\index{scalar field CFT!Euclidean adjoint!vacuum}%
This implies that the Virasoro operators \eqref{cft:eq:scalar-Ln} are Hermitian:
\begin{equation}
	\adj{L_n}
		= L_{-n},
\end{equation}
as expected since $T(z)$ is self-adjoint for a free scalar field.

As a consequence of \eqref{cft:eq:scalar-modes-adj}, the adjoint of the vacuum $\ket{k}$ follows from \eqref{cft:eq:scalar-vac-k}:
\begin{equation}
	\bra{k}
		= \eadj{\ket{k}}
		= \bra{0} \e^{- \I \epsilon k x},
	\qquad
	\bra{k} p
		= \bra{k} k.
\end{equation}

\index{scalar field CFT!BPZ conjugate!modes}%
The BPZ conjugate \eqref{cft:eq:modes-bpz} of the mode $\alpha_n$ is:
\begin{equation}
	\alpha_n^t
		= - (\pm 1)^{n} \alpha_{-n},
	\qquad
\end{equation}
where the sign depends on the choice of $I^\pm$ in \eqref{cft:eq:modes-bpz}.
\index{scalar field CFT!BPZ conjugate!vacuum}%
Using \eqref{cft:eq:scalar-pw-a0}, this implies that the momentum operator gets a minus sign:\footnotemark{}
\footnotetext{%
	Be careful that $\ket{k}$ is not the state associated to the operator $p$ through the state--operator correspondence.
	Instead, they are associated to $V_k$, see \eqref{cft:eq:scalar-vac-k}.
	This explains why $\bra{k} \neq (\ket{k})^t$ as in \eqref{cft:eq:state-bpz-phi}.
}%
\begin{equation}
	p^t = - p,
	\qquad
	\bra{-k}
		= \ket{k}^t.
\end{equation}

\index{scalar field CFT!vacuum!conjugate}%
\index{scalar field CFT!inner product}%
The inner product between two vacua $\ket{k}$ and $\ket{k'}$ is normalized as:
\begin{equation}
	\bracket{k}{k'} = 2\pi \, \delta(k - k')
\end{equation}
such that the conjugate state \eqref{cft:eq:bpz-conj-state} of the vacuum reads
\begin{equation}
	\bra{k^c}
		= \frac{1}{2\pi} \, \bra{k}.
\end{equation}
The Hermitian and BPZ conjugate states are related as:
\begin{equation}
	\eadj{\ket{k}} = - \ket{k}^t,
\end{equation}
which can be interpreted as a reality condition on $\ket{k}$.

\begin{draft}

\section{Free scalar: specific cases}

\subsection{Non-compact scalar}

\subsection{Compact scalar}

\subsection{Twisted scalar (orbifold)}

\subsection{Unoriented scalar}

\end{draft}

\begin{draft}

\section{Coulomb gas (linear dilaton)}
\label{cft:sec:systems:coulomb-gas}

The Coulomb gas is the most useful CFT system in string theory because all free theories (scalar, fermion, ghosts and superghosts) can be described in terms of it.

\begin{equation}
	L_0 = - m^2 \ell^2 + N, \qquad
	- m^2 = \epsilon\, a (Q - a)
		= \epsilon \left(\frac{Q^2}{4} + p^2 \right).
\end{equation}

\subsection{Compact Coulomb gas}

\end{draft}

\section{First-order \texorpdfstring{$bc$}{bc} ghost system}
\label{cft:sec:systems:ghosts}

\index{first-order CFT}%
First-order systems describe two free fields called ghosts which have a first-order action and whose conformal weights sum to $1$.
Commuting (resp.\ anti-commuting) fields are often denoted by $\beta$ and $\gamma$ (resp.\ $b$ and $c$) and correspondingly first-order systems are also called $\beta\gamma$ or $bc$ systems.
We will introduce a sign $\epsilon = \pm 1$ to denote the Grassmann parity of the fields and always write them as $b$ and $c$.
In string theory, first-order systems describe the Faddeev--Popov ghosts associated to reparametrizations and supersymmetries (\Cref{bos:sec:ws-int:ghosts,sws:chap:worldsheet}).

\subsection{Covariant action}

\index{first-order CFT!action}%
A first-order system is defined by two symmetric and traceless fields $b_{\mu_1 \cdots \mu_\lambda}$ and $c^{\mu_1 \cdots \mu_{\lambda-1}}$ called ghosts.
For fields of integer spins, the dynamics is governed by the first-order action
\begin{equation}
	\label{cft:eq:1storder-action}
	S
		= \frac{1}{4\pi} \int \dd^2 x \sqrt{g} \,
			g^{\mu\nu} \, b_{\mu \mu_1 \cdots \mu_{\lambda-1}}
			\grad_{\nu} c^{\mu_1 \cdots \mu_{\lambda-1}}
\end{equation}
after taking into account the symmetries of the field indices.
Obviously, for $\lambda = 2$, one recovers the reparametrization ghost action \eqref{bos:eq:action-ghost}.
The action \eqref{cft:eq:1storder-action} is invariant under Weyl transformations (the fields and covariant derivatives are inert) such that it describes a CFT on flat space.

When the fields have half-integer spins (and often denoted as $\beta$ and $\gamma$ in this case), they carry a spinor index.
In this case, the action contains a Dirac matrix, and the covariant derivative a spin connection.

\index{first-order CFT!$\mathrm{U}(1)$ symmetry}%
The ghost action \eqref{cft:eq:1storder-action} is invariant under a global $\group{U}(1)$ symmetry
\begin{equation}
	\label{cft:eq:1storder-U1-global}
	b_{\mu_1 \cdots \mu_n} \longrightarrow \e^{- \I \theta} b_{\mu_1 \cdots \mu_n},
	\qquad
	c^{\mu_1 \cdots \mu_{n-1}} \longrightarrow \e^{\I \theta} c^{\mu_1 \cdots \mu_{n-1}}.
\end{equation}

\subsection{Action on the complex plane}

The simplest description of the system is on the complex plane.
Due to the conditions imposed on the fields, they have only two independent components for all $n$, and the equations of motion imply that one is holomorphic, and the other anti-holomorphic:
\index{first-order CFT!complex components}%
\begin{equation}
	b(z)
		:= b_{z \cdots z}(z),
	\qquad
	\bar b(\bar z)
		:= b_{\bar z \cdots \bar z}(\bar z),
	\qquad
	c(z)
		:= c^{z \cdots z}(\bar z),
	\qquad
	\bar c(\bar z)
		:= c^{\bar z \cdots \bar z}(z).
\end{equation}
In this language, the action becomes
\index{first-order CFT!action}%
\begin{equation}
	\label{cft:eq:1storder-action-plane}
	S = \frac{1}{2\pi} \int \dd^2 z \big( b \bar\pd c + \bar b \pd \bar c).
\end{equation}
\index{first-order CFT!equation of motion}%
This action gives the correct equations of motion
\begin{equation}
	\pd \bar b = 0,
	\qquad
	\bar\pd b = 0,
	\qquad
	\pd \bar c = 0,
	\qquad
	\bar\pd c = 0.
\end{equation}
Since the fields split into holomorphic and anti-holomorphic sectors, it is convenient to study only the holomorphic sector as usual.
This system is even simpler than the scalar field because the zero-modes don't couple both sectors.\footnotemark{}
\footnotetext{%
	For the scalar field, the coupling of both sectors happened because of the periodicity condition \eqref{cft:eq:scalar-nc-periodic-pLR}.
}%
All formulas for the anti-holomorphic sector are directly obtained from the holomorphic one by adding bars on quantities, except for conserved charges which have an index $L$ or $R$ and are both written explicitly.

\index{first-order CFT!weight}%
The action describes a CFT, and the weight of the fields are given by
\begin{equation}
	\label{cft:eq:1storder-weight}
	h(b) = \lambda,
	\qquad
	h(c) = 1 - \lambda,
	\qquad
	h(\bar b) = \lambda,
	\qquad
	h(\bar c) = 1 - \lambda,
\end{equation}
where $\lambda = n$ if the fields are in a tensor representation, and $\lambda = n + 1/2$ if they are in a spinor-tensor representation.
The holomorphic energy--momentum reads
\index{first-order CFT!energy--momentum tensor}%
\begin{subequations}
\label{cft:eq:1storder-T}
\begin{align}
	T
		&= - \lambda \, \norder{b \pd c} + (1 - \lambda) \, \norder{\pd b \, c}
		\\
		&= - \lambda \, \norder{\pd (b c)} + \norder{\pd b \, c}
		\\
		&= (1 - \lambda) \, \norder{\pd (b c)} - \norder{b \, \pd c}.
\end{align}
\end{subequations}
Normal ordering is taken with respect to the $\group{SL}(2, \C)$ vacuum \eqref{cft:eq:vacuum-conformal}.

Finally, both fields can be classically commuting or anticommuting (see below for the quantum commutators):
\begin{equation}
	b(z) c(w)
		= - \epsilon \, c(w) b(z),
	\qquad
	b(z) b(w)
		= - \epsilon \, b(w) b(z),
	\qquad
	c(z) c(w)
		= - \epsilon \, c(w) c(z),
\end{equation}
where $\epsilon$ denotes the Grassmann parity
\begin{equation}
	\epsilon =
	\begin{cases}
		+ 1 & \text{anticommuting}, \\
		- 1 & \text{commuting}.
	\end{cases}
\end{equation}
Sometimes, if $\epsilon = +1$, one denotes $b$ and $c$ respectively by $\beta$ and $\gamma$.
If $b$ and $c$ are ghosts arising from Faddeev--Popov gauge fixing, then $\epsilon = 1$ if $\lambda$ is integer; and $\epsilon = - 1$ if $\lambda$ is half-integer (“wrong” spin--statistics assignment).

\index{first-order CFT!$\mathrm{U}(1)$ symmetry}%
The $\group{U}(1)$ global symmetry \eqref{cft:eq:1storder-U1-global} reads infinitesimally
\begin{equation}
	\delta b = - \I b,
	\qquad
	\delta c = \I c,
	\qquad
	\delta \bar b = - \I \bar b,
	\qquad
	\delta \bar c = \I \bar c.
\end{equation}
\index{first-order CFT!$\mathrm{U}(1)$ ghost current}%
It is generated by the conserved \emph{ghost current} with components:
\begin{equation}
	\label{cft:eq:1storder-current}
	j(z)
		= - \norder{b(z) c(z)},
	\qquad
	\bar \jmath(\bar z)
		= - \norder{\bar b(\bar z) \bar c(\bar z)}
\end{equation}
and the associated charge is called the ghost number
\index{first-order CFT!ghost number}%
\begin{equation}
	\label{cft:eq:1storder-number}
	N_{\text{gh}}
		= N_{\text{gh},L} + N_{\text{gh},R},
	\qquad
	N_{\text{gh},L}
		= \oint \frac{\dd z}{2\pi \I} \, j(z),
	\qquad
	N_{\text{gh},R}
		= - \oint \frac{\dd \bar z}{2\pi \I} \, \bar \jmath(\bar z).
\end{equation}
This charge counts the number of $c$ ghosts minus the number of $b$ ghosts, such that
\begin{equation}
	\label{cft:eq:1storder-number-bc-plane}
	N_{\text{gh}}(c) = 1,
	\qquad
	N_{\text{gh}}(b) = - 1,
	\qquad
	N_{\text{gh}}(\bar c) = 1,
	\qquad
	N_{\text{gh}}(\bar b) = - 1.
\end{equation}

\index{first-order CFT!propagator}%
The propagator can be derived from the path integral
\begin{equation}
	\int \dd' b \, \dd' c \;
			\frac{\delta}{\delta b(z)} \left[ b(w) \e^{- S[b, c]} \right]
		= 0
\end{equation}
which gives the differential equation
\begin{equation}
	\delta^{(2)}(z - w) + \frac{1}{2\pi} \, \mean{b(w) \bar\pd c(z)} = 0.
\end{equation}
Using \eqref{app:eq:pd-z-inv}, the solution is easily found to be
\begin{equation}
	\label{cft:eq:1storder-propagator}
	\mean{c(z) b(w)} = \frac{1}{z - w}.
\end{equation}

\begin{remark}
	The propagator is constructed with the path integral.
	For convenience, the zero-modes are removed from the measure: reintroducing them, one finds that the propagator is computed not in the conformal vacuum (which has no operator insertion), but in a state with ghost insertions.
	This explains why the propagator \eqref{cft:eq:1storder-propagator} is not of the form \eqref{cft:eq:corr-func-2pt}.
	However, this form is sufficient to extract the OPE as changing the vacuum does not introduce singular terms.
\end{remark}

\subsection{OPE}

\index{first-order CFT!OPE|(}%

The OPEs between the $b$ and $c$ fields are found from the propagator \eqref{cft:eq:1storder-propagator}:
\begin{subequations}
\label{cft:eq:1storder-ope-bc}
\begin{gather}
	c(z) b(w)
		\sim \frac{1}{z - w},
	\qquad
	b(z) c(w)
		\sim \frac{\epsilon}{z - w},
	\\
	b(z) b(w)
		\sim 0,
	\qquad
	c(z) c(w)
		\sim 0.
\end{gather}
\end{subequations}

The OPE of each ghost with $T$ confirms the conformal weights in \eqref{cft:eq:1storder-weight}:
\begin{subequations}
\label{cft:eq:1storder-ope-bc-T}
\begin{gather}
	\label{cft:eq:1storder-ope-b-T}
	T(z) b(w)
		\sim \lambda \, \frac{b(w)}{(z - w)^2} + \frac{\pd b(w)}{z - w},
	\\
	\label{cft:eq:1storder-ope-c-T}
	T(z) c(w)
		\sim (1 - \lambda) \, \frac{c(w)}{(z - w)^2} + \frac{\pd c(w)}{z - w}.
\end{gather}
\end{subequations}
The OPE of $T$ with itself is
\begin{equation}
	\label{cft:eq:1storder-ope-T-T}
	T(z) T(w)
		\sim \frac{c_\lambda/2}{(z - w)^4} + \frac{2 T(w)}{(z - w)^2} + \frac{\pd T(w)}{z - w},
\end{equation}
\index{first-order CFT!central charge}%
where the central charge is:
\begin{equation}
	\label{cft:eq:1storder-central-charge}
	c_\lambda
		= 2 \epsilon (- 1 + 6 \lambda - 6 \lambda^2)
		= - 2 \epsilon \big( 1 + 6 \lambda (\lambda - 1) \big).
\end{equation}
\index{first-order CFT!ghost charge}%
Introducing the ghost charge:
\begin{equation}
	\label{cft:eq:1storder-charge}
	q_\lambda = \epsilon (1 - 2 \lambda),
\end{equation}
the central charge can also be written as
\begin{equation}
	c_\lambda
		= \epsilon (1 - 3 q_\lambda^2 ).
\end{equation}
This parameter will appear many times in this section and its meaning will become clearer as we proceed.

The OPE between the ghost current \eqref{cft:eq:1storder-current} and the $b$ and $c$ ghosts read
\begin{subequations}
\label{cft:eq:1storder-ope-bc-j}
\begin{gather}
	\label{cft:eq:1storder-ope-b-j}
	j(z) b(w)
		\sim - \frac{b(w)}{z - w},
	\\
	\label{cft:eq:1storder-ope-c-j}
	j(z) c(w)
		\sim \frac{c(w)}{z - w}.
\end{gather}
\end{subequations}
The coefficients of the $(z - w)^{-1}$ terms correspond to the ghost number of the $b$ and $c$ fields \eqref{cft:eq:1storder-number-bc-plane}.
More generally, the ghost number $N_{\text{gh}}(\mc O)$ of any operator $\mc O(z)$ is defined by
\begin{equation}
	j(z) \mc O(w)
		\sim N_{\text{gh}}(\mc O) \, \frac{\mc O(w)}{z - w}.
\end{equation}

The OPE for $j$ with itself is
\begin{equation}
	\label{cft:eq:1storder-ope-j-j}
	j(z) j(w)
		\sim \frac{\epsilon}{(z - w)^2}.
\end{equation}
\begin{check}
This will be interpreted later in the context of bosonization.
\end{check}

Finally, the OPE of the current with $T$ reads:
\begin{equation}
	\label{cft:eq:1storder-ope-T-j}
	T(z) j(w)
		\sim \frac{q_\lambda}{(z - w)^3}
			+ \frac{j(w)}{(z - w)^2}
			+ \frac{\pd j(w)}{z - w}.
\end{equation}
Due to the presence of the $z^{-3}$ term, the current $j(z)$ is not a primary field if $q_\lambda \neq 0$, that is, if $\lambda \neq 1/2$.
In that case, its transformation under changes of coordinates gets an anomalous contribution:
\index{first-order CFT!$\mathrm{U}(1)$ ghost current!transformation law}%
\begin{equation}
	j(z)
		= \frac{\dd w}{\dd z} \, j'(w)
			+ \frac{q_\lambda}{2} \, \frac{\dd}{\dd z} \ln \frac{\dd w}{\dd z}
		= \frac{\dd w}{\dd z} \, j'(w)
			+ \frac{q_\lambda}{2} \, \frac{\pd_z^2 w}{\pd_z w}.
\end{equation}
\index{first-order CFT!cylinder}%
This implies in particular that the currents on the plane and on the cylinder ($w = \ln z$) are related by:
\begin{equation}
	j(z)
		= \frac{\dd w}{\dd z} \left(
			j^{\text{cyl}}(w)
			- \frac{q_\lambda}{2}
			\right),
\end{equation}
which leads to the following relation between the ghost numbers on the plane and on the cylinder:
\index{first-order CFT!ghost number!cylinder}%
\begin{equation}
	\label{cft:eq:1storder-number-cyl}
	N_{\text{gh}}
		= N_{\text{gh}}^{\text{cyl}} - q_\lambda,
	\qquad
	N_{\text{gh},L}
		= N_{\text{gh},L}^{\text{cyl}} - \frac{q_\lambda}{2},
	\qquad
	N_{\text{gh},R}
		= N_{\text{gh},R}^{\text{cyl}} - \frac{q_\lambda}{2}.
\end{equation}
For this reason, it is important to make clear the space with respect to which is given the ghost number: if not explicitly stated, ghost numbers in this \revname{} are given on the plane.\footnotemark{}
\footnotetext{%
	Other references, especially old ones, give it on the cylinder.
	This can be easily recognized if some ghost numbers in the holomorphic sector are half-integers: for the reparametrization ghosts, $q_\lambda$ is an integer such that the shift in \eqref{cft:eq:1storder-number-cyl} is a half-integer.
}%
Due to this anomaly, one finds that the ghost number is not conserved on a curved space:
\begin{equation}
	N^c - N^b
		= - \frac{\epsilon \, q_\lambda}{2} \, \chi_g
		= (1 - 2 \lambda) (g - 1),
\end{equation}
where $\chi_g$ is the Euler characteristics \eqref{bos:eq:chi-g}, $N^b$ and $N^c$ are the numbers of $b$ and $c$ operators.
In string theory, where the only ghost insertions are zero-modes, this translates into a statement on the number of zero-modes to be inserted.
Hence, this can be interpreted as a generalization of \eqref{bos:eq:P1-Riemann-Roch}.
For a proof, see for example~\cite[p.~397]{Blumenhagen:2014:BasicConceptsString}.

\begin{computation}[cft:eq:1storder-ope-b-T]
	\begin{align*}
		T(z) b(w)
			&= \Big( - \lambda \, \norder{b(z) \pd c(z)} + (1 - \lambda) \, \norder{\pd b(z) \, c(z)} \Big) b(w)
			\\
			&\sim - \lambda \, \wick{\norder{b(z) \pd \c c(z)} \, \c b(w)}
				+ (1 - \lambda) \, \wick{ \norder{\pd b(z) \, \c c(z)} \, \c b(w) }
			\\
			&\sim - \lambda \, b(z) \pd_z \frac{1}{z - w}
				+ (1 - \lambda) \, \pd b(z) \, \frac{1}{z - w}
			\\
			&\sim \lambda \, \big( b(w) + \cancel{(z - w) \pd b(w)} \big) \, \frac{1}{(z - w)^2}
				+ (1 - \cancel{\lambda}) \, \frac{\pd b(w)}{z - w}.
	\end{align*}
\end{computation}

\begin{computation}[cft:eq:1storder-ope-c-T]
	\begin{align*}
		T(z) c(w)
			&= \Big( - \lambda \, \norder{b(z) \pd c(z)} + (1 - \lambda) \, \norder{\pd b(z) c(z)} \Big) c(w)
			\\
			&\sim \epsilon \lambda \, \wick{\norder{\pd c(z) \c b(z)} \, \c c(w)}
				- \epsilon (1 - \lambda) \, \wick{ \norder{c(z) \pd \c b(z)} \, \c c(w) }
			\\
			&\sim \lambda \, \frac{\pd c(z)}{z - w}
				- (1 - \lambda) \, c(z) \, \pd_z \frac{1}{z - w}
			\\
			&\sim \lambda \, \frac{\pd c(w)}{z - w}
				+ (1 - \lambda) \, \big( c(w) + (z - w) \pd c(w) \big) \, \frac{1}{(z - w)^2}
			\\
			&\sim (1 - \lambda) \, \frac{c(w)}{(z - w)^2}
				+ \frac{\pd c(w)}{(z - w)^2}.
	\end{align*}
\end{computation}

\begin{computation}[cft:eq:1storder-ope-b-j]
	\[
		j(z) b(w)
			= - \norder{b(z) c(z)} \, b(w)
			\sim - \wick{\norder{b(z) \c c(z)} \, \c b(w)}
			\sim - \frac{b(z)}{z - w}
			\sim - \frac{b(w)}{z - w}.
	\]
\end{computation}

\begin{computation}[cft:eq:1storder-ope-c-j]
	\[
		j(z) c(w)
			= - \norder{b(z) c(z)} \, c(w)
			\sim \epsilon \, \wick{\norder{c(z) \c b(z)} \, \c c(w)}
			\sim \frac{c(z)}{z - w}
			\sim \frac{c(w)}{z - w}.
	\]
\end{computation}

\begin{computation}[cft:eq:1storder-ope-j-j]
	\begin{align*}
		j(z) j(w)
			&= \norder{b(z) c(z)} \, \norder{b(w) c(w)}
			\\
			&\sim \wick{ \norder{\c2 b(z) \c1 c(z)} \, \norder{\c1 b(w) \c2 c(w)} }
				+ \wick{ \norder{b(z) \c c(z)} \, \norder{\c b(w) c(w)} }
				+ \wick{ \norder{\c b(z) c(z)} \, \norder{b(w) \c c(w)} }
			\\
			&\sim \frac{\epsilon}{(z - w)^2}
				+ \frac{\epsilon \, \norder{c(z) b(w)}}{z - w}
				+ \frac{\norder{b(z) c(w)}}{z - w}
			\sim \frac{\epsilon}{(z - w)^2}.
	\end{align*}
\end{computation}

\index{first-order CFT!OPE|)}%

\subsection{Mode expansions}

\index{first-order CFT!mode expansion}%
The $b$ and $c$ ghosts are expanded as
\begin{equation}
	\label{cft:eq:1storder-exp-bc}
	b(z)
		= \sum_{n \in \Z + \lambda + \nu} \frac{b_n}{z^{n + \lambda}},
	\qquad
	c(z)
		= \sum_{n \in \Z + \lambda + \nu} \frac{c_n}{z^{n + 1 - \lambda}},
\end{equation}
where $\nu = 0, 1/2$ depends on $\epsilon$ and on the periodicity of the fields, see \eqref{cft:eq:mode-range}.
The modes are extracted with the contour formulas
\begin{equation}
	\label{cft:eq:1storder-modes-int}
	b_n
		= \oint \frac{\dd z}{2\pi \I} \, z^{n + \lambda - 1} b(z),
	\qquad
	c_n
		= \oint \frac{\dd z}{2\pi \I} \, z^{n - \lambda} c(z).
\end{equation}

\index{first-order CFT!boundary condition}%
Ghosts with $\lambda \in \Z$ have integer indices and $\nu = 0$ (we don't consider ghosts with twisted boundary conditions).
On the other hand, ghosts with $\lambda \in \Z + 1/2$ have integer indices and $\nu = 1/2$ in the R sector, and half-integer indices and $\nu = 0$ in the NS sector (see \Cref{cft:chap:plane:radial:modes}).
The choices in the boundary conditions arise from the $\Z_2$ symmetry of the action:
\begin{equation}
	b \longrightarrow - b,
	\qquad
	c \longrightarrow - c.
\end{equation}

\index{first-order CFT!number operator}%
The number operators $N^b_n$ and $N^c_n$ are defined to count the numbers of excitations above the $\group{SL}(2, \C)$ vacuum of $b$ and $c$ ghosts at level $n$:
\begin{equation}
	\label{cft:eq:1storder-Nn}
	N^b_n
		= \norder{b_{-n} c_n},
	\qquad
	N^c_n
		= \epsilon \, \norder{c_{-n} b_n}.
\end{equation}
The definitions follow from the commutators \eqref{cft:eq:1storder-com-bn-cn}.
\index{first-order CFT!level operator}%
Then, the level operators $N^b$ and $N^c$ are obtained by summing over $n$:
\begin{equation}
	\label{cft:eq:1storder-N}
	N^b
		= \sum_{n > 0} n \, N^b_n,
	\qquad
	N^c
		= \sum_{n > 0} n \, N^c_n.
\end{equation}

\index{first-order CFT!Virasoro operators}%
\index{first-order CFT!energy--momentum tensor!mode expansion}%
The Virasoro operators are
\begin{equation}
	\label{cft:eq:1storder-Ln}
	L_m
		= \sum_{n} \big( n - (1 - \lambda) m \big) \, \norder{b_{m-n} c_n}
		= \sum_{n} (\lambda m - n) \, \norder{b_{n} c_{m-n}}.
\end{equation}
\index{first-order CFT!$L_0$}%
Of particular importance is the zero-mode
\begin{equation}
	\label{cft:eq:1storder-L0}
	L_0
		= - \sum_{n} n\, \norder{b_{n} c_{-n}}
		= \sum_{n} n\, \norder{b_{-n} c_{n}}.
\end{equation}
We will give the expression of $L_0$ in terms of the level operators below, see \eqref{cft:eq:1storder-L0-ev}.
To do this, we will first need to change the normal ordering, which first requires to study the Hilbert space.

\index{first-order CFT!$\mathrm{U}(1)$ ghost current!mode expansion}%
The modes of the ghost current are
\begin{equation}
	\label{cft:eq:1storder-jn}
	j_m
		= - \sum_{n} \norder{b_{m-n} c_{n}}
		= - \sum_{n} \norder{b_{n} c_{m-n}}.
\end{equation}
Note that the zero-mode of the current also equals the ghost number
\begin{equation}
	\label{cft:eq:1storder-modes-Ngh}
	N_{\text{gh},L}
		= j_0
		= - \sum_{n} \norder{b_{-n} c_{n}}.
\end{equation}

When both the holomorphic and anti-holomorphic sectors enter, it is convenient to introduce the combinations
\begin{equation}
	\label{cft:eq:bn-cn-pm}
	b_n^\pm = b_n \pm \bar b_n,
	\qquad
	c_n^\pm = \frac{1}{2} \, (c_n \pm \bar c_n).
\end{equation}
The normalization of $b_m^\pm$ is chosen to match the one of $L_m^\pm$ \eqref{cft:eq:Ln-pm}, and the one of $c_m^\pm$ such that \eqref{cft:eq:1storder-com-bc-pm} holds.
Note the following useful identities:
\begin{equation}
	b_n^- b_n^+
		= 2 b_n \bar b_n,
	\qquad
	c_n^- c_n^+
		= \frac{1}{2} \, c_n \bar c_n.
\end{equation}

\begin{computation}[cft:eq:1storder-Ln]
	\begin{align*}
		T &= - \lambda \, \norder{b \pd c} + (1 - \lambda) \, \norder{\pd b c}
			\\
			&= \sum_{m,n} \left(
				\lambda \, \norder{b_m c_n} \, \frac{n + 1 - \lambda}{z^{m + \lambda} z^{m + 2 - \lambda}}
				- (1 - \lambda) \, \norder{b_m c_n} \, \frac{m + \lambda}{z^{m + \lambda + 1} z^{m + 1 - \lambda}}
				\right)
			\\
			&= \sum_{m,n} \Big(
					\lambda \, (n + 1 - \lambda)
					- (1 - \lambda) (m + \lambda)
				\Big)
				\frac{\norder{b_m c_n}}{z^{m - n + 2}}
			\\
			&= \sum_{m,n} \Big(
					\lambda \, (n + 1 - \lambda)
					- (1 - \lambda) (m - n + \lambda)
				\Big) \,
				\frac{\norder{b_{m-n} c_n}}{z^{m + 2}}
			\\
			&= \sum_{m,n} (n - m + \lambda m) \,
				\frac{\norder{b_{m-n} c_n}}{z^{m + 2}}
			= \sum_{m} \frac{L_m}{z^{m + 2}}.
	\end{align*}
	The fourth line follows from shifting $m \to m - n$.
	The second equality in \eqref{cft:eq:1storder-Ln} follows by shifting $n \to m - n$.
\end{computation}

\begin{computation}[cft:eq:1storder-jn]
	\[
		j = - \norder{b c}
			= \sum_{m,n} \frac{\norder{b_{m} c_{n}}}{z^{m + \lambda} z^{n + 1 - \lambda}}
			= \sum_{m,n} \frac{\norder{b_{m-n} c_{n}}}{z^{m + 1}}
			= \sum_{m} \frac{j_m}{z^{m + 1}}.
	\]
\end{computation}

\subsection{Commutators}

\index{first-order CFT!commutator}%
The (anti)commutators between the modes $b_n$ and $c_n$ read:
\begin{equation}
	\label{cft:eq:1storder-com-bn-cn}
	\com{b_m}{c_n}_\epsilon
		= \delta_{m+n,0},
	\qquad
	\com{b_m}{b_n}_\epsilon
		= 0,
	\qquad
	\com{c_m}{c_n}_\epsilon
		= 0.
\end{equation}
Therefore, the modes with $n < 0$ are creation operators and the modes with $n > 0$ are annihilation operators:
\begin{itemize}
	\item a $b$ ghost excitation at level $n > 0$ is created by $b_{-n}$ and annihilated by $c_n$;
	\item a $c$ ghost excitation at level $n > 0$ is created by $c_{-n}$ and annihilated by $b_n$.
\end{itemize}
In terms of $b_m^\pm$ and $c_m^\pm$ \eqref{cft:eq:bn-cn-pm}, we have:
\begin{equation}
	\label{cft:eq:1storder-com-bc-pm}
	\com{b_m^+}{c_n^+}_\epsilon
		= \delta_{m+n},
	\qquad
	\com{b_m^-}{c_n^-}_\epsilon
		= \delta_{m+n}.
\end{equation}

The commutators of the number operators with the modes are:
\begin{equation}
	\com{N^b_m}{b_{-n}}
		= b_{-n} \delta_{m,n},
	\qquad
	\com{N^c_m}{c_{-n}}
		= c_{-n} \delta_{m,n},
\end{equation}
while those between the $L_n$ and the ghost modes are:
\begin{equation}
	\label{cft:eq:com-Ln-bn-cn}
	\com{L_m}{b_n}
		= \big( m (\lambda - 1) - n \big) b_{m+n},
	\qquad
	\com{L_m}{c_n}
		= - (m \lambda + n) c_{m+n},
\end{equation}
in agreement with \eqref{cft:eq:com-Ln-phin}.
If $n \in \Z$, each ghost field has zero-modes $b_0$ and $c_0$ which commutes with $L_0$
\begin{equation}
	\label{cft:eq:com-L0-b0-c0}
	\com{L_0}{b_0}
		= 0,
	\qquad
	\com{L_0}{c_0}
		= 0.
\end{equation}

The commutator of the current modes reads
\begin{equation}
	\com{j_m}{j_n}
		= m \, \delta_{m+n,0}.
\end{equation}
Then, the commutator with the Virasoro operators are
\begin{equation}
	\com{L_m}{j_n}
		= - n j_{m+n} + \frac{q_\lambda}{2} \, m (m + 1) \delta_{m+n,0}.
\end{equation}
Finally, the commutators of the ghost number operator with the ghosts are:
\begin{equation}
	\label{cft:eq:1storder-com-bc-Ngh}
	\com{N_{\text{gh}}}{b(w)} = - b(w),
	\qquad
	\com{N_{\text{gh}}}{c(w)} = c(w).
\end{equation}

\begin{computation}[cft:eq:1storder-com-bn-cn]
	\begin{align*}
		\com{b_m}{c_n}_\epsilon
			&= \epsilon \oint_{C_0} \frac{\dd w}{2\pi \I} \, w^{-1} \oint_{C_w} \frac{\dd z}{2\pi \I} \, z^{-1} \, w^{n + \lambda} z^{m - \lambda + 1} b(z) c(w)
			\\
			&\sim \epsilon \oint_{C_0} \frac{\dd w}{2\pi \I} \, w^{-1} \oint_{C_w} \frac{\dd z}{2\pi \I} \, z^{-1} \, w^{n + \lambda} z^{m - \lambda + 1} \frac{\epsilon}{z - w}
			\\
			&= \oint_{C_0} \frac{\dd w}{2\pi \I} \, w^{m + n - 1}
			= \delta_{m+n,0}.
	\end{align*}
\end{computation}

\begin{computation}[cft:eq:1storder-com-bc-Ngh]
	\[
		\com{N_{\text{gh}}}{b(w)}
			= \oint \frac{\dd z}{2\pi \I} \, j(z) b(w)
			\sim - \oint \frac{\dd z}{2\pi \I} \, \frac{b(w)}{z - w}
			= - b(w).
	\]
	The computation for $c$ is similar.
\end{computation}

\subsection{Hilbert space}
\label{cft:sec:systems:1st:hilbert}

\index{first-order CFT!vacuum!$\mathrm{SL}(2, \C)$}%
The $\group{SL}(2, \C)$ vacuum $\ket{0}$ \eqref{cft:eq:vacuum-conformal} is defined by:
\begin{equation}
	\label{cft:eq:1storder-vac-ann}
	\forall n > - \lambda:
		\quad
		b_n \ket{0} = 0,
	\qquad
	\forall n > \lambda - 1:
		\quad
		c_n \ket{0} = 0.
\end{equation}
If $\lambda > 1$, there are positive modes which do not annihilate the vacuum.

To simplify the notation, we consider the case $\lambda \in \Z$, the half-integer case following by shifting the indices by $1/2$.
Since the modes $\{ c_1, \ldots, c_{\lambda - 1} \}$ do not annihilate $\ket{0}$, one can create states
\begin{equation}
	\ket{ n_1, \ldots, n_{\lambda-1} }
		= c_1^{n_1} \cdots c_{\lambda - 1}^{n_{\lambda-1}} \ket{0}
\end{equation}
which have negative energies:
\begin{equation}
	L_0 \ket{ n_1, \ldots, n_{\lambda-1} }
		= - \left( \sum_{j=1}^{\lambda - 1} j \, n_j \right)
			\ket{ n_1, \ldots, n_{\lambda-1} },
\end{equation}
where \eqref{cft:eq:com-Ln-bn-cn} has been used.
Moreover, this state is degenerate due to the existence of zero-modes since they commute with the Hamiltonian -- see \eqref{cft:eq:com-L0-b0-c0}.
As a consequence, it must be in a representation of the zero-mode algebra.

If the ghosts are commuting ($\epsilon = - 1$), then it seems hard to make sense of the theory since one can find a state of arbitrarily negative energy since $n_i \in \N$.
The zero-modes make the problem even worse.
The appropriate interpretation of these states will be discussed in the context of the superstring theory for $\lambda = 3/2$ (superconformal ghosts).

In the rest of this section, we focus on the Grassmann odd case $\epsilon = 1$.

\subsubsection{Energy vacuum (Grassmann odd)}

\index{first-order CFT!vacuum!energy|(}%
Since $n_i = 0$ or $n_i = 1$ for anticommuting ghosts ($\epsilon = 1$), there is a state of lowest energy.
This is the \emph{energy vacuum} \eqref{cft:eq:vacuum-energy}.
Since the zero-modes $b_0$ and $c_0$ commute with $L_0$, it is doubly degenerate.
A convenient basis is
\begin{equation}
	\big\{ \ket{\downarrow}, \ket{\uparrow} \big\},
\end{equation}
where
\begin{equation}
	\label{cft:eq:ghost-vacuum-ud}
	\ket{\downarrow}
		:= c_1 \cdots c_{\lambda - 1} \ket{0},
	\qquad
	\ket{\uparrow}
		:= c_0 c_1 \cdots c_{\lambda - 1} \ket{0}.
\end{equation}
A general vacuum is a linear combination of the two basis vacua:
\begin{equation}
	\ket{\Omega}
		= \omega_{\downarrow} \ket{\downarrow}
			+ \omega_{\uparrow} \ket{\uparrow},
	\qquad
	\omega_{\downarrow}, \omega_{\uparrow} \in \C.
\end{equation}

The algebra of these vacua is the one of a two-state system:
\begin{equation}
	\label{cft:eq:1storder-evac-algebra-hol}
	b_0 \ket{\uparrow} = \ket{\downarrow},
	\qquad
	c_0 \ket{\downarrow} = \ket{\uparrow},
	\qquad
	b_0 \ket{\downarrow} = 0,
	\qquad
	c_0 \ket{\uparrow} = 0.
\end{equation}
Hence, for the vacuum $\ket{\downarrow}$ (resp.\ $\ket{\uparrow}$), $b_0$ (resp.\ $c_0$) acts as an annihilation operator, and conversely $c_0$ (resp.\ $b_0$) acts as a creation operator.
Finally, both states are annihilated by all positive modes:
\begin{equation}
	\label{cft:eq:1storder-evac-ann}
	\forall n > 0:
		\qquad
		b_n \ket{\downarrow}
			= b_n \ket{\uparrow}
			= 0,
		\qquad
		c_n \ket{\downarrow}
			= b_n \ket{\downarrow}
			= 0.
\end{equation}
Note that the $\group{SL}(2, \C)$ vacuum can be recovered by acting with $b_{-n}$ with $n < \lambda$:
\begin{equation}
	\ket{0}
		= b_{1-\lambda} \cdots b_{-1} \ket{\downarrow}
		= b_{1-\lambda} \cdots b_{-1} b_0 \ket{\uparrow}.
\end{equation}

\index{first-order CFT!zero-point energy}%
The zero-point energy \eqref{cft:eq:0pt-energy} of these states is the conformal weight of the vacuum:
\begin{equation}
	L_0 \ket{\downarrow}
		= a_\lambda \ket{\downarrow},
	\qquad
	L_0 \ket{\uparrow}
		= a_\lambda \ket{\uparrow},
\end{equation}
where $a_\lambda$ can be written in various forms:
\begin{equation}
	\label{cft:eq:1storder-zero-energy}
	a_\lambda
		= - \sum_{n=1}^{\lambda-1} n
		= - \frac{\lambda (\lambda - 1)}{2}
		= \frac{c_\lambda}{24} + \frac{2}{24}.
\end{equation}

\medskip

Taking into account the anti-holomorphic sector leads to a four-fold degeneracy.
The basis
\begin{equation}
	\big\{ \ket{\downarrow\downarrow}, \ket{\uparrow\downarrow}, \ket{\downarrow\uparrow}, \ket{\uparrow\uparrow} \big\},
\end{equation}
is built as follows:
\begin{equation}
	\label{cft:eq:1storder-vacuum-LR}
	\begin{gathered}
	\ket{\downarrow\downarrow}
		:= c_1 \bar c_1 \cdots c_{\lambda-1} \bar c_{\lambda-1} \ket{0},
	\\
	\ket{\uparrow\downarrow}
		:= c_0 \ket{\downarrow\downarrow},
	\qquad
	\ket{\downarrow\uparrow}
		:= \bar c_0 \ket{\downarrow\downarrow},
	\qquad
	\ket{\uparrow\uparrow}
		:= c_0 \bar c_0 \ket{\downarrow\downarrow}.
	\end{gathered}
\end{equation}
The modes $b_0$ and $\bar b_0$ can be used to flip the arrows downward, leading to the following algebra:
\begin{subequations}
\label{cft:eq:1storder-evac-algebra}
\begin{equation}
	\begin{gathered}
	c_0 \ket{\downarrow\downarrow}
		= \ket{\uparrow\downarrow},
	\qquad
	\bar c_0 \ket{\downarrow\downarrow}
		= \ket{\downarrow\uparrow},
	\qquad
	c_0 \ket{\downarrow\uparrow}
		= - \bar c_0 \ket{\uparrow\downarrow}
		= \ket{\uparrow\uparrow},
	\\
	b_0 \ket{\uparrow\uparrow}
		= \ket{\downarrow\uparrow},
	\qquad
	\bar b_0 \ket{\uparrow\uparrow}
		= - \ket{\uparrow\downarrow},
	\qquad
	b_0 \ket{\uparrow\downarrow}
		= \bar b_0 \ket{\downarrow\uparrow}
		= \ket{\downarrow\downarrow},
	\end{gathered}
\end{equation}
The vacua are annihilated by different combinations of the zero-modes:
\begin{equation}
	\begin{gathered}
	b_0 \ket{\downarrow\downarrow}
		= \bar b_0 \ket{\downarrow\downarrow}
		= 0,
	\qquad
	c_0 \ket{\uparrow\downarrow}
		= \bar b_0 \ket{\uparrow\downarrow}
		= 0,
	\\
	b_0 \ket{\downarrow\uparrow}
		= \bar c_0 \ket{\downarrow\uparrow}
		= 0,
	\qquad
	c_0 \ket{\uparrow\uparrow}
		= \bar c_0 \ket{\uparrow\uparrow}
		= 0.
	\end{gathered}
\end{equation}
\end{subequations}
In these manipulations, one has to be careful to correctly anti-commute the modes with the ones hidden in the definitions of the vacua.

There is a second basis which is more natural when using the zero-modes $c_0^\pm$ and $b_0^\pm$ \eqref{cft:eq:bn-cn-pm}:
\begin{equation}
	\label{cft:eq:1storder-evac-basis-pm}
	\big\{ \ket{\downarrow\downarrow}, \ket{+}, \ket{-}, \ket{\uparrow\uparrow} \big\},
\end{equation}
where the two vacua $\ket{\pm}$ are combinations of the $\ket{\downarrow\uparrow}$ and $\ket{\uparrow\downarrow}$ vacua:
\begin{equation}
	\ket{\pm}
		= \ket{\uparrow\downarrow} \pm \ket{\downarrow\uparrow}.
\end{equation}
The different vacua are naturally related by acting with $c_0^\pm$ and $b_0^\pm$ which act as raising and lowering operators:
\begin{subequations}
\label{cft:eq:1storder-evac-algebra-pm}
\begin{equation}
	\begin{gathered}
	c_0^\pm \ket{\downarrow\downarrow}
		= \frac{1}{2} \, \ket{\pm},
	\qquad
	c_0^\mp \ket{\pm}
		= \pm \ket{\uparrow\uparrow},
	\\
	b_0^\pm \ket{\pm}
		= \pm 2 \ket{\downarrow\downarrow},
	\qquad
	b_0^\mp \ket{\uparrow\uparrow}
		= \pm \ket{\pm}.
	\end{gathered}
\end{equation}
From the previous relations, it follows that the different vacua are annihilated by the zero-modes as follow:
\begin{equation}
	\begin{gathered}
	b_0^+ \ket{\downarrow\downarrow}
		= b_0^- \ket{\downarrow\downarrow}
		= 0,
	\qquad
	c_0^- \ket{-}
		= b_0^+ \ket{-}
		= 0,
	\\
	c_0^+ \ket{+}
		= b_0^- \ket{+}
		= 0
	\qquad
	c_0^+ \ket{\uparrow\uparrow}
		= c_0^- \ket{\uparrow\uparrow}
		= 0,
	\end{gathered}
\end{equation}
This also means that we have
\begin{equation}
	c_0^- c_0^+ \ket{\downarrow\downarrow}
		= \frac{1}{2} \, \ket{\uparrow\uparrow},
	\qquad
	b_0^+ b_0^- \ket{\uparrow\uparrow}
		= 2 \ket{\downarrow\downarrow}.
\end{equation}
\end{subequations}

\begin{computation}[cft:eq:1storder-evac-algebra-pm]
	\begin{gather*}
		2 \, c_0^+ \ket{\pm}
			= (c_0 + \bar c_0) \ket{\uparrow\downarrow} \pm (c_0 + \bar c_0) \ket{\downarrow\uparrow}
			= \bar c_0 \ket{\uparrow\downarrow} \pm c_0 \ket{\downarrow\uparrow}
			= (- 1 \pm 1) \ket{\uparrow\uparrow}
		\\
		b_0^+ \ket{\pm}
			= (b_0 + \bar b_0) \ket{\uparrow\downarrow} \pm (b_0 + \bar b_0) \ket{\downarrow\uparrow}
			= b_0 \ket{\uparrow\downarrow} \pm \bar b_0 \ket{\downarrow\uparrow}
			= (1 \pm 1) \ket{\downarrow\downarrow}
		\\
		2 \, c_0^\pm \ket{\downarrow\downarrow}
			= (c_0 \pm \bar c_0) \ket{\downarrow\downarrow}
			= c_0 \ket{\downarrow\downarrow} \pm \bar c_0 \ket{\downarrow\downarrow}
			= \ket{\uparrow\downarrow} \pm \ket{\downarrow\uparrow}
			= \ket{\pm}
		\\
		b_0^\pm \ket{\uparrow\uparrow}
			= (b_0 \pm \bar b_0) \ket{\uparrow\uparrow}
			= b_0 \ket{\uparrow\uparrow} \pm \bar b_0 \ket{\uparrow\uparrow}
			= \ket{\downarrow\uparrow} \mp \ket{\uparrow\downarrow}
			= \mp \ket{\mp}
	\end{gather*}
\end{computation}

\index{first-order CFT!vacuum!energy|)}%

\subsubsection{Energy normal ordering (Grassmann odd)}

\index{first-order CFT!energy normal ordering}%
We now turn towards the definition of the energy normal ordering \eqref{cft:eq:energy-normal-order}.
Ultimately, it will be found that $\ket{\downarrow}$ is the physical vacuum in string theory.
For this reason, the energy normal ordering $\norderv{\cdots}$ is associated to the vacuum $\ket{\downarrow}$ in order to resolve the ambiguity of the zero-modes.
In particular, $b_0$ is an annihilation operator in this case, while $c_0$ is a creation operator.
In the rest of this section, we translate the normal ordering of expressions from the conformal vacuum to the energy vacuum.

The Virasoro operators $L_n$ for $n \neq 0$ have no ordering problems since the modes which compose them commute.
\index{first-order CFT!$L_0$}%
The expression of $L_0$ \eqref{cft:eq:1storder-L0} in the energy ordering becomes
\begin{equation}
	\label{cft:eq:1storder-L0-ev}
	L_0
		= \sum_{n} n \, \norderv{b_{-n} c_n} + a_\lambda
		= N^b + N^c + a_\lambda
\end{equation}
where $a_\lambda$ is the zero-point energy \eqref{cft:eq:1storder-zero-energy} and $N^b$ and $N^c$ are the ghost mode numbers \eqref{cft:eq:1storder-N}.
The contribution of the non-zero modes is denoted by:
\begin{equation}
	\label{cft:eq:1storder-L0-hat}
	\what L_0
		= N^b + N^c.
\end{equation}
\index{first-order CFT!Virasoro operators}%
The expression can be rewritten to encompass all modes:
\begin{equation}
	\label{cft:eq:1storder-Lm-ev}
	L_m
		= \sum_{n} \big(n - (1 - \lambda) m \big) \, \norderv{b_{m-n} c_n}
			+ a_\lambda \, \delta_{m,0}
\end{equation}
\index{first-order CFT!ghost number}%
Similarly, the expression of the ghost number is
\begin{subequations}
\label{cft:eq:1storder-jn0-ev}
\begin{align}
	N_{\text{gh},L}
		&
		= j_0
		= \sum_{n} \norderv{b_{-n} c_n}
			- \left( \frac{q_\lambda}{2} + \frac{1}{2} \right)
		\\ &
		= \sum_{n > 0} \big(N^c_n - N^b_n \big)
			+ \frac{1}{2} \, \big(N^c_0 - N^b_0 \big)
			- \frac{q_\lambda}{2},
\end{align}
\end{subequations}
and thus:
\index{first-order CFT!$\mathrm{U}(1)$ ghost current}%
\begin{equation}
	\label{cft:eq:1storder-jnm-ev}
	j_m
		= \sum_{n} \, \norderv{b_{m-n} c_n}
			- \left( \frac{q_\lambda}{2} + \frac{1}{2} \right) \, \delta_{m,0}.
\end{equation}
It is useful to define the ghost number without ghost zero-modes:
\begin{equation}
	\label{cft:eq:1storder-modes-Ngh-hat}
	\what N_{\text{gh}, L}
		:= \sum_{n > 0} \big(N^c_n - N^b_n \big).
\end{equation}

One can straightforwardly compute the ghost number of the vacua:
\begin{subequations}
\begin{gather}
	j_0 \ket{\downarrow}
		= (\lambda - 1) \ket{\downarrow}
		= \left( - \frac{q_\lambda}{2} - \frac{1}{2} \right) \ket{\downarrow},
	\\
	j_0 \ket{\uparrow}
		= \lambda \ket{\uparrow}
		= \left(- \frac{q_\lambda}{2} + \frac{1}{2} \right) \ket{\uparrow}.
\end{gather}

\end{subequations}
This confirms that the $\group{SL}(2, \C)$ vacuum has vanishing ghost number since $\ket{\downarrow}$ contains exactly $\lambda - 1$ ghosts:
\begin{equation}
	j_0 \ket{0}
		= 0.
\end{equation}
Using \eqref{cft:eq:1storder-number-cyl} allows to write the ghost numbers on the cylinder:
\begin{equation}
	j^{\text{cyl}}_0 \ket{\downarrow} = - \frac{1}{2} \ket{\downarrow},
	\qquad
	j^{\text{cyl}}_0 \ket{\uparrow} = \frac{1}{2} \ket{\uparrow}.
\end{equation}
That both ghost numbers have same magnitude but opposite signs could be expected: since the ghost number changes as $N_{\text{gh}} \to - N_{\text{gh}}$ when $b \leftrightarrow c$, the mean value of the ghost number should be zero.

\begin{remark}[Ghost number conventions]
	\index{conventions!ghost number}%

	Since the ghost number is an additive quan\-tum number, it is always possible to shift its definition by a constant.
	This can be used to set the ghost numbers of the vacua to some other values.
	For example,~\cite[p.~116]{Blumenhagen:2014:BasicConceptsString} adds $q_\lambda / 2$ to the ghost number in order to get $N_{\text{gh}} = \pm 1/2$ on the plane (instead of the cylinder).
	We do not follow this convention in order to keep the symmetry between the vacuum ghost numbers on the cylinder.
\end{remark}

\begin{computation}[cft:eq:1storder-L0-ev]
	Start with \eqref{cft:eq:1storder-L0} and use \eqref{cft:eq:modes-normal-order}:
	\begin{align*}
		L_0 &= - \sum_{n} n \, \norder{b_{n} c_{-n}}
			= - \sum_{n \le - \lambda} n \, b_{n} c_{-n}
				+ \epsilon \sum_{n > - \lambda} n \, c_{-n} b_{n}
			\\
			&= \sum_{n \ge \lambda} n \, b_{-n} c_{n}
				+ \epsilon \sum_{n > - \lambda} n \, c_{-n} b_{n}
			\\
			&= \sum_{n \ge \lambda} n \, b_{-n} c_{n}
				+ \epsilon \sum_{n > 0} n \, c_{-n} b_{n}
				+ \epsilon \sum_{n = - \lambda + 1}^{0} n \, c_{-n} b_{n}
			\\
			&= \sum_{n \ge \lambda} n \, b_{-n} c_{n}
				+ \epsilon \sum_{n > 0} n \, c_{-n} b_{n}
				+ \epsilon \sum_{n = 0}^{\lambda - 1} n \, b_{-n} c_{n}
				+ a_\lambda
			\\
			&= \sum_{n > 0} n \, b_{-n} c_{n}
				+ \epsilon \sum_{n > 0} n \, c_{-n} b_{n}
				+ a_\lambda,
			\\
			&= \sum_{n} n \, \norderv{b_{-n} c_n}
				+ a_\lambda,
	\end{align*}
	using that
	\[
		\sum_{n = - \lambda + 1}^{0} \, c_{-n} b_{n}
			= - \sum_{n = 0}^{\lambda - 1} n \, c_{n} b_{-n}
			= - \sum_{n = 0}^{\lambda - 1} n \, (- \epsilon \, b_{-n} c_{n} + 1)
			= \epsilon \sum_{n = 0}^{\lambda - 1} n \, b_{-n} c_{n} + a_\lambda.
	\]
	The result also follows from \eqref{cft:eq:modes-relations-cno-eno}.
\end{computation}

\begin{computation}[cft:eq:1storder-jn0-ev]
	\begin{align*}
		j_0 &= - \sum_{n} \, \norder{b_{-n} c_{n}}
			= - \sum_{n \ge \lambda} b_{-n} c_{n}
				+ \epsilon \sum_{n > - \lambda} \, c_{-n} b_{n}
			\\
			&= - \sum_{n \ge \lambda} b_{-n} c_{n}
				+ \epsilon \sum_{n > 0} \, c_{-n} b_{n}
				+ \epsilon \sum_{n=1}^{\lambda-1} \, c_{n} b_{-n}
				+ \epsilon \, c_{0} b_{0}
			\\
			&= - \sum_{n \ge \lambda} b_{-n} c_{n}
				+ \epsilon \sum_{n > 0} \, c_{-n} b_{n}
				- \sum_{n=1}^{\lambda-1} \, b_{-n} c_{n}
				+ \epsilon (\lambda - 1)
				+ \epsilon \, c_{0} b_{0}
			\\
			&= - \sum_{n > 0} b_{-n} c_{n}
				+ \epsilon \sum_{n > 0} \, c_{-n} b_{n}
				+ \epsilon (\lambda - 1)
				+ \epsilon \, c_{0} b_{0}.
	\end{align*}
	Finally, one can write
	\begin{equation}
		\epsilon (\lambda - 1) = - \frac{q_\lambda}{2} - \frac{\epsilon}{2}.
	\end{equation}
	The result also follows from \eqref{cft:eq:modes-relations-cno-eno}.
	The second expression is obtained by symmetrizing the last term such that
	\begin{align*}
		\epsilon \, c_{0} b_{0} + \epsilon (\lambda - 1)
			&= \frac{\epsilon}{2} \, c_{0} b_{0}
				+ \frac{1}{2} (- b_{0} c_{0} + \epsilon)
				+ \epsilon (\lambda - 1)
			\\
			&= \frac{1}{2} \, (\epsilon \, c_{0} b_{0} - b_{0} c_{0})
				+ \epsilon \left(\lambda - \frac{1}{2} \right).
	\end{align*}
\end{computation}

\subsubsection{Structure of the Hilbert space (Grassmann odd)}

\index{first-order CFT!zero-mode decomposition}%
Since the zero-modes commute with the Hamiltonian and with all other negative- and positive-frequency modes, the Hilbert space is decomposed in several subspaces, each associated to a zero-mode.\footnotemark{}
\footnotetext{%
	Due to the specific structure of the inner product defined below, these subspaces are not orthonormal to each other.
}%

\index{first-order CFT!Fock space}%

Starting with the holomorphic sector only, the Hilbert space $\mc H_{\text{gh}}$ is:
\begin{equation}
	\label{cft:eq:1storder-Hilbert}
	\mc H_{\text{gh}}
		= \mc H_{\text{gh},0}
			\oplus c_0 \mc H_{\text{gh},0},
	\qquad
	\mc H_{\text{gh},0}
		:= \mc H_{\text{gh}} \cap \ker b_0,
\end{equation}
which follows from the $2$-state algebra \eqref{cft:eq:1storder-evac-algebra-hol}.
Obviously, one has $c_0 \mc H_{\text{gh},0}
= \mc H_{\text{gh}} \cap \ker c_0$.
\index{first-order CFT!Hilbert space!holomorphic}%
The oscillator basis of the Hilbert space $\mc H_{\text{gh},0}$ is generated by applying the negative-frequency modes and has the structure of a fermionic Fock space without zero-modes:
\begin{subequations}
\begin{gather}
	\mc H_{\text{gh},0}
		= \Span \Big\{ \Ket{\downarrow ; \{ N^b_n \} ; \{ N^c_n \}} \Big\},
	\\
	\Ket{\downarrow ; \{ N^b_n \} ; \{ N^c_n \}}
		= \prod_{n \ge 1} (b_{-n})^{N^b_n} (c_{-n})^{N^c_n} \ket{\downarrow},
	\qquad
	N^b_n, N^c_n \in \N^*
\end{gather}
\end{subequations}
(again, number operators and their eigenvalues are not distinguished).
This means that $\mc H_{\text{gh},0}$ can also be regarded as a Fock space built on the vacuum $\ket{\downarrow}$, for which $c_0$ and $b_0$ are respectively creation and annihilation operators.
Conversely, $c_0$ and $b_0$ are respectively annihilation and creation operators for $c_0 \mc H_{\text{gh},0}$.

In particular, this means that any state can be written as the sum of two states
\begin{equation}
	\psi
		= \psi_\downarrow + \psi_\uparrow,
	\qquad
	\psi_\downarrow \in \mc H_{\text{gh},0},
	\qquad
	\psi_\uparrow \in c_0 \mc H_{\text{gh},0},
\end{equation}
with $\psi_{\downarrow}$ and $\psi_{\uparrow}$ built respectively on top of the $\ket{\downarrow}$ and $\ket{\uparrow}$ vacua.

\bigskip

\index{first-order CFT!Hilbert space!full}%
This pattern generalizes when considering both the holomorphic and anti-holomorphic sectors.
In that case, the Hilbert space is decomposed in four subspaces:\footnotemark{}
\footnotetext{%
	The reader should not get confused by the same symbol $\mc H_{\text{gh},0}$ as in the case of the holomorphic sector.
}%
\begin{equation}
	\begin{gathered}
	\mc H_{\text{gh}}
		= \mc H_{\text{gh},0}
			\oplus c_0 \mc H_{\text{gh},0}
			\oplus \bar c_0 \mc H_{\text{gh},0}
			\oplus c_0 \bar c_0 \mc H_{\text{gh},0},
	\\
	\mc H_{\text{gh},0}
		:= \mc H_{\text{gh}} \cap \ker b_0 \cap \ker \bar b_0.
	\end{gathered}
\end{equation}
Basis states of the Hilbert space $\mc H_{\text{gh},0}$ are:
\begin{equation}
	\begin{gathered}
	\Ket{\downarrow\downarrow ; \{ N^b_n \} ; \{ N^c_n \} ; \{ \bar N^b_n \} ; \{ \bar N^c_n \}}
		= \prod_{n \ge 1} (b_{-n})^{N^b_n} (\bar b_{-n})^{\bar N^b_n}
			(c_{-n})^{N^c_n} (\bar c_{-n})^{\bar N^c_n}
			\ket{\downarrow\downarrow},
	\\
	N^b_n, \bar N^b_n, N^c_n, \bar N^c_n \in \N^*.
	\end{gathered}
\end{equation}
A general state of $\mc H_{\text{gh}}$ can be decomposed as
\begin{equation}
	\psi
		= \psi_{\downarrow\downarrow} + \psi_{\uparrow\downarrow}
			+ \psi_{\downarrow\uparrow} + \psi_{\uparrow\uparrow},
\end{equation}
where each state is built by acting with negative-frequency modes on the corresponding vacuum.

In terms of the second basis \eqref{cft:eq:1storder-evac-basis-pm}, the Hilbert space admits a second decomposition:
\begin{equation}
	\label{cft:eq:1storder-Hilbert-LR}
	\begin{gathered}
	\mc H_{\text{gh}}
		= \mc H_{\text{gh},0}
			\oplus c_0^+ \mc H_{\text{gh},0}
			\oplus c_0^- \mc H_{\text{gh},0}
			\oplus c_0^- c_0^+ \mc H_{\text{gh},0},
	\\
	\mc H_{\text{gh},0}
		:= \mc H_{\text{gh}} \cap \ker b_0^- \cap \ker b_0^+.
	\end{gathered}
\end{equation}
In view of applications to string theory, it is useful to introduce two more subspaces:
\begin{equation}
	\label{cft:eq:1storder-Hpm}
	\mc H_{\text{gh},\pm}
		:= \mc H_{\text{gh}} \cap \ker b_0^\pm
		= \mc H_{\text{gh},0} \oplus c_0^\mp \mc H_{\text{gh},0},
\end{equation}
and the associated decomposition
\begin{equation}
	\mc H_{\text{gh}}
		= \mc H_{\text{gh},\pm}
			\oplus c_0^\pm \mc H_{\text{gh},\pm}.
\end{equation}
In off-shell closed string theory, the principal Hilbert space will be $\mc H_{\text{gh}}^-$ due to the level-matching condition.
In this case, $\mc H_{\text{gh}}^-$ has the same structure as $\mc H_{\text{gh}}$ in the pure holomorphic sector, and $c_0^+$ plays the same role as $c_0$.
A state in $\mc H_{\text{gh}}^-$ is built on top of the vacua $\ket{\downarrow\downarrow}$ and $\ket{+}$.

\subsection{Euclidean and BPZ conjugates}

\index{first-order CFT!Euclidean adjoint!modes}%
In order for the Virasoro operators to be Hermitian, the $b_n$ and $c_n$ must satisfy the following conditions:
\begin{equation}
	\label{cft:eq:1storder-adj-bn-cn}
	\adj{b_n} = \epsilon b_{-n},
	\qquad
	\adj{c_n} = c_{-n}.
\end{equation}
Hence, $b_n$ is anti-Hermitian if $\epsilon = - 1$.
\index{first-order CFT!BPZ conjugate!modes}%
The BPZ conjugates of the modes are:
\begin{equation}
	b_n^t
		= (- 1)^{\lambda} \, b_{-n},
	\qquad
	c_n^t
		= (- 1)^{1 - \lambda} \, c_{-n},
\end{equation}
using $I^+(z)$ with \eqref{cft:eq:modes-bpz}.

In the rest of this section, we consider only the case $\epsilon = 1$ and $\lambda \in \N$.
\index{first-order CFT!Euclidean adjoint!vacuum}%
The adjoints of the vacuum read:
\begin{equation}
	\eadj{\ket{\downarrow}}
		= \bra{0} c_{1 - \lambda} \cdots c_{-1},
	\qquad
	\eadj{\ket{\uparrow}}
		= \bra{0} c_{1 - \lambda} \cdots c_{-1} c_0.
\end{equation}
\index{first-order CFT!BPZ conjugate!vacuum}%
The BPZ conjugates of the vacua are:
\begin{equation}
	\begin{aligned}
	\bra{\downarrow}
		&
		:= \ket{\downarrow}^t
		= (- 1)^{(1 - \lambda)^2} \bra{0} c_{-1} \cdots c_{1 - \lambda},
	\\
	\bra{\uparrow}
		&
		:= \ket{\uparrow}^t
		= (- 1)^{\lambda (1 - \lambda)} \bra{0} c_0 c_{-1} \cdots c_{1 - \lambda}.
	\end{aligned}
\end{equation}
The signs are inconvenient but will disappear when considering both the left and right vacua together as in \eqref{cft:eq:1storder-vacuum-LR}.
We have the following relations:
\begin{equation}
	\label{cft:eq:1storder-vac-conj-relations}
	\bra{\downarrow}
		= (- 1)^{a_\lambda + (1 - \lambda) (2 - \lambda)}
			\eadj{\ket{\downarrow}},
	\qquad
	\bra{\uparrow}
		= (- 1)^{a_\lambda}
			\eadj{\ket{\uparrow}},
\end{equation}
where $a_\lambda$ is the zero-point energy \eqref{cft:eq:1storder-zero-energy}.

\begin{computation}[cft:eq:1storder-vac-conj-relations]
	To prove the relation, we can start from the BPZ conjugate $\bra{\downarrow}$ and reorder the modes to bring them in the same order as the adjoint:
	\[
		\bra{\downarrow}
			= (- 1)^{(1 - \lambda)^2 + \frac{1}{2} (2 - \lambda) (1 - \lambda)}
				\eadj{\ket{\downarrow}}
			= (- 1)^{- a_\lambda + (1 - \lambda) (2 - \lambda)}
				\eadj{\ket{\downarrow}}
	\]
	The reordering gives a factor $(-1)$ to the power:
	\[
		\sum_{i=1}^{\lambda - 2} i
			= \frac{1}{2} (2 - \lambda) (1 - \lambda)
			= - a_\lambda + 1 - \lambda.
	\]
	Similarly, for the second vacuum:
	\[
		\bra{\uparrow}
			= (- 1)^{\lambda (1 - \lambda) - \frac{1}{2} \lambda (1 - \lambda)}
				\eadj{\ket{\uparrow}}
			= (- 1)^{\frac{1}{2} \lambda (1 - \lambda)}
				\eadj{\ket{\uparrow}}.
	\]
	We can identify the power with \eqref{cft:eq:1storder-zero-energy}.
\end{computation}

Then, we have the following relations:
\begin{equation}
	\bra{\uparrow} b_0
		= \bra{\downarrow},
	\qquad
	\bra{\downarrow} c_0
		= \bra{\uparrow},
	\qquad
	\bra{\downarrow} b_0
		= 0,
	\qquad
	\bra{\uparrow} c_0
		= 0.
\end{equation}

\index{first-order CFT!inner product}%
There is a subtlety in defining the inner product because the vacuum is degenerate.
If we write the two vacua as vectors
\begin{equation}
	\ket{\downarrow}
		=
		\begin{pmatrix}
			0 \\ 1
		\end{pmatrix}
		,
	\qquad
	\ket{\uparrow}
		=
		\begin{pmatrix}
			1 \\ 0
		\end{pmatrix}
		,
\end{equation}
then the zero-modes have the following matrix representation:
\begin{equation}
	b_0
		=
		\begin{pmatrix}
			0 & 0
			\\
			1 & 0
		\end{pmatrix}
		,
	\qquad
	c_0
		=
		\begin{pmatrix}
			0 & 1
			\\
			0 & 0
		\end{pmatrix}
		.
\end{equation}
These matrices are not Hermitian as required by \eqref{cft:eq:1storder-adj-bn-cn}: since Hermiticity follows from the choice of an inner product, it means that the vacua cannot form an orthonormal basis.
An appropriate choice for the inner products is:\footnotemark{}
\footnotetext{%
	To avoid confusions, let us note that the adjoint in \eqref{cft:eq:1storder-vac-conj-relations} are defined only through the adjoint of the modes \eqref{cft:eq:modes-adj} but not with respect to the inner product given here, which would lead to exchanging $\eadj{\ket{\downarrow}} \sim \bra{\uparrow}$ and $\eadj{\ket{\uparrow}} \sim \bra{\downarrow}$.
}%
\begin{equation}
	\label{cft:eq:1storder-vacuum-product}
	\begin{gathered}
	\bracket{\downarrow}{\downarrow}
		= \bracket{\uparrow}{\uparrow}
		= 0,
	\\
	\bracket{\uparrow}{\downarrow}
		= \bra{\downarrow} c_0 \ket{\downarrow}
		= \bra{0} c_{1 - \lambda} \cdots c_{-1} c_0 c_1 \cdots c_{\lambda-1} \ket{0}
		= 1.
	\end{gathered}
\end{equation}
The effect of changing the definition of the inner product or to consider a non-orthonormal basis is represented by the insertion of $c_0$.
\index{first-order CFT!vacuum!conjugate}%
The last condition implies that the conjugate state \eqref{cft:eq:bpz-conj-state} to the $\group{SL}(2, \C)$ vacuum is:
\begin{equation}
	\bra{0^c}
		= \bra{0} c_{1 - \lambda} \cdots c_{-1} c_0 c_1 \cdots c_{\lambda-1},
	\qquad
	\bra{\downarrow^c}
		= \bra{\uparrow}.
\end{equation}

\begin{draft}

\subsection{Bosonization}

\end{draft}

\subsection{Summary}

In this section we summarize the values of the parameters for different theories of interest (\Cref{cft:tab:ghost-summary}).
The $(\eta, \xi)$ system will be introduced in \Cref{part:superstring} in the bosonization of the super-reparametrization $(\beta, \gamma)$ ghosts.
The $\psi^\pm$ system can be used to describe spin-$1/2$ fermions.

\index{first-order CFT!summary}%

\begin{table}[ht]
	\centering
	\begin{tabular}{c|ccccc}
		&
			$\epsilon$ &
			$\lambda$ &
			$q_\lambda$ &
			$c_\lambda$ &
			$a_\lambda$
		\\
		\hline
		$b, c$ (diff.) &
			$1$ &
			$2$ &
			$- 3$ &
			$- 26$ &
			$- 1$
		\\
		$\beta, \gamma$ (susy.) &
			$-1$ &
			$3/2$ &
			$2$ &
			$11$ &
			$3/8$
		\\
		$\psi^\pm$ &
			$1$ &
			$1/2$ &
			$0$ &
			$1$ &
			$0$
		\\
		$\eta, \xi$ &
			$1$ &
			$1$ &
			$- 1$ &
			$- 2$ &
			$0$
	\end{tabular}
	\caption{%
		Summary of the first-order systems.
		Remember that $h(b) = \lambda$ and $h(c) = 1 - \lambda$.
	}
	\label{cft:tab:ghost-summary}
\end{table}

\refchapter

\begin{itemize}
	\item Free scalar: general references~\cites[sec.~4.1.3, 4.3, 4.6.2]{Tong:2009:LecturesStringTheory}[sec.~5.3.1, 6.3]{DiFrancesco:1999:ConformalFieldTheory}[sec.~4.2]{Blumenhagen:2014:BasicConceptsString}{Polchinski:2005:StringTheory-1}{Kiritsis:2007:StringTheoryNutshell}, topological current and winding~\cites{Hull:2007:DoubledGeometryTFolds}[sec.~17.2--3]{Zwiebach:2009:FirstCourseString}.

	\item First-order system: general references~\cites[chap.~5, sec.~13.1]{Blumenhagen:2014:BasicConceptsString}[sec.~4.15]{Kiritsis:2007:StringTheoryNutshell}[sec.~2.5]{Polchinski:2005:StringTheory-1}, ghost vacua~\cite[sec.~15.3]{Lawrie:2012:UnifiedGrandTour}.
\end{itemize}

\chapter{BRST quantization}
\label{cft:chap:brst}

\introchapter

The BRST quantization can be introduced either by following the standard QFT treatment (outlined in \Cref{bos:sec:ws-int:brst}), or by translating it in the CFT language.
One can then use all the CFT techniques to extract information on the spectrum, which makes this approach more powerful.
Moreover, this also provides an elegant description of states and string fields.
In this chapter, we set the stage of the BRST quantization using the CFT language and we apply it to string theory.
The main results of this chapter are a proof of the no-ghost theorem and a characterization of the BRST cohomology (physical states).

\section{BRST for reparametrization invariance}

The BRST symmetry we are interested in results from gauge fixing the reparametrization invariance.
In this chapter, we focus on the holomorphic sector: since both sectors are independent, most results follow directly, except those concerning the zero-modes.
We consider a generic matter CFT coupled to reparametrization ghosts:
\begin{enumerate}
	\item matter:
	central charge $c_m$, energy--momentum tensor $T_m$ and Hilbert space $\mc H_m$;

	\item reparametrization ghosts:
	$bc$ ghost system (\Cref{bos:sec:ws-int:faddeev-popov,cft:sec:systems:ghosts}) with $\epsilon = +1$ and $\lambda = 2$, $c_{\text{gh}} = - 26$, energy--momentum tensor $T^{\text{gh}}$ and Hilbert space $\mc H_{\text{gh}}$.
\end{enumerate}
The formulas for the reparametrization ghosts are summarized in \Cref{app:chap:formulas:cft:bc}.
For modes, the system ($m$, gh, $b$ or $c$) is indicated as a superscript to not confuse it with the mode index.
The total central charge, energy--momentum tensor and Hilbert space are denoted by:
\begin{equation}
	c
		= c_m + c_{\text{gh}}
		= c_m - 26,
	\qquad
	T(z)
		= T^{m}(z) + T^{\text{gh}}(z),
	\qquad
	\mc H
		= \mc H_m \otimes \mc H_{\text{gh}}.
\end{equation}

\index{BRST cohomology}%
The goal is to find the physical states in the cohomology, that is, which are BRST closed
\begin{equation}
	Q_B \ket{\psi}
		= 0
\end{equation}
but non exact (\Cref{bos:sec:ws-int:brst}): the latter statement can be understood as an equivalence between closed states under shift by exact states:
\begin{equation}
	\ket{\psi}
		\sim \ket{\psi} + Q_B \ket{\Lambda}.
\end{equation}

We introduce the BRST current and study its CFT properties.
Then, we give a computation of the BRST cohomology when the matter CFT contains at least two scalar fields.

\section{BRST in the CFT formalism}
\label{cft:sec:brst:cft}

\index{BRST current}%
The BRST current can be found from \eqref{bos:eq:sym-brst-B} to be~\cite{Polchinski:2005:StringTheory-1}:
\begin{subequations}
\begin{align}
	j_B(z)
		&= \norder{c(z) \left( T^m(z) + \frac{1}{2} \, T^{\text{gh}}(z) \right) } + \kappa \, \pd^2 c(z)
		\\
		&= c(z) T^m(z) + \norder{b(z) c(z) \pd c(z)} + \kappa \, \pd^2 c(z),
\end{align}
\end{subequations}
and similarly for the anti-holomorphic sector.
This can be derived from \eqref{bos:eq:transf-inf-brst-psi-c-b}: the generator of infinitesimal changes of coordinates (given by the Lie derivative) is the energy--momentum tensor.
The factor of $1/2$ comes from the expression \eqref{cft:eq:1storder-T} of the ghost energy--momentum tensor: the second term does not contribute while the first has a factor of $2$.
Since the transformation of $c$ in \eqref{bos:eq:transf-inf-brst-psi-c-b} has no factor, the $1/2$ is necessary to recover the correct normalization.
Finally, one finds that the transformation of $b$ is reproduced.
The different computations can be checked using the OPEs given below.
The last piece is a total derivative and does not contribute to the charge: for this reason, it cannot be derived from \eqref{bos:eq:transf-inf-brst-psi-c-b}, its coefficient will be determined below.
Note that it is the only total derivative of dimension $1$ and of ghost number $1$.

\index{BRST operator}%
The BRST charge is then obtained by the contour integral:
\begin{equation}
	Q_B
		= Q_{B,L} + Q_{B,R},
	\qquad
	Q_{B,L}
		= \oint \frac{\dd z}{2\pi \I} \, j_B(z),
	\qquad
	Q_{B,R}
		= \oint \frac{\dd \bar z}{2\pi \I} \, \bar \jmath_B(\bar z).
\end{equation}
As usual, $Q_B \sim Q_{B,L}$ when considering only the holomorphic sectors such that we generally omit the index.

\subsection{OPE}

\index{BRST current!OPE|(}%

The OPE of the BRST current with $T$ is
\begin{equation}
	\label{cft:eq:brst-ope-T-jb-anomalous}
	T(z) j_B(w)
		\sim \left( \frac{c_m}{2} - 4 - 6 \kappa \right) \frac{c(w)}{(z - w)^4}
			+ (3 - 2 \kappa) \frac{\pd c(w)}{(z - w)^3}
			+ \frac{j_B(w)}{(z - w)^2}
			+ \frac{\pd j_B(w)}{z - w}.
\end{equation}
\index{critical dimension}%
Hence, the BRST current is a primary operator only if
\begin{equation}
	c_m
		= 26,
	\qquad
	\kappa
		= \frac{3}{2}.
\end{equation}
The BRST current must be primary, otherwise, the BRST symmetry is anomalous, which means that the theory is not consistent.
This provides another derivation of the critical dimension.
In this case, the OPE becomes
\begin{equation}
	\label{cft:eq:brst-ope-T-jb}
	T(z) j_B(w)
		\sim \frac{j_B(w)}{(z - w)^2}
			+ \frac{\pd j_B(w)}{z - w}.
\end{equation}

\begin{remark}[Critical dimension in $2d$ gravity]
	\index{2d@$2d$ gravity}%

	The value $c_m = 26$ (critical dimension) was obtained in \Cref{bos:sec:ws-int:faddeev-popov} by requiring that the Liouville field decouples from the path integral.
	In $2d$ gravity, where this condition is not necessary, (nor even desirable) the Liouville field is effectively part of the matter, such that $c_L + c_m = 26$.
	One can also study the BRST cohomology in this case.
\end{remark}

The OPE of $j_B(z)$ with the ghosts are
\begin{subequations}
\label{cft:eq:brst-ope-jb-fields}
\begin{gather}
	\label{cft:eq:brst-ope-jb-b}
	j_B(z) b(w) \sim \frac{2 \kappa}{(z - w)^3}
		+ \frac{j(w)}{(z - w)^2}
		+ \frac{T(w)}{z - w},
	\\
	\label{cft:eq:brst-ope-jb-c}
	j_B(z) c(w) \sim \frac{\norder{c(w) \pd c(w)}}{z - w}.
\end{gather}
Similarly, the OPE with any matter weight $h$ primary field $\phi$ is
\begin{equation}
	\label{cft:eq:brst-ope-jb-phi}
	j_B(z) \phi(w)
		\sim h \, \frac{c(w) \phi(w)}{(z - w)^2}
			+ \frac{\norder{h \, \pd c(w) \phi(w) + c(w) \pd \phi(w)}}{z - w},
\end{equation}
\end{subequations}
using that $c(w)^2 = 0$ to cancel one term.

The OPE with the ghost current is
\begin{equation}
	j_B(z) j(w) \sim \frac{2 \kappa + 1}{(z - w)^3}
		- \frac{2 \pd c(w)}{(z - w)^2}
		- \frac{j_B(w)}{z - w},
\end{equation}
while the OPE with itself is (for $\kappa = 3/2$)
\begin{equation}
	j_B(z) j_B(w) \sim - \frac{c_m - 18}{2} \, \frac{\norder{c(w) \pd c(w)}}{(z - w)^3}
		- \frac{c_m - 18}{4} \, \frac{\norder{c(w) \pd^2 c(w)}}{(z - w)^2}
		- \frac{c_m - 26}{12} \, \frac{\norder{c(w) \pd^3 c(w)}}{z - w}.
\end{equation}
There is no first order pole if $c_m = 26$: as we will see shortly, this implies that the BRST charge is nilpotent.

\index{BRST current!OPE|)}%

\subsection{Mode expansions}

\index{BRST operator!mode expansion}%
The mode expansion of the BRST charge can be written equivalently
\begin{subequations}
\begin{align}
	Q_B
		&= \sum_{m} \norder{c_m \left( L^{m}_{-m} + \frac{1}{2} \, L^{\text{gh}}_{-m} \right)}
		\\
		&= \sum_{m} c_{-m} L^m_{m}
			+ \frac{1}{2} \sum_{m,n} (n - m)\, \norder{c_{-m} c_{-n} b_{m+n}}
\end{align}
\end{subequations}
In the energy ordering, this expression becomes
\begin{subequations}
\label{cft:eq:brst-charge-expr}
\begin{align}
	Q_B
		&= \sum_m \norderv{c_{m} \left( L^m_{-m} + \frac{1}{2}\, L^{\text{gh}}_{-m} \right)}
			- \frac{c_0}{2}
		\\
		&= \sum_n c_{m} L^m_{-m}
			+ \frac{1}{2} \sum_{m,n} (n - m)\, \norderv{c_{-m} c_{-n} b_{m+n}}
			- c_0,
\end{align}
\end{subequations}
where the ordering constant is the same as in $L^{\text{gh}}_0$ (as can be checked by comparing both sides of the anticommutator).
The simplest derivation of this term is to use the algebra and to ensure that it is consistent.
The only ambiguity is in the second term, when one $c$ does not commute with the $b$: this happens for $- n + (m + n) = 0$, such that the ordering ambiguity is proportional to $c_0$.
Then, one finds that it is equal to $a_{\text{gh}} = - 1$.

\index{BRST operator!zero-mode decomposition}%
The BRST operator can be decomposed on the ghost zero-modes as
\begin{subequations}
\begin{equation}
	\label{brst:eq:splitting-Q}
	Q_B
		= c_0 L_0 - b_0 M + \what Q_B
\end{equation}
where
\begin{gather}
	\label{brst:eq:Q-hat}
	\what Q_B
		= \sum_{m \neq 0} c_{-m} L^m_m
			- \frac{1}{2} \sum_{\substack{m,n \neq 0 \\ m + n \neq 0}}
				(m - n)\, \norderv{c_{-m} c_{-n} b_{m+n}}\ ,
	\\
	M
		= \sum_{m \neq 0} m \, c_{-m} c_m
\end{gather}
\end{subequations}
The interest of this decomposition is that $L_0$, $M$ and $\what Q$ do not contain $b_0$ or $c_0$, which make it very useful to act on states decomposed according to the zero-modes \eqref{cft:eq:1storder-Hilbert}.
The nilpotency of the BRST operator implies the relations
\begin{equation}
	\com{L_0}{M}
		= \com{\what Q_B}{M}
		= \com{\what Q_B}{L_0}
		= 0,
	\qquad
	\what Q_B^2
		= L_0 M.
\end{equation}
Moreover, one has $N_{\text{gh}}(\what Q_B) = 1$ and $N_{\text{gh}}(M) = 2$.

\subsection{Commutators}

\index{BRST operator!commutator}%
From the various OPEs, one can compute the (anti-)commutators of the BRST charge with the other operators.
For the ghosts and a weight $h$ primary field $\phi$, one finds
\begin{subequations}
\label{cft:eq:com-Q}
\begin{gather}
	\label{cft:eq:brst-com-T-b}
	\anticom{Q_B}{b(z)}
		= T(z),
	\\
	\label{cft:eq:brst-com-T-c}
	\anticom{Q_B}{c(z)}
		= c(z) \pd c(z),
	\\
	\label{cft:eq:brst-com-T-phi}
	\com{Q_B}{\phi(z)}
		= h \, \pd c(z) \phi(z) + c(z) \pd \phi(z).
\end{gather}
\end{subequations}
This reproduces correctly \eqref{bos:eq:transf-inf-brst-psi-c-b}.

Two facts will be useful in string theory.
First, \eqref{cft:eq:brst-com-T-phi} is a total derivative for $h = 1$:
\begin{equation}
	\com{Q_B}{\phi(z)}
		= \pd\big(c(z) \phi(z) \big).
\end{equation}
Second, $c(z) \phi(z)$ is closed if $h = 1$
\begin{equation}
	\anticom{Q_B}{c(z) \phi(z)}
		= (1 - h) c(z) \pd c(z) \phi(z).
\end{equation}

The commutator with the ghost current is
\begin{equation}
	\com{Q_B}{j(z)}
		= - j_B(z),
\end{equation}
which confirms that the BRST charge increases the ghost number by $1$
\begin{equation}
	\com{N_{\text{gh}}}{Q_B}
		= Q_B.
\end{equation}

One finds that the BRST charge is nilpotent
\begin{equation}
	\anticom{Q_B}{Q_B} = 0
\end{equation}
and commutes with the energy--momentum tensor
\begin{equation}
	\com{Q_B}{T(z)}
		= 0
\end{equation}
\index{critical dimension}%
only if the matter central charge corresponds to the critical dimension:
\begin{equation}
	c_m = 26.
\end{equation}

The most important commutator for the modes is
\begin{equation}
	\label{cft:eq:brst-com-QB-bn}
	L_n
		= \anticom{Q_B}{b_n}.
\end{equation}
Nilpotency of $Q_B$ then implies that $Q_B$ commutes with $L_n$:
\begin{equation}
	\label{cft:eq:brst-com-QB-Ln}
	\com{Q_B}{L_n}
		= 0.
\end{equation}

\section{BRST cohomology: two flat directions}
\label{cft:sec:brst:lc}

\index{BRST cohomology!two flat directions|(}%

\index{bosonic string CFT}%
The simplest case for studying the BRST cohomology is when the target spacetime has at least two non-compact flat directions represented by two free scalar fields $(X^0, X^1)$ (\Cref{cft:sec:systems:free-scalar}).
The remaining matter fields are arbitrary as long as the critical dimension $c_m = 26$ is reached.
The reason for introducing two flat directions is that the cohomology is easily worked out by introducing light-cone (or complex) coordinates in target spacetime.

The field $X^0$ can be spacelike or timelike $\epsilon_0 = \pm 1$, while we consider $X^1$ to be always spacelike, $\epsilon_1 = 1$.
The oscillators are denoted by $\alpha^0_m$ and $\alpha^1_m$, and the momenta of the Fock vacua by $k_\| = (k^0, k^1)$ such that
\begin{equation}
	k_\|^2
		= \epsilon_0 (k^0)^2 + (k^1)^2.
\end{equation}
The rest of the matter sector, called the transverse sector $\perp$, is an arbitrary CFT with energy--momentum tensor $T^\perp$, central charge $c_\perp = 24$ and Hilbert space $\mc H_\perp$.
The ghost together with the two scalar fields form the longitudinal sector $\|$.
The motivation for the names longitudinal and transverse will become clear later: they will be identified with the light-cone and perpendicular directions in the target spacetime (and, correspondingly, with unphysical and physical states).

\index{bosonic string CFT!Hilbert space}%
The Hilbert space of the theory is decomposed as
\begin{equation}
	\mc H
		:= \mc H_\|
			\otimes \mc H_\perp,
	\qquad
	\mc H_\|
		:= \int \dd k^0 \, \mc F_0(k^0)
			\otimes \int \dd k^1 \, \mc F_1(k^1)
			\otimes \mc H_{\text{gh}},
\end{equation}
where $\mc F_0(k^0)$ and $\mc F_1(k^1)$ are the Fock spaces \eqref{cft:eq:scalar-fock} of the scalar fields $X^0$ and $X^1$, and $\mc H_{\text{gh}}$ is the ghost Hilbert space \eqref{cft:eq:1storder-Hilbert}.
As a consequence, a generic state of $\mc H$ reads
\begin{equation}
	\ket{\psi}
		= \ket{\psi_\|} \otimes \ket{\psi_\perp},
\end{equation}
where $\psi_\perp$ is a generic state of the transverse matter CFT $\mc H_\perp$ and $\psi_\|$ is built by acting with oscillators on the Fock vacuum of $\mc H_\|$:
\begin{equation}
	\begin{gathered}
	\ket{\psi_\|}
		= c_0^{N_0^c}
			\prod_{m > 0} (\alpha^0_{-m})^{N_m^0} (\alpha^1_{-m})^{N_m^1} \,
			(b_{-m})^{N_m^b} (c_{-m})^{N_m^c} \,
			\ket{k^0, k^1, \downarrow}
	\\
	\ket{k^0, k^1, \downarrow}
		:= \ket{k^0} \otimes \ket{k^1} \otimes \ket{\downarrow},
	\qquad
	N_m^0, N_m^1
		\in \N,
	\qquad
	N_m^b, N_m^c
		= 0, 1.
	\end{gathered}
\end{equation}
Since the Virasoro modes commute with the ghost number, eigenstates of the Virasoro operators without zero-modes $\what L_0$, given by the sum of \eqref{cft:eq:scalar-L0-hat} and \eqref{cft:eq:1storder-L0-hat}, can also be taken to be eigenstates of $N_{\text{gh}}$.
It is also useful to define the Hilbert space of states lying in the kernel of $b_0$:
\begin{equation}
	\mc H_0
		= \mc H \cap \ker b_0
\end{equation}
such that
\begin{equation}
	\label{cft:eq:Hilbert-split-open}
	\mc H
		= \mc H_0 \oplus c_0 \mc H_0.
\end{equation}

\index{bosonic string CFT!$L_0$}%
The full $L_0$ operator reads
\begin{equation}
	\label{cft:eq:full-L0}
	L_0
		= L^m_0 + L^{\text{gh}}_0
		= (L^m_0 - 1) + N^b + N^c,
\end{equation}
using \eqref{cft:eq:1storder-L0} for $L^{\text{gh}}_0$.
A more useful expression is obtained by separating the two sectors and by extracting the zero-modes using \eqref{cft:eq:scalar-L0}:
\begin{equation}
	\label{brst:eq:L0-total}
	L_0
		= \big( L_0^\perp - m_{\|,L}^2 \ell^2 - 1 \big)
			+ \what L_0^\|,
\end{equation}
\index{bosonic string CFT!level operator}%
where the longitudinal mass and total level operator are:
\begin{equation}
	m_{\|,L}^2
		= - p_{\|,L}^2,
	\qquad
	\what L_0^\|
		= N^0 + N^1 + N^b + N^c \in \N.
\end{equation}
A state $\ket{\psi}$ is said to be on-shell if it is annihilated by $L_0$:
\begin{equation}
	\label{brst:eq:on-shell-condition}
	\text{on-shell:}
	\qquad
	L_0 \ket{\psi}
		= 0.
\end{equation}

\index{BRST cohomology!absolute}%
The absolute BRST cohomology $\mc H_{\text{abs}}(Q_B)$ defines the physical states (\Cref{bos:sec:ws-int:brst}) and is given by the states $\psi \in \mc H$ that are $Q_B$-closed but not exact:
\begin{equation}
	\mc H_{\text{abs}}(Q_B)
		:= \Big\{
			\ket{\psi} \in \mc H
			\bigm| Q_B \ket{\psi} = 0,
			\nexists \ket{\chi} \in \mc H
			\bigm| \ket{\psi} = Q_B \ket{\chi}
			\Big\}.
\end{equation}
Since $Q_B$ commutes with $L_0$, \eqref{cft:eq:brst-com-QB-Ln}, the cohomology subspace is preserved under time evolution.

\index{contracting homotopy operator}%
Before continuing, it is useful to outline the general strategy for studying the cohomology of a BRST operator $Q$ in the CFT language.
The idea is to find an operator $\Delta$ -- called contracting homotopy operator -- which, if it exists, trivializes the cohomology.
Conversely, this implies that the cohomology is to be found within states which are annihilated by $\Delta$ or for which $\Delta$ is not defined.
Then, it is possible to restrict $Q$ on these subspaces: this is advantageous when the restriction of the BRST charge on these subspaces is a simpler.
In fact, we will find that the reduced operator is itself a BRST operator, for which one can search for another contracting homotopy operator.\footnotemark{}
\footnotetext{%
	A similar strategy shows that there is no open string excitation for the open SFT in the tachyon vacuum.
}

Given a BRST operator $Q$, a contracting homotopy operator $\Delta$ for $Q$ is an operator such that
\begin{equation}
	\label{cft:eq:com-Q-D}
	\anticom{Q}{\Delta}
		= 1.
\end{equation}
\index{propagator}%
Interpreting $Q$ as a derivative operator, $\Delta$ corresponds to the Green function or propagator.
The existence of a well-defined $\Delta$ with empty kernel implies that the cohomology is empty because all closed states are exact.
Indeed, consider a state $\ket{\psi} \in \mc H$ which is an eigenstate of $\Delta$ and closed $Q_B \ket{\psi} = 0$.
Inserting \eqref{cft:eq:com-Q-D} in front of the state gives:
\begin{equation}
	\ket{\psi}
		= \anticom{Q_B}{\Delta} \ket{\psi}
		= Q_B \big(\Delta \ket{\psi} \big).
\end{equation}
If $\Delta$ is well-defined on $\ket{\psi}$ and $\ket{\psi} \notin \ker \Delta$, then $\Delta \ket{\psi}$ is another state in $\mc H$, which implies that $\ket{\psi}$ is exact.
Hence, the BRST cohomology has to be found inside the subspaces $\ker \Delta$ or on which $\Delta$ is not defined.

\subsection{Conditions on the states}
\label{cft:sec:brst:lc:cond}

\index{propagator}%
In this subsection, we apply explicitly the strategy just discussed to get conditions on the states.
A candidate contracting homotopy operator for $Q_B$ is
\begin{equation}
	\label{cft:eq:brst-prop}
	\Delta
		:= \frac{b_0}{L_0}
\end{equation}
thanks to \eqref{cft:eq:brst-com-QB-bn}:
\begin{equation}
	\label{ghost:eq:L0-anticom}
	L_0
		= \anticom{Q_B}{b_0}.
\end{equation}
Indeed, suppose that $\ket{\psi}$ is an eigenstate of $L_0$, and that it is closed but not on-shell:
\begin{equation}
	Q_B \ket{\psi} = 0,
	\qquad
	L_0 \ket{\psi} \neq 0.
\end{equation}
One can use \eqref{ghost:eq:L0-anticom} in order to write:
\begin{equation}
	\label{brst:eq:off-shell-psi}
	\ket{\psi}
		= Q_B \left( \frac{b_0}{L_0} \ket{\psi} \right).
\end{equation}
The operator inside the parenthesis is $\Delta$ defined above in \eqref{cft:eq:brst-prop}.
The formula \eqref{brst:eq:off-shell-psi} breaks down if $\psi$ is in the kernel of $L_0$ since the inverse is not defined.
\index{on-shell condition}%
This implies that a necessary condition for a $L_0$-eigenstate $\ket{\psi}$ to be in the BRST cohomology is to be on-shell \eqref{brst:eq:on-shell-condition}.
Considering explicitly the subset of states annihilated by $b_0$ is not needed at this stage since $\ker b_0 \subset \ker L_0$ for $Q_B$-closed states, according to \eqref{cft:eq:brst-com-QB-bn}.
Hence, we conclude:
\begin{equation}
	\mc H_{\text{abs}}(Q_B)
		\subset \ker L_0.
\end{equation}

Note that this statement holds only at the level of vector spaces, i.e.\ when considering equivalence classes of states $\ket{\psi} \sim \ket{\psi} + Q \ket{\Lambda}$.
This means that there exists a representative state of each equivalence class inside $\ker L_0$, but a generic state is not necessarily in $\ker L_0$.
For example, consider a state $\ket{\psi} \in \ker L_0$ and closed.
Then, $\ket{\psi'} = \ket{\psi} + Q_B \ket{\Lambda}$ with $\ket{\Lambda} \notin \ker L_0$ is still in $\mc H_{\text{abs}}(Q_B)$ but $\ket{\psi'} \notin \ker L_0$ since $\com{L_0}{Q_B} = 0$.

\begin{computation}[brst:eq:off-shell-psi]
For $L_0 \ket{\psi} \neq 0$, one has:
	\[
		\ket{\psi}
			= \frac{L_0}{L_0} \ket{\psi}
			= \frac{1}{L_0}\, \anticom{Q_B}{b_0} \ket{\psi}
			= \frac{1}{L_0}\, Q_B \big( b_0 \ket{\psi} \big)
	\]
	where the fact that $\ket{\psi}$ is closed has been used to cancel the second term of the anticommutator.
	Note that $L_0$ commutes with both $Q_B$ and $b_0$ such that it can be moved freely.
\end{computation}

This shows that $\Delta = b_0 / L_0$ given by \eqref{cft:eq:brst-prop} is not a contracting homotopy operator.
\index{projector!on-shell, $\ker L_0$}%
A proper definition involves the projector $P_0$ on the kernel of $L_0$:
\begin{equation}
	\ket{\psi} \in \ker L_0:
	\quad
	P_0 \ket{\psi}
		= \ket{\psi},
	\qquad
	\ket{\psi} \in (\ker L_0)^\perp:
	\quad
	P_0 \ket{\psi}
		= 0.
\end{equation}
Then, the appropriate contracting homotopy operator reads $\Delta (1 - P_0)$ and \eqref{cft:eq:com-Q-D} is changed to:
\begin{equation}
	\Anticom{Q_B}{\Delta (1 - P_0)}
		= (1 - P_0).
\end{equation}
This parallels completely the definition of the Green function in presence of zero-modes, see \eqref{app:eq:green-eq}.
By abuse of language, we will also say that $\Delta$ is a contracting homotopy operator, remembering that this statement is correct only when multiplying with $(1 - P_0)$.

\index{free SFT}%
We will revisit these aspects later from the SFT perspective.
In fact, we will find that $Q_B$ is the kinetic operator of the gauge invariant theory, while $\Delta$ is the gauge fixed propagator in the Siegel gauge.
This is expected from experience with standard gauge theories: the inverse of the kinetic operator (Green function) is not defined when the gauge invariance is not fixed.

The on-shell condition \eqref{brst:eq:on-shell-condition} is already a good starting point.
In order to simplify the analysis further, one can restrict the question of computing the cohomology on the subspace:
\begin{equation}
	\mc H_0
		:= \mc H \cap \ker b_0
		= \mc H_m \otimes \mc H_{\text{gh},0},
\end{equation}
where $\mc H_{\text{gh},0} = \mc H_{\text{gh}} \cap \ker b_0$ was defined in \eqref{cft:sec:systems:1st:hilbert}.
This subspace contains all states $\ket{\psi}$ such that:
\begin{equation}
	\ket{\psi} \in \mc H_0
	\quad \Longrightarrow \quad
	b_0 \ket{\psi}
		= 0.
\end{equation}
In this subspace, there is no exact state $\ket{\psi}$ with $L_0 \ket{\psi} \neq 0$ such that $b_0 \ket{\psi} = Q_B \ket{\psi} = 0$.
Indeed, assuming these conditions, \eqref{brst:eq:off-shell-psi} leads to a contraction:
\begin{equation}
	b_0 \ket{\psi}
		= Q_B \ket{\psi}
		= 0,
	\quad
	L_0 \ket{\psi}
		\neq 0
	\quad \Longrightarrow \quad
	\ket{\psi} = 0.
\end{equation}
Note that the converse statement is not true: there are on-shell states such that $b_0 \ket{\psi} \neq 0$.
This also makes sense because the ghost Hilbert space can be decomposed with respect to the ghost zero-modes.
\index{BRST cohomology!relative}%
The cohomology of $Q_B$ in the subspace $\mc H_0$ is called the relative cohomology:
\begin{equation}
	\mc H_{\text{rel}}(Q_B)
		:= \mc H_0(Q_B)
		= \Big\{
			\ket{\psi} \in \mc H_0
			\bigm| Q_B \ket{\psi} = 0,
			\nexists \ket{\chi} \in \mc H
			\bigm| \ket{\psi} = Q_B \ket{\chi}
			\Big\}.
\end{equation}

The advantage of the subspace $b_0 = 0$ is to precisely pick the representative of $\mc H_{\text{abs}}$ which lies in $\ker L_0$.
In particular, the operator $L_0$ is simple and has a direct physical interpretation as the worldsheet Hamiltonian.
\index{Siegel gauge}%
This condition is also meaningful in string theory because these states are also mass eigenstates, which have a nice spacetime interpretation, and it will later be interpreted in SFT as fixing the Siegel gauge.
Moreover, it is implied by the choice of $\Delta$ in \eqref{cft:eq:brst-prop} as the contracting homotopy operator, which is particularly convenient to work with to derive the cohomology.
However, there are other possible choices, which are interpreted as different gauge fixings.

After having built this cohomology, we can look for the full cohomology by relaxing the condition $b_0 = 0$.
In view of the structure of the ghost Hilbert space \eqref{cft:eq:1storder-Hilbert}, one can expect that $\mc H_{\text{abs}}(Q_B) = \mc H_{\text{rel}}(Q_B) \oplus c_0 \mc H_{\text{rel}}(Q_B)$, which is indeed the correct answer.
But, we will see (building on \Cref{bos:sec:ws-int:brst:states}) that, in fact, it is this cohomology which contains the physical states in string theory, instead of the absolute cohomology.

As a summary, we are looking for $Q_B$-closed non-exact states annihilated by $b_0$ and $L_0$:
\begin{equation}
	Q_B \ket{\psi}
		= 0,
	\qquad
	L_0 \ket{\psi}
		= 0,
	\qquad
	b_0 \ket{\psi}
		= 0.
\end{equation}

\subsection{Relative cohomology}

In \eqref{brst:eq:splitting-Q}, the BRST operator was decomposed as:
\begin{equation}
	Q_B
		= c_0 L_0 - b_0 M + \what Q_B,
	\qquad
	\what Q_B^2 = L_0 M.
\end{equation}
This shows that, on the subspace $L_0 = b_0 = 0$, $\what Q_B$ is nilpotent and equivalent to $Q_B$:
\begin{equation}
	\ket{\psi} \in \mc H_0 \cap \ker L_0
	\quad \Longrightarrow \quad
	Q_B \ket{\psi}
		= \what Q_B \ket{\psi},
	\qquad
	\what Q_B^2 \ket{\psi}
		= 0.
\end{equation}
Hence, this implies that $\what Q_B$ is a proper BRST operator and the relative cohomology of $Q_B$ is isomorphic to the cohomology of $\what Q_B$:
\begin{equation}
	\mc H_0(Q_B)
		= \mc H_0(\what Q_B).
\end{equation}

Next, we introduce light-cone coordinates in the target spacetime.
While it does not allow to write Lorentz covariant expressions, it is helpful mathematically because it introduces a grading of the Hilbert space, for which powerful theorems exist (even if we will need only basic facts for our purpose).

\subsubsection{Light-cone parametrization}

\index{bosonic string CFT!complex parametrization}%
\index{bosonic string CFT!light-cone parametrization}%
The two scalar fields $X^0$ and $X^1$ are combined in a light-cone (if $\epsilon_0 = - 1$) or complex (if $\epsilon_0 = 1$) fashion:
\begin{equation}
	X_L^\pm
		= \frac{1}{\sqrt{2}} \, \left(
			X_L^0
			\pm \frac{\I}{\sqrt{\epsilon_0}} \, X_L^1
			\right).
\end{equation}
The modes of $X^\pm$ are found by following \eqref{cft:eq:scalar-exp-XLR}:\footnotemark{}
\footnotetext{%
	For $\epsilon_0 = 1$, this convention matches the ones from~\cite{Bouwknegt:1992:BRSTAnalysisPhysical} for $X^0 = X$ and $X^1 = \phi$.
	For $\epsilon = - 1$, this convention matches~\cite{Polchinski:2005:StringTheory-1}.
}%
\begin{subequations}
\begin{gather}
	\alpha^\pm_n
		= \frac{1}{\sqrt{2}} \, \left(
			\alpha^0_n
			\pm \frac{\I}{\sqrt{\epsilon_0}} \, \alpha^1_n
			\right),
		\qquad
		n \neq 0,
	\\
	x_L^\pm
		= \frac{1}{\sqrt{2}} \, \left(
			x_L^0
			\pm \frac{\I}{\sqrt{\epsilon_0}} \, x_L^1
			\right),
	\qquad
	p_L^\pm
		= \frac{1}{\sqrt{2}} \, \left(
			p_L^0
			\pm \frac{\I}{\sqrt{\epsilon_0}} \, p_L^1
			\right),
\end{gather}
\end{subequations}
The non-zero commutation relations are:
\begin{equation}
	\label{cft:eq:scalar-lc-modes-com}
	\com{\alpha^+_m}{\alpha^-_n}
		= \epsilon_0\, m \, \delta_{m+n,0},
	\qquad
	\com{x_L^\pm}{p_L^\mp}
		= \I \epsilon_0.
\end{equation}
This implies that negative-frequency (creation) modes $\alpha_{-n}^\pm$ are canonically conjugate to positive-frequency (annihilation) modes $\alpha_{n}^\mp$.
Note the similarity with the first-order system \eqref{cft:eq:1storder-com-bn-cn}.

For later purposes, it is useful to note the following relations:
\begin{subequations}
\label{cft:eq:scalar-lc-modes-products}
\begin{gather}
	2 \, p_L^+ p_L^-
		= (p_L^0)^2 + \epsilon_0 (p_L^1)^2
		= \epsilon_0 \, p_{\|,L}^2,
	\\
	x^+ p^- + x^- p^+
		= x^0 p^0 + \epsilon_0 \, x^1 p^1,
	\\
	\sum_{n} \alpha_n^+ \alpha_{m-n}^-
		= \frac{1}{2} \sum_n \big(
			\alpha^0_n \alpha^0_{m-n}
			+ \epsilon_0 \, \alpha^1_n \alpha^1_{m-n}
			\big).
\end{gather}
\end{subequations}

In view of the commutators \eqref{cft:eq:scalar-lc-modes-com}, the appropriate definitions of the light-cone number $N_n^\pm$ and level operators $N^\pm$ are:
\begin{equation}
	N^\pm_n
		= \frac{\epsilon_0}{n}\, \alpha_{-n}^\pm \alpha_n^\mp,
	\qquad
	N^\pm
		= \sum_{n > 0} n\, N^\pm_n.
\end{equation}
The insertion of $\epsilon_0$ follows \eqref{cft:eq:scalar-Nn}.
Then, one finds the following relation:
\begin{equation}
	\label{cft:eq:scalar-lc-level-rel}
	N^+ + N^-
		= N^0 + N^1.
\end{equation}

\index{bosonic string CFT!$L_0$}%
Using these definitions, the variables appearing in $L_0$ \eqref{brst:eq:L0-total}
\begin{equation}
	L_0
		= \big( L_0^\perp - m_{\|,L}^2 \ell^2 - 1 \big)
			+ \what L_0^\|
\end{equation}
can be rewritten as:
\index{bosonic string CFT!level operator}%
\begin{equation}
	\label{brst:eq:brst-long-mass-level-lc}
	m_{\|,L}^2
		= - 2 \epsilon_0 \, p_L^+ p_L^-,
	\qquad
	\what L_0^\|
		= N^+ + N^- + N^b + N^c.
\end{equation}
The expression for the sum of the Virasoro operators \eqref{cft:eq:scalar-Ln} easily follows from \eqref{cft:eq:scalar-lc-modes-products}:
\begin{equation}
	L^0_m + L^1_m
		= \epsilon_0 \sum_{n} \norder{\alpha^+_n \alpha^-_{m-n}}
		= \epsilon_0 \sum_{n \neq 0, m} \norder{\alpha^+_n \alpha^-_{m-n}}
			+ \epsilon_0\, \big( \alpha^-_0 \alpha^+_m + \alpha^+_m \alpha^-_m \big).
\end{equation}

\begin{computation}[cft:eq:scalar-lc-modes-com]
	For the modes $\alpha_m^\pm$, we have:
	\begin{align*}
		\com{\alpha^+_m}{\alpha^\pm_n}
			&
			= \frac{1}{2} \,
				\Com{\left(
					\alpha^0_m
					+ \frac{\I}{\sqrt{\epsilon_0}} \, \alpha^1_m
					\right)
				}{\left(
					\alpha^0_n
					\pm \frac{\I}{\sqrt{\epsilon_0}} \, \alpha^1_n
					\right)
				}
			\\ &
			= \frac{1}{2} \left(
				\com{\alpha^0_m}{\alpha^0_n}
				\mp \frac{1}{\epsilon_0} \com{\alpha^1_m}{\alpha^1_n}
				\right)
			= \frac{\epsilon_0}{2} \, m \,
				\delta_{m+n,0} (1 \mp 1),
	\end{align*}
	where we used \eqref{cft:eq:scalar-com-pw}.
	The other commutators follow similarly from \eqref{cft:eq:scalar-com-xpLR}, for example:
	\begin{align*}
		\com{x_L^-}{p_L^\pm}
			&
			= \frac{1}{2} \,
				\Com{\left(
					x_L^0
					- \frac{\I}{\sqrt{\epsilon_0}} \, x_L^1
					\right)
				}{\left(
					p_L^0
					\pm \frac{\I}{\sqrt{\epsilon_0}} \, p_L^1
					\right)
				}
			\\ &
			= \frac{1}{2} \left(
				\com{x_L^0}{p_L^0}
				\pm \epsilon_0 \com{x_L^1}{p_L^1}
				\right)
			= \frac{\epsilon_0}{2} \, (1 \pm 1).
	\end{align*}
\end{computation}

\begin{computation}[cft:eq:scalar-lc-modes-products]
	For the modes $\alpha_m^\pm$, we have:
	\begin{align*}
		\sum_{n} \alpha_n^+ \alpha_{m-n}^-
			&
			= \frac{1}{2} \sum_{n}
				\left(
					\alpha^0_n
					+ \frac{\I}{\sqrt{\epsilon_0}} \, \alpha^1_n
					\right)
				\left(
					\alpha^0_{m-n}
					- \frac{\I}{\sqrt{\epsilon_0}} \, \alpha^1_{m-n}
					\right)
			\\ &
			= \frac{1}{2} \sum_n \left(
				\alpha^0_n \alpha^0_{m-n}
				+ \epsilon_0 \, \alpha^1_n \alpha^1_{m-n}
				+ \frac{\I}{\sqrt{\epsilon_0}}
					(\alpha^0_{m-n} \alpha^1_n - \alpha^0_n \alpha^1_{m-n})
				\right).
	\end{align*}
	The last two terms in parenthesis cancel as can be seen by shifting the sum $n \to m - n$ in one of the term.
	Note that, for $m \neq 2n$, there is no cross-term only after summing over $n$.

	The relations for the zero-modes follow simply by observing that expressions in both coordinates can be rewritten in terms of the $2$-dimensional (spacetime) flat metric.
\end{computation}

\begin{computation}[cft:eq:scalar-lc-level-rel]
	Using \eqref{cft:eq:scalar-lc-modes-products}, one finds:
	\begin{align*}
		N^0 + N^1
			= \sum_{n} n \big( N^0_n + N^1_n \big)
			= \sum_{n} n \big( N^+_n + N^-_n \big)
			= N^+ + N^-.
	\end{align*}
\end{computation}

\subsubsection{Reduced cohomology}

In terms of the light-cone variables, the reduced BRST operator $\what Q_B$ reads:
\begin{equation}
	\what Q_B
		= \sum_{m \neq 0} c_{-m} \left(
				L^\perp_m
				+ \epsilon_0 \sum_{n} \alpha^+_n \alpha^-_{m-n}
				\right)
			+ \frac{1}{2} \sum_{m,n} (n - m)\, \norder{c_{-m} c_{-n} b_{m+n}}.
\end{equation}
This operator can be further decomposed.
Introducing the degree
\begin{equation}
	\deg
		:= N^+ - N^- + \what N^c - \what N^b
\end{equation}
such that
\begin{equation}
	\forall m \neq 0:
	\qquad
	\deg(\alpha_m^+)
		= \deg(c_m)
		= 1,
	\qquad
	\deg(\alpha_m^-)
		= \deg(b_m)
		= - 1,
\end{equation}
and $\deg = 0$ for the other variables, the operator $\what Q_B$ is decomposed as:\footnotemark{}
\footnotetext{%
	The general idea behind this decomposition is the notion of filtration, nicely explained in~\cites[sec.~3]{Asano:2000:NoghostTheoremString}{Chow:2006:YouCouldHave}.
}%
\begin{subequations}
\begin{equation}
	\label{brst:eq:splitting-hat-Q}
	\what Q_B
		= Q_0 + Q_1 + Q_2,
	\qquad
	\deg(Q_j)
		= j,
\end{equation}
where
\begin{equation}
	\begin{gathered}
	Q_1
		= \sum_{m \neq 0} c_{-m} L^\perp_m
			+ \sum_{\substack{m,n \neq 0 \\ m + n \neq 0}}
				\norderv{c_{-m} \left(
					\epsilon_0 \, \alpha^+_n \alpha^-_{m-n}
					+ \frac{1}{2}\, (m - n)\, c_{-m} b_{m+n}
					\right)},
	\\
	Q_0
		= \sum_{n \neq 0} \alpha_0^+ \, c_{-n} \alpha^-_n,
	\qquad
	Q_2
		= \sum_{n \neq 0} \alpha_0^- \, c_{-n} \alpha^+_n.
	\end{gathered}
\end{equation}
\end{subequations}
The nilpotency of $\what Q_B$ implies the following conditions on the $Q_j$:
\begin{equation}
	\label{brst:eq:conditions-Qj}
	Q_0^2
		= Q_2^2
		= 0,
	\qquad
	\anticom{Q_0}{Q_1}
		= \anticom{Q_1}{Q_2} = 0,
	\qquad
	Q_1^2 + \anticom{Q_0}{Q_2}
		= 0.
\end{equation}
Hence, $Q_0$ and $Q_2$ are both nilpotent and define a cohomology.

\index{BRST cohomology!relative}%
One can show that the cohomologies of $\what Q_B$ and $Q_0$ are isomorphic\footnotemark{}
\footnotetext{%
	The role of $Q_0$ and $Q_2$ can be reversed by changing the sign in the definition of the degree and the role of $P^\pm_n$.
}%
\begin{equation}
	\mc H_0(\what Q_B)
		\simeq \mc H_0(Q_0)
\end{equation}
under general conditions~\cite{Bouwknegt:1992:BRSTAnalysisPhysical}, in particular, if the cohomology is ghost-free (i.e.\ all states have $N_{\text{gh}} = 1$).

\index{contracting homotopy operator}%
The contracting homotopy operator for $Q_0$ is
\begin{equation}
	\label{cft:eq:brst-B}
	\what \Delta
		:= \frac{B}{\what L_0^\|},
	\qquad
	B
		:= \epsilon_0 \sum_{n \neq 0}
			\frac{1}{\alpha_0^+}\, \alpha^+_{-n} b_n.
\end{equation}
Indeed, it is straightforward to check that
\begin{equation}
	\label{cft:eq:brst-com-Q0-B}
	\what L_0^\|
		= \anticom{Q_0}{B}
	\quad \Longrightarrow \quad
	\anticom{Q_0}{\what \Delta}
		= 1.
\end{equation}

As a consequence, a necessary condition for a closed $\what L_0^\|$-eigenstate $\ket{\psi}$ to be in the cohomology of $Q_0$ is to be annihilated by $\what L_0^\|$:
\begin{equation}
	\what L_0^\| \ket{\psi}
		= 0,
	\quad \Longrightarrow \quad
	N^\pm \ket{\psi}
		= N^c \ket{\psi}
		= N^b \ket{\psi}
		= 0,
\end{equation}
since $\what L_0^\|$ is a sum of positive integers.
This means that the state $\psi$ contains no ghost or light-cone excitations $\alpha^\pm_{-n}$, $b_{-n}$ and $c_{-n}$, and lies in the ground state of the Fock space $\mc H_{\|, 0}$.

Then, we need to prove that this condition is sufficient: states with $\what L_0^\| = 0$ are closed.
First, note that a state $\ket{\psi} \in \mc H_0$ with $\what L_0^\|$ has ghost number $1$ since there are no ghost excitations on top of the vacuum $\ket{\downarrow}$, which has $N_{\text{gh}} = 1$.
Second, $\what L_0$ and $Q_0$ commute, such that:
\begin{equation}
	0
		= Q_0 \what L_0^\| \ket{\psi}
		= \what L_0^\| Q_0 \ket{\psi}.
\end{equation}
Since $Q_0$ increases the ghost number by $1$, one can invert $\what L_0^\| = N^b + N^c + \cdots$ in the last term since $\what L_0^\| \neq 0$ in this subspace.
This gives:
\begin{equation}
	Q_0 \ket{\psi} = 0.
\end{equation}
Hence, the condition $\what L_0^\| \ket{\psi} = 0$ is sufficient for $\ket{\psi}$ to be in the cohomology.
This has to be contrasted with \Cref{cft:sec:brst:lc:cond} where the condition $L_0^\| = 0$ is necessary but not sufficient.

\index{on-shell condition}%
In this case, the on-shell condition \eqref{brst:eq:L0-total} reduces to
\begin{equation}
	\label{brst:eq:L0-cohom-std}
	L_0
		= L_0^\perp - m_{\|,L}^2 \ell^2 - 1
		= 0.
\end{equation}

But, additional states can be found in $\ker B$ or in a subspace of $\mc H$ on which $B$ is singular.
We have $\ker B = \ker \what L_0^\|$ such that nothing new can be found there.
However, the operator $B$ is not defined for states with vanishing momentum $\alpha_0^+ \propto p_L^+ = 0$.
In fact, one must also have $\alpha_0^- \propto p_L^- = 0$ (otherwise, the contracting operator for $Q_2$ is well-defined and can be used instead).
But, these states do not satisfy the on-shell condition (except for massless states with $L_0^\perp = 1$), as it will be clear later (see~\cite[sec.~2.2]{Thorn:1989:StringFieldTheory} for more details).
For this reason, we assume that states have a generic non-zero momentum and that there is no pathology.

\subsubsection{Full relative cohomology}

\index{BRST cohomology!relative}%
This section aims to construct states in $\mc H_0(\what Q_B)$ from states in $\mc H(Q_0)$.
We follow the construction from~\cite{Bouwknegt:1992:BRSTAnalysisPhysical}.

Given a state $\ket{\psi_0} \in \mc H_0(Q_0)$, the state $Q_1 \ket{\psi_0}$ is $Q_0$-closed since $Q_0$ and $Q_1$ anticommute \eqref{brst:eq:conditions-Qj}:
\begin{equation}
	\anticom{Q_0}{Q_1} \ket{\psi_0}
		= 0
	\quad \Longrightarrow \quad
	Q_0 \big( Q_1 \ket{\psi_0} \big)
		= 0.
\end{equation}
Since $Q_1 \ket{\psi_0}$ is not in $\ker \what L_0^\|$ (because $Q_1$ increases the ghost number by $1$), the state $Q_1 \ket{\psi_0}$ is $Q_0$-exact and can be written as $Q_0$ of another state $\ket{\psi_1}$:
\begin{equation}
	\label{brst:eq:def-psi1}
	Q_1 \ket{\psi_0}
		=: - Q_0 \ket{\psi_1}
	\quad \Longrightarrow \quad
	\ket{\psi_1}
		= - \frac{B}{\what L_0^\|} \, Q_1 \ket{\psi_0}.
\end{equation}

\begin{computation}[brst:eq:def-psi1]
	Start from the definition and insert \eqref{cft:eq:brst-com-Q0-B} since $\what L_0$ is invertible:
	\[
		Q_1 \ket{\psi_0}
			= \Anticom{Q_0}{\frac{B}{\what L_0^\|}} Q_1 \ket{\psi_0}
			= Q_0 \left( \frac{B}{\what L_0^\|} \, Q_1 \ket{\psi_0} \right).
	\]
	The state $\ket{\psi_1}$ is identified with minus the state inside the parenthesis (up to a BRST exact state).
\end{computation}

As for $\ket{\psi_0}$, apply $\anticom{Q_0}{Q_1}$ on $\psi_1$:
\begin{equation}
	\label{brst:eq:com-Q0-Q1-psi1}
	\anticom{Q_0}{Q_1} \ket{\psi_1}
		= Q_0 \big(Q_1 \ket{\psi_1} + Q_2 \ket{\psi_0} \big).
\end{equation}
This implies that the combination in parenthesis is $Q_0$-closed and, for the same reason as above, it is exact:
\begin{equation}
	Q_1 \ket{\psi_1} + Q_2 \ket{\psi_0}
		= Q_0 \ket{\psi_2},
	\qquad
	\ket{\psi_2}
		= - \frac{B}{\what L_0^\|} \big(
			Q_1 \ket{\psi_1}
			+ Q_2 \ket{\psi_0}
			\big).
\end{equation}

\begin{computation}[brst:eq:com-Q0-Q1-psi1]
	\[
		\anticom{Q_0}{Q_1} \ket{\psi_1}
			= Q_0 Q_1 \ket{\psi_1} - Q_1^2 \ket{\psi_0}
			= Q_0 Q_1 \ket{\psi_1} + \anticom{Q_0}{Q_2} \ket{\psi_0}.
	\]
	The first equality follows from \eqref{brst:eq:def-psi1}, the second by using \eqref{brst:eq:conditions-Qj}.
	The final result is obtained after using that $\ket{\psi_0}$ is $Q_0$-closed.
\end{computation}

Iterating this procedure leads to a series of states:
\begin{equation}
	\ket{\psi_{k+1}}
		= - \frac{B}{\what L_0^\|} \big(
			Q_1 \ket{\psi_k}
			+ Q_2 \ket{\psi_{k-1}}
			\big).
\end{equation}
We claim that a state in the relative cohomology $\ket{\psi} \in \mc H_0(\what Q_B)$ is built by summing all these states:
\begin{equation}
	\ket{\psi} = \sum_{k \in \N} \ket{\psi_k}.
\end{equation}
Indeed, it is easy to check that $\ket{\psi}$ is $\what Q_B$-closed:
\begin{equation}
	\label{cft:eq:brst-sum-psik-closed}
	\what Q_B \ket{\psi}
		= 0.
\end{equation}
We leave aside the proof that $\psi$ is not exact (see~\cite{Bouwknegt:1992:BRSTAnalysisPhysical}).
Note that $\psi$ and $\psi_0$ have the same ghost numbers
\begin{equation}
	N_{\text{gh}}(\psi)
		= N_{\text{gh}}(\psi_0)
		= 1
\end{equation}
since $N_{\text{gh}}(B Q_j) = 0$.

In fact, since $\psi_0$ does not contain longitudinal modes, it is annihilated by $Q_1$ and $Q_2$ (these operators contain either a ghost creation operator together with a light-cone annihilation operator, or the reverse):
\begin{equation}
	Q_1 \ket{\psi_0}
		= Q_2 \ket{\psi_0}
		= 0.
\end{equation}
As a consequence, one has $\psi_k = 0$ for $k \ge 1$ and $\psi = \psi_0$.

\begin{computation}[cft:eq:brst-sum-psik-closed]
	\begin{align*}
		\what Q_B \ket{\psi}
			&
			= \sum_{k \in \N} \what Q_B \ket{\psi_k}
			\\ &
			= Q_0 \ket{\psi_0}
				+ \underbrace{Q_1 \ket{\psi_0} + Q_0 \ket{\psi_1}}_{= 0}
				+ \underbrace{Q_2 \ket{\psi_0} + Q_1 \ket{\psi_1} + Q_0 \ket{\psi_2}}_{= 0}
				+ \cdots
			\\ &
			= 0.
	\end{align*}
\end{computation}

\subsection{Absolute cohomology, states and no-ghost theorem}
\label{cft:sec:brst:lc:abs}

\index{BRST cohomology!absolute}%
The absolute cohomology is constructed from the relative cohomology:
\begin{equation}
	\label{brst:eq:Habs-decomposition-Hrel}
	\mc H_{\text{abs}}(Q_B)
		= \mc H_{\text{rel}}(Q_B) \oplus c_0 \, \mc H_{\text{rel}}(Q_B).
\end{equation}
The interested reader is refereed to~\cite{Bouwknegt:1992:BRSTAnalysisPhysical} for the proof.
A simple motivation is that the Hilbert space is decomposed in terms of the ghost zero-modes as in \eqref{cft:eq:1storder-Hilbert}.
Since the zero-modes commute with $\what Q_0$, linear combination of states in $\mc H_{\text{rel}}(Q_B)$ and $c_0 \mc H_{\text{rel}}(Q_B)$ are expected to be in the cohomology.
Obviously, one has to work out the other terms of $Q_B$ and prove that there are no other states.

It looks like there is a doubling of the physical states, one built on $\ket{\downarrow}$ and one on $\ket{\uparrow}$.
The remedy is to impose the condition $b_0 = 0$ on the states (see also \Cref{bos:sec:ws-int:brst:states} and~\cite[sec.~2.2]{Thorn:1989:StringFieldTheory} for more details).
As already pointed out, states in $\mc H_{\text{abs}}$ form equivalence class under $\ket{\psi} \sim \ket{\psi} + Q_B \ket{\Lambda}$, and it is necessary to select a single representative.
This is what the condition $b_0 = 0$ achieves.
Obviously, it is always possible to add BRST exact states to write another representative (for example, to restore the Lorentz covariance).

The last step is to discuss the no-ghost theorem: the latter states that there is no negative-norm states in the BRST cohomology of string theory.
This follows straightforwardly from the condition $\what L_0 = 0$: it implies that there are no ghost and no light-cone excitations.
The ghosts and the time direction (if $X^0$ is timelike) are responsible for negative-norm states.
Hence, the cohomology has no negative-norm states if the transverse CFT is unitary (which implies that all states in $\mc H_\perp$ have a positive-definite inner-product).

\index{on-shell condition}%
Physical states $\ket{\psi} \in \mc H_{\text{rel}}(Q_B)$ are thus of the form:
\begin{subequations}
\begin{gather}
	\label{cft:eq:brst-state-cohom}
	\ket{\psi}
		= \ket{k^0, k^1, \downarrow} \otimes \ket{\psi_\perp},
	\qquad
	\ket{\psi_\perp}
		\in \mc H_\perp,
	\\
	\big( L_0^\perp - m_{\|,L}^2 \ell^2 - 1 \big) \ket{\psi}
		= 0,
	\qquad
	p_{L,\|}^2
		= - m_{\|,L}^2 \ell^2.
\end{gather}
\end{subequations}
\index{old covariant quantization (OCQ)}%
This form can be made covariant: taking a state of the form $\ket{\psi} \otimes \ket{\downarrow}$ with $\ket{\psi} \in \mc H_m$, acting with $Q_B$ implies the equivalence with the old covariant quantization:
\begin{equation}
	(L_0^m - 1) \ket{\psi}
		= 0,
	\qquad
	\forall n > 0:
		\quad
		L_n^m \ket{\psi}
			= 0.
\end{equation}
This means that $\psi$ must be a weight $1$ primary field of the matter CFT.

\begin{remark}[Open string]
	The results of this section provide, in fact, the cohomology for the open string after taking $p_L = p$ (instead of $p_L = p / 2$ for the closed string).
\end{remark}

\subsection{Cohomology for holomorphic and anti-holomorphic sectors}
\label{cft:sec:brst:lc:closed}

It remains to generalize the computation of the cohomology when considering both the holomorphic and anti-holomorphic sectors.

\index{BRST operator!full!zero-mode decomposition}%
In this case, the BRST operator is
\begin{equation}
	Q_B
		= c_0 L_0 - b_0 M + \what Q_B
			+ \bar c_0 \bar L_0 - \bar b_0 \bar M + \what{\overline Q}_B.
\end{equation}
It is useful to rewrite this expression in terms of $L_0^\pm$, $b_0^\pm$ and $c_0^\pm$:
\begin{equation}
	\label{cft:eq:brst-decomp-LR-pm}
	Q_B
		= c_0^+ L_0^+ - b_0^+ M^+
			+ c_0^- L_0^- - b_0^- M^-
			+ \what Q_B^+,
\end{equation}
where
\begin{equation}
	L_0^+
		= \left( L_0^{\perp +} - \frac{m_{\|}^2 \ell^2}{2} - 2 \right)
			+ \what L_0^{\| +},
	\qquad
	L_0^-
		= L_0^{\perp -} + \what L_0^{\| -}
\end{equation}
and
\begin{equation}
	M^\pm
		:= \frac{1}{2} (M \pm \bar M).
\end{equation}

\index{on-shell condition}%
\index{level-matching condition}%
Because of the relations $L_0^\pm = \anticom{Q_B}{b_0^\pm}$, we find that states in the cohomology must be on-shell $L_0^+ = 0$ and must satisfy the level-matching condition $L_0^- = 0$:\footnotemark{}
\footnotetext{%
	In the current case, the propagator is less easily identified.
	We will come back on its definition later.
}%
\begin{equation}
	L_0^+ \ket{\psi}
		= L_0^- \ket{\psi}
		= 0.
\end{equation}

\index{BRST cohomology!semi-relative}%
\index{BRST cohomology!relative}%
Again, it is possible to reduce the cohomology by imposing conditions on the zero-modes such that the above conditions are automatically satisfied (see also \Cref{bos:sec:ws-int:brst:states}).
Imposing first the condition $b_0^- = 0$ defines the semi-relative cohomology.
The relative cohomology is found by imposing $b_0^\pm = 0$ and in fact corresponds to the physical space (see~\cite[sec.~2.3]{Thorn:1989:StringFieldTheory} for more details).
The rest of the derivation follows straightforwardly because the two sectors commute: we find that the cohomology is ghost-free and has no light-cone excitations:
\begin{equation}
	\what L_0^{\| \pm}
		= N^0 \pm \bar N^0
			+ N^1 \pm \bar N^1
			+ N^b \pm \bar N^b
			+ N^c \pm \bar N^c
		= 0.
\end{equation}

In general, it is simpler to work with a covariant expression and to impose the necessary conditions.
Taking a state $\ket{\psi} \otimes \ket{\downarrow\downarrow}$ with $\ket{\psi} \in \mc H_m$, we find that $\psi$ is a weight $(1, 1)$ primary field of the matter CFT:
\begin{equation}
	\begin{gathered}
	(L_0^m + \bar L_0^m - 2) \ket{\psi}
		= 0,
	\qquad
	(L_0^m - \bar L_0^m) \ket{\psi}
		= 0,
	\\
	\forall n > 0:
		\quad
		L_n^m \ket{\psi}
			= \bar L_n^m \ket{\psi}
			= 0.
	\end{gathered}
\end{equation}
An important point is that the usual mass-shell condition $k^2 = - m^2$ is provided by the first condition only.
This also shows that states in the cohomology naturally appears with $c \bar c$ insertion since
\begin{equation}
	\ket{\downarrow\downarrow}
		= c(0) \bar c(0) \ket{0}
		= c_1 \bar c_1 \ket{0}.
\end{equation}
This hints at rewriting of scattering amplitudes in terms of unintegrated states \eqref{bos:eq:vertex-unintegrated} only.

A state is said to be of level $(\ell, \bar\ell)$ and denoted as $\psi_{\ell, \bar \ell}$ if it satisfies:
\begin{equation}
	\what L_0 \ket{\psi_{\ell, \bar \ell}}
		= \ell \ket{\psi_{\ell, \bar \ell}},
	\qquad
	\what{\bar L}_0 \ket{\psi_{\ell, \bar \ell}}
		= \bar \ell \ket{\psi_{\ell, \bar \ell}}.
\end{equation}

\begin{example}[Closed string tachyon]
	\index{closed string states!tachyon $T$!physical}%

	As an example, let's construct the state $\psi_{0, 0}$ with level zero for a spacetime with $D$ non-compact dimensions.
	In this case, the transverse CFT contains $D - 2$ free scalars which combine with $X^0$ and $X^1$ into $D$ scalars $X^\mu$.
	The Fock space is built on the vacuum $\ket{k}$ and we define the mass such that on-shell condition reduces to the standard QFT expression:
	\begin{equation}
		k^2
			= - m^2,
		\qquad
		m^2
			:= \frac{2}{\ell^2} \, (N + \bar N - 2),
	\end{equation}
	where $N$ and $\bar N$ are the matter level operators.
	The state in the remaining transverse CFT (without the $D - 2$ scalars) is the $\group{SL}(2, \C)$ vacuum with $L_0^\perp = \bar L_0^\perp = 0$ (this is the state with the lowest energy for a unitary CFT).
	In this case, the on-shell condition reads
	\begin{equation}
		m^2 \ell^2
			= - 4 < 0.
	\end{equation}
	Since the mass is negative, this state is a tachyon.
	The vertex operator associated to the state reads:
	\begin{equation}
		\scr V(k, z, \bar z)
			= c(z) \bar c(\bar z) \e^{\I k \cdot X(z, \bar z)}.
	\end{equation}
	\begin{check}
	This also illustrates that the closed string states are product of open string states, up to the exponential term.
	\end{check}
\end{example}

\index{BRST cohomology!two flat directions|)}%

\begin{draft}

\section{BRST cohomology: one flat direction}

A more general approach to the BRST cohomology has been proposed in~\cite{Asano:2000:NoghostTheoremString}.
However, this approach are less explicit is less explicit and does not construct the states.

\end{draft}

\section{Summary}

In this chapter, we have described the BRST quantization from the CFT point of view.
We have first considered only the holomorphic sector (equivalently, the open string).
We proved that the cohomology does not contain negative-norm states and we provided an explicit way to construct the states.
Finally, we glued together both sectors and characterized the BRST cohomology of the closed string.

What is the next step?
We could move to computations of on-shell string amplitudes, but this falls outside the scope of this \revname{}.
We can also start to consider string field theory.
Indeed, the BRST equation $Q_B \ket{\psi} = 0$ and the equivalence $\ket{\psi} \sim \ket{\psi} + Q_B \ket{\Lambda}$ completely characterize the states.
In QFT, states are solutions of the linearized equations of motion: hence, the BRST equation can provide a starting point for building the action.
This is the topic of \Cref{bsft:chap:free-brst}.

\refchapter

\begin{itemize}
	\item The general method to construct the absolute cohomology follows~\cite{Bouwknegt:1992:BRSTAnalysisPhysical, Polchinski:2005:StringTheory-1}.
	Other works and reviews include~\cite{Itoh:1990:BRSTQuantizationPolyakovs, Mukhi:1991:ExtraStatesC1, Bouwknegt:1992:BRSTAnalysisPhysicalSugra, Bilal:1992:RemarksBRSTcohomologycM, Itoh:1992:SpectrumTwoDimensionalSuperGravity, Ohta:1992:DiscreteStatesTwoDimensional, Distler:1992:NewDiscreteStates}.

	\item String states are discussed in~\cites[sec.~3.3, ]{Blumenhagen:2014:BasicConceptsString}[sec.~4.1]{Polchinski:2005:StringTheory-1}.
\end{itemize}

\part{String field theory}
\label{part:bosonic-sft}
\label{part:spacetime-sft}

\chapter{String field}
\label{bsft:chap:string-field}

In this chapter, we introduce general concepts about the string field.
The goal is to give an idea of which type of object it is and of the different possibilities for describing it.
We will see that the string field is a functional and, for this reason, it is more convenient to work with the associated ket field, which can itself be represented in momentum space.
We focus on what to expect from a free field, taking inspiration from the worldsheet theory.
The interpretation becomes more difficult when taking into account the interactions.

\section{Field functional}

\index{string field!functional}%
A string field, after quantization, is an operator which creates or destroys a string at a given time.
Since a string is a $1$-dimension extended object, the string field $\Psi$ must depend on the spatial positions of each point of the string denoted collectively as $X^\mu$.
Hence, the string field is a functional $\Psi[X^\mu]$.
The fact that it is a functional rather than a function makes the construction of a field theory much more challenging: it asks for revisiting all concepts we know in point-particle QFT without any prior experience with a simple model.\footnotemark{}
\footnotetext{%
	The problem is not in working with the wordline formalism and writing a BRST field theory, but really to take into account the spatial extension of the objects.
	In fact, generalizing further to functionals of extended $(p > 1)$-dimensional objects -- branes -- shows that SFT is the simplest of such field theories.
}%

It is important that the dependence is only on the shape and not on the parametrization.
However, it is simpler to first work with a specific parametrization $X(\sigma)$ and make sure that nothing depends on it at the end (equivalent to imposing the invariance under reparametrization of the worldsheet).
This leads to work with a functional $\Psi[X(\sigma)]$ of fields on the worldsheet (at fixed time).
To proceed, one should first determine the degrees of freedom of the string, and then to find the interactions.
The simplest way to achieve the first step is to perform a second-quantization of the string wave-functional: the string field is written as a linear combination of first-quantized states with spacetime wave functionals as coefficients.\footnotemark{}
\footnotetext{%
	The description of the first-quantized states depends on the CFT used to describe the theory.
	This explains the lack of manifest background independence of SFT.
	Unfortunately, no better approach has been found until now.
}%
This provides a free Hamiltonian; trying to add interactions perturbatively does not work well.

It is not possible to go very far with this approach and one is lead to choose a specific gauge, breaking the manifest invariance under reparametrizations.
The simplest is the light-cone gauge since one works only with the physical degrees of freedom of the string.
While this approach is interesting to gain some intuitions and to show that, in principle, it is possible to build a string field theory, it requires making various assumptions and ends up with problems (especially for superstrings).\footnotemark{}
\footnotetext{%
	While this approach has been mostly abandoned, recent results show that it can still be used when defined with a proper regularization~\cite{Baba:2009:LightConeGaugeString, Ishibashi:2013:MultiloopAmplitudesLightcone, Ishibashi:2016:WorldsheetTheoryLightcone, Ishibashi:2018:MultiloopAmplitudesLightcone-1, Ishibashi:2018:MultiloopAmplitudesLightcone-2}.
}%

Since worldsheet reparametrization invariance is just a kind of gauge symmetry -- maybe less familiar than the non-Abelian gauge symmetries in Yang--Mills, but still a gauge symmetry --, one may surmise that it should be possible to gauge fix this symmetry and to introduce a BRST symmetry in its place.
This is the program of the BRST (or covariant) string field theory in which the string field depends not only of the worldsheet (at fixed time), but also on the ghosts: $\Psi[X(\sigma), c(\sigma)]$.
There is no dependence on the $b$ ghost because the latter is the conjugate momentum of the $c$ ghost: in the operator language, $b(\sigma) \sim \frac{\delta}{\delta c(\sigma)}$.

The BRST formalism has the major advantage to allow to move easily from $D = 26$ dimensions -- described by $X^\mu$ scalars ($\mu = 0, \ldots, 25$) -- to a (possibly curved) $D$-dimensional spacetime and a string with some internal structure -- described by a more general CFT, in which $D$ scalars $X^\mu$ represent the non-compact dimensions and the remaining system with central charge $26 - D$ describes the compactification and structure.
It is sufficient to consider the string field as a general functional of all the worldsheet fields.
For simplicity, we will continue to write $X$ in the functional dependence, keeping the other matter fields implicit.

It is complicated to find an explicit expression for the string field as a functional of $X(\sigma)$ and $c(\sigma)$.
\index{string field!position representation}%
In fact, the field written in this way is in the \emph{position representation} and, as usual in quantum mechanics, one can choose to work with the representation independent ket $\ket{\Psi}$:
\index{string field!ket representation}%
\begin{equation}
	\Psi[X(\sigma), c(\sigma)]
		:= \Bracket{X(\sigma), c(\sigma)}{\Psi}.
\end{equation}
It is often more convenient to work with $\ket{\Psi}$ (which we will also denote simply as $\Psi$, not distinguishing between states and operators).
The latter will be the basic object of SFT in most of this \revname{}.

\bigskip

Writing a field theory in terms of $\ket{\Psi}$ may not be intuitive since in point-particle QFT, one is used to work with the position or momentum representation.
In fact, there is a very simple way to recover a formulation in terms of spacetime point-particle fields, which can be used almost whenever there is a doubt about what is going on.
Indeed, as is well-known from standard worldsheet string theory, the string states behave like a collection of particles.
This is because the modes of the CFT fields (like $\alpha_n^\mu$) carry spacetime indices (Lorentz, group representation…) such that the states themselves carries indices.
Indeed, these quantum numbers classify eigenstates of the operators $L_0$ and $\bar L_0$.
On the other hand, positions and shapes are not eigenstates of any simple CFT operator.

\section{Field expansion}

\index{string field!momentum expansion}%
It follows that the second-quantized string field can be written as a linear combination of first-quantized off-shell states $\ket{\phi_\alpha(k)} = \scr V_\alpha(k; 0, 0) \ket{0}$ (which form a basis of the CFT Hilbert space $\mc H$):
\begin{equation}
	\label{bsft:eq:field-exp-kj}
	\ket{\Psi}
		= \sum_\alpha \int \frac{\dd^D k}{(2\pi)^D} \, \psi_\alpha(k) \ket{\phi_\alpha(k)},
\end{equation}
where $k$ is the $D$-dimensional momentum of the string (conjugated to the position of the centre-of-mass) and $\alpha$ is a collection of discrete quantum numbers (Lorentz indices, group representation…).
When inserting this expansion inside the action, we find that it reduces to a standard field theory with an infinite number of particles described by the spacetime fields $\psi_\alpha(k)$ (momentum representation).
The fields can also be written in the position representation by Fourier transforming only the momentum $k$ to the centre-of-mass $x$:
\begin{equation}
	\psi_\alpha(x)
		= \int \frac{\dd^D k}{(2\pi)^D} \, \e^{\I k \cdot x} \psi_\alpha(k).
\end{equation}
However, we will see that it is often not convenient because the action is non-local in position space (including for example exponentials of derivatives).

The physical intuition is that the string is a non-local object in spacetime.
It can be expressed in momentum space through a Fourier transformation: variables dual to non-compact (resp.\ compact) dimensions are continuous (discrete).
As a consequence, the momentum is continuous since the centre-of-mass move in the non-compact spacetime, while the string itself has a finite extension and the associated modes are discrete but still not bounded (and similarly for compact dimensions).
This indicates that the spectrum is the collection of a set of continuous and discrete modes.
Hence, the non-locality of the string (due to the spatial extension) is traded for an infinite number of modes which behave like standard particles.
In this description, the non-locality arises: 1) in the infinite number of fields, 2) in the coupling between the modes, 3) as a complicated momentum-dependence of the action.

When we are not interested in the spacetime properties, we will write a generic basis of the Hilbert space $\mc H$ as $\{ \phi_r \}$:
\index{string field!expansion}%
\begin{equation}
	\label{bsft:eq:string-field-expansion}
	\ket{\Psi}
		= \sum_r \psi_r \ket{\phi_r}.
\end{equation}
The sum over $r$ includes discrete and continuous labels.

\begin{check}
\begin{remark}[Classical fields and quantum states]
	The states $\ket{\phi_\alpha(k)}$, being eigenstates of the CFT Hamiltonian, are (first-quantized) quantum states.
	A classical field $\Psi$ can then be written as a linear combination of such states with coefficients $\psi_\alpha(k)$ corresponding to wave functions.
	A specific choice of these functions gives a profile to the classical field.
	As long as the coefficients are functions, the field $\Psi$ is classical: (second-)quantization replaces the functions by operators, and, at this point, the field $\Psi$ is promoted to a quantum operator.
	The fact that one uses first-quantized states to describe the field configuration does not mean that the field itself is quantized: this is just a particularly useful description since one knows already a basis of possible configurations.
\end{remark}
\end{check}

\begin{example}[Scalar field]
	In order to illustrate the notations for a point-particle, consider a scalar field $\phi(x)$.
	It can be expanded in Fourier modes as:
	\begin{equation}
		\phi(x) = \int \frac{\dd^D k}{(2\pi)^D} \, \phi(k) \e^{\I k \cdot x}.
	\end{equation}
	The corresponding ket $\ket{\phi}$ is found by expanding on a basis $\{ \ket{k} \}$:
	\begin{equation}
		\ket{\phi}
			= \int \frac{\dd^D k}{(2\pi)^D} \, \phi(k) \ket{k},
		\qquad
		\phi(k)
			= \bracket{k}{\phi}.
	\end{equation}
	Similarly, the position space field is defined from the basis $\{ \ket{x} \}$ such that:
	\begin{equation}
		\phi(x)
			= \bracket{x}{\phi}
			= \int \frac{\dd^D k}{(2\pi)^D} \, \bracket{x}{k} \bracket{k}{\phi},
		\qquad
		\bracket{x}{k}
			= \e^{\I k \cdot x}.
	\end{equation}
\end{example}

\section{Summary}

In this chapter, we introduced general ideas about what a string field is.
We now need to write an action.
In general, one proceeds in two steps:
\begin{enumerate}
	\item build the kinetic term (free theory):
	\begin{enumerate}
		\item equations of motion → physical states
		\item equivalence relation → gauge symmetry
	\end{enumerate}

	\item add interactions and deform the gauge transformation
\end{enumerate}
We consider the first point in the next chapter, but we will have to introduce more machinery in order to discuss interactions.

\refchapter

\begin{itemize}
	\item General discussions of the string field and of the ideas of string field theory can be found in~\cites[sec.~4]{Polchinski:1994:WhatStringTheory}{Zwiebach:1993:ClosedStringFieldIntro}.

	\item Light-cone SFT is reviewed in~\cites{Thorn:1989:StringFieldTheory}[chap.~6]{Kaku:1999:IntroductionSuperstringsMTheory}[chap.~9]{Kaku:1999:StringsConformalFields}.
\end{itemize}

\chapter{Free BRST string field theory}
\label{bsft:chap:free-brst}

\introchapter

In this chapter, we construct the BRST (or covariant) free bosonic string field theories.
It is useful to first ignore the interactions in order to introduce some general tools and structures in a simpler setting.
Moreover, the free SFT is easily constructed and does not require as much input as the interactions.
In this chapter, we discuss mostly the open string, keeping the closed string for the last section.
We start by describing the classical theory: equations of motion, action, gauge invariance and gauge fixing.
Then, we perform the path integral quantization and compute the action in terms of spacetime fields for the first two levels (tachyon and gauge field).

\section{Classical action for the open string}
\label{bsft:sec:free-brst:classical}

Contrary to most of this \revname{}, we will exemplify the discussion with the open string.
The reason is that most computations are the same in both the open and closed string theories, but the latter requires twice more writing.
There are also a few subtleties which can be more easily explained once the general structure is understood.
Everything needed for the open string for this chapter can be found in \Cref{cft:chap:brst}: in fact, describing the open string (at this level) is equivalent to consider only the holomorphic sector of the CFT and to set $p_L = p$ (instead of $p/2$).
We consider a generic matter CFT in addition to the ghost system and we denote as $\mc H$ the space of states.
The open and closed string fields are denoted respectively by $\Phi$ and $\Psi$, such that it is clear which theory is studied.

An action can be either constructed from first principles, or it can be derived from the equations of motion.
Since the fundamental structure of string field theory is not (really) known, one needs to rely on the second approach.
But do we already know the (free) equations of motion for the string field?
The answer is yes.
But, before showing how these can be found from the worldsheet formalism, we will study the case of the point-particle to fix ideas and notations.

\subsection{Warm-up: point-particle}

The free (or linearized) equation of motion for a scalar particle reads:
\begin{equation}
	( - \lap + m^2) \phi(x)
		= 0.
\end{equation}
Solutions to this equation provides one-particle state of the free theory: a convenient basis is $\{ \e^{\I k x} \}$, where each state satisfies the on-shell condition
\begin{equation}
	k^2 = - m^2.
\end{equation}

The field $\phi(x)$ is decomposed on the basis as
\begin{equation}
	\phi(x)
		= \int \dd k \, \phi(k) \e^{\I k x},
\end{equation}
where $\phi(k)$ are the coefficients of the expansion.
Since the field is off-shell, the condition $k^2 = - m^2$ is not imposed.
Following \Cref{bsft:chap:string-field}, the field can also be represented as a ket:
\begin{equation}
	\phi(x)
		= \bracket{x}{\phi},
	\qquad
	\phi(k)
		= \bracket{k}{\phi},
\end{equation}
or, conversely:
\begin{equation}
	\ket{\phi}
		= \int \dd x \, \phi(x) \ket{x}
		= \int \dd k \, \phi(k) \ket{k}.
\end{equation}

Writing the kinetic operator as a kernel:
\begin{equation}
	K(x, x')
		:= \bra{x} K \ket{x'}
		= \delta(x - x') \, (- \lap_x + m^2),
\end{equation}
the equations of motion reads
\begin{equation}
	\int \dd x' \, K(x, x') \phi(x')
		= 0
	\quad \Longleftrightarrow \quad
	K \ket{\phi}
		= 0.
\end{equation}
An action can easily be found from the equation of motion by multiplying with $\phi(x)$ and integrating:
\begin{equation}
	S
		= \frac{1}{2} \int \dd x \, \phi(x) (- \lap + m^2) \phi(x)
		= \frac{1}{2} \int \dd x \dd x' \, \phi(x) K(x, x') \phi(x').
\end{equation}
It is straightforward to write the action in terms of the ket:
\begin{equation}
	S
		= \frac{1}{2} \, \bra{\phi} K \ket{\phi}.
\end{equation}

There is one hidden assumption in the previous lines: the definition of a scalar product.
A natural inner product is provided in the usual quantum mechanics by associating a bra to a ket.
Similarly, integration provides another definition of the inner product when working with functions.
We will find that the definition of the inner product requires more care in closed SFT.
To summarize, to write the kinetic term of the action, one needs the linearized equation of motion and an appropriate inner product on the space of states.

\subsection{Open string action}

The worldsheet equation which yields precisely all the string physical states $\ket{\psi}$ is the BRST condition:
\begin{equation}
	Q_B \ket{\psi}
		= 0.
\end{equation}
\index{string field!open bosonic}%
Considering the open string field $\Phi$ to be a linear combination of all possible one-string states $\ket{\psi}$
\begin{equation}
	\Phi \in \mc H,
\end{equation}
\index{free covariant SFT!open bosonic!equation of motion}%
the equation of motion is:
\begin{equation}
	\label{bsft:eq:open-eom}
	Q_B \ket{\Phi}
		= 0.
\end{equation}
Moving away from the physical state condition, the string field $\Phi$ is off-shell and is expanded on a general basis $\{ \phi_r \}$ of $\mc H$.
This presents a first difficulty because the worldsheet approach -- and the description of amplitudes -- looks ill-defined for off-shell states: extending the usual formalism will be the topic of \Cref{bos:chap:offshell}.
However, this is not necessary for the free theory and we can directly proceed.

\index{covariant SFT!open bosonic!inner product}%
Next, we need to find an inner product $\mean{\cdot, \cdot}$ on the Hilbert space $\mc H$.
A natural candidate is the BPZ inner product since it is not degenerate
\begin{equation}
	\mean{A, B}
		:= \bracket{A}{B},
\end{equation}
where $\bra{A} = \ket{A}^t$ is the BPZ conjugate \eqref{cft:eq:conj-bpz} of $\ket{A}$, using $I^-$.
This leads to the action:
\index{free covariant SFT!open bosonic!classical action}%
\begin{equation}
	\label{bsft:eq:open-action}
	S
		= \frac{1}{2} \, \mean{\Phi, Q_B \Phi}
		= \frac{1}{2} \, \bra{\Phi} Q_B \ket{\Phi}.
\end{equation}
Due to the definition of the BPZ product, the action is equivalent to a $2$-point correlation function on the disk.

The inner product satisfies the following identities:
\begin{equation}
	\label{bsft:eq:open-inner-identities}
	\mean{A, B}
		= (- 1)^{\abs{A} \abs{B}} \mean{B, A},
	\qquad
	\mean{Q_B A, B}
		= - (-1)^{\abs{A}} \mean{A, Q_B B},
\end{equation}
where $\abs{A}$ denotes the Grassmann parity of the operator $A$.

\index{string field!open bosonic!classical}%
A first consistency check is to verify that the ghost number of the string can be defined such that the action is not vanishing.
Indeed, the ghost number anomaly on the disk implies that the total ghost number must be $N_{\text{gh}} = 3$.
Since physical states have $N_{\text{gh}} = 1$, it is reasonable to take the string field to satisfy the same condition, even off-shell:
\begin{equation}
	 N_{\text{gh}}(\Phi)
		= 1.
\end{equation}
This condition means that there is no ghost at the classical level beyond the one of the energy vacuum $\ket{\downarrow}$, which has $N_{\text{gh}} = 1$.
Moreover, the BRST charge has $N_{\text{gh}}(Q_B) = 1$, such that the action has ghost number $3$.

\index{string field!open bosonic!parity}%
One needs to find the Grassmann parity of the string field.
Using the properties of the BPZ inner product, the string field should be Grassmann odd
\begin{equation}
	\label{bsft:eq:open-string-parity}
	\abs{\Phi}
		= 1
\end{equation}
for the action to be even.
This is in agreement with the fact that the string field has ghost number $1$ and that the ghosts are Grassmann odd.
\index{string field!open bosonic!reality condition}%
One must impose a reality condition on the string field (a complex field would behave like two real fields and have too many states).
The appropriate reality condition identifies the Euclidean and BPZ conjugates:
\begin{equation}
	\eadj{\ket{\Phi}}
		= \ket{\Phi}^t.
\end{equation}
That this relation is correct will be checked a posteriori for the tachyon field in \Cref{bsft:sec:free-brst:spacetime:class}.

\begin{computation}[bsft:eq:open-string-parity]
	\begin{align*}
		\mean{\Phi, Q_B \Phi}
			&= (- 1)^{\abs{\Phi} (\abs{Q_B \Phi})} \mean{Q_B \Phi, \Phi}
			= (- 1)^{\abs{\Phi} (1 + \abs{\Phi})} \mean{Q_B \Phi, \Phi}
			\\
			&= \mean{Q_B \Phi, \Phi}
			= - (- 1)^{\abs{\Phi}} \mean{\Phi, Q_B \Phi},
	\end{align*}
	where both properties \eqref{bsft:eq:open-inner-identities}, together with the fact that $\abs{\Phi} (1 + \abs{\Phi})$ is necessarily even.
	In order for the bracket to be non-zero, one must have $\abs{\Phi} = 1$.
\end{computation}

\index{free covariant SFT!open bosonic!zero-mode decomposition}%
Since the Hilbert space splits as $\mc H = \mc H_0 \oplus c_0 \mc H_0$ with $\mc H_0 = \mc H \cap \ker b_0$, see \eqref{cft:eq:Hilbert-split-open}, it is natural to split the field as (this is discussed further in \Cref{bsft:sec:free-brst:basis}):
\begin{equation}
	\label{bsft:eq:open-field-decomposition-b0-c0}
	\ket{\Phi}
		= \ket{\Phi_{\downarrow}} + c_0 \ket{\wtilde\Phi_{\downarrow}},
\end{equation}
where
\begin{equation}
	\Phi_{\downarrow}, \wtilde\Phi_{\downarrow}
		\in \mc H_0
	\quad \Longrightarrow \quad
	b_0 \ket{\Phi_{\downarrow}}
		= b_0 \ket{\wtilde\Phi_{\downarrow}}
		= 0.
\end{equation}
The ghost number of each component is
\begin{equation}
	N_{\text{gh}}(\Phi_{\downarrow})
		= 1,
	\qquad
	N_{\text{gh}}(\wtilde\Phi_{\downarrow})
		= 0.
\end{equation}
Remembering the decomposition \eqref{brst:eq:splitting-Q} of the BRST operator
\begin{equation}
	Q_B
		= c_0 L_0 - b_0 M + \what Q_B,
\end{equation}
inserting the decomposition \eqref{bsft:eq:open-field-decomposition-b0-c0} in the action \eqref{bsft:eq:open-action} gives:
\index{free covariant SFT!open bosonic!classical action}%
\begin{equation}
	\label{bsft:eq:open-action-components}
	S
		= \frac{1}{2} \bra{\Phi_{\downarrow}} c_0 L_0 \ket{\Phi_{\downarrow}}
			+ \frac{1}{2} \bra{\wtilde\Phi_{\downarrow}} c_0 M \ket{\wtilde\Phi_{\downarrow}}
			+ \bra{\wtilde\Phi_{\downarrow}} c_0 \what Q_B \ket{\Phi_{\downarrow}}.
\end{equation}
The equations of motion are obtained by varying the different fields:
\begin{equation}
	\label{bsft:eq:open-eom-components}
	0
		= - M \ket{\wtilde\Phi_\downarrow}
			+ \what Q_B \ket{\Phi_\downarrow},
	\qquad
	0
		= c_0 L_0 \ket{\Phi_\downarrow}
			+ c_0 \what Q_B \ket{\wtilde \Phi_\downarrow}.
\end{equation}

 \begin{computation}[bsft:eq:open-action-components]
	Let's introduce the projector $\Pi_s = b_0 c_0$ on the space $\mc H_0 = \mc H \cap \ker b_0$ and the orthogonal projector $\bar \Pi_s = c_0 b_0$ such that
	\begin{equation}
		\ket{\Phi}
			= \ket{\Phi_\downarrow} + \ket{\Phi_\uparrow},
		\qquad
		\ket{\Phi_\downarrow}
			= \Pi_s \ket{\Phi},
		\qquad
		\ket{\Phi_\uparrow}
			= \bar \Pi_s \ket{\Phi}.
	\end{equation}
	We then have:
	\begin{equation}
		\Pi_s Q_B \ket{\Phi}
			= - b_0 M \ket{\Phi_\uparrow}
				+ \what Q_B \ket{\Phi_\downarrow},
		\qquad
		\bar \Pi_s Q_B \ket{\Phi}
			= c_0 L_0 \ket{\Phi_\downarrow}
				+ \what Q_B \ket{\Phi_\uparrow},
	\end{equation}
	using
	\begin{equation}
		\com{\Pi_s}{\what Q_B}
			= \com{\Pi_s}{M}
			= \com{\Pi_s}{L_0}
			= 0.
	\end{equation}
	Then, we need the fact that $\adj{\Pi_s} = \bar \Pi_s$. to compute the action:
 	\begin{align*}
 	S
 		&
 		= \frac{1}{2} \, \mean{\Phi, Q_B \Phi}
 		\\ &
 		= \frac{1}{2} \, \mean{\Pi_s \Phi + \bar \Pi_s \Phi, Q_B \Phi}
 		\\ &
 		= \frac{1}{2} \, \mean{\Pi_s \Phi, \bar \Pi_s Q_B \Phi}
 			+ \frac{1}{2} \, \mean{\bar \Pi_s \Phi, \Pi_s Q_B \Phi}
 		\\ &
 		= \frac{1}{2} \, \mean{
 				\Phi_\downarrow,
 				c_0 L_0 \Phi_\downarrow + \what Q_B \Phi_\uparrow
 				}
 			+ \frac{1}{2} \, \mean{
 				\Phi_\uparrow,
 				- b_0 M \Phi_\uparrow + \what Q_B \Phi_\downarrow
				}
 		\\ &
 		= \frac{1}{2} \, \mean{\Phi_\downarrow, c_0 L_0 \Phi_\downarrow}
 			+ \frac{1}{2} \, \mean{\Phi_\downarrow, \what Q_B \Phi_\uparrow}
 			- \frac{1}{2} \, \mean{\Phi_\uparrow, b_0 M \Phi_\uparrow}
 			+ \frac{1}{2} \, \mean{\Phi_\uparrow, \what Q_B \Phi_\downarrow}.
 	\end{align*}
 	The result follows by setting $\ket{\Phi_\uparrow} = c_0 \ket{\wtilde{\Phi}}$, using \eqref{bsft:eq:open-inner-identities} and that the BPZ conjugate of $c_0$ is $- c_0$.
 \end{computation}

\subsection{Gauge invariance}

In writing the action, only the condition that the states are BRST closed has been used.
\index{free covariant SFT!open bosonic!gauge transformation}%
One needs to interpret the condition that the state are not BRST-exact, or phrased differently that two states differing by a BRST exact state are equivalent:
\begin{equation}
	\ket{\phi}
		\sim \ket{\psi} + Q_B \ket{\lambda}.
\end{equation}
Uplifting this condition to the string field, the most direct interpretation is that it corresponds to a gauge invariance:
\begin{equation}
	\label{bsft:eq:open-gauge-transf}
	\ket{\Phi}
	\longrightarrow
	\ket{\Phi'}
		= \ket{\Phi} + \delta_\Lambda \ket{\Phi},
	\qquad
	\delta_\Lambda \ket{\Phi}
		= Q_B \ket{\Lambda}
	\qquad
	N_{\text{gh}}(\Lambda)
		= 0.
\end{equation}
In order for the ghost numbers to match, the gauge parameter has vanishing ghost number.
The action \eqref{bsft:eq:open-action} is obviously invariant since the BRST charge is nilpotent.

\subsection{Siegel gauge}
\label{bsft:sec:free-brst:open:siegel}

In writing the action \eqref{bsft:eq:open-action}, the condition $b_0 \ket{\psi} = 0$ has \emph{not} been imposed on the string field.
In \Cref{bos:sec:ws-int:brst:states}, this condition was found by restricting the BRST cohomology, projecting out states built on the ghost vacuum $\ket{\uparrow}$, as required by the behaviour of the on-shell scattering amplitudes.
In \Cref{cft:chap:brst}, we obtained it by finding that the absolute cohomology contains twice more states as necessary.
This was also understood as a way to work with a specific representative of the BRST cohomology.
Since the field is off-shell and since the action computes off-shell Green functions, these arguments cannot be used, which explains why we did not use this condition earlier.

\index{Siegel gauge}%
On the other hand, the condition
\begin{equation}
	\label{bsft:eq:open-siegel-gauge}
	b_0 \ket{\Phi}
		= 0
\end{equation}
can be interpreted as a gauge fixing condition, called \emph{Siegel gauge}.
It can be reached from any field through a gauge transformation \eqref{bsft:eq:open-gauge-transf} with
\begin{equation}
	\label{bsft:eq:open-siegel-param}
	\ket{\Lambda}
		= - \Delta \ket{\Phi},
	\qquad
	\Delta
		= \frac{b_0}{L_0},
\end{equation}
where $\Delta$ was defined in \eqref{cft:eq:brst-prop} and will be identified with the propagator.
\index{propagator!open bosonic}%
Note that $b_0 = 0$ does not imply $L_0 = 0$ since the string field is not BRST closed.

This gauge choice is well-defined and completely fixes the gauge symmetry off-shell, meaning that no solution of the equation of motion is pure gauge after the gauge fixing.
This is shown as follows: assume that $\ket{\psi} = Q_B \ket{\chi}$ is an off-shell pure-gauge state with $L_0 \neq 0$, then, because it is also annihilated by $b_0$, one finds:
\begin{equation}
	0
		= \anticom{Q_B}{b_0} \ket{\psi}
		= L_0 \ket{\psi}
\end{equation}
which yields a contradiction.

The gauge fixing condition breaks down for $L_0 = 0$, but this does not pose any problem when working with Feynman diagrams since they are not physical by themselves (nor are the off-shell and on-shell Green functions).
Only the sum giving the scattering amplitudes (truncated on-shell Green functions) is physical: in this case, the singularity $L_0 = 0$ corresponds to the on-shell condition and it is well-known how such infrared divergences for intermediate states are removed (through the LSZ prescription, mass renormalization and tadpole cancellation).
\begin{draft}
These problems will be analysed on their own later.
\end{draft}

\begin{computation}[bsft:eq:open-siegel-param]
	Performing a gauge transformation gives
	\begin{equation}
		b_0 \ket{\Phi'}
			= b_0 \ket{\Phi} + b_0 Q_B \ket{\Lambda}
			= 0.
	\end{equation}
	Then, one writes
	\begin{equation}
		b_0 \ket{\Phi}
			= b_0 \anticom{Q_B}{\Delta} \ket{\Phi}
			= b_0 Q_B \Delta \ket{\Phi},
	\end{equation}
	using the relation \eqref{cft:eq:com-Q-D}, the expression \eqref{cft:eq:brst-prop} and the fact that $b_0^2 = 0$.
	Plugging this back in the first equation gives:
	\begin{equation}
		b_0 Q_B \left( \Delta \ket{\Phi} + \ket{\Lambda} \right)
			= 0.
	\end{equation}
	The factor of $b_0$ can be removed by multiplying with $c_0$, and the parenthesis should vanish (since it is not identically closed), which means that \eqref{bsft:eq:open-siegel-param} holds up to a BRST exact state.
\end{computation}

\begin{example}[Gauge fixing and singularity]
	In Maxwell theory, the gauge transformation
	\begin{equation}
		A'_\mu
			= A_\mu + \pd_\mu \lambda,
	\end{equation}
	is used to impose the Lorentz condition
	\begin{equation}
		\pd^\mu A'_\mu
			= 0
		\quad \Longrightarrow \quad
		\lap \lambda
			= - \pd^\mu A_\mu.
	\end{equation}
	In momentum space, the parameter reads
	\begin{equation}
		\lambda
			= - \frac{k^\mu}{k^2} \, A_\mu.
	\end{equation}
	It is singular when $k$ is on-shell, $k^2 = 0$.
	However, this does not prevent from computing Feynman diagrams.
\end{example}

To understand the effect of the gauge fixing on the string field components, decompose the field as \eqref{bsft:eq:open-field-decomposition-b0-c0} $\ket{\Phi} = \ket{\Phi_{\downarrow}} + c_0 \ket{\wtilde\Phi_{\downarrow}}$.
Then, imposing the condition \eqref{bsft:eq:open-siegel-gauge} yields
\begin{equation}
	\ket{\wtilde\Phi_{\downarrow}}
		= 0
	\quad \Longrightarrow \quad
	\ket{\Phi}
		= \ket{\Phi_{\downarrow}}.
\end{equation}
This has the expected effect of dividing by two the number of states and show that they are not physical.

\index{free covariant SFT!open bosonic!gauge fixed action}%
Plugging this condition in the action \eqref{bsft:eq:open-action-components} leads to gauge fixed action
\begin{equation}
	\label{bsft:eq:open-action-siegel}
	S
		= \frac{1}{2} \, \bra{\Phi} c_0 L_0 \ket{\Phi},
\end{equation}
\index{free covariant SFT!closed bosonic!gauge fixed equation of motion}%
for which the equation of motion is
\begin{equation}
	\label{bsft:eq:open-eom-siegel}
	L_0 \ket{\Phi}
		= 0.
\end{equation}
But, note that this equation contains much less information than the original \eqref{bsft:eq:open-eom}: as $\ket{\wtilde\Phi_{\downarrow}}$ is truncated from \eqref{bsft:eq:open-action-siegel}, a part of the equations of motion is lost.
The missing equation can be found by setting $\ket{\wtilde \Phi} = 0$ in \eqref{bsft:eq:open-eom-components} and must be imposed on top of the action:
\index{out-of-Siegel gauge constraint}%
\begin{equation}
	\label{bsft:eq:open-constraints-siegel}
	\what Q_B \ket{\Phi}
		= 0.
\end{equation}
It is called \emph{out-of-Siegel gauge constraint} and is equivalent to the Gauss constraint in electromagnetism: the equations of motion for pure gauge states contain also the physical fields, thus, when one fixes a gauge, these relations are lost and must be imposed on the side of the action.
This procedure mimics what happens in the old covariant theory, where the Virasoro constraints are imposed after choosing the flat gauge (if $\Phi$ contains no ghost on top of $\ket{\downarrow}$, then $\what Q_B = 0$ implies $L_n = 0$, see \Cref{cft:sec:brst:lc:abs}).
Moreover, the states which do not satisfy the condition $b_0 = 0$ do not propagate: this restricts the external states to be considered in amplitudes.

\begin{remark}
	Another way to derive \eqref{bsft:eq:open-action-siegel} is to insert $\anticom{b_0}{c_0} = 1$ in the action:
	\begin{align*}
		S
			&= \frac{1}{2} \, \bra{\Phi} Q_B \anticom{c_0}{b_0} \ket{\Phi}
			= \frac{1}{2} \, \bra{\Phi} Q_B b_0 c_0 \ket{\Phi}
			\\
			&= \frac{1}{2} \, \bra{\Phi} \anticom{b_0}{Q_B} c_0 \ket{\Phi}
				- \frac{1}{2} \, \bra{\Phi} b_0 Q_B c_0 \ket{\Phi}
			\\
			&= \frac{1}{2} \, \bra{\Phi} c_0 L_0 \ket{\Phi}.
	\end{align*}
	The drawback of this computation is that it does not show directly how the constraints \eqref{bsft:eq:open-constraints-siegel} arise.
\end{remark}

\begin{remark}[Generalized gauge fixing]
	It is possible to generalize the Siegel gauge, in the same way that the Feynman gauge generalizes the Lorentz gauge.
	This has been studied in~\cite{Asano:2007:NewCovariantGauges, Asano:2009:GeneralLinearGauges}.
\end{remark}

In this section, we have motivated different properties and adopted some normalizations.
The simplest way to check that they are consistent is to derive the action in terms of the spacetime fields and to check that it has the expected properties from standard QFT.
This will be the topic of \Cref{bsft:sec:free-brst:spacetime}.

\section{Open string field expansion, parity and ghost number}
\label{bsft:sec:free-brst:basis}

\index{string field!open bosonic!expansion}%
A basis for the off-shell Hilbert space $\mc H$ is denoted by $\{ \phi_r \}$, where the ghost numbers and parity of the states are written as:
\begin{equation}
	n_r
		:= N_{\text{gh}}(\phi_r),
	\qquad
	\abs{\phi_r}
		= n_r \mod 2.
\end{equation}
The corresponding basis of dual (or conjugate) states $\{ \phi_r^c \}$ is defined by \eqref{cft:eq:bpz-conj-state}:
\begin{equation}
	\bracket{\phi_r^c}{\phi_s}
		= \delta_{rs}.
\end{equation}
The basis states can be decomposed according to the ghost zero-modes
\begin{equation}
	\label{bsft:eq:basis-decomposition-b0-c0}
	\ket{\phi_r}
		= \ket{\phi_{\downarrow, r}} + \ket{\phi_{\uparrow, r}},
	\qquad
	b_0 \ket{\phi_{\downarrow, r}}
		= c_0 \ket{\phi_{\uparrow, r}}
		= 0.
\end{equation}
Finally, each state $\psi_{\uparrow} \in c_0 \mc H$ can be associated to a state $\wtilde\psi$:
\begin{equation}
	\ket{\psi_{\uparrow}}
		= c_0 \ket{\wtilde\psi_{\downarrow}},
	\qquad
	b_0 \ket{\wtilde\psi_{\downarrow}}
		= 0,
	\qquad
	N_{\text{gh}}(\psi_{\uparrow})
		= N_{\text{gh}}(\wtilde\psi_{\downarrow}) + 1.
\end{equation}
More details can be found in \Cref{bos:sec:offshell:states}.

Any field $\Phi$ can be expanded as
\begin{equation}
	\label{bsft:eq:open-basis-decomposition}
	\ket{\Phi}
		= \sum_r \psi_r \ket{\phi_r},
\end{equation}
where the $\psi_r$ are spacetime fields (remembering that $r$ denotes collectively the continuous and discrete quantum numbers).\footnotemark{}
\footnotetext{%
	The notation is slightly ambiguous: from \eqref{bsft:eq:basis-decomposition-b0-c0}, it looks like both components of $\phi_r$ have the same coefficient $\psi_r$.
	But, in fact, one sums over all linearly independent states: in terms of the components of $\phi_r$, different basis can be considered; for example $\{ \phi_{\downarrow, r}, \phi_{\uparrow, r} \}$, or $\{ \phi_{\downarrow, r} \pm \phi_{\uparrow, r} \}$.
	A more precise expression can be found in \eqref{bsft:eq:field-decomposition-b0-c0} and \eqref{bsft:eq:open-basis-exp-ud}.
}%

Obviously, the coefficients do not carry a ghost number since they are not worldsheet operators.
However, they can be Grassmann even or odd such that each term of the sum has the same parity, so that the field has a definite parity:
\begin{equation}
	\forall r:
	\qquad
	\abs{\Phi}
		= \abs{\psi_r} \, \abs{\phi_r}.
\end{equation}
If the field is Grassmann odd (resp.\ even) then the coefficients $\psi_r$ and the basis states must have opposite (resp.\ identical) parities, such that $\abs{\Phi} = 1$.

\index{spacetime ghost number!open string}%
Since the parity results from worldsheet ghosts and since there would be Grassmann odd states even in a purely bosonic theory, it suggests that the parity of the coefficients $\psi_r$ is also related to a \emph{spacetime ghost number} $G$ defined as:
\begin{equation}
	G(\psi_r)
		= 1 - n_r.
\end{equation}
The normalization is chosen such that the component of a classical string field ($N_{\text{gh}} = 1$) are classical spacetime fields with $G = 0$ (no ghost).
We will see later that this definition makes sense.

\index{string field!open bosonic!quantum}%
A quantum string field $\Phi$ generally contains components $\Phi_n$ of all worldsheet ghost numbers $n$:
\begin{equation}
	\Phi
		= \sum_{n \in \Z} \Phi_n,
	\qquad
	N_{\text{gh}}(\Phi_n)
		= n.
\end{equation}
The projections on the positive and negative (cylinder) ghost numbers are denoted by $\Phi_{\pm}$:
\begin{equation}
	\label{bsft:eq:projection-pos-neg-Ngh}
	\Phi
		= \Phi_+ + \Phi_-,
	\qquad
	\Phi_+
		= \sum_{n > 1} \Phi_n,
	\qquad
	\Phi_-
		= \sum_{n \le 1} \Phi_n.
\end{equation}
The shift in the indices is explained by the relation \eqref{app:eq:ghost-number-cyl} between the cylinder and plane ghost numbers.

For a field $\Phi_n$ of fixed ghost number, coefficients of the expansion vanish whenever the ghost number of the basis state does not match the one of the field:
\begin{equation}
	\forall n_r \neq n:
	\quad
	\psi_r
		= 0.
\end{equation}
Another possibility to define the field $\Phi_n$ is to insert a delta function:
\begin{equation}
	\ket{\Phi_n}
		= \delta(N_{\text{gh}} - n) \ket{\Psi}
		= \sum_r \delta(n_r - n) \, \psi_r \ket{\phi_r}.
\end{equation}

According to \eqref{bsft:eq:basis-decomposition-b0-c0}, a string field $\Phi$ can also be separated in terms of the ghost zero-modes:
\begin{subequations}
\label{bsft:eq:field-decomposition-b0-c0}
\begin{gather}
	\ket{\Phi}
		= \ket{\Phi_{\downarrow}} + \ket{\Phi_{\uparrow}}
		= \ket{\Phi_{\downarrow}} + c_0 \ket{\wtilde\Phi_{\downarrow}},
	\\
	\ket{\Phi_{\uparrow}}
		= c_0 \ket{\wtilde\Phi_{\downarrow}},
	\qquad
	\ket{\wtilde\Phi_{\downarrow}}
		= b_0 \ket{\Phi_{\uparrow}},
\end{gather}
\end{subequations}
where the components satisfy the constraints
\begin{equation}
	b_0 \ket{\Phi_{\downarrow}}
		= 0,
	\qquad
	c_0 \ket{\Phi_{\uparrow}}
		= 0,
	\qquad
	b_0 \ket{\wtilde\Phi_{\downarrow}}
		= 0.
\end{equation}
The fields $\ket{\Phi_{\downarrow}}$ and $\ket{\Phi_{\uparrow}}$ (or $\ket{\wtilde\Phi_{\downarrow}}$) are called the down and top components and they can be expanded as:
\begin{equation}
	\label{bsft:eq:open-basis-exp-ud}
	\ket{\Phi_{\downarrow}}
		= \sum_r \psi_{\downarrow, r} \ket{\phi_{\downarrow, r}},
	\qquad
	\ket{\Phi_{\uparrow}}
		= \sum_r \psi_{\uparrow, r} \ket{\phi_{\uparrow, r}}.
\end{equation}

\section{Path integral quantization}
\label{bsft:sec:free-brst:path-integral}

\index{string field path integral!free covariant open bosonic string}%
The string field theory can be quantized with a path integral:
\begin{equation}
	Z
		= \int \dd \Phi_{\text{cl}} \, \e^{- S[\Phi_{\text{cl}}]}
		= \int \dd \Phi_{\text{cl}} \, \e^{- \frac{1}{2} \bra{\Phi_{\text{cl}}} Q_B \ket{\Phi_{\text{cl}}}}.
\end{equation}
An index has been added to the field to emphasize that it is the classical field (no spacetime ghosts).
The simplest way to define the measure is to use the expansion \eqref{bsft:eq:string-field-expansion} such that
\begin{equation}
	Z
		= \int \prod_s \dd \psi_s \, \e^{- S[\{\psi_r\}]}.
\end{equation}

\subsection{Tentative Faddeev--Popov gauge fixing}

\index{string field path integral!Faddeev--Popov gauge fixing}%
The action can be gauge fixed using the Faddeev--Popov formalism.
The gauge fixing condition is
\begin{equation}
	F(\Phi_{\text{cl}})
		:= b_0 \ket{\Phi_{\text{cl}}}
		= 0.
\end{equation}
Its variation under a gauge transformation \eqref{bsft:eq:open-gauge-transf} reads
\begin{equation}
	\delta F
		= b_0 Q_B \ket{\Lambda_{\text{cl}}},
\end{equation}
which implies that the Faddeev--Popov determinant is
\begin{equation}
	\det \frac{\delta F}{\delta \Lambda_{\text{cl}}}
		= \det b_0 Q_B.
\end{equation}
This determinant is rewritten as a path integral by introducing a ghost $C$ and an antighost $B'$ string fields (the prime on $B'$ will become clear below):
\begin{equation}
	\label{bsft:eq:free-open-FP-1}
	\det b_0 Q_B
		= \int \dd B' \dd C \, \e^{- S_{\text{FP}}},
	\qquad
	S_{\text{FP}}
		= - \bra{B'} b_0 Q_B \ket{C}.
\end{equation}
The ghost numbers are attributed by selecting the same ghost number for the $C$ ghost and for the gauge parameter, and then requiring that the Faddeev--Popov action is non-vanishing:
\begin{equation}
	N_{\text{gh}}(B')
		= 3,
	\qquad
	N_{\text{gh}}(C)
		= 0.
\end{equation}
The ghosts can be expanded as
\begin{equation}
	\ket{B'}
		= \delta(N_{\text{gh}} - 3) \sum_r b'_r \ket{\phi_r},
	\qquad
	\ket{C}
		= \delta(N_{\text{gh}}) \sum_r c_r \ket{\phi_r},
\end{equation}
where the coefficients $b_r$ and $c_r$ are Grassmann odd in order for the determinant formula to make sense:
\begin{equation}
	\abs{b_r}
		= \abs{c_r}
		= 1.
\end{equation}
Then, since the basis states appearing in $B'$ and $C$ are respectively odd and even, this implies
\begin{equation}
	\abs{B'}
		= 0,
	\qquad
	\abs{C}
		= 1.
\end{equation}

However, there is a redundancy in the gauge fixing because the Faddeev--Popov action is itself invariant under two independent transformations:
\begin{subequations}
\begin{gather}
	\label{bsft:eq:open-transf-C}
	\delta \ket{C}
		= Q_B \ket{\Lambda_{-1}},
	\qquad
	N_{\text{gh}}(\Lambda_{-1})
		= - 1,
	\\
	\label{bsft:eq:open-transf-Bp}
	\delta \ket{B'}
		= b_0 \ket{\Lambda'},
	\qquad
	N_{\text{gh}}(\Lambda')
		= 4.
\end{gather}
\end{subequations}
This residual invariance arises because not all $\ket{\Lambda_{\text{cl}}}$ generate a gauge transformation.
Indeed, if
\begin{equation}
	\ket{\Lambda}
		= \ket{\Lambda_0} + Q_B \ket{\Lambda_{-1}},
\end{equation}
the field transforms as
\begin{equation}
	\ket{\Phi'_{\text{cl}}}
	\longrightarrow
	\ket{\Phi_{\text{cl}}} + Q_B \ket{\Lambda_0}
\end{equation}
and there is no trace left of $\ket{\Lambda_{-1}}$, so it should not be counted.

The second invariance \eqref{bsft:eq:open-transf-Bp} is not problematic because $b_0$ is an algebraic operator (the Faddeev--Popov action associated to the determinant has no dynamics).
The decompositions of the gauge parameter $\Lambda'$ and the $B'$ field into components \eqref{bsft:eq:field-decomposition-b0-c0} read:
\begin{subequations}
\label{bsft:eq:free-Bp-decomp}
\begin{align}
	\ket{B'}
		&
		= \ket{B'_{\downarrow}} + c_0 \ket{B},
	\qquad
	\ket{B}
		:= \ket{\wtilde B'_{\downarrow}},
	\\
	\ket{\Lambda'}
		&
		= \ket{\Lambda'_{\downarrow}} + c_0 \ket{\wtilde\Lambda'_{\downarrow}}.
\end{align}
\end{subequations}
The gauge transformations act on the components as:
\begin{equation}
	\label{bsft:eq:open-transf-Bp-comp}
	\delta \ket{B'_{\downarrow}}
		= \ket{\wtilde\Lambda'_{\downarrow}},
	\qquad
	\delta \ket{B}
		= 0.
\end{equation}
This shows that $B$ is gauge invariant and $B'_{\downarrow}$ can be completely removed by the gauge transformation.
This makes sense because $B'_{\downarrow}$ does not appear in the action \eqref{bsft:eq:free-open-FP-1}.
The gauge transformation \eqref{bsft:eq:open-transf-Bp} can be used to fix the gauge:
\begin{equation}
	\ket{F'}
		= c_0 \ket{B'}
		= 0
	\quad \Longrightarrow \quad
	\ket{B'_{\downarrow}}
		= 0.
\end{equation}
This fixes completely the gauge invariance since the field $B$ is restricted to satisfy $b_0 \ket{B} = 0$, and the component form \eqref{bsft:eq:open-transf-Bp-comp} of the gauge transformation shows that no transformation is allowed.
Moreover, there is no need to introduce a Faddeev--Popov determinant for this gauge fixing because the corresponding ghosts would not couple to the other fields (and this would continue to hold even in the presence of interactions, see \Cref{bsft:rem:decoupling-ghosts}).
Indeed, from the absence of derivatives in the gauge transformation, one finds that the determinant is constant and thus a ghost-representation is not necessary:
\begin{equation}
	\det \frac{\delta F'}{\delta \Lambda'}
		= \det c_0 b_0
		= \det c_0 \det b_0
		= \frac{1}{2} \, \det \anticom{b_0}{c_0}
		= \frac{1}{2}.
\end{equation}
Then, redefining the measure, the partition function and action reduce to
\begin{equation}
	\label{bsft:eq:open-action-ghosts-02}
	\Delta_{\text{FP}}
		= \int \dd B \, \dd C \, \e^{- S_{\text{FP}}[B, C]},
	\qquad
	S_{\text{FP}}
		= \bra{B} Q_B \ket{C}.
\end{equation}
Note that the field $B$ satisfies
\begin{equation}
	b_0 \ket{B} = 0,
	\qquad
	N_{\text{gh}}(B) = 2,
	\qquad
	\abs{B} = 1.
\end{equation}
Since both fields are Grassmann odd, the action can be rewritten in a symmetric way:
\begin{equation}
	S_{\text{FP}}
		= \frac{1}{2} \Big( \bra{B} Q_B \ket{C} + \bra{C} Q_B \ket{B} \Big).
\end{equation}

\begin{remark}[Ghost and anti-ghost definitions]
	The definition of the anti-ghost $B$ and ghost $C$ is appropriate because the worldsheet and spacetime ghost numbers are related by a minus sign (and a shift of one unit).
	In the BV formalism, we will see that the fields contain the matter and ghost fields, while the antifields contain the anti-ghosts.
	These two sets are respectively defined with $N_{\text{gh}} \le 1$ and $N_{\text{gh}} > 1$.
\end{remark}

The constraint $b_0 \ket{B} = 0$ can be lifted by adding a top component:
\begin{equation}
	\ket{B} = \ket{B_{\downarrow}} + c_0 \ket{\wtilde B_{\downarrow}}
\end{equation}
together with the gauge invariance
\begin{equation}
	\label{bsft:eq:open-transf-B}
	\delta \ket{B} = Q_B \ket{\Lambda_1}.
\end{equation}
Note the difference with \eqref{bsft:eq:free-Bp-decomp}: while $B = \wtilde B'_{\downarrow}$ was the top component of the $B'$ field, here, it is defined to be the down component, such that $\ket{B_{\downarrow}} = \ket{\wtilde B'_{\downarrow}}$.
However, for the moment, we keep $B$ to satisfy $b_0 \ket{B} = 0$.

\begin{remark}[Decoupling of the ghosts]
	\label{bsft:rem:decoupling-ghosts}

	Since the theory is free the Faddeev--Popov action \eqref{bsft:eq:open-action-ghosts-02} could be ignored and absorbed in the normalization because it does not couple to the field.
	On the other hand, when interactions are included, the gauge transformation is modified and the ghosts couple to the matter fields.
	But this is true only for the $C$ transformation \eqref{bsft:eq:open-transf-C}, not for \eqref{bsft:eq:open-transf-Bp}.
	Then it means that ghosts introduced for gauge fixing \eqref{bsft:eq:open-transf-Bp} will never couple to the matter and other ghosts.
\end{remark}

The invariance \eqref{bsft:eq:open-transf-C} is a gauge invariance for $C$ and must be treated in the same way as \eqref{bsft:eq:open-gauge-transf}.
Then, following the Faddeev--Popov procedure, one is lead to introduce new ghosts for the ghosts.
But, the same structure appears again.
This leads to a residual gauge invariance, which has the same form.
This process continues recursively and one finds an infinite tower of ghosts.

\subsection{Tower of ghosts}

In order to simplify the notations, all the fields are denoted by $\Phi_n$ where $n$ gives the ghost number:
\begin{itemize}
	\item $\Phi_1 := \Phi_{\text{cl}}$ is the original physical field
	\item $\Phi_0 := C$ and, more generally, $\Phi_n$ with $n < 1$ are ghosts
	\item $\Phi_2 := B$ and, more generally, $\Phi_n$ with $n > 1$ are anti-ghosts
\end{itemize}
The recipe is that each pair of ghost fields $(\Phi_{n+2}, \Phi_{-n})$ is associated to a gauge parameter $\Lambda_{-n-1}$ with $n \ge 0$.
It is then natural to gather all the fields in a single field
\begin{equation}
	\ket{\Phi}
		= \sum_n \ket{\Phi_n}
\end{equation}
satisfying the gauge fixing constraint:
\begin{equation}
	b_0 \ket{\Phi}
		= 0
	\quad \Longrightarrow \quad
	b_0 \ket{\Phi_n}
		= 0.
\end{equation}
For $n \le 1$, these constraints are gauge fixing conditions for the invariance $\delta \ket{\Phi_n} = Q_B \Lambda_{n}$.
For $n > 1$, they arise by considering only the top component of the $B$ field.

\index{free covariant SFT!open bosonic!BV action}%
Finally, the gauge fixing condition can be incorporated inside the action by using a Lagrange multiplier $\beta$, which is an auxiliary string field containing also components of all ghost numbers:
\begin{equation}
	\ket{\beta} = \sum_{n \in \Z} \ket{\beta_n}.
\end{equation}
The path integral then reads
\begin{equation}
	Z = \int \dd \Phi \dd \beta \, \e^{- S[\Phi, \beta]},
\end{equation}
where
\begin{subequations}
\label{bsft:eq:open-quantum-action-beta}
\begin{align}
	S[\Phi, \beta]
		&
		= \frac{1}{2} \bra{\Phi} Q_B \ket{\Phi} + \bra{\beta} b_0 \ket{\Phi}
		\\ &
		= \sum_{n \in \Z} \left(
			\frac{1}{2} \bra{\Phi_{2-n}} Q_B \ket{\Phi_{n}}
			+ \bra{\beta_{4-n}} b_0 \ket{\Phi_{n}} \right).
\end{align}
\end{subequations}
The first term of the action has the same form as the classical action \eqref{bsft:eq:open-action}, but now includes fields at every ghost number.
The complete BV analysis is relegated to the interacting theory.

\begin{draft}
The action \eqref{bsft:eq:open-quantum-action-beta} is invariant under two BRST symmetries with parameters $\eta$ and $\eta'$:
\begin{equation}
	\delta \Phi_{+}
		= \eta (b_0 \Phi)_{+} + \eta' (Q_B \Phi)_{+},
	\qquad
	\delta \Phi_{-}
		= \eta (Q_B \Phi)_{-} + \eta' (b_0 \Phi)_{-},
	\qquad
	\delta \beta
		= 0,
\end{equation}
where the signs denote the projection \eqref{bsft:eq:projection-pos-neg-Ngh} on the positive and negative ghost numbers.
\end{draft}

Removing the auxiliary field $\beta = 0$, one finds that the action is invariant
under the extended gauge transformation
\index{free covariant SFT!open bosonic!gauge transformation}%
\begin{equation}
	\delta \ket{\Phi}
		= Q_B \ket{\Lambda},
\end{equation}
where the gauge parameter has also components of all ghost numbers:
\begin{equation}
	\ket{\Lambda}
		= \sum_{n \in \Z} \ket{\Lambda_n}.
\end{equation}

\section{Spacetime action}
\label{bsft:sec:free-brst:spacetime}

In order to make the string field action more concrete, and as emphasized in \Cref{bsft:chap:string-field}, it is useful to expand the string field in spacetime fields and to write the action for the lowest modes.
This also helps to check that the normalization chosen until here correctly reproduces the standard QFT normalizations.
For simplicity we focus on the open bosonic string in $D = 26$.

\begin{check}
\subsection{Classical action}
\end{check}
\label{bsft:sec:free-brst:spacetime:class}

\index{spacetime level-truncated action!open bosonic}%
We build the string from the vacuum $\ket{k, \downarrow}$ (\Cref{cft:chap:brst}) by acting with the ghost positive-frequency modes $b_{-n}$ and $c_{-n}$, the zero-mode $c_0$, and from the scalar oscillators $\I \alpha_{-n}^\mu$.

Up to level $\ell = 1$, the classical open string field can be expanded as
\begin{equation}
	\ket{\Phi}
		= \frac{1}{\sqrt{\alpha'}} \int \frac{\dd^{D} k}{(2\pi)^{D}}
			\Big(
				T(k)
				+ A_\mu(k) \alpha^\mu_{-1}
				+ \I \sqrt{\frac{\alpha'}{2}} \, B(k) b_{-1} c_0
				+ \cdots
				\Big) \ket{k, \downarrow}
\end{equation}
before gauge fixing.
The spacetime fields are $T(k)$, $A_\mu(k)$ and $B(k)$: their roles will be interpreted below.
The first two terms are part of the $\ket{\Phi_{\downarrow}}$ component, while the last term is part of the $\ket{\Phi_{\uparrow}}$ component.
All terms are correctly Grassmann even and they have vanishing spacetime ghost numbers.
The normalizations are chosen in order to retrieve the canonical normalization in QFT.
The factor of $\I$ in front of $B$ is needed for the field $B$ to be real (as can be seen below, this leads to the expected factor $\I k_\mu$ which maps to $\pd_\mu$ in position space).

The equation \eqref{bsft:eq:open-eom} leads to the following equations of motion of the spacetime fields:
\begin{equation}
	\label{bsft:eq:open-eom-TAB}
	\begin{gathered}
	(\alpha' k^2 - 1) T(k)
		= 0,
	\qquad
	k^2 A_\mu(k) + \I k_\mu B(k)
		= 0,
	\\
	k^\mu A_\mu(k) + \I B(k)
		= 0.
	\end{gathered}
\end{equation}
Moreover, plugging the last equation into the second one gives
\begin{equation}
	k^2 A_\mu(k) - k_\mu k \cdot A(k)
		= 0.
\end{equation}
After Fourier transformation, the equations in position space read:
\begin{equation}
	\left(\alpha' \lap + 1 \right) T
		= 0,
	\qquad
	B
		= \pd^\mu A_\mu,
	\qquad
	\lap A_\mu
		= \pd_\mu B.
\end{equation}
This shows that $T(k)$ is a tachyon with mass $m^2 = - 1 / \alpha'$ and $A_\mu(k)$ is a massless gauge field.
\index{string field!open bosonic!Nakanishi--Lautrup auxiliary field}%
The field $B(k)$ is the Nakanishi--Lautrup auxiliary field: it is completely fixed once $A_\mu$ is known since its equation has no derivative.
\index{Siegel gauge}%
Siegel gauge imposes $B = 0$ which shows that it generalizes the Feynman gauge to the string field.

\begin{computation}[bsft:eq:open-eom-TAB]
	Keeping only the levels $0$ and $1$ terms in the string field, it is sufficient to truncate the BRST operator as
	\begin{equation}
		\begin{gathered}
			Q_B
				= c_0 L_0 - b_0 M + \what Q_B,
			\\
			M
				\sim 2 c_{-1} c_{1},
			\qquad
			\what Q_B
				\sim c_{1} L^m_{-1} + c_{-1} L^m_{1},
			\\
			L^m_{1}
				\sim \alpha_0 \cdot \alpha_{1},
			\qquad
			L^m_{-1}
				\sim \alpha_0 \cdot \alpha_{-1}.
		\end{gathered}
	\end{equation}
	Acting on the string field gives
	\begin{align*}
		Q_B \ket{\Phi}
			&= \frac{1}{\sqrt{\alpha'}} \int \frac{\dd^D k}{(2\pi)^{D}}
				\bigg( T(k) c_0 L_0 \ket{k, \downarrow}
					+ A_\mu(k) \Big(
						c_0 L_0
						+ \eta_{\nu\rho} c_{-1} \alpha^\nu_{1} \alpha^\rho_0
						\Big) \alpha^\mu_{-1} \ket{k, \downarrow}
					\\
					&\hspace{3cm}
					+ \I \sqrt{\frac{\alpha'}{2}} \, B(k) \Big(
						- 2 b_0 c_{-1} c_{1}
						+ \eta_{\nu\rho} c_{1} \alpha^\nu_{-1} \alpha^\rho_0
						\Big) b_{-1} c_0 \ket{k, \downarrow}
					\bigg)
			\\
			&= \frac{1}{\sqrt{\alpha'}} \int \frac{\dd^D k}{(2\pi)^{D}}
				\bigg(
					T(k) (\alpha' k^2 - 1) c_0 \ket{k, \downarrow}
					\\
					&\hspace{3cm}
					+ A_\mu(k) \left(
						\alpha' k^2 c_0 \alpha^{\mu}_{-1}
						+ \sqrt{2 \alpha'} \eta_{\nu\rho} \eta^{\mu\nu} \, k^\rho \, c_{-1}
						\right) \ket{k, \downarrow}
					\\
					&\hspace{3cm}
					+ \I \sqrt{\frac{\alpha'}{2}} \, B(k) \Big(
						2 c_{-1}
						+ \sqrt{2 \alpha'} \eta_{\nu\rho} k^\rho \alpha^\nu_{-1} c_0
						\Big) \ket{k, \downarrow}
					\bigg)
			\\
			&= \frac{1}{\sqrt{\alpha'}} \int \frac{\dd^D k}{(2\pi)^{D}}
				\bigg(
					T(k) (\alpha' k^2 - 1) c_0 \ket{k, \downarrow}
					\\
					&\hspace{3cm}
					+ \alpha' \Big(
						A_\mu(k) k^2
						+ \I k^\mu B(k
						\Big) c_0 \alpha^{\mu}_{-1} \ket{k, \downarrow}
					\\
					&\hspace{3cm}
					+ \sqrt{2 \alpha'} \Big(
						k^\mu A_\mu(k)
						+ \I B(k)
						\Big) c_{-1} \ket{k, \downarrow}
					\bigg).
	\end{align*}
	One needs to be careful when anticommuting the ghosts and we used that $p_L = k$ and $\alpha_0 = \sqrt{2 \alpha'} k$ for the open string.
	It remains to require that the coefficient of each state vanishes.
\end{computation}

\index{spacetime level-truncated action!open bosonic!gauge transformation}%
In order to confirm that $A_\mu$ is indeed a gauge field, we must study the gauge transformation.
The gauge parameter is expanded at the first level:
\begin{equation}
	\ket{\Lambda}
		= \frac{\I}{\sqrt{2} \alpha'} \int \frac{\dd^D k}{(2\pi)^{D}}
			\big( \lambda(k) \, b_{-1} \ket{k, \downarrow}
				+ \cdots
				\big).
\end{equation}
Note that $b_{-1} \ket{\downarrow}$ is the $\group{SL}(2, \C)$ ghost vacuum.
Since
\begin{equation}
	Q_B \ket{\Lambda}
		= \frac{\I}{\sqrt{\alpha'}} \int \frac{\dd^D k}{(2\pi)^D} \,
			\lambda(k) \left(
			- \sqrt{\frac{\alpha'}{2}} \, k^2 \, b_{-1} c_0
			+ k_\mu \alpha^\mu_{-1}
			\right) \ket{k, \downarrow},
\end{equation}
matching the coefficients in \eqref{bsft:eq:open-gauge-transf} gives
\begin{equation}
	\delta A_\mu
		= - \I k_\mu \lambda,
	\qquad
	\delta B
		= k^2 \lambda.
\end{equation}
This is the appropriate transformation for a $\group{U}(1)$ gauge field.

Finally, one can derive the action; for simplicity, we work in the Siegel gauge.
We consider only the tachyon component:
\begin{equation}
	\ket{T}
		= \int \frac{\dd^D k}{(2\pi)^D} \, T(k) c_{1} \ket{k, 0},
\end{equation}
with $c_1 \ket{0} = \ket{\downarrow}$.
The BPZ conjugate and Hermitian conjugates are respectively:
\begin{subequations}
\begin{align}
	\bra{T}
		&
		= \int \frac{\dd^D k}{(2\pi)^D} \, T(k) \bra{- k, 0} c_{-1},
	\\
	\bra{\eadj{T}}
		&
		= \int \frac{\dd^D k}{(2\pi)^D} \, \conj{T(k)} \bra{k, 0} c_{-1}.
\end{align}
\end{subequations}
since $c_1^t = c_{-1}$ when using the operator $I^-$ in \eqref{cft:eq:modes-bpz}.
Imposing equality of both leads to the reality condition
\begin{equation}
	\conj{T(k)}
		= T(-k),
\end{equation}
which agrees with the fact that the tachyon is real (the integration measure changes as $\dd^D k \to - \dd^D k$, but the contour is reversed).

Then, the action reads:
\begin{equation}
	\label{bfst:eq:open-action-tachyon}
	S[T]
		= \frac{1}{2} \int \frac{\dd^D k}{(2\pi)^D} \,
			T(-k) \left( k^2 - \frac{1}{\alpha' } \right) T(k).
\end{equation}
This shows that the action is canonically normalized as it should for a real scalar field.
Similarly, one can compute the action for the gauge field:
\begin{equation}
	S[A]
		= \frac{1}{2} \int \frac{\dd^D k}{(2\pi)^D} \, A_\mu(-k) k^2 A^\mu(k).
\end{equation}
The correct normalization of the tachyon (real scalar field of negative mass) gives a justification a posteriori for the normalization of the action \eqref{bsft:eq:open-action}.
Typically, string field actions are normalized in this way, by requiring that the first physical spacetime fields has the correct normalization.
Note how this implies the correct normalization for all the others physical fields.
Generalizing this computation for higher-levels, one always find the kinetic term to be:
\begin{equation}
	\frac{L_0^+}{2}
		= \frac{1}{2} (k^2 + m^2),
\end{equation}
which is the canonical normalization.

\begin{computation}[bfst:eq:open-action-tachyon]
	\begin{align*}
		\bra{T} c_0 L_0 \ket{T}
			&
			= \frac{1}{\alpha'} \int \frac{\dd^D k}{(2\pi)^D} \, \frac{\dd^D k'}{(2\pi)^D} \,
				T(k) T(k') \bra{- k', 0} c_{-1} c_0 L_0 c_1 \ket{k, 0}
			\\ &
			= \frac{1}{\alpha'} \int \frac{\dd^D k}{(2\pi)^D} \, \frac{\dd^D k'}{(2\pi)^D} \,
				T(k) T(k') (\alpha' k^2 - 1) \bra{- k', 0} c_{-1} c_0 c_1 \ket{k, 0}
			\\ &
			= \frac{1}{\alpha'} \int \frac{\dd^D k}{(2\pi)^D} \, \dd^D k' \,
				T(k) T(k') (\alpha' k^2 - 1) \, \delta^{(D)}(k + k'),
	\end{align*}
	where we used $\bra{0} c_{-1} c_0 c_1 \ket{0} = 1$ and $\bracket{k'}{k} = (2\pi)^D \delta^{(D)}(k + k')$.
\end{computation}

\begin{draft}

\subsection{Quantum action}

The quantum open string field can be expanded as
\begin{equation}
	\begin{aligned}
		\ket{\Phi}
			= \int \frac{\dd^D k}{(2\pi)^{D}}
				\bigg(& T(p)
					+ \I A_\mu(p) \alpha^\mu_{-1}
					+ B(p) b_{-1}
					+ C(p) c_{-1}
					\\
					&+ T'(p) c_0
					+ A'_\mu(p) \alpha^\mu_{-1} c_0
					+ B'(p) b_{-1} c_0
					+ C'(p) c_{-1} c_0
					+ \cdots
					\bigg) \ket{k, \downarrow}.
	\end{aligned}
\end{equation}
The fields of the second line (with a prime) are unphysical: they can either be gauged away or they are auxiliary fields (they satisfy algebraic instead of differential equations).
They are all absent in the Siegel gauge.
For example, the fields $C(p)$ and $B(p)$ are ghosts for the $\group{U}(1)$ spacetime symmetry.

\end{draft}

\section{Closed string}
\label{bsft:sec:free-brst:closed}

\index{free covariant SFT!closed bosonic!equation of motion}%
The derivation of the BRST free action for the closed string is very similar.
The starting point is the equation of motion
\begin{equation}
	\label{bsft:eq:closed-eom}
	Q_B \ket{\Psi} = 0
\end{equation}
for the closed string field $\ket{\Psi}$.
The difference with \eqref{bsft:eq:open-eom} is that the BRST charge $Q_B$ now includes both the left- and right-moving sectors.
In the case of the open string, the field $\Phi$ was free of any constraint: we will see shortly that this is not the case for the closed string.

The next step is to find an inner product $\mean{\cdot, \cdot}$ to write the action:
\begin{equation}
	\label{bsft:eq:closed-action-om}
	S
		= \frac{1}{2} \, \mean{\Psi, Q_B \Psi}.
\end{equation}
\index{string field!closed bosonic}%
Following the open string, it seems logical to give the string field $\Psi$ the same ghost number as the states in the cohomology:
\begin{equation}
	N_{\text{gh}}(\Psi)
		= 2.
\end{equation}
\index{covariant SFT!closed bosonic!inner product}%
In this case, the ghost number of the arguments of $\mean{\cdot, \cdot}$ in \eqref{bsft:eq:closed-action-om} is $N_{\text{gh}} = 5$.
The ghost number anomaly requires the total ghost number to be $6$, that is:
\begin{equation}
	N_{\text{gh}}(\mean{\cdot, \cdot})
		= 1.
\end{equation}
There is no other choice because $N_{\text{gh}}(\Psi)$ must be integer.
The simplest solution is to insert one $c$ zero-mode $c_0$ or $\bar c_{0}$, or a linear combination.
The BRST operator $Q_B$ contains both $L_0^\pm$ (see the decomposition \eqref{cft:eq:brst-decomp-LR-pm}): the natural expectation (and by analogy with the open string) is that the gauge fixed equation of motion (to be discussed below) should be equivalent to the on-shell equation $L_0^+ = 0$ (see also \Cref{cft:sec:brst:lc:closed}).
This is possible only if the insertion is $c_0^-$.
\index{covariant SFT!closed bosonic!inner product}%
With this insertion, $\mean{\cdot, \cdot}$ can be formed from the BPZ product:
\begin{equation}
	\mean{A, B}
		= \bra{A} c_0^- \ket{B}.
\end{equation}
\index{free covariant SFT!closed bosonic!classical action}%
Then, the action reads:
\begin{equation}
	\label{bsft:eq:closed-action}
	S
		= \frac{1}{2} \, \bra{\Psi} c_0^- Q_B \ket{\Psi}.
\end{equation}

\index{free covariant SFT!closed bosonic!}%
However, the presence of $c_0^-$ has a drastic effect because it annihilates part of the string field.
Decomposing the Hilbert space as in \eqref{cft:eq:1storder-Hilbert-LR}
\begin{equation}
	\mc H
		= \mc H^- \oplus c_0^- \mc H^-,
	\qquad
	\mc H^-
		:= \mc H \cap \ker b_0^-,
\end{equation}
the string field reads:
\begin{equation}
	\ket{\Psi}
		= \ket{\Psi_-} + c_0^- \ket{\wtilde\Psi_-},
	\qquad
	\Psi_-, \wtilde \Psi_- \in \mc H^-,
\end{equation}
such that
\begin{equation}
	c_0^- \ket{\Psi}
		= c_0^- \ket{\Psi_-}.
\end{equation}
The problem in such cases is that the kinetic term may become non-invertible.
This motivates to project out the component $\wtilde \Psi_-$ by imposing the following constraint on the string field:
\begin{equation}
	\label{bsft:eq:constraint-b0m}
	b_0^- \ket{\Psi}
		= 0.
\end{equation}

\index{level-matching condition}%
The constraint \eqref{bsft:eq:constraint-b0m} is stronger than the constraint $L_0^- = 0$ for states in the cohomology (\Cref{cft:sec:brst:lc:cond}), so there is no information lost on-shell by imposing it.
\index{level-matching condition}%
For this reason, we will also impose the level-matching condition:
\begin{equation}
	L_0^- \ket{\Psi}
		= 0,
\end{equation}
such that
\begin{equation}
	\Psi \in \mc H^- \cap \ker L_0^-.
\end{equation}
This will later be motivated by studying the propagator and the off-shell scattering amplitudes.
To avoid introducing more notations, we will not use a new symbol for this space and keep implicit that $\Psi \in \ker L_0^-$.

The necessity of this condition can be understood differently.
We had found that it is necessary to ensure that the closed string parametrization is invariant under translations along the string (\Cref{bos:sec:ws-int:brst:states}).
Since there is no BRST symmetry associated to this symmetry, one needs to keep the constraint.\footnotemark{}
\footnotetext{%
	Yet another reason can be found in \Cref{bos:sec:ws-int:brst:states} (see also \Cref{cft:sec:brst:lc:closed}): to motivate the need of the $b_0^+$ condition, we could take the on-shell limit from off-shell states because $L_0^+$ is continuous.
	However, the $L_0^-$ operator is discrete and there is no such limit we can consider~\cite{Thorn:1989:StringFieldTheory}.
	So we must always impose this condition, both off- and on-shell.
}%
This suggests that one may enlarge further the gauge symmetry and interpret \eqref{bsft:eq:constraint-b0m} as a gauge fixing condition.
This would be quite desirable: one could argue that a fundamental field should be completely described by the Lagrangian (if such a description exists) and that it should not be necessary to supplement it with constraints imposed by hand.
While this can be achieved at the free level, this idea runs into problems in the presence of interactions (\Cref{bos:sec:offshell:geometry:hat-Pgn}) and the interpretation is not clear.\footnotemark{}
\footnotetext{%
	A recent proposal can be found in~\cite{Okawa:2018:ClosedStringField}.
}%

\index{free covariant SFT!closed bosonic!gauge transformation}%
The action \eqref{bsft:eq:closed-action} is gauge invariant under:
\begin{equation}
	\label{bsft:eq:closed-gauge-transf}
	\ket{\Psi}
	\longrightarrow
	\ket{\Psi'}
		= \ket{\Psi} + \delta_\Lambda \ket{\Psi},
	\qquad
	\delta_\Lambda \ket{\Psi}
		= Q_B \ket{\Lambda},
\end{equation}
where the gauge parameter has ghost number $1$ and also lives in $\mc H^- \cap \ker L_0^-$:
\begin{equation}
	N_{\text{gh}}(\Lambda)
		= 1,
	\qquad
	L_0^- \ket{\Lambda}
		= 0,
	\qquad
	b_0^- \ket{\Lambda}
		= 0.
\end{equation}

\index{Siegel gauge}%
As for the open string, the gauge invariance \eqref{bsft:eq:closed-gauge-transf} can be gauge fixed in the Siegel gauge:
\begin{equation}
	\label{bsft:eq:closed-siegel}
	b_0^+ \ket{\Psi}
		= 0.
\end{equation}
\index{free covariant SFT!closed bosonic!gauge fixed action}%
Then, the action reduces to:
\begin{equation}
	\label{bsft:eq:closed-action-siegel}
	S
		= \frac{1}{2} \, \bra{\Psi} c_0^- c_0^+ L_0^+ \ket{\Psi}
		= \frac{1}{4} \, \bra{\Psi} c_0 \bar c_0 L_0^+ \ket{\Psi}.
\end{equation}
\index{free covariant SFT!closed bosonic!gauge fixed equation of motion}%
The equation of motion is equivalent to the on-shell condition as expected:
\begin{equation}
	\label{bsft:eq:closed-eom-siegel}
	L_0^+ \ket{\Psi}
		= 0.
\end{equation}
Additional constraints must be imposed to ensure that only the physical degrees of freedom propagate.

\begin{computation}[bsft:eq:closed-action-siegel]
	\[
		c_0^- Q_B
			= (c_0 - \bar c_0) (c_0 L_0 + \bar c_0 \bar L_0)
				= c_0 \bar c_0 (L_0 + \bar L_0).
	\]
\end{computation}

\section{Summary}

In this chapter, we have shown how the BRST conditions defining the cohomology can be interpreted as an equation of motion for a string field together with a gauge invariance.
We found a subtlety for the closed string due to the ghost number anomaly and because of the level-matching condition.
Then, we studied several basic properties in order to prove that the free action has the expected properties.

The next step is to add the interactions to the action, but we don't know first principles to write them.
For this reason, we need to take a detour and to consider off-shell amplitudes.
By introducing a factorization of the amplitudes, it is possible to rewrite them as Feynman diagrams, where fundamental interactions are connected by propagators (which we will find to match the one in the Siegel gauge).
This can be used to extract the interacting terms of the action.

\refchapter

\begin{itemize}
	\item The free BRST string field theory is discussed in details in~\cite{Thorn:1989:StringFieldTheory} (see also~\cites[chap.~7]{Kaku:1999:IntroductionSuperstringsMTheory}[chap.~9]{Kaku:1999:StringsConformalFields}[chap.~11]{Siegel:2001:IntroductionStringField}).
	Shorter discussions can be found in~\cites{Zwiebach:1993:ClosedStringFieldIntro}[sec.~4]{Polchinski:1994:WhatStringTheory}{Witten:1986:NoncommutativeGeometryString}{Asano:2007:NewCovariantGauges}{Thorn:1987:PerturbationTheoryQuantized}.

	\item Spacetime fields and actions are discussed in~\cites{Taylor:2004:DBranesTachyonsString}[sec.~4]{Polchinski:1994:WhatStringTheory}.

	\item Gauge fixing~\cites{Labastida:1987:BRSTQuantizationSiegel}[sec.~6.5, 7.2, 7.4]{Taylor:2004:DBranesTachyonsString}{Asano:2007:NewCovariantGauges}[sec.~2.1]{Kroyter:2012:OpenSuperstringField}{Asano:2009:GeneralLinearGauges}{Bochicchio:1987:StringFieldTheory}.

	\item General properties of string field (reality, parity, etc.)~\cite{Zwiebach:1993:ClosedStringField, Asano:2007:NewCovariantGauges}.
\end{itemize}

\chapter{Introduction to off-shell string theory}
\label{bos:chap:offshell}

\introchapter

In this chapter, we introduce a framework to describe off-shell amplitudes in string theory.
We first start by motivating various concepts -- in particular, local coordinates and factorization -- by focusing on the $3$- and $4$-point amplitudes.
We then prepare the stage for a general description of off-shell amplitudes.
We focus again on the closed bosonic string only.

\section{Motivations}
\label{bos:sec:offshell:motivations}

\subsection{3-point function}

\index{closed string amplitude!tree-level!3-point}%
The tree-level $3$-point amplitude of $3$ weight $h_i$ vertex operators\footnotemark{} $\scr V_i$ is given by
\footnotetext{%
	The quantum number $(k, j)$ of the vertex operator is mostly irrelevant for the discussion of the current and next chapters, and they are omitted.
	We will distinguish them by a number and reintroduce the momentum $k$ when necessary.
	We also omit the overall normalization of the amplitudes.
}%
\begin{equation}
	A_{0,3}
		= \Mean{\prod_{i=1}^3 \scr V_i(z_i)}_{S^2}
		\propto (z_1 - z_2)^{h_3 - h_1 - h_2} \times \text{perms} \times \cc
\end{equation}
There is no integration since $\dim \mc M_{0,3} = 0$.

\index{string amplitude!conformal invariance}%
The amplitude is independent of the $z_i$ only if the matter state is on-shell, $h_i = 0$, for example if $\scr V_i = c \bar c V_i$ with $h(V_i) = 1$.
Indeed, if $h_i \neq 0$, then $A_{0,3}$ is not invariant under conformal transformations \eqref{cft:eq:sl2C-transf-fg}:
\begin{equation}
	\label{bos:eq:sl2C-transf-fg}
	z
	\longrightarrow
	f_g(z)
		= \frac{a z + b}{c z + d}
		\in \group{SL}(2, \C)
\end{equation}
(it transforms covariantly).
This is a consequence of the punctures: the presence of the latter modifies locally the metric, since they act as sources of negative curvature.
When performing a conformal transformation, the metric around the punctures changes in a different way as away from them.
This implies that the final result depends on the metric chosen around the punctures.
This looks puzzling because the original path integral derivation (\Cref{chap:bos:ws-int-amp}) indicates that the $3$-point amplitude should not depend on the locations of the operators because its moduli space is empty (hence, all choices of $z_i$ should be equivalent).

\index{local coordinates}%
The solution is to introduce local coordinates $w_i$ with a flat metric $\abs{\dd w_i}^2$ around each puncture conventionally located at $w_i = 0$.
The local coordinates are defined by the maps:
\begin{equation}
	z
		= f_i(w_i),
	\qquad
	z_i
		= f_i(0).
\end{equation}
This is also useful to characterize in a simpler way the dependence of off-shell amplitudes rather than using the metric around the punctures (computations may be more difficult with a general metric).

The expression of a local operator in the local coordinate system is found by applying the corresponding change of coordinates \eqref{cft:eq:transf-coord-circ}:
\begin{equation}
	f \circ \scr V(w)
		= f'(w)^{h} \overline{f'(w)}^{\bar h} \, \scr V\big(f(w)\big).
\end{equation}
\index{off-shell closed string amplitude!tree-level!3-point}%
The amplitude reads then
\begin{subequations}
\begin{align}
	A_{0,3}
		&
		= \Mean{\prod_{i=1}^3 f_i \circ \scr V_i(0)}_{S^2}
		= \left( \prod_{i=1}^3 f'_i(0)^{h_i} \overline{f'_i(0)}^{\bar h_i} \right)
			\Mean{\prod_{i=1}^3 \scr V_i\big(f_i(0)\big)}_{S^2}
		\\ &
		\propto \left( \prod_{i=1}^3 f'_i(0)^{h_i} \overline{f'_i(0)}^{\bar h_i} \right)
			\big(f_1(0) - f_2(0) \big)^{h_3 - h_1 - h_2}
			\times \text{perms} \times \cc
\end{align}
\end{subequations}
The amplitude depends on the local coordinate choice $f_i$, but not on the metric around the punctures.
It is also invariant under $\group{SL}(2, \C)$: the transformation \eqref{bos:eq:sl2C-transf-fg} written in terms of the local coordinates is
\begin{equation}
		f_i
			\longrightarrow
			\frac{a f_i + b}{c f_i + d}
\end{equation}
from which we get:
\begin{equation}
	f'_i
		\longrightarrow
		\frac{f'_i}{(c f_i + d)^2},
	\qquad
	f_i - f_j
		\longrightarrow
		\frac{f_i - f_j}{(c f_i + d) (c f_j + d)}.
\end{equation}
All together, this implies the invariance of the $3$-point amplitude since the factors in the denominator cancel.
When the states are on-shell $h_i = 0$, the dependence in the local coordinate cancels, showing that the latter is non-physical.
\index{off-shell string amplitude!conformal invariance}%

\bigskip

\index{string Feynman diagram}%
One can ask how Feynman graphs can be constructed in string theory.
By definition, an amplitude is the sum of Feynman graphs contributing at that order in the loop expansion and for the given number of external legs.
The Feynman graphs are themselves built from a set of Feynman rules.
These correspond to the data of the fundamental interactions together with the definition of a propagator.
\index{closed string fundamental vertex!tree-level!3-point}%
Since a tree-level cubic interaction is the interaction of the lowest order, it makes sense to promote it to a fundamental cubic vertex\footnotemark{} $\mc V_{0,3}$:
\footnotetext{%
	The notation will become clear later, and should not be confused with the vertex operators.
}%
\begin{equation}
	\label{bos:eq:vertex-03-intro}
	\mc V_{0,3}(\scr V_1, \scr V_2, \scr V_3)
		:= \vcenter{\hbox{\includegraphics[scale=1]{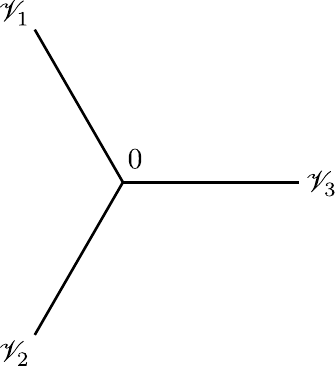}}}
		\quad
		= A_{0,3}(\scr V_1, \scr V_2, \scr V_3).
\end{equation}
The index $0$ reminds that it is a tree-level interaction.

\subsection{4-point function}
\label{bos:sec:offshell:motivations:4pt}

\index{closed string amplitude!tree-level!4-point}%
The tree-level $4$-point amplitude is expressed as
\begin{equation}
	A_{0,4}
		= \int \dd^2 z_4 \Mean{\prod_{i=1}^3 c \bar c V_i(z_i) \, V_4(z_4)}_{S^2}.
\end{equation}
The conformal weights are denoted by $h(V_i) = h_i$.
For on-shell states, $h_i = 1$: while there is no dependence on the positions $z_1$, $z_2$ and $z_3$, there are divergences for
\begin{equation}
	z_4 \longrightarrow z_1, z_2, z_3,
\end{equation}
corresponding to collisions of punctures in the integration process.
Moreover, the expression does not look symmetric: it would me more satisfactory if all the insertions were accompanied by ghost insertions and if all the puncture locations were treated on an equal footing.

\begin{example}[Tachyons]
	Given tachyon states $V_i = \e^{\I k_i \cdot X}$, the amplitude reads:
	\begin{equation}
		A_{0,4}
			\propto \prod_{\substack{i, j = 1 \\ i < j}}^3 \abs{z_i - z_j}^{2 + k_i \cdot k_j}
				\int \dd^2 z_4 \, \prod_{i=1}^3 \abs{z_4 - z_i}^{k_i \cdot k_4}.
	\end{equation}
	The integral diverges for $z_4 \to z_i$ if $k_i \cdot k_4 \le 0$.
	This can happen for physical values of the momenta $k_i$.
\end{example}

The idea is to cut out regions around $z_1$, $z_2$ and $z_3$ in the $z_4$-plane and to change the interpretation of these contributions.
First, we consider the case $z_4 \to z_3$, which corresponds to cutting a region around $z_3$.
Writing $z_4 = q y_4$ with $y_4 \in \C$ fixed, the contribution of this region to the amplitude is denoted by $\mc F_{0,4}^{(s)}$.
For simplicity, we take $z_3 = 0$.
The contribution reads:
\begin{equation}
	\mc F_{0,4}^{(s)}
		= \int \frac{\dd^2 q}{\abs{q}^2} \,
			\mean{c \bar c V_1(z_1) c \bar c V_2(z_2) c \bar c V_3(0) \abs{q y_4}^2 V_4(q y_4)}.
\end{equation}
The implicit radial ordering pushes $V_3$ to the left of $V_4$ and using the OPE between the $b$ and $c$ ghosts gives:
\begin{equation}
	\mc F_{0,4}^{(s)}
		= - \int \frac{\dd^2 q}{\abs{q}^2}
			\Mean{
				c \bar c V_1(z_1) c \bar c V_2(z_2)
				\oint_{\mathrlap{\abs{w} = \abs{q}^{1/2}}} \quad \dd w \, w \, b(w)
				\oint_{\mathrlap{\abs{w} = \abs{q}^{1/2}}} \quad \dd \bar w \, \bar w \, \overline{b(w)} \;
				c \bar c V_4(q y_4) c \bar c V_3(0)
				}.
\end{equation}
The sign arises by anti-commuting $c$ and $\bar b$.
The integration variable $q$ can be removed from the argument of $V_4$ using the $L_0$ and $\bar L_0$ operators:
\begin{equation}
	\label{bos:eq:G04-int-q-s}
	\mc F_{0,4}^{(s)}
		= - \int \frac{\dd^2 q}{\abs{q}^2}
			\Mean{c \bar c V_1(z_1) c \bar c V_2(z_2)
				\oint \dd w \, w \, b(w)
				\oint \dd \bar w \, \bar w \, \overline{b(w)} \,
				q^{L_0} \bar q^{\bar L_0} c \bar c V_4(y_4) c \bar c V_3(0)}.
\end{equation}
This expression is more satisfactory because all vertex operators are accompanied with $c$-ghost insertions and none of the arguments is integrated over.
But, in fact, even better can be achieved.

Inserting two complete sets of states $\{ \phi_r \}$ (see \Cref{bos:sec:offshell:states}) inside this expression gives (restoring a generic $z_3$-dependence):
\begin{equation}
	\mc F_{0,4}^{(s)}
		= \mean{c \bar c V_1(z_1) c \bar c V_2(z_2) \phi_r(0)} \,
			\mean{c \bar c V_3(z_3) c \bar c V_4(y_4) \phi_s(0)}
			\int \frac{\dd^2 q}{\abs{q}^2} \, \Mean{\phi_r^c q^{L_0} \bar q^{\bar L_0} b_0 \bar b_0 \phi_s^c},
\end{equation}
where the sum over $r$ and $s$ is implicit.
The conjugate states $\phi_r^c$ are defined by $\bracket{\phi_r^c}{\phi_s} = \delta_{rs}$.
The first two terms are cubic interactions \eqref{bos:eq:vertex-03-intro}, and the last term connects both.
\index{propagator!closed string|(}%
It is then tempting to identify the latter with a propagator $\Delta$
\begin{equation}
	\Delta\big(\phi_r^c, \phi_s^c\big)
		:= \bra{\phi_r^c} \Delta \ket{\phi_s^c}
		:= - \int \frac{\dd^2 q}{\abs{q}^2} \, \Mean{\phi_r^c q^{L_0} \bar q^{\bar L_0} b_0 \bar b_0 \phi_s^c},
\end{equation}
such that:
\index{string amplitude!factorization}%
\begin{equation}
	\begin{aligned}
	\mc F_{0,4}^{(s)}
		&
		= \mc V_{0,3}\big(c \bar c V_1(z_1), c \bar c V_2(z_2), \phi_r(0)\big)
			\times \Delta\big(\phi_r^c, \phi_s^c\big)
			\times \mc V_{0,3}\big(c \bar c V_3(z_3), c \bar c V_4(y_4), \phi_s(0)\big)
		\\ &
		= \vcenter{\hbox{\includegraphics[scale=1]{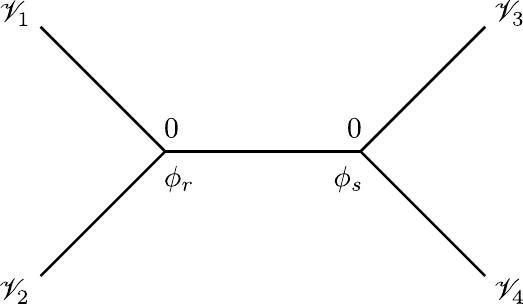}}}
	\end{aligned}
\end{equation}

To make this more precise, change the coordinates as:
\begin{equation}
	q
		= \e^{- s + \I \theta},
	\qquad
	s \in \R_+,
	\qquad
	\theta \in [0, 2\pi),
\end{equation}
such that the integral becomes:
\begin{equation}
	\int \frac{\dd^2 q}{\abs{q}^2} \, q^{L_0} \bar q^{\bar L_0}
		= 2 \int_0^\infty \dd s \int_0^{2\pi} \dd\theta \,
			\e^{- s (L_0 + \bar L_0)} \e^{\I \theta (L_0 - \bar L_0)}
		= \frac{2}{L_0 + \bar L_0} \, \delta_{L_0, \bar L_0}.
\end{equation}
This shows that the propagator can be rewritten as
\begin{equation}
	\Delta
		= - \frac{2 b_0 \bar b_0}{L_0 + \bar L_0} \, \delta_{L_0, \bar L_0}
		= \frac{b_0^+}{L_0^+} \, b_0^- \delta_{L_0^-, 0},
\end{equation}
where $L_0^\pm = L_0 \pm \bar L_0$ and $b_0^\pm = b_0 \pm \bar b_0$.
The sign is added by anticipating the normalization to be derived later.
Its properties will be studied in details in \Cref{bos:sec:amp-prop:propagator}.

Taking the basis states $\phi_r := \phi_\alpha(k)$ to be eigenstates of $L_0$ and $\bar L_0$
\begin{equation}
	L_0 \ket{\phi_\alpha(k)}
		= \bar L_0 \ket{\phi_\alpha(k)}
		= \frac{\alpha'}{4} \, (k^2 + m_\alpha^2) \, \ket{\phi_\alpha(k)}
\end{equation}
allows to rewrite the last term of $\mc F_{0,4}^{(s)}$ as
\begin{equation}
	\Delta_{\alpha\beta}(k)
		= \int \frac{\dd^2 q}{\abs{q}^2} \,
			\Mean{\phi_\alpha^c(k) q^{L_0} \bar q^{\bar L_0} b_0 \bar b_0 \phi_\beta^c(-k)}
		= \frac{M_{\alpha\beta}(k)}{k^2 + m_\alpha^2}.
\end{equation}
The finite-dimensional matrix $M_{\alpha\beta}$ gives the overlap of states of identical masses:
\begin{equation}
	M_{\alpha\beta}(k)
		:= \frac{2}{\alpha'} \, \bra{\phi_\alpha^c(k)} b_0^+ b_0^- \ket{\phi_\beta^c(-k)}.
\end{equation}
The propagator depends only on one momentum because $\bracket{k}{k'} \sim \delta^{(D)}(k - k')$.
This is exactly the standard propagator one finds in QFT and this justifies the above claim.
The contribution $\mc F_{0,4}^{(s)}$ to the amplitude can be seen as a $s$-channel Feynman graph obtained by gluing two cubic fundamental vertices with a propagator.
We will see later the interpretation in terms of Riemann surfaces.
\index{propagator!closed string|)}%

The same procedure can be followed by considering $z_4 \sim z_2$ and $z_4 \sim z_1$.
This leads to contributions $\mc F_{0,4}^{(t)}$ and $\mc F_{0,4}^{(u)}$ corresponding to $t$- and $u$-channel Feynman graphs:
\begin{equation}
	\mc F_{0,4}^{(t)}
		= \vcenter{\hbox{\includegraphics[scale=1]{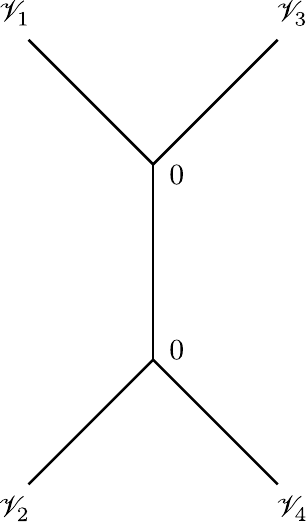}}}
	\hspace{1cm}
	\mc F_{0,4}^{(u)}
		= \vcenter{\hbox{\includegraphics[scale=1]{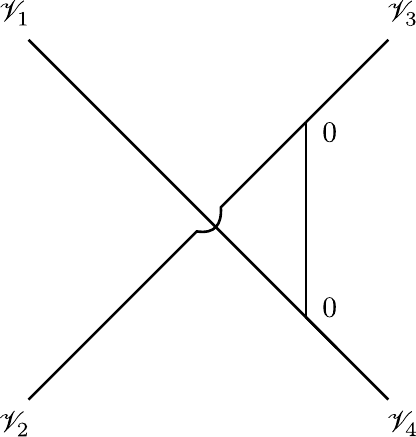}}}
\end{equation}

In general, the sum of the three contributions $\mc F_{0,4}^{(s,t,u)}$ does not reproduce the full amplitude $A_{0,4}$.
Said differently, the regions cut in the $z_4$-plane does not cover it completely.
\index{closed string fundamental vertex!tree-level!4-point}%
It is then natural to interpret the remaining part as a fundamental tree-level quartic interaction denoted by
\begin{equation}
	\mc V_{0,4}
		= \vcenter{\hbox{\includegraphics[scale=1]{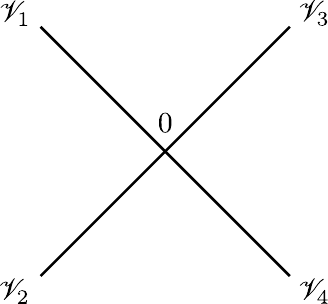}}}
\end{equation}
such that
\begin{equation}
	A_{0,4}
		= \mc F_{0,4}^{(s)} + \mc F_{0,4}^{(t)} + \mc F_{0,4}^{(u)} + \mc V_{0,4}.
\end{equation}

Up to \eqref{bos:eq:G04-int-q-s} it was sufficient to consider on-shell states, but the insertion of the complete basis requires to consider also off-shell states since on-shell states do not form a basis of the Hilbert space.
As discussed for the $3$-point functions, it is necessary to introduce local coordinates to describe off-shell states properly.

With the $3$- and $4$-point functions, we motivated the use of off-shell states and introduced the two important ideas of local coordinates and amplitude factorization.
We also indicated that amplitudes can be written in a more symmetric way (see also the discussion at the end of \Cref{bos:sec:ws-int:amp:gauge-fixing}).
In the rest of this chapter, we give additional ideas on off-shell string theory.

\begin{remark}[Riemann surface interpretation]
	The interpretation of the insertion of a propagator in terms of Riemann surface consists in gluing two of them thanks to the plumbing fixture procedure (\Cref{bos:sec:offshell:geometry:plumbing}).
\end{remark}

\section{Off-shell states}
\label{bos:sec:offshell:states}

\index{string states|(}%

\index{string states!off-shell}%
An off-shell state is a generic state of the CFT Hilbert space
\begin{equation}
	\label{bos:eq:hilbert-H}
	\mc H
		= \mc H_m \otimes \mc H_{\text{gh}}.
\end{equation}
without any constraint.
A basis for the off-shell states is denoted by
\begin{equation}
	\label{bos:eq:hilbert-H-basis}
	\mc H
		= \Span \{ \ket{\phi_r} \}.
\end{equation}
Since there is no constraint on the states, the ghost number of $\phi_r$ are arbitrary and denoted as:
\begin{equation}
	n_r
		:= N_{\text{gh}}(\phi_r) \in \Z
\end{equation}
(the ghost number is restricted for states in the cohomology of $Q_B$).
The Grassmann parity of a state $\phi_r$ is denoted as $\abs{\phi_r}$.
When there is no fermions in the matter sector (usually the case for the bosonic string), only ghosts are odd.
Then, the Grassmann parity of a state is odd (resp.\ even) if its ghost number is odd (resp.\ even):
\begin{equation}
	\abs{\phi_r}
		:= N_{\text{gh}}(\phi_r) \mod 2
		=
		\begin{cases}
			0 & \text{$N_{\text{gh}}(\phi_r)$ even},
			\\
			1 & \text{$N_{\text{gh}}(\phi_r)$ odd}.
		\end{cases}
\end{equation}

\index{string states!dual}%
The dual basis $\{ \ket{\phi_r^c} \}$ is defined from the BPZ inner product:
\begin{equation}
	\bracket{\phi_r^c}{\phi_s}
		= \delta_{rs}.
\end{equation}
Denoting the ghost numbers of the dual states by
\begin{equation}
	n_r^c
		:= N_{\text{gh}}(\phi_r^c),
\end{equation}
the product is non-vanishing if
\ifbook
\begin{equation}
	n_r^c + n_r
		= 6,
\end{equation}
\fi
due to the ghost number anomaly on the sphere.
This condition cannot be satisfied if the dual state $\phi_r^c$ is simply taken to be the BPZ conjugate $\phi_r^t$ since the BPZ conjugation does not change the ghost number.
This implies that
\begin{equation}
	\bracket{\phi_r}{\phi_s}
		= 0
\end{equation}
from the ghost anomaly for every state, except for the closed string states with $N_{\text{gh}} = 3$ (in fact, the inner product of these states is also zero as can be seen after investigation).
One can show that
\begin{equation}
	\bracket{\phi_r}{\phi_s^c}
		= (- 1)^{\abs{\phi_r}} \, \delta_{rs}.
\end{equation}
\index{string states!resolution of identity}%
Hence, the resolution of the identity can be written in the two equivalent ways
\begin{equation}
	\label{bos:eq:identity-offshell}
	1
		= \sum_{r} \ket{\phi_r} \bra{\phi_r^c}
		= \sum_{r} (- 1)^{\abs{\phi_r}} \ket{\phi_r^c} \bra{\phi_r}.
\end{equation}

\begin{check}
\subsection{Open string}

\index{open string states!zero-mode decomposition}%
The Hilbert space $\mc H$ can be separated according to the ghost zero-modes (\Cref{cft:sec:systems:1st:hilbert}):
\begin{equation}
	\label{bos:eq:open-hilbert-H-decomposition}
	\mc H
		\sim \mc H_0 \oplus c_0 \mc H_0,
\end{equation}
where the relative Hilbert space is defined as:
\begin{equation}
	\label{bos:eq:hilbert-open-H0}
	\mc H_0
		:= \mc H \cap \ker b_0
		= \big\{ \ket{\phi} \in \mc H \mid
			b_0 \ket{\phi} = 0 \big\}
		\sim b_0 \mc H.
\end{equation}
Given a state $\psi \in c_0 \mc H$, it is mapped to another state $\widetilde\psi \in \mc H_0$ under the isomorphism
\begin{equation}
	\ket{\psi}
		= c_0 \ket{\widetilde\psi},
	\qquad
	\ket{\widetilde\psi}
		= b_0 \ket{\psi}.
\end{equation}

The basis states of the open string Hilbert space are decomposed as
\begin{equation}
	\label{bos:eq:basis-decomposition-b0-c0}
	\phi_r
		= \phi_{\downarrow, r} + \phi_{\uparrow, r},
	\qquad
	b_0 \ket{\phi_{\downarrow, r}}
		= 0,
	\qquad
	c_0 \ket{\phi_{\uparrow, r}}
		= 0.
\end{equation}
Each state $\phi_{\uparrow, r} \in c_0 \mc H_0$ can be associated to a state $\widetilde\phi_{\downarrow, r} \in \mc H_0$ (the arrow is changed to indicate properly in which subspace the state lies):
\begin{equation}
	\begin{gathered}
	\ket{\phi_{\uparrow, r}}
		= c_0 \ket{\widetilde\phi_{\downarrow, r}},
	\qquad
	\ket{\widetilde\phi_{\downarrow, r}}
		= b_0 \ket{\phi_{\uparrow, r}},
	\\
	b_0 \ket{\widetilde\phi_{\downarrow, r}}
		= 0,
	\qquad
	N_{\text{gh}}(\phi_{\uparrow, r})
		= N_{\text{gh}}(\widetilde\phi_{\downarrow, r}) + 1.
	\end{gathered}
\end{equation}

\subsection{Closed string}

\end{check}

\index{closed string states!zero-mode decomposition}%
Following \eqref{cft:eq:1storder-Hpm}, the Hilbert space can be decomposed as:
\begin{equation}
	\label{bos:eq:hilbert-closed}
	\mc H
		= \mc H_\pm
			\oplus c_0^\pm \mc H_\pm,
\end{equation}
where
\begin{equation}
	\label{bos:eq:hilbert-closed-Hpm0}
	\mc H_\pm
		:= \mc H \cap \ker b_0^\pm
		= \mc H_0 \oplus c_0^\mp \mc H_0,
	\qquad
	\mc H_0
		:= \mc H \cap \ker b_0^- \cap \ker b_0^+.
\end{equation}

\index{level-matching condition}%
In fact, we will find that a consistent description of the off-shell amplitudes for the closed string requires to impose some conditions on the states even at the off-shell level.
The off-shell states will have to satisfy the level-matching condition and to be annihilated by $b_0^-$:
\begin{equation}
	L_0^- \ket{\phi}
		= 0,
	\qquad
	b_0^- \ket{\phi}
		= 0.
\end{equation}
This implies that the off-shell states will be elements of $\mc H^- \cap \ker L_0^-$.
This will appear as consistency conditions on the geometry of the moduli space and by studying the propagator.
In general, we shall work with $\mc H$ and indicate when necessary the restriction to $\mc H^-$ (keeping the condition $\ker L_0^-$ implicit to avoid new notations).

The Hilbert space $\mc H$ can be separated according to the ghost zero-modes
\begin{equation}
	\label{bos:eq:hilbert-H-decomposition}
	\mc H
		\sim \mc H_{\downarrow\downarrow} \oplus \mc H_{\downarrow\uparrow} \oplus \mc H_{\uparrow\downarrow} \oplus \mc H_{\uparrow\uparrow},
\end{equation}
with the following definitions:
\begin{equation}
	\mc H_{\downarrow\uparrow} \sim \mc H_0,
	\qquad
	\mc H_{\downarrow\uparrow}
		\sim \bar c_0 \mc H_{\downarrow\downarrow}
	\qquad
	\mc H_{\uparrow\downarrow}
		\sim c_0 \mc H_{\downarrow\downarrow}
	\qquad
	\mc H_{\uparrow\uparrow}
		\sim c_0 \bar c_0 \mc H_{\downarrow\downarrow}.
\end{equation}
Accordingly, every basis state can be split as
\begin{equation}
	\label{bos:eq:closed-basis-expansion}
	\phi_r
		= \phi_{\downarrow\downarrow, r}
			+ \phi_{\downarrow\uparrow, r}
			+ \phi_{\uparrow\downarrow, r}
			+ \phi_{\uparrow\uparrow, r}
\end{equation}
such that
\begin{equation}
	\begin{gathered}
	b_0 \ket{\phi_{\downarrow\downarrow, r}}
		= \bar b_0 \ket{\phi_{\downarrow\downarrow, r}}
		= 0,
	\qquad
	b_0 \ket{\phi_{\downarrow\uparrow, r}}
		= \bar c_0 \ket{\phi_{\downarrow\uparrow, r}}
		= 0,
	\\
	c_0 \ket{\phi_{\uparrow\downarrow, r}}
		= \bar b_0 \ket{\phi_{\uparrow\downarrow, r}}
		= 0,
	\qquad
	c_0 \ket{\phi_{\uparrow\uparrow, r}}
		= \bar c_0 \ket{\phi_{\uparrow\uparrow, r}}
		= 0.
		\end{gathered}
\end{equation}
Moreover, the basis can be indexed such that
\begin{equation}
	\ket{\phi_{\downarrow\uparrow, r}}
		= \bar c_0 \ket{\phi_{\downarrow\downarrow, r}}
	\qquad
	\ket{\phi_{\uparrow\downarrow, r}}
		= c_0 \ket{\phi_{\downarrow\downarrow, r}}
	\qquad
	\ket{\phi_{\uparrow\uparrow, r}}
		= c_0 \bar c_0 \ket{\phi_{\downarrow\downarrow, r}}.
\end{equation}

A dual state $\phi_r^c$ is also expanded:
\begin{equation}
	\phi_r^c
		= \phi_{\downarrow\downarrow, r}^c
			+ \phi_{\downarrow\uparrow, r}^c
			+ \phi_{\uparrow\downarrow, r}^c
			+ \phi_{\uparrow\uparrow, r}^c
\end{equation}
and the components satisfy
\begin{equation}
	\begin{gathered}
	\bra{\phi_{\downarrow\downarrow, r}^c} c_0
		= \bra{\phi_{\downarrow\downarrow, r}^c} \bar c_0
		= 0,
	\qquad
	\bra{\phi_{\downarrow\uparrow, r}^c} c_0
		= \bra{\phi_{\downarrow\uparrow, r}^c} \bar b_0
		= 0,
	\\
	\bra{\phi_{\uparrow\downarrow, r}^c} b_0
		= \bra{\phi_{\uparrow\downarrow, r}^c} \bar c_0
		= 0,
	\qquad
	\bra{\phi_{\uparrow\uparrow, r}^c} b_0
		= \bra{\phi_{\uparrow\uparrow, r}^c} \bar b_0
		= 0.
	\end{gathered}
\end{equation}
The indexing of the basis is chosen such that:
\begin{equation}
	\bra{\phi_{\downarrow\uparrow, r}^c}
		= \bra{\phi_{\downarrow\downarrow, r}^c} \bar b_0
	\qquad
	\bra{\phi_{\uparrow\downarrow, r}^c}
		= \bra{\phi_{\downarrow\downarrow, r}^c} b_0
	\qquad
	\bra{\phi_{\uparrow\uparrow, r}^c}
		= \bra{\phi_{\downarrow\downarrow, r}^c} \bar b_0 b_0.
\end{equation}
such that
\begin{equation}
	\bracket{\phi_{x,r}^c}{\phi_{y,s}}
		= \delta_{xy} \delta_{rs},
\end{equation}
where $x, y = \downarrow\downarrow, \uparrow\downarrow, \downarrow\uparrow, \uparrow\uparrow$.

Consider the Hilbert space $\mc H^-$, then a basis state must satisfy
\begin{equation}
	b_0^- \ket{\phi_r}
		= 0
	\quad \Longrightarrow \quad
	b_0 \ket{\phi_r}
		= \bar b_0 \ket{\phi_r}.
\end{equation}
The expansion \eqref{bos:eq:closed-basis-expansion} gives the relation
\begin{equation}
	\phi_{\uparrow\downarrow, r} + \phi_{\uparrow\uparrow, r}
		= \phi_{\downarrow\downarrow, r} + \phi_{\downarrow\uparrow, r}
\end{equation}
such that
\begin{equation}
	\phi_r = 2 (\phi_{\downarrow\downarrow, r}
		+ \phi_{\downarrow\uparrow, r}).
\end{equation}
For convenience, the factor of $2$ can be omitted (which amounts to rescaling the basis states).

\index{string states|)}%

\section{Off-shell amplitudes}
\label{bos:sec:offshell:amplitudes-intro}

\index{off-shell string amplitude}%

In this section, we provide a guideline of what we need to look for in order to write an off-shell amplitude.
The geometrical tools will be described in the next chapter, and the construction of off-shell amplitudes in the following one.

\subsection{Amplitudes from the marked moduli space}
\label{bos:sec:offshell:amp-intro:marked-moduli}

In \Cref{chap:bos:ws-int-amp}, the scattering amplitudes were written as an integral over the moduli space $\mc M_{g}$ of the Riemann surface $\Sigma_{g}$.
As a consequence, the moduli of $\mc M_{g}$ and the positions of the vertex operators are not treated on an equal footing.
Moreover, the insertions of operators is not symmetric since some are integrated, and others have factors of $c$.
These problems can be solved by reinterpreting the scattering amplitudes in a more geometrical way.

The key is to consider the punctures where vertex operators are inserted as part of the geometry and not as external data added on top of the Riemann surface $\Sigma_{g}$.

\index{punctured Riemann surface}%
Then, the worldsheet with the external states is described as a punctured (or marked) Riemann surface $\Sigma_{g,n}$, which is a Riemann surface $\Sigma_g$ with $n$ punctures (marked points) $z_i$.
\index{punctured Riemann surface!Euler characteristics}%
The Euler number of such a surface was given in \eqref{bos:eq:chi-gn}:
\begin{equation}
	\label{offb:eq:chi-gn}
	\chi_{g,n}
		:= \chi(\Sigma_{g,n})
		= 2 - 2 g - n.
\end{equation}
This makes sense since punctures can be interpreted as disks (boundaries).
Note that the punctures are labeled and are thus distinguishable.

Since the marked points are distinguished, marked Riemann surfaces with identical $g$ and $n$ but with punctures located at different points are seen as different (this statement requires some care for $g = 0$ and $g = 1$ due to the presence of CKV).
\index{moduli space!with punctures}%
The corresponding moduli space is denoted by $\mc M_{g,n}$, and it can be viewed as a fibre bundle with $\mc M_{g}$ as the base and the puncture positions as the fibre.
The dimension of $\mc M_{g,n}$ is
\begin{equation}
	\label{bos:eq:dim-Mgn}
	\M_{g,n}
		:= \dim_{\R} \mc M_{g,n}
		= 6 g - 6 + 2 n,
	\qquad
	\text{for}
	\quad
	\begin{cases}
		g \ge 2, \\
		g = 1, n \ge 1, \\
		g = 0, n \ge 3.
	\end{cases}
\end{equation}
These cases are equivalent to $\chi_{g,n} < 0$, that is, when the surfaces have a negative curvature.
The corresponding coordinates are denoted by $t_\lambda$, $\lambda = 1, \ldots, \M_{g,n}$.
Comparing with \eqref{bos:eq:dim-Mg} and \eqref{bos:eq:dim-Kg}, this corresponds to the situation where $\Sigma_{g,n}$ has no CKV left unfixed.

\begin{example}[$4$-punctured sphere $\Sigma_{0,4}$]
	The positions of the punctures are denoted by $z_i$ with $i = 1, \ldots, 4$.
	Since there are three CKV, the positions of three punctures (say $z_1$, $z_2$ and $z_3$) can be fixed, leaving only one position which characterizes $\Sigma_{0,4}$.
	Hence, the moduli space has dimension $\M_{0,4} = 2$ and $\mc M_{0,4}$ is parametrized by $\{ z_4 \}$.
\end{example}

\begin{example}[$2$-punctured torus $\Sigma_{1,2}$]
	The positions of the punctures are denoted by $z_i$ with $i = 1, 2$.
	One puncture can be fixed using the single CKV of the surface, which leaves one position.
	Together with the moduli parameter $\tau$ of the torus, this gives $\M_{1,2} = 4$ and the coordinates of $\mc M_{1,2}$ are $\{ z_2, \tau \}$.
\end{example}

\index{closed string amplitude!on punctured moduli space}%
The $g$-loop $n$-point scattering amplitude with external states $\{ \scr V_i \}$ can be written as an integral over $\mc M_{g,n}$ of some $\M_{g,n}$-form $\omega^{(g,n)}_{\M_{g,n}}$:
\begin{equation}
	\label{bos:eq:Ggn-Mgn}
	A_{g,n}(\scr V_1, \ldots, \scr V_n)
		= \int_{\mc M_{g,n}} \omega^{g,n}_{\M_{g,n}}(\scr V_1, \ldots, \scr V_n).
\end{equation}
The integration over $\M_{g,n}$ has the correct dimension to reproduce the formulas from \Cref{bos:sec:ws-int:amp}.

While it is possible to derive this amplitude from the path integral (see the comments at the end of \Cref{bos:sec:ws-int:amp:gauge-fixing}), we will make only use of the properties of CFT on Riemann surfaces in the next chapter.
This provides an alternative point of view on the computation of scattering amplitudes and how to derive the formulas, which can be helpful when the manipulation of the path integral is more complicated (for example, with the superstring).

The expression of the form $\omega^{g,n}_{\M_{g,n}}$ must 1) provide a measure on the moduli space and 2) extract a function of the moduli from the states $\scr V_i$.
It is natural to achieve the second point by computing a correlation function on the Riemann surfaces $\Sigma_{g,n}$.
Moreover, \Cref{chap:bos:ws-int-vac,chap:bos:ws-int-amp} indicate that the ghosts are part of the definition of the measure.
Hence, one can expect the $\omega^{g,n}_{\M_{g,n}}$ to have the form:
\begin{equation}
	\omega^{g,n}_{\M_{g,n}}(\scr V_1, \ldots, \scr V_n)
		= \Mean{\text{ghosts} \times \prod_{i=1}^n \scr V_i}_{\mathrlap{\Sigma_{g,n}}}
			\times \;
			\bigwedge_{\lambda=1}^{\M_{g,n}} \dd t_\lambda.
\end{equation}
We will motivate an expression in \Cref{bos:chap:feynman} before checking that it has the correct properties.
The ghost insertions are necessary to saturate the number of zero-modes to obtain a non-vanishing result.
By convention, the ghosts are inserted on the left: while this does not make difference for on-shell closed states, this will for off-shell states and for the other types of strings (open and supersymmetric) since the operators can be Grassmann odd.

\subsection{Local coordinates}

\index{local coordinates}%

\begin{figure}[htp]
	\centering
	\subcaptionbox{Original surface with three punctures.}{%
		\centering
		\includegraphics[scale=1]{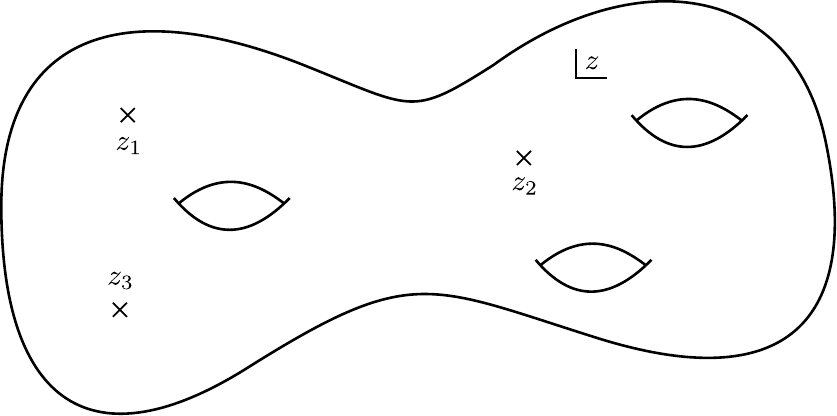}
	}%
	\\
	\bigskip
	\subcaptionbox{Disks delimiting the local coordinate patches.}{%
		\centering
		\includegraphics[scale=1]{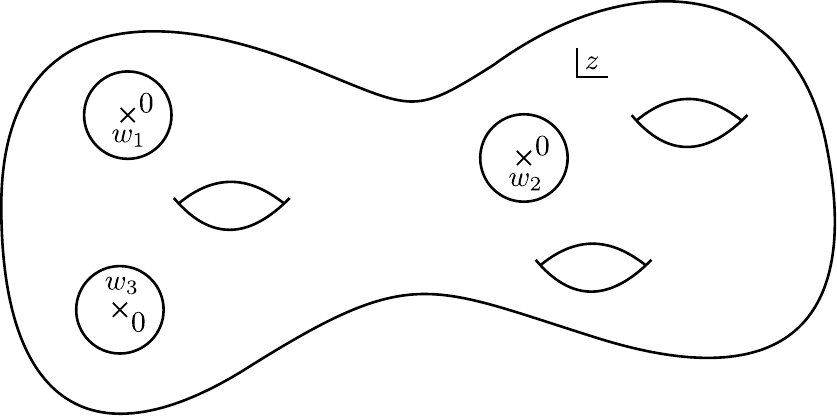}
	}%
	\\
	\bigskip
	\subcaptionbox{Complex plane mapped to disks centred around the previous puncture location.}{%
		\centering
		\includegraphics[scale=1]{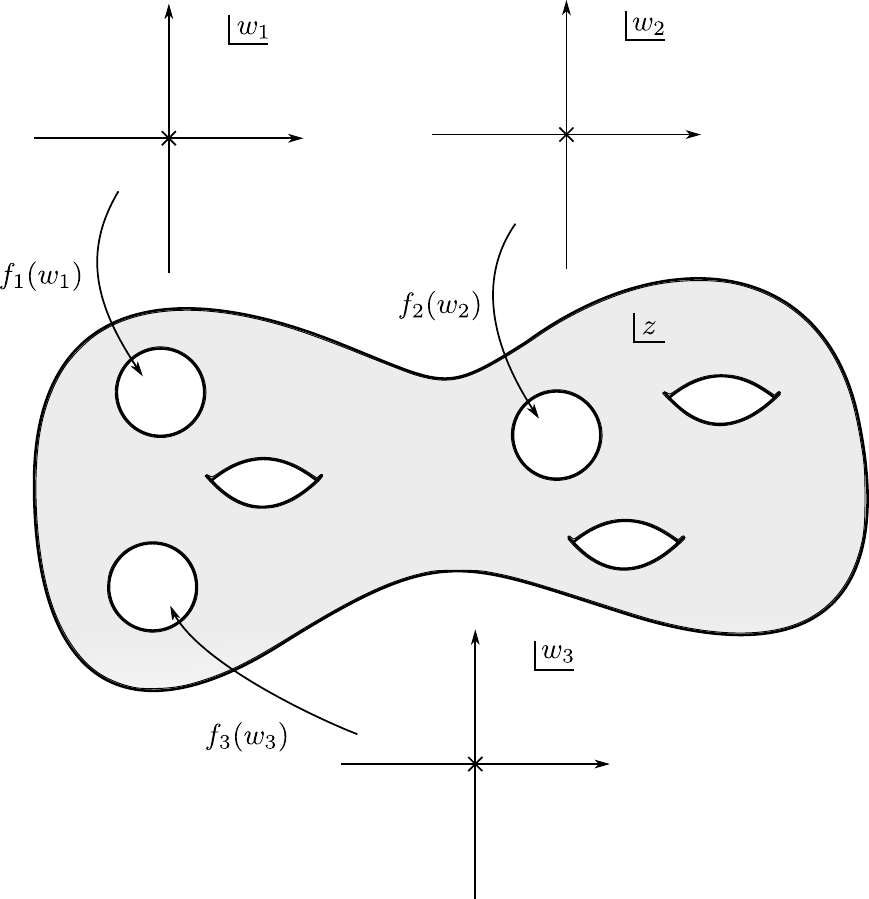}
	}%
	\caption{Usage of local coordinates for a Riemann surface.}
	\label{bos:fig:Sgn-local-coord}
\end{figure}

The next step is to consider off-shell states $\scr V_i \in \mc H$.
As motivated previously, one needs to introduce local coordinates defined by the maps:
\begin{equation}
	z
		= f_i(w_i),
	\qquad
	z_i
		= f_i(0).
\end{equation}
There is one local coordinate for each operator, which is inserted at the origin.
Local coordinates on the surfaces can be seen in two different fashions (\Cref{bos:fig:Sgn-local-coord}).
Either as describing patches on the surface, in which case the maps $f_i$ correspond to transition functions.
Or, one can interpret them by cutting disks centred at the punctures and whose interiors are mapped to complex planes, and the maps $f_i$ tell how to insert the plane inside the disk.

When the amplitude $A_{g,n}$ is defined in terms of local coordinates, it will depend on the maps $f_i$ and one needs to ensure that this cancels when the $A_i$ are on-shell.
But, the choice of the maps $f_i$ is arbitrary: selecting a specific set hides that all choices are physically equivalent and should lead to the same results on-shell.
For this reason, the geometry can be enriched with the local coordinates, in the same way that the puncture locations were added as a fibre to the moduli space $\mc M_{g}$ to get the marked moduli space $\mc M_{g,n}$.
Hence, the fundamental geometrical object is the fibre bundle $\mc P_{g,n}$ with $\mc M_{g,n}$ being the base and the local coordinates the fibre.
\index{Pgn@$\mc P_{g,n}$ space}%
Since there is an infinite number of functions, the fibre is infinite-dimensional, and so is the space $\mc P_{g,n}$.

\begin{figure}[htp]
	\centering
	\includegraphics[scale=1.2]{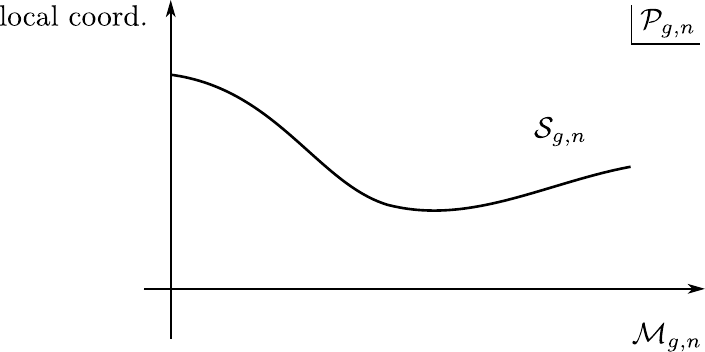}
	\caption{Section $\mc S_{g,n}$ of the fibre bundle $\mc P_{g,n}$, the latter having $\mc M_{g,n}$ as a base the local coordinates as a fibre.}
	\label{bos:fig:Pgn-section}
\end{figure}

Every point of $\mc P_{g,n}$ corresponds to a genus-$g$ Riemann surface with $n$ punctures together with a choice of local coordinates around the punctures.
\index{Pgn@$\mc P_{g,n}$ space!section}%
The form $\omega^{g,n}_{\M_{g,n}}$ is defined in this bigger space and the integration giving the off-shell amplitude \eqref{bos:eq:Ggn-Mgn} is performed over a $\M_{g,n}$-dimensional section $\mc S_{g,n} \subset \mc P_{g,n}$ (\Cref{bos:fig:Pgn-section}):
\index{off-shell closed string amplitude}%
\begin{equation}
	\label{bos:eq:Ggn-Mgn-section}
	A_{g,n}(\scr V_1, \ldots, \scr V_n)_{\mc S_{g,n}}
		= \int_{\mc S_{g,n}} \omega^{g,n}_{\M_{g,n}}(\scr V_1, \ldots, \scr V_n)\big|_{\mc S_{g,n}}.
\end{equation}
The subscript in the LHS indicates that the amplitudes depend on $\mc S_{g,n}$ through the choice of local coordinates.
The on-shell independence of $A_{g,n}$ on the local coordinates translate into the independence on the choice of the section:
\begin{equation}
	\forall \mc S_{g,n}:
	\quad
	A_{g,n}(\scr V_1, \ldots, \scr V_n)_{\mc S_{g,n}}
		= A_{g,n}(\scr V_1, \ldots, \scr V_n)
	\qquad
	\text{(on-shell)}.
\end{equation}
The section is taken to be continuous, which means that two neighbouring surfaces of the moduli space must have close local coordinates.

In order to define the amplitude, one needs to find the expression of the $\M_{g,n}$-form $\omega^{g,n}_{\M_{g,n}}$ on $\mc P_{g,n}$.
It is in fact simpler to define general $p$-forms $\omega^{g,n}_p$ on $\mc P_{g,n}$, in particular, for proving general properties about the forms and the amplitudes.
Given a manifold, a $p$-form is an element of the cotangent space and it can be defined through its contraction with vectors (tangent space).
Since vectors correspond to small variation of the manifold coordinates, it is necessary to find a parametrization of $\mc P_{g,n}$.
The geometry of $\mc P_{g,n}$ -- and of its relevant subspaces and tangent space -- is studied in the next chapter.
Then, we will come back on the construction of the amplitudes in \Cref{bos:chap:offshell-amp}.

\refchapter

\begin{itemize}
	\item General references on off-shell string theory include~\cites[sec.~7]{Zwiebach:1993:ClosedStringField}[sec.~2]{Sen:2015:OffshellAmplitudesSuperstring}{deLacroix:2017:ClosedSuperstringField}{Polchinski:2005:StringTheory-1}.

	\item Interpretation of local coordinates~\cite[sec.~5.2]{Polchinski:2005:StringTheory-1}.
\end{itemize}

\chapter{Geometry of moduli spaces and Riemann surfaces}
\label{bos:sec:offshell:geometry}

\introchapter

In this chapter, we describe how to parametrize the moduli space $\mc M_{g,n}$ and the local coordinates which together form the fibre bundle $\mc P_{g,n}$ introduced in the previous chapter.
Then, we can characterize the tangent space which we will need in the next chapter to write the $p$-forms on $\mc P_{g,n}$ necessary to write the amplitudes.
Finally, we introduce the notion of plumbing fixture, an operation which glue together punctures located on the same or different surfaces.

\section{Parametrization of \texorpdfstring{$\mc P_{g,n}$}{P(g,n)}}
\label{bos:sec:offshell:geometry:Pgn-param}

\index{punctured Riemann surface!parametrization|(}%
The first step is to find a parametrization of the Riemann surfaces.
As we have seen (\Cref{bos:chap:offshell}), the dependence of the surface on the punctures can be described by local coordinates, that is, transition functions.
The patch is defined by cutting a disk around each puncture and the transition functions are defined on the circle given by the intersection of the disk with the rest of the surface.
The number of disks is simply:
\begin{equation}
	\# \text{disks}
		= n.
\end{equation}

It makes sense to look for a similar description of the other moduli (associated to the genus) by introducing additional coordinate patches.
One can imagine that all the dependence of the moduli and punctures will reside in the transition functions between patches if the different patches are isomorphic to a surface without any moduli: the $3$-punctured sphere $\Sigma_{0,3}$.
Hence, one can look for a decomposition of the surface by cutting disks such that one is left with $3$-punctured spheres only, and transition functions are defined on the circles at the intersections of the spheres.

\begin{figure}[htp]
	\centering
	\includegraphics[scale=1]{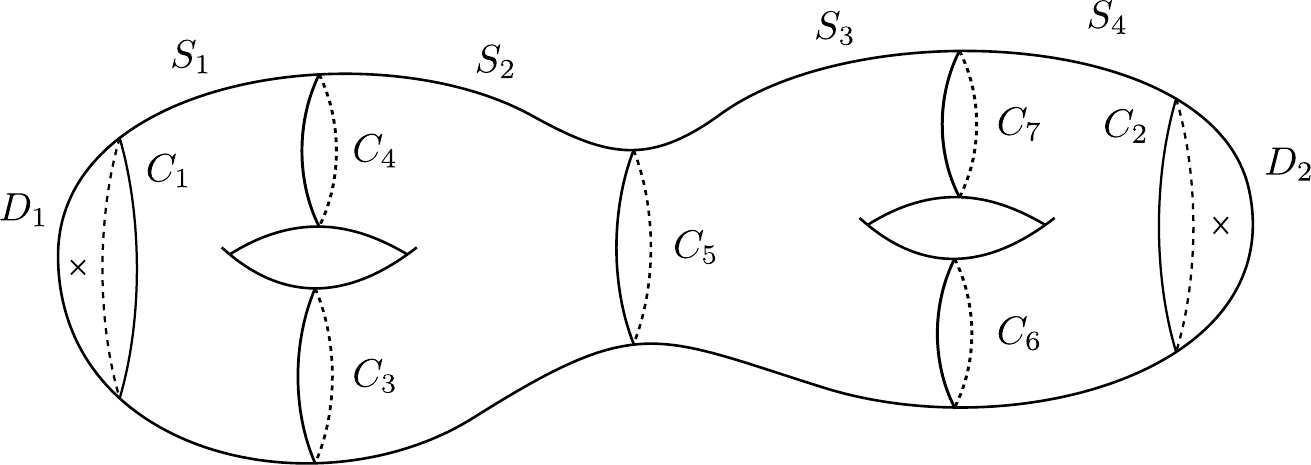}
	\caption{Parametrization of $\Sigma_{2,2}$.}
	\label{bos:fig:S22-param}
\end{figure}

Next, we need to find the number of spheres with $3$ holes (or punctures).
We start first with $\Sigma_{0,n}$: in this case, it is straightforward to find that there will be $n - 2$ spheres.
Indeed, for each additional puncture beyond $n = 3$, an additional sphere is created by cutting a circle.
For $g \ge 1$, natural places to split the surface are handles: two circles can be cut for each of them.
By inspection, one finds that it leads to $2$ spheres for each handle (one on the right and one on the left).\footnotemark{}
\footnotetext{%
	The simplest way to find this result is to consider $\Sigma_{g,2}$ and to write one puncture at each side of the surface (as in \Cref{bos:fig:S22-param}).
	To generalize further, one can consider a generic $n$ and put all punctures but one on one side of the surfaces.
}%
This shows that the number of spheres is:
\begin{equation}
	\# \text{spheres}
		= 2 g - 2 + n.
\end{equation}

The number of circles corresponds to the number of boundaries divided by two since the boundaries are glued pairwise: each disk has one boundary and each sphere has $3$, which leads to:
\begin{equation}
	\# \text{circles}
		= \frac{n + 3 (2 g - 2 + n)}{2}
		= 3 g - 3 + 2 n.
\end{equation}

The idea of the construction is to split the surface into elementary objects (spheres and disks) such that the full surface is seen as the union of all of them (gluing along circles), and no information is left in the individual geometries.
This parametrization is particularly useful because there are simple coordinate systems on spheres and disks and these surfaces are easy to visualize and to work with.
For example, they can be easily mapped to the complex plane.

\begin{figure}[htp]
	\centering
	\includegraphics[scale=1]{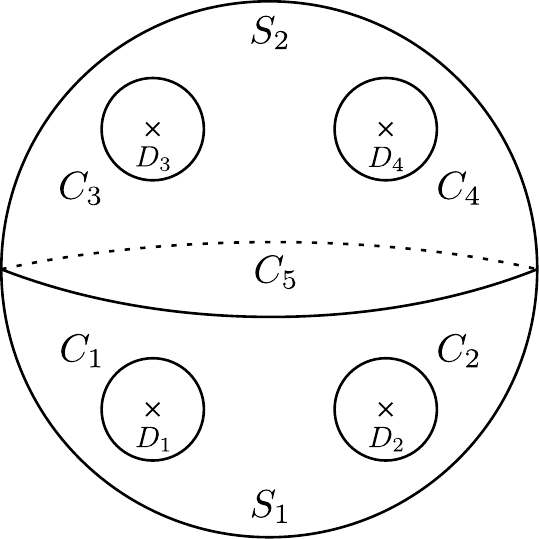}
	\caption{Parametrization of $\Sigma_{0,4}$.}
	\label{bos:fig:S04-param}
\end{figure}

To conclude, a genus-$g$ Riemann surface $\Sigma_{g,n}$ with $n$ punctures can be seen as the collection of:
\begin{itemize}
	\item $2g - 2 + n$ three-punctured spheres $\{ S_a \}$ with coordinates $z_a$

	\item $n$ disks $\{ D_i \}$ with coordinates $w_i$ around each puncture

	\item $3g - 3 + 2n$ circles $\{ C_\alpha \}$ at the intersections of the spheres and disks
\end{itemize}
Examples for $\Sigma_{0,4}$ and $\Sigma_{2,2}$ are given in \Cref{bos:fig:S04-param,bos:fig:S22-param}.

\index{punctured Riemann surface!parametrization|)}%

There are two types of circles: respectively, the ones at the overlap between two spheres, and between a disk and a $3$-sphere:
\begin{equation}
	C_{\Lambda(ab)}
		:= S_a \cap S_b,
	\qquad
	C_{i(a)}
		:= S_a \cap D_i,
	\qquad
	\{ C_\alpha \}
		= \{ C_{\Lambda}, C_{i} \},
\end{equation}
where $\Lambda$ counts in fact the number of moduli:
\begin{equation}
	\Lambda
		= 1, \ldots, \M_{g,n}^c,
	\qquad
	\M_{g,n}^c
		= 3 g - 3 + n.
\end{equation}

\index{local coordinates!transition function}%
On the overlap circles, the coordinate systems are related by transitions functions:
\begin{equation}
	\label{bos:eq:trans-functions-za-wi}
	\begin{aligned}
	\text{on $C_{\Lambda(ab)}$:}&
	\qquad
	z_a
		= F_{ab}(z_b),
	\\
	\text{on $C_{i(a)}$:}&
	\qquad
	z_a
		= f_{ai}(w_i).
	\end{aligned}
\end{equation}
Then, the set of functions $\{ F_{ab}, f_{ai} \}$ completely specifies the Riemann surface $\Sigma_{g,n}$ together with the choice of the local coordinate systems around the punctures.
The transition functions can thus be used to parametrize the moduli space $\mc M_{g,n}$ and the fibre bundle $\mc P_{g,n}$, but it is highly redundant because many different functions lead to the same Riemann surface.
A unique characterization of the different spaces is obtained by making identifications up to symmetries.

\index{local coordinates!global phase}%
In the previous chapter, we have seen that the metric in the local coordinate system is flat, $\dd s^2 = \abs{\dd w}^2$.
This means that two systems differing by a global phase rotation
\begin{equation}
	w_i
	\longrightarrow
	\tilde w_i
			= \e^{\I \alpha_i} w_i
\end{equation}
lead to surfaces with local coordinates which cannot be distinguished.
Correspondingly, the two maps $f_i$ and $\tilde f_i$ which relate the local coordinates $w_i$ and $\tilde w_i$ to the coordinate $z$
\begin{equation}
	z
		= f_i(w_i),
	\qquad
	z
		= \tilde f_i(\tilde w_i)
\end{equation}
are related as:
\begin{equation}
	\label{bos:eq:equiv-f-phase}
	f_i(w_i)
		= \tilde f_i(\e^{\I \alpha_i} w_i).
\end{equation}
\index{level-matching condition}%
\index{local coordinates!constraints}%
\index{Pgn@$\hat{\mc P}_{g,n}$}%
Hence, this motivates to consider the smaller space
\begin{equation}
	\label{bos:eq:def-Pgn-hat}
	\hat{\mc P}_{g,n}
		= \mc P_{g,n} / \group{U}(1)^n,
\end{equation}
where the action of each $\group{U}(1)$ is defined by the equivalence \eqref{bos:eq:equiv-f-phase}.
The necessity to consider this subspace will be strengthen further later, and will correspond to the level-matching condition.
Below, global phase rotations are also interpreted in terms of the plumbing fixture, see \eqref{bos:eq:coord-s-th-rescaling}.

\index{local coordinates!reparametrization}%
The different spaces which we need are parametrized by the transition functions up to the following identifications:
\begin{itemize}
	\item $\mc P_{g,n} = \{ F_{ab}, f_{ai} \}$ modulo reparametrizations of $z_a$

	\item $\hat{\mc P}_{g,n} = \{ F_{ab}, f_{ai} \}$ modulo reparametrizations of $z_a$ and phase rotations of $w_i$

	\item $\mc M_{g,n} = \{ F_{ab}, f_{ai} \}$ modulo reparametrizations of $z_a$ and of $w_i$ keeping the points $w_i = 0$ fixed

	\item $\mc M_{g} = \{ F_{ab}, f_{ai} \}$ modulo reparametrizations of $z_a$ and $w_i$
\end{itemize}
At each step, the dimension of the space is reduced because one divides by bigger and bigger groups.
The highest reduction occurs when dividing by the reparametrizations of $w_i$ which form an infinite-dimensional group (phase rotations form a finite-dimensional subgroup of them).

\index{local coordinates!transition function}%
For concreteness, it is useful to introduce explicit coordinates $x_s$ on $\mc P_{g,n}$ ($s \in \N$ since the space is infinite-dimensional).
The transition functions on the Riemann surface depend on the $x_s$ which explains why they can be used to parametrize the moduli spaces.
Describing the spheres $S_a$ by complex planes with punctures located at $z_{a,1}$, $z_{a,2}$ and $z_{a,3}$, the transition functions on the $\M_{g,n}$ circles $C_{\Lambda(ab)} = S_a \cap S_b$ for all $a < b$ can be taken to be:
\begin{equation}
	\text{on $C_{\Lambda(ab)}$:}
	\qquad
	z_a - z_{a,m}
		= \frac{q_\Lambda}{z_b - z_{b,n}},
\end{equation}
where $z_{a,m}$ and $z_{b,n}$ denote the punctures of $S_a$ and $S_b$ lying in $C_\Lambda$.
Then, the complex parameters $q_\Lambda$ with $\Lambda = 1, \ldots, \mathsf{M}_{g,n}^c$ are coordinates on the moduli space $\mc M_{g,n}$.
On the remaining $n$ circles $C_{i(a)} = S_a \cap D_i$, the transition functions can be expanded in series
\begin{equation}
	\text{on $C_{i(a)}$:}
	\qquad
	z_a - z_{a,m}
		= w_i + \sum_{N=1}^\infty p_{i,N} w_i^N,
\end{equation}
where $z_{a,m}$ is the puncture of $S_a$ lying in $C_i$.
There is no negative index in the series because the RHS must vanish for $w_i = 0$ which maps to $z_a = z_{a,m}$ (puncture location).
The complex coefficients of the series $p_{i,N}$ ($i = 1, \ldots, n$ and $N \in \N^*$) provide coordinates for the fibre.
\index{Pgn@$\mc P_{g,n}$ space!coordinates}%
Thus, coordinates for $\mc P_{g,n}$ are:
\begin{equation}
	\label{bos:eq:Pgn-coord}
	\{ x_s \}
		= \{ q_\Lambda, p_{i,N} \}.
\end{equation}
As usual, derivatives with respect to $x_s$ are abbreviated by $\pd_s$.
\begin{draft}
This set of coordinates may not cover completely $\mc P_{g,n}$ and it may be necessary to use different patches of coordinates, but locally they are sufficient.
For simplicity, we will always work with only one patch, the extension to several patches being straightforward.
\end{draft}

When the dependence in the $x_s$ must be stressed, the transition functions \eqref{bos:eq:trans-functions-za-wi} are denoted by:
\begin{equation}
	z_a
		= F_{ab}(z_b; x_s),
	\qquad
	z_a
		= f_{ai}(w_i; x_s).
\end{equation}
In this coordinate system, each parameter appears in only one transition function and it looks like one can separate the fibre from the basis.
But, this is not an invariant statement as this would not hold in other coordinate systems.
For example, one can rescale the coordinates to lump all dependence on $q_\Lambda$ in a single circle.

Since both cases are formally identical, it is convenient to fix the orientation of each $C_{\alpha}$ and to denote by $\sigma_\alpha$ (resp.\ $\tau_\alpha$) the coordinate on the left (resp.\ right) of the contour, such that the transition functions reads:
\begin{equation}
	\text{on $C_{\alpha}$:}
		\qquad
		\sigma_\alpha
			= F_\alpha(\tau_\alpha; x_s).
\end{equation}

Now that we have coordinates on $\mc P_{g,n}$, it is possible to construct tangent vectors.

\section{Tangent space}
\label{bos:sec:offshell:geometry:Pgn-tangent}

\index{Pgn@$\mc P_{g,n}$ space!vector|(}%

A tangent vector $V_s \in T \mc P_{g,n}$ corresponds to an infinitesimal variation of the coordinates on the manifold
\begin{equation}
	\delta x_s
		= \epsilon \, V_s,
\end{equation}
where $\epsilon$ is a small parameter, such that functions of $x_s$ vary as:
\begin{equation}
	\epsilon \, V_s\, \pd_s f
		= f(x_s + \epsilon \, V_s) - f(x_s).
\end{equation}

\index{local coordinates!transition function}%
The transition functions $F_\alpha$ provide an equivalent (but redundant) set of coordinates for $\mc P_{g,n}$.
Hence, vectors in $T \mc P_{g,n}$ can also be obtained by considering small variations of the transition functions $F_\alpha$:
\begin{equation}
	F_\alpha
		\longrightarrow F_\alpha + \epsilon \, \delta F_\alpha.
\end{equation}
Considering an overlap circle $C_{\alpha}$, a deformation of the transition function
\begin{equation}
	\sigma_\alpha
		= F_\alpha(\tau_\alpha)
\end{equation}
for fixed $\tau_\alpha$ can be interpreted as a change of the coordinate $\sigma_\alpha$:
\begin{equation}
	\sigma_\alpha'
		= F_\alpha(\tau_\alpha) + \epsilon \, \delta F_\alpha(\tau_\alpha)
		= \sigma_\alpha + \epsilon \, \delta F_\alpha(\tau_\alpha)
		= \sigma_\alpha + \epsilon \, \delta F_\alpha\big(F_\alpha^{-1}(\sigma_\alpha)\big).
\end{equation}
This transformation is generated by a vector field $v^{(\alpha)}$ on the Riemann surface $\Sigma_{g,n}$:
\begin{equation}
	\sigma_\alpha'
		= \sigma_\alpha + \epsilon v^{(\alpha)}(\sigma_\alpha),
	\qquad
	v^{(\alpha)}
		= \delta F_\alpha \circ F_\alpha^{-1}.
\end{equation}
The situation is symmetrical and one can obviously fix $\sigma_\alpha$ and vary $\tau_\alpha$.
The vector field is regular around the circle $C_\alpha$ (to have a well-defined changes of coordinates) but it can have singularities away from the circle $C_{\alpha}$.
Hence, the vector field $v^{(\alpha)}$ together with the circle $C_\alpha$ define a vector of $\mc P_{g,n}$:
\begin{equation}
	V^{(\alpha)}
		\sim \big( v^{(\alpha)}, C_{(\alpha)} \big).
\end{equation}

This provides a basis of $T \mc P_{g,n}$.
This is sufficient when using the coordinate system \eqref{bos:eq:Pgn-coord}, but, in more general situations, one needs to consider linear combinations.
For example, if a modulus appears in several transitions functions, then the associated vector field will be defined on the corresponding circles.
A general vector $V$ is described by a vector field $v$ with support on a subset $C$ of the circles $C_\alpha$:
\begin{equation}
	\label{bos:eq:Pgn-vector}
	V
		\sim \big( v, C \big),
	\qquad
	C
		\subseteq \bigcup_{\alpha} C_\alpha,
\end{equation}
and the restriction of $v$ on the various circles is written as:
\begin{equation}
	v|_{C_{\alpha}}
		= v^{(\alpha)}.
\end{equation}
Note that the vector field $v^{(\alpha)}$ and its complex conjugate $\bar v^{(\alpha)}$ are independent and are associated to different tangent vectors.
This construction is called the Schiffer variation.

The simplest tangent vectors $\pd_s$ are given by varying one coordinate of $\mc P_{g,n}$ while keeping the other fixed:
\begin{equation}
	x_s
		\longrightarrow x_s + \epsilon \, \delta x_s.
\end{equation}
On each circle $C_{\alpha}$, this gives a deformation of the transition functions
\begin{equation}
	C_\alpha:
	\quad
	F_\alpha
		\longrightarrow F_{\alpha} + \epsilon \, \delta F_{\alpha},
	\qquad
	\delta F_{\alpha}
		= \frac{\pd F_{\alpha}}{\pd x_s} \, \delta x_s
\end{equation}
(no sum over $s$), such that the change of coordinates reads
\begin{equation}
	\sigma_\alpha'
		= \sigma_\alpha + \epsilon \, v^{(\alpha)}_s(\sigma_\alpha) \, \delta x_s,
	\qquad
	v_s^{(\alpha)}(\sigma_\alpha)
		= \frac{\pd F_\alpha}{\pd x_s}\big(F_\alpha^{-1}(\sigma_\alpha)\big).
\end{equation}
If the $x_s$ are given by \eqref{bos:eq:Pgn-coord}, the vectors have support in only one circle.

There is, however, a redundancy in these vectors.
Not all of them leads to a motion in $\mc P_{g,n}$ because some modifications can be absorbed with a reparametrization of the $z_a$.
For example, if a given $v^{(\alpha)}$ can be extended holomorphically outside the circle $C_\alpha$ in the neighbour sphere, then its effect can be undone by reparametrizing the corresponding coordinate.
A similar discussion holds for the other spaces and relations can be found by restricting the vector on subspaces.
A non-trivial vector $(v^{(i)}, C_i)$ of $\mc P_{g,n}$ becomes trivial on $\mc M_{g,n}$ if it can be cancelled with a reparametrization of $w_i$ which leaves the origin fixed.

\index{Pgn@$\mc P_{g,n}$ space!vector|(}%

\section{Plumbing fixture}
\label{bos:sec:offshell:geometry:plumbing}

\index{plumbing fixture}%

The plumbing fixture is a way to glue together two Riemann surfaces (separating case) or two parts of the same surface (non-separating case) together, in order to build a surface with a higher number of holes and punctures.
This geometric operation will correspond precisely to the concept of gluing two Feynman graphs with a propagator in Siegel gauge.

The plumbing fixture depends on a (complex) one-parameter, which leads to a family of surfaces.
This provides the correct number of moduli for the surface obtained after gluing.
This brings to the question of describing the moduli spaces $\mc M_{g,n}$ in terms of the moduli spaces with lower genus and number of punctures.

\subsection{Separating case}
\label{bos:sec:geometry:plumbing:sep}

\index{plumbing fixture!separating|(}%

Consider two Riemann surfaces $\Sigma_{g_1,n_1}$ and $\Sigma_{g_2,n_2}$ with local coordinates $w_1^{(1)}$, …, $w_{n_1}^{(1)}$ and $w_1^{(2)}$, …, $w_{n_2}^{(2)}$.

\begin{figure}[htp]
	\centering
	\includegraphics[scale=1]{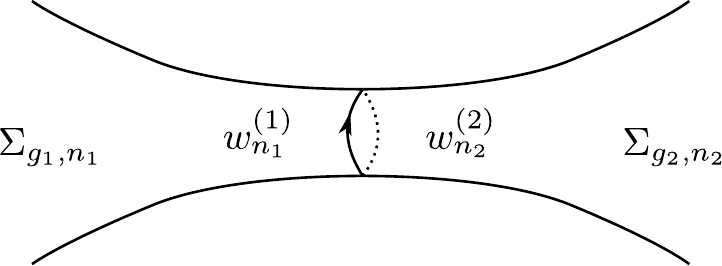}
	\caption{Integration contour on the circle between the two local coordinates which are glued together.}
	\label{bos:fig:gluing-contour-Bq}
\end{figure}

\begin{figure}[htp]
	\centering
	\includegraphics[scale=1]{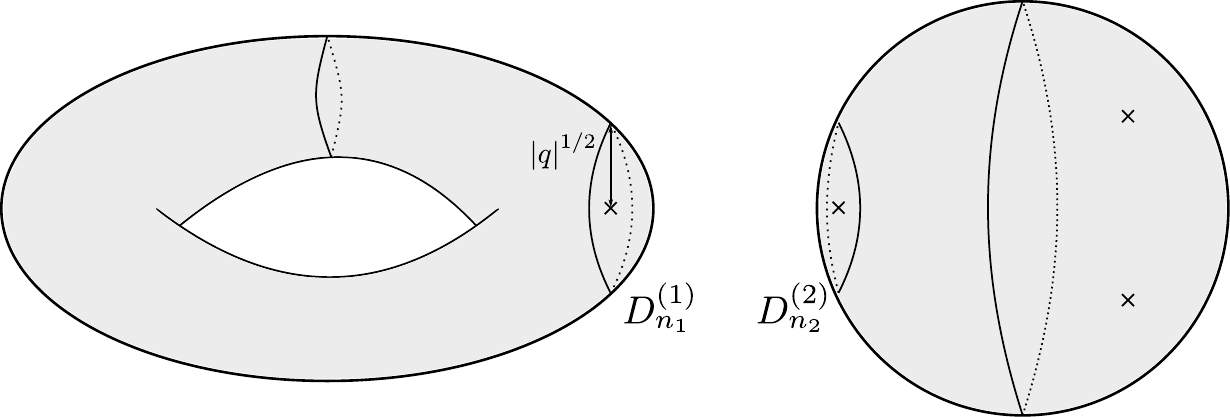}
	\caption{Disks around one puncture of the surfaces $\Sigma_{1,1}$ and $\Sigma_{0,3}$.
		The disks appear as a cap because it is on top of the surface, which is curved.}
	\label{bos:fig:gluing-sep-S11-S03-disks}
\end{figure}

The first step is to cut two disks $D_q^{(1)}$ and $D_q^{(2)}$ of radius $\abs{q}^{1/2}$ around a puncture on each surface, taken to be the $n_1$-th and $n_2$-th for definiteness:
\begin{equation}
	D_q^{(1)}
		= \big\{ \abs{w_{n_1}^{(1)}} \le \abs{q}^{1/2} \big\},
	\qquad
	D_q^{(2)}
		= \big\{ \abs{w_{n_2}^{(2)}} \le \abs{q}^{1/2} \big\},
\end{equation}
where $q \in \C$ is fixed (\Cref{bos:fig:gluing-sep-S11-S03-disks}).\footnotemark{}
\footnotetext{%
	The disks $D_q^{(i)}$ should be equal or smaller than the disks $D_{n_1}^{(1)}$ and $D_{n_2}^{(2)}$.
}%
\index{$\#$}%
Then, both surfaces can be glued (indicated by the binary operation
$\#$) together into a new surface
\begin{equation}
	\label{bos:eq:Sgn-gluing-sep-op}
	\Sigma_{g,n} = \Sigma_{g_1,n_1} \# \Sigma_{g_2,n_2},
	\qquad
	\begin{cases}
	g = g_1 + g_2,
	\\
	n = n_1 + n_2 - 2
	\end{cases}
\end{equation}
by removing the disks $D_q^{(1)}$ and $D_q^{(2)}$ and by identifying the circles $\pd D_q^{(1)}$ and $\pd D_q^{(2)}$.
\index{plumbing fixture}%
At the level of the coordinates, this is achieved by the \emph{plumbing fixture} operation:
\begin{equation}
	\label{bos:eq:plumbing}
	w_{n_1}^{(1)} w_{n_2}^{(2)}
		= q,
	\qquad
	\abs{q} \le 1,
\end{equation}
The restriction on $q$ arises because we have $\abs{w_{n_1}^{(1)}}, \abs{w_{n_2}^{(2)}} \le 1$ (for a discussion, see~\cite{Erler:2020:FourLecturesClosed}).
This case is called \emph{separating} because cutting the new tube splits the surface in two components.
Locally, the new surface looks like \Cref{bos:fig:gluing-contour-Bq}.
It is also convenient to parametrize $q$ as
\begin{equation}
	\label{bos:eq:q-s-th}
	q
		= \e^{- s + \I\theta},
	\qquad
	s \in \R_+,
	\qquad
	\theta \in [0, 2\pi).
\end{equation}
\index{plumbing fixture!moduli}%
The parameters $s$ and $\theta$ are interpreted below as moduli of the Riemann surface.

\begin{figure}[htp]
	\begin{adjustwidth}{-1cm}{-1cm}
	\centering
	\subcaptionbox{Direct gluing of circles.\label{bos:fig:gluing-sep-S11-S03-sharp}}{%
		\centering
		\includegraphics[scale=1]{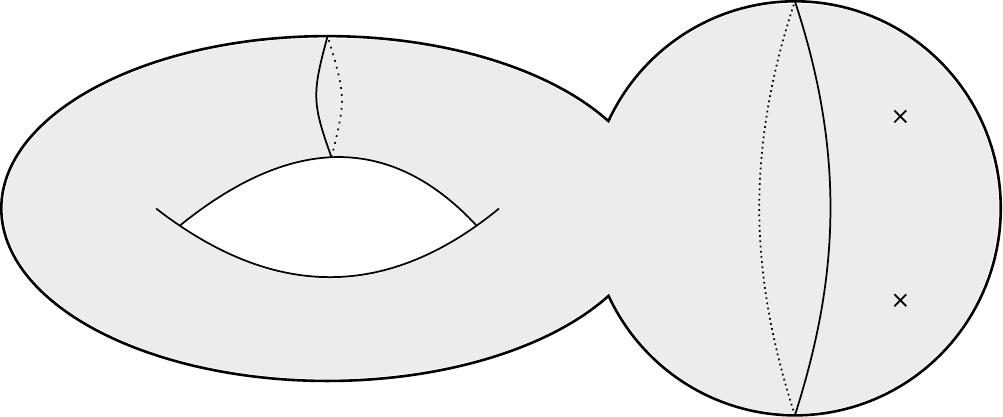}
	}%
	\\
	\subcaptionbox{Connection by a long tube.\label{bos:fig:gluing-sep-S11-S03-tube}}{%
		\centering
		\includegraphics[scale=1]{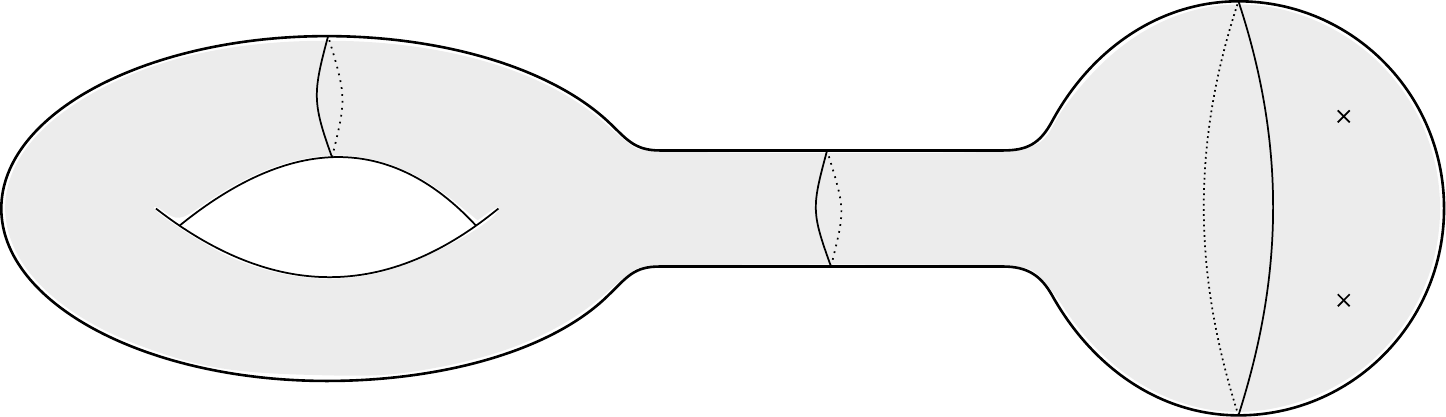}
	}%
	\\
	\subcaptionbox{Insertion of the second surface into the first one.\label{bos:fig:gluing-sep-S11-S03-insert-LR}}{%
		\centering
		\includegraphics[scale=1]{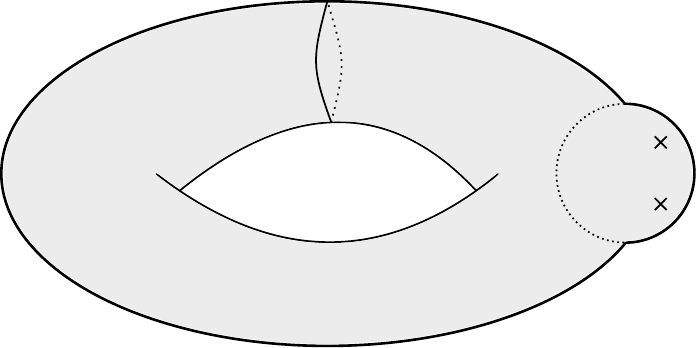}
	}%
	\hspace{1.5cm}
	\subcaptionbox{Insertion of the first surface into the second one.\label{bos:fig:gluing-sep-S11-S03-insert-RL}}{%
		\centering
		\includegraphics[scale=1]{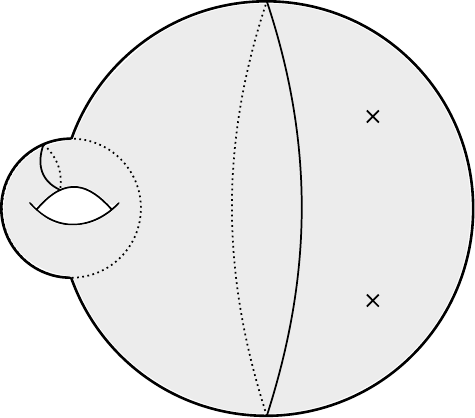}
	}%
	\end{adjustwidth}
	\caption{Different representations of the surface $\Sigma_{1,2}$ obtained after gluing $\Sigma_{1,1}$ and $\Sigma_{0,3}$ through the plumbing fixture.}
	\label{bos:fig:gluing-sep-S11-S03}
\end{figure}

\noindent
The geometry of the new surface can be viewed in three different ways:
\begin{enumerate}
	\item both surfaces $\Sigma_{g_1,n_1}$ and $\Sigma_{g_2,n_2}$ (with the disks removed) are connected directly at their boundaries (\Cref{bos:fig:gluing-sep-S11-S03-sharp});

	\item both surfaces $\Sigma_{g_1,n_1}$ and $\Sigma_{g_2,n_2}$ (with the disks removed) are connected by a cylinder of finite size (\Cref{bos:fig:gluing-sep-S11-S03-tube});

	\item the surface $\Sigma_{g_2,n_2}$ is inserted inside the disk $D^{(1)}_q$, or conversely $\Sigma_{g_1, n_1}$ inside $D^{(2)}_q$ (\Cref{bos:fig:gluing-sep-S11-S03-insert-LR,bos:fig:gluing-sep-S11-S03-insert-RL}).
\end{enumerate}

\begin{figure}[htp]
	\centering
	\includegraphics[scale=1]{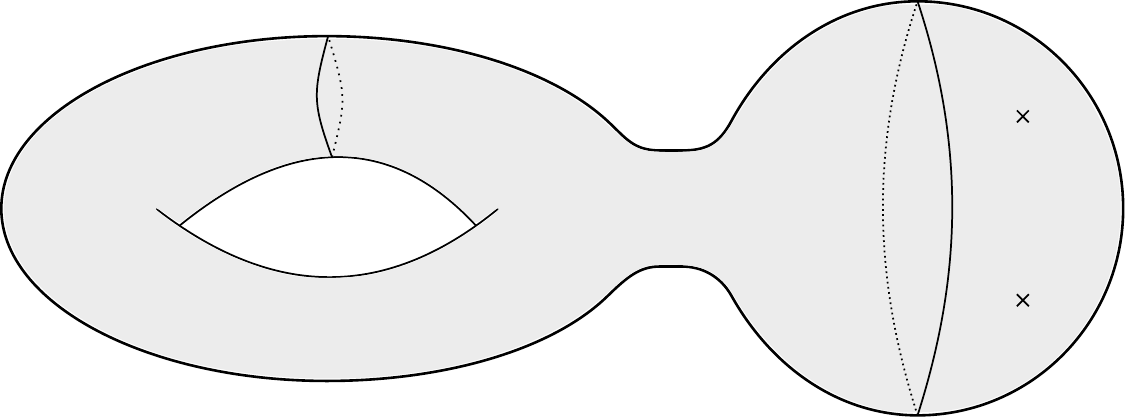}
	\caption{Smoothed connection between both surfaces.}
	\label{bos:fig:gluing-sep-S11-S03-smooth}
\end{figure}

The first interpretation is the most direct one: the disks are simply removed and the boundaries are glued together by a small tube of radius $\abs{w_1^{(1)}} < \abs{q}^{1/2}$.
The connection between both surfaces can be smoothed (and figures are often drawn in this way -- for example \Cref{bos:fig:gluing-sep-S11-S03-smooth}), but this is not necessary (the smoothing is achieved by cutting disks of size $\abs{q}^{1/2} - \epsilon$ and gluing the boundaries to the disks of radius $\abs{q}^{1/2}$).

In the second interpretation, one rescales the local coordinates in order to bring the radius of the disk to $1$ instead of $\abs{q}^{1/2}$.
In terms of these new coordinates, the surfaces are connected by a tube of length $s = - \ln \abs{q}$ after a Weyl transformation (see~\cite[sec.~9.3]{Polchinski:2005:StringTheory-1} for a longer discussion).

\begin{draft}
The radius of the tube is of order $\abs{q}^{1/2}$.
It can be set to be of order $1$ by rescaling the coordinate (now describing $\Sigma_{g_2, n_2}$) $w_{n_1}^{(1)} = q^{1/2} w'^{(1)}_{n_1}$.
Remembering that the metric is taken to be flat around the puncture $\abs{\dd w^{(1)}_{n_1}}^2$, one obtains a new metric $\abs{q} \, \abs{\dd w'^{(1)}_{n_1}}^2$.
The conformal factor can be removed through a Weyl transformation $\e^{2 \omega}$ with $\omega = - \ln \abs{q}^{1/2}$.
The effect is to also multiply the length of the tube, which becomes of order $s = - \ln \abs{q}$.
Another way to find this result is by changing coordinates to
\begin{equation}
	w_{n_1}^{(1)}
		= \e^{r + \I \phi},
	\qquad
	w_{n_2}^{(2)}
		= q \, \e^{- r - \I \phi},
\end{equation}
with $r \in \R$ and $\phi \in [0, 2\pi)$.
Then, if the original coordinates are valid for $\abs{w_{n_1}^{(1)}} < 1$ and $\abs{w_{n_2}^{(2)}} < 1$, the range of $r$ changes as:
\begin{equation}
	s
		= \ln \abs{q} < r < 0.
\end{equation}
Thus, the coordinates $r$ and $\phi$ describe a tube of length $s$ and of circumference $2\pi$.
\end{draft}

The last interpretation is obtained by performing a conformal mapping of the second case: the region $\abs{w_{n_1}^{(1)}} < \abs{q}^{1/2}$ is mapped to the region
\begin{equation}
	\abs{w_{n_2}^{(2)}}
		= \frac{\abs{q}}{\abs{w_{n_1}^{(1)}}}
		> \abs{q}^{1/2},
\end{equation}
and conversely.
The idea is that the disk $D_q^{(1)}$ of $\Sigma_{g_1,n_1}$ is removed and replaced by the complement of $D_q^{(2)}$ in $\Sigma_{g_2,n_2}$, i.e.\ the full surface $\Sigma_{g_2,n_2} - D_q^{(2)}$ is glued inside $D_q^{(1)}$.
While it is clear geometrically, this statement may look confusing from the coordinate point of view because the local coordinates $w_{n_1}^{(1)}$ and $w_{n_2}^{(2)}$ do not cover completely the Riemann surfaces, but their relation still encodes information about the complete surface.
The reason is that one can always use transition functions to relate the coordinates on the two surfaces.

\begin{example}
	Denote by $S^{(1)}_a$ and $S^{(2)}_b$ the spheres sharing a boundary with $D_{n_1}^{(1)}$ and $D_{n_2}^{(2)}$, and write the corresponding coordinates by $z_a^{(1)}$ and $z_b^{(2)}$ such that the transition functions are
	\begin{equation}
		z_a^{(1)}
			= f^{(1)}_{a n_1}(w_{n_1}^{(1)}),
		\qquad
		z_b^{(2)}
			= f^{(2)}_{b n_2}(w_{n_2}^{(2)}).
	\end{equation}
	Then the coordinates $z_a$ and $z_b$ are related by
	\begin{equation}
		z_a^{(1)}
			= f^{(1)}_{a n_1}\big(w_{n_1}^{(1)} \big)
			= f^{(1)}_{a n_1}\left( \frac{q}{w_{n_2}^{(2)}} \right)
			= f^{(1)}_{a n_1}\left( \frac{q}{f^{(2)-1}_{b n_2}\big(z_b^{(2)}\big)} \right)
	\end{equation}
	such that the new transition function reads
	\begin{equation}
		z_a
			= F_{ab}(z_b),
		\qquad
		F_{ab}
			= f^{(1)}_{a n_1} \circ (q \cdot I) \circ f^{(2)-1}_{b n_2},
	\end{equation}
	where $I$ is the inversion (the superscript on the coordinates $z_a$ and $z_b$ has been removed to indicate that they are now seen as coordinates on the same surface $\Sigma_{g,n}$).
\end{example}

\index{moduli space!plumbing fixture decomposition}%
The Riemann surface $\Sigma_{g,n}$ is a point of $\mc M_{g,n}$.
By varying the moduli parameters of $\Sigma_{g_1, n_1}$ and $\Sigma_{g_2, n_2}$, one obtains other surfaces in $\mc M_{g,n}$.
But the number of parameters furnished by $\Sigma_{g_1,n_1}$ and $\Sigma_{g_2,n_2}$ does not match the dimension \eqref{bos:eq:dim-Mgn} of $\mc M_{g,n}$:
\begin{equation}
	\mathsf{M}_{g_1,n_1} + \mathsf{M}_{g_2,n_2}
		= 6 g_1 - 6 + 2 n_1
			+ 6 g_2 - 6 + 2 n_2
		= \mathsf{M}_{g,n} - 2.
\end{equation}
This means that the subspace of $\mc M_{g,n}$ obtained by gluing all the possible surfaces in $\mc M_{g_1, n_1}$ and $\mc M_{g_2, n_2}$ is of codimension $2$.
The missing complex parameter is $q$: in writing the plumbing fixture, it was taken to be fixed, but it can be varied to generate a $2$-parameter family of Riemann surfaces in $\mc M_{g,n}$, with the moduli of the original surfaces held fixed.

\index{local coordinates!constraints}%
\index{gluing compatibility}%
The surface $\Sigma_{g,n}$ is equipped with local coordinates inherited from the original surfaces $\Sigma_{g_1,n_1}$ and $\Sigma_{g_2,n_2}$.
Hence, the plumbing fixture of points in $\mc P_{g_1, n_1}$ and $\mc P_{g_2, n_2}$ automatically leads to a point of $\mc P_{g,n}$.
The fact that the local coordinates are inherited from lower-order surfaces is called \emph{gluing compatibility}.
It is also not necessary to add parameters to describe the fibre direction.

\index{plumbing fixture!separating|)}%

\subsection{Non-separating case}
\label{bos:sec:geometry:plumbing:nonsep}

\index{plumbing fixture!non-separating|(}%

In the previous section, the plumbing fixture was used to glue punctures on two different surfaces.
\index{$\#$}%
In fact, one can also glue two punctures on the same surface to get a new surface with an additional handle:
\begin{equation}
	\label{bos:eq:Sgn-gluing-nonsep-op}
	\Sigma_{g,n}
		= \# \Sigma_{g_1,n_1},
	\qquad
	\begin{cases}
		g = g_1 + 1,
		\\
		n = n_1 - 2,
	\end{cases}
\end{equation}
defining $\#$ as a unary operator.
This gluing is called \emph{non-separating} because there is a single surface before the identification of the disks.

\begin{draft}
\begin{remark}[Plumbing fixture operation $\#$]
	The operation $\#$ is binary (resp.\ unary) for the separating (non-separating) plumbing fixture.
	Parenthesis will be used to remove ambiguity in case where $\#$ could be interpreted either as binary or unary.
	In topology, the connected sum $\#$ is defined similarly, except that one does not consider punctures on surfaces.
\end{remark}
\end{draft}

In terms of the local coordinates, the gluing relation reads
\begin{equation}
	w_{n_1-1}^{(1)} w_{n_1}^{(1)}
		= q,
\end{equation}
where we consider the last two punctures for definiteness.

\index{moduli space!plumbing fixture decomposition}%
The dimensions of both moduli spaces are related by
\begin{equation}
	\mathsf{M}_{g_1,n_1}
		= \mathsf{M}_{g,n} - 2.
\end{equation}
Again, the two missing parameters are provided by varying $q$ and we obtain a $\M_{g,n}$-dimensional subspace of $\mc M_{g,n}$.

\index{plumbing fixture!non-separating|)}%

\begin{example}
	Here are some examples of surfaces obtained by gluing:
	\begin{multicols}{2}
	\begin{itemize}
		\item $\Sigma_{0,4} = \Sigma_{0,3} \# \Sigma_{0,3}$
		\item $\Sigma_{0,5} = \Sigma_{0,3} \# \Sigma_{0,3} \# \Sigma_{0,3}, \Sigma_{0,3} \# \Sigma_{0,4}$
		\item $\Sigma_{1,1} = \# \Sigma_{0,3}$
		\item $\Sigma_{1,2} = \# \Sigma_{0,4}, \Sigma_{1,1} \# \Sigma_{0,3}$
	\end{itemize}
	\end{multicols}
	Note that the moduli on the LHS and RHS are fixed (we will see later that not all surfaces can be obtained by gluing).
\end{example}

\subsection{Decomposition of moduli spaces and degeneration limit}
\label{bos:sec:geometry:plumbing:decomposition}

\index{moduli space!plumbing fixture decomposition}%
We have seen that the separating and non-separating plumbing fixtures yield a family of surfaces in $\mc M_{g,n}$ described in terms of lower-dimensional moduli spaces.
The question is whether all points in $\mc M_{g,n}$ can be obtained in this way by looking at all the possible gluing (varying $g_1$, $n_1$, $g_2$ and $n_2$).
It turns out that this is not possible, which is at the core of the difficulties to construct a string field theory.

\index{degeneration limit}%
Which surfaces are obtained from this construction?
In order to interpret the regions of $\mc M_{g,n}$ covered by the plumbing fixture, the parametrization \eqref{bos:eq:q-s-th} is the most useful.
Previously, we explained that $s$ gives the size of the tube connecting the two surfaces.
Since the latter is like a sphere with two punctures, it corresponds to a cylinder (interpreted as an intermediate closed string propagating).
The angle $\theta$ in \eqref{bos:eq:q-s-th} is the twist of the cylinder connecting both components.
This amounts to start with $\theta = 0$, then to cut the cylinder, to twist it by an angle $\theta$ and to glue again.

\index{degeneration limit}%
The limit $s \to \infty$ ($\abs{q} \to 0$) is called the \emph{degeneration limit}: the degenerate surface $\Sigma_{g,n}$ reduces to $\Sigma_{g_1, n_1}$ and $\Sigma_{g_2, n_2}$ connected by a very long tube attached to two punctures (separating case), or to $\Sigma_{g-1, n+2}$ with a very long handle (non-separating case).
So it means that the family of surfaces described by the plumbing fixture are “close” to degeneration.
Another characterization (for the separating case) is that the punctures on $\Sigma_{g_1, n_1}$ are closer (according to some distance, possibly after a conformal transformation) to each other than to the punctures on $\Sigma_{g_2, n_2}$.

Conversely, there are surfaces which cannot be described in this way: the plumbing fixture does not cover all the possible values of the moduli.
For a given $\mc M_{g,n}$, we denote the surfaces which cannot be obtained by the plumbing fixture by $\mc V_{g,n}$.
This space does not contain any surface arbitrarily close to degeneration (i.e.\ with long handles or tubes).
In terms of punctures, it also means that there is no conformal frame where the punctures split in two sets.

\begin{figure}[htp]
	\centering
	\subcaptionbox{Degeneration $1 2 \to 3 4$}{%
		\includegraphics[scale=0.7]{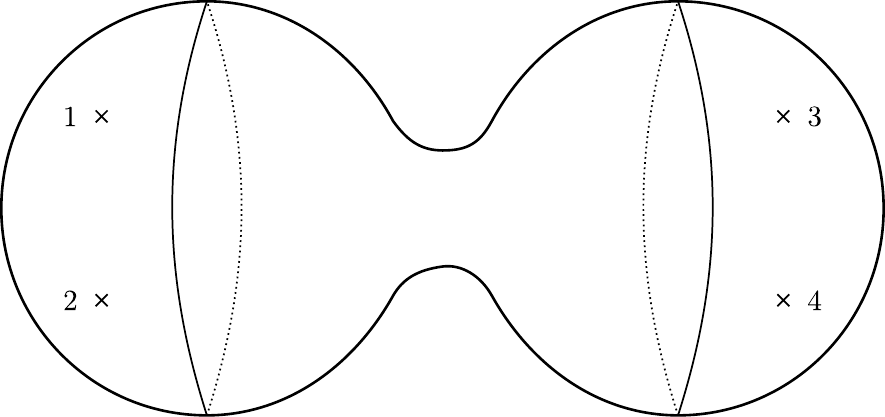}
	}%
	\qquad
	\subcaptionbox{Degeneration $1 3 \to 2 4$}{%
		\includegraphics[scale=0.7]{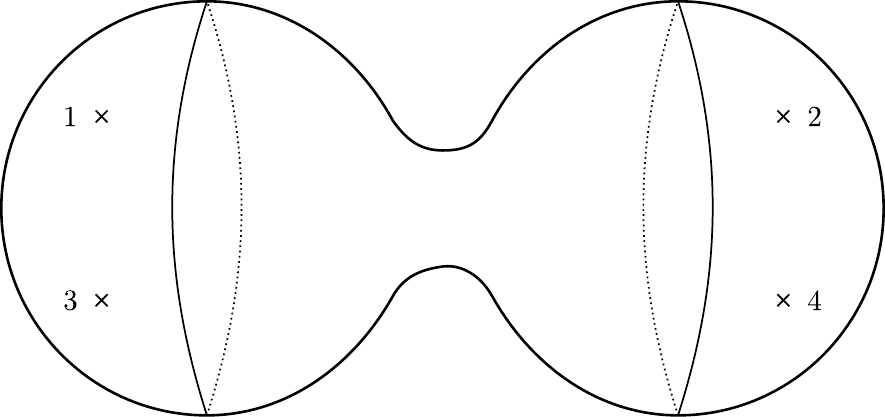}
	}%
	\\
	\medskip
	\subcaptionbox{Degeneration $1 4 \to 2 3$}{%
		\includegraphics[scale=0.7]{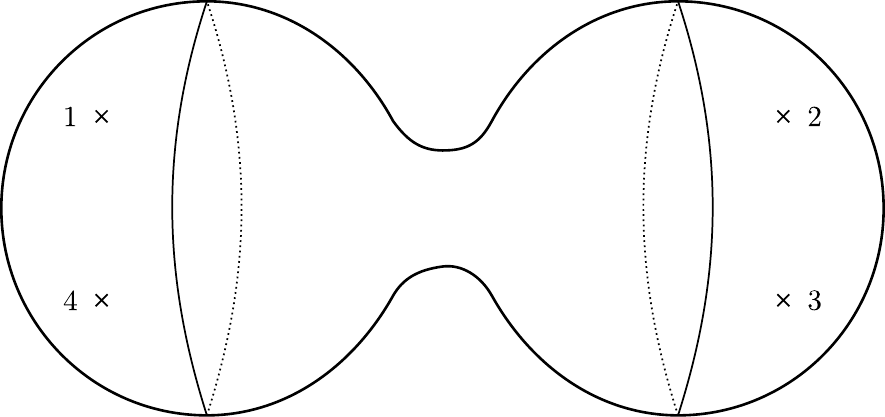}
	}%
	\caption{Permutations of punctures while gluing two spheres: they correspond to different (disconnected) parts of $\mc M_{0,4}$.}
	\label{bos:fig:gluing-S03-S03-perms}
\end{figure}

In the previous subsection, we considered two specific punctures, but any other punctures could be chosen.
Hence, there are many ways to split $\Sigma_{g,n}$ in two surfaces $\Sigma_{g_1,n_1}$ and $\Sigma_{g_2,n_2}$ (with fixed $g_1$, $g_2$, $n_1$ and $n_2$): every partition of the punctures and holes in two sets lead to different degeneration limits (because they are associated to different moduli -- \Cref{bos:fig:gluing-S03-S03-perms}).
Since each puncture is described by a modulus, choosing different punctures for gluing give different set of moduli for $\Sigma_{g,n}$, such that each possibility covers a different subspace of $\mc M_{g,n}$.
The part of the moduli space $\mc M_{g,n}$ covered by the plumbing fixture of all surfaces $\Sigma_{g_1,n_1}$ and $\Sigma_{g_2,n_2}$ (with fixed $g_1$, $g_2$, $n_1$, $n_2$) is denoted by $\mc M_{g_1,n_1} \# \mc M_{g_2,n_2}$:
\begin{equation}
	\mc M_{g_1, n_1} \# \mc M_{g_2, n_2}
		\subset \mc M_{g,n},
\end{equation}
where the operation $\#$ includes the plumbing fixture for all values of $q$ and all pairs of punctures.
Similarly, the part covered by the non-separating plumbing fixture is written as $\# \mc M_{g_1,n_1}$:
\begin{equation}
	\# \mc M_{g_1,n_1}
		\subset \mc M_{g,n}.
\end{equation}
Importantly, the regions covered by the plumbing fixture depend on the choice of the local coordinates because \eqref{bos:eq:plumbing} is written in terms of local coordinates.
The subspaces $\mc M_{g_1,n_1} \# \mc M_{g_2,n_2}$ and $\# \mc M_{g_1,n_1}$ are not necessarily connected (in the topological sense).

\index{moduli space!plumbing fixture decomposition|(}%

\begin{figure}
	\centering
	\includegraphics[scale=1.2]{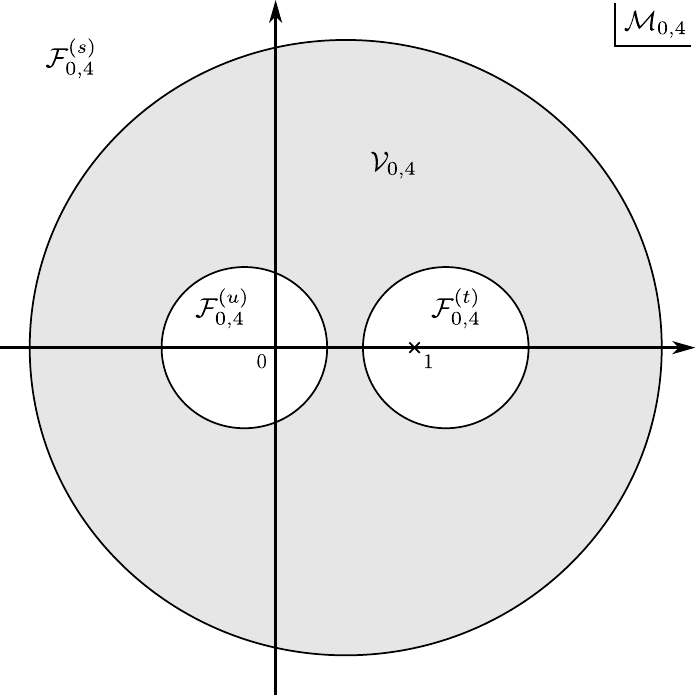}
	\caption{%
		In white are the subspaces of the moduli space $\mc M_{0,4}$ covered by the plumbing fixture.
		The three different regions correspond to the three different ways to pair the punctures (see \Cref{bos:fig:gluing-S03-S03-perms}).
		In grey is the fundamental vertex region $\mc V_{0,4}$.
	}%
	\label{bos:fig:M04-covering}
\end{figure}

\begin{figure}[htp]
	\centering
	\includegraphics[scale=1.2]{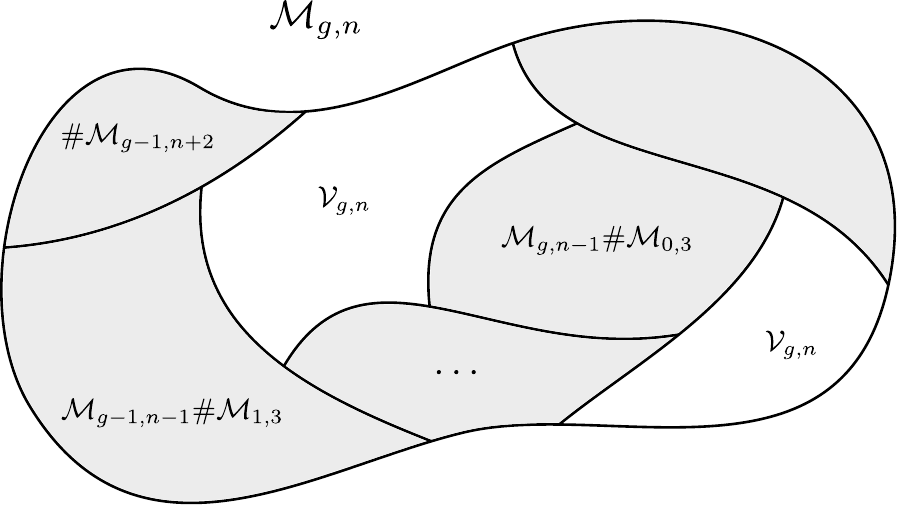}
	\caption{%
		Schematic illustration of the covering of $\mc M_{g,n}$ from the plumbing fixture of lower-dimensional spaces.
		The fundamental region $\mc V_{g,n}$ (usually disconnected) is not covered by the plumbing fixture.
	}%
	\label{bos:fig:Mgn-covering-gluing}
\end{figure}

\index{fundamental vertex!region}%
\index{propagator region}%
The moduli space $\mc M_{g,n}$ cannot be completely covered by the plumbing fixture of lower-dimensional surfaces.
We define the propagator and fundamental vertex regions $\mc F_{g,n}$ and $\mc V_{g,n}$ as the subspaces which can and cannot be described by the plumbing fixture:
\begin{subequations}
\begin{align}
	\label{bos:eq:def-Fgn-space}
	\mc F_{g,n}
		&
		:=
			\# \mc M_{g - 1, n + 2}
			\quad {\textstyle \bigcup} \quad
			\bigg(
				\bigcup_{\cramped{\substack{
					n_1 + n_2 = n + 2 \\
					g_1 + g_2 = g }}}
				\mc M_{g_1,n_1} \# \mc M_{g_2,n_2}
			\bigg),
	\\
	\label{bos:eq:def-Vgn-space}
	\mc V_{g,n}
		&
		:= \mc M_{g,n}
			- \mc F_{g,n},
\end{align}
\end{subequations}
In the RHS, it is not necessary to consider multiple non-separating plumbing fixtures for the first term because $\# \mc M_{g - 2, n + 4} \subset \mc M_{g - 1, n + 2}$, etc.
For the same reason, it is sufficient to consider a single separating plumbing fixture.
Note that $\mc V_{g,n}$ and $\mc F_{g,n}$ are in general not connected subspaces.
A simple illustration is given in \Cref{bos:fig:Mgn-covering-gluing}.
The actual decomposition of $\mc M_{0,4}$ is given in \Cref{bos:fig:M04-covering}.
Importantly, $\mc F_{g,n}$ and $\mc V_{g,n}$ depend on the choice of the local coordinates for all $\mc V_{g',n'}$ appearing in the RHS.

\index{1PI vertex!region}%
\index{1PR region}%
It is also useful to define the subspaces $\mc F_{g,n}^{\text{1PR}}$ and $\mc V_{g,n}^{\text{1PI}}$ of $\mc M_{g,n}$ which can and cannot be described with the separating plumbing fixture only:
\begin{subequations}
\begin{align}
	\label{bos:eq:def-Fgn-1PR-space}
	\mc F_{g,n}^{\text{1PR}}
		&
		:= \bigcup_{\cramped{\substack{
				n_1 + n_2 = n + 2 \\
				g_1 + g_2 = g }}}
			\mc M_{g_1,n_1} \# \mc M_{g_2,n_2},
	\\
	\label{bos:eq:def-Vgn-1PI-space}
	\mc V_{g,n}^{\text{1PI}}
		&
		:= \mc M_{g,n}
			- \mc F_{g,n}^{\text{1PR}}.
\end{align}
\end{subequations}
1PR (1PI) stands for $1$-particle (ir)reducible, a terminology which will become clear later.
Note the relation:
\begin{equation}
	\mc V_{g,n}^{\text{1PI}}
		= \mc V_{g,n}
			\quad {\textstyle \bigcup} \quad
			\bigg(
				\bigcup_{g'}
				\# \mc M_{g-g', n+g'}
			\bigg).
\end{equation}

The two plumbing fixtures behave as follow:
\begin{itemize}
	\item separating: increases both $n$ and $g$ (if both surfaces have a non-vanishing $g$);

	\item non-separating plumbing: increases $g$ but decreases $n$.
\end{itemize}
The construction is obviously recursive: starting from the lowest-dimensional moduli space, which is $\mc M_{0,3}$ (no moduli), one has:
\begin{equation}
	\mc V_{0,3}
		= \mc M_{0,3},
	\qquad
	\mc F_{0,3}
		= \emptyset.
\end{equation}
Next, the subspace of $\mc M_{0,4}$ obtained from the plumbing fixture is:
\begin{equation}
	\mc F_{0,4}
		= \mc V_{0,3} \# \mc V_{0,3},
\end{equation}
and $\mc V_{0,4}$ is characterized as the remaining region.
Then, one has:
\begin{equation}
	\begin{aligned}
	\mc F_{0,5}
		&
		= \mc M_{0,4} \# \mc M_{0,3}
		\\ &
		= \mc F_{0,4} \# \mc V_{0,3}
			+ \mc V_{0,4} \# \mc V_{0,3}
		= \mc V_{0,3} \# \mc V_{0,3} \# \mc V_{0,3}
			+ \mc V_{0,4} \# \mc V_{0,3},
	\end{aligned}
\end{equation}
and $\mc V_{0,5}$ is what remains of $\mc M_{0,5}$.
The pattern continues for $g = 0$.
The same story holds for $g \ge 1$: the first such space is
\begin{equation}
	\mc F_{1,1}
		= \# \mc V_{0,3},
\end{equation}
and $\mc V_{1,1} = \mc M_{1,1} - \mc F_{1,1}$.
The gluing of a $3$-punctured sphere and the addition of a handle are the two most elementary operations.

\index{index!Riemann surface}%
To keep track of which moduli spaces can contribute, it is useful to find a function of $\Sigma_{g,n}$, called the index, which increases by $1$ for each of the two elementary operations:
\begin{equation}
	r(\Sigma_{g_1,n_1} \# \Sigma_{0, 3})
		= r(\Sigma_{g_1,n_1}) + 1,
	\qquad
	r(\# \Sigma_{g_1,n_1})
		= r(\Sigma_{g_1,n_1}) + 1.
\end{equation}
An appropriate function is
\begin{equation}
	\label{bos:eq:Sgn-index}
	r(\Sigma_{g,n})
		= 3 g + n - 2
		\in \N^*.
\end{equation}
which is normalized such that:
\begin{equation}
	r(\Sigma_{0,3})
		= 1.
\end{equation}
For a generic separating plumbing fixture, we find:
\begin{equation}
	r(\Sigma_{g_1,n_1} \# \Sigma_{g_2,n_2})
		= r(\Sigma_{g_1,n_1}) + r(\Sigma_{g_2,n_2}).
\end{equation}
Since the index increases, surfaces with a given $r$ can be obtained by considering all the gluings of surfaces with $r' < r$.

\index{moduli space!plumbing fixture decomposition|)}%

\subsection{Stubs}
\label{bos:sec:offshell:geometry:plumbing:stubs}

\begin{figure}
	\centering
	\includegraphics[scale=1.2]{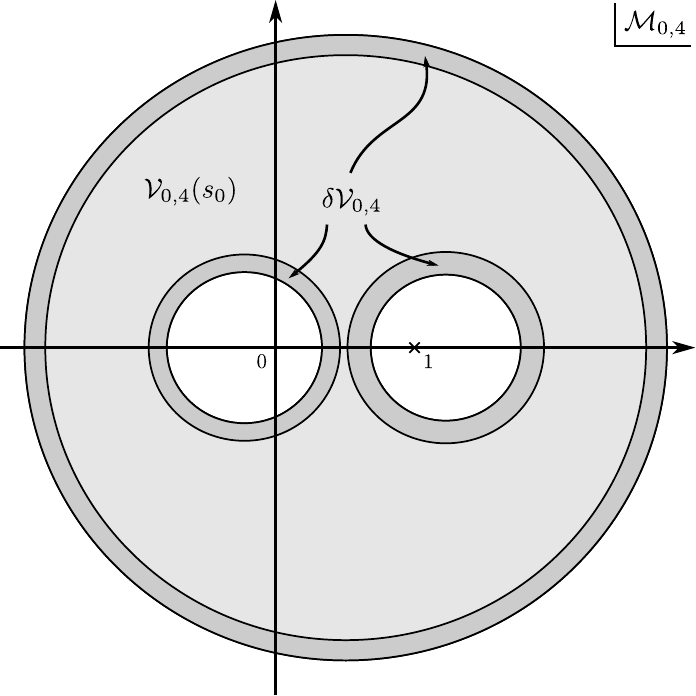}
	\caption{%
		In light grey is the subspace covered by the $\mc V_{0,4}(s_0)$ as in \Cref{bos:fig:M04-covering}.
		In dark grey is the difference $\delta \mc V_{0,4} = \mc V_{0,4}(s_0 + \delta s_0) - \mc V_{0,4}(s_0)$ with $\delta s_0 > 0$.
	}%
	\label{bos:fig:M04-covering-stub}
\end{figure}

\index{stub|(}%
To conclude this chapter, we introduce the concept of \emph{stubs}.
Previously in \eqref{bos:eq:q-s-th}, the range of the parameter $s$ was the complete line of positive numbers, $s \in \R_+$.
\index{stub!parameter}%
This means that tubes of all lengths were considered to glue surfaces.
But, we could also introduce a minimal length $s_0 > 0$, called the \emph{stub parameter}, for the tube.
In this case, the plumbing fixture parameter is generalized to:
\begin{equation}
	\label{bos:eq:q-s-th-stub}
	q
		= \e^{- s + \I\theta},
	\qquad
	s \in [s_0, \infty),
	\qquad
	\theta \in [0, 2\pi),
	\qquad
	s_0 \ge 0.
\end{equation}
What is the effect on the subspaces $\mc F_{g,n}(s_0)$ and $\mc V_{g,n}(s_0)$?
Obviously, less surfaces can be described by the plumbing fixture if $s_0 > 0$ than if $s_0 = 0$, since the plumbing fixture cannot describe anymore surfaces which contain a tube of length less than $s_0$.
\index{moduli space!plumbing fixture decomposition|)}%
Equivalently, the values of the moduli described by the plumbing fixture is more restricted when $s_0 > 0$.
More generally, one has:
\begin{equation}
	s_0 < s_0':
	\qquad
	\mc F_{g,n}(s_0')
		\subset \mc F_{g,n}(s_0)
	\qquad
	\mc V_{g,n}(s_0)
		\subset \mc V_{g,n}(s_0').
\end{equation}
This is illustrated on \Cref{bos:fig:M04-covering-stub}.
Even if $s_0$ is very large, $\mc V_{g,n}$ still does not include surfaces arbitrarily close to degeneracy.
In general, we omit the dependence in $s_0$ except when it is necessary.

To interpret the stub parameter, consider two local coordinates $w_1$ and $w_2$ and rescale them by $\lambda \in \C$ with $\Re \lambda > 0$:
\begin{equation}
	w_1
		= \lambda \, \tilde w_1,
	\qquad
	w_2
		= \lambda \, \tilde w_2.
\end{equation}
Then, the plumbing fixture \eqref{bos:eq:plumbing} becomes
\begin{equation}
	\tilde w_1 \tilde w_2
		= \e^{- \tilde s + \I \tilde\theta}.
\end{equation}
with
\begin{equation}
	\label{bos:eq:coord-s-th-rescaling}
	\tilde s
		= s + 2 \ln \abs{\lambda},
	\qquad
	\tilde \theta
		= \theta + \I \ln \frac{\lambda}{\bar \lambda}.
\end{equation}
If $s \in \R_+$, the corresponding range of $\tilde s$ is
\begin{equation}
	\tilde s
		\in [s_0, \infty),
	\qquad
	s_0
		:= 2 \ln \abs{\lambda}.
\end{equation}
\index{local coordinates!global rescaling}%
This shows that rescaling the local coordinates by a constant parameter is equivalent to change the stub parameter.

\index{stub|)}%

Note also how performing a global phase rotation in \eqref{bos:eq:coord-s-th-rescaling} is equivalent to shift the twist parameter.
Working in $\hat{\mc P}_{g,n}$ forces to take $\lambda \in \R_+$.

\begin{draft}

\section{BV structure}

\end{draft}

\section{Summary}

In this chapter, we have explained how to parametrize the fibre bundle $\mc P_{g,n}$, that is, appropriate coordinates for the moduli space and the local coordinate systems.
This was realized by introducing different coordinate patches and encoding all the informations of $\mc P_{g,n}$ in the transition functions.
Then, this description lead to a simple description of the tangent vectors through the Schiffer variation.

In the next chapter, we will continue the program by building the $p$-forms required to describe off-shell amplitudes.

\refchapter

\begin{itemize}
	\item Plumbing fixture~\cite[sec.~9.3]{Polchinski:2005:StringTheory-1}.
\end{itemize}

\chapter{Off-shell amplitudes}
\label{bos:chap:offshell-amp}

\introchapter

While the previous chapter was purely geometrical, this one makes contact with string theory through the worldsheet CFT.
We continue the description of $\mc P_{g,n}$ by constructing $p$-forms.
The reason why we need to consider the CFT is that ghosts are necessary to build the $p$-forms: this can be understood from \Cref{chap:bos:ws-int-vac}, where we found that the ghosts must be interpreted as part of the measure on the moduli space.
Then, we build the off-shell amplitudes and discuss some properties.

\section{Cotangent spaces and amplitudes}
\label{bos:sec:offshell:geometry:Pgn-cotangent}

In this section, we construct the $p$-forms on $\mc P_{g,n}$ which are needed for the amplitudes.
We first motivate the expressions from general ideas, and check later that they have the correct properties.

\subsection{Construction of forms}

\index{Pgn@$\mc P_{g,n}$ space!$p$-form}%

A $p$-form $\omega_p^{(g,n)} \in \bigwedge^p T^* \mc P_{g,n}$ is a multilinear antisymmetric map from $\bigwedge^p T \mc P_{g,n}$ to a function of the moduli parameters.
The superscript on the form is omitted when there is no ambiguity about the space considered.
The components $\omega_{i_1 \cdots i_p}$ of the $p$-form are defined by inserting $p$ basis vectors $\pd_{s_1}, \ldots, \pd_{s_p}$
\begin{equation}
	\omega_{i_1 \cdots i_p}
		:= \omega_p(\pd_{s_1}, \ldots, \pd_{s_p}),
\end{equation}
where $\pd_s = \frac{\pd}{\pd x_s}$ and $x_s$ are the coordinates \eqref{bos:eq:Pgn-coord}.
It is antisymmetric in any pair of two indices
\begin{equation}
	\omega_{i_1 i_2 \cdots i_p}
		= - \omega_{i_2 i_1 \cdots i_p},
\end{equation}
and multilinearity implies that
\begin{equation}
	\omega_p\big(V^{(1)}, \ldots, V^{(p)}\big)
		= \omega_p\big(V^{(1)}_{s_1} \pd_{s_1}, \ldots, V^{(p)}_{s_p} \pd_{s_p}\big)
		= \omega_{i_1 \cdots i_p} V^{(1)}_{s_1} \cdots V^{(p)}_{s_p},
\end{equation}
given vectors $V^{(\alpha)} = V^{(\alpha)}_s \pd_s$.

The $p$-forms which are needed to define off-shell amplitudes depend on the external states $\scr V_i$ ($i = 1, \ldots, n$) inserted at the punctures $z_i$.
They are maps from $\bigwedge^p T \mc P_{g,n} \times \mc H^n$ to a function on $\mc P_{g,n}$.
The dependence on the states is denoted equivalently as
\begin{equation}
	\label{bos:eq:omega-p-states-dep}
	\omega_p(\scr V_1, \ldots, \scr V_n)
		:= \omega_p(\otimes_i \scr V_i).
\end{equation}
The simplest way to get a function on $\mc P_{g,n}$ from the states $\scr V_i$ is to compute a CFT correlation function of the operators inserted at the points $z_i = f_i(0)$ on the surface $\Sigma_{g,n}$ described by the point in $\mc M_{g,n}$.

\index{Pgn@$\mc P_{g,n}$ space!0-form}%
The $0$-form is just a function and is defined by:
\begin{equation}
	\omega_0
		= (2\pi \I)^{- \M_{g,n}^c}
			\Mean{\prod_{i=1}^n f_i \circ \scr V_i(0)}_{\mathrlap{\Sigma_{g,n}}}.
\end{equation}
For simplicity, the dependence in the local coordinates $f_i$ is kept implicit in the rest of the chapter.

\index{Pgn@$\mc P_{g,n}$ space!1-form|(}%

A natural approach for constructing $p$-forms is to build them from elementary $1$-forms and to use ghosts to enforce the antisymmetry.
Remembering the Beltrami differentials found in \Cref{chap:bos:ws-int-vac}, the contour integral of ghosts $b(z)$ weighted by some vector field is a good starting point.
In the current language, it is defined by its contraction with a vector $V = (v, C) \in T \mc P_{g,n}$ defined in \eqref{bos:eq:Pgn-vector}:
\begin{equation}
	\label{bos:eq:B-form-V}
	B(V)
		:= \oint_{C} \frac{\dd z}{2\pi \I} \, b(z) v(z)
			+ \oint_{C} \frac{\dd \bar z}{2\pi \I} \, \bar b(\bar z) \bar v(\bar z),
\end{equation}
where $b(z)$ and $\bar b(\bar z)$ are the $b$-ghost components, and $v$ is the vector field on $\Sigma_{g,n}$ defining $V$.
The contours run anti-clockwise.
If the contour $C$ includes several circles ($C = \cup_\alpha C_\alpha$), $B(V)$ is defined as the sum of the contour integral on each circle:
\begin{equation}
	B(V)
		:= \sum_\alpha \oint_{C_\alpha} \frac{\dd z}{2\pi \I} \, b(z) v(z)
			+ \cc
\end{equation}

It is also useful to define another object built from the energy--momentum tensor:
\begin{equation}
	\label{bos:eq:T-form-V}
	T(V)
		:= \oint_{C} \frac{\dd z}{2\pi \I} \, T(z) v(z)
			+ \oint_{C} \frac{\dd \bar z}{2\pi \I} \, \bar T(\bar z) \bar v(\bar z),
\end{equation}
where $T$ and $\bar T$ are the components of the energy--momentum tensor.
It is defined such that
\begin{equation}
	\label{bos:eq:T-form-V-Qb-B}
	T(V)
		= \anticom{Q_B}{B(V)}.
\end{equation}

Considering the coordinate system \eqref{bos:eq:Pgn-coord}, the Beltrami form can be decomposed as:
\begin{subequations}
\label{bos:eq:beltrami-form}
\begin{gather}
	B
		= B_s \dd x_s,
	\qquad
	B_s
		:= B(\pd_s),
	\\
	B_s
		= \sum_\alpha \oint_{C_\alpha}
			\frac{\dd \sigma_\alpha}{2\pi \I} \,
				b(\sigma_\alpha) \,
				\frac{\pd F_\alpha}{\pd x_s}\big(F_\alpha^{-1}(\sigma_\alpha)\big)
			+ \sum_\alpha \oint_{C_\alpha} \frac{\dd \bar\sigma_\alpha}{2\pi \I} \,
				\bar b(\bar\sigma_\alpha) \,
				\frac{\pd \bar F_\alpha}{\pd x_s}\big(\bar F_\alpha^{-1}(\bar\sigma_\alpha)\big),
\end{gather}
\end{subequations}
where the contour orientations are defined by having the $\sigma_\alpha$ coordinate system on the left.

\index{Pgn@$\mc P_{g,n}$ space!1-form|)}%

\index{Pgn@$\mc P_{g,n}$ space!$p$-form}%
We define the $p$-form contracted with a set of vectors $V^{(1)}$, …, $V^{(p)}$ by
\begin{equation}
	\label{bos:eq:omega-p-V}
	\omega_p\big(V^{(1)}, \ldots, V^{(p)}\big)(\scr V_1, \ldots, \scr V_n)
		:= (2\pi \I)^{- \M_{g,n}^c}
			\Mean{B(V^{(1)}) \cdots B(V^{(p)}) \prod_{i=1}^n \scr V_i}_{\mathrlap{\Sigma_{g,n}}},
\end{equation}
and the corresponding $p$-form reads
\begin{subequations}
\label{bos:eq:omega-p}
\begin{align}
	\omega_p
		&= \omega_{p,s_1 \cdots s_p} \,
			\dd x_{s_1} \wedge \cdots \wedge \dd x_{s_p}
		\\
		&= (2\pi \I)^{- \M_{g,n}^c}
			\Mean{B_{s_1} \dd x_{s_1} \wedge \cdots \wedge B_{s_p} \dd x_{s_p} \prod_{i=1}^n \scr V_i}_{\mathrlap{\Sigma_{g,n}}}.
\end{align}
\end{subequations}
In this expression, the form contains an infinite numbers of components $\omega_{p,s_1 \cdots s_p}$ since there is an infinite number of coordinates.
Note that the normalization is independent of $p$.

In practice, one is not interested in $\mc P_{g,n}$, but rather in a subspace of it.
Given a $q$-dimensional subspace $\mc S$ of $\mc P_{g,n}$ parametrized by $q$ real coordinates $t_1$, …, $t_q$
\begin{equation}
	x_s
		= x_s(t_1, \ldots, t_q),
\end{equation}
the restriction of a $p$-form to this subspace is obtained by the chain rule:
\begin{equation}
	\label{bos:eq:omega-p-restriction}
	\begin{gathered}
		\forall p \le q:
			\quad
			\omega_p|_{\mc S}
				= (2\pi \I)^{- \M_{g,n}^c}
					\Mean{B_{r_1} \frac{\pd x_{s_1}}{\pd t_{r_1}} \, \dd t_{r_1} \wedge \cdots \wedge B_{r_p} \frac{\pd x_{s_p}}{\pd t_{r_p}} \, \dd t_{r_p}
					\prod_{i=1}^n \scr V_i}_{\mathrlap{\Sigma_{g,n}}},
		\\
		\forall p > q:
			\quad
			\omega_p|_{\mc S} = 0.
	\end{gathered}
\end{equation}
We will often write the expression directly in terms of the coordinates of $\mc S$ and abbreviate the notation as:
\begin{equation}
	B_r
		:= \frac{\pd x_s}{\pd t_r} \, B_s.
\end{equation}

\subsection{Amplitudes and surface states}
\label{bos:sec:offshell:geometry:surface-states}

It is now possible to write the amplitude more explicitly.
An on-shell amplitude is defined as an integral over $\mc M_{g,n}$.
Off-shell, one needs to consider local coordinates around each puncture, that is, a point of the fibre for each point of the base $\mc M_{g,n}$.
This defines a $\M_{g,n}$-dimensional section $\mc S_{g,n}$ of $\mc P_{g,n}$ (\Cref{bos:fig:Pgn-section}).
\index{off-shell closed string amplitude}%
The $g$-loop $n$-point off-shell amplitude of the states $\scr V_1, \ldots, \scr V_n$ reads:
\begin{subequations}
\label{bos:eq:Agn-offshell-Sgn}
\begin{gather}
	A_{g,n}(\scr V_1, \ldots, \scr V_n)_{\mc S_{g,n}}
		:= \int_{\mc S_{g,n}} \omega^{g,n}_{\M_{g,n}}(\scr V_1, \ldots, \scr V_n)\big|_{\mc S_{g,n}},
	\\
	\omega^{g,n}_{\M_{g,n}}(\scr V_1, \ldots, \scr V_n)\big|_{\mc S_{g,n}}
		= (2\pi \I)^{- \M_{g,n}^c}
			\Mean{ \bigwedge_{\lambda=1}^{\M_{g,n}} B_{s} \frac{\pd x_s}{\pd t_\lambda} \, \dd t_\lambda
			\prod_{i=1}^n f_i \circ \scr V_i(0)}_{\mathrlap{\Sigma_{g,n}}},
\end{gather}
\end{subequations}
where the choice of the $f_i$ is dictated by the section $\mc S_{g,n}$.
From now on, we stop to write the restriction of the form to the section.
We also restrict to the cases where $\chi_{g,n} = 2 - 2 g - n < 0$.

The complete (perturbative) $n$-point amplitude is the sum of contributions from all loops:
\begin{equation}
	\label{bos:eq:An-offshell}
	A_{n}(\scr V_1, \ldots, \scr V_n)
		:= \sum_{g \ge 0} A_{g,n}(\scr V_1, \ldots, \scr V_n).
\end{equation}

\index{off-shell closed string amplitude!contribution from subspace}%
More generally, we define the integral over a section $\mc R_{g,n}$ which projection on the base is a subspace of $\mc M_{g,n}$ (and not the full space as for the amplitude) as:
\begin{equation}
	\mc R_{g,n}(\scr V_1, \ldots, \scr V_n)
		:= \int_{\mc R_{g,n}} \omega^{g,n}_{\M_{g,n}}(\scr V_1, \ldots, \scr V_n),
\end{equation}
For simplicity, we will sometimes use the same notation for the section of $\mc P_{g,n}$ and its projection on the base $\mc M_{g,n}$.
For this reason, the reader should assume that some choice of local coordinates around the punctures is made except otherwise stated.

Given sections $\mc R_{g,n}$, the sum over all genus contribution is written formally as
\begin{equation}
	\label{bos:eq:int-omega-An}
	\mc R_{n}
		:= \sum_{g \ge 0} \mc R_{g,n},
\end{equation}
such that
\begin{equation}
	\mc R_{n}(\scr V_1, \ldots, \scr V_n)
		:= \sum_{g \ge 0} \mc R_{g,n}(\scr V_1, \ldots, \scr V_n)
		= \sum_{g \ge 0} \int_{\mc R_{g,n}} \omega^{g,n}_{\M_{g,n}}(\scr V_1, \ldots, \scr V_n).
\end{equation}

\index{surface state}%
A \emph{surface state} is defined as a $n$-fold bra which reproduces the expression of a given function when contracted with $n$ states $A_i$.
The surface $\bra{\Sigma^{g,n}}$, form $\bra{\omega^{g,n}}$, section $\bra{\mc R_{g,n}}$ and amplitude $\bra{A^{g,n}}$ $n$-fold states are defined by the following expressions:
\begin{subequations}
\begin{gather}
	\label{bos:eq:state-surface}
	\bra{\Sigma_{g,n}} B_{s_1} \cdots B_{s_p} \ket{\otimes_i \scr V_i}
		:= \omega_{s_1 \cdots s_p}(\scr V_1, \ldots, \scr V_n),
	\\
	\label{bos:eq:state-form}
	\bracket{\omega_p^{g,n}}{\otimes_i \scr V_i}
		:= \omega_{p}(\scr V_1, \ldots, \scr V_n),
	\\
	\label{bos:eq:state-amplitude}
	\bracket{A_{g,n}}{\otimes_i \scr V_i}
		:= A_{g,n}(\scr V_1, \ldots, \scr V_n).
\end{gather}
The last relation is generalized to any section $\mc R_{g,n}$:
\begin{equation}
	\label{bos:eq:state-section}
	\bracket{\mc R_{g,n}}{\otimes_i \scr V_i}
		:= \mc R_{g,n}(\scr V_1, \ldots, \scr V_n).
\end{equation}
\end{subequations}
The reason for introducing these objects is that the form \eqref{bos:eq:omega-p} is a linear map from $\mc H^{\otimes n}$ to a form on $\mc M_{g,n}$ -- see \eqref{bos:eq:omega-p-states-dep}.
Thus, there is always a state $\bra{\Sigma_{g,n}}$ such that its BPZ product with the states reproduces the form.
In particular, the state $\bra{\Sigma_{g,n}}$ contains all the information about the local coordinates and the moduli (the dependence is kept implicit).
The definition of the other states follow similarly.
These states are defined as bras, but they can be mapped to kets.

One finds the obvious relations:
\begin{equation}
	\bra{\omega_p^{g,n}}
		= \bra{\Sigma_{g,n}} B_{s_1} \dd x^{s_1} \cdots B_{s_p} \dd x^{s_p},
	\qquad
	\bra{A_{g,n}}
		= \int_{\mc M_{g,n}} \bra{\omega_p^{g,n}}.
\end{equation}

The surface states don't contain information about the matter CFT: they collect the universal data (like local coordinates) needed to describe amplitudes.
Hence, it is an important step in the description of off-shell string theory to characterize this data.
However, note that the relation between a surface state and the corresponding form \emph{does depend} on the CFT.

\begin{example}[On-shell amplitude $A_{0,4}$]

	\index{closed string amplitude!tree-level!4-point}%

	The transition functions are given by (see \Cref{bos:fig:S04-param}):
	\begin{equation}
		\begin{aligned}
			C_1:
				w_1 &= z_1 - y_1, &
			\qquad
			C_3:
				w_3 &= z_2 - y_3,
			\qquad
			C_5:
				z_1 = z_2,
			\\
			C_2:
				w_2 &= z_1 - y_2, &
			\qquad
			C_4:
				w_4 &= z_2 - y_4.
		\end{aligned}
	\end{equation}
	Three of the parameters ($y_1$, $y_2$ and $y_3$) are fixed while the single complex modulus of $\mc M_{0,4}$ is taken to be $y_4$.
	Since we are interested in the on-shell amplitude, it is not necessary to introduce local coordinates and the associated parameters.

	A variation of the modulus
	\begin{equation}
		y_4
			\longrightarrow y_4 + \delta y_4,
		\qquad
		\bar y_4
			\longrightarrow \bar y_4 + \delta \bar y_4
	\end{equation}
	is equivalent to a change in the transition function of $C_4$.
	This translates in turn into a transformation of $z_2$:
	\begin{equation}
		z_2'
			= z_2 + \delta y_4,
		\qquad
		\bar z_2'
			= \bar z_2 + \delta \bar y_4.
	\end{equation}
	Then, the tangent vector $V = \pd_{y_4}$ is associated to the vector field
	\begin{equation}
		v = 1,
		\qquad
		\bar v = 0,
	\end{equation}
	with support on $C_4$.
	For $V = \pd_{\bar y_4}$, one finds
	\begin{equation}
		v = 0,
		\qquad
		\bar v = 1.
	\end{equation}
	The Beltrami $1$-form for the unit vectors are
	\begin{equation}
		B(\pd_{y_4})
			= \oint_{C_4} \dd z_2 \, b(z_2) (+ 1),
		\qquad
		B(\pd_{\bar y_4})
			= \oint_{C_4} \dd \bar z_2 \, \bar b(\bar z_2) (+ 1),
	\end{equation}
	with both contours running anti-clockwise.

	The components of the $2$-form reads
	\begin{align*}
		\omega_2(\pd_{y_4}, \pd_{\bar y_4})
			&
			= \frac{1}{2\pi \I} \Mean{B(\pd_{y_4}) B(\pd_{\bar y_4})
				\prod_{i=1}^4 \scr V_i}_{\Sigma_{0,4}}
			\\ &
			= \frac{1}{2\pi \I} \Mean{
				\oint_{C_4} \dd z_2 \, b(z_2) \oint_{C_4} \dd \bar z_2 \, \bar b(\bar z_2)
				\prod_{i=1}^4 \scr V_i}_{\mathrlap{\Sigma_{0,4}}}.
	\end{align*}
	For on-shell states $\scr V_i = c \bar c V_i(y_i, \bar y_i)$, this becomes
	\[
		\omega_2(\pd_{y_4}, \pd_{\bar y_4})
			= \frac{1}{2\pi \I} \Mean{
				\prod_{i=1}^3 c \bar c V_i(y_i, \bar y_i)
				\oint_{C_4} \dd z_2 \, b(z_2) \oint_{C_4} \dd \bar z_2 \, \bar b(\bar z_2)
				\bar c(\bar y_4) c(y_4) V_4(y_4, \bar y_4)}_{\mathrlap{\Sigma_{0,4}}}.
	\]
	The first three operators could be moved to the left because they are not encircled by the integration contour.
	Note the difference with the example discussed in \Cref{bos:sec:offshell:motivations:4pt}: here, the contour encircles $z_3$, while it was encircling $y_3$ for the $s$-channel.

	Using the OPE
	\begin{equation}
		\oint_{C_4} \dd z_2 \, b(z_2) c(y_4)
			\sim \oint_{C_4} \dd z_2 \, \frac{1}{z_2 - y_4}
	\end{equation}
	to simplify the product of $b$ and $c$ gives the amplitude
	\begin{equation}
		A_{0,4}
			= \frac{1}{2\pi \I} \int \dd y_4 \wedge \dd \bar y_4 \,
				\Mean{\prod_{i=1}^3 c \bar c V_i(y_i) \, V_4(y_4)}_{\mathrlap{\Sigma_{0,4}}}.
	\end{equation}
	This is the standard formula for the $4$-point function derived from the Polyakov path integral.
\end{example}

\section{Properties of forms}
\label{bos:sec:offshell-amp:form-prop}

\index{Pgn@$\mc P_{g,n}$ space!$p$-form!properties|(}%

In this section, we check that the form \eqref{bos:eq:omega-p} has the correct properties:
\begin{itemize}
	\item antisymmetry under exchange of two vectors;

	\item given a trivial vector of (a subspace of) $\mc P_{g,n}$ (\Cref{bos:sec:offshell:geometry:Pgn-param}), its contraction with the form vanishes: $\omega_p(V^{(1)}, \ldots, V^{(p)}) = 0$ if any of the $V^{(i)}$ generates:
	\begin{itemize}
		\item reparametrizations of $z_a$ for $V^{(i)} \in T \mc P_{g,n}$,
		\item rotation $w_i \to (1 + \I \alpha_i) w_i$ for $V^{(i)} \in T \hat{\mc P}_{g,n}$,
		\item reparametrizations of $w_i$ keeping $w_i = 0$ if the states are on-shell for $V^{(i)} \in T \mc M_{g,n}$;
	\end{itemize}

	\item BRST identity, which is necessary to prove several properties of the amplitudes.
\end{itemize}

The first property is obvious.
Indeed, the form is correctly antisymmetric under the exchange of two vectors $V^{(i)}$ and $V^{(j)}$ due to the ghost insertions.

\subsection{Vanishing of forms with trivial vectors}
\label{bos:sec:offshell-amp:form-prop:sym}

\paragraph{Reparametrization of $z_a$}

\begin{figure}[htp]
	\centering
	\includegraphics[scale=1]{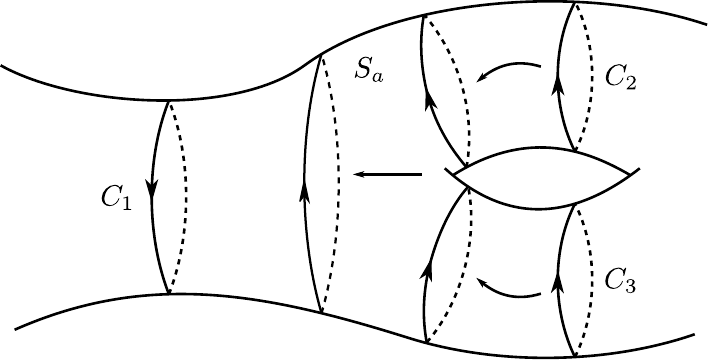}
	\caption{%
		Deformation of the contour of integration defining the Beltrami form for a reparametrization of $z_a$.
		The figure is drawn for two circles at a hole, but the proof is identical for other types of circles.
	}
	\label{bos:fig:form-reparam-z-contour}
\end{figure}

\index{local coordinates!reparametrization}%
Consider the sphere $S_a$ with coordinate $z_a$, and denote by $C_1$, $C_2$ and $C_3$ the three boundaries.
Then, a reparametrization
\begin{equation}
	z_a
		\longrightarrow z_a + \phi(z_a)
\end{equation}
is generated by a vector field $\phi(z)$ which is regular on $S_a$.
This transformation modifies the transition functions on the three circles and is thus associated to a tangent vector $V$ described by a vector field $v$ with support on the three circles:
\begin{equation}
	C_i:
		\quad
		v^{(i)} = \phi|_{C_i}.
\end{equation}
The Beltrami form then reads
\begin{equation}
	B(V)
		= \sum_{i=1}^3 \oint_{C_i} \dd z_a \, b(z_a) \phi(z_a) + \cc
\end{equation}
where the orientations of the contours are such that $S_a$ is on the left.
Since the vector field $\phi$ is regular in $S_a$, two of the contours can be deformed until they merge together.
The resulting orientation is opposite to the one of the last contour (\Cref{bos:fig:form-reparam-z-contour}).
As a consequence, both cancel and the integral vanishes.

\paragraph{Rotation of $w_i$}

\index{local coordinates!global phase}%
Consider an infinitesimal phase rotation of the local coordinate $w_i$ in the disk $D_i$:
\begin{equation}
	w_i
		\longrightarrow (1 + \I \alpha_i) w_i,
	\qquad
	\bar w_i
		\longrightarrow (1 - \I \alpha_i) \bar w_i,
\end{equation}
with $\alpha_i \in \R$.
The tangent vector is defined by the circle $C_i$ and the vector field by
\begin{equation}
	v
		= \I w_i,
	\qquad
	\bar v
		= - \I \bar w_i.
\end{equation}
The Beltrami form for this vector is
\begin{equation}
	B(V)
		= \I \oint_{C_i} \dd w_i \, w_i \, b(w_i)
			- \I \oint_{C_i} \dd \bar w_i \, \bar w_i \, \bar b(\bar w_i),
\end{equation}
where $D_i$ is kept to the left.

In the $p$-form \eqref{bos:eq:omega-p-V}, the $i$th operator $\scr V_i$ is inserted in $D_i$ and encircled by $C_i$.
Because there is no other operator inside $D_i$, the contribution of this disk to the form is
\begin{equation}
	B(V) \scr V_i(0)
		= \I \oint_{C_i} \dd w_i \, w_i \, b(w_i) \scr V_i(0)
			- \I \oint_{C_i} \dd \bar w_i \, \bar w_i \, \bar b(\bar w_i) \scr V_i(0),
\end{equation}
The state--operator correspondence allows to rewrite this result as
\begin{equation}
	\I (b_0 - \bar b_0) \ket{\scr V_i},
\end{equation}
since the contour integral picks the zero-modes of $b$ and of $\bar b$.
\index{level-matching condition}%
Requiring that the form vanishes implies the ghost counter-part of the level-matching condition:
\index{level-matching condition}%
\begin{equation}
	b_0^- \ket{\scr V_i}
		= 0.
\end{equation}
Hence, consistency of off-shell amplitudes imply that
\begin{equation}
	\scr V_i \in \mc H^-,
\end{equation}
where $\mc H^-$ is defined in \eqref{bos:eq:hilbert-closed-Hpm0}.

\paragraph{Reparametrization of $w_i$}

\index{local coordinates!reparametrization}%
A reparametrization of the local coordinate $w_i$ keeping the origin of $D_i$ fixed reads:
\begin{equation}
	w_i
		\longrightarrow f(w_i),
	\qquad
	f(0)
		= 0.
\end{equation}
The function can be expanded in series:
\begin{equation}
	f(w_i) = \sum_{m \ge 0} p_m w_i^{m+1}.
\end{equation}
Because the transformation is holomorphic, it can be extended on $C_i$.
Each parameter $p_m$ provides a coordinate of $\mc P_{g,n}$ and whose deformation corresponds to a vector field:
\begin{equation}
	v_m
		= w_i^{m+1},
	\qquad
	\bar v_m
		= 0.
\end{equation}
The corresponding Beltrami differential is
\begin{equation}
	B(\pd_{p_m})
		= \oint_{C_i} \dd w_i \, b(w_i) w_i^{m+1}.
\end{equation}
Since only the operator $\scr V_i$ is inserted in the disk, the state--operator correspondence gives $b_m \ket{\scr V_i}$.
Requiring that the form vanishes on $\mc M_{g,n}$ for all $m$ and also for the anti-holomorphic vectors gives the conditions:
\begin{equation}
	\forall m \ge 0:
	\qquad
	b_m \ket{\scr V_i}
		= 0,
	\qquad
	\bar b_m \ket{\scr V_i}
		= 0.
\end{equation}
This holds automatically for on-shell states $\scr V_i = c \bar c V_i$.

\subsection{BRST identity}

\index{Pgn@$\mc P_{g,n}$ space!$p$-form!BRST identity}%
The BRST identity for the $p$-form \eqref{bos:eq:omega-p} reads
\begin{equation}
	\label{bos:eq:omega-p-brst-identity}
	\omega_p\Big(\sum_i Q_B^{(i)} \otimes_i \scr V_i \Big)
		= (-1)^p \dd\omega_{p-1}(\otimes \scr V_i),
\end{equation}
using the notation \eqref{bos:eq:omega-p-states-dep}.
The BRST operator acting on the $i$th Hilbert space is written as
\begin{equation}
	Q_B^{(i)} = 1_{i-1} \otimes Q_B \otimes 1_{n-i}
\end{equation}
and acts as
\begin{equation}
	Q_B \scr V_i(z, \bar z)
		= \frac{1}{2\pi \I} \oint \dd w \, j_B(w) \scr V_i(z, \bar z)
			+ \cc
\end{equation}
More explicitly, the LHS corresponds to
\begin{equation}
	\begin{multlined}
		\omega_p\Big(\sum_i Q_B^{(i)} \otimes_i \scr V_i \Big)
			= \omega_p(Q_B \scr V_1, \scr V_2, \ldots, \scr V_n)
				+ (-1)^{\abs{\scr V_1}} \omega_p(\scr V_1, Q_B \scr V_2, \ldots, \scr V_n)
				\\
				+ \cdots
				+ (-1)^{\abs{\scr V_1} + \cdots + \abs{\scr V_{n-1}}} \omega_p(\scr V_1, \scr V_2, \ldots, Q_B \scr V_n).
	\end{multlined}
\end{equation}
We give just an hint of this identity, the complete proof can be found in~\cites[pp.~85--89]{Zwiebach:1993:ClosedStringField}[sec.~2.5]{Sen:2015:OffshellAmplitudesSuperstring}.

The contour of the BRST current around each puncture can be deformed, picking singularities due to the presence of the Beltrami forms.
Using \eqref{bos:eq:T-form-V-Qb-B}, we find that anti-commuting the BRST charge with the Beltrami form $B_s$ leads to an insertion of
\begin{equation}
	T_s
		= \anticom{Q_B}{B_s}.
\end{equation}
The energy--momentum tensor generates changes of coordinates.
Hence, $T_s = T_{\pd_s}$ is precisely the generator associated to an infinitesimal change of the coordinate $x_s$ on $\mc P_{g,n}$.
The latter is given by the vector $\pd_s$.
For this reason, one can write:
\begin{equation}
	\dd x_s \, \anticom{Q_B}{B_s}
		= \dd x_s \, T_s
		= \dd x_s\, \pd_s
		= \dd,
\end{equation}
where $\dd$ is the exterior derivative on $\mc P_{g,n}$.
The minus signs arise if the states $\scr V_i$ are Grassmann odd.

\index{Pgn@$\mc P_{g,n}$ space!$p$-form!properties|)}%

\section{Properties of amplitudes}
\label{bos:sec:amp-prop:properties}

\index{string amplitude!properties|(}%

\index{string amplitude!ghost number}%
\index{ghost number!anomaly}%
In order for the $p$-form \eqref{bos:eq:omega-p} to be non-vanishing, its total ghost number should match the ghost number anomaly:
\begin{equation}
	\label{bos:eq:Ngh-omega}
	N_{\text{gh}}\big(\omega_p(\scr V_1, \ldots, \scr V_n)\big)
		= \sum_{i=1}^n N_{\text{gh}}(\scr V_i) - p
		= 6 - 6 g,
\end{equation}
using $N_{\text{gh}}(B) = - 1$.
For an amplitude, one has $p = \M_{g,n} = 6 g - 6 + 2n$ and thus:
\begin{equation}
	N_{\text{gh}}(\omega_{\M_{g,n}})
		= 6 - 6 g
	\quad \Longrightarrow \quad
	\sum_{i=1}^n N_{\text{gh}}(\scr V_i)
		= 2 n.
\end{equation}
This condition holds automatically for on-shell states since $N_{\text{gh}}(c \bar c V_i) = 2$.

\subsection{Restriction to \texorpdfstring{$\hat{\mc P}_{g,n}$}{hat P(g,n)}}
\label{bos:sec:offshell:geometry:hat-Pgn}

\index{local coordinates!global phase}%
The goal of this section is to explain why amplitudes must be described in terms of a section of $\hat{\mc P}_{g,n}$ \eqref{bos:eq:def-Pgn-hat} instead of $\mc P_{g,n}$.
This means that one should identify local coordinates differing by a global phase rotation.

The off-shell amplitudes \eqref{bos:eq:Agn-offshell-Sgn} are multi-valued on $\mc P_{g,n}$.
Indeed, the amplitude depends on the local coordinates\footnotemark{} and changes by a factor under a global phase rotation of any local coordinate $w_i \to \e^{\I \alpha} w_i$.
\footnotetext{%
	The current argument does not apply for on-shell amplitudes.
}%
However, such a global rotation leaves the surface unchanged, since the flat metric $\abs{\dd w_i}^2$ is invariant.
This means that the same surface leads to different values for the amplitude.
To prevent this multi-valuedness of the amplitudes, it is necessary to identify local coordinates differing by a constant phase.

\begin{figure}[ht]
	\centering
	\subcaptionbox{Closed curve in $\mc M_{g,n}$.}[0.4\linewidth]{%
		\centering\includegraphics[scale=1]{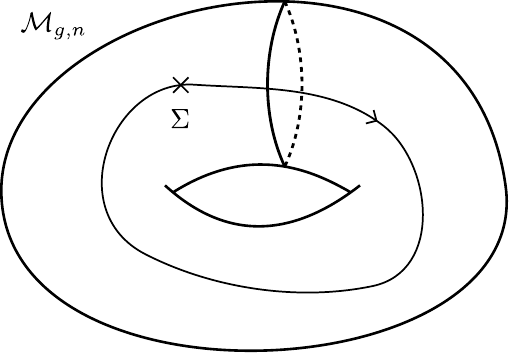}
	}%
	\hspace{1cm}
	\subcaptionbox{Change in the phase of $w_i$.}[0.4\linewidth]{%
		\includegraphics[scale=1]{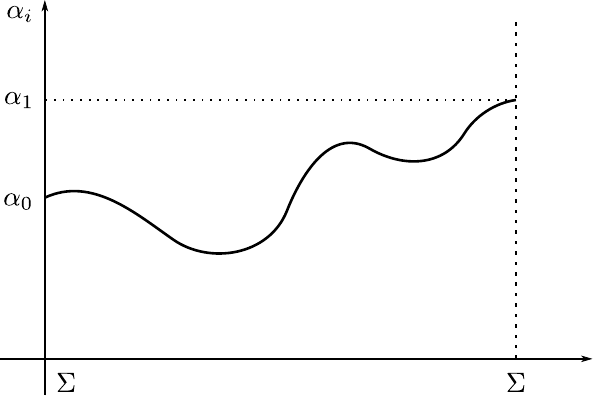}
	}%
	\caption{%
		Schematic plot of the change in the phase of the local coordinate $w_i$ as one follows a closed curve in $\mc M_{g,n}$.
		If the original phase at $\Sigma$ is $\alpha_0$ and if the phase varies continuously along the path, then $\alpha_1 \neq \alpha_0$ when returning back to $\Sigma$ by continuity.
	}%
	\label{bos:fig:Pgn-curve}
\end{figure}

A second way to obtain this condition is to require that the section $\mc S_{g,n}$ is globally defined: every point of the section should correspond to a single point of the moduli space $\mc M_{g,n}$.
However, there is a topological obstruction which prevents finding a global section in $\mc P_{g,n}$ in general.
One hint~\cites[sec.~2]{Distler:1991:TopologicalCouplingsContact}[sec.~3]{Erler:2020:FourLecturesClosed} is to exhibit a nowhere vanishing $1$-form if $\mc S_{g,n}$ is globally defined: this leads to a contradiction since such a $1$-form does not generally exist (see for example~\cite[sec.~6.3.2, ch.~7]{Donaldson:2011:RiemannSurfaces}).
Then, consider a closed curve in the moduli space (such curves exist since $\mc M_{g,n}$ is compact).
Starting at a given point $\Sigma$ of the curve, one finds that the local coordinates typically change by a global phase when coming back to the point $\Sigma$ (\Cref{bos:fig:Pgn-curve}), since this describes the same surface and there is no reason to expect the phase to be invariant.
Up to this identification, it is possible to find a global section.
The latter corresponds to a section of $\hat{\mc P}_{g,n}$.

\begin{remark}[Degeneracy of the antibracket]
	It is possible to define a BV structure on Riemann surfaces~\cite{Sen:1994:QuantumBackgroundIndependence, Sen:1996:BackgroundIndependentAlgebraic}.
	The antibracket is degenerate in $\mc P_{g,n}$ but not in $\hat{\mc P}_{g,n}$~\cite{Sen:1996:BackgroundIndependentAlgebraic}.
\end{remark}

Global phase rotations of the local coordinates are generated by $L_0^-$.
Hence, identifying the local coordinates $w_i \to \e^{\I \alpha_i} w_i$ amounts to require that the amplitude is invariant under $L_0^-$.
This is equivalent to imposing the level-matching condition
\index{level-matching condition}%
\begin{equation}
	\label{bos:eq:constraint-L0m-Ai}
	L_0^- \ket{\scr V_i}
		= 0
\end{equation}
on the off-shell states.
This condition was interpreted in \Cref{bos:sec:ws-int:brst:states} as a gauge-fixing condition for translations along the $S^1$ of the string.
This shows, in agreement with earlier comments, that the level-matching condition should also be imposed off-shell because no gauge symmetry is introduced for the corresponding transformation.

If the generator $L_0^-$ is trivial, this means that the ghost associated to the corresponding tangent vector must be decoupled.
According to \Cref{bos:sec:offshell-amp:form-prop:sym}, this corresponds to the constraint:
\begin{equation}
	b_0^- \ket{\scr V_i}
		= 0.
\end{equation}
This can be interpreted as a gauge fixing condition (\Cref{bsft:sec:free-brst:closed}), which could in principle be relaxed.
However, the decoupling of physical states (equivalent to gauge invariance in SFT) happens only after integrating over the moduli space.
This requires having a globally defined section.

As a consequence, off-shell states are elements of the semi-relative Hilbert space
\begin{equation}
	\scr V_i \in \mc H^- \cap \ker L_0^-,
\end{equation}
and the amplitudes are defined by integrating the form $\omega_{\M_{g,n}}$ over a section $\mc S_{g,n} \subset \hat{\mc P}_{g,n}$.

\begin{computation}[bos:eq:constraint-L0m-Ai]
	The operator associated to the state through $\ket{A_i} = A_i(0) \ket{0}$ transforms as
	\begin{equation}
		\scr V_i(0)
			\longrightarrow (\e^{\I \alpha_i})^{h} (\e^{-\I \alpha_i})^{\bar h} \scr V_i(0)
	\end{equation}
	which translates into
	\begin{equation}
		\ket{\scr V_i}
			\longrightarrow \e^{\I \alpha_i (L_0 - \bar L_0)} \ket{\scr V_i}
	\end{equation}
	for the state, using the fact that the vacuum is invariant under $L_0$ and $\bar L_0$.
	Then, requiring the invariance of the state leads to \eqref{bos:eq:constraint-L0m-Ai}.
\end{computation}

\subsection{Consequences of the BRST identity}

Two important properties of the on-shell amplitudes can be deduced from the BRST identity \eqref{bos:eq:omega-p-brst-identity}: the independence of physical results on the choice of local coordinates and the decoupling of pure gauge states.

\begin{figure}[htp]
	\centering
	\includegraphics[scale=1]{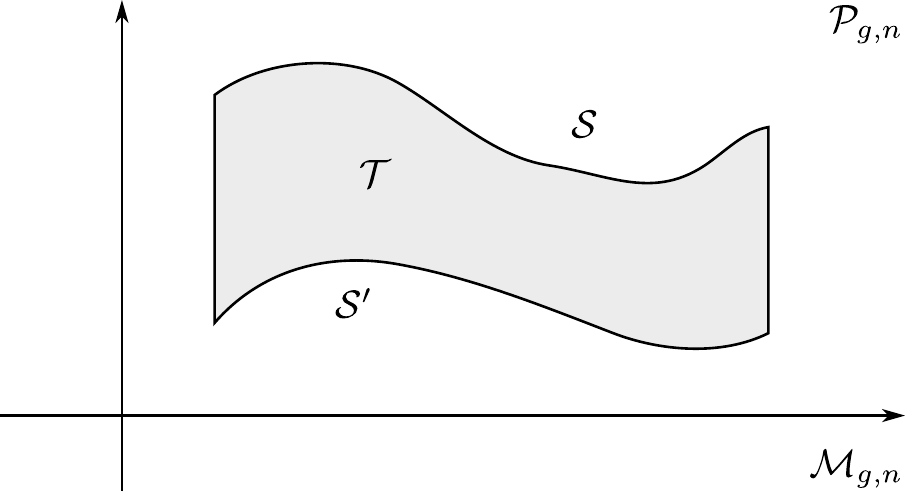}
	\caption{Two sections $\mc S$ and $\mc S'$ of $\mc P_{g,n}$ delimiting a surface $\mc T$.}
	\label{bos:fig:Ggn-onshell-section-indep}
\end{figure}

\index{string amplitude!section independence}%
Given BRST closed states, the LHS of \eqref{bos:eq:omega-p-brst-identity} vanishes identically
\begin{equation}
	\forall i:
		\quad
		Q_B \ket{\scr V_i}
			= 0
	\quad \Longrightarrow \quad
	\dd \omega_{p-1}(\scr V_1, \ldots, \scr V_n)
		= 0.
\end{equation}
Using this result, one can compare the on-shell amplitudes computed for two different sections $\mc S$ and $\mc S'$:
\begin{equation}
	\int_{\mc S} \omega_{\M_{g,n}} - \int_{\mc S'} \omega_{\M_{g,n}}
		= \int_{\pd \mc T} \omega_{\M_{g,n}-1}
		= \int_{\mc T} \dd \omega_{\M_{g,n}-1}
		= 0,
\end{equation}
using Stokes' theorem and where $\mc T$ is the surface delimited by the two sections (\Cref{bos:fig:Ggn-onshell-section-indep}).
This implies that on-shell amplitudes do not depend on the section, and thus on the local coordinates.
In obtaining the result, one needs to assume that the vertical segments do not contribute.
The latter correspond to boundary contributions of the moduli space.
In general, many statements hold up to this condition, which we will not comment more in this book.

\index{string amplitude!pure gauge states decoupling}%
Next, we consider a pure gauge state together with BRST closed states:
\begin{equation}
	\ket{\scr V_1}
		= Q_B \ket{\Lambda},
	\qquad
	Q_B \ket{\scr V_i}
		= 0.
\end{equation}
The BRST identity \eqref{bos:eq:omega-p-brst-identity} reads:
\begin{equation}
	\omega_{\M_{g,n}}(Q_B \Lambda, \scr V_2, \ldots, \scr V_n)
		= \dd \omega_{\M_{g,n}-1}(\Lambda, \scr V_2, \ldots, \scr V_n),
\end{equation}
which gives the amplitude
\begin{equation}
	\int_{\mc S} \omega_{\M_{g,n}}(Q_B \Lambda, \scr V_2, \ldots, \scr V_n)
		= \int_{\mc S} \dd \omega_{\M_{g,n}-1}(\Lambda, \scr V_2, \ldots, \scr V_n)
		= \int_{\pd\mc S} \omega_{\M_{g,n}-1}(\Lambda, \scr V_2, \ldots, \scr V_n)
\end{equation}
where the last equality follows from Stokes' theorem.
Assuming again that there is no boundary contribution, this vanishes:
\begin{equation}
	\int_{\mc S} \omega_{\M_{g,n}}(Q_B \Lambda, \scr V_2, \ldots, \scr V_n)
		= 0.
\end{equation}
This implies that pure gauge states decouple from the physical states.

\index{string amplitude!properties|)}%

\refchapter

\begin{itemize}
	\item Definition of the forms~\cite{Zwiebach:1993:ClosedStringField, Sen:2015:OffshellAmplitudesSuperstring, Erler:2017:VerticalIntegrationLarge, Erler:2020:FourLecturesClosed}.

	\item Global phase rotation of local coordinates~\cites[sec.~2]{Distler:1991:TopologicalCouplingsContact}[sec.~3]{Erler:2020:FourLecturesClosed}{Nelson:1989:CovariantInsertionGeneral}[p.~54]{Zwiebach:1993:ClosedStringField}.
\end{itemize}

\chapter{Amplitude factorization and Feynman diagrams}
\label{bos:chap:feynman}

\introchapter

In the previous chapter, we built the off-shell amplitudes by integrating forms on sections of $\mc P_{g,n}$.
Studying their factorizations lead to rewrite them in terms of Feynman diagrams, which allows to identify the fundamental interactions vertices.
We will then be able to write the SFT action in the next chapter.

\section{Amplitude factorization}
\label{bos:sec:amp-prop:factorization}

We have seen how to write off-shell amplitudes.
The next step is to rewrite them as a sum of Feynman diagrams through factorization of amplitudes.

\index{string amplitude!factorization}%
Factorization consists in writing a $g$-loop $n$-point amplitude in terms of lower-order amplitudes in both $g$ and $n$ connected by propagators.
Since an amplitude corresponds to a sum over all possible processes, which corresponds to integrating over the moduli space, it is natural to associate Feynman diagrams to different subspaces of the moduli space.
\index{plumbing fixture}%
One can expect that the plumbing fixture (\Cref{bos:sec:offshell:geometry:plumbing}) is the appropriate translation of the factorization at the level of Riemann surfaces.
We will assume that it is the case and check that it is correct a posteriori.

To proceed, we consider the contribution to the amplitude $A_{g,n}$ of the family of surfaces obtained by the plumbing fixture of two surfaces (separating case) or a surface with itself (non-separating case).

\subsection{Separating case}
\label{bos:sec:amp-prop:factorization:sep}

In this section, we consider the separating plumbing fixture where part of the moduli space $\mc M_{g,n}$ is covered by $\mc M_{g_1, n_1} \# \mc M_{g_2, n_2}$ with $g = g_1 + g_2$ and $n = n_1 + n_2 - 2$ (\Cref{bos:sec:geometry:plumbing:sep}).
The local coordinates read $w^{\cramped{(1)}}_i$ and $w^{\cramped{(2)}}_j$ for $i = 1, \ldots, n_1$ and $j = 1, \ldots, n_2$.
\index{plumbing fixture!separating}%
By convention, the last coordinate of each set is used for the plumbing fixture:
\begin{equation}
	\label{bos:eq:fact-plumbing-sep}
	w^{(1)}_{n_1} w^{(2)}_{n_2}
		= q.
\end{equation}

The $g$-loop $n$-point amplitude with external states $\{ \scr V_1^{\cramped{(1)}}, \ldots, \scr V_{n_1-1}^{\cramped{(1)}}, \scr V_1^{\cramped{(2)}}, \ldots, \scr V_{n_2-1}^{\cramped{(2)}} \}$ is denoted as:
\begin{equation}
	\label{bos:eq:fact-amp-sep}
	A_{g,n}
		= \int_{\mc S_{g,n}} \omega^{g,n}_{\M_{g,n}}\big(\scr V_1^{(1)}, \ldots, \scr V_{n_1-1}^{(1)}, \scr V_1^{(2)}, \ldots, \scr V_{n_2-1}^{(2)}\big).
\end{equation}

\begin{figure}[htp]
	\centering
	\includegraphics[scale=0.9]{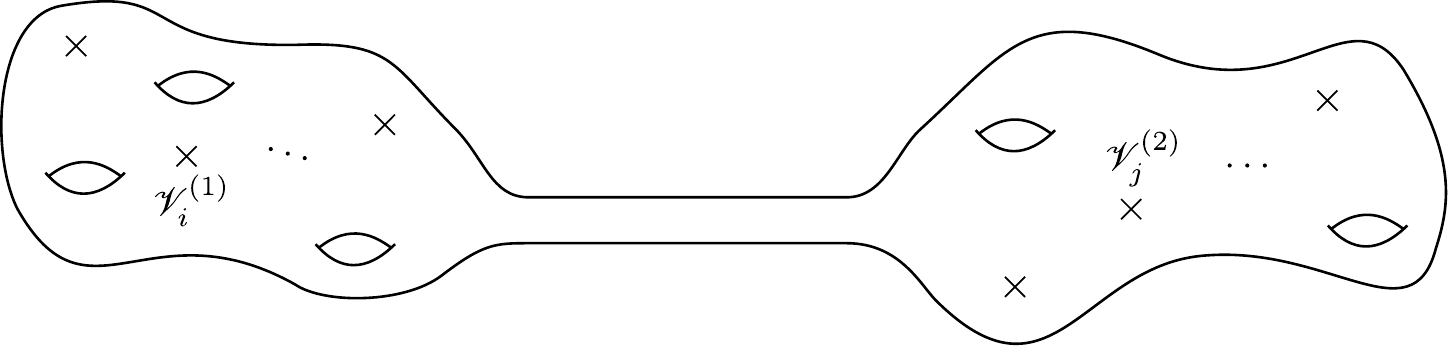}
	\caption{Degeneration limit of $\Sigma_{g,n}$ where the punctures $\scr V_i^{(1)}$ and $\scr V_j^{(2)}$ move apart from each other.}
	\label{bos:fig:factorization-sep-setup}
\end{figure}

We need to study the form $\omega^{g,n}_{\M_{g,n}}$ on $\mc M_{g_1, n_1} \# \mc M_{g_2, n_2}$, which means to rewrite it in terms of the data from $\mc M_{g_1,n_1}$ and from $\mc M_{g_2,n_2}$.
This corresponds to the degeneration limit where the two groups of punctures denoted by $\scr V_i^{(1)}$ and $\scr V_j^{(2)}$ ($i = 1, \ldots, n_1 - 1$, $j = 1, \ldots, n_2 - 1$) together with $g_1$ and $g_2$ holes move apart from each other (\Cref{bos:fig:factorization-sep-setup}).

Since $q$ is a coordinate of $\mc P_{g,n}$, its variation is associated with a tangent vector and a Beltrami $1$-form.
The latter has to be inserted inside $\omega^{g,n}_{\M_{g,n}}$.
A change $q \to q + \delta q$ translates into a change of coordinate
\begin{equation}
	\label{bos:eq:plumbing-dq}
	w'^{(1)}_{n_1}
		= w^{(1)}_{n_1} + \frac{w^{(1)}_{n_1}}{q} \, \delta q,
\end{equation}
where $w_{n_2}^{(2)}$ is kept fixed (obviously, this choice is conventional as explained in \Cref{bos:sec:offshell:geometry:Pgn-tangent}).
\index{plumbing fixture!vector field}%
\index{plumbing fixture!ghost 1-form}%
Thus, the vector field and the Beltrami form are
\begin{equation}
	\label{bos:eq:beltrami-Bq}
	v_q
		= \frac{w^{(1)}_{n_1}}{q},
	\qquad
	B_q
		= \frac{1}{q} \oint_{C_q} \dd w^{(1)}_{n_1} \, b\big(w^{(1)}_{n_1}\big) w^{(1)}_{n_1}.
\end{equation}

\begin{computation}[bos:eq:plumbing-dq]
	Starting from \eqref{bos:eq:fact-plumbing-sep}, vary $q \to q + \delta q$ while keeping $w^{(2)}_{n_2}$ fixed:
	\begin{align*}
		w'^{(1)}_{n_1} w^{(2)}_{n_2}
			&
			= q + \delta q
		\\
		w'^{(1)}_{n_1}
			&
			= \frac{w^{(1)}_{n_1}}{q} (q + \delta q)
			= w^{(1)}_{n_1} + \frac{q}{w^{(1)}_{n_1}} \, \delta q.
	\end{align*}
	The second line follows by replacing $w^{(2)}_{n_2}$ using \eqref{bos:eq:fact-plumbing-sep}.
\end{computation}

\index{plumbing fixture!$p$-form}%
The $\M_{g,n}$-form for the moduli described by the plumbing fixture can be expressed as:
\begin{align}
	\label{bos:eq:omega-Mgn-data-M1-M2}
	\omega_{\M_{g,n}}\big( & V_1^{(1)}, \ldots, V_{\M_{g_1,n_1}}^{(1)}, \pd_q, \pd_{\bar q}, V_1^{(2)}, \ldots, V_{\M_{g_2,n_2}}^{(2)} \big)
	\\
	\nonumber
		&
		= (2\pi \I)^{-\M_{g,n}^c}
			\Mean{
				\;\;
				\prod_{\lambda=1}^{\mathclap{\M_{g_1,n_1}}} B\big(V_{\lambda}^{(1)}\big)
				B(\pd_q) B(\pd_{\bar q})
				\prod_{\kappa=1}^{\mathclap{\M_{g_2,n_2}}} B\big(V_{\kappa}^{(2)}\big)
				\prod_{i=1}^{\mathclap{n_1-1}} \scr V_i^{(1)}
				\prod_{j=1}^{\mathclap{n_2-1}} \scr V_j^{(2)}
				}_{\mathrlap{\Sigma_{g,n}}}.
\end{align}

\index{surface state}%
We introduce the surface states $\Sigma_{n_1}$ and $\Sigma_{n_2}$ such that the BPZ inner product with the new states $\scr V_{n_1}^{(1)}$ and $\scr V_{n_2}^{(2)}$ reproduce the $\M_{g_1,n_1}$- and $\M_{g_2,n_2}$-forms:
\begin{subequations}
\begin{align}
	\bracket{\Sigma_{n_1}}{\scr V_{n_1}^{(1)}}
		&
		:= \omega_{\M_{g_1,n_1}}(\scr V_1^{(1)}, \ldots, \scr V_{n_1}^{(1)})
		= (2\pi \I)^{-\M_{g_1,n_1}^c}
			\Mean{
				\prod_{\lambda=1}^{\M_{g_1,n_1}} B\big(V_{\lambda}^{(1)}\big)
				\prod_{i=1}^{n_1-1} \scr V_i^{(1)}}_{\mathrlap{\Sigma_{g_1,n_1}}},
	\\
	\bracket{\Sigma_{n_2}}{\scr V_{n_2}^{(2)}}
		&
		:= \omega_{\M_{g_2,n_2}}(\scr V_1^{(2)}, \ldots, \scr V_{n_2}^{(2)})
		= (2\pi \I)^{-\M_{g_2,n_2}^c}
			\Mean{
				\prod_{\lambda=1}^{\M_{g_1,n_1}} B\big(V_{\lambda}^{(2)}\big)
				\prod_{j=1}^{n_2-1} \scr V_j^{(2)}}_{\mathrlap{\Sigma_{g_2,n_2}}}.
\end{align}
\end{subequations}
As described in \Cref{bos:sec:offshell:geometry:surface-states}, these states exist since the $p$-form are linear in each of the external state and the BPZ inner-product is non-degenerate.
Each of the surface states corresponds to an operator
\begin{equation}
	\bra{\Sigma_{n_1}}
		= \bra{0} I \circ \Sigma_{n_1}(0),
	\qquad
	\bra{\Sigma_{n_2}}
		= \bra{0} I \circ \Sigma_{n_2}(0),
\end{equation}
defined from \eqref{cft:eq:state-bpz-phi}.
Then, the forms can be interpreted as $2$-point functions on the complex plane:
\begin{equation}
	\bracket{\Sigma_{n_1}}{\scr V_{n_1}^{(1)}}
		= \mean{I \circ \Sigma_{n_1}(0) \scr V_{n_1}(0)}_{w_{n_1}^{(1)}},
	\qquad
	\bracket{\Sigma_{n_2}}{\scr V_{n_2}^{(2)}}
		= \mean{I \circ \Sigma_{n_2}(0) \scr V_{n_2}(0)}_{w_{n_2}^{(2)}}.
\end{equation}
All the complexity of the amplitudes has been lumped into the definitions of the surface states which contain information about the surface moduli (including the ghost insertions) and about the $n_1 - 1$ remaining states (including the local coordinate systems).
The local coordinates around $\scr V_{n_1}^{(1)}$ and $\scr V_{n_2}^{(2)}$ are denoted respectively as $w_{n_1}^{(1)}$ and $w_{n_2}^{(2)}$.
Correspondingly, the surface operators are inserted in the local coordinates $w_1$ and $w_2$ which are related to $w_{n_1}^{(1)}$ and $w_{n_2}^{(2)}$ through the inversion:
\begin{equation}
	w_1
		= I\big(w_{n_1}^{(1)}\big),
	\qquad
	w_2
		= I\big(w_{n_2}^{(2)}\big).
\end{equation}

In order to rewrite \eqref{bos:eq:omega-Mgn-data-M1-M2} in terms of $\Sigma_1$ and $\Sigma_2$, it is first necessary to express all operators in one coordinate system, for example $w_{n_1}^{(1)}$.
Hence, we need to find its relation to $w_2$.
Using the plumbing fixture \eqref{bos:eq:fact-plumbing-sep}, the relation between $w_{n_1}^{(1)}$ and $w_2$ is:
\begin{equation}
	w_{n_1}^{(1)}
		= \frac{q}{w_{n_2}^{(2)}}
		= \frac{q}{I(w_2)}
		= q w_2
		:= f(w_2).
\end{equation}
Then, the form \eqref{bos:eq:omega-Mgn-data-M1-M2} becomes
\begin{equation}
	\omega_{\M_{g,n}}
		= \frac{1}{2\pi\I} \, \mean{I \circ \Sigma_{n_1}(0) B_q B_{\bar q} \, f \circ \Sigma_{n_2}(0)}_{w_{n_1}^{(1)}}
		= \frac{1}{2\pi\I} \, \bra{\Sigma_{n_1}} B_q B_{\bar q} \, q^{L_0} \bar q^{\bar L_0} \ket{\Sigma_2},
\end{equation}
using that $\Sigma_2$ has a well-defined scaling dimension.
The factor of $2\pi\I$ arises by comparing the contribution from $\Sigma_{n_1}$ and $\Sigma_{n_2}$ with the factor in \eqref{bos:eq:omega-Mgn-data-M1-M2}.
The expression can be simplified by using the relation
\begin{equation}
	\bra{\Sigma_{n_1}} B_q B_{\bar q} \ket{\scr V_{n_1}^{(1)}}
		= \frac{1}{q \bar q} \, \bra{\Sigma_{n_1}} b_0 \bar b_0 \ket{\scr V_{n_1}^{(1)}}
\end{equation}
using the expression \eqref{bos:eq:beltrami-Bq} for $B_q$ and the state--operator correspondence:
\begin{equation}
	B_q \scr V_{n_1}^{(1)}(z, \bar z)
		= \frac{1}{q} \oint_{C_q} \dd w^{(1)}_{n_1} \, b\big(w^{(1)}_{n_1}\big) w^{(1)}_{n_1} \scr V_{n_1}^{(1)}(z, \bar z)
		\longrightarrow
		\frac{1}{q} \, b_0 \ket{\scr V_{n_1}^{(1)}}.
\end{equation}
Ultimately, the form \eqref{bos:eq:omega-Mgn-data-M1-M2} reads
\begin{equation}
	\label{bos:eq:omega-Mgn-states-M1-M2}
	\omega_{\M_{g,n}}
		= \frac{1}{2\pi\I} \, \frac{1}{q \bar q} \, \bra{\Sigma_{n_1}} b_0 \bar b_0 \, q^{L_0} \bar q^{\bar L_0} \ket{\Sigma_{n_2}}.
\end{equation}

It is important to remember that the plumbing fixture describes only a patch of the moduli space, and the form defined in this way is valid only locally.
As a consequence, the integration over all moduli of $\mc M_{g_1, n_1} \# \mc M_{g_2, n_2}$ does \emph{not} describe $\mc M_{g,n}$, but only a part of it (\Cref{bos:sec:geometry:plumbing:decomposition}).
Every degeneration limit with a different puncture distribution in two different groups contributes to a different part of the amplitude.

\index{string Feynman diagram!1PR diagram}%
We denote the contribution to the total amplitude \eqref{bos:eq:fact-amp-sep} from the region of the moduli space connected to this degeneration limit as:
\begin{equation}
	\mc F_{g,n}\big(\scr V_i^{(1)} | \scr V_j^{(2)}\big)
		:= \frac{1}{2\pi\I} \int
			\bigwedge_{\lambda=1}^{\M_{g_1,n_1}} \dd t_{\lambda}^{(1)}
			\bigwedge_{\kappa=1}^{\M_{g_2,n_2}} \dd t_{\kappa}^{(2)}
			\wedge \frac{\dd q}{q} \wedge \frac{\dd \bar{q}}{\bar{q}} \,
			\bra{\Sigma_{n_1}} b_0 \bar b_0 \, q^{L_0} \bar q^{\bar L_0} \ket{\Sigma_{n_2}}.
\end{equation}
To proceed, we introduce a basis $\{ \phi_\alpha(k) \}$ of eigenstates of $L_0$ and $\bar L_0$, where $k^\mu$ is the $D$-dimensional momentum and $\alpha$ denotes the remaining quantum number.
Then, introducing twice the resolution of the identity \eqref{bos:eq:identity-offshell} gives:
\begin{align}
	\label{bos:eq:Ggn-degen-M1-M2}
	\nonumber
	\mc F_{g,n}\big(\scr V_i^{(1)} | \scr V_j^{(2)}\big)
		= \frac{1}{2\pi\I} & \int \frac{\dd^D k}{(2\pi)^D} \, \frac{\dd^D k'}{(2\pi)^D} \,
			(-1)^{\abs{\phi_\alpha}}
			\\ &
			\times \int \frac{\dd q}{q} \wedge \frac{\dd \bar{q}}{\bar{q}} \,
				\bra{\phi_\alpha(k)^c} b_0 \bar b_0 \, q^{L_0} \bar q^{\bar L_0} \ket{\phi_\beta(k')^c}
			\\ &
			\nonumber
			\times \int \bigwedge_{\lambda=1}^{\M_{g_1,n_1}} \! \dd t_{\lambda}^{(1)}
				\Bracket{\Sigma_{n_1}}{\phi_\alpha(k)}
			\int \bigwedge_{\kappa=1}^{\M_{g_2,n_2}} \! \dd t_{\kappa}^{(2)}
				\Bracket{\phi_\beta(k')}{\Sigma_{n_2}}
\end{align}
(with implicit sums over $\alpha$ and $\beta$).
In the last line, one recognizes the expressions of the $g_1$-loop $n_1$-point amplitude with external states $\{ \scr V_1^{\cramped{(1)}}, \ldots, \scr V_{n_1-1}^{\cramped{(1)}}, \phi_\alpha \}$ and of the $g_2$-loop and $n_2$-point amplitudes with external states $\{ \scr V_1^{\cramped{(2)}}, \ldots, \scr V_{n_2-1}^{\cramped{(2)}}, \phi_\beta \}$:
\begin{subequations}
\begin{gather}
	\begin{aligned}
	A_{g_1,n_1}\big(\scr V_1^{(1)}, \ldots, \scr V_{n_1-1}^{(1)}, \phi_\alpha(k)\big)
		&
		= \int_{\mc S_{g_1,n_1}} \omega_{\M_{g_1,n_1}}\big(\scr V_1^{(1)}, \ldots, \scr V_{n_1-1}^{(1)}, \phi_\alpha(k)\big)
		\\ &
		= \int_{\mc S_{g_1,n_1}} \bigwedge_{\lambda=1}^{\M_{g_1,n_1}} \dd t_{\lambda}^{(1)} \Bracket{\Sigma_{n_1}}{\phi_\alpha(k)},
	\end{aligned}
	\\
	\begin{aligned}
	A_{g_2,n_2}\big(\scr V_1^{(2)}, \ldots, \scr V_{n_2-1}^{(2)}, \phi_\beta(k')\big)
		&
		= \int_{\mc S_{g_2,n_2}} \omega_{\M_{g_2,n_2}}\big(\scr V_1^{(2)}, \ldots, \scr V_{n_2-2}^{(2)}, \phi_\beta(k')\big)
		\\ &
		= \int_{\mc S_{g_2,n_2}} \bigwedge_{\lambda=1}^{\M_{g_2,n_2}} \dd t_{\lambda}^{(2)} \Bracket{\Sigma_{n_2}}{\phi_\beta(k')}.
	\end{aligned}
\end{gather}
\end{subequations}
The property \eqref{app:eq:cft-bpz-product-prop} has been used to reverse the order of the BPZ product for the second Riemann surface, and this cancels the factor $(-1)^{\abs{\phi_\alpha}}$.

\begin{figure}[htp]
	\centering
	\includegraphics[scale=1]{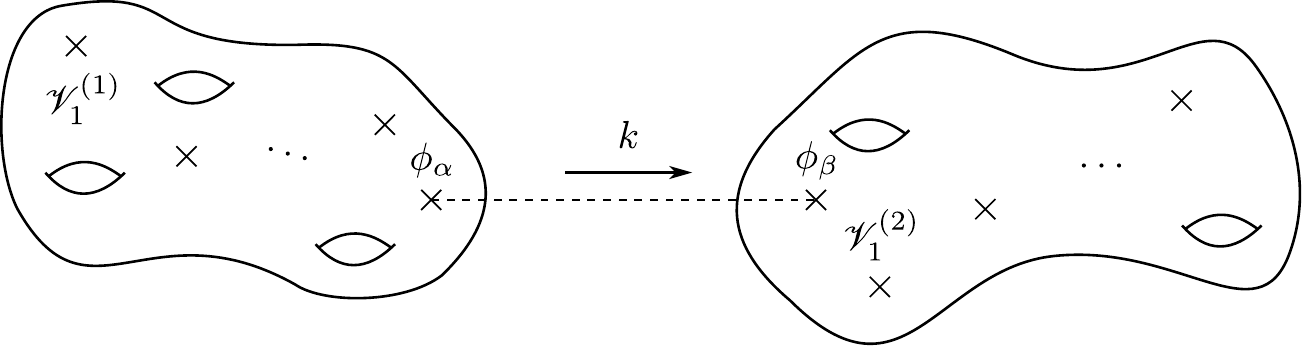}
	\caption{%
		Factorization of the amplitude into two sub-amplitudes connected by a propagator (dashed line).
	}%
	\label{bos:fig:factorization-sep-prop}
\end{figure}

Defining the second line of \eqref{bos:eq:Ggn-degen-M1-M2} as
\begin{equation}
	\label{bos:eq:fact-prop}
	\Delta_{\alpha\beta}(k, k')
		:= \Delta\big(\phi_\alpha(k)^c, \phi_\beta(k')^c\big)
		:= \frac{1}{2\pi \I} \int \frac{\dd q}{q} \wedge \frac{\dd \bar{q}}{\bar{q}} \,
			\bra{\phi_\alpha(k)^c} b_0 \bar b_0 \, q^{L_0} \bar q^{\bar L_0} \ket{\phi_\beta(k')^c},
\end{equation}
one has:
\begin{equation}
	\begin{aligned}
		\mc F_{g,n}\big(\scr V_i^{(1)} | \scr V_j^{(2)}\big)
			= \int \frac{\dd^D k}{(2\pi)^D} \frac{\dd^D k'}{(2\pi)^D} \,
				& A_{g_1,n_1}\big(\scr V_1^{(1)}, \ldots, \scr V_{n_1-1}^{(1)}, \phi_\alpha(k)\big) \,
				\Delta_{\alpha\beta}(k, k')
				\\
				&\times A_{g_2,n_2}\big(\scr V_1^{(2)}, \ldots, \scr V_{n_2-1}^{(2)}, \phi_\beta(k')\big).
	\end{aligned}
\end{equation}
We recover the expressions from \Cref{bos:sec:offshell:motivations:4pt}, but for a more general amplitude.
\index{propagator!closed bosonic}%
We had found that $\Delta$ corresponds to the propagator: its properties are studied further in \Cref{bos:sec:amp-prop:propagator}.
Hence, the object \eqref{bos:eq:Ggn-degen-M1-M2} corresponds to the product of two amplitudes connected by a propagator (\Cref{bos:fig:factorization-sep-prop}).

There are several points to mention about this amplitude:
\begin{itemize}
	\item We will find that the propagator depends only on one momentum because $\bracket{k}{k'} \sim \delta^{(D)}(k + k')$, which removes one of the integral.
	Then, both amplitudes $A_{g_1,n_1}$ and $A_{g_2,n_2}$ contain a delta function for the momenta:
	\begin{equation}
		A_{g_1,n_1}
			\sim \delta^{(D)}\big(k_1^{(1)} + \cdots + k_{n_1-1}^{(1)} + k\big),
		\qquad
		A_{g_2,n_2}
			\sim \delta^{(D)}\big(k_1^{(2)} + \cdots + k_{n_2-1}^{(2)} + k'\big).
	\end{equation}
	As a consequence, the second momentum integral can be performed and yields a delta function:
	\begin{equation}
		\mc F_{g,n}
			\sim \delta^{(D)}\big(k_1^{(1)} + \cdots + k_{n_1-1}^{(1)} + k_1^{(2)} + \cdots + k_{n_2-1}^{(2)}\big).
	\end{equation}
	Hence, the momentum flowing in the internal line is fixed and this ensures the overall momentum conservation as expected.
	\index{string Feynman diagram!intermediate states!momentum}%

	\item The ghost numbers of the states $\phi_\alpha$ and $\phi_\beta$ are also fixed (in terms of the external states).
	Indeed, because of the ghost number anomaly, the amplitudes on $\mc M_{g_1, n_1}$ and $\mc M_{g_2, n_2}$ are non-vanishing only if the ghost numbers of these states satisfy:
	\begin{equation}
		N_{\text{gh}}(\phi_\alpha)
			= 2 n_1 - \sum_{i=1}^{n_1-1} N_{\text{gh}}\big(\scr V_i^{(1)}\big),
		\qquad
		N_{\text{gh}}(\phi_\beta)
			= 2 n_2 - \sum_{j=1}^{n_2-1} N_{\text{gh}}\big(\scr V_j^{(2)}\big).
	\end{equation}
	The non-vanishing of $\mc F_{g, n}$ also gives another relation:
	\begin{equation}
		N_{\text{gh}}(\phi_\alpha)
				+ N_{\text{gh}}(\phi_\beta)
			= 4.
	\end{equation}
	In particular, if the external states are on-shell with $N_{\text{gh}} = 2$, we find:
	\begin{equation}
		N_{\text{gh}}(\phi_\alpha)
			= N_{\text{gh}}(\phi_\beta)
			= 2.
	\end{equation}
	As indicated in \Cref{bsft:chap:free-brst}, such states are appropriate at the classical level since they do not contain spacetime ghosts.
	\index{string Feynman diagram!intermediate states!ghost number}%

	\item The sum over $\alpha$ and $\beta$ is over an infinite number of states and could diverge.
	In fact, the sum can be made convergent by tuning the stub parameter (\Cref{bos:sec:amp-prop:stubs}).

\end{itemize}
Properties of Feynman graphs and amplitudes in the momentum space will be discussed further in \Cref{sft:chap:momentum-sft}.

\subsection{Non-separating case}
\label{bos:sec:amp-prop:factorization:nonsep}

Next, we consider the non-separating plumbing fixture (\Cref{bos:sec:geometry:plumbing:nonsep}).
The computations are almost identical to the separating case, thus we outline only the general steps.

Part of the moduli space $\mc M_{g,n}$ is covered by $\# \mc M_{g_1, n_1}$, with $g = g_1 + 1$ and $n = n_1 - 2$.
\index{plumbing fixture!non-separating}%
The local coordinates are denoted as $w_i$ for $i = 1, \ldots, n_1$ and the plumbing fixture reads:
\begin{equation}
	\label{bos:eq:fact-plumbing-nonsep}
	w_{n_1 - 1} w_{n_1}
		= q.
\end{equation}
The $g$-loop $n$-point amplitude with external states $\{ \scr V_1^{(1)}, \ldots, \scr V_{n_1-2}^{(1)} \}$ is denoted as:
\begin{equation}
	\label{bos:eq:fact-amp-nonsep}
	A_{g,n}
		= \int_{\mc S_{g,n}} \omega^{g,n}_{\M_{g,n}}\big(\scr V_1^{(1)}, \ldots, \scr V_{n_1-2}^{(1)}\big).
\end{equation}

\begin{figure}[htp]
	\centering
	\includegraphics[scale=1]{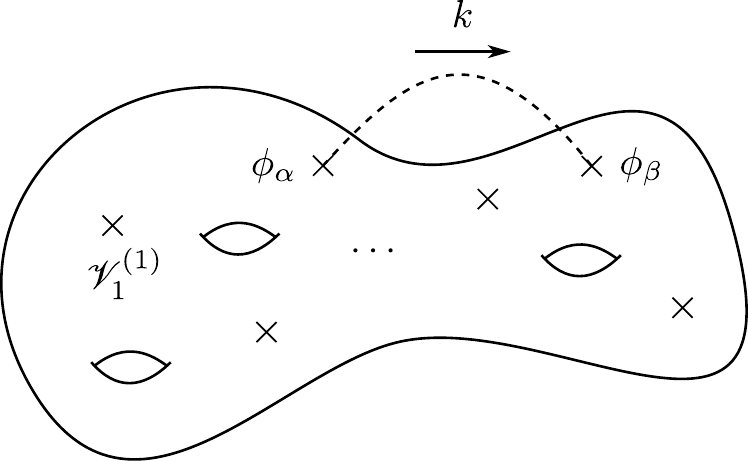}
	\caption{%
		Factorization of the amplitude into two sub-amplitudes connected by a propagator (dashed line).
		The propagator connects two punctures of the same surface, which is equivalent to a loop.
	}%
	\label{bos:fig:factorization-nonsep}
\end{figure}

When the $n_1 - 2$ punctures and $g_1 = g - 1$ holes move lose to each other, the form can be written as:
\begin{equation}
	\label{bos:eq:omega-Mgn-data-M1}
	\omega_{\M_{g,n}}\big( V_1^{(1)}, \ldots, V_{\M_{g_1,n_1}}^{(1)}, \pd_q, \pd_{\bar q} \big)
		= (2\pi \I)^{-\M_{g,n}^c}
			\Mean{
				\prod_{\lambda=1}^{\M_{g_1,n_1}} B\big(V_{\lambda}^{(1)}\big)
				B(\pd_q) B(\pd_{\bar q})
				\prod_{i=1}^{n_1-2} \scr V_i^{(1)}
				}_{\mathrlap{\Sigma_{g,n}}}.
\end{equation}
To proceed, one needs to introduce the surface state $\Sigma_{n_1-1,n_1}$:
\begin{equation}
	\bracket{\Sigma_{n_1-1,n_1}}{\scr V_{n_1-1}^{(1)} \otimes \scr V_{n_1}^{(1)}}
		:= \omega_{\M_{g_1,n_1}}(\scr V_{1}^{(1)}, \ldots, \scr V_{n_1}^{(1)}).
\end{equation}
\index{string Feynman diagram!loop diagram}%
Following the same step as in the previous section leads to:
\begin{equation}
	\mc F_{g,n}\big(\scr V_i^{(1)}|\big)
		= \int \frac{\dd^D k}{(2\pi)^D} \frac{\dd^D k'}{(2\pi)^D} \,
			A_{g_1,n_1}\big(\scr V_1^{(1)}, \ldots, \scr V_{n_1-2}^{(1)}, \phi_\alpha(k), \phi_\beta(k')\big) \,
			\Delta_{\alpha\beta}(k, k'),
\end{equation}
where the propagator is given in \eqref{bos:eq:fact-prop}.
This is equivalent to an amplitude for which two external legs are glued together with a propagator, giving a loop (\Cref{bos:fig:factorization-nonsep}).

\index{string Feynman diagram!intermediate states!ghost number}%
Since both states $\phi_\alpha$ and $\phi_\beta$ are inserted on the same surface, their ghost numbers are not fixed, even if the external states are physical.
The non-vanishing of $\mc F_{g,n}$ only leads to the constraint:
\begin{equation}
	N_{\text{gh}}(\phi_\alpha) + N_{\text{gh}}(\phi_\beta)
		= 2 n_1 - \sum_{i=1}^{n_1-2} N_{\text{gh}}\big(\scr V_i^{(1)}\big)
		= 4.
\end{equation}
As a consequence, loop diagrams force to introduce states of every ghost number.
Internal states with $N_{\text{gh}} \neq 2$ correspond to spacetime ghosts.

\index{string Feynman diagram!intermediate states!momentum}%
Since the propagator contains a delta function $\delta^{(D)}(k - k')$, the integral over $k'$ can be removed by setting $k' = - k$.
However, the integral over $k$ remains since
\begin{equation}
	A_{g_1,n_1}\big(\scr V_1^{(1)}, \ldots, \scr V_{n_1-2}^{(1)}, \phi_\alpha(k), \phi_\beta(-k)\big)
		\sim \delta^{(D)}\big(k_1^{(1)} + \cdots + k_{n_1-2}^{(1)}\big).
\end{equation}
Hence, the loop momentum $k$ is not fixed, as expected in QFT.

\begin{remark}
	Not all values of the moduli associated to the holes can be associated to loops in Feynman diagrams.
	Only the values close to the degeneration limit can be interpreted in this way, the other being just standard (quantum) vertices.
\end{remark}

\section{Feynman diagrams and Feynman rules}
\label{bos:sec:amp-prop:feynman}

\index{string Feynman diagram!Feynman rules}%
In the standard QFT approach, Feynman graphs compute Green functions, and scattering amplitudes are obtained by amputating the external propagators through the LSZ prescription.
For connected tree-level processes, this requires $n \ge 3$ (corresponding to $\chi_{0, n} < 0$).

Given a theory, there is a minimal set of Feynman diagrams -- the Feynman rules -- from which every other diagram can be constructed.
These rules include the definitions of the fundamental vertices -- the fundamental interactions -- and of the propagator -- how states propagate between two interactions (or, how to glue vertices together).
In this section, we describe these different elements.

\subsection{Feynman graphs}
\label{bos:sec:amp-prop:feynman:graphs}

\index{string amplitude!factorization}%
The amplitude factorization described in \Cref{bos:sec:amp-prop:factorization} gives a natural separation of amplitudes into several contributions.
Considering all the possible degeneration limits lead to a set of diagrams with amplitudes of lower order connected by propagators (\Cref{bos:fig:factorization-sep-prop}, \Cref{bos:fig:factorization-nonsep}).
This corresponds exactly to the idea behind Feynman graphs.
Then, the goal is to find the Feynman rules of the theory: since the propagator has already been identified (further studied in \Cref{bos:sec:amp-prop:propagator}), it is sufficient to find the interaction vertices.

Let's make this more precise by considering an amplitude $A_{g,n}(\scr V_1, \ldots, \scr V_n)$.
\index{index!Riemann surface}%
\index{index!amplitude}%
The index of an amplitude is defined to be the index \eqref{bos:eq:Sgn-index} of the corresponding Riemann surfaces
\begin{equation}
	r(A_{g,n})
		:= r(\Sigma_{g,n})
		= 3 g + n - 2.
\end{equation}
Contributions to an amplitude with a given $r(A_{g,n})$ can be described in terms of amplitudes $A_{g',n'}$ with $r(A_{g',n'}) < r(A_{g,n})$.
But, the moduli space $\mc M_{g,n}$ cannot (generically) be completely covered with the plumbing fixture of lower-dimensional moduli spaces, i.e.\ with $r(\mc M_{g',n'}) < r(\mc M_{g,n})$ (\Cref{bos:sec:geometry:plumbing:decomposition}).
Then, the same must be true for the amplitudes, such that $A_{g,n}$ cannot be uniquely expressed in terms of amplitudes $A_{g',n'}$.

\index{closed string fundamental vertex}%
The $g$-loop $n$-point fundamental vertex is defined by:
\begin{equation}
	\label{bos:eq:Vgn-vertex}
	\mc V_{g,n}(\scr V_1, \ldots, \scr V_n)
		:= \quad
			\vcenter{\hbox{\includegraphics{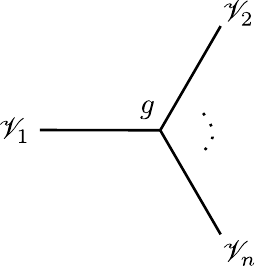}}}
			\quad
		:= \int_{\mc R_{g,n}} \omega^{g,n}_{\M_{g,n}}(\scr V_1, \ldots, \scr V_n),
\end{equation}
\index{Pgn@$\mc P_{g,n}$ space!section}%
The form defined in \eqref{bos:eq:Agn-offshell-Sgn} is integrated over a sub-section $\mc R_{g,n} \subset \mc S_{g,n}$ of $\hat{\mc P}_{g,n}$.
Its projection on the base is the region $\mc V_{g,n} \subset \mc M_{g,n}$ which cannot be described by the plumbing fixture, see \eqref{bos:eq:def-Vgn-space}.
In general, we will keep the choice of local coordinates implicit and always write $\mc V_{g,n}$ to avoid surcharging the notations.

\index{closed string amplitude!Feynman diagram decomposition}%
It corresponds to the remaining contribution of the amplitude once all graphs containing propagators have been taken into account:
\begin{equation}
	\label{bos:eq:amplitude-vertex-offshell}
	\begin{aligned}
	\vcenter{\hbox{\includegraphics[scale=1]{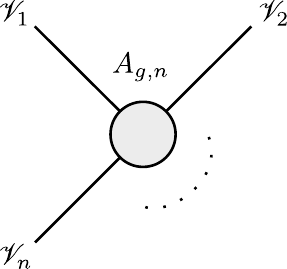}}}
		&
		= \sum_{\substack{0 \le h \le g \\ 0 \le m < n - 1}}
		\vcenter{\hbox{\includegraphics[scale=1]{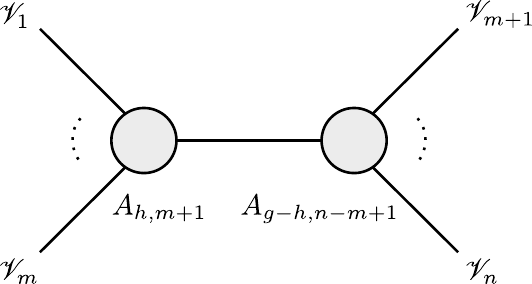}}}
		\\ &
		+ \quad
		\vcenter{\hbox{\includegraphics[scale=1]{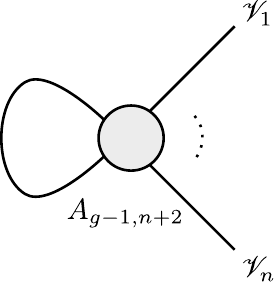}}}
		+ \text{ perms }
		+ \quad \vcenter{\hbox{\includegraphics[scale=1]{bosonic/vertex_Vgn}}}
	\end{aligned}
\end{equation}
where the permutations are taken over all legs exiting the amplitudes in the first two terms (this includes the two legs glued together in the second term), including if necessary a weight to avoid overcounting.
In the RHS, the amplitudes $A_{g, 1}$ are tadpoles and have no external vertices $\scr V_i$ (from $A_{g,n}$); this corresponds to the terms for $m = 0$ and $m = n - 2$.

In general, the fundamental vertex is non-vanishing for every value $g, n \in \N$ such that $\chi_{g,n} < 0$.
For this reason, the index $g$ helps to distinguish between graphs with identical values of $n$.
It may look strange that one needs vertices at every loop: the interpretation will be made clearer when translating this in the language of string field theory (\Cref{bsft:chap:bv-sft}).
We stress again that the definition of the fundamental vertex (and the region covered) depends on the choice of local coordinates for \emph{all} lower-order vertices $\mc V_{g',n'}$ such that $r(\mc V_{g',n'}) < r(\mc V_{g,n})$.

\begin{remark}
	There are different alternative notations for \eqref{bos:eq:Vgn-vertex}:
	\begin{equation}
		\mc V_{g,n}(\scr V_1, \ldots, \scr V_n)
			:= \mc V_{g,n}(\otimes_i \scr V_i)
			:= \{ \scr V_1, \ldots, \scr V_n \}_g.
	\end{equation}
\end{remark}

\begin{example}[Scalar QFT]
	Consider a scalar field theory with a cubic and a quartic interaction.
	The $4$-point amplitude contains four contributions, three from gluing $3$-point vertices with a propagator, and one from the fundamental quartic vertex.
	The mismatch between the amplitude and the three graphs with a propagator hints at the existence of the quartic interactions.
	This example gives an idea of how one can identify the fundamental interactions recursively.
\end{example}

\index{Green function}%
The definition \eqref{bos:eq:amplitude-vertex-offshell} of the vertex shows that it can also be interpreted as an amputated Green function without internal propagator (i.e.\ there is no propagator at all).
This definition is expected from the definition of the interactions vertices from an action, as will be exemplified in \Cref{bsft:chap:bv-sft}.
Before describing (and generalizing) the vertices, we describe first the properties of the propagator.

\subsection{Propagator}
\label{bos:sec:amp-prop:propagator}

\index{propagator!closed bosonic|(}%

The propagator has been defined in \eqref{bos:eq:fact-prop}:
\begin{equation}
	\label{bos:eq:fact-prop-op}
	\Delta
		= \frac{1}{2\pi \I} \int \frac{\dd q}{q} \wedge \frac{\dd \bar{q}}{\bar{q}} \,
			b_0 \bar b_0 \, q^{L_0} \bar q^{\bar L_0}.
\end{equation}
The plumbing modulus $q$ is parametrized by \eqref{bos:eq:q-s-th}
\begin{equation}
	\label{bos:eq:q-s-th-fact}
	q
		= \e^{- s + \I \theta},
	\qquad
	s \in \R_+,
	\qquad
	\theta \in [0, 2\pi),
\end{equation}
such that the integration measure becomes
\begin{equation}
	\frac{\dd q}{q} \wedge \frac{\dd \bar{q}}{\bar{q}}
		= - 2 \I \, \dd s \wedge \dd \theta.
\end{equation}

Using the variables $L_0^\pm = L_0 \pm \bar L_0$ and $b_0^\pm = b_0 \pm \bar b_0$, the propagator can be recast as:
\begin{equation}
	\Delta
		= \frac{1}{2\pi} \, b_0^+ b_0^-
			\int_{0}^{\infty} \dd s \, \e^{- s L_0^+}
			\int_{0}^{2\pi} \dd \theta \, \e^{\I \theta L_0^-}.
\end{equation}
\index{propagator!Schwinger parametrization}%
The form of the first integral is recognized as the Schwinger parametrization of the propagator, while the second is the Fourier transformation of the discrete delta function:
\begin{equation}
	\int_{0}^{\infty} \dd s \, \e^{- s L_0^+}
		= \frac{1}{L_0^+},
	\qquad
	\int_{0}^{2\pi} \dd \theta \, \e^{\I \theta L_0^-}
		= 2\pi \, \delta_{L_0^-, 0}.
\end{equation}
In fact, the first integral converges only if $L_0^+ > 0$.
As argued in the introduction, divergences for $L_0^+ \le 0$ are either non-physical or IR divergences which can be cured by renormalization.
For this reason, we take the RHS as a definition of the integral, which would be the correct result if one starts with a field theory action instead of a first-quantized formalism.

In this case, the propagator becomes
\begin{equation}
	\label{bos:eq:offshell-prop-bpm}
	\Delta
		= \frac{b_0^+}{L_0^+} \;
			b_0^- \delta_{L_0^-, 0}.
\end{equation}
This is the standard expression for the propagator.
For completeness, the form in terms of the holomorphic and anti-holomorphic components is:
\begin{equation}
	\Delta
		= - 2 b_0 \bar b_0 \,
			\frac{1}{L_0 + \bar L_0} \,
			\delta_{L_0, \bar L_0}.
\end{equation}
\index{level-matching condition}%
The delta function restricts the amplitude to states satisfying the level-matching condition, that is, annihilated by $L_0^-$.

Considering a basis $\{ \phi_\alpha(k) \}$ of eigenstates of both $L_0$ and $\bar L_0$:
\begin{equation}
	L_0^+ \ket{\phi_\alpha(k)}
		= \frac{\alpha'}{2} \, (k^2 + m_\alpha^2) \, \ket{\phi_\alpha(k)},
	\qquad
	L_0^- \ket{\phi_\alpha(k)}
		= 0
\end{equation}
leads to the following momentum-space kernel for the propagator:
\begin{subequations}
\label{bos:eq:offshell-prop-k}
\begin{gather}
	\Delta_{\alpha\beta}(k, k')
		:= \bra{\phi_\alpha(k)^c} \Delta \ket{\phi_\beta(k')^c}
		:= (2\pi)^D \delta^{(D)}(k + k') \, \Delta_{\alpha\beta}(k),
	\\
	\Delta_{\alpha\beta}(k)
		:= \frac{M_{\alpha\beta}(k)}{k^2 + m_\alpha^2},
	\qquad
	M_{\alpha\beta}(k)
		:= \frac{2}{\alpha'} \, \bra{\phi_\alpha^c(k)} b_0^+ b_0^- \ket{\phi_\beta^c(-k)},
\end{gather}
\end{subequations}
with $M_{\alpha\beta}$ a finite-dimensional matrix giving the overlap of states of identical masses (because the number of states at a given level is finite).

For the propagator to be well-defined, it must be invertible (in particular, to define a kinetic term).
The propagator \eqref{bos:eq:offshell-prop-bpm} is non-vanishing if the states it acts on satisfy:
\begin{equation}
	b_0^+ \ket{\phi_\alpha^c}
		\neq 0,
	\qquad
	b_0^- \ket{\phi_\alpha^c}
		\neq 0.
\end{equation}
Necessary and sufficient conditions for this to be true are
\begin{equation}
	c_0^+ \ket{\phi_\alpha^c}
		= 0,
	\qquad
	c_0^- \ket{\phi_\alpha^c}
		= 0.
\end{equation}
Indeed, decomposing the state on the ghost zero-modes
\begin{equation}
	\ket{\phi_\alpha^c}
		= \ket{\phi_1} + b_0^\pm \ket{\phi_2},
	\qquad
	c_0^\pm \ket{\phi_1}
		= c_0^\pm \ket{\phi_2}
		= 0
\end{equation}
gives
\begin{equation}
	c_0^\pm \ket{\phi_\alpha^c}
		= 0
	\quad \Longrightarrow \quad
	\ket{\phi_2}
		= 0,
\end{equation}
and one has correctly $b_0^\pm \ket{\phi_1} \neq 0$.

These conditions are given for the dual states: translating them on the normal states reverses the roles of $b_0$ and $c_0$.
Hence, the states must satisfy the conditions:
\begin{equation}
	b_0^+ \ket{\phi_\alpha}
		= 0,
	\qquad
	b_0^- \ket{\phi_\alpha}
		= 0.
\end{equation}
The second condition is satisfied automatically because the Hilbert space is $\mc H^-$ when working with $\hat{\mc P}_{g,n}$ (\Cref{bos:sec:offshell:geometry:hat-Pgn}).
However, the first condition further restricts the states which propagate in internal lines.
This leads to postulate that the external states should also be taken to satisfy this condition
\begin{equation}
	b_0^+ \ket{\scr V_i}
		= 0,
\end{equation}
since external states are usually a subset of the internal states.
This provides another motivation of the statement in \Cref{bos:sec:ws-int:brst:states} that scattering amplitudes for the states not annihilated by $b_0^+$ must be trivial.
A field interpretation of this condition is given in \Cref{bsft:chap:free-brst,bsft:chap:bv-sft}.

Under these constraints on the states, the propagator can be inverted:
\begin{equation}
	\Delta^{-1}
		= c_0^+ c_0^- L_0^+ \delta_{L_0^-, 0}.
\end{equation}

\index{propagator!closed bosonic|)}%

\subsection{Fundamental vertices}
\label{bos:sec:amp-prop:feynman:vertices}

\index{closed string fundamental vertex!recursive construction}%
The vertices \eqref{bos:eq:Vgn-vertex} can be constructed recursively assuming that all amplitudes are known.
The starting point is the tree-level cubic amplitude $A_{0,3}$: since it does not contain any internal propagator, it is equal to the fundamental vertex $\mc V_{0,3}$.

The fist thing to extract from the recursion relations are the background independent data.
This amounts to find local coordinates and a characterization of the subspaces $\mc V_{g,n} \subset \mc M_{g,n}$, starting with $\mc P_{0,3}$ and iterating.

In the rest of this section, we show how this works schematically.

\subsubsection{Recursive definition: tree-level vertices}

The description of tree-level amplitudes $A_{0,n}$ is the simplest since only the separating plumbing fixture is used and Feynman graphs are trees.
The possible factorizations of the amplitude correspond basically to all the partitions of the set $\{ \scr V_i \}$ into subsets.

\paragraph{Tree-level cubic vertex}

\begin{figure}[htp]
	\centering
	\subcaptionbox{%
		A section $\mc S_{0,3}$ over $\mc P_{0,3}$ reduces to a point.
		\label{bos:fig:P03-section}
	}[0.4\linewidth]{%
		\centering
		\includegraphics[scale=1.4]{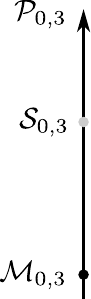}
	}%
	\hspace{2cm}
	\subcaptionbox{%
		Fundamental cubic vertex.
		\label{bos:fig:feynman-V03}
	}[0.4\linewidth]{%
		\centering
		\includegraphics[scale=1]{bosonic/feynman_V03}

	}%
	\caption{Section of $\mc P_{0,3}$ and cubic vertex.}
\end{figure}

Since $\M_{0,3} = 0$, the moduli space of the $3$-punctured sphere $\Sigma_{0,3}$ reduces to a point, and so does the section $\mc S_{0,3}$ of $\mc P_{0,3}$ (\Cref{bos:fig:P03-section}):
\begin{equation}
	\mc V_{0,3}(\scr V_1, \scr V_2, \scr V_3)
		:= A_{0,3}(\scr V_1, \scr V_2, \scr V_3)
		= \omega_{0}^{0,3}(\scr V_1, \scr V_2, \scr V_3).
\end{equation}
The corresponding graph is indicated in \Cref{bos:fig:feynman-V03}.

\paragraph{Tree-level quartic vertex}

\begin{figure}[htp]
	\centering
	\subcaptionbox{$s$-channel.\label{bos:fig:feynman-G04-s}}{%
		\centering
		\includegraphics[scale=1]{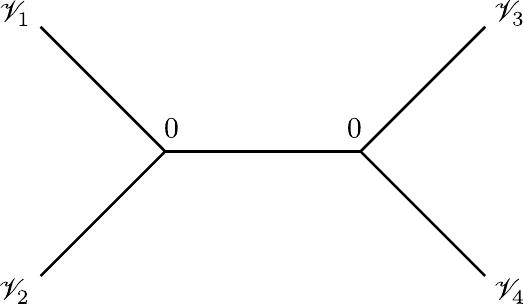}
	}%
	\\
	\medskip
	\subcaptionbox{$t$-channel.\label{bos:fig:feynman-G04-t}}{%
		\centering
		\includegraphics[scale=1]{bosonic/feynman_G04_t}
	}%
	\hspace{2cm}
	\subcaptionbox{$u$-channel.\label{bos:fig:feynman-G04-u}}{%
		\centering
		\includegraphics[scale=1]{bosonic/feynman_G04_u}
	}%
	\caption{Factorization of the quartic amplitude $A_{0,4}$ in the $s$-, $t$- and $u$-channels.}
	\label{bos:fig:feynman-G04-stu}
\end{figure}

\begin{figure}[htp]
	\centering
	\includegraphics[scale=1.2]{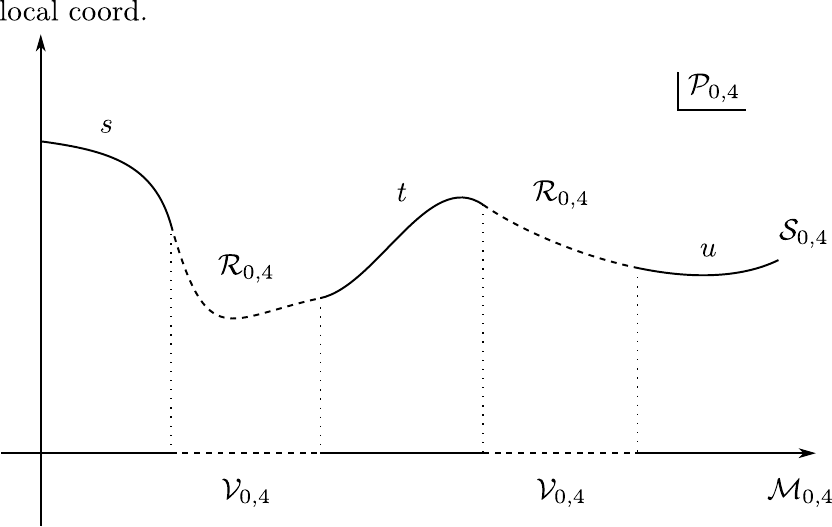}
	\caption{%
		A section $\mc S_{0,4}$ over $\mc P_{0,4}$, the contribution from the $s$-, $t$- and $u$-channels (\Cref{bos:fig:feynman-G04-stu}) are indicated by the corresponding indices.
		The fundamental vertex is defined by the section $\mc V_{0,4}$.
	}
	\label{bos:fig:P04-section-stu-R04}
\end{figure}

\begin{figure}[htp]
	\centering
	\includegraphics[scale=1]{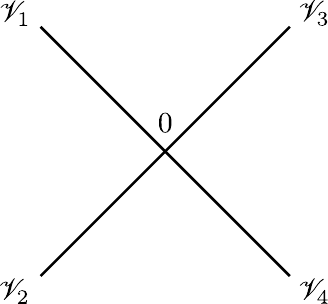}
	\caption{Fundamental quartic vertex.}
	\label{bos:fig:feynman-V04}
\end{figure}

Part of the contributions to the $4$-point amplitude $A_{0,4}$ with external states $\scr V_i$ ($i = 1, \ldots, 4$) comes from gluing two cubic vertices.
Because there are four external states, there are three different partitions $2|2$ which are described in \Cref{bos:fig:feynman-G04-stu} (see also \Cref{bos:fig:gluing-S03-S03-perms}).
The sum of these three diagrams does not reproduce $A_{0,4}$: the moduli space $\mc M_{0,4}$ is not completely covered by the three amplitudes.
Equivalently, the projection of the section over $\mc P_{0,4}$ does not cover all of $\mc M_{0,4}$.
The missing contribution is defined by the quartic vertex (\Cref{bos:fig:feynman-V04})
\begin{equation}
	\mc V_{0,4}(\scr V_1, \scr V_2, \scr V_3, \scr V_4)
		:= \int_{\mc R_{0,4}} \omega_2^{0,4}(\scr V_1, \ldots, \scr V_4),
\end{equation}
and the corresponding section is denoted by $\mc R_{0,4}$ (\Cref{bos:fig:P04-section-stu-R04}).
Denoting by $\mc F_{0,4}^{(s,t,u)}$ the graphs~\ref{bos:fig:feynman-G04-stu} in the $s$-, $t$- and $u$-channels, one has the relation
\index{off-shell closed string amplitude!tree-level!4-point}%
\begin{equation}
	A_{0,4}
		= \mc F_{0,4}^{(s)} + \mc F_{0,4}^{(t)} + \mc F_{0,4}^{(u)}
			+ \mc V_{0,4}.
\end{equation}

\paragraph{Tree-level quintic vertex}

\begin{figure}[htp]
	\centering
	\subcaptionbox{%
		Factorization $12 | 3 | 45$.
		\label{bos:fig:feynman-G05-12-3-45}
	}{%
		\centering
		\includegraphics[scale=1]{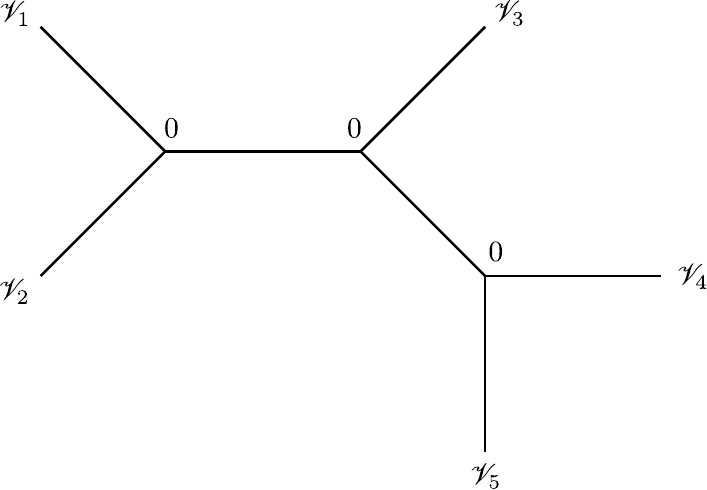}
	}%
	\\
	\medskip
	\subcaptionbox{%
		Factorization $12 | 345$.
		\label{bos:fig:feynman-G05-12-345}
	}{%
		\centering
		\includegraphics[scale=1]{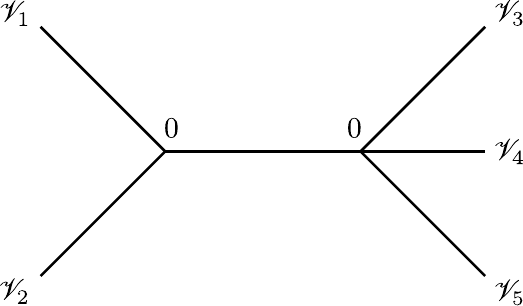}
	}%
	\hspace{1cm}
	\subcaptionbox{%
		Fundamental vertex.
		\label{bos:fig:feynman-V05}
	}{%
		\centering
		\includegraphics[scale=1]{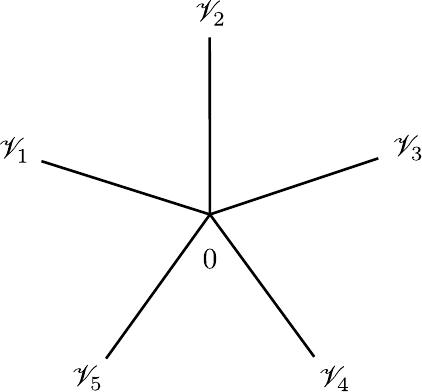}
	}%
	\caption{%
		Factorization of the amplitude $G_{0,5}$ in channels and fundamental quintic vertex.
		Only the cases where $\scr V_1$ and $\scr V_2$ factorize on one side is indicated, the other cases follow by permutations of the external states.
	}%
	\label{bos:fig:feynman-G05}
\end{figure}

The amplitude $A_{0,5}$ can be factorized in a greater number of channels, the two types being $2 | 2 | 1$ and $2 | 3$.
The possible Feynman graphs are built either from three cubic vertices and two propagators (\Cref{bos:fig:feynman-G05-12-3-45} and permutations), or from one cubic and one quartic vertices together with one propagator (\Cref{bos:fig:feynman-G05-12-345} and permutations).
The remaining contribution is the fundamental vertex (\Cref{bos:fig:feynman-V05}):
\begin{equation}
	\mc V_{0,5}(\scr V_1, \ldots, \scr V_5)
		:= \int_{\mc R_{0,5}} \omega_4^{0,5}(\scr V_1, \ldots, \scr V_5).
\end{equation}

The construction to higher-order follows exactly this scheme.

\subsubsection{Recursive definition: general vertices}

\begin{figure}[htp]
	\centering
	\subcaptionbox{Internal loop.}[0.4\linewidth]{%
		\centering
		\includegraphics[scale=1]{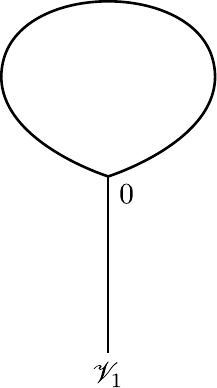}
		\label{bos:fig:feynman-G11-glued}
	}%
	\hspace{1cm}
	\subcaptionbox{Fundamental vertex.}[0.4\linewidth]{%
		\centering
		\includegraphics[scale=1]{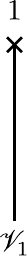}
		\label{bos:fig:feynman-V11}
	}%
	\caption{%
		Factorization of the amplitude $G_{1,1}$ and fundamental tadpole at $1$-loop.
	}%
	\label{bos:fig:feynman-G11}
\end{figure}

Next, one needs to consider Feynman diagrams with loops.
The first amplitude which can be considered is the one-loop tadpole $A_{1,1}(\scr V_1)$.
The factorization region corresponds to the graph obtained by gluing two legs of the cubic vertex (\Cref{bos:fig:feynman-G11-glued}).
\index{closed string fundamental vertex!1-loop!1-point}%
The remaining contribution is the fundamental tadpole vertex $\mc V_{1,1}(\scr V_1)$ (\Cref{bos:fig:feynman-V11}) -- note the index $g = 1$ on the vertex, indicating that it is a $1$-loop effect.

Next, the $1$-loop $2$-point amplitude can be obtained using the cubic and quartic tree-level vertices $\mc V_{0,3}$ and $\mc V_{0,4}$, but also the one-loop tadpole $\mc V_{1,1}$.
Iterating, the number of loops can be increased either by gluing together two external legs of a graph, or by gluing two different graphs with loops together.

\index{closed string fundamental vertex!$g$-loop!0-point}%
For $g \ge 2$, the recursion implies the existence of vertices with no external states $\mc V_{g,0}$: they should be interpreted as loop corrections to the vacuum energy density.

It is important to realize that, in this language, a handle in the Riemann surface is not necessarily mapped to a loop in the Feynman graph: only handles described by the region $\mc F_{g,n} = \mc M_{g,n} - \mc V_{g,n}$ do.
The higher-order vertices -- corresponding to surfaces with small handles only and described by $\mc V_{g,n}$ -- should be regarded as quantum fundamental interactions.
In \Cref{bsft:chap:bv-sft}, it will be explained that they really correspond to (finite) counter-terms: the measure is not invariant under the gauge symmetry of the theory and these terms must be introduced to restore it.

\subsubsection{Other vertices}

\index{closed string fundamental vertex!special vertices|(}%

The definition given at the end of \eqref{bos:sec:amp-prop:feynman:graphs} suggests to introduce additional vertices.
The previous recursive definition gives only vertices with $\chi_{g,n} = 2 - 2 g - n < 0$, but, in fact, it makes sense to consider the additional cases: $g = 0$ and $n = 0, 1, 2$, and $g = 1$, $n = 0$.

The definition of the vertices as amputated Green function without internal propagators provides a hint for the tree-level quadratic vertex $\mc V_{0,2}$.
\index{closed string fundamental vertex!tree-level!2-point}%
\index{Green function!tree-level 2-point}%
We define the latter as the amputated tree-level $2$-point Green function:
\begin{equation}
	\label{bos:eq:V02-vertex}
	\mc V_{0,2}
		:= \Delta^{-1} \Delta \Delta^{-1}
		= \Delta^{-1}.
\end{equation}
Hence, we have
\begin{equation}
	\mc V_{0,2}(\scr V_1, \scr V_2)
		:= \bra{\scr V_1} c_0^+ c_0^- L_0^+ \delta_{L_0^-,0} \ket{\scr V_2}.
\end{equation}
Note that $\mc V_{0,2}$ is \emph{not} the $2$-point scattering amplitude.

\index{closed string fundamental vertex!tree-level!1-point}%
\index{closed string fundamental vertex!tree-level!0-point}%
We denote the tree-level $1$-point and $0$-point vertices as $\mc V_{0,1}(\scr V_1)$ and $\mc V_{0,0}$.
The first can be interpreted as a classical source in the action, while the second is a classical vacuum energy.
They are set to zero in most applications and can be safely ignored.
However, they appear when formulating the theory on a background which does not solve the equation of motion~\cite{Zwiebach:1996:BuildingStringField}.

\index{closed string fundamental vertex!1-loop!0-point}%
Finally, the $1$-loop vacuum energy $\mc V_{1,0}$ can also be defined as the partition function of the worldsheet CFT integrated over the torus modulus.

This allows to define the vertices $\mc V_{g,n}$ for all $g, n \in \N$.
We define the sum of all loop contributions for a fixed $n$ as:
\begin{equation}
	\label{bos:eq:Vn-vertex}
	\mc V_{n}(\scr V_1, \ldots, \scr V_n)
		:= \sum_{g \ge 0} (\hbar g_s^2)^g \,
			\mc V_{g,n}(\scr V_1, \ldots, \scr V_n).
\end{equation}

\index{closed string fundamental vertex!special vertices|)}%

\subsection{Stubs}
\label{bos:sec:amp-prop:stubs}

\index{stub!Feynman diagram|(}%

In \Cref{bos:sec:offshell:geometry:plumbing:stubs}, we have indicated that the plumbing fixture can be modified by adding stubs or, equivalently, by rescaling the local coordinates.
This amounts to introduce a cut-off \eqref{bos:eq:q-s-th-stub} on the variable $s$ such that
\begin{equation}
	\label{bos:eq:q-s-th-stub-fact}
	q
		= \e^{- s + \I \theta},
	\qquad
	s \in [s_0, \infty),
	\qquad
	\theta \in [0, 2\pi).
\end{equation}
instead of \eqref{bos:eq:q-s-th-fact}.
In this case, the $s$-integral in the propagator \eqref{bos:eq:fact-prop-op} is modified to
\begin{equation}
	\int_{s_0}^{\infty} \dd s \, \e^{- s L_0^+}
		= \frac{\e^{- s_0 L_0^+}}{L_0^+}.
\end{equation}
\index{propagator!closed bosonic!with stub}%
This leads to a new expression for the propagator:
\begin{equation}
	\label{bos:eq:offshell-prop-bpm-stub}
	\Delta(s_0)
		= b_0^+ \, \frac{\e^{- s_0 L_0^+}}{L_0^+} \;
			b_0^- \delta_{L_0^-, 0}.
\end{equation}
In momentum space, this reads
\begin{equation}
	\label{bos:eq:offshell-prop-k-stub}
	\Delta_{\alpha\beta}(k)
		:= \frac{\e^{- \frac{\alpha' s_0}{2} (k^2 + m_\alpha^2)}}{k^2 + m_\alpha^2} \, M_{\alpha\beta}(k).
\end{equation}

It is more convenient to work with the canonical propagator \eqref{bos:eq:offshell-prop-bpm}.
This can be achieved by absorbing $\e^{- \frac{s_0}{2} L_0^+}$ in the interaction vertex: a $n$-point interaction will get $n$ such factors.\footnotemark{}
\footnotetext{%
	To make this identification precise for vertices involving external states, one has to consider the non-amputated Green functions.
}%

\index{string Feynman diagram!change of stub parameter}%
Since $s_0$ changes the local coordinates, this means that it also changes the region $\mc V_{g,n}$ (\Cref{bos:fig:M04-covering-stub}).
The freedom in the choice of $s_0$ translates into a freedom to choose which part of the amplitude is described by propagator graphs $\mc F_{g,n}(s_0)$, and which part is described by a fundamental vertex $\mc V_{g,n}(s_0)$.
The amplitude $A_{g,n}$ is independent of $s_0$ since it is described in terms of the complete moduli space $\mc M_{g,n}$.
This also means that the parameter $s_0$ must disappear when summing over the contributions from $\mc V_{g,n}(s_0)$ and $\mc F_{g,n}(s_0)$.
This indicates that the value of $s_0$ is not relevant, even off-shell: it can be taken to any convenient value.

\index{momentum-space SFT!finiteness!UV divergence}%
\index{momentum-space SFT!finiteness!infinite number of states}%
The possibility of adding stubs solves the problem that the sum over all states could diverge (see \Cref{bos:sec:amp-prop:factorization:sep}).
Indeed, the expression \eqref{bos:eq:offshell-prop-k-stub} in momentum space shows that the propagator includes an exponential suppression for very massive particle propagating as intermediate states.
Since the mass of a particle increases with the level, this shows that the sum converges for a sufficiently large value of $s_0$ thanks to the factor $\e^{- \alpha' s_0 m^2}$.
A second interesting aspect is the exponential momentum suppression $\e^{- \alpha' s_0 k^2}$: this is responsible for the nice UV behaviour of string theory.
Since the value of $s_0$ is not physical, this means that all Feynman graphs must share these properties.
These two points will be made more precise in \Cref{sft:chap:momentum-sft}.

\index{stub!Feynman diagram|)}%

\subsection{1PI vertices}
\label{bos:sec:amp-prop:1PI-vertices}

\index{plumbing fixture!separating}%
We can follow the same procedure as before, but considering only the separating plumbing fixture.
In this case, the Feynman diagrams are all 1PR ($1$-particle reducible): if the propagator line is cut, then the graphs split in two disconnected components.
\index{string Feynman diagram!1PR diagram}%
The region of the moduli space covered by these graphs is written as $\mc F_{g,n}^{\text{1PR}}$ \eqref{bos:eq:def-Fgn-1PR-space}.
\index{1PI vertex}%
The complement defines the 1PI region $\mc V_{g,n}^{\text{1PI}}$ \eqref{bos:eq:def-Vgn-1PI-space}.
Then, the 1PI $g$-loop $n$-point fundamental vertices are defined as:
\begin{equation}
	\label{bos:eq:Vgn-vertex-1PI}
	\mc V_{g,n}^{\text{1PI}}(\scr V_1, \ldots, \scr V_n)
		:= \quad
			\vcenter{\hbox{\includegraphics{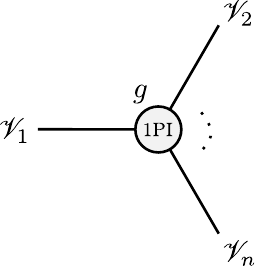}}}
		:= \int_{\mc R_{g,n}^{\text{1PI}}} \omega^{g,n}_{\M_{g,n}}(\scr V_1, \ldots, \scr V_n),
\end{equation}
where $\mc R_{g,n}^{\text{1PI}}$ is a section of $\mc P_{g,n}$ which projection on the base is $\mc V_{g,n}^{\text{1PI}}$.

\section{Properties of fundamental vertices}
\label{bos:sec:amp-prop:vertices}

\index{closed string fundamental vertex!properties|(}%

\subsection{String product}

\index{vertex state}%
Following the definition of surfaces states (\Cref{bos:sec:offshell:geometry:surface-states}), the vertex state is defined as:
\begin{equation}
	\label{bos:eq:state-vertex}
	\bracket{\mc V^{g,n}}{\otimes_i \scr V_i}
		:= \mc V_{g,n}(\otimes_i \scr V_i).
\end{equation}

The vertex is a map $\mc V_{g,n}: \mc H^{\otimes n} \to \C$ where $\C \simeq \mc H^{\otimes 0}$.
\index{closed string product}%
We will find very useful to introduce the string products $\ell_{g,n}: \mc H^{\otimes n} \to \mc H$ through the closed string inner product:
\begin{equation}
	\label{bos:eq:lgn-product}
	\mc V_{g,n+1}(\scr V_0, \scr V_1, \ldots, \scr V_n)
		:= \bra{\scr V_0} c_0^- \Ket{\ell_{g,n}(\scr V_1, \ldots, \scr V_n)}.
\end{equation}
An alternative notation is:
\begin{equation}
	\ell_{g,n}(\scr V_1, \ldots, \scr V_n)
		:= [\scr V_1, \ldots, \scr V_n]_g
\end{equation}
The advantage of the second notation is to show that the products with $n \ge 3$ are direct generalization of the $2$-product, which is very similar to a super-Lie bracket.
These products play a central role in SFT -- in fact, the description of SFT is more natural using the $\ell_{g,n}$ rather than the $\mc V_{g,n}$.

Note that the products with $n = 0$ are maps $\C \to \mc H$, which means that they correspond to a particular fixed state.
\begin{equation}
	\label{bos:eq:lg0-state}
	\ell_{g,0}
		:= [\cdot]_g
		\in \mc H.
\end{equation}

\index{closed string product!ghost number}%
The ghost number of the product \eqref{bos:eq:lgn-product} is
\begin{equation}
	\label{bos:eq:Ngh-bracket-state}
	N_{\text{gh}}\big(\ell_{g,n}(\scr V_1, \ldots, \scr V_n)\big)
		= 3 - 2 n + \sum_{i=1}^n N_{\text{gh}}(\scr V_i)
		= 3 + \sum_{i=1}^n \big( N_{\text{gh}}(\scr V_i) - 2 \big),
\end{equation}
and it is independent of the genus $g$.
As a consequence, the parity of the product is
\begin{equation}
	\Abs{\ell_{g,n}(\scr V_1, \ldots, \scr V_n)}
		= 1 + \sum_{i=1}^n \abs{\scr V_i} \mod 2,
\end{equation}
and the string product itself is always odd.

\index{closed string vertex!fundamental identity}%
The vertices satisfy the following identity for $g \ge 0$ and $n \ge 1$~\cite[pp.~41--42]{Zwiebach:1993:ClosedStringField}
\begin{equation}
	\label{bos:eq:vertex-identity}
	0
		= \sum_{\substack{g_1, g_2 \ge 0 \\ g_1 + g_2 = g}}
				\sum_{\substack{n_1, n_2 \ge 0 \\ n_1 + n_2 = n}}
				\frac{n!}{n_1! \, n_2!} \mc V_{g_1, n_1+1}\big(\Psi^{n_1}, \ell_{g_2, n_2}(\Psi^{n_2})\big)
			+ (-1)^{\abs{\phi_s}} \mc V_{g-1, n+2}\big(\phi_s, b_0^{-} \phi_s^c, \Psi^n\big).
\end{equation}
The last term is absent for $g = 0$.
It is a consequence of the definition of the vertices as the missing region from gluing lower-order vertices.

\index{closed string fundamental vertex!properties|)}%

\subsection{Feynman graph interpretation}

The vertices must satisfy a certain number of conditions to be interpreted as Feynman diagrams.
The first is that they must be symmetric under permutations of the states.
\index{section of $\mc P_{g,n}$!generalized}%
Not every choice of local coordinates satisfies this requirement:
this can be solved by defining the vertex over a generalized section.
In this case, the vertex is defined as the average of the integrals over $N$ sections $\mc S^{(a)}_{g,n}$ of $\mc P_{g,n}$:
\begin{equation}
	\label{bos:eq:vertex-offshell-generalized}
	\mc V_{g,n}(\scr V_1, \ldots, \scr V_n)
		= \frac{1}{N} \sum_{a=1}^N
			\int_{\mc R^{(a)}_{g,n}} \omega^{g,n}_{\M_{g,n}}(\scr V_1, \ldots, \scr V_n).
\end{equation}

\begin{example}[$3$-point vertex]
	The cubic vertex must be symmetric under permutations
	\begin{equation}
		\mc V_{0,3}(\scr V_1, \scr V_2, \scr V_3)
			= \mc V_{0,3}(\scr V_3, \scr V_1, \scr V_2)
			+ \cdots
	\end{equation}
	Taking the vertex to be given by a section $\mc S_{0,3}$ with local coordinates $f_i$
	\begin{equation}
		\mc V_{0,3}(\scr V_1, \scr V_2, \scr V_3)
			= \omega_{0}^{0,3}(\scr V_1, \scr V_2, \scr V_3)|_{\mc S_{0,3}}
			= \mean{f_1 \circ \scr V_1(0) f_2 \circ \scr V_2(0) f_3 \circ \scr V_3(0)},
	\end{equation}
	one finds that a permutation looks different
	\begin{equation}
		\mc V_{0,3}(\scr V_3, \scr V_1, \scr V_2)
			= \mean{f_1 \circ \scr V_3(0) f_2 \circ \scr V_1(0) f_3 \circ \scr V_2(0)}
			\neq \mc V_{0,3}(\scr V_1, \scr V_2, \scr V_3),
	\end{equation}
	unless the local coordinates satisfy special properties (remember that the local coordinates are specified by the vertex state $\mc V$ and not by the external states $\scr V_i$, so a permutation of them does not permute the local maps).
	Obviously, both amplitudes agree on-shell since the dependence in the local coordinates cancel (equivalently one can rotate the punctures using $\group{SL}(2, \C)$).

	Writing $z_i = f_i(0)$, there is a $\group{SL}(2, \C)$ transformation $g(z)$ such that
	\begin{equation}
		g(z_1) = z_2,
		\qquad
		g(z_2) = z_3,
		\qquad
		g(z_3) = z_1
	\end{equation}
	such that
	\begin{equation}
		\mc V_{0,3}(\scr V_3, \scr V_1, \scr V_2)
			= \mean{g \circ f_1 \circ \scr V_3(0) g \circ f_2 \circ \scr V_1(0) g \circ f_3 \circ \scr V_2(0)}.
	\end{equation}
	While the state $\scr V_i$ is correctly inserted at the puncture $z_i$ in this expression, this is not sufficient to guarantee the equality of the amplitudes.
	Indeed the fibre is defined by the complete functions $f_i(w)$ and not only by their values at $w = 0$.
	For this reason the amplitudes can be equal only if
	\begin{equation}
		g \circ f_1 = f_2,
		\qquad
		g \circ f_2 = f_3,
		\qquad
		g \circ f_3 = f_1.
	\end{equation}
	This provides constraints on the functions $f_i$, but it is often not possible to solve them.

	If the constraints cannot be solved, then one must introduce a general section.
	In this case a generalized section will be made of $6$ sections $\mc S^{(a)}$ ($a = 1, \ldots, 6$) because there are $6$ permutations.
	Then the amplitude reads
	\begin{equation}
		\mc V_{0,3}(\scr V_1, \scr V_2, \scr V_3)
			= \frac{1}{6} \sum_{a=1}^{6} \omega_{0}^{0,3}(\scr V_1, \scr V_2, \scr V_3)|_{\mc S^{(a)}_{0,3}}.
	\end{equation}

\end{example}

\begin{figure}[htp]
	\centering
	\includegraphics[scale=1.4]{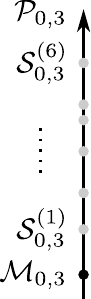}
	\caption{%
		A generalized section $\{ \mc S^{(a)}_{0,3} \}$ ($a = 1, \ldots, 6$) of $\mc P_{0,3}$ for the $3$-point vertex.
		This is to be compared with \Cref{bos:fig:P03-section}.
	}%
	\label{bos:fig:P03-section-generalized}
\end{figure}

\index{section of $\mc P_{g,n}$!overlap}%
When computing the Feynman graphs by gluing lower-dimensional amplitudes, it is possible that parts of the section overlap, meaning that several graphs cover the same part of the moduli space.
In this case, the fundamental vertex should be defined as a negative contribution in the overlap region.
This procedure is perfectly well-defined since all graphs are finite and there is no ambiguity.
In practice, it is always simpler to work with non-overlapping sections (i.e.\ a single covering of the moduli space).
A simple way to prevent overlaps is to tune the stub parameter $s_0$ to a large value.

By construction, the integral over $\mc V_{g,n}$ should be finite.
If this is not the case, it means that the propagator graphs also diverge and that the parametrization is not good.
This can also be solved by considering a sufficiently large value of the stub parameter $s_0$.

\refchapter

\begin{itemize}
	\item Plumbing fixture and amplitude factorization~\cites[sec.~9.3, 9.4]{Polchinski:2005:StringTheory-1}[sec.~6]{Witten:2012:SuperstringPerturbationTheory}.
\end{itemize}

\chapter{Closed string field theory}
\label{bsft:chap:bv-sft}

\introchapter

We bring together the elements from the previous chapters in order to write the closed string field action.
We first study the gauge fixed theory before reintroducing the gauge invariance.
We then prove that the action satisfies the BV master equation meaning that closed SFT is completely consistent at the quantum level.
Finally, we describe the 1PI effective action.

\begin{draft}
Since SFT is off-shell, we will need the results of \Cref{bos:chap:offshell,bos:chap:offshell-amp,bos:chap:feynman}, and the reader not familiar with off-shell amplitudes is invited to read first these chapters.
We focus on the closed string theory.
\end{draft}

\section{Closed string field expansion}

\index{string field!closed bosonic!expansion}%
In \Cref{bos:chap:offshell,bos:chap:offshell-amp,bos:chap:feynman}, constraints on the external and internal states were found to be necessary.
But, to provide another perspective and decouples the properties of the field from the ones of the state, we assume that the string field does not obey any constraint.
They will be derived later in order to reproduce the scattering amplitudes from the action and to make the latter well-defined.

The string field is expanded on a basis $\{ \phi_r \}$ of the CFT Hilbert space $\mc H$ (see \Cref{bos:sec:offshell:states} for more details)
\begin{equation}
	\label{bsft:eq:field-exp}
	\ket{\Psi}
		= \sum_r \psi_r \ket{\phi_r}.
\end{equation}
Using the decomposition \eqref{bos:eq:hilbert-H-decomposition} of the Hilbert space according to the ghost zero-modes, the string field can also be expanded as
\begin{equation}
	\ket{\Psi}
		= \sum_r \Big( \psi_{\downarrow\downarrow, r} \ket{\phi_{\downarrow\downarrow, r}}
			+ \psi_{\downarrow\uparrow, r} \ket{\phi_{\downarrow\uparrow, r}}
			+ \psi_{\uparrow\downarrow, r} \ket{\phi_{\uparrow\downarrow, r}}
			+ \psi_{\uparrow\uparrow, r} \ket{\phi_{\uparrow\uparrow, r}}
			\Big),
\end{equation}
where we recall that the basis states satisfy
\begin{equation}
	\begin{gathered}
		b_0 \ket{\phi_{\downarrow\downarrow, r}}
			= \bar b_0 \ket{\phi_{\downarrow\downarrow, r}}
			= 0,
		\qquad
		b_0 \ket{\phi_{\downarrow\uparrow, r}}
			= \bar c_0 \ket{\phi_{\downarrow\uparrow, r}}
			= 0,
		\\
		c_0 \ket{\phi_{\uparrow\downarrow, r}}
			= \bar b_0 \ket{\phi_{\uparrow\downarrow, r}}
			= 0,
		\qquad
		c_0 \ket{\phi_{\uparrow\uparrow, r}}
			= \bar c_0 \ket{\phi_{\uparrow\uparrow, r}}
			= 0.
		\end{gathered}
\end{equation}
We recall the definition of the dual basis $\{ \phi_r^c \}$ through the BPZ inner product
\begin{equation}
	\bracket{\phi_r^c}{\phi_s}
		= \delta_{rs}.
\end{equation}
In terms of the ghost decomposition, the components of the dual states satisfy:
\begin{equation}
	\begin{gathered}
		\bra{\phi_{\downarrow\downarrow, r}^c} c_0
			= \bra{\phi_{\downarrow\downarrow, r}^c} \bar c_0
			= 0,
		\qquad
		\bra{\phi_{\downarrow\uparrow, r}^c} c_0
			= \bra{\phi_{\downarrow\uparrow, r}^c} \bar b_0
			= 0,
		\\
		\bra{\phi_{\uparrow\downarrow, r}^c} b_0
			= \bra{\phi_{\uparrow\downarrow, r}^c} \bar c_0
			= 0,
		\qquad
		\bra{\phi_{\uparrow\uparrow, r}^c} b_0
			= \bra{\phi_{\uparrow\uparrow, r}^c} \bar b_0
			= 0,
	\\
	\bracket{\phi_{x,r}^c}{\phi_{y,s}}
		= \delta_{xy} \delta_{rs},
	\end{gathered}
\end{equation}
where $x, y = \downarrow\downarrow, \uparrow\downarrow, \downarrow\uparrow, \uparrow\uparrow$.
\index{spacetime ghost number!closed string}%
The spacetime ghost number of the fields $\psi_r$ is defined by
\begin{equation}
	\label{bsft:eq:closed-spacetime-Ngh}
	G(\psi_r) = 2 - n_r.
\end{equation}
Remember that the ghost number of the basis states are denoted by
\begin{equation}
	\label{bsft:eq:closed-states-Ngh}
	n_r = N_{\text{gh}}(\phi_r),
	\qquad
	n_r^c = N_{\text{gh}}(\phi_r^c)
		= 6 - n_r.
\end{equation}

\section{Gauge fixed theory}
\label{bsft:sec:bv-sft:gauge-fixed}

Having built the kinetic term (\Cref{bsft:chap:string-field}), one needs to construct the interactions.
For the same reason -- our ignorance of SFT first principles -- that forced us to start with the free equation of motion to derive the quadratic action (\Cref{bsft:chap:free-brst}), we also need to infer the interactions from the scattering amplitudes.
Preparing the stage for this analysis was the goal of \Cref{bos:chap:feynman}, where we introduced the factorization of amplitudes to derive the fundamental interactions.

Scattering amplitudes are expressed in terms of gauge fixed states since only them are physical.
This allows to give an alternative derivation of the kinetic term by defining it as the inverse of the propagator, which is well-defined for gauge fixed states.\footnotemark{}
\footnotetext{%
	This step is not necessary because the propagator corresponding to the plumbing fixture (\Cref{bos:sec:amp-prop:propagator}) matches the one found in \Cref{bsft:sec:free-brst:closed} by considering the simplest gauge fixing.
	However, this would have been necessary if the factorization had given another propagator, or if the structure of the theory was more complicated, for example for the superstring.
}%
The price to pay by constructing interactions in this way is that the SFT action itself is gauge fixed.
To undercover its deeper structure it is necessary to release the gauge fixing condition.
In view of the analysis of the quadratic action in \Cref{bsft:chap:free-brst}, we can expect that the BV formalism is required.
Another possibility is to consider directly the 1PI action.

In this section, we first derive the kinetic term by inverting the propagator.
For this to be possible, the string field must obey some constraints: we will find that they correspond to the level-matching and Siegel gauge conditions.
Then, we introduce the interactions into the action.

\subsection{Kinetic term and propagator}

In \Cref{bos:chap:feynman}, it was found that the propagator reads \eqref{bos:eq:offshell-prop-bpm}:
\begin{equation}
	\label{bsft:eq:propagator-fixed}
	\Delta
		= b_0^+ b_0^- \, \frac{1}{L_0^+} \, \delta_{L_0^-, 0},
	\qquad
	\Delta_{rs}
		= \bra{\phi_r^c} b_0^+ b_0^- \, \frac{1}{L_0^+} \, \delta_{L_0^-, 0} \ket{\phi_s^c}.
\end{equation}
The most natural guess for the kinetic term is
\begin{equation}
	\label{bsft:eq:closed-action-fixed-02}
	S_{0,2}
		= \frac{1}{2} \, \bra{\Psi} K \ket{\Psi}
		= \frac{1}{2} \, \psi_r K_{rs} \psi_s
\end{equation}
where
\begin{equation}
	\label{bsft:eq:closed-fixed-kinetic-op}
	K
		= c_0^- c_0^+ L_0^+ \delta_{L_0^-, 0}
	\qquad
	K_{rs}
		= \bra{\phi_r} c_0^- c_0^+ L_0^+ \delta_{L_0^-, 0} \ket{\phi_s}.
\end{equation}
Indeed, it looks like $K \Delta = 1$ using the identities $c_0^\pm b_0^\pm \sim 1$ and it matches \eqref{bsft:eq:closed-action-siegel}.
In terms of the holomorphic and anti-holomorphic modes, we have
\begin{equation}
	K
		= \frac{1}{2} \, c_0 \bar c_0 L_0^+ \delta_{L_0^-, 0}.
\end{equation}

But, when writing $c_0^\pm b_0^\pm \sim 1$, the second part of the anti-commutator $\anticom{b_0^\pm}{c_0^\pm} = 1$ is missing.
The relation $c_0^\pm b_0^\pm \sim 1$ is correct only when acting on basis dual states annihilated by $c_0^\pm$.
The problem stems from the fact that $\Psi$ is not yet subject to any constraint.
Moreover, some of the string field components will not appear in the expression since they are annihilated by the ghost zero-mode.
As a consequence, the kinetic operator in \eqref{bsft:eq:closed-fixed-kinetic-op} (or equivalently the propagator) is not invertible in the Hilbert space $\mc H$ because its kernel is not empty:
\begin{equation}
	\ker K|_{\mc H} \neq \emptyset.
\end{equation}
This can be seen by writing $\phi_r$ as a $4$-vector and $K_{rs}$ as a $4 \times 4$-matrix:
\begin{equation}
	K_{rs}
		= \frac{1}{2}
			\begin{pmatrix}
				\bra{\phi_{\downarrow\downarrow, r}} \\
				\bra{\phi_{\downarrow\uparrow, r}} \\
				\bra{\phi_{\uparrow\downarrow, r}} \\
				\bra{\phi_{\uparrow\uparrow, r}}
			\end{pmatrix}^t
			\begin{pmatrix}
				c_0 \bar c_0 L_0^+ & 0 & 0 & 0 \\
				0 & 0 & 0 & 0 \\
				0 & 0 & 0 & 0 \\
				0 & 0 & 0 & 0
			\end{pmatrix}
			\begin{pmatrix}
				\ket{\phi_{\downarrow\downarrow, s}} \\
				\ket{\phi_{\downarrow\uparrow, s}} \\
				\ket{\phi_{\uparrow\downarrow, s}} \\
				\ket{\phi_{\uparrow\uparrow, s}}
			\end{pmatrix}.
\end{equation}
The matrix is mostly empty because the states $\phi_{x,r}$ with different $x = \downarrow\downarrow, \uparrow\downarrow, \downarrow\uparrow, \uparrow\uparrow$ are orthogonal (no non-diagonal terms) and the states with $x \neq \downarrow\downarrow$ are annihilated by $c_0$ or $\bar c_0$.
The same consideration applies for the delta-function: if the field does not satisfy $L_0^- = 0$, then the kinetic operator is non-invertible.

\index{level-matching condition}%
To summarize the string field must satisfy three conditions in order to have an invertible kinetic term
\begin{equation}
	\label{bsft:eq:field-constraints}
	L_0^- \ket{\Psi}
		= 0,
	\qquad
	b_0^- \ket{\Psi}
		= 0,
	\qquad
	b_0^+ \ket{\Psi}
		= 0.
\end{equation}
This means that the string field is expanded on the $\mc H_0 \cap \ker L_0^-$ Hilbert space:
\begin{equation}
	\label{bsft:eq:field-exp-dd}
	\ket{\Psi}
		= \sum_r \psi_{\downarrow\downarrow, r} \ket{\phi_{\downarrow\downarrow, r}}.
\end{equation}
Ill-defined kinetic terms are expected in the presence of a gauge symmetry: this was already discussed in \Cref{bsft:sec:free-brst:open:siegel,bsft:sec:free-brst:closed} for the free theory, and this will be discussed further later in this chapter for the interacting case.

\begin{computation}
	Let's check that $K_{rs}$ is correctly the inverse of $\Delta_{rs}$ when $\Psi$ is restricted to $\mc H_0$:
	\begin{align*}
		K_{rs} \Delta_{st}
			&= \bra{\phi_r} c_0^- c_0^+ L_0^+ \delta_{L_0^-, 0} \ket{\phi_s}
				\bra{\phi_s^c} b_0^+ b_0^- \, \frac{1}{L_0^+} \, \delta_{L_0^-, 0} \ket{\phi_t^c}
			\\
			&= \bra{\phi_r} c_0^- c_0^+ L_0^+ \delta_{L_0^-, 0} b_0^+ b_0^- \, \frac{1}{L_0^+} \, \delta_{L_0^-, 0} \ket{\phi_t^c}
			\\
			&= \bra{\phi_r} \anticom{c_0^-}{b_0^-} \anticom{c_0^+}{b_0^+} \ket{\phi_t^c}
			\\
			&= \bracket{\phi_r}{\phi_t^c}
			= \delta_{rt}.
	\end{align*}
	The second equality follows from the resolution of the identity \eqref{bos:eq:identity-offshell}: due to the zero-mode insertions, the resolution of the identity collapses to a sum over the $\downarrow\downarrow$ states
	\begin{equation}
		1
			= \sum_{r} \ket{\phi_r} \bra{\phi_r^c}
			= \sum_{r} \ket{\phi_{\downarrow\downarrow,r}} \bra{\phi_{\downarrow\downarrow,r}^c}.
	\end{equation}
	The third equality uses that $L_0^+$ commutes with the ghost modes, that $\phi_r$ is annihilated by $b_0^\pm$, and that $(\delta_{L_0^-,0})^2 = \delta_{L_0^-, 0} = 1$ on states with $L_0^- = 0$.
\end{computation}

\index{free covariant SFT!closed bosonic!gauge fixed action}%
Finally, we find that the kinetic term matches the classical quadratic vertex $\mc V_{0,2}$ defined in \eqref{bos:eq:V02-vertex} such that
\begin{equation}
	\label{bsft:eq:action-S02}
	S_{0,2}
		= \frac{1}{2} \, \mc V_{0,2}(\Psi^2)
		= \frac{1}{2} \, \bra{\Psi} c_0^- c_0^+ L_0^+ \delta_{L_0^-, 0} \ket{\Psi}.
\end{equation}

\subsection{Interactions}

The second step to build the action is to write the interaction terms from the Feynman rules.
Before proceeding to SFT, it is useful to remember how this works for a standard QFT.

\begin{example}[Feynman rules for a scalar field]
	\label{bsft:ex:feynman-qft}

	Consider a scalar field with a standard kinetic term and a $n$-point interaction:
	\begin{equation}
		S
			= \int \dd^D x \left( \frac{1}{2} \, \phi(x) (- \pd^2 + m^2) \phi(x) + \frac{\lambda}{n!} \, \phi(x)^n \right).
	\end{equation}
	First, one needs to find the physical states, which correspond to solutions of the linearised equation of motion.
	In the current case, they are plane-waves (in momentum representation):
	\begin{equation}
		\phi_k(x)
			= \e^{\I k \cdot x}.
	\end{equation}
	Then, the vertex (in momentum representation) $V_n(k_1, \ldots, k_n)$ is found by replacing in the interaction each occurrence of the field by a different state, and summing over all the different contributions.
	Here, this means that one considers states $\phi_{k_i}(x)$ with different momenta:
	\begin{equation}
		\begin{aligned}
		V_n(k_1, \ldots, k_n)
			&
			= \frac{\lambda}{n!} \int \dd^D x \, n! \prod_{i=1}^n \phi_{k_i}(x)
			= \lambda \int \dd^D x \, \e^{\I (k_1 + \cdots + k_n) x}
			\\ &
			= \lambda (2\pi)^D \, \delta^{(D)}(k_1 + \cdots + k_n).
		\end{aligned}
	\end{equation}
	The factor $n!$ comes from all the permutations of the $n$ states in the monomial of order $n$.
	Reversing the argument, one sees how to move from the vertex $V_n(k_1, \ldots, k_n)$ written in terms of states to the interaction in the action in terms of the field.

	Obviously, if the field has more states (for example if it has a spin or if it is in a representation of a group), then one needs to consider all the different possibilities.
	The above prescription also yields directly the insertion of the momentum necessary if the interaction contains derivatives.
\end{example}

In \Cref{bos:sec:amp-prop:feynman}, the Feynman rule for a $g$-loop $n$-point fundamental vertex of states $(\scr V_1, \ldots, \scr V_n)$ was found to be given by \eqref{bos:eq:Vgn-vertex}:
\begin{equation}
	\label{bsft:eq:Vgn-vertex}
	\mc V_{g,n}(\scr V_1, \ldots, \scr V_n)
		= \int_{\mc R_{g,n}} \omega^{g,n}_{\mathsf{M}_{g,n}}(\scr V_1, \ldots, \scr V_n)
		= \quad \vcenter{\hbox{\includegraphics{bosonic/vertex_Vgn}}}
\end{equation}
where $\mc R_{g,n}$ is a section over the fundamental region $\mc V_{g,n} \subset \mc M_{g,n}$ \eqref{bos:eq:def-Vgn-space} which cannot be covered from the plumbing fixture of lower-dimensional surfaces.

From the example \Cref{bsft:ex:feynman-qft}, it should be clear that the $g$-loop $n$-point contribution to the action can be obtained simply by replacing every state with a string field in $\mc V_{g,n}$:
\begin{equation}
	\label{bsft:eq:action-Sgn}
	S_{g,n}
		= \hbar^g \, \frac{g_s^{2 g - 2 + n}}{n!} \, \mc V_{g,n}(\Psi^n).
\end{equation}
where $\Psi^n := \Psi^{\otimes n}$.
The power of the coupling constant has been reinstated: it can be motivated by the fact that it should have the same power as the corresponding amplitude (\Cref{bos:sec:ws-int-amp:Mg:vertex}).
Note that the interactions are defined only when the power of $g_s$ is positive: $\chi_{g,n} = 2 - 2 g - n < 0$.
We have also written explicitly the power of $\hbar$, which counts the number of loops.

Before closing this section, we need to comment on the effect of the constraints \eqref{bsft:eq:field-constraints} on the interactions.
Building a Feynman graph by gluing two $m$- and $n$-point interactions with a propagator, one finds that the states proportional to $\phi_{x,r}$ for $x \neq \downarrow\downarrow$ do not propagate inside internal legs
\begin{multline}
	\mc V_{g,m}(\scr V_1, \ldots, \scr V_{m-1}, \phi_r)
			\bra{\phi_r^c} b_0^+ b_0^- \, \frac{1}{L_0^+} \ket{\phi_s^c}
			\mc V_{g',n}(\scr W_1, \ldots, \scr W_{n-1}, \phi_s)
		\\
		= \mc V_{g,m}(\scr V_1, \ldots, \scr V_{m-1}, \phi_{\downarrow\downarrow,r})
			\bra{\phi_{\downarrow\downarrow,r}^c} b_0^+ b_0^- \, \frac{1}{L_0^+} \ket{\phi_{\downarrow\downarrow,s}^c}
			\mc V_{g',n}(\scr W_1, \ldots, \scr W_{n-1}, \phi_{\downarrow\downarrow,s}).
\end{multline}
Thus, they do not contribute to the final result even if the interactions contain them.
While the conditions $L_0^- = b_0^- = 0$ were found to be necessary for defining off-shell amplitudes, the condition $b_0^+ = 0$ does not arise from any consistency requirement.
But, it is also consistent with the interactions, since only fundamental vertices have a chance to give a non-vanishing result for states which do not satisfy \eqref{bsft:eq:field-constraints}.
Hence, the interactions \eqref{bsft:eq:action-Sgn} are compatible with the definition of the kinetic term and the restriction of the string field.

\subsection{Action}

The interacting gauge-fixed action is built from the kinetic term $\mc V_{0,2}$ \eqref{bsft:eq:action-S02} and from the interactions $\mc V_{g,n}$ \eqref{bsft:eq:action-Sgn} with $\chi_{g,n} < 0$.
However, this is not sufficient: we have seen in \Cref{bos:sec:amp-prop:vertices} that it makes sense to consider the vertices with $\chi_{g,n} \ge 0$.
First, we should consider the $1$-loop cosmological constant $\mc V_{1,0}$.
Then, we can also add the classical source $\mc V_{0,1}$ and the tree-level cosmological constant $\mc V_{0,0}$.
With all the terms together, the action reads:
\index{covariant SFT!closed bosonic!gauge fixed action}%
\begin{equation}
	\label{bsft:eq:closed-action-gf}
	\begin{aligned}
	S
		&
		= \sum_{g, n \ge 0} \,
			\hbar^{g} \, \frac{g_s^{2 g - 2 + n}}{n!} \, \mc V_{g,n}(\Psi^n)
		\\ &
		:= \frac{1}{2} \, \bra{\Psi} c_0^- c_0^+ L_0^+ \delta_{L_0^-, 0} \ket{\Psi}
			+ \sideset{}{'}\sum_{g, n \ge 0}
				\hbar^{g} \, \frac{g_s^{2 g - 2 + n}}{n!} \, \mc V_{g,n}(\Psi^n).
	\end{aligned}
\end{equation}
where $\mc V_n$ was defined in \eqref{bos:eq:Vn-vertex}.
A prime on the sum indicates that the term $g = 0, n = 2$ is removed, such that one can single out the kinetic term.
We will often drop the delta function imposing $L_0^- = 0$ because the field are taken to satisfy this constraint.

Rewriting the vertices in terms of the products $\ell_{g,n}$ defined in \eqref{bos:eq:lgn-product}
\begin{equation}
	\mc V_{g,n}(\Psi^n)
		:= \bra{\Psi} c_0^- \Ket{\ell_{g,n-1}(\Psi^{n-1})}
\end{equation}
leads to the alternative form
\begin{equation}
	\label{bsft:eq:closed-action-gf-prod}
	S
		= \sum_{g, n \ge 0} \,
			\hbar^{g} \, \frac{g_s^{2 g - 2 + n}}{n!} \, \bra{\Psi} c_0^- \Ket{\ell_{g,n-1}(\Psi^{n-1})}.
\end{equation}
\index{closed string product!g0n1@$g = 0, n = 1$}%
The definition \eqref{bos:eq:V02-vertex} leads to the following explicit expression for $\ell_{0,1}$:
\begin{equation}
	\ell_{0,1}(\Psi_{\text{cl}})
		= c_0^+ L_0^+ \ket{\Psi_{\text{cl}}}.
\end{equation}

In most cases, the terms $g = 0, n = 0, 1$ vanish such that the action reads:
\begin{equation}
	S
		= \sum_{\mathclap{\substack{g, n \ge 0 \\ \chi_{g,n} \le 0}}} \,
			\hbar^{g} \, \frac{g_s^{2 g - 2 + n}}{n!} \, \mc V_{g,n}(\Psi^n).
\end{equation}
However, we will often omit the condition $\chi_{g,n} \le 0$ to simplify the notation, except when the distinction is important, and the reader can safely assumes $\mc V_{0,0} = \mc V_{0,1} = 0$ if not otherwise stated.
The classical action is obtained by setting $\hbar = 0$:
\begin{equation}
	\label{bsft:eq:closed-action-gf-cl}
	S_{\text{cl}}
		= \frac{1}{2} \, \bra{\Psi_{\text{cl}}} c_0^- c_0^+ L_0^+ \ket{\Psi_{\text{cl}}}
			+ \sum_{n \ge 3} \frac{g_s^{n}}{n!} \, \mc V_{0,n}(\Psi_{\text{cl}}^n).
\end{equation}

\index{covariant SFT!closed bosonic!normalization}%
\index{covariant SFT!closed bosonic!gauge fixed action}%
Rescaling the string field by $g_s^{-1}$ gives the more canonical form of the action (using the same symbol):
\begin{equation}
	\label{bsft:eq:closed-action-gf-gs}
	\begin{aligned}
	S
		&
		= \sum_{g, n \ge 0} \,
			\hbar^{g} g_s^{2 g - 2} \, \frac{1}{n!} \, \mc V_{g,n}(\Psi^n)
		\\ &
		:= \frac{1}{2 g_s^2} \, \bra{\Psi} c_0^- c_0^+ L_0^+ \delta_{L_0^-, 0} \ket{\Psi}
			+ \frac{1}{g_s^2} \sideset{}{'}\sum_{g, n \ge 0}
				\frac{(\hbar g_s^2)^{g}}{n!} \, \mc V_{g,n}(\Psi^n).
	\end{aligned}
\end{equation}
\index{covariant SFT!parameters}%
In the path integral, the action is divided by $\hbar$ such that
\begin{equation}
	\frac{S}{\hbar}
		= \sum_{g, n \ge 0} \,
				(\hbar g_s^2)^{g - 1} \, \frac{1}{n!} \, \mc V_{g,n}(\Psi^n).
\end{equation}
This shows that there is a single coupling constant $\hbar g_s^2$, instead of two ($\hbar$ and $g_s$ separately) as it looks at the first sight.
This makes sense because $g_s$ is in fact the expectation value of the dilaton field \eqref{bos:eq:gs-dilaton} and its value can be changed by deforming the background with dilatons~\cite{Belopolsky:1996:WhoChangesString, Rahman:1996:VacuumVerticesGhostdilaton, Bergman:1995:DilatonTheoremClosed}.

\index{covariant SFT!closed bosonic!normalization}%
The previous remark also allows to easily change the normalization of the action, for example, to perform a Wick rotation, to normalize canonically the action in terms of spacetime fields, or reintroduce $\hbar$.
Rescaling the action by $\alpha$ is equivalent to rescale $g_s^2$ by $\alpha^{-1}$:
\begin{equation}
	S \to \alpha \, S
	\quad \Longrightarrow \quad
	g_s^2 \to \frac{g_s^2}{\alpha}.
\end{equation}

\index{free covariant SFT!closed bosonic!gauge fixed equation of motion}%
The linearized equation of motion is:
\begin{equation}
	L_0^+ \ket{\Psi} = 0,
\end{equation}
which corresponds to the Siegel gauge equation of motion of the free theory \eqref{bsft:eq:closed-eom-siegel}.
Hence, this equation is not sufficient to determine the physical states (cohomology of the BRST operator, \Cref{cft:chap:brst}), as discussed in \Cref{bsft:chap:free-brst}, and additional constraints must be imposed.
One can interpret this by saying that the action \eqref{bsft:eq:closed-action-gf} provides only the Feynman rules, not the physical states.
Removing the gauge fixing will be done in \Cref{bsft:sec:bv-sft:classical,bsft:sec:bv-sft:bv}.

\index{fundamental vertex!interpretation}%
\index{covariant SFT!renormalization}%
\index{string Feynman diagram!IR divergence}%
The action \eqref{bsft:eq:closed-action-gf} looks overly more complicated than a typical QFT theory: instead of few interaction terms for low $n$ ($n \le 4$ in $d = 4$ renormalizable theories), it has contact interactions of all orders $n \in \N$.
The terms with $g \ge 1$ are associated to quantum corrections as indicate the power of $\hbar$, which means that they can be interpreted as counter-terms.
But, how is it that one needs counter-terms despite the claim that every Feynman graphs (including the fundamental vertices) in SFT are finite?
The role of renormalization is not only to cure UV divergences, but also IR divergences (due to vacuum shift and mass renormalization).
Equivalently, this can be understood by the necessity to correct the asymptotic states of the theory, or to consider renormalized instead of bare quantities.
Indeed, the asymptotic states obtained from the linearized classical equations of motion are idealization: turning on interactions modify the states.
In typical QFTs, these corrections are infinite and renormalization is crucial to extract a number; however, even if the effect is finite, it is needed to describe correctly the physical quantities~\cite[p.~411]{Weinberg:2005:QuantumTheoryFields-1}.
There is a second reason for these additional terms: when relaxing the gauge fixing condition, the path integral is anomalous under the gauge symmetry, and the terms with $g > 0$ are necessary to cancel the anomaly (this will be discussed more precisely in \Cref{bsft:sec:bv-sft:bv}).
It may thus seem that SFT cannot be predictive because of the infinite number of counter-terms.
Fortunately, this is not the case: the main reason for the loss of predictability in non-renormalizable theory is that the renormalization procedure introduces an infinite number\footnotemark{} of arbitrary parameters (and thus making a prediction would require to have already made an infinite number of observations to determine all the parameters).
\footnotetext{%
	In practice, this number does not need to be infinite to wreck predictability, it is sufficient that it is very large.
}%
These parameters come from the subtraction of two infinities: there is no unique way to perform it and thus one needs to introduce a new parameter.
The case of SFT is different: since every quantity is finite, the renormalization has no ambiguity because one subtracts two finite numbers, and the result is unambiguous.
As a consequence, renormalization does not introduce any new parameter and there is a unique coupling constant $g_s$ in the theory, which is determined by the tree-level cubic interaction.
The coupling constants of higher-order and higher-loop interactions are all determined by powers of $g_s$, and thus a unique measurement is sufficient to make predictions.

Another important point is that the action \eqref{bsft:eq:closed-action-gf} is not uniquely defined.
\index{stub parameter}%
The definition of the vertices depends on the choice of the local coordinates and of the stub parameter $s_0$.
Changing them modifies the vertices, and thus the action.
But, one can show that the different theories are related by field redefinitions and are thus equivalent.

\section{Classical gauge invariant theory}
\label{bsft:sec:bv-sft:classical}

In the previous section, we have found the gauge fixed action \eqref{bsft:eq:closed-action-gf}.
Since the complete gauge invariant quantum action has a complicated structure, it is instructive to first focus on the classical action \eqref{bsft:eq:closed-action-gf-cl}.
The full action is discussed in \Cref{bsft:sec:bv-sft:bv}.

\index{string field!closed bosonic!classical}%
The gauge fixing is removed by relaxing the $b_0^+ = 0$ constraint on the field (the other constraints must be kept in order to have well-defined the interactions).
The classical field $\Psi_{\text{cl}}$ is then defined by:
\begin{equation}
	\Psi_{\text{cl}} \in \mc H^- \cap \ker L_0^-,
	\qquad
	N_{\text{gh}}(\Psi_{\text{cl}}) = 2.
\end{equation}
The restriction on the ghost number translates the condition that the field is classical, i.e.\ that there are no spacetime ghosts at the classical level.
The relation \eqref{bsft:eq:closed-spacetime-Ngh} implies that all components have vanishing spacetime ghost number.

\index{free covariant SFT!closed bosonic!action}%
In the free limit, the gauge invariant action should match \eqref{bsft:eq:closed-action}
\begin{equation}
	S_{0,2}
		= \frac{1}{2} \, \bra{\Psi} c_0^- Q_B \ket{\Psi}.
\end{equation}
and lead to the results from \Cref{bsft:sec:free-brst:closed}.
A natural guess is that the form of the interactions is not affected by the gauge fixing (the latter usually modifies the propagator but not the interactions).
This leads to the gauge invariant classical action:
\index{covariant SFT!closed bosonic!classical action}%
\begin{equation}
	\label{bsft:eq:closed-action-cl}
	S_{\text{cl}}
		= \frac{1}{2} \, \bra{\Psi_{\text{cl}}} c_0^- Q_B \ket{\Psi_{\text{cl}}}
			+ \frac{1}{g_s^2} \sum_{n \ge 3} \frac{g_s^{n}}{n!} \, \mc V_{0,n}(\Psi_{\text{cl}}^n),
\end{equation}
where the vertices $\mc V_{0,n}$ with $n \ge 3$ are the ones defined in \eqref{bos:eq:Vgn-vertex} (we consider the case where $\mc V_{0,0} = \mc V_{0,1} = 0$).
\index{closed string fundamental vertex!tree-level!2-point}%
It is natural to generalize the definition of $\mc V_{0,2}$ as:
\begin{equation}
	\label{bos:eq:V02-vertex-QB}
	\mc V_{0,2}(\Psi_{\text{cl}}^2)
		:= \bra{\Psi_{\text{cl}}} c_0^- Q_B \ket{\Psi_{\text{cl}}}
\end{equation}
such that
\begin{equation}
	S_{\text{cl}}
		= \frac{1}{g_s^2} \sum_{n \ge 2} \frac{g_s^{n}}{n!} \, \mc V_{0,n}(\Psi_{\text{cl}}^n)
		= \frac{1}{g_s^2} \sum_{n \ge 2} \frac{g_s^{n}}{n!} \, \bra{\Psi_{\text{cl}}} c_0^- \Ket{\ell_{0,n-1}(\Psi_{\text{cl}}^{n-1})},
\end{equation}
\index{closed string product!g0n1@$g = 0, n = 1$}%
where \eqref{bos:eq:V02-vertex-QB} implies:
\begin{equation}
	\ell_{0,1}(\Psi_{\text{cl}})
		= Q_B \ket{\Psi_{\text{cl}}}.
\end{equation}

\index{covariant SFT!closed bosonic!classical equation of motion}%
The equation of motion is
\begin{equation}
	\label{bsft:eq:closed-eom-cl}
	\mc F_{\text{cl}}(\Psi_{\text{cl}})
		:= \sum_{n \ge 1} \frac{g_s^{n-1}}{n!} \, \ell_{0,n}(\Psi_{\text{cl}}^{n})
		= Q_B \ket{\Psi_{\text{cl}}}
			+ \sum_{n \ge 2} \frac{g_s^{n-1}}{n!} \, \ell_{0,n}(\Psi_{\text{cl}}^{n})
		= 0.
\end{equation}

\begin{computation}[bsft:eq:closed-eom-cl]
	\begin{equation}
		\delta S_{\text{cl}}
			= \frac{1}{g_s^2} \sum_{n \ge 2} \frac{g_s^{n}}{n!} \, n \{ \delta\Psi_{\text{cl}}, \Psi_{\text{cl}}^{n-1} \}_0
			= \frac{1}{g_s^2} \sum_{n \ge 2} \frac{g_s^{n}}{(n-1)!} \, \bra{\delta \Psi_{\text{cl}}} c_0^- \Ket{\ell_{0,n-1}(\Psi_{\text{cl}}^{n-1}}.
	\end{equation}
	The first equality follows because the vertex is completely symmetric.
	Simplifying and shifting $n$, one obtains $c_0^- \ket{\mc F_{\text{cl}}}$.
	The factor $c_0^-$ is invertible because of the constraint $b_0^- = 0$ imposed on the field.
\end{computation}

The action is invariant
\begin{equation}
	\label{bsft:eq:closed-int-action-invariance}
	\delta_\Lambda S_{\text{cl}}
		= 0
\end{equation}
under the gauge transformation
\index{covariant SFT!closed bosonic!classical gauge transformation}%
\begin{equation}
	\label{bsft:eq:closed-int-sym}
	\delta_\Lambda \Psi_{\text{cl}}
		= \sum_{n \ge 0} \frac{g_s^{n}}{n!} \, \ell_{0,n+1}(\Psi_{\text{cl}}^n, \Lambda)
		= Q_B \ket{\Lambda}
			+ \sum_{n \ge 1} \frac{g_s^{n}}{n!} \, \ell_{0,n+1}(\Psi_{\text{cl}}^n, \Lambda).
\end{equation}
\begin{subequations}
\index{covariant SFT!closed bosonic!classical gauge algebra}%
The gauge algebra is~\cite[sec.~4]{Zwiebach:1993:ClosedStringField}:
\begin{equation}
	\label{bsft:eq:closed-gauge-alg-cl}
	\com{\delta_{\Lambda_2}}{\delta_{\Lambda_1}} \Psi_{\text{cl}}
		= \delta_{\Lambda(\Lambda_1, \Lambda_2, \Psi_{\text{cl}})} \ket{\Psi_{\text{cl}}}
			+ \sum_{n \ge 0} \frac{g_s^{n+2}}{n!} \, \ell_{0,n+3}\big(\Psi_{\text{cl}}^n, \Lambda_2, \Lambda_1, \mc F_{\text{cl}}(\Psi_{\text{cl}})\big),
\end{equation}
where $\mc F_{\text{cl}}$ is the equation of motion \eqref{bsft:eq:closed-eom-cl}, and $\Lambda(\Lambda_1, \Lambda_2, \Psi_{\text{cl}})$ is a field-dependent gauge parameter:
\begin{equation}
	\begin{aligned}
	\Lambda(\Lambda_1, \Lambda_2, \Psi_{\text{cl}})
		&
		= \sum_{n \ge 0} \frac{g_s^{n+1}}{n!} \, \ell_{0,n+2}(\Lambda_1, \Lambda_2, \Psi_{\text{cl}}^n)
		\\ &
		= g_s \, \ell_{0,2}(\Lambda_1, \Lambda_2)
			+ \sum_{n \ge 1} \frac{g_s^{n+1}}{n!} \, \ell_{0,n+2}(\Lambda_1, \Lambda_2, \Psi_{\text{cl}}^n).
	\end{aligned}
\end{equation}
\end{subequations}
The classical gauge algebra is complicated which explains why a direct quantization (for example through the Faddeev--Popov procedure) cannot work: the second term in \eqref{bsft:eq:closed-gauge-alg-cl} indicates that the algebra is open (it closes only on-shell), while the first term is a gauge transformation with a field-dependent parameter.
As reviewed in \Cref{app:sec:qft:bv}, both properties require using the BV formalism for the quantization, and the latter is performed in \Cref{bsft:sec:bv-sft:bv}.
An important point is that if the theory had only cubic interactions, i.e.\ if
\begin{equation}
	\forall n \ge 4:
		\quad
		\mc V_{0,4}(\scr V_1, \ldots, \scr V_n)
			= 0,
		\qquad
		\ell_{g,n-1}(\scr V_1, \ldots, \scr V_{n-1})
			= 0,
	\qquad
	\text{(cubic theory)},
\end{equation}
then the algebra closes off-shell and $\Lambda(\Lambda_1, \Lambda_2, \Psi_{\text{cl}})$ becomes field-independent.

\begin{computation}[bsft:eq:closed-int-action-invariance]
	\begin{align*}
		\delta_\Lambda S_{\text{cl}}
			&
			= \sum_{n \ge 2} \frac{g_s^{n-2}}{n!} \,
				n \mc V_{0,n}(\delta\Psi_{\text{cl}}, \Psi_{\text{cl}}^{n-1})
			= \sum_{m,n \ge 0} \frac{g_s^{m+n-1}}{m! \, n!} \,
				\mc V_{0,n+1}\big(\ell_{0,m+1}(\Psi_{\text{cl}}^m, \Lambda), \Psi_{\text{cl}}^n\big)
			\\ &
			= \sum_{m \ge 0} \sum_{n=0}^{m} \frac{g_s^{m-1}}{(m-n)! \, n!} \,
				\Bra{\ell_{m-n+1}(\Psi_{\text{cl}}^{m-n}, \Lambda)} c_0^- \Ket{\ell_{0,n}(\Psi_{\text{cl}}^n)}.
	\end{align*}
	For simplicity we have extended the sum up to $n = 0$ and $m = 0$ by using the fact that lower-order vertices vanish.
	The bracket can be rewritten as
	\begin{align*}
			&
			= \Bra{\ell_{0,n}(\Psi_{\text{cl}}^n)} c_0^- \Ket{\ell_{0,m-n+1}(\Psi_{\text{cl}}^{m-n}, \Lambda)}
			\\ &
			= \mc V_{0,m-n+2}\big(\ell_{0,n}(\Psi_{\text{cl}}^n), \Psi_{\text{cl}}^{m-n}, \Lambda\big)
			\\ &
			= - \mc V_{0,m-n+2}\big(\Lambda, \ell_{0,n}(\Psi_{\text{cl}}^n), \Psi_{\text{cl}}^{m-n}\big)
			\\ &
			= \Bra{\Lambda} c_0^- \Ket{\ell_{0,m-n+1}\big(\ell_{0,n}(\Psi_{\text{cl}}^n), \Psi_{\text{cl}}^{m-n}\big)}.
	\end{align*}
	Then, one needs to use the identity (defined for all $m \ge 0$)
	\begin{equation}
		0
			= \sum_{n=0}^{m} \frac{m!}{(m - n)! \, n!} \,
				\ell_{0,m-n+1}\big(\ell_{0,n}(\Psi_{\text{cl}}^n), \Psi_{\text{cl}}^{m-n}\big),
	\end{equation}
	which comes from \eqref{bos:eq:vertex-identity}.
	Multiplying this by $g_s^{m-1} / m!$ and summing over $m \ge 0$ proves \eqref{bsft:eq:closed-int-action-invariance}.
\end{computation}

\begin{remark}[$L_\infty$ algebra]
	\index{L@$L_\infty$ algebra}%

	The identities satisfied by the products $\ell_{0,n}$ from the gauge invariance of the action implies that they form a $L_\infty$ homotopy algebra~\cite{Zwiebach:1993:ClosedStringField, Muenster:2014:HomotopyClassificationBosonic, Erler:2014:NSNSSectorClosed} (for more general references, see~\cite{Lada:1993:IntroductionShLie, Lada:1994:StronglyHomotopyLie, Hohm:2017:LinftyAlgebrasField, Hohm:2017:GeneralConstructionsLinfty}).
	The latter can also be mapped to a BV structure, which explains why the BV quantization \Cref{bsft:sec:bv-sft:bv} is straightforward.
	This interplay between gauge invariance, covering of the moduli space, BV and homotopy algebra is particularly beautiful.
	It has also been fruitful in constructing super-SFT.
\end{remark}

\section{BV theory}
\label{bsft:sec:bv-sft:bv}

As indicated in the previous section (\Cref{bsft:sec:bv-sft:classical}), the classical gauge algebra is open and has field-dependent structure constants.
The BV formalism (\Cref{app:sec:qft:bv}) is necessary to define the theory.

\index{Batalin--Vilkovisky formalism}%
In the BV formalism, the classical action for the physical fields is extended to the quantum master action by solving the quantum master equation \eqref{qft:eq:master-equation-quantum}.
It is generically difficult to build this action exactly, but the discussion of \Cref{bsft:sec:free-brst:path-integral} can serve as a guide: it was found that the free quantum action (with the tower of ghosts) has exactly the same form as the free classical action (without ghosts).
Hence, this motivates the ansatz that it should be of the same form as the classical action \eqref{bsft:eq:closed-action-cl} to which are added the counter-terms from \eqref{bsft:eq:closed-action-gf}:
\index{covariant SFT!closed bosonic!quantum action}%
\begin{subequations}
\label{bsft:eq:closed-action-bv}
\begin{align}
	S
		&
		= \frac{1}{g_s^2} \sum_{g \ge 0} \hbar^{g} g_s^{2g}
			\sum_{n \ge 0} \frac{g_s^{n}}{n!} \, \mc V_{g,n}(\Psi^n)
		\\ &
		= \frac{1}{2} \, \bra{\Psi} c_0^- Q_B \ket{\Psi}
			+ \sideset{}{'}\sum_{g, n \ge 0} \frac{\hbar^g g_s^{2g - 2 + n}}{n!} \, \mc V_{g,n}(\Psi^n)
		\\ &
		= \frac{1}{g_s^2} \sum_{g, n \ge 0} \frac{\hbar^g g_s^{2g - 2 + n}}{n!} \, \bra{\Psi} c_0^- \Ket{\ell_{g,n-1}(\Psi^{n-1})},
\end{align}
\end{subequations}
\index{string field!closed bosonic!quantum}%
but without any constraint on the ghost number of $\Psi$:
\begin{equation}
	\Psi \in \mc H^- \cap \ker L_0^-.
\end{equation}

\begin{draft}

\index{covariant SFT!closed bosonic!quantum equation of motion}%
The equation of motion for \eqref{bsft:eq:closed-action-bv} is
\begin{equation}
	\label{bsft:eq:closed-eom-bv}
	\mc F(\Psi)
		:= \sum_{g, n \ge 0} \frac{g_s^{2g - n - 1}}{n!} \ell_{g,n}(\Psi^{n})
		= 0.
\end{equation}

\end{draft}

\index{covariant SFT!closed bosonic!quantum BV master equation}%
In order to show that \eqref{bsft:eq:closed-action-bv} is a consistent quantum master action, it is necessary to show that it solves the master BV equation \eqref{qft:eq:master-equation-quantum}:
\begin{equation}
	\psp{S}{S} - 2 \hbar \lap S
		= 0.
\end{equation}
The first step is to introduce the fields and antifields.
In fact, because the CFT ghost number induces a spacetime ghost number, there is a natural candidate set.

\index{string field!closed bosonic!expansion}%
The string field is expanded as \eqref{bsft:eq:field-exp}
\begin{equation}
	\ket{\Psi}
		= \sum_r \psi_r \ket{\phi_r},
\end{equation}
where the $\{ \phi_r \}$ forms a basis of $\mc H^-$.
The string field can be further separated as:
\begin{equation}
	\Psi
		= \Psi_+ + \Psi_-,
\end{equation}
where $\Psi_-$ ($\Psi_+$) contains only states which have negative (positive) cylinder ghost numbers (this gives an offset of $3$ when using the plane ghost number):
\begin{equation}
	\Psi_-
		= \sum_r \sum_{n_r \le 2} \ket{\phi_r} \psi^r,
	\qquad
	\Psi_+
		= \sum_r \sum_{n_r^c > 2} b_0^- \ket{\phi_r^c} \psi_r^*.
\end{equation}
The order of the basis states and coefficients matter if they anti-commute.
The sum in $\Psi_+$ can be rewritten as a sum over $n_r \le 2$ like the first term since $n_r + n_r^c = 6$.
Correspondingly, the spacetime ghost numbers \eqref{bsft:eq:closed-spacetime-Ngh} for the coefficients in $\Psi_-$ ($\Psi_+$) are positive (negative)
\begin{equation}
	G(\psi^r) \ge 0,
	\qquad
	G(\psi_r^*) < 0.
\end{equation}
Moreover, one finds that the ghost numbers of $\psi^r$ and $\psi_r^*$ are related as:
\begin{equation}
	\label{bsft:eq:bv-relation-Ngh}
	G(\psi_r^*) = - 1 - G(\psi^r),
\end{equation}
which also implies that they have opposite parity.
\index{covariant SFT!closed bosonic!fields and antifields}%
Comparing with \Cref{app:sec:qft:bv}, this shows that the $\psi^r$ ($\psi_r^*$) contained in $\Psi_-$ ($\Psi_+$) can be identified with the fields (antifields).

\begin{computation}[bsft:eq:bv-relation-Ngh]
	\begin{align*}
		G(\psi_r^*)
			&
			= 2 - N_{\text{gh}}(b_0^- \phi_r^c)
			= 2 + 1 - n_r^c
			\\ &
			= 3 - (6 - n_r)
			= - 3 + (2 - G(\psi^r))
			= - 1 - G(\psi^r).
	\end{align*}
\end{computation}

In terms of fields and antifields, the master action is
\begin{equation}
	\frac{\pd_R S}{\pd \psi^r} \frac{\pd_L S}{\pd \psi_r^*}
			+ \hbar \, \frac{\pd_R \pd_L S}{\pd \psi^r \pd \psi_r^*}
		= 0.
\end{equation}
Plugging the expression \eqref{bsft:eq:closed-action-bv} of $S$ inside and requiring that the expression vanishes order by order in $g$ and $n$ give the set of equations:
\begin{equation}
	\sum_{\substack{g_1, g_2 \ge 0 \\ g_1 + g_2 = g}}
				\sum_{\substack{n_1, n_2 \ge 0 \\ n_1 + n_2 = n}}
				\frac{\pd_R S_{g_1,n_1}}{\pd \psi^r} \frac{\pd_L S_{g_2,n_2}}{\pd \psi_r^*}
			+ \hbar \, \frac{\pd_R \pd_L S_{g-1,n}}{\pd \psi^r \pd \psi_r^*}
		= 0,
\end{equation}
where $S_{g,n}$ was defined in \eqref{bsft:eq:action-Sgn}.
This holds true due to the identity \eqref{bos:eq:vertex-identity} (the complete proof can be found in~\cite[pp.~42--45]{Zwiebach:1993:ClosedStringField}).
The fact that the second term is not identically zero means that the measure is not invariant under the classical gauge symmetry (anomalous symmetry): corrections need to be introduced to cancel the anomaly.
It is a remarkable fact that one can construct directly the quantum master action in SFT and that it takes the same form as the classical action.

\begin{draft}

Finally, let's discuss the BRST transformation.
The action of $s$ on the string field is proportional to its (quantum) equation of motion \eqref{bsft:eq:closed-eom-bv}
\begin{equation}
	s \ket{\Psi} = \mc F(\Psi).
\end{equation}

\end{draft}

\section{1PI theory}
\label{bsft:sec:bv-sft:1pi}

The BV action is complicated: instead, it is often simpler and sufficient to work with the 1PI effective action.
The latter incorporates all the quantum corrections in 1PI vertices such that scattering amplitudes are expressed only in terms of tree Feynman graphs (there are no loops in diagrams since they correspond to quantum effects, already included in the definitions of the vertices).

A 1PI graph is a Feynman graph which stays connected if one cuts any single internal line.
On the other hand, a 1PR graph splits in two disconnected by cutting one of the line.
The scattering amplitudes $A_{g,n}$ are built by summing all the different ways to connect two 1PI vertices with a propagator: diagrams connecting two legs of the same 1PI vertex are forbidden by definition.

The $g$-loop $n$-point 1PR and 1PI Feynman diagrams are associated to some regions of the moduli space $\mc M_{g,n}$.
Comparing the previous definitions with the gluing of Riemann surfaces (\Cref{bos:sec:offshell:geometry:plumbing}), 1PR diagrams are obtained by gluing surfaces with the separating plumbing fixture (\Cref{bos:sec:amp-prop:factorization:sep}).
\index{1PI vertex}%
Thus, the 1PR and 1PI regions $\mc F_{g,n}^{\text{1PR}}$ and $\mc V_{g,n}^{\text{1PI}}$ can be identified with the regions defined in \eqref{bos:eq:def-Fgn-1PR-space} and \eqref{bos:eq:def-Vgn-1PI-space}.
In particular, the $n$-point 1PI interaction is the sum over $g$ of the $g$-loop $n$-point 1PI interactions \eqref{bos:eq:Vgn-vertex-1PI}:
\begin{equation}
	\label{bsft:eq:vertex-1PI}
	\begin{gathered}
	\mc V_{n}^{\text{1PI}}(\scr V_1, \ldots, \scr V_n)
		:= \vcenter{\hbox{\includegraphics{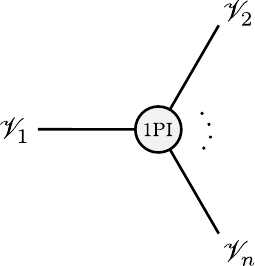}}}
		:= \sum_{g \ge 0} (\hbar g_s^2)^{g} \,
			\mc V_{g,n}^{\text{1PI}}(\scr V_1, \ldots, \scr V_n),
	\\
	\mc V_{g,n}^{\text{1PI}}(\scr V_1, \ldots, \scr V_n)
		:= \int_{\mc R_{g,n}^{\text{1PI}}} \omega^{g,n}_{\M_{g,n}}(\scr V_1, \ldots, \scr V_n),
	\end{gathered}
\end{equation}
where $\mc R_{g,n}^{\text{1PI}}$ is a section of $\mc P_{g,n}$ over $\mc V_{g,n}^{\text{1PI}}$.

Given the interactions vertices, it is possible to follow the same reasoning as in \Cref{bsft:sec:bv-sft:gauge-fixed,bsft:sec:bv-sft:classical}.

\index{covariant SFT!closed bosonic!1PI action}%
The gauge fixed 1PI effective action reads:
\begin{equation}
	\label{bsft:eq:closed-action-fixed-1PI}
	S_{\text{1PI}}
		= \frac{1}{g_s^2} \sum_{n \ge 0} \frac{g_s^{n}}{n!} \,
			\mc V_{n}^{\text{1PI}}(\Psi^n)
		:= \frac{1}{2} \, \bra{\Psi} c_0^- c_0^+ L_0^+ \ket{\Psi}
			+ \frac{1}{g_s^2} \sideset{}{'}\sum_{n \ge 0}
				\frac{g_s^{n}}{n!} \, \mc V_{n}^{\text{1PI}}(\Psi^n).
\end{equation}
Here, the prime means again that the term $g = 0, n = 2$ is excluded from the definition of $\mc V_{2}^{\text{1PI}}$.
The action has the same form as the classical gauge fixed action \eqref{bsft:eq:closed-action-gf-cl}, which is logical since it generates only tree-level Feynman graphs.
For this reason the vertices $\mc V_{n}^{\text{1PI}}$ have exactly the same properties as the brackets $\mc V_{0,n}$.
This fact can be used to write the 1PI gauge invariant action:
\begin{equation}
	\label{bsft:eq:closed-action-1PI}
	S_{\text{1PI}}
		= \frac{1}{2} \, \bra{\Psi} c_0^- Q_B \ket{\Psi}
			+ \frac{1}{g_s^2} \sideset{}{'}\sum_{n \ge 0}
				\frac{g_s^{n}}{n!} \, \mc V_{n}^{\text{1PI}}(\Psi^n),
\end{equation}
which mirrors the classical gauge invariant action \eqref{bsft:eq:closed-action-cl}.
\index{covariant SFT!closed bosonic!1PI gauge symmetry}%
Then it is straightforward to see that it enjoys the same gauge symmetry upon replacing the tree-level vertices by the 1PI vertices.
But, since this action incorporates all quantum corrections this also proves that the quantum theory is correctly invariant under a quantum gauge symmetry.

\begin{remark}
	The 1PI action \eqref{bsft:eq:closed-action-1PI} can also be directly constructed from the BV action \eqref{bsft:eq:closed-action-bv}.
\end{remark}

\refchapter

\begin{itemize}
	\item Gauge fixed and classical gauge invariant closed SFT~\cite{Zwiebach:1993:ClosedStringField} (see also~\cite{Kugo:1989:NonpolynomialClosedString, Kugo:1990:NonpolynomialClosedString}).

	\item BV closed SFT~\cite{Zwiebach:1993:ClosedStringField} (see also~\cite{Thorn:1989:StringFieldTheory}).

	\item Construction of the open--closed BV SFT~\cite{Zwiebach:1998:OrientedOpenClosedString}.

	\item 1PI SFT~\cites{Sen:2015:GaugeInvariant1PI-NS, Sen:2015:GaugeInvariant1PI-R}[sec.~4.1, 5.2]{deLacroix:2017:ClosedSuperstringField}.
\end{itemize}

\chapter{Background independence}
\label{bsft:chap:background}

\introchapter

Spacetime background independence is a fundamental property of any candidate quantum gravity theory.
In this chapter, we outline the proof of background independence for the closed SFT by proving that the equations of motion of two background related by a marginal deformation are equivalent after a field redefinition.

\section{The concept of background independence}

\index{background independence}
Background independence means that the formalism does not depend on the background -- if any -- used to write the theory.
A dependence in the background would imply that there is a distinguished background among all possibilities, which seems in tension with the dynamics of spacetime and the superposition principle from quantum mechanics.
Moreover, one would expect a fundamental theory to tell which backgrounds are consistent and that they could be derived instead of postulated.
Background independence allows spacetime to emerge as a consequence of the dynamics of the theory and of its defining fundamental laws.

Background independence can be manifest or not.
In the second case, one needs to fix a background to define the theory, but the dynamics on different backgrounds are physically equivalent.\footnotemark{}
\footnotetext{%
	This does not mean that the physics in all backgrounds are identical, but that the laws are.
	Hence, a computation made in one specific background can be translated into another background.
}%
This implies that two theories with different backgrounds can be related, for example by a field redefinition.

While fields other than the metric can also be expanded around a background, no difficulty is expected in this case.
Indeed, the topic of background independence is particularly sensible only for the metric because it provides the frame for all other computations -- and in particular for the questions of dynamics and quantization.
Generally, these questions are subsumed into the problem of the emergence of time in a generally covariant theory.
In the previous language, QFTs without gravity are (generically) manifestly background independent after minimal coupling.\footnotemark{}
\footnotetext{%
	However, non-minimal coupling terms may be necessary to make the theory physical.
}%
For example, a classical field theory is defined on a fixed Minkowski background and a well-defined time is necessary to perform its quantization and to obtain a QFT, but it is not needed to choose a background for the other fields.
For this reason, the extension of a QFT on a curved background is generally possible if the spacetime is hyperbolic, implying that there is a distinguished time direction.
But the coupling to gravity is difficult and restricted to a (semi-)classical description.

\index{string theory!background independence}%
What is the status of background independence in string theory?
The worldsheet formulation requires to fix a background (usually Minkowski) to quantize the theory and to compute scattering amplitudes.
Thus, the quantum theory is at least not manifestly background independent.
On the other hand, the worldsheet action can be modified to a generic CFT including a generic non-linear sigma model describing an arbitrary target spacetime.
Conformal invariance reproduces (at leading order) Einstein equations coupled to various matter and gauge field equations of motion.
From this point of view, the classical theory can be written as a manifestly background independent theory, and this provides hopes that the quantum theory may be also background independent, even if non-manifestly.
\index{AdS/CFT}%
This idea is supported by other definitions of string theory (e.g.\ through the AdS/CFT conjecture -- and other holographic realizations -- or through matrix models) which provide, at least partially, background independent formulations.

\index{covariant SFT!background independence}%
Ultimately, the greatest avenue to establish the background independence is string field theory.
Indeed, the form of the SFT action and of its properties (gauge invariance, equation of motion…) are identical irrespective of the background~\cite{Sen:1996:BackgroundIndependentAlgebraic}.
This provides a good starting point.
The background dependence enters in the precise definition of the string products (BRST operator and vertices).
The origin of this dependence lies in the derivation of the action (\Cref{bos:chap:feynman,bsft:chap:bv-sft}): one begins with a particular CFT describing a given background (spacetime compactifications, fluxes, etc.) and defines the vertices from correlation functions of vertex operators, and the Hilbert space from the CFT operators.
As a consequence, even though it is clear that no specific property of the background has been used in the derivation -- and that the final action describes SFT for any background --, this is not sufficient to establish background independence.
Since the theory assumes implicitly a background choice, one cannot guarantee that the physical quantities have no residual dependence in the background, even if the action looks superficially background independent.
Background independence in SFT is thus the statement that theories characterized by different CFTs can be related by a field redefinition.

In this chapter, we will sketch the proof of background independence for backgrounds related by marginal deformations.\footnotemark{}
\footnotetext{%
	An alternative approach based on morphism of $L_\infty$ algebra is followed in~\cite{Muenster:2014:HomotopyClassificationBosonic}.
}%
It is possible to prove it at the level of the action~\cite{Sen:1994:ProofLocalBackground, Sen:1994:QuantumBackgroundIndependence}, or at the level of the equations of motion~\cite{Sen:2018:BackgroundIndependenceClosed}.
The advantage of the second approach is that one can use the 1PI theory, which simplifies vastly the analysis.
It also generalizes directly to the super-SFT.

\begin{remark}[Field theory on the CFT space]
	\index{field theory space}%

	As mentioned earlier, the string field is defined as a functional on the state space of a given CFT and not as a functional on the field theory space (off-shell states would correspond to general QFTs, only on-shell states are CFTs).
	In this case, background independence would amount to reparametrization invariance of the action in the theory space, and would thus almost automatically hold.
	A complete formulation of SFT following this line is currently out of reach, but some ideas can be found in~\cite{Witten:1992:BackgroundIndependentOpenString}.
\end{remark}

\section{Problem setup}

Given a SFT on a background, there are two ways to describe it on another background:
\begin{itemize}
	\item deform the worldsheet CFT and express the SFT on the new background;

	\item expand the original action around the infinitesimal classical solution (to the linearised equations of motion) corresponding to the deformation.
\end{itemize}
Background independence amounts to the equivalence of both theories up to a field redefinition.
The derivation can be performed at the level of the action or of the equations of motion.
To prove the background independence at the quantum level, one needs to take into account the changes in the path integral measure or to work with the 1PI action.

\index{marginal deformation!action}%
The simplest case is when the two CFTs are related by an infinitesimal marginal deformation
\begin{equation}
	\label{bsft:eq:bg-op-int}
	\delta S_{\text{cft}}
		= \frac{\lambda}{2\pi} \int \dd^2 z \, \varphi(z, \bar z),
\end{equation}
with $\varphi$ a $(1, 1)$-primary operator and $\lambda$ infinitesimal.
The two CFTs are denoted by $\text{CFT}_1$ and $\text{CFT}_2$, and quantities associated to each CFTs is indexed with the appropriate number.

Establishing background independence in this case also implies it for finite marginal deformation since they can be built from a series of successive deformations.
In the latter case, the field redefinition may be singular, which reflects that the parametrization of one CFT is not adapted for the other (equivalently, the coordinate systems for the string field breaks down), which is expected if both CFTs are far in the field theory space.

Remember the form of the 1PI action \eqref{bsft:eq:closed-action-1PI}:
\begin{equation}
	S_1[\Psi_1]
		= \frac{1}{g_s^2} \left(
			\frac{1}{2} \, \bra{\Psi_1} c_0^- Q_B \ket{\Psi_1}
			+ \sideset{}{'}\sum_{n \ge 0} \frac{1}{n!} \, \mc V_{n}^{\text{1PI}}(\Psi_1^n) \right),
\end{equation}
where the prime indicates that vertices with $n < 3$ do not include contributions from the sphere.
In all this chapter, we remove the index 1PI to lighten the notations.
The equation of motion is:
\begin{equation}
	\label{bsft:eq:bg-eom-1PI}
	\mc F_1(\Psi_1)
		= Q_B \ket{\Psi_1}
			+ \sum_{n} \frac{1}{n!} \, \ell_{n}(\Psi_1^n)
		= 0.
\end{equation}

\section{Deformation of the CFT}

\index{marginal deformation!action}%
Consider the case where the theory $\text{CFT}_1$ is described by an action $S_{\text{cft},1}[\psi_1]$ given in terms of fields $\psi_1$.
Then, the deformation of this action by \eqref{bsft:eq:bg-op-int} gives an action for $\text{CFT}_2$:
\begin{equation}
	S_{\text{cft},2}[\psi_1]
		= S_{\text{cft},1}[\psi_1]
			+ \frac{\lambda}{2\pi} \int \dd^2 z \, \varphi(z, \bar z).
\end{equation}
\index{marginal deformation!correlation function}%
Correlation functions on a Riemann surface $\Sigma$ in both theories can be related by expanding the action to first order in $\lambda$ in the path integral:
\begin{subequations}
\begin{align}
	\Mean{\prod_i \mc O_i(z_i, \bar z_i)}_2
		&= \Mean{\exp\left( - \frac{\lambda}{2\pi} \int \dd^2 z \, \varphi(z, \bar z) \right) \prod_i \mc O_i(z_i, \bar z_i)}_1
		\\
		&\approx \Mean{\prod_i \mc O_i(z_i, \bar z_i)}_1
			- \frac{\lambda}{2\pi} \int_{\Sigma} \dd^2 z
				\Mean{\varphi(z, \bar z) \prod_i \mc O_i(z_i, \bar z_i)}_1,
\end{align}
\end{subequations}
where the $\mc O_i$ are operators built from the matter fields $\psi_1$.
This expression presents two obvious problems.
First, the correlation function may diverge when $\varphi$ collides with one of the insertions, i.e.\ when $z = z_i$ in the integration.
Second, there is an inherent ambiguity: the correlation functions are written in terms of operators in the Hilbert space of $\text{CFT}_1$, which is different from the $\text{CFT}_2$ Hilbert space, and there is no canonical isomorphism between both spaces.

\index{field theory space!Hilbert space bundle}%
Seeing the Hilbert space as a vector bundle over the CFT theory space, the second problem can be solved by introducing a connection on this bundle.
This allows to relate Hilbert spaces of neighbouring CFTs.
In fact, the choice of a non-singular connection also regularizes the divergences.

\index{field theory space!connection}%
The simplest definition of a connection corresponds to cut unit disks around each operator insertions~\cite{Sen:1990:BackgroundIndependenceString, Campbell:1991:StressTensorPerturbations, Sonoda:1993:ConnectionTheorySpace, Ranganathan:1993:NearbyCFTsOperator, Ranganathan:1994:ConnectionsStateSpaceConformal}.
This amounts to define the variation between the two correlation functions as:
\begin{equation}
	\delta \Mean{\prod_i \mc O_i(z_i, \bar z_i)}_1
		= - \frac{\lambda}{2\pi} \int_{\mathrlap{\Sigma - \cup_i D_i}} \qquad \dd^2 z \,
			\Mean{\varphi(z, \bar z) \prod_i \mc O_i(z_i, \bar z_i)}_1.
\end{equation}
The integration is over $\Sigma$ minus the disks $D_i = \{ \abs{w_i} \le 1 \}$ where $w_i$ is the local coordinate for the insertion $\mc O_i$.
The divergences are cured because $\varphi$ never approaches another operator since the corresponding regions have been removed.
The changes in the correlation functions induce a change in the string vertices denoted by $\delta \mc V_{n}(\scr V_1, \ldots, \scr V_n)$.

The next step consists in computing the deformations of the operator modes.
Since it involves only a matter operator, the modes in the ghost sector are left unchanged.
The Virasoro generators change as:
\begin{equation}
	\delta L_n
		= \lambda \oint_{\abs{z} = 1} \frac{\dd \bar{z}}{2\pi\I} \,
			z^{n+1} \varphi(z, \bar z),
	\qquad
	\delta \bar L_n
		= \lambda \oint_{\abs{z} = 1} \frac{\dd z}{2\pi\I} \,
			\bar z^{n+1} \varphi(z, \bar z).
\end{equation}
As a consequence, the BRST operator changes as
\begin{equation}
	\delta Q_B
		= \lambda \oint_{\abs{z} = 1} \frac{\dd \bar{z}}{2\pi\I} \,
				c(z) \varphi(z, \bar z)
			+ \lambda \oint_{\abs{z} = 1} \frac{\dd \bar{z}}{2\pi\I} \,
				\bar c(\bar z) \varphi(z, \bar z).
\end{equation}
One can prove that
\begin{equation}
	\anticom{Q_B}{\delta Q_B} = O(\lambda^2)
\end{equation}
such that the BRST charge $Q_B + \delta Q_B$ in $\text{CFT}_2$ is correctly nilpotent if $Q_B$ is nilpotent in $\text{CFT}_1$.

For the deformation to provide a consistent SFT, the conditions $b_0^- = 0$ and $L_0^- = 0$ must be preserved.
The first is automatically satisfied since the ghost modes are not modified.
Considering an weight-$(h, h)$ operator $\mc O$, one finds
\begin{equation}
	\delta L_0^- \ket{\mc O}
		= \lambda \oint_{\abs{z} = 1} \frac{\dd \bar{z}}{2\pi\I} \, z \sum_{p,q} z^{p-1} \bar z^{q-1} \ket{\mc O_{p,q}}
			- \lambda \oint_{\abs{z} = 1} \frac{\dd \bar{z}}{2\pi\I} \, \sum_{p,q} z^{p-1} \bar z^{q-1} \ket{\mc O_{p,q}},
\end{equation}
where $\mc O_{p,q}$ are the fields appearing in the OPE with $\varphi$:
\begin{equation}
	\varphi(z, \bar z) \mc O(0, 0)
		= \sum_{p,q} z^{p-1} \bar z^{q-1} \mc O_{p,q}(0, 0).
\end{equation}
The terms with $p \neq q$ vanish because the contour integrals are performed around circles of unit radius centred at the origin.
Moreover, the terms $p = q$ are identical and cancel with each other, showing that $\delta L_0^- = 0$ when acting on states satisfying $L_0^- = 0$.

The SFT action $S_2[\Psi_1]$ in the new background reads
\begin{equation}
	S_2[\Psi_1]
		= S_1[\Psi_1] + \delta S_1[\Psi_1]
\end{equation}
where the change $\delta S_1$ in the action is induced by the changes in the string vertices:
\begin{equation}
	\delta S_1[\Psi_1]
		= \frac{1}{g_s^2} \left(
			\frac{1}{2} \, \bra{\Psi_1} c_0^- \delta Q_B \ket{\Psi_1}
			+ \sum_{n \ge 0} \frac{1}{n!} \, \delta \mc V_n(\Psi_1^n) \right).
\end{equation}
The equation of motion is:
\begin{equation}
	\label{bsft:eq:bg-eom-cft-deform}
	\mc F_2(\Psi_1)
		= \mc F_1(\Psi_1) + \lambda \, \delta \mc F_1(\Psi_1)
		= 0,
\end{equation}
where $\mc F_1$ is given in \eqref{bsft:eq:bg-eom-1PI} and
\begin{equation}
	\lambda \, \delta \mc F_1(\Psi_1)
		= \delta Q_B \ket{\Psi_1}
			+ \sum_{n} \frac{1}{n!} \, \delta \ell_n(\Psi_1^n).
\end{equation}

\section{Expansion of the action}

\index{classical solution!marginal deformation}%
Given a $(1, 1)$ primary $\varphi$, a BRST invariant operator is $c \bar c \varphi$.
Hence the field
\begin{equation}
	\ket{\Psi_1}
		= \lambda \ket{\Psi_0},
	\qquad
	\ket{\Psi_0}
		= c_1 \bar c_1(0) \ket{\varphi}
\end{equation}
is a classical solution to first order in $\lambda$ since the interactions on the sphere are at least cubic.

Separating the string field as the contribution from the (fixed) background and a fluctuation $\Psi'$
\begin{equation}
	\ket{\Psi_1}
		= \lambda \ket{\Psi_0} + \ket{\Psi'},
\end{equation}
the action expanded to first order in $\lambda$ reads:
\begin{equation}
	S_1[\Psi_1]
		= S_1[\Psi_0] + S'[\Psi'],
\end{equation}
where
\begin{equation}
	S'[\Psi']
		= \frac{1}{g_s^2} \left(
			\frac{1}{2} \, \bra{\Psi'} c_0^- Q_B \ket{\Psi'}
			+ \sum_{n} \frac{1}{n!} \, \big( \mc V_n(\Psi'^n)
				+ \lambda \, \mc V_{n+1}(\Psi_0, \Psi'^n) \big)
			\right).
\end{equation}
The equation of motion is:
\begin{equation}
	\label{bsft:eq:bg-eom-sft-vac}
	\mc F'(\Psi')
		:= \mc F_1(\Psi') + \lambda \, \delta \mc F'(\Psi')
		= 0,
\end{equation}
where $\mc F_1$ is given in \eqref{bsft:eq:bg-eom-1PI} and
\begin{equation}
	\delta \mc F'(\Psi')
		= \sum_{n} \frac{1}{n!} \, \ell_{n+1}(\Psi_0, \Psi'^n).
\end{equation}

\section{Relating the equations of motion}

In the previous section, we have derived the equations of motion for two different descriptions of a SFT obtained after shifting the background: \eqref{bsft:eq:bg-eom-cft-deform} arises by deforming the CFT and computing the changes in the BRST operator and string products, while \eqref{bsft:eq:bg-eom-sft-vac} arises by expanding the SFT action around the new background.
The theory is background independent if both sets of equations \eqref{bsft:eq:bg-eom-cft-deform} and \eqref{bsft:eq:bg-eom-sft-vac} are related by a (possibly field-dependent) linear transformation $\mc M(\Psi')$ after a field redefinition of $\Psi_1 = \Psi_1(\Psi')$:
\begin{subequations}
\begin{gather}
	\mc F_1(\Psi_1) + \lambda \, \delta \mc F_1(\Psi_1)
		= \big( 1 + \lambda \mc M(\Psi') \big) \big( \mc F_1(\Psi') + \lambda \, \delta \mc F'(\Psi') \big),
	\\
	\ket{\Psi_1} = \ket{\Psi'} + \lambda \ket{\delta\Psi'}.
\end{gather}
\end{subequations}
The zero-order equation is automatically satisfied.
To first order, this becomes
\begin{equation}
	\frac{\dd}{\dd\lambda} \mc F_1(\Psi' + \lambda \delta\Psi') \bigg|_{\lambda=0}
			+ \delta \mc F_1(\Psi_1)
			- \delta \mc F'(\Psi')
		= \mc M(\Psi') \mc F_1(\Psi').
\end{equation}
Taking $\Psi'$ to be a solution of the original action removes the RHS, such that:
\begin{equation}
	\begin{aligned}
	\lambda\, Q_B \ket{\delta\Psi'}
		&+ \lambda\, \sum_{n} \frac{1}{n!} \, \ell_{n+1}(\delta\Psi', \Psi'^n)
		+ \delta Q_B \ket{\Psi'}
		\\ &
		+ \sum_{n} \frac{1}{n!} \, \delta \ell_n(\Psi'^n)
			- \lambda \sum_{n} \frac{1}{n!} \, \ell_{n+1}(\Psi_0, \Psi'^n)
		= 0.
	\end{aligned}
\end{equation}
To simplify the computations, it is simpler to consider the inner product of this quantity with an arbitrary state $A$ (assumed to be even):
\begin{equation}
	\label{bsft:eq:bg-delta-eq}
	\begin{aligned}
	\Delta
		:= \lambda\, \bra{A} c_0^- Q_B \ket{\delta\Psi'}
			&
			+ \lambda\, \sum_{n} \frac{1}{n!} \, \mc V_{n+2}(A, \delta\Psi', \Psi'^n)
			+ \bra{A} c_0^- \delta Q_B \ket{\Psi'}
			\\ &
			+ \sum_{n} \frac{1}{n!} \, \delta \mc V_{n+1}(A, \Psi'^n)
			- \lambda \sum_{n} \frac{1}{n!} \, \mc V_{n+2}(A, \Psi_0, \Psi'^n).
	\end{aligned}
\end{equation}
The goal is to prove the existence of $\delta\Psi'$ such that $\Delta = 0$ up to the zero-order equation of motion $\mc F_1(\Psi') = 0$.

\section{Idea of the proof}

In this section, we give an idea of how the proof ends, referring to~\cite{Sen:2018:BackgroundIndependenceClosed} for the details.

The first step is to introduce new vertices $\mc V'_{0,3}$ and $\mc V'_{n}$ parametrizing the variations of the string vertices:
\begin{equation}
	\bra{A} c_0^- \delta Q_B \ket{B}
		= \lambda \, \mc V'_{0,3}(\Psi_0, B, A),
	\qquad
	\delta \mc V_n(\Psi'^n)
		= \lambda \, \mc V'_{n+1}(\Psi_0, \Psi'^n),
\end{equation}
where the notation \eqref{bos:eq:int-omega-An} has been used.
Each subspace $\mc V'_{g,n}$ is defined such that the LHS is recovered upon integrating the appropriate $\omega_{g,n}$ over this section segment.
Next, the field redefinition $\delta \Psi'$ is parametrized as:
\begin{equation}
	\bra{A} c_0^- \ket{\delta\Psi'}
		= \sum_{n} \frac{1}{n!} \, \mc B_{n+2}(\Psi_0, \Psi'^n, A).
\end{equation}
The objective is to prove the existence (and if possible the form) of the subspaces $\mc B_{n+2}$.
Both the vertices $\mc V'_n$ and $\mc B_n$ admit a genus expansion:
\begin{equation}
	\mc V'_{n}
		= \sum_{g \ge 0} \mc V'_{g, n},
	\qquad
	\mc B_{n}
		= \sum_{g \ge 0} \mc B_{g, n}.
\end{equation}

Plugging the new expressions in \eqref{bsft:eq:bg-delta-eq} give:
\begin{equation}
	\begin{multlined}
	\Delta
		= - \sum_{n} \frac{1}{n!} \, \mc B_{n+2}(\Psi_0, \Psi'^n, Q_B A)
			+ \sum_{m,n} \frac{1}{m! n!} \, \mc B_{n+2}(\Psi_0, \Psi'^m, \ell_{n+1}(A, \Psi'^n))
			\\
			+ \sum_{n} \frac{1}{n!} \, \mc V'_{n+2}(A, \Psi_0, \Psi'^n)
			- \sum_{n} \frac{1}{n!} \, \mc V_{n+2}(A, \Psi_0, \Psi'^n).
	\end{multlined}
\end{equation}
Next, the BRST identity \eqref{bos:eq:omega-p-brst-identity} and the equation of motion $\mc F_1(\Psi') = 0$ allow to rewrite the first term as:
\begin{subequations}
\begin{align}
	\mc B_{n+2}(\Psi_0, \Psi'^n, Q_B A)
		&= \pd\mc B_{n+2}(\Psi_0, \Psi'^n, A)
			+ n \, \mc B_{n+2}(\Psi_0, \Psi'^{n-1}, Q_B \Psi', A)
		\\
		&= \pd\mc B_{n+2}(\Psi_0, \Psi'^n, A)
			- \sum_{m} \frac{n}{m!} \, \mc B_{n+2}(\Psi_0, \Psi'^{n-1}, \ell_m(\Psi'^m), A).
\end{align}
\end{subequations}
In the second term, the sum over $n$ is shifted.
Combining everything together gives:
\begin{equation}
	\begin{aligned}
	\Delta
		= \sum_{n}& \frac{1}{n!} \, \pd\mc B_{n+2}(\Psi_0, \Psi'^n, A)
			- \sum_{m,n} \frac{1}{m! n!} \, \mc B_{n+3}(\Psi_0, \Psi'^n, \ell_m(\Psi'^m), A)
			\\
			&+ \sum_{m,n} \frac{1}{m! n!} \, \mc B_{n+2}(\Psi_0, \Psi'^m, \ell_{n+1}(A, \Psi'^n))
			+ \sum_{n} \frac{1}{n!} \, \mc V'_{n+2}(A, \Psi_0, \Psi'^n)
			\\
			&- \sum_{n} \frac{1}{n!} \, \mc V_{n+2}(A, \Psi_0, \Psi'^n).
	\end{aligned}
\end{equation}
Solving for $\Delta = 0$ requires that each term with a different power of $\Psi'$ vanishes independently:
\begin{equation}
	\begin{aligned}
		\pd\mc B_{n+2}(\Psi_0, \Psi'^n, A)
			= &- \mc V'_{n+2}(A, \Psi_0, \Psi'^n)
			+ \mc V_{n+2}(A, \Psi_0, \Psi'^n)
			\\
			&+ \sum_{\mathclap{\substack{m_1,m_2 \\ m_1 + m_2 = n}}} \quad
				\frac{n!}{m_1! m_2!} \,
				\mc B_{m_1+3}(\Psi_0, \Psi'^{m_1}, \ell_{m_2}(\Psi'^{m_2}), A)
			\\
			&- \sum_{\mathclap{\substack{m_1,m_2 \\ m_1 + m_2 = n}}} \quad
				\frac{n!}{m_1! m_2!} \,
				\mc B_{m_1+2}\big(\Psi_0, \Psi'^{m_1}, \ell_{m_2+1}(A, \Psi'^{m_2})\big).
	\end{aligned}
\end{equation}

In order to proceed, one needs to perform a genus expansion of the various spaces: this allows to solve recursively for all $\mc B_{g,n}$ starting from $\mc B_{0,3}$.
One can then build $\ket{\delta\Psi'}$ recursively, which provides the field redefinition.
Indeed, the RHS of this equation contains only $\mc B_{g',n'}$ for $g' < g$ or $n' < n$ and the equation for $\mc B_{0,3}$ contains no $\mc B_{g,n}$ in the RHS.
It should be noted that the field redefinition is not unique, but there is the freedom of performing (infinite-dimensional) gauge transformations.
Finding an obstruction to solve these equations mean that the field redefinition does not exist, and thus that the theory is not background independent

The form of the equation
\begin{equation}
	\pd\mc B_{0,3}
		= \mc V_{0,3} - \mc V'_{0,3}
\end{equation}
suggests to use homology theory.
The interpretation of $\mc B_{0,3}$ is that it is a space interpolating between $\mc V_{0,3}$ and $\mc V'_{0,3}$.
A preliminary step is to check that there is no obstruction: since the LHS is already a boundary one has $\pd^2 \mc B_{0,3} = 0$ and one should check that $\pd(\text{RHS}) = 0$ as well.
It can be shown that it is indeed true.
It was proved in~\cite{Sen:2018:BackgroundIndependenceClosed} that this equation admits a solution and that the equations for higher $g$ and $n$ can all be solved.
Hence, there exists a field redefinition and SFT is background independent.

\refchapter

\begin{itemize}
	\item Proof of the background independence under marginal deformations~\cite{Sen:1994:QuantumBackgroundIndependence, Sen:1994:ProofLocalBackground, Sen:2018:BackgroundIndependenceClosed} (see also~\cite{Sen:1990:BackgroundIndependenceString, Sen:1990:BackgroundIndependenceString-2, Sen:1993:BackgroundIndependenceString-3} for earlier results laying foundations for the complete proof).

	\item $L_\infty$ perspective~\cite[sec.~4]{Muenster:2014:HomotopyClassificationBosonic} (see also~\cites{Muenster:2013:QuantumOpenClosedHomotopy}[sec.~III.B]{Muenster:2013:HomotopyAlgebrasQuantum}.

	\item Connection on the space of CFTs~\cite{Sen:1990:BackgroundIndependenceString, Campbell:1991:StressTensorPerturbations, Sonoda:1993:ConnectionTheorySpace, Ranganathan:1993:NearbyCFTsOperator, Ranganathan:1994:ConnectionsStateSpaceConformal}.
\end{itemize}

\chapter{Superstring}
\label{part:superstring}

\introchapter

Superstring theory is generally the starting point for physical model building.
It has indeed several advantages over the bosonic string, most importantly, the removal of the tachyon and the inclusion of fermions in the spectrum.
The goal of this chapter is to introduce the most important concepts needed to generalize the bosonic string to the superstring, both for off-shell amplitudes and string field theory.
We refer to the review~\cite{deLacroix:2017:ClosedSuperstringField} for more details.

\section{Worldsheet superstring theory}
\label{sws:chap:worldsheet}

There are five different superstring theories with spacetime supersymmetry: the types I, IIA and IIB, and the $\group{E}_8 \times \group{E}_8$ and $\group{SO}(32)$ heterotic models.

\index{superstring}%
In the Ramond--Neveu--Schwarz formalism (RNS), the left- and right-moving sectors of the superstring worldsheet are described by a two-dimensional super-conformal field theory (SCFT), possibly with different numbers of supersymmetries.
The prototypical example is the heterotic string with $N = (1, 0)$ and we will focus on this case: only the left-moving sector is supersymmetric, while the right-moving is given by the same bosonic theory as in the other chapters.
Up to minor modifications, the type II theory follows by duplicating the formulas of the left-moving sector to the right-moving one.

\subsection{Heterotic worldsheet}

\index{superstring!heterotic (-)}%
The ghost super-CFT is characterized by anti-commuting ghosts $(b, c)$ (left-moving) and $(\bar b, \bar c)$ (right-moving) with central charge $c = (-26, -26)$, associated to diffeomorphisms, and by commuting ghosts $(\beta, \gamma)$ with central charge $c = (11, 0)$, associated to local supersymmetry.
As a consequence the matter SCFT must have a central charge $c = (15, 26)$.
If spacetime has $D$ non-compact dimensions, then the matter CFT is made of:
\begin{itemize}
	\item a free theory of $D$ scalars $X^\mu$ and $D$ left-moving fermions $\psi^\mu$ ($\mu = 0, \ldots, D - 1$) such that $c_{\text{free}} = 3 D / 2$ and $\bar c_{\text{free}} = D$;

	\item an internal theory with $c_{\text{int}} = 15 - 3 D / 2$ and $\bar c_{\text{int}} = 26 - D$.
\end{itemize}
\index{critical dimension!superstring}%
The critical dimension is reached when $c_{\text{int}} = 0$ which corresponds to $D = 10$.

The diffeomorphisms are generated by the energy--momentum tensor $T(z)$; correspondingly, supersymmetry is generated by its super-partner $G(z)$ (sometimes also denoted by $T_F$).
\index{operator product expansion!$GG$}%
\index{operator product expansion!$TG$}%
The OPEs of the algebra formed by $T(z)$ and $G(z)$ is:
\begin{subequations}
\begin{align}
	T(z) T(w)
		&
		\sim \frac{c / 2}{(z - w)^4} + \frac{2 T(w)}{(z - w)^2} + \frac{\pd T(w)}{z - w},
	\\
	G(z) G(w)
		&
		\sim \frac{2 c / 3}{(z - w)^3} + \frac{2 T(w)}{(z - w)},
	\\
	T(z) G(w)
		&
		\sim \frac{3}{2} \, \frac{G(w)}{(z - w)^2} + \frac{\pd G(w)}{(z - w)}.
\end{align}
\end{subequations}

\index{superconformal $\beta\gamma$ ghosts}%
The superconformal ghosts form a first-order system (see \Cref{cft:sec:systems:ghosts}) with $\epsilon = - 1$ and $\lambda = 3/2$.
\index{superconformal $\beta\gamma$ ghosts!conformal weights}%
Hence, they have conformal weights
\begin{equation}
	h(\beta)
		= \left( \frac{3}{2}, 0 \right),
	\qquad
	h(\gamma)
		= \left( - \frac{1}{2}, 0 \right)
\end{equation}
and OPEs
\index{superconformal $\beta\gamma$ ghosts!OPE}%
\begin{equation}
	\gamma(z) \beta(w)
		\sim \frac{1}{z - w},
	\qquad
	\beta(z) \gamma(w)
		\sim - \frac{1}{z - w}.
\end{equation}
\index{superconformal $\beta\gamma$ ghosts!energy--momentum tensor}%
The expressions of the ghost energy--momentum tensors are
\begin{equation}
	T^{\text{gh}}
		= - 2 b \, \pd c + c \pd b,
	\qquad
	T^{\beta\gamma}
		= \frac{3}{2} \, \beta \pd\gamma + \frac{1}{2} \, \gamma \, \pd\beta.
\end{equation}
The ghost numbers of the different fields are
\begin{equation}
	N_{\text{gh}}(b)
		= N_{\text{gh}}(\beta)
		= - 1,
	\qquad
	N_{\text{gh}}(c)
		= N_{\text{gh}}(\gamma)
		= 1.
\end{equation}

The worldsheet scalars satisfy periodic boundary conditions.
On the other hand, fermions can satisfy anti-periodic or periodic conditions: this leads to two different sectors, called Neveu--Schwarz (NS) and Ramond (R) respectively.

\subsubsection{\texorpdfstring{$\beta\gamma$}{Beta-gamma} system}

\index{superconformal $\beta\gamma$ ghosts!bosonization}%
\index{e@$\eta\xi$ ghosts}%
The $\beta\gamma$ system can be bosonized as
\begin{equation}
	\gamma
		= \eta \, \e^{\phi},
	\qquad
	\beta
		= \pd\xi \, \e^{- \phi},
\end{equation}
where $(\xi, \eta)$ are fermions with conformal weights $0$ and $1$ (this is a first-order system with $\epsilon = 1$ and $\lambda = 1$), and $\phi$ is a scalar field with a background charge (Coulomb gas).
This provides an alternative representation of the delta functions:
\begin{equation}
	\delta(\gamma)
		= \e^{-\phi},
	\qquad
	\delta(\beta)
		= \e^{\phi}.
\end{equation}
Introducing these operators is necessary to properly define the path integral with bosonic zero-modes.
They play the same role as the zero-modes insertions for fermionic fields needed to obtain a finite result (see also \Cref{sec:form:path-integrals:zero-mode}):
\begin{equation}
	\int \dd c_0
		= 0
	\quad \Longrightarrow \quad
	\int \dd c_0 \, c_0
		= 1,
\end{equation}
because $c_0 = \delta(c_0)$.
For a bosonic path integral, one needs a delta function:
\begin{equation}
	\int \dd\gamma_0
		= \infty
	\quad \Longrightarrow \quad
	\int \dd\gamma_0 \, \delta(\gamma_0)
		= 1.
\end{equation}

By definition of the bosonization, one has:
\begin{equation}
	T^{\beta\gamma}
		= T^{\eta\xi} + T^{\phi},
\end{equation}
where
\begin{equation}
	T^{\eta\xi}
		= - \eta \, \pd \xi,
	\qquad
	T^{\phi}
		= - \frac{1}{2} \, (\pd\phi)^2 - \pd^2 \phi.
\end{equation}
The OPE between the new fields are:
\begin{equation}
	\xi(z) \eta(w)
		\sim \frac{1}{z - w},
	\quad
	\e^{q_1 \phi(z)} \e^{q_2 \phi(w)}
		\sim \frac{\e^{(q_1 + q_2) \phi(w)}}{(z - w)^{q_1 q_2}},
	\quad
	\pd\phi(z) \pd\phi(w)
		\sim - \frac{1}{(z - w)^2}.
\end{equation}
The simplest attribution of ghost numbers to the new fields is:
\begin{equation}
	N_{\text{gh}}(\eta)
		= 1,
	\qquad
	N_{\text{gh}}(\xi)
		= - 1,
	\qquad
	N_{\text{gh}}(\phi)
		= 0.
\end{equation}

\index{picture number}%
To the scalar field $\phi$ is associated another $\group{U}(1)$ symmetry whose quantum number is called the picture number $N_{\text{pic}}$.
The picture number of $\eta$ and $\xi$ are assigned\footnotemark{} such that $\beta$ and $\gamma$ have $N_{\text{pic}} = 0$:
\footnotetext{%
	Any linear combination of both $\group{U}(1)$ could have been used.
	The one given here is conventional, but also the most convenient.
}%
\begin{equation}
	N_{\text{pic}}(\e^{q \phi})
		= q,
	\qquad
	N_{\text{pic}}(\xi)
		= 1,
	\qquad
	N_{\text{pic}}(\eta)
		= -1.
\end{equation}
Because of the background charge, this symmetry is anomalous and correlation functions are non-vanishing if the total picture number (equivalently the number of $\phi$ zero-modes) is:
\index{picture number!anomaly}%
\begin{equation}
	N_{\text{pic}}
		= 2 (g - 1)
		= - \chi_g.
\end{equation}
For the same reason, the vertex operators $\e^{q \phi}$ are the only primary operators:
\begin{equation}
	h(\e^{q \phi})
		= - \frac{q}{2} (q + 2),
\end{equation}
and the Grassmann parity of these operators is $(-1)^{q}$.
Special values are
\begin{equation}
	h(\e^{\phi})
		= \frac{3}{2},
	\qquad
	h(\e^{- \phi})
		= \frac{1}{2}.
\end{equation}

\index{GSO symmetry}%
The superstring theory features a $\Z_2$ symmetry called the GSO symmetry.
All fields are taken to be GSO even, except $\beta$ and $\gamma$ which are GSO odd and $\e^{q \phi}$ whose parity is $(-1)^{q}$.
Physical states in the NS sector are restricted to be GSO even: it is required to remove the tachyon of the spectrum and to get a spacetime with supersymmetry.
In type II, the Ramond sector can be projected in two different ways, leading to the type IIA and type IIB theories.

\index{superstring!BRST current}%
The components of the BRST current are:
\begin{subequations}
\begin{gather}
	j_B
		= c ( T^{\text{m}} + T^{\beta\gamma} )
			+ \gamma G + b c \pd c
			- \frac{1}{4} \, \gamma^2 b,
	\\
	\bar\jmath_B
		= \bar c \bar T^{\text{m}}
			+ \bar b \bar c \bar\pd \bar c.
\end{gather}
\end{subequations}
From there, it is useful to define the picture changing operator (PCO):
\begin{equation}
	\mc X(z)
		= \anticom{Q_B}{\xi(z)}
		= c \pd \xi + \e^{\phi} G
			- \frac{1}{4} \, \pd\eta \, \e^{2\phi} \, b
			- \frac{1}{4} \, \pd(\eta \, \e^{2\phi} b ),
\end{equation}
which is a weight-$(0, 0)$ primary operator which carries a unit picture number.
It is obviously BRST exact.
This operator will be necessary to saturate the picture number condition: the naive insertion of $\e^{\phi} \sim \delta(\beta)$ breaks the BRST invariance.
The PCO zero-mode is obtained from the contour integral:
\begin{equation}
	\mc X_0
		= \frac{1}{2\pi \I} \oint \frac{\dd z}{z} \, \mc X(z).
\end{equation}
It can be interpreted as delocalizing a PCO insertion from a point to a circle, which decreases the risk of divergence.

\subsection{Hilbert spaces}

\index{superstring!Hilbert space}%
The description in terms of the $(\eta, \xi, \phi)$ fields leads to a subtlety: the bosonization involves only the derivative $\pd \xi$ and not the field $\xi$ itself, meaning that the zero-mode $\xi_0$ is absent from the original Hilbert space defined from $(\beta, \gamma)$.
\index{superstring!Hilbert space!small}%
In the bosonized language, the Hilbert space without the $\xi$ zero-mode is called the \emph{small Hilbert space} and is made of state annihilated by $\eta_0$ (the $\eta$ zero-mode)
\begin{equation}
	\mc H_{\text{small}}
		= \big\{ \ket{\psi}
			\mid \eta_0 \ket{\psi} = 0 \big\}.
\end{equation}
Removing this condition leads to the \emph{large Hilbert space}:\footnotemark{}
\footnotetext{%
	The relation between the small and large Hilbert spaces is similar to the one between the $\mc H$ and $\mc H_0 = b_0 \mc H$ Hilbert space from the open string since the $(b, c)$ and $(\eta, \xi)$ are both fermionic first-order systems.
}%
\begin{equation}
	\mc H_{\text{small}}
		= \mc H_{\text{large}} \cap \ker \eta_0 .
\end{equation}
\index{superstring!Hilbert space!large}%
A state in $\mc H_{\text{small}}$ contains $\xi$ with at least one derivative acting on it.

A correlation function defined in terms of the $(\eta, \xi, \phi)$ system is in the large Hilbert space and will vanish since there is no $\xi$ factor to absorb the zero-mode of the path integral.
As a consequence, correlation functions (and the inner product) are defined with a $\xi_0$ insertion (by convention at the extreme left) or, equivalently, $\xi(z)$.
The position does not matter since only the zero-mode contribution survives, and the correlation function is independent of $z$.
Sometimes it is more convenient to work in the large Hilbert space and to restrict later to the small Hilbert space.

\index{superstring!normalization}%
The $\group{SL}(2, \C)$ invariant vacuum is normalized as
\begin{equation}
	\bra{k} c_{-1} \bar c_{-1} c_0 \bar c_0 c_1 \bar c_1 \, \e^{- 2\phi(z)} \ket{k'}
		= (2\pi)^D \cdirac[D](k + k').
\end{equation}

\begin{remark}[Normalization in type II]
	In type II theory, the $\group{SL}(2, \C)$ is normalised as:
	\begin{equation}
		\bra{k} c_{-1} \bar c_{-1} c_0 \bar c_0 c_1 \bar c_1 \, \e^{- 2\phi(z)} \e^{- \bar\phi(\bar w)} \ket{k'}
			= - (2\pi)^D \cdirac[D](k + k').
	\end{equation}
	The sign difference allows to avoid sign differences between type II and heterotic string theories in most formulas~\cite{deLacroix:2017:ClosedSuperstringField}.
\end{remark}

\index{superstring!Hilbert space!picture number}%
The Hilbert space of GSO even states satisfying the $b_0^- = 0$ and $L_0^- = 0$ conditions is denoted by $\mc H_T$ (ghost and picture numbers are arbitrary).
This Hilbert space is the direct sum of the NS and R Hilbert spaces:
\begin{equation}
	\mc H_T
		= \mc H_{\text{NS}} \oplus \mc H_{\text{R}}.
\end{equation}
The subspace of states with picture number $N_{\text{pic}} = n$ is written $\mc H_n$.
The picture number of NS and R states are respectively integer and half-integer.
Two special subspaces of $\mc H_T$ play a distinguished role:
\begin{equation}
	\what{\mc H}_T
		= \mc H_{-1} \oplus \mc H_{-1/2},
	\qquad
	\wtilde{\mc H}_T
		= \mc H_{-1} \oplus \mc H_{-3/2}.
\end{equation}
To understand this, consider the vacuum $\ket{p}$ of the $\phi$ field with picture number $p$:
\begin{equation}
	\ket{p}
		= \e^{p \phi}(0) \ket{0}.
\end{equation}
Then, acting on the vacuum with the $\beta_n$ and $\gamma_n$ modes implies
\begin{equation}
	\begin{aligned}
		\forall n \ge - p - \frac{1}{2} &:
			\qquad
			\beta_n \ket{p} = 0,
		\\
		\forall n \ge p + \frac{3}{2} &:
			\qquad
			\gamma_n \ket{p} = 0.
	\end{aligned}
\end{equation}
For $p = - 1$, all positive modes (starting with $n = 1/2$) annihilate the vacuum in the NS sector.
This is a positive asset because positive modes which do not annihilate the vacuum can create states with arbitrary negative energy (since it is bosonic).\footnotemark{}
\footnotetext{%
	This is not a problem on-shell since the BRST cohomology is independent of the picture number.
	However, this matters off-shell since such states would propagate in loops and make the theory inconsistent.
}%
For $p = - 1/2$ or $p = - 3/2$, the vacuum is annihilated by all positive modes, but not by one of the zero-mode $\gamma_0$ or $\beta_0$.
Nonetheless, one can show that the propagator in the R sector allows to propagate only a finite number of states if one chooses $\mc H_{-1/2}$; the role of $\mc H_{-3/2}$ will become apparent when discussing how to build the superstring field theory.

Basis states are introduced as in the bosonic case:
\begin{equation}
	\what{\mc H}_T
		= \Span \{ \ket{\phi_r} \},
	\qquad
	\wtilde{\mc H}_T
		= \Span \{ \ket{\phi_r^c} \}
\end{equation}
such that
\begin{equation}
	\bracket{\phi_r^c}{\phi_s}
		= \delta_{rs}.
\end{equation}
The completeness relations are
\begin{equation}
	1
		= \sum_r \ket{\phi_r} \bra{\phi_r^c}
\end{equation}
$\what{\mc H}_T$, and
\begin{equation}
	1
		= \sum_{r} (- 1)^{\abs{\phi_r}} \ket{\phi_r^c} \bra{\phi_r}
\end{equation}
on $\wtilde{\mc H}_T$.

Finally, the operator $\mc G$ is defined as:
\begin{equation}
	\label{sws:eq:def-G-PCO}
	\mc G
		=
		\begin{cases}
			1 & \text{NS sector},
			\\
			\mc X_0 & \text{R sector}.
		\end{cases}
\end{equation}
Note the following properties
\begin{equation}
	\com{\mc G}{L_0^\pm}
		= \com{\mc G}{b_0^\pm}
		= \com{\mc G}{Q_B}
		= 0.
\end{equation}
It will be appear in the propagator and kinetic term of the superstring field theory.

\section{Off-shell superstring amplitudes}
\label{sws:chap:amplitudes}

In this section, we are going to build the scattering amplitudes.
The procedure is very similar to the bosonic case, except for the PCO insertions and of the Ramond sector.
For this reason, we will simply state the result and motivate the modifications with respect to the bosonic case.

\subsection{Amplitudes}
\label{sws:sec:amplitudes:amplitudes}

External states can be either NS or R: the Riemann surface corresponding to the $g$-loop scattering of $m$ external NS states and $n$ external R states is denoted by $\Sigma_{g,m,n}$.
R states must come in pairs because they correspond to fermions.
As in the bosonic case, the amplitude is written as the integration of an appropriate $p$-form $\Omega^{(g,m,n)}_p$ over the moduli space $\mc M_{g,m,n}$ (or, more precisely, of a section of a fibre bundle with this moduli space as a basis).
From the geometric point of view, nothing distinguishes the punctures and thus:
\begin{equation}
	\M_{g,m,n}
		:= \dim \mc M_{g,m,n}
		= 6 g - 6 + 2 m + 2 n.
\end{equation}
The form $\Omega_{\M_{g,m,n}}$ is defined as a SCFT correlation function of the physical vertex operators together with ghost and PCO insertions.

\begin{remark}
	A simple way to avoid making errors with signs is to multiply every Grassmann odd external state with a Grassmann odd number.
	These can be removed at the end to read the sign.
\end{remark}

\index{ghost number!anomaly}%
\index{picture number!anomaly}%
The two conditions from the $\group{U}(1)$ anomalies on the scattering amplitude are:
\begin{equation}
	N_{\text{gh}}
		= 6 - 6 g,
	\qquad
	N_{\text{pic}}
		= 2 g - 2.
\end{equation}
\index{off-shell superstring amplitude!PCO insertions}%
Given an amplitude with $m$ NS states $\scr V^{\text{NS}}_i \in \mc H_{-1}$ and $n$ R states $\scr V^{\text{R}}_j \in \mc H_{-1/2}$, the above picture number can be reached by introducing a certain number of PCO $\mc X(y_A)$:
\begin{equation}
	n_{\text{pco}}
		:= 2 g - 2 + m + \frac{n}{2}.
\end{equation}
These PCO are inserted at various positions: while the amplitude does not depend on these locations on-shell, off-shell it will (because the vertex operators are not BRST invariant).
\index{off-shell superstring amplitude!consistency}%
\index{spurious pole}%
The choices of PCO locations are arbitrary except for several consistency conditions:
\begin{enumerate}
	\item avoid spurious poles (\Cref{sws:sec:amplitudes:spurious-poles});
	\item consistent with factorization (each component of the surface in the degeneration limits must saturate the picture number condition).
\end{enumerate}

\index{Pgn@$\wtilde{\mc P}_{g,m,n}$ space}%
This parallels the discussion of the choices of local coordinates: as a consequence, the natural object is a fibre bundle $\wtilde{\mc P}_{g,m,n}$ with the local coordinate choices (up to global phase rotations) and the PCO locations as fibre, and the moduli space $\mc M_{g,m,n}$ as base.
Forgetting about the PCO locations leads to a fibre bundle $\what{\mc P}_{g,m,n}$ which is a generalization of the one found in the bosonic case.
The coordinate system of the fibre bundle presented in the bosonic case is extended by including the PCO locations $\{ y_A \}$.

\index{off-shell superstring amplitude}%
With these information, the amplitude can be written as:
\begin{subequations}
\begin{equation}
	\label{sws:eq:Agmn}
	A_{g,m,n}(\scr V^{\text{NS}}_i, \scr V^{\text{R}}_j)
		= \int_{\mc S_{g,m,n}} \Omega_{\M_{g,m,n}}(\scr V^{\text{NS}}_i, \scr V^{\text{R}}_j),
\end{equation}
where
\index{Pgn@$\wtilde{\mc P}_{g,m,n}$ space!$p$-form}%
\begin{equation}
	\Omega_{\M_{g,m,n}}
		= (- 2\pi \I)^{- \M_{g,m,n}^c}
			\Mean{ \bigwedge_{\lambda=1}^{\M_{g,m,n}} \mc B_{\lambda} \, \dd t_{\lambda}
				\prod_{A=1}^{n_{\text{pco}}} \mc X(y_A)
				\prod_{i=1}^m \scr V^{\text{NS}}_i
				\prod_{j=1}^n \scr V^{\text{R}}_j
				}_{\mathrlap{\Sigma_{g,n}}}.
\end{equation}
\end{subequations}
where $\mc S_{g,m,n}$ is a $\M_{g,m,n}$-dimensional section of $\wtilde{\mc P}_{g,m,n}$ parametrized by coordinates $t_\lambda$.
The $1$-form $\mc B$ corresponds to a generalization of the bosonic $1$-form.
\index{Pgn@$\wtilde{\mc P}_{g,m,n}$ space!1-form}%
It has ghost number $1$ and includes a correction to compensate the variation of the PCO locations in terms of the moduli parameters:
\begin{equation}
	\begin{aligned}
		\mc B_\lambda
			= \sum_\alpha \oint_{C_{\alpha}} \frac{\dd \sigma_\alpha}{2\pi \I} \, b(\sigma_\alpha) \, \frac{\pd F_\alpha}{\pd t_{\lambda}}\big(F_\alpha^{-1}(\sigma_\alpha)\big)
				&+ \sum_\alpha \oint_{C_{\alpha}} \frac{\dd \bar\sigma_\alpha}{2\pi \I} \, \bar b(\bar\sigma_\alpha) \, \frac{\pd \bar F_\alpha}{\pd t_{\lambda}}\big(\bar F_\alpha^{-1}(\bar\sigma_\alpha)\big)
				\\
				&- \sum_{A} \frac{1}{\mc X(y_A)} \, \frac{\pd y_A}{\pd t_{\lambda}} \, \pd \xi(y_A).
	\end{aligned}
\end{equation}
The last factor amounts to consider the combination
\begin{equation}
	\mc X(y_A) - \pd \xi(y_A) \, \dd y_A
\end{equation}
for each PCO insertion:\footnotemark{} the correction is necessary to ensure that the BRST identity \eqref{bos:eq:omega-p-brst-identity} holds.
\footnotetext{%
	The sum is formal since it is composed of $0$- and $1$-forms.
}%
This can be understood as follows: the derivative acting on the PCO gives a term $\dd \mc X(z) = \pd \mc X(z) \dd z$ which must be cancelled.
This is achieved by the second term since $\anticom{Q_B}{\pd \xi(z)} = \pd \mc X(z)$.

\begin{remark}
	While it is sufficient to work with $\mc M_{g,n}$ for on-shell bosonic amplitudes, on-shell superstring amplitudes are naturally expressed in $\wtilde{\mc P}_{g,m,n}$ (with the local coordinate removed) since the positions of the PCO must be specified even on-shell.
\end{remark}

\begin{remark}[Amplitudes on the supermoduli space]
	\index{off-shell superstring amplitude!supermoduli space}%

	Following Polyakov's appro\-ach from \Cref{chap:bos:ws-int-vac,chap:bos:ws-int-amp} to the superstring would lead to replace the moduli space by the supermoduli space.
	The latter includes Grassmann-odd moduli parameters in addition to the moduli parameters from $\mc M_{g,m+n}$ (in the same way the superspace includes odd coordinates $\theta$ along with spacetime coordinates $x$).
	The natural question is whether it is possible to split the integration over the even and odd moduli, and to integrate over the latter such that only an integral over $\mc M_{g,m+n}$ remains.
	In view of \eqref{sws:eq:Agmn}, the answer seems positive.
	\index{holomorphic factorization}%
	However, this is incorrect: it was proven in~\cite{Donagi:2013:SupermoduliSpaceNot} that there is no global holomorphic projection of the supermoduli space to the moduli space.
	This is related to the problem of spurious poles described below.
	But, this does not prevent to do it locally: in that case, implementing the procedure carefully should give the rules of vertical integration~\cite{Sen:2015:OffshellAmplitudesSuperstring, Sen:2015:FillingGapsPCOs, Erler:2017:VerticalIntegrationLarge}.
\end{remark}

\subsection{Factorization}
\label{sws:sec:amplitudes:factorization}

\index{off-shell superstring amplitude!factorization|(}%

\index{plumbing fixture}%
The plumbing fixture of two Riemann surfaces $\Sigma_{g_1,m_1,n_1}$ and $\Sigma_{g_2,m_2,n_2}$ can be performed in two different ways since two NS or two R punctures can be glued.

If two NS punctures are glued, the resulting Riemann surface is $\Sigma^{(\text{NS})}_{g_1+g_2, m_1+m_2-2, n_1+n_2}$.
The number of PCO inherited from the two original surfaces is
\begin{equation}
	n_{\text{pco}}^{(1)} + n_{\text{pco}}^{(2)}
		= 2 (g_1 + g_2) - 2 + (m_1 + m_2 - 2) + \frac{n_1 + n_2}{2}
		= n_{\text{pco}}^{\text{(NS)}},
\end{equation}
which is the required number for a non-vanishing amplitude.
\index{propagator!NS sector}%
As a consequence, the propagator is the same as in the bosonic case:
\begin{equation}
	\label{sws:eq:propagator-ns}
	\Delta_{\text{NS}}
		= b_0^+ b_0^- \, \frac{1}{L_0 + \bar L_0} \, \delta(L_0^-).
\end{equation}

If two R punctures are glued, the numbers of PCO do not match by one unit:
\begin{equation}
	n_{\text{pco}}^{(1)} + n_{\text{pco}}^{(2)}
		= 2 (g_1 + g_2) - 2 + (m_1 + m_2) + \frac{n_1 + n_2 - 2}{2} - 1
		= n_{\text{pco}}^{\text{(R)}} - 1.
\end{equation}
This means that an additional PCO must be inserted in the plumbing fixture procedure: the natural place for it is in the propagator since this is the only way to keep both vertices symmetric as required for a field theory interpretation.
Another way to see the need of this modification is to study the propagator \eqref{sws:eq:propagator-ns} for Ramond states: since Ramond states carry a picture number $-1/2$, the conjugate states have $N_{\text{pic}} = - 3/2$ and thus the propagator has a total picture number $- 3$ instead of $- 2$ (the propagator graph is equivalent to a sphere).
\index{propagator!R sector}%
Then, to avoid localizing the PCO at a point of the propagator, one inserts the zero-mode which corresponds to smear the PCO:
\begin{equation}
	\Delta_{\text{R}}
		= b_0^+ b_0^- \, \frac{\mc X_0}{L_0 + \bar L_0} \, \delta(L_0^-).
\end{equation}
Delocalizing the PCO amounts to average the amplitude over an infinite number of points (i.e.\ to consider a generalized section): this is necessary to preserve the $L_0^-$ eigenvalue since $\mc X_0$ is rotationally invariant while $\mc X(z)$ is not.
Note that the zero-mode can be written equivalently as a contour integral around one of the two glued punctures:
\begin{equation}
	\mc X_0
		= \frac{1}{2\pi\I} \oint \frac{\dd w_n^{(1)}}{w_n^{(1)}} \, \mc X\big(w_n^{(1)}\big)
		= \frac{1}{2\pi\I} \oint \frac{\dd w_n^{(2)}}{w_n^{(2)}} \, \mc X\big(w_n^{(2)}\big).
\end{equation}
The equality of both expressions holds because $\mc X(z)$ has conformal weight $0$.

Using the operator $\mc G$ \eqref{sws:eq:def-G-PCO}, the propagator can be written generically as
\begin{equation}
	\label{sws:eq:propagator}
	\Delta
		= b_0^+ b_0^- \, \frac{\mc G}{L_0 + \bar L_0} \, \delta(L_0^-).
\end{equation}

\begin{remark}[Propagators]
	NS and R states correspond respectively to bosonic and fermionic fields: the operators $L_0^+$ and $\mc X_0$ can be interpreted as the (massive) Laplacian and Dirac operators, such that both propagators can be written
	\begin{equation}
		\normalfont
		\Delta_{\text{NS}}
			\sim \frac{1}{k^2 + m^2},
		\qquad
		\Delta_{\text{R}}
			\sim \frac{\I \slashed\pd + m}{k^2 + m^2}.
	\end{equation}
	To motivate the identification of $\mc X_0$ with the Dirac operator, remember that $\mc X(z)$ contains a term $\e^{\phi(z)} G(z)$ (this is the only term which contributes on-shell), where $G(z)$ in turn contains $\psi_\mu \pd X^\mu$.
	But, the zero-modes of $\psi_\mu$ and $\pd X^\mu$ correspond respectively to the gamma matrix $\gamma^\mu$ and momentum $k^\mu$ when acting on a state.
\end{remark}

\index{off-shell superstring amplitude!consistency}%
The PCO zero-mode insertion inside the propagator has another virtue.
It was noted previously that states with $N_{\text{pic}} = - 3/2$ are infinitely degenerate since one can apply $\beta_0$ an arbitrary number of time.
These states have large negative ghost numbers.
Considering a loop amplitude, all these states would appear in the sum over the states and lead to a divergence.
The problem is present only for loops because the ghost number is not fixed: in a tree propagator, the ghost number is fixed and only a finite number of $\beta_0$ can be applied.
But, the PCO insertion turns these states into $N_{\text{pic}} = - 1/2$ states.
In this picture number, one cannot create an arbitrarily large negative ghost number since $\gamma_0^n$ can only increase the ghost number.

\index{off-shell superstring amplitude!factorization|)}%

\subsection{Spurious poles}
\label{sws:sec:amplitudes:spurious-poles}

\index{spurious pole|(}%

A spurious pole corresponds to a singularity of the amplitude which cannot be interpreted as the degeneration limit of Riemann surfaces.
As a consequence, they do not correspond to infrared divergences and don't have any physical meaning; they must be avoided in order to define a consistent theory.
To achieve this, the section $\mc S_{g,m,n}$ must be chosen such that it avoids all spurious poles.
However, while it is always possible to avoid these poles locally, it is not possible globally (this is related to the results from~\cite{Donagi:2013:SupermoduliSpaceNot}).
Poles can be avoided using vertical integration: two methods have been proposed, in the small (Sen--Witten)~\cite{Sen:2015:OffshellAmplitudesSuperstring, Sen:2015:FillingGapsPCOs} and large (Erler--Konopka)~\cite{Erler:2017:VerticalIntegrationLarge} Hilbert spaces respectively.
Before describing the essence of both approaches, we review the origin of spurious poles.

\index{vertical integration}%

\subsubsection{Origin}

Spurious poles arise in three different ways:
\begin{itemize}
	\item two PCOs collide;
	\item one PCO and one matter vertex collide;
	\item other singularities of the correlation functions.
\end{itemize}
The last source is the less intuitive one and we focus on it.

A general correlation function of $(\eta, \xi, \phi)$ on the torus\footnotemark{} (satisfying the ghost number condition) reads
\footnotetext{%
	The discussion generalizes directly to higher-genus Riemann surfaces.
}%
\begin{equation}
	\begin{aligned}
		\mc C(x_i, y_j, z_q)
			&
			= \Mean{
				\prod_{i=1}^{n+1} \xi(x_i)
				\prod_{j=1}^{n} \eta(y_j)
				\prod_{k=1}^{m} \e^{q_k \phi(z_k)}
				}
			\\ &
			= \frac{\prodl_{j'=1}^{n} \vartheta_{\delta}\Big(- y_{j'} + \suml_i x_i - \suml_j y_j + \suml_k q_k z_k \Big)}{\prodl_{i'=1}^{n+1} \vartheta_{\delta}\Big(- x_{i'} + \suml_i x_i - \suml_j y_j + \suml_k q_k z_k \Big)}
				\times \frac{\prodl_{i < i'} E(x_i, x_{i'}) \prodl_{j < j'} E(y_j, y_{j'})}{\prodl_{i,j} E(x_i, y_j) \prodl_{k, \ell} E(z_k, z_\ell)^{q_k q_\ell}}.
	\end{aligned}
\end{equation}
The additional $\xi$ insertion is necessary since it provides the $\xi$ zero-mode, the correlation function being defined in the large Hilbert space.
On the torus, the picture numbers must add to zero and thus the charges $q_k$ satisfy
\begin{equation}
	\sum_k q_k
		= 0.
\end{equation}
The function $E(x, y)$ is called the prime form and is a generalization of the function $x - y$ on te torus:
\begin{equation}
	E(x, y)
		= \frac{\vartheta_1(x - y)}{\vartheta_1'(0)}
		\sim_{x \to y} x - y.
\end{equation}
Its presence ensures that $\mc C$ vanishes or diverges appropriately when the operators collide (i.e.\ that the zeros and poles of $\mc C$ are the expected ones from the OPE).
The theta functions are used to make sure that the correlation function satisfies the appropriate boundary conditions (specified by the spin structure $\delta$) for each cycle of the surface.

However, theta functions can also vanish and the ones in the denominator lead to additional singularities (not implied by any OPE) for the correlation function.
Since $x_1$ can be chosen arbitrarily, the only theta function which can have poles is
\begin{equation}
	\vartheta_\delta\Big(\suml_{i=2}^{n+1} x_i - \sum_{j=1}^{n} y_j + \sum_{k=1}^{m} q_k z_k \Big)
		= 0.
\end{equation}
This defines a complex codimension $1$ curve in $\wtilde{P}_{g,m,n}$, depending on the vertex and PCO locations, but also on the moduli parameters (appearing in the definition of the theta function).
On the other hand, it does not depend on the local coordinate choice.
If the section $\mc S_{g,m,n}$ intersects this curve, it will be ill-defined (even on-shell).

From this formula, several comments can be made.
If an operator inserted at $z$ contains $n_\xi$ fields $\pd\xi$, $n_\eta$ fields $\eta$ and a factor $\e^{p \phi}$, then the dependence in $z$ of the theta function is of the form $(n_\xi - n_\eta + p) z = N_{\text{pic}} z$.
Then, if the PCO locations are chosen as to avoid spurious poles for a given operator, this will also avoid them for any operator of the same picture number.
This also implies that insertion of $\beta$ and $\gamma$ cannot lead to spurious poles since they have $N_{\text{pic}} = 0$; this is important since they appear in the BRST current, thus insertion of the latter cannot lead to new poles.

Since it is always possible to choose locally a distribution of PCO to avoid spurious poles, the idea is to discretize the moduli space in small pieces.
But, since the PCO cannot be distributed continuously along the different components of the moduli space, correction terms are required.
These can be generated in two different ways in both the small and large Hilbert spaces.
The second is more general while the first may be more adapted since it keeps the amplitude in the small Hilbert space.

\subsubsection{Vertical integration: large Hilbert space}

Consider the $n$-point amplitude state $\bra{A^{(p)}}$ which produces an amplitude with $p$ PCO when contracted with $n$ external states are specified.
The BRST identity implies that the amplitude state is closed (i.e.\ gauge invariant)
\begin{equation}
	\bra{A^{(p)}} Q
		= 0,
\end{equation}
where
\begin{equation}
	Q
		= Q_B \otimes 1^{\otimes n - 1} + \cdots + 1^{\otimes n - 1} \otimes Q_B.
\end{equation}
Moreover, this state is in the small Hilbert space which implies that it is in the kernel of $\eta_0$:
\begin{equation}
	\bra{A^{(p)}} \eta
		= 0,
\end{equation}
where
\begin{equation}
	\eta
		= \eta_0 \otimes 1^{\otimes n - 1} + \cdots + 1^{\otimes n - 1} \otimes \eta_0.
\end{equation}

The BRST cohomology is trivial in the large Hilbert space: thus, if $\bra{A^{(p)}}$ is closed, it must be exact in this space:
\begin{equation}
	\bra{A^{(p)}}
		= \bra{\alpha^{(p)}} Q,
\end{equation}
where the state $\alpha^{(p)}$ (called gauge amplitude) must be in the large Hilbert space.
This is consistent with $\bra{A^{(p)}} \eta = 0$ only if
\begin{equation}
	\bra{\alpha^{(p)}} \eta Q
		= 0
\end{equation}
($Q$ and $\eta$ anti-commute);
It is then natural to interpret the state on which $Q$ acts as an amplitude with one less PCO
\begin{equation}
	\bra{A^{(p-1)}}
		= \bra{\alpha^{(p)}} \eta
\end{equation}
since $N_{\text{pic}}(\eta) = - 1$.

Continuing this procedure leads to an amplitude $\bra{A^{(0)}}$ without any PCO insertion, and thus without spurious singularities.
Consistency with the picture number anomaly requires the external state to have non-canonical picture numbers.
But, this should not be a puzzle since the amplitude states should be viewed as intermediate object to obtain the final amplitude.

Hence, the amplitude $\bra{A^{(p)}}$ can be constructed by starting with $\bra{A^{(0)}}$: inserting $\xi(z)$ in the amplitude leads to the gauge amplitude $\bra{\alpha^{(1)}}$, whose BRST variation yields $\bra{A^{(2)}}$.
Continuing recursively helps to construct the desired amplitude.
Moreover, $Q$ and $\xi$ insertions automatically take care of the corrections at the interfaces of the components.

Showing that the amplitude is independent of the non-physical data (i.e.\ gauge invariance) is trivial since it is expressed as a BRST exact expression.

\subsubsection{Vertical integration: small Hilbert space}

The section of $\wtilde{\mc P}_{g,m,n}$ is given by a series of discontinuous components linked by vertical segments.
On the vertical segment, the PCO configuration interpolates continuously between the components and the integrand can encounter a spurious pole.
Since the integrand is not a total derivative in terms of the fibre coordinates, its integration over a segment depends on the path followed and not only on the end points.
This implies that it diverges when it encounters the spurious pole.
However, there is a specific prescription which avoids these problems.
When only one PCO varies, the integrand is a total derivative of the PCO location and can thus be integrated directly, giving a difference in the two end points.
In this case, the result is independent from the specific path and from the presence of the spurious pole.

To be more concrete, given a PCO insertion $\mc X(y_1)$, the variation of its location inserts a factor $- \pd \xi(y_1)$.
Integrating this term between two components labelled by $i$ and $j$ -- keeping everything else fixed -- leads to a factor $\xi(y_1^{(j)}) - \xi(y_1^{(j)})$.

When several PCOs are involved, it is not sufficient to integrate the vertical segment along a path where only one PCO varies at a time.
Indeed, because a hole is left in the process of the vertical integration.
Additional segments must be added and integrated over.
This avoids the spurious poles and one can show that it yields a well-defined amplitude.
Moreover, it agrees with the large Hilbert space approach.

Finally, it remains to address the question of the Feynman diagrams construction.
In this case, every graph obtained by plumbing fixture inherits its PCO locations from the lower-dimensional surfaces, and there is no control on the resulting distribution.
It can be shown that no spurious singularity is generated in the gluing process if the lower-dimensional graphs have no spurious poles.
Hence, it is sufficient to ensure that the fundamental graphs have no spurious poles.

\index{spurious pole|)}%

\section{Superstring field theory}
\label{ssft:chap:field-theory}

\index{super-SFT}%
The construction of super-SFT has proceeded along different directions (for reviews, see~\cite{deLacroix:2017:ClosedSuperstringField, Okawa:2018:ConstructionSuperstringField, Erler:2020:FourLecturesClosed}).
There are two main strategies for constructing the superstring vertices:
\begin{enumerate}
	\item brute-force construction: build the vertices recursively from amplitude factorization;

	\item dress the bosonic products with superconformal ghosts.
\end{enumerate}
While the second approach is simpler and preferred for explicit construction, the first allows to derive the general structure as was done for the bosonic string.
\index{L@$L_\infty$ algebra}%
There are two main strategies for dressing the vertices:
\begin{enumerate}
	\item Munich construction (homotopy algebra bootstrap): use the $L_\infty$ and $A_\infty$ structures to derive the superstring vertices from the bosonic vertices (small Hilbert space).

	\item Berkovits' construction (WZW action): generalize Witten's cubic bosonic open SFT (NS / R in large / small Hilbert space).

\end{enumerate}
As indicated in parenthesis, a super-SFT can be written in the small or large Hilbert space (or a combination).
The different approaches have been shown to be equivalent at the classical level.

The main difficulty in building a super-SFT is to properly describe the Ramond sector.
\index{superstring field!Ramond field}%
This can be done following two different approaches:
\begin{itemize}
	\item constraining the Ramond string field;

	\item using an auxiliary string field.
\end{itemize}
\index{Berkovits' super-SFT}%
Berkovits' original SFT cannot describe the Ramond sector in the large Hilbert space, but it is possible to couple Berkovits' action for the NS field in the large Hilbert space to a Ramond field in the small Hilbert space.
Another limitation of Berkovits' approach is that it works only for the open and heterotic superstrings (but not for type II).

We assume that the problems with PCO are absent (in Berkovits' and supermoduli constructions) or that they have been defined using vertical integration.

In the rest of this chapter, we will discuss the kinetic term for each of the first three approaches.
At the level of the free action, the open and heterotic super-SFT differs only in the bosonic factors as in \Cref{bsft:chap:free-brst}.

\subsection{String field and propagator}

\index{superstring field!small Hilbert space}%
As in the bosonic case, it is natural to consider a string field gathering all possible states
\begin{equation}
	\Psi
		= \Psi_{-1} + \Psi_{-1/2},
\end{equation}
where $\Psi_{-1}$ and $\Psi_{-1/2}$ are respectively the NS and R string fields.
If the field is in the small Hilbert space, it satisfies:
\begin{equation}
	\eta_0 \ket{\Psi}
		= 0.
\end{equation}

\index{propagator!superstring}%
The propagator was found in \eqref{sws:eq:propagator} to be
\begin{equation}
	\Delta
		= b_0^+ b_0^- \, \frac{\mc G}{L_0 + \bar L_0} \, \delta(L_0^-),
	\qquad
	\mc G =
		\begin{cases}
			1 & \text{NS},
			\\
			\mc X_0 & \text{R}.
		\end{cases}
\end{equation}
As for the bosonic case, the constraints
\begin{equation}
	b_0^- \ket{\Psi}
		= L_0^- \ket{\Psi}
		= 0,
	\qquad
	b_0^+ \ket{\Psi}
		= 0
\end{equation}
must be imposed on the field to ensure that the propagator is invertible.

For similar reasons, the PCO insertion implies that the propagator is not invertible since $\mc X_0$ has zero-modes: this means equivalently that it has a non-empty kernel off-shell or that it contains derivatives.
Two different solutions can be chosen to address this issue: imposing constraints as for the level-matching condition, or introducing auxiliary fields.

\subsection{Constraint approach}

\index{picture changing operator}%
Two new PCO operators must be introduced:
\begin{equation}
	X
		= G_0 \, \delta(\beta_0) + b_0 \, \delta'(\beta_0),
	\qquad
	Y
		= - c_0 \, \delta'(\gamma_0).
\end{equation}
The first operator commutes with the BRST operator
\begin{equation}
	\com{Q_B}{X} = 0.
\end{equation}
The product of these operators is a projector
\begin{equation}
	X Y X
		= X.
\end{equation}
\index{superstring field!constrained Ramond field}%
Then, the R string field is constrained to satisfy
\begin{equation}
	X Y \ket{\Psi_{-1/2}}
		= \ket{\Psi_{-1/2}}.
\end{equation}
A state satisfying this condition is said to be in the restricted Hilbert space.
It can be shown that it reproduces the cohomology of $Q_B$ on-shell.

\begin{remark}
	Since $G_0$ contains derivatives, the restriction is not purely algebraic as in the bosonic case.
	It prevents the degeneration due to $\gamma_0^n$.
\end{remark}

\begin{remark}[Comparison with level-matching]
	\index{level-matching condition}%

	The conditions $b_0^- = L_0^- = 0$ can be rephrased as the statement that the string field $\Psi$ is invariant under the action of the projector $B c_0^-$
	\begin{equation}
		B c_0^- \ket{\Psi}
			= \ket{\Psi},
	\end{equation}
	where
	\begin{equation}
		B
			= b_0^- \int_{0}^{2\pi} \frac{\dd\theta}{2\pi} \, \e^{\I \theta L_0^-}
			= \delta(b_0^-) \delta(L_0^-).
	\end{equation}
\end{remark}

\index{free super-SFT!action}%
The kinetic term (after unfixing the gauge) reads:
\begin{equation}
	S_{0,2}
		= - \frac{1}{2} \, \bra{\Psi_{-1}} c_0^- Q_B \ket{\Psi_{-1}}
			- \frac{1}{2} \, \bra{\Psi_{-1/2}} c_0^- Y Q_B \ket{\Psi_{-1/2}}.
\end{equation}
\index{free super-SFT!gauge transformation}%
The action is invariant under the gauge transformation
\begin{equation}
	\delta \ket{\Psi} = Q_B \ket{\Lambda}
\end{equation}
where
\begin{equation}
	\Lambda
		= \Lambda_{-1} + \Lambda_{-1/2}.
\end{equation}
Each gauge parameter satisfies the same conditions as the associated field (in particular, $\Lambda_{-1/2}$ is in the restricted Hilbert space).

\subsection{Auxiliary field approach}

The disadvantage of the constraint approach is two-fold.
First, it treats both components of the field on a different footing.
Second, the constraint must be imposed by hand and does not follow from any fundamental principle.
Another possibility is to embed the propagator in a higher-dimensional field space by introducing additional fields: in this way, the propagator can be inverted without introducing the inverse of $\mc X_0$.

\index{superstring field!auxiliary field}%
Let's introduce the new field
\begin{equation}
	\wtilde{\Psi}
		= \wtilde{\Psi}_{-1} + \wtilde{\Psi}_{-3/2}
\end{equation}
which satisfies the same conditions as $\Psi$:
\begin{equation}
	b_0^- \ket{\wtilde{\Psi}}
		= L_0^- \ket{\wtilde{\Psi}}
		= 0,
	\qquad
	b_0^+ \ket{\wtilde{\Psi}}
		= 0.
\end{equation}

\index{free super-SFT!action}%
A tentative kinetic term is then:
\begin{equation}
	S_{0,2}
		= \frac{1}{2} \, \bra{\wtilde{\Psi}} c_0^- c_0^+ L_0^+ \mc G \ket{\wtilde{\Psi}}
			- \bra{\wtilde{\Psi}} c_0^- c_0^+ L_0^+ \ket{\Psi}.
\end{equation}
The kinetic operator in matrix form for $(\wtilde{\Psi}, \Psi)$ reads
\begin{equation}
	K
		= c_0^- c_0^+ L_0^+
			\begin{pmatrix}
				- \mc G & 1 \\
				1 & 0
			\end{pmatrix}
\end{equation}
and its inverse is
\begin{equation}
	\Delta
		= b_0^- b_0^+ \frac{1}{L_0^+}
			\begin{pmatrix}
				0 & 1 \\
				1 & \mc G
			\end{pmatrix}.
\end{equation}
This reproduces the expected propagator for $(\Psi, \Psi)$ without needing to invert $\mc X_0$.

What is the interpretation of the additional fields?
\index{free super-SFT!gauge transformation}%
The gauge invariance of the action is
\begin{equation}
	\delta \ket{\Psi}
		= Q_B \ket{\Lambda},
	\qquad
	\delta \ket{\wtilde{\Psi}}
		= Q_B \ket{\wtilde{\Lambda}},
\end{equation}
where $\Lambda$ satisfies the same constraints as $\Psi$ (in particular, it contains more components than the $\Lambda$ of the previous section).
Then, the equations of motion are
\begin{equation}
	Q_B \ket{\Psi}
		= 0,
	\qquad
	Q_B \ket{\wtilde{\Psi}}
		= 0.
\end{equation}
This shows that both fields are free and decoupled and that the spectrum is doubled.
To push the interpretation further, one needs to consider the interactions.

Amplitudes involve only the states contained in $\Psi$ and thus the interactions are built solely in terms of $\Psi$.
Then, the equations of motion have the form:
\begin{equation}
	Q_B \big( \ket{\Psi} - \mc G \ket{\wtilde{\Psi}} \big)
		= 0,
	\qquad
	Q_B \ket{\wtilde{\Psi}}
		= \ket{J(\Psi)},
\end{equation}
where $J(\Psi)$ is a source term due to the interactions.
An equation for $\Psi$ only is obtained by multiplying the second with $\mc G$
\begin{equation}
	Q_B \ket{\Psi} = \mc G \ket{J(\Psi)}.
\end{equation}
Once $\Psi$ is determined by solving this equation, the auxiliary field $\wtilde{\Psi}$ is completely fixed by the second equation up to free field solutions.
This shows that $\wtilde{\Psi}$ describes only free fields even when $\Psi$ is interacting.
Note that this implies that the degrees of freedom contained in $\wtilde{\Psi}$ do not even couple to the gravitational field!
This can also be shown at the level of Feynman diagrams.

\begin{remark}
	The field $\wtilde{\Psi}$ is not an auxiliary field strictly speaking since it is propagating (its equation of motion is not algebraic).
\end{remark}

\subsection{Large Hilbert space}

\index{superstring field!large Hilbert space}%
The last formulation of the kinetic term considers the NS string field to be in the large Hilbert space, i.e.\ $\eta_0 \neq 0$.
The Ramond field must be described with one of the two previous approach.

\index{free super-SFT!action}%
Writing the action requires to use a NS field $\Psi_0$ with picture number $0$.
The kinetic term becomes
\begin{equation}
	S_{0,2}
		= - \frac{1}{2} \, \psbrbr{\Psi_{0}}{\eta_0 Q_B \Psi_{0}},
\end{equation}
where $\psbrbr{\cdot}{\cdot}$ is the inner product in the large Hilbert space (contains a $\xi_0$ insertion).
\index{free super-SFT!gauge transformation}%
This action has an enlarged gauge invariance:
\begin{equation}
	\delta \ket{\Psi_{0}}
		= Q_B \ket{\Lambda_{0}} + \eta_0 \ket{\Omega_{1}},
\end{equation}
and the equation of motion reads
\begin{equation}
	Q_B \eta_0 \ket{\Psi_{0}}
		= 0.
\end{equation}

The $\eta_0$ gauge invariance can be fixed with the condition
\begin{equation}
	\xi_0 \ket{\Psi_{0}}
		= 0,
\end{equation}
and one can introduce a new field $\Psi_{-1}$ such that
\begin{equation}
	\ket{\Psi_{0}}
		= \xi_0 \ket{\Psi_{-1}}
\end{equation}
to satisfy automatically the condition.
The equation of motion becomes
\begin{equation}
	Q_B \ket{\Psi_{-1}}
		= 0,
\end{equation}
and one recovers the small Hilbert space formulation.

\refchapter

\begin{itemize}
	\item General reviews~\cites{deLacroix:2017:ClosedSuperstringField}[sec.~6]{Erler:2020:FourLecturesClosed}.

	\item Spurious poles and vertical integration:
	\begin{itemize}
		\item small Hilbert space~\cites[app.~C, D]{deLacroix:2017:ClosedSuperstringField}{Sen:2015:OffshellAmplitudesSuperstring}{Sen:2015:FillingGapsPCOs}

		\item large Hilbert space~\cite{Erler:2017:VerticalIntegrationLarge}
	\end{itemize}

	\item Constructions of super-SFT:
	\begin{itemize}
		\item “Sen's” amplitude factorization construction~\cite{Sen:2015:GaugeInvariant1PI-NS, Sen:2015:GaugeInvariant1PI-R, Sen:2015:SupersymmetryRestorationSuperstring, Sen:2016:BVMasterAction, deLacroix:2017:ClosedSuperstringField, Pius:2018:QuantumClosedSuperstring}

		\item “Munich” homotopy algebra bootstrap~\cite{Erler:2014:ResolvingWittensSuperstring, Erler:2014:NSNSSectorClosed, Erler:2015:RamondEquationsMotion, Erler:2016:CompleteActionOpen, Konopka:2016:OpenSuperstringField}

		\item Berkovits' SFT~\cite{Berkovits:1995:SuperPoincareInvariantSuperstring, Kroyter:2012:OpenSuperstringField, Kunitomo:2016:CompleteActionOpen, Erler:2016:CompleteActionOpen, Matsunaga:2016:CommentsCompleteActions, Erler:2017:SuperstringFieldTheory}

		\item supermoduli space~\cite{Ohmori:2018:OpenSuperstringField, Takezaki:2019:OpenSuperstringField}

		\item democratic SFT~\cite{Kroyter:2009:SuperstringFieldTheory, Kroyter:2011:DemocraticSuperstringField}

		\item light-cone SFT~\cite{Ishibashi:2018:MultiloopAmplitudesLightcone-1, Ishibashi:2018:MultiloopAmplitudesLightcone-2}

	\end{itemize}

	\item Relations between different constructions~\cite{Iimori:2014:BerkovitsFormulationWitten, Erler:2015:RelatingBerkovitsAinftySmall, Erler:2015:RelatingBerkovitsAinftyLarge, Erler:2015:AinftyStructureBerkovits, Erler:2016:CompleteActionOpen}.

	\item Ramond string field:
	\begin{itemize}
		\item constrained field~\cite{Kunitomo:2016:CompleteActionOpen, Erler:2016:CompleteActionOpen, Erler:2017:SuperstringFieldTheory, Takezaki:2019:OpenSuperstringField}

		\item auxiliary field~\cite{Sen:2015:GaugeInvariant1PI-R, Sen:2016:BVMasterAction, Erler:2016:CompleteActionOpen, deLacroix:2017:ClosedSuperstringField}
	\end{itemize}

\end{itemize}

\chapter{Momentum-space SFT}
\label{sft:chap:momentum-sft}

\introchapter

In this chapter, we describe the general properties of SFT actions in the momentum space.
This allows to make SFT more intuitive, but also to use standard QFT methods to prove various properties of string theory.
We explain how the Wick rotation is generalized for theories with vertices diverging at infinite real energies (Lorentzian signature).
This allows to prove important properties of string theory, such as unitarity or crossing symmetry.

\section{General form}
\label{sft:sec:momentum-sft:general}

Since the explicit expressions of the string vertices are not known, it is not possible to write explicitly the SFT action.
However, the general properties of the vertices are known: then, one can write a general QFT which contains SFT as a subcase.
This is sufficient to already extract a lot of informations.
The other advantage is that the QFT language is more familiar and intuitive in many situations.
Hence, one can use this general form to built intuition before translating the results in a more stringy language.
\index{momentum-space SFT!properties}%
In a nutshell, SFT is a QFT:
\begin{itemize}
	\item with an infinite number of fields (of all spins);

	\item with an infinite number of interactions;

	\item with non-local interactions $\propto \e^{- \# k^2}$;

	\item which reproduces the worldsheet amplitudes (if the latter are well-defined).
\end{itemize}

\index{non-locality}%
The non-locality of the interactions is the most salient property of SFT, beyond the infinite number of fields.
This has a number of consequences:
\begin{itemize}
	\item the Wick rotation is ill-defined;

	\item the position representation cannot be used, nor any property relying on it (micro-causality, largest time equation…);

	\item standard assumptions from local QFT (in particular, from the constructive S-matrix program, such as micro-causality) break down.
\end{itemize}
Together, these points imply that the usual arguments from QFTs must be improved.
This has been an active topic in the recent years and the results will be summarized in \Cref{sft:sec:momentum-sft:consistency}.

\index{momentum-space SFT!string field expansion}%
\index{string field!momentum expansion}%
We expand the string field in Fourier space using a basis $\{ \phi_\alpha(k) \}$ as (\Cref{bsft:chap:string-field}):
\begin{equation}
	\ket{\Psi}
		= \sum_j \int \frac{\dd^D k}{(2\pi)^D} \, \psi_\alpha(k) \ket{\phi_\alpha(k)},
\end{equation}
where $k$ is the $D$-dimensional momentum and $\alpha$ the discrete indices (Lorentz indices, group representation, KK modes…) of the spacetime fields $\psi_\alpha(k)$.
\index{momentum-space SFT!action}%
The action in momentum space takes the form (in Lorentzian signature):
\begin{equation}
	\begin{aligned}
	S
		= &\ - \int \dd^D k \, \psi_\alpha(k) K_{\alpha\beta}(k) \psi_{\beta}(-k)
		\\ &
		- \sum_{n \ge 0} \int \dd^D k_1 \cdots \dd^D k_n \,
			V^{(n)}_{\alpha_1 \cdots \alpha_n}(k_1, \ldots, k_n) \,
			\psi_{\alpha_1}(k_1) \cdots \psi_{\alpha_n}(k_n).
	\end{aligned}
\end{equation}
The kinetic matrix $K_{\alpha\beta}$ is usually quadratic in the momentum.
In the direct Fourier expansion of the SFT action \eqref{bsft:eq:closed-action-gf}, it describes only the classical kinetic term: the quantum corrections are found in the vertex $V^{(2)}$.

\index{momentum-space SFT!Feynman rules}
From the action, we can write the Feynman rules (for the path integral weight $\e^{\I S}$ and S-matrix $S = 1 + \I T$).
The propagator reads:
\begin{equation}
	\label{sft:eq:momentum-prop}
	\vcenter{\hbox{\includegraphics{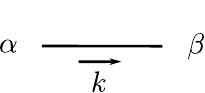}}}
		\quad
		= K_{\alpha\beta}(k)^{-1}
		= \frac{- \I \, M_{\alpha\beta}}{k^2 + m_\alpha^2} \, Q_\alpha(k),
\end{equation}
where $M_{\alpha\beta}$ is mixing matrix for states of equal mass and $Q_\alpha$ a polynomial in $k$ (there is no sum over $\alpha$).
The interactions are obtained by plugging the basis states $\{ \phi_\alpha \}$ inside the vertices $\mc V_{n}$ \eqref{bos:eq:Vn-vertex}:
\index{momentum-space SFT!interaction vertex}%
\begin{equation}
	\label{sft:eq:momentum-vertex}
	\begin{aligned}
	\vcenter{\hbox{\includegraphics{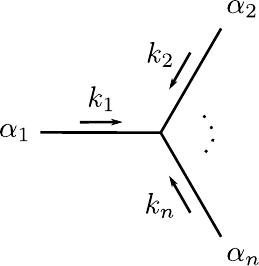}}}
		&
		= \I \, V^{(n)}_{\alpha_1 \cdots \alpha_n}(k_1, \ldots, k_n)
		:= \I \, \mc V_n\big(\phi_{\alpha_1}(k_1), \ldots, \phi_{\alpha_n}(k_n)\big)
		\\ &
		= \I \int \dd t \,
			\e^{- \cramped{g_{ij}^{\{\alpha_k\}}(t)} \, k_i \cdot k_{j} - \lambda \sum\limits_{\alpha} m_\alpha^2}
			P_{\alpha_1, \ldots, \alpha_n}\big(k_1, \ldots, k_n; t \big),
	\end{aligned}
\end{equation}
where $t$ denotes collectively the moduli parameters, $P_{\{ \alpha_i \}}$ is a polynomial in $k$, $g_{ij}$ is a positive-definite matrix, $\lambda > 0$ is a number.
There is an implicit sum over the momentum indices.

\index{stub parameter}%
The terms quadratic in the momenta inside the exponential arise from two sources:
\begin{itemize}
	\item The correlation functions of the vertex operators $\mean{\prod_i \e^{\I k_i \cdot X(z_i)}}$ is proportional to $\e^{- k_i \cdot k_j G(z_i, z_j)}$, where $G$ is the Green function.
	Additional factors like $\pd X$ contribute to the polynomial $P_{\alpha_1, \ldots, \alpha_n}$.

	\item It is possible to add stubs to the vertices.
	The effect is to multiply each leg by a factor $\e^{- \lambda (k_i^2 + m_i^2)}$ with $\lambda > 0$ (we take $\lambda$ to be the same for all vertices for simplicity).
	The first term of the exponential contributes to the diagonal of the matrix $g_{ij}$.
	By taking $\lambda$ sufficiently large, one can enforce that all eigenvalues are positive.
\end{itemize}
\index{momentum-space SFT!finiteness!UV divergence}%
\index{momentum-space SFT!finiteness!infinite number of states}%
Finally, the exponential term with the masses $m_\alpha^2$ ensures that the sum over all intermediate states converge despite an infinite number of states.
Indeed, the number of states of mass $m_\alpha$ grows as $\e^{c m_\alpha}$, which is dominated by $\e^{- \lambda m_\alpha^2}$ for sufficiently large $\lambda$.
Hence, the addition of stubs make explicit the absence of divergences in SFT.\footnotemark{}
\footnotetext{%
	Remember that $\lambda$ is not a physical parameter and disappears on-shell.
	This means that the cancellation of the divergences is independent of $\lambda$ and must always happen on-shell.
}%

The vertices have no singularity for $k_i \in \C$ finite.
As the energy becomes infinite $\abs{k_i^0} \to \infty$, they behave as:
\begin{equation}
	\lim\limits_{k^0 \to \pm \I \infty} V^{(n)}
		= 0,
	\qquad
	\lim\limits_{k^0 \to \pm \infty} V^{(n)}
		= \infty.
\end{equation}
The first property is responsible for the soft UV behaviour of string theory in Euclidean signature, while the second prevents from performing the Wick rotation (indeed, the pole at infinity implies that the arcs closing the contour contribute).

\index{momentum-space SFT!Green function}%
The $g$-loop $n$-point amputated Green functions are sums of Feynman diagrams, each of the form:
\begin{equation}
	\label{sft:eq:momentum-feynman-graph}
	\begin{multlined}
	F_{g,n}(p_1, \ldots, p_n)
		\sim \int \dd T \prod_s \dd^D \ell_s \,
			\e^{- G_{rs}(T) \, \ell_r \cdot \ell_s - 2 H_{ri}(T) \, \ell_r \cdot p_i - F_{ij}(T) \, p_i \cdot p_j}
			\\
			\times \prod_a \frac{1}{k_a^2 + m_a^2} \,
			\mc P(p_i, \ell_r; T),
	\end{multlined}
\end{equation}
where $\{ p_i \}$ are the external momenta, $\{ \ell_r \}$ the loop momenta and $\{ k_i \}$ the internal momenta, with the latter given by a linear combination of the other.
Moreover, $T$ denotes the dependence in the moduli parameters of all the internal vertices, and $\mc P$ is a polynomial in $(p_i, \ell_r)$.
The matrix $G_{rs}$ is positive definite, which implies that:
\begin{itemize}
	\item integrations over spatial loop momenta $\vec \ell_r$ converge;

	\item integrations over loop energies $\ell_r^0$ diverge.
\end{itemize}
As a consequence, the Feynman diagrams in Lorentzian signature are ill-defined: we will explain in the next section how to fix this problem.

\section{Generalized Wick rotation}
\label{sft:sec:momentum-sft:consistency}

We have seen that loop integrals in Lorentzian signature are divergent because of the large energy behaviour of the interactions.
But, this is not different from the usual QFT, where the loop integrals are also ill-defined in Lorentzian signature.
Indeed, poles of the propagators sit on the real axis and also give divergent loop integrals (note that the same problem arise also here).
\index{Wick rotation}%
\index{ie@$\I\varepsilon$-prescription}%
In that case, the strategy is to define the Feynman diagrams in Euclidean space and to perform a Wick rotation: the latter matches the expressions in Lorentzian signature up to the $\I\varepsilon$-prescription.
The goal of the latter is to move slightly the poles away from the real axis.

\begin{figure}[ht]
	\centering
	\includegraphics[scale=1.5]{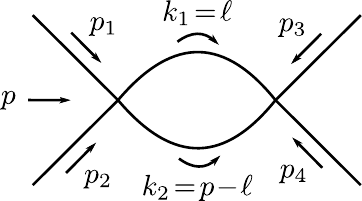}
	\caption{$1$-loop $4$-point function for a scalar field theory.}
	\label{sft:fig:ex-F14-scalar-graph}
\end{figure}

\begin{figure}[ht]
	\centering
	\includegraphics[scale=1]{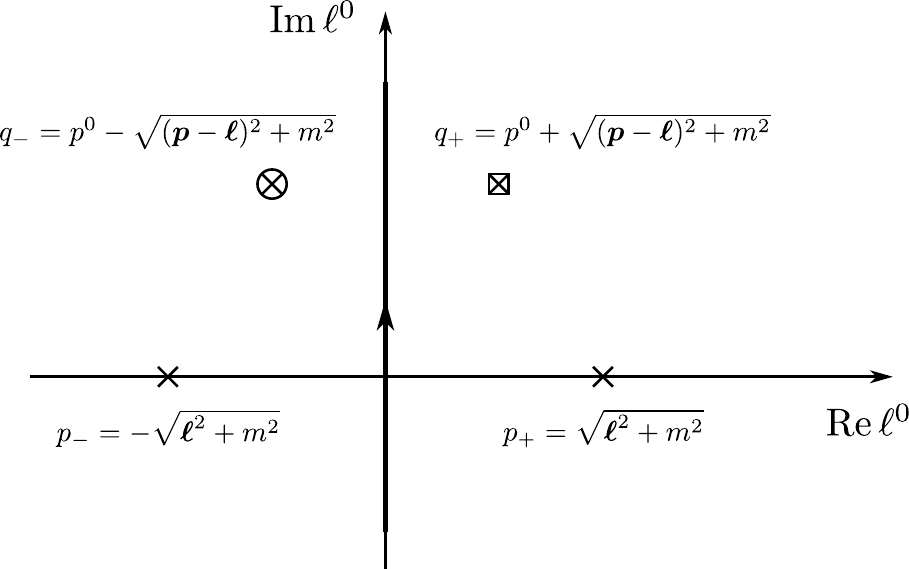}
	\caption{Integration contour for external Euclidean momenta.}
	\label{sft:fig:ex-F14-scalar-poles-E-Im}
\end{figure}

\begin{figure}[ht]
	\centering
	\includegraphics[scale=1]{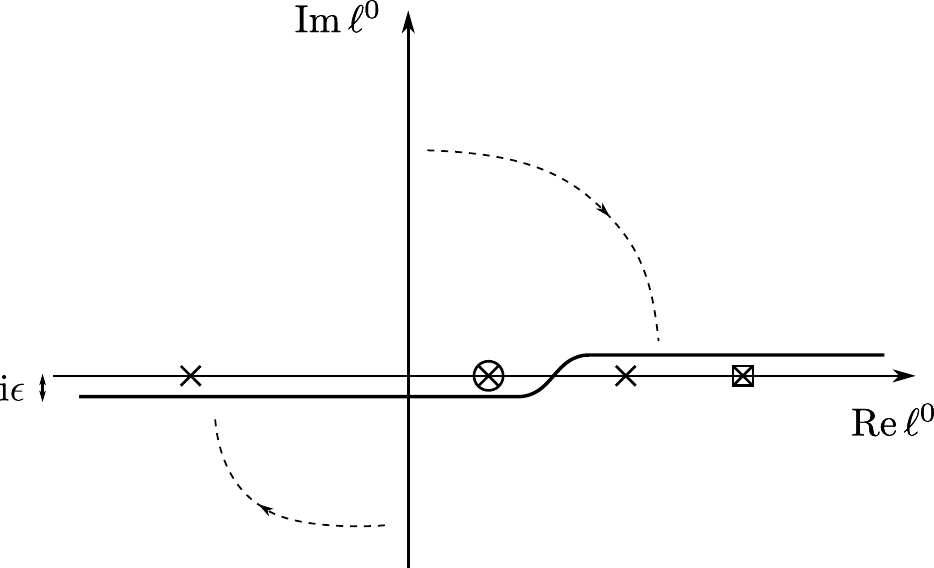}
	\caption{Integration contour for external Lorentzian momenta after Wick rotation (regular vertices).}
	\label{sft:fig:ex-F14-scalar-poles-E-wick}
\end{figure}

\begin{example}[Scalar field]
	Consider a scalar field of mass $m$ with a quartic interaction.
	The $1$-loop $4$-point Feynman diagram is given in \Cref{sft:fig:ex-F14-scalar-graph}.
	The external momenta are $p_i$, $i = 1, \ldots, 4$.
	There are one loop momentum $\ell$ and two internal momenta $k_1 = \ell$ and $k_2 = p - \ell$, where $p = p_1 + p_2$.
	The poles in the loop energy $\ell^0$ are located at:
	\begin{equation}
		p_{\pm}
			= \pm \sqrt{\vec\ell^2 + m^2},
		\qquad
		q_{\pm}
			= p^0 \pm \sqrt{(\vec p - \vec\ell)^2 + m^2}.
	\end{equation}

	The graph is first defined in Euclidean signature, where the external and loop energies are pure imaginary, $p_i^0, \ell^0 \in \I \R$.
	The poles are shown in \Cref{sft:fig:ex-F14-scalar-poles-E-Im}.
	Then, the external momenta are analytically continued to real values, $p_i^0 \in \R$.
	At the same time, the integration contour is also analytically continued thanks to the Wick rotation (\Cref{sft:fig:ex-F14-scalar-poles-E-wick}).
	The contour is closed with arcs, but they don't contribute since there is no poles in the upper-right and lower-left quadrants, and no poles at infinity.
	However, one cannot continue the contour such that $\ell^0 \in \R$ because of the poles on the real axis.
	The Wick rotation is possible for $\ell^0$ in the upper-right quadrant, $\Re \ell^0 \ge 0, \Im \ell^0 > 0$, which leads to the $\I\varepsilon$-prescription $\ell^0 \in \R + \I \varepsilon$.
\end{example}

Since the Feynman diagram \eqref{sft:eq:momentum-feynman-graph} is not defined in Lorentzian signature because of the poles at $\ell_r^0 \to \pm \infty$, it is also necessary to start with Euclidean momenta.
However, the same behaviour at infinity prevents from using the Wick rotation since the contribution from the arcs does not vanish.
\index{Wick rotation!generalized}%
It is then necessary to find another prescription for defining the Feynman diagrams in SFT starting from the Euclidean Green functions.
This is given by the following \emph{generalized Wick rotation} (Pius--Sen~\cite{Pius:2016:CutkoskyRulesSuperstring}):
\index{Wick rotation!generalized}%
\begin{enumerate}
	\item Define the Green functions for Euclidean internal and external momenta.

	\item Perform an analytic continuation of the external energies and of the integration contour such that:
	\begin{itemize}
		\item keep poles on the same side;

		\item keep the contour ends fixed at $\pm \I \infty$.
	\end{itemize}
\end{enumerate}
One can show~\cite{Pius:2016:CutkoskyRulesSuperstring} that the Green functions are analytic in the upper-right quadrant $\Im p_a^0 > 0, \Re p_a^0 \ge 0$, for $\vec p_a \in \R$, $p_a^0$.
Moreover, the result is independent of the contour chosen as long as it satisfies the conditions described above.
In fact, this generalized Wick rotation is valid even for normal QFT, which raises interesting questions.
For example, it seems that the internal and external set of states have no intersection, which can be puzzling when trying to interpret the Cutkosky rules.
Nonetheless, everything works as expected.

\begin{figure}[htpb]
	\centering
	\subcaptionbox{}{\includegraphics[scale=1]{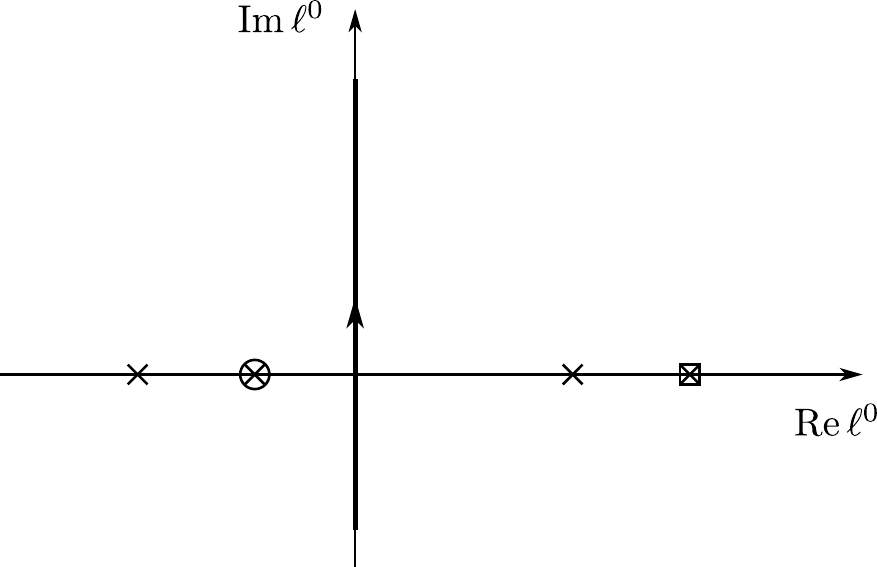}}
	\subcaptionbox{}{\includegraphics[scale=1]{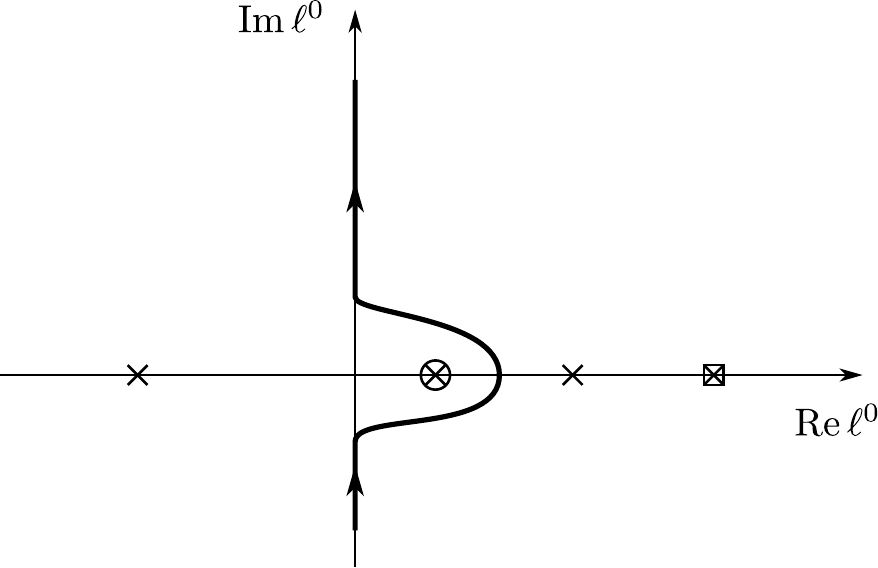}}
	\subcaptionbox{}{\includegraphics[scale=1]{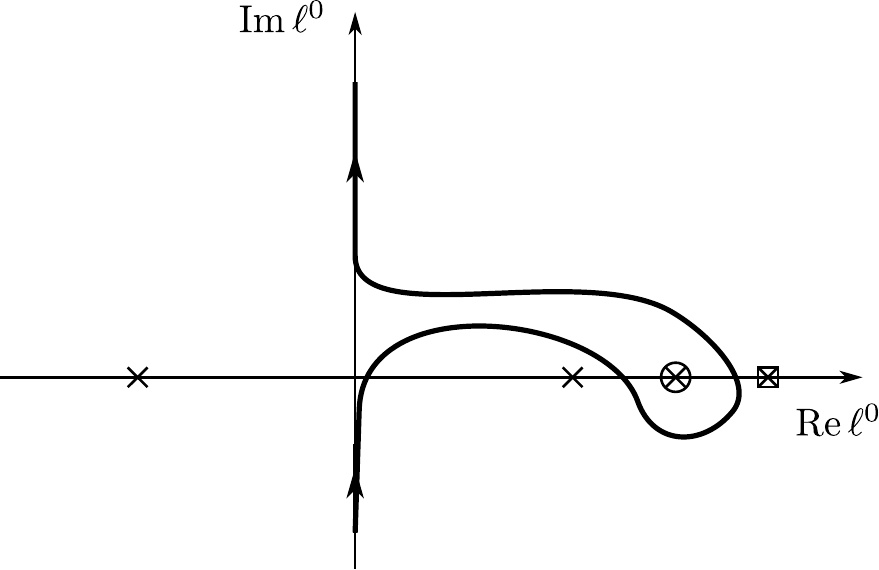}}
	\caption{%
		Integration contour after analytic continuation to external Lorentzian momenta.
		Depending on the values of the external momenta, different cases can happen.
	}%
	\label{sft:fig:ex-F14-scalar-poles-E}
\end{figure}

\begin{remark}[Timelike Liouville theory]
	\index{Liouville!theory}%

	It has been shown in~\cite{Bautista:2019:QuantumGravityTimelike} that this generalized Wick rotation is also the correct way for defining the timelike Liouville theory.
\end{remark}

\index{momentum-space SFT!consistency}%
\index{string theory!consistency}%
The fact that the amplitude is analytic only when the imaginary parts of the momenta are not zero, $\Im p_a^0 > 0$, is equivalent to the usual $\I\varepsilon$-prescription for QFT.
Moreover, it has been shown~\cite{Sen:2017:EquivalenceTwoContour} to be equivalent to the moduli space $\I\varepsilon$-prescription from~\cite{Witten:2013:FeynmaniEpsilon}.
Then, it has also been used to prove several important properties of string theory shared by local QFTs: Cutkosky rules~\cite{Pius:2016:CutkoskyRulesSuperstring, Pius:2018:UnitarityBoxDiagram}, unitarity~\cite{Sen:2016:RealitySuperstringField, Sen:2016:UnitaritySuperstringField}, analyticity in a subset of the primitive domain and crossing symmetry~\cite{deLacroix:2019:AnalyticityCrossingSymmetry}.
Finally, general soft theorems for string theory (and, in fact, any theory of quantum gravity) have been proven in~\cite{Sen:2017:SoftTheoremsSuperstring, Sen:2017:SubleadingSoftGraviton, Laddha:2017:SubsubleadingSoftGraviton, Chakrabarti:2017:SubleadingSoftTheorem}.
All together, these properties establish string theory as a very strong candidate for a consistent theory of everything.
The next main question is how to obtain an expression of SFT which is amenable to explicit computations.
This will certainly require to understand even better the deep structure of SFT, a goal which this \revname{} will hopefully help the reader to achieve.

\refchapter

\begin{itemize}
	\item SFT momentum space action~\cite{Sen:2016:WilsonianEffectiveAction, Pius:2016:CutkoskyRulesSuperstring, deLacroix:2017:ClosedSuperstringField}.

	\item Consistency properties of string theory~\cite{deLacroix:2017:ClosedSuperstringField}:
	\begin{itemize}
		\item generalized Wick rotation, Cutkosky rules and unitarity~\cite{Pius:2016:CutkoskyRulesSuperstring, Sen:2016:OneLoopMass, Sen:2016:RealitySuperstringField, Sen:2016:UnitaritySuperstringField, Sen:2017:EquivalenceTwoContour, Pius:2018:UnitarityBoxDiagram}.

		\item analyticity and crossing symmetry~\cite{deLacroix:2019:AnalyticityCrossingSymmetry}.

		\item soft theorems~\cite{Sen:2017:SoftTheoremsSuperstring, Sen:2017:SubleadingSoftGraviton, Laddha:2017:SubsubleadingSoftGraviton, Chakrabarti:2017:SubleadingSoftTheorem}.
	\end{itemize}

\end{itemize}

\appendix

\chapter{Conventions}

Most of the \revname{} uses natural units where $c = \hbar = 1$, but the string length $\ell_s$ (or Regge slope $\alpha')$ are kept.

A bar is used to denote both complex conjugation and the anti-holomorphic operators.
The symbol $:=$ (resp.\ $=:$) means that the LHS (RHS) is defined by the expression in the RHS (LHS).

\section{Coordinates}

The number of spacetime (target-space) dimensions is denoted by $D = d + 1$, where $d$ is the number of spatial dimensions.
The corresponding spacetime and spatial coordinates are written with Greek and Latin indices:
\begin{equation}
	x^\mu = (x^0, x^i),
	\qquad
	\mu = 0, \ldots, D - 1 = d
	\qquad
	i = 1, \ldots, d
\end{equation}
When time is singled out, one writes $x^0 = t$ in Lorentzian signature and $x^0 = t_E$ in Euclidean signature (or $x^0 = \tau$ when there is no ambiguity with the worldsheet time).

A $p$-brane is a $(p+1)$-dimensional object whose worldvolume is parametrized by coordinates:
\begin{equation}
	\sigma^a = (\sigma^0, \sigma^\alpha),
	\qquad
	a = 0, \ldots, p - 1,
	\qquad
	\alpha = 1, \ldots, p.
\end{equation}
The time coordinate can also be singled out as $\sigma^0 = \tau_M$ in Lorentzian signature and as $\sigma^0 = \tau$ in Euclidean signature.
For the string, the index $\alpha$ is omitted since it takes only one value.

The Lorentzian signature is taken to be mostly plus and the flat Minkowski metric reads
\begin{equation}
	\eta_{\mu\nu} = \diag(-1, \underbrace{1, \ldots, 1}_{d}).
\end{equation}
The flat Euclidean metric is
\begin{equation}
	\delta_{\mu\nu} = \diag(\underbrace{1, \ldots, 1}_{D}).
\end{equation}
Similar notations hold for the worldvolume metrics $\eta_{ab}$ and $\delta_{ab}$.
The Levi--Civita (completely antisymmetric) tensor is normalized by
\begin{equation}
	\epsilon_{01}
		= - \epsilon^{01}
		= 1.
\end{equation}

\index{Wick rotation}%
Wick rotation from Lorentzian time $t$ to Euclidean time $\tau$ (either worldsheet or target spacetime) is defined by
\begin{equation}
	\label{app:eq:wick-rot-time}
	t = - \I \tau.
\end{equation}
Accordingly, contravariant (covariant) vector transforms with the same (opposite) factor:
\begin{equation}
	\label{app:eq:wick-rot-vector}
	V_M^0 = - \I V_E^0,
	\qquad
	V_{M,0} = \I V_{E,0}.
\end{equation}
Most computations are performed with both spacetime and worldsheet Euclidean signatures.
Expressions are Wick rotated when needed.

Light-cone coordinates are defined by
\begin{equation}
	x^{\pm} = x^0 \pm x^1.
\end{equation}
\index{left/right-moving sectors}%
A function depending only on $x^+$ ($x^-$) is said to be left-moving (right-moving) by analogy with the displacement of a wave.
\index{holomorphic/anti-holomorphic sectors}%
Under analytic continuation, the left-moving (right-moving) coordinate is mapped to the holomorphic\footnotemark{} (anti-holomorphic) coordinate $z$ ($\bar z$).
\footnotetext{%
	The terms of \emph{holomorphic} is simply used to indicate that the object depends only on $z$, but not on $\bar z$.
	Typically, the objects have singularities and are really \emph{meromorphic} in $z$.
}%
In chiral theories, the left-moving value is written first.

The worldsheet coordinates $(\tau, \sigma)$ on the cylinder are defined by
\begin{equation}
	\tau \in \R,
	\qquad
	\sigma \in [0, L),
	\qquad
	\sigma \sim \sigma + L,
\end{equation}
where typically $L = 2\pi$.
The integration over the spatial coordinate is normalized such that the perimeter of spatial slice is normalized to $1$ if $L = 2\pi$:
\begin{equation}
	\mathscr L
		= \frac{1}{2\pi} \int_0^L \dd \sigma
		= \frac{L}{2\pi}.
\end{equation}
This implies that $2d$ action, conserved charges, etc.\ are divided by an extra factor of $2\pi$.

The coordinates can be written in terms of complex coordinates
\begin{equation}
	w = \tau + \I \sigma,
	\qquad
	\bar w = \tau - \I \sigma
\end{equation}
such that the flat metric is
\begin{equation}
	\dd s^2 = \dd \tau^2 + \dd \sigma^2
		= \dd w \dd \bar{w}.
\end{equation}
Under Wick rotation, the complex coordinates are mapped to light-cone coordinates as follows:
\begin{equation}
	w
		= \I \sigma^+,
	\qquad
	\bar w
		= \I \sigma^-.
\end{equation}
The cylinder can be mapped to the complex plane through
\begin{equation}
	z = \e^{2\pi w / L},
	\qquad
	\bar z = \e^{2\pi \bar{w} / L}.
\end{equation}

The definition of the Levi--Civita tensor includes the $\sqrt{g}$ factor, such that
\begin{equation}
	\epsilon_{z \bar z}
		= \frac{\I}{2},
	\qquad
	\epsilon^{z \bar z}
		= - 2 \I
\end{equation}
on the complex plane with flat metric.

\begin{table}[ht]
	\begin{adjustwidth}{-2cm}{-2cm}
	\centering
	\begin{tabular}{l|cccc}
		refs &
			cylinder &
			plane &
			light-cone &
			left-moving
		\\
		\hline
		\makecell[l]{here, Di Francesco et al. \\
			\cite{Zwiebach:2009:FirstCourseString, DiFrancesco:1999:ConformalFieldTheory, Blumenhagen:2009:IntroductionConformalField, Sen:StringTheory1, Kiritsis:2007:StringTheoryNutshell, Kaku:1999:IntroductionSuperstringsMTheory, West:2012:IntroductionStringsBranes, Ketov:1995:ConformalFieldTheory}
			}
			&
			$w = \tau + \I \sigma$ &
			$z = \e^{w}$ &
			$w = \I \sigma^+$, $\bar w = i \sigma^-$ &
			holomorphic
		\\
		\makecell[l]{Blumenhagen et al. \\
			\cite{Blumenhagen:2014:BasicConceptsString, Becker:2006:StringTheoryMTheory, Johnson:2006:DBranes, Schomerus:2017:PrimerStringTheory}
			}
			&
			$w = \tau - \I \sigma$ &
			$z = \e^{w}$ &
			$w = \I \sigma^-$, $\bar w = i \sigma^+$ &
			anti-holomorphic
		\\
		Polchinski~\cite{Polchinski:2005:StringTheory-1, Tong:2009:LecturesStringTheory, Ohmori:2001:ReviewTachyonCondensation} &
			$w = \sigma + \I \tau$ &
			$z = \e^{- \I w}$ &
			$w = - \sigma^-$, $\bar w = \sigma^+$ &
			anti-holomorphic\footnotemark{}
	\end{tabular}
	\end{adjustwidth}
	\caption{Conventions for the coordinates.%
		The notations are the following (they can slightly vary depending on the references): the Euclidean time is obtained by the analytic continuation $\tau = \I t$ (denoted also by $\tau = \sigma^0 = \sigma^2$) the spatial direction is $\sigma = \sigma^1$, and the light-cone coordinates are $\sigma^\pm = t \pm \sigma$.
		\index{conventions!complex coordinates}%
	}%
	\label{cft:tab:conventions-coord}
\end{table}

\footnotetext{%
	In fact, the terms of “left”- and “right”-moving are interchanged in~\cite[p.~34]{Polchinski:2005:StringTheory-1} to get agreement with the literature.
	But, it means that the spatial axis orientation is reversed.

	Moreover, concerning~\cite{DiFrancesco:1999:ConformalFieldTheory}, the first definition agrees with (6.1) but not with (6.53) since the definition of $\xi$ (our $w$) is modified in-between.
	This explains why the definitions of left- and right-moving~\cite[ p.~161]{DiFrancesco:1999:ConformalFieldTheory} do not agree with the one given in the table.
}%

\section{Operators}

Commutators and anti-commutators are denoted by
\begin{equation}
	\com{A}{B}
		:= \com{A}{B}_-
		= A B - B A,
	\qquad
	\anticom{A}{B}
		:= \com{A}{B}_+
		= A B + B A.
\end{equation}

The Grassmann parity of a field $A$ is denoted by $\abs{A}$
\begin{equation}
	\abs{A} =
	\begin{cases}
		+1 & \text{Grassmann odd},
		\\
		0 & \text{Grassmann even}.
	\end{cases}
\end{equation}
Two (anti-)commuting operators satisfy
\begin{equation}
	A B = (- 1)^{\abs{A} \abs{B}} B A.
\end{equation}

\section{QFT}

Energy is defined as the first component of the momentum vector
\begin{equation}
	p^\mu := (E, p^i).
\end{equation}

\medskip

The following notations are used to denote the number of supersymmetries:
\begin{equation}
	(N_L, N_R),
	\qquad
	N = N_L + N_R,
\end{equation}
where $N_L$ and $N_R$ are the numbers of left- and right-chirality supersymmetries.
The last form is used when it is not important to know the chirality of the supercharges.

\medskip

The variation of a field $\phi(x)$ is defined by
\begin{equation}
	\delta \phi(x) = \phi'(x) - \phi(x).
\end{equation}
\index{conserved current}%
Given an internal symmetry with parameters $\alpha^a$, the Noether current in Lorentzian signature is given by:
\begin{equation}
	\label{app:eq:current}
	J_a^\mu
		= \lambda \, \frac{\pd \mc L}{\pd(\pd_\mu \phi)} \, \frac{\delta \phi}{\delta \alpha^a},
	\qquad
	\grad_\mu J_a^\mu = 0,
\end{equation}
where $\mc L$ is the Lagrangian which does not include the factor $\sqrt{g}$ for curved spaces and $\lambda$ is some normalization.\footnotemark{}
\footnotetext{%
	Including the $\sqrt{g}$ would give the current density $\sqrt{g} J_a^\mu$.
	The simple derivative of the latter vanishes $\pd_\mu (\sqrt{g} J_a^\mu) = 0$ in view of the identity \eqref{app:eq:div-vector}.
}%
The conserved charges $Q_a$ associated to the currents $J_a^\mu$ for a fixed spatial slice $t = \text{cst}$ are
\index{conserved charge}%
\begin{equation}
	\label{app:eq:charge}
	Q_a = \frac{1}{\lambda} \oint_{\Sigma} \dd^{D-1} x \sqrt{h} \, J_a^0,
\end{equation}
where $\Sigma$ is a spatial slice and $h$ is the induced metric.
One sets $\lambda = 2\pi$ in two dimensions, otherwise $\lambda = 1$.
The variation of a field under a transformation generated by $Q$ is
\begin{equation}
	\label{app:eq:variation}
	\delta_{\alpha^a} \phi(x)
		= \I \alpha^a \com{Q_a}{\phi(x)}
\end{equation}
In Euclidean signature, the current and variation are:
\begin{subequations}
\begin{gather}
	\label{app:eq:current-euc}
	J_a^\mu
		= \I \lambda \, \frac{\pd \mc L}{\pd(\pd_\mu \phi)} \, \frac{\delta \phi}{\delta \alpha^a},
	\\
	\label{app:eq:variation-euc}
	\delta_{\alpha^a} \phi(x)
		= - \alpha^a \com{Q_a}{\phi(x)}.
\end{gather}
\end{subequations}
Note that the charge is still given by \eqref{app:eq:charge}.
The factor of $\I$ in \eqref{app:eq:current-euc} can be understood as follows.\footnotemark{}
\footnotetext{%
	We stress that these formulas and arguments do not apply to the energy--momentum tensor.
}%
First, the time component $J_a^0$ of the current transforms like time such that $J_a^0 \to \I J_a^0$, which implies that the charge also gets a factor $\I$, $Q_a \to \I Q_a$.
This explains the minus sign in \eqref{app:eq:variation-euc}.
Then, one needs to make this consistent with the formula \eqref{app:eq:charge-surface} for the charge associated to a general surface.
Given a spacelike $n_\mu$, the integration measure includes the time which transforms with a factor of $\I$: one can interpret it as coming from the spatial components of the current, $J_a^i \to \I J_a^i$, while working with a Euclidean region.
Another way to understand this factor for the spatial vector is by considering the electromagnetic case, where $\vec J$ contains a time derivative.

\medskip

\index{zero-mode}%
The term “zero-mode” has two (related) meanings:
\begin{enumerate}
	\item given an operator $D$ acting on a space of fields $\psi(z)$, zero-modes $\psi_{0,i}(z)$ of the operator are all fields with zero eigenvalue $D \psi_{0,i}(z) = 0$, $i = 1, \ldots, \dim \ker D$

	\item the zero-mode of a field expansion $\psi = \sum_n \psi_n z^{-n-h}$ is the mode $\psi_0$ for $n = 0$: on the cylinder, it corresponds to the constant term of the Fourier expansion on the cylinder (hence, a zero-mode of $\pd_z$ according to the previous definition)
\end{enumerate}
A prime indicates that the zero-modes are excluded.
For example, $\det' D$ is the product of non-zero eigenvalues, $\phi'$ is a field without zero-mode and $\dd' \phi$ the corresponding integration measure.

\section{Curved space and gravity}

The covariant derivative is defined by
\begin{equation}
	\grad_\mu = \pd_\mu + \Gamma_\mu
\end{equation}
where $\Gamma_\mu$ is the connection.
For example, one has for a vector field
\begin{equation}
	\grad_\mu A^\nu
		= \pd_\mu A^\nu + \tensor{\Gamma}{_{\mu\rho}^\nu} A^\rho.
\end{equation}
The negative-definite Laplacian (or Laplace--Beltrami operator) is defined by
\begin{equation}
	\label{app:eq:laplacian}
	\lap
		= g^{\mu\nu} \grad_\mu \grad_\nu
		= \frac{1}{\sqrt{g}} \, \grad_\mu \big( \sqrt{g} g^{\mu\nu} \grad_\nu).
\end{equation}
Note that $\grad_\mu$ does not contain the Christoffel symbol for the index $\nu$ because of the identity \eqref{app:eq:div-vector} (but it contains a connection for any other index of the field).
For a scalar field, both derivatives become simple derivatives.

The energy--momentum tensor is defined by
\begin{equation}
	T_{\mu\nu}
		= - \frac{2\lambda}{\sqrt{g}} \, \frac{\delta S}{\delta g^{\mu\nu}},
\end{equation}
where $\lambda = 2\pi$ for $D = 2$ and $\lambda = 1$ otherwise.

\section{List of symbols}

\noindent
General:
\begin{itemize}
	\item $D$:
	number of non-compact spacetime dimensions

	\item $g$:
	loop order for a scattering amplitude
	\item $n$:
	number of external closed string states

	\item $x^\mu$:
	spacetime non-compact coordinates
	\item $\sigma^a = (t, \sigma)$:
	worldsheet coordinates

	\item $g_s$:
	closed string coupling

	\item $Z_g = A_{g,0}$: genus-$g$ vacuum amplitude

	\item $A_{g,n}(k_1, \ldots, k_n)_{\alpha_1, \ldots, \alpha_n} :=
	A_{g,n}(\{ k_i \})_{\{ \alpha_i \}}$:
	$g$-loop $n$-point scattering amplitude for states with quantum numbers $\{ k_i, \alpha_i \}$ (if connected, amputated Green functions for $n \ge 3$)

	\item $G_{g,n}(k_1, \ldots, k_n)_{\alpha_1, \ldots, \alpha_n}$:
	$g$-loop $n$-point Green function for states with quantum numbers $\{ k_i, \alpha_i \}$

	\item $T_{ab}^\perp$:
	traceless symmetric tensor or traceless component of the tensor $T_{ab}$

	\item $\Psi$:
	generic (set of) matter field(s)
\end{itemize}

\noindent
Hilbert spaces:
\begin{itemize}
	\item $\mc H$:
	generic Hilbert space (in general, Hilbert space of the matter plus ghost CFT)
	\item $\mc H_\pm = \mc H \cap \ker b_0^\pm$
	\item $\mc H_0 = \mc H \cap \ker b_0^- \cap \ker b_0^+$

	\item $\mc H(Q_B)$:
	absolute cohomology of the operator $Q_B$ inside the space $\mc H$
	\item $\mc H_-(Q_B) = \mc H(Q_B) \cap \mc H^-$:
	semi-relative cohomology of the operator $Q_B$ inside the space $\mc H$
	\item $\mc H_0(Q_B) = \mc H(Q_B) \cap \mc H^0$:
	relative cohomology of the operator $Q_B$ inside the space $\mc H$

	\item $A$:
	Grassmann parity of the operator or state $A$
\end{itemize}

\noindent
Riemann surfaces:
\begin{itemize}

	\item $g$:
	Riemann surface genus (number of holes / handles)
	\item $n$:
	number of bulk punctures / marked points

	\item $\Sigma_{g,n}$:
	genus-$g$ Riemann surface with $n$ punctures
	\item $\Sigma_g = \Sigma_{g,0}$:
	genus-$g$ Riemann surface

	\item $\mc M_{g,n}$:
	moduli space of genus-$g$ Riemann surfaces with $n$ punctures
	\item $\mc M_{g} = \mc M_{g,0}$:
	moduli space of genus-$g$ Riemann surfaces

	\item $\mathsf{M}_{g,n} = \dim \mc M_{g,n}$
	\item $\mathsf{M}_{g,n}^c = \dim_\C \mc M_{g,n}$

	\item $\mathsf{M}_{g} = \mathsf{M}_{g,0} = \dim \mc M_g = \dim \ker \adj{P_1}$
	\item $\mathsf{M}_{g}^c = \mathsf{M}_{g,0}^c = \dim_\C \mc M_g$

	\item $\mc K_{g,n}$:
	conformal Killing vector group of genus-$g$ Riemann surfaces with $n$ punctures
	\item $\mc K_{g} = \mc K_{g,0} = \ker P_1$:
	conformal Killing vector group of genus-$g$ Riemann surfaces

	\item $\mathsf{K}_{g,n} = \dim \mc K_{g,n}$
	\item $\mathsf{K}_{g,n}^c = \dim_\C \mc K_{g,n} = \dim_\C \ker P_1$

	\item $\mathsf{K}_{g} = \mathsf{K}_{g,0} = \dim \ker P_1$
	\item $\mathsf{K}_{g}^c \mathsf{K}_{g,0}^c = \dim_\C \ker P_1$

	\item $\psi_i$:
	real basis of $\ker P_1$, CKV
	\item $\phi_i$:
	real basis of $\ker \adj{P_1}$, real quadratic differentials
	\item $(\psi_K, \bar\psi_K)$:
	complex basis of $\ker P_1$, (anti-)holomorphic CKV
	\item $(\phi_I, \bar\phi_I)$:
	complex basis of $\ker \adj{P_1}$, (anti-)holomorphic quadratic differentials

	\item $t_\lambda \in \mc M_{g,n}$:
	real moduli of $\mc M_{g,n}$
	\item $m_\Lambda \in \mc M_{g,n}$:
	complex moduli of $\mc M_{g,n}$

	\item $t_i \in \mc M_g$:
	real moduli of $\mc M_g$
	\item $m_I \in \mc M_g$:
	complex moduli of $\mc M_g$

	\item $z$:
	coordinate on the Riemann surface
	\item $w_i$:
	local coordinates around punctures
	\item $z_a$:
	local coordinates away from punctures
	\item $f_i(w_i)$:
	transition functions from $w_i$ to $z$
	\item $\sigma_\alpha$:
	coordinate system on the left of the contour $C_{\alpha}$
	\item $\tau_\alpha$:
	coordinate system on the right of the contour $C_{\alpha}$
\end{itemize}

\noindent
CFT:
\begin{itemize}
	\item $V_\alpha(k; \sigma^a) := V_{k,\alpha}(\sigma^a)$:
	matter vertex operator with\footnotemark{} momentum $k$ and quantum numbers $\alpha$ inserted at position $\sigma^a = (z, \bar z)$
	\footnotetext{%
		When the momentum and/or quantum numbers are not relevant, we remove them or simply index the operators by a number.
	}%

	\item $\scr V_\alpha(k; \sigma^a)$:
	unintegrated vertex operator with momentum $k$ and quantum numbers $\alpha$ inserted at position $\sigma^a$

	\item $V_\alpha(k) = \int \dd^2 \sigma \sqrt{g} \, V_\alpha(k; \sigma)$: integrated vertex operator

	\item on-shell (closed bosonic string):
	$\scr V_\alpha(k; \sigma^a) = c \bar c V_\alpha(k; \sigma^a)$ is a $(0, 0)$-primary, with $V_\alpha(k; \sigma^a)$ a $(1,1)$-primary matter operator

	\begin{draft}
	\item $V_\alpha(k; \sigma) = v_\alpha(k; \sigma) \, \e^{\I k \cdot X(\sigma)}$

	\item $V_\alpha(k; \sigma) = \what V_\alpha(k; \sigma) \, \e^{\I k \cdot x}$
	($x$ is the zero-mode)
	\end{draft}

	\item $\what{\mc O}$:
	operator $\mc O$ with zero-modes removed

	\item $\adj{\mc O}$:
	Hermitian adjoint
	\item $\eadj{\mc O}$:
	Euclidean adjoint
	\item $\mc O^t$:
	BPZ conjugation

	\item $\bracket{\mc O_1}{\mc O_2}$:
	BPZ inner-product
	\item $\bracket{\eadj{\mc O_1}}{\mc O_2}$:
	Hermitian inner-product

	\item $\ket{0}$:
	$\group{SL}(2, \C)$ (conformal) vacuum
	\item $\ket{\Omega}$:
	energy vacuum (lowest energy state)

	\item $\norder{\mc O}$\ :
	conformal normal ordering (with respect to $\group{SL}(2, \C)$ vacuum $\ket{0}$)
	\item $\norderv{\mc O}$\ :
	energy normal ordering (with respect to energy vacuum $\ket{\Omega}$)
\end{itemize}

\noindent
SFT:
\begin{itemize}
	\item $\Psi$:
	closed string field

	\begin{check}
	\item $\Phi$:
	open string field
	\end{check}

	\item $\{ \phi_r \} = \{ \phi_\alpha(k) \}$:
	basis of $\mc H$ (or some subspace)
\end{itemize}

\noindent
Indices:
\begin{itemize}
	\item $\mu = 0, \ldots, D - 1$:
	non-compact spacetime dimensions
	\item $a = 0, \ldots, p$:
	worldvolume coordinates ($p = 1$: worldsheet)

	\item $i = 1, \ldots, n$: external states, local coordinates

	\item $\lambda = 1, \ldots, \mathsf{M}_{g,n}$:
	real moduli of $\mc M_{g,n}$
	\item $\Lambda = 1, \ldots, \mathsf{M}_{g,n}^c$:
	complex moduli of $\mc M_{g,n}$

	\item $i = 1, \ldots, \mathsf{M}_{g}$:
	real moduli of $\mc M_{g}$
	\item $I = 1, \ldots, \mathsf{M}_{g}^c$:
	complex moduli of $\mc M_{g}$

	\item $i = 1, \ldots, \mathsf{K}_{g}$:
	real CKV of $\mc K_{g}$
	\item $K = 1, \ldots, \mathsf{K}_{g}^c$:
	complex CKV of $\mc K_{g}$

	\item $r = (k, \alpha)$:
	index for basis state of $\mc H$ (or some subspaces), $\alpha$: non-momentum indices
\end{itemize}

\begin{check}
Superstring:
\begin{itemize}
	\item $\mc M_{g,m,n}$

	\item $\M_{g,m,n}$

	\item $A = 1, \ldots, n_{\text{pco}}$:
	number of PCO
\end{itemize}
\end{check}

\chapter{Summary of important formulas}
\label{app:chap:formulas}

This appendix summarizes formulas which appear in the \revname{} or which are needed but assumed to be known to the reader (such as formulas from QFT and general relativity).

\section{Complex analysis}

The Cauchy--Riemann formula is
\begin{equation}
	\label{app:eq:cauchy-riemann}
	\oint_{C_z} \frac{\dd w}{2\pi \I} \, \frac{f(w)}{(w - z)^n}
		= \frac{f^{(n-1)}(z)}{(n - 1)!},
\end{equation}
where $f(z)$ is an holomorphic function.

One has
\begin{equation}
	\label{app:eq:pd-z-inv}
	\bar\pd \frac{1}{z} = 2\pi \, \delta^{(2)}(z).
\end{equation}

\section{QFT, curved spaces and gravity}

The Green function $G$ of a differential operator $D$ is defined by
\begin{equation}
	\label{app:eq:green-eq}
	D_x G(x, y)
		= \frac{\delta(x - y)}{\sqrt{g}} - P(x, y),
\end{equation}
where $P$ is the projector on the zero-modes of $D$.

The covariant divergence of a vector can be rewritten in terms of a simple derivative:
\begin{equation}
	\label{app:eq:div-vector}
	\grad_\mu v^\mu
		= \frac{1}{\sqrt{g}} \, \pd_\mu (\sqrt{g} v^\mu).
\end{equation}

Under an infinitesimal change of coordinates
\begin{equation}
	\delta x^\mu = \xi^\mu,
\end{equation}
the metric transforms as
\begin{equation}
	\label{app:eq:var-inf-diffeo-metric}
	\delta g_{\mu\nu}
		= \mc L_\xi g_{\mu\nu}
		= \grad_\mu \xi_\nu + \grad_\nu \xi_\mu.
\end{equation}

Stokes' theorem reads
\begin{equation}
	\label{app:eq:stokes-thm}
	\int_V \dd^D x \, \grad_\mu v^\mu
		= \oint_{\pd V} \dd \Sigma_\mu v^\mu,
	\qquad
	\dd \Sigma_\mu
		:= \epsilon \, n_\mu \, \dd^{D-1} \Sigma,
\end{equation}
where $V$ is a spacetime region, $S = \pd V$ its boundary and $\dd^{D-1} \Sigma$ the induced integration measure.
The vector $n_\mu$ normal to $S$ points outward and $\epsilon := n_\mu n^\mu = 1$ ($-1$) if $S$ is timelike (spacelike).
If the surface is defined by $x^0 = \cst$, then
\begin{equation}
	\dd^{D-1} \Sigma
		= \sqrt{g} \, \dd^{D-1} x,
	\qquad
	n_\mu = \delta_\mu^0.
\end{equation}

\index{conserved charge}%
We can write a generalization of \eqref{app:eq:charge} for a charge associated to a general surface $S$:
\begin{equation}
	\label{app:eq:charge-surface}
	Q_S
		= \frac{1}{\lambda} \int_{S} \dd \Sigma_\mu \, J_a^\mu.
\end{equation}
If the current $J_a^\mu$ is conserved, $\grad_\mu J_a^\mu = 0$ (no source), Stokes' theorem \eqref{app:eq:stokes-thm} shows that the charge vanishes $Q_S = 0$ if $S$ is a closed surface and that it is conserved $Q_{S_1} = - Q_{S_2}$ for two spacelike surfaces $S_1$ and $S_2$ extending to infinity (if $J_a^\mu$ vanishes at infinity)
(see~\cites[chap.~3]{Poisson:2007:RelativistsToolkitMathematics}[sec.~8.4]{Zwiebach:2009:FirstCourseString} for more details).

\subsection{Two dimensions}

Stokes' theorem \eqref{app:eq:stokes-thm} on flat space reads
\begin{equation}
	\label{app:eq:stokes-thm-2d}
	\int \dd^2 x \, \pd_\mu v^\mu
		= \oint \epsilon_{\mu\nu} \, \dd x^\nu v^\mu
		= \oint (v^0 \dd \sigma - v^1 \dd \tau),
\end{equation}
since $\dd \Sigma_\mu = \epsilon_{\mu\nu} \dd x^\nu$.

The integral of the curvature is a topological invariant
\begin{equation}
	\label{app:eq:euler-number-gb}
	\begin{aligned}
	\chi_{g;b}
		&
		:= \frac{1}{4\pi} \int \dd^2 \sigma \sqrt{g}\, R + \frac{1}{2\pi} \oint \dd s \, k
		\\ &
		= 2 - 2 g - b,
	\end{aligned}
\end{equation}
called the Euler characteristics and where $g$ is the number of holes and $b$ the number of boundaries.

\section{Conformal field theory}
\label{app:chap:formulas:cft}

In two dimensions, the energy--momentum tensor is defined by
\begin{equation}
	T_{ab}
		= - \frac{4\pi}{\sqrt{g}} \, \frac{\delta S}{\delta g^{ab}}.
\end{equation}

\subsection{Complex plane}

Defining the real coordinates $(x, y)$ from the complex coordinate on the complex plane
\begin{equation}
	z = x + \I y,
	\qquad
	z = x - \I y,
\end{equation}
we have the formulas:
\begin{subequations}
\begin{gather}
	\dd s^2
		= \dd x^2 + \dd y^2
		= \dd z \dd \bar{z},
	\qquad
	g_{z \bar z}
		= \frac{1}{2},
	\qquad
	g_{zz}
		= g_{\bar z \bar z}
		= 0,
	\\
	\epsilon_{z \bar z}
		= \frac{\I}{2},
	\qquad
	\epsilon^{z \bar z}
		= - 2 \I,
	\\
	\pd := \pd_{z}
		= \frac{1}{2} \, (\pd_x - \I \pd_y),
	\qquad
	\bar\pd := \pd_{\bar z}
		= \frac{1}{2} \, (\pd_x + \I \pd_y),
	\\
	V^{z} = V^x + \I V^y,
	\qquad
	V^{\bar z} = V^x - \I V^y,
	\\
	\dd^2 x = \dd x \dd y
		= \frac{1}{2} \, \dd^2 z,
	\qquad
	\dd^2 z = \dd z \dd \bar{z},
	\\
	\delta(z) = \frac{1}{2} \, \delta^{(2)}(x),
	\qquad
	1
		= \int \dd^2 z \, \delta^{(2)}(z)
		= \int \dd^2 x \, \delta^{(2)}(x),
	\\
	\label{app:eq:stokes-thm-2d-complex}
	\int_R \dd^2 z \, (\pd_z v^z + \pd_{\bar z} v^{\bar z})
		= - \I \oint_{\pd R} \big( \dd z \, v^{\bar z} - \dd \bar z v^{z} \big)
		= - 2 \I \oint_{\pd R} (v_z \dd z - v_{\bar z} \dd \bar z).
\end{gather}
\end{subequations}

\subsection{General properties}

A primary holomorphic field $\phi(z)$ of weight $h$ transforms as
\begin{equation}
	f \circ \phi(z)
		= \left( \frac{\dd f}{\dd z} \right)^{h} \phi\big( f(z) \big)
\end{equation}
for any local change of coordinates $f$.
A quasi-primary operator transforms like this only for $f \in \group{SL}(2, \C)$.
Its mode expansion reads
\begin{equation}
	\phi(z) = \sum_{n} \frac{\phi_n}{z^{n + h}},
	\qquad
	\phi_n = \oint_{C_0} \frac{\dd z}{2\pi \I} \, z^{n + h - 1} \phi(z),
\end{equation}
where the integration is counter-clockwise around the origin.

The $\group{SL}(2, \C)$ vacuum $\ket{0}$ is defined by
\begin{equation}
	\forall n \ge - h + 1:
	\quad
	\phi_n \ket{0} = 0.
\end{equation}
Its BPZ conjugate $\bra{0}$ satisfies:
\begin{equation}
	\forall n \le h - 1:
		\quad
		\bra{0} \phi_n = 0.
\end{equation}

The state--operator correspondence associates a state $\ket{\phi}$ to each operator $\phi(z)$:
\begin{equation}
	\ket{\phi}
		:= \phi(0) \ket{0}
		= \phi_{-h} \ket{0}.
\end{equation}
The operator corresponding to the vacuum is the identity $1$.\footnotemark{}
\footnotetext{%
	Exceptionally, the state $\ket{0}$ and the operator $1$ does not have the same symbol.
}%
The Hermitian and BPZ conjugated states are
\begin{equation}
	\bra{\eadj{\phi}}
		:= \bra{0} I \circ \adj\phi(0)
		= \lim_{z \to \infty} z^{2h} \bra{0} \adj\phi(z),
	\qquad
	\bra{\phi}
		:= \bra{0} I^\pm \circ \phi(0)
		= (\pm 1)^h \lim_{z \to \infty} z^{2h} \bra{0} \phi(z).
\end{equation}

The energy--momentum tensor is a quasi-primary operator of weight $h = 2$
\begin{equation}
	T(z) = \sum_{n} \frac{L_n}{z^{n + 2}}.
\end{equation}

The OPE between $T$ and a primary operator $h$ of weight $h$ is
\begin{equation}
	T(z) \phi(w) \sim \frac{h \, \phi(w)}{(z - w)^2} + \frac{\pd \phi(w)}{z - w}.
\end{equation}
The OPE of $T$ with itself defines the central charge $c$
\begin{equation}
	T(z) T(w) \sim \frac{c / 2}{(z - w)^4} + \frac{2 T(w)}{(z - w)^2} + \frac{\pd T(w)}{z - w}.
\end{equation}

\subsection{Hermitian and BPZ conjugations}

Both conjugations do not change the ghost number of a state.

\subsubsection{Hermitian}

The Hermitian conjugate of a general state built from $n$ operators $A_i$ and a complex number $\lambda$ is
\begin{equation}
	\adj{(\lambda \, A_1 \cdots A_n \ket{0})}
		= \conj{\lambda} \, \bra{0} \adj{A_n} \cdots \adj{A_1}.
\end{equation}

\begin{check}
The Hermitian conjugation defines an anti-linear inner product
\begin{equation}
	\psp{A}{B} = \bracket{\eadj{A}}{B},
	\qquad
	\adj{\psp{A}{B}} = - \psp{B}{A}.
\end{equation}
It has the properties
\begin{equation}
	\psp{A}{B + \lambda C}
		= \psp{A}{B} + \lambda \psp{A}{C},
	\qquad
	\psp{B + \lambda C}{A}
		= \psp{B}{A} + \conj{\lambda} \psp{C}{A}.
\end{equation}

The Hermitian adjoint of an operator can then be defined as
\begin{equation}
	\psp{A}{\mc O B} = \psp{\adj{\mc O} A}{B}.
\end{equation}
\end{check}

\subsubsection{BPZ}

The BPZ conjugate of modes is
\begin{equation}
	\phi_n^t = (I^\pm \circ \phi)_n
		= (- 1)^{h} (\pm 1)^{n} \phi_{-n},
\end{equation}
where $I^\pm(z) = \pm 1/z$.
The plus sign is usually used for the closed string, and the minus sign for the open string.
Given a general state built from $n$ operators $n$ and a complex number $\lambda$, the conjugation does not change the order of the operators and does not conjugate complex numbers:
\begin{equation}
	(\lambda \, A_1 \cdots A_n \ket{0})^t
		= \lambda \, \bra{0} (A_1)^t \cdots (A_n)^t.
\end{equation}
However, it reverses radial ordering such that operators must be (anti-)commuted in radial ordered expressions.

The BPZ product satisfies
\begin{equation}
	\label{app:eq:cft-bpz-product-prop}
	\mean{A, B}
		= (-1)^{\abs{A} \abs{B}} \mean{B, A}.
\end{equation}
Moreover the inner product is non-degenerate, so
\begin{equation}
	\forall A: \quad
		\bracket{A}{B} = 0
	\quad \Longrightarrow \quad
	\ket{B} = 0.
\end{equation}

Denoting by $\{ \ket{\phi_r} \}$ a complete basis of states, then the conjugate basis $\{ \bra{\phi_r^c} \}$ is defined by the BPZ product as
\begin{equation}
	\bracket{\phi_r^c}{\phi_s} = \delta_{rs}.
\end{equation}
We have
\begin{equation}
	\bracket{\phi_r}{\phi_s^c} = (-1)^{\abs{\phi_r}} \delta_{rs}.
\end{equation}

\subsection{Scalar field}

The simplest matter CFT is a set of $D$ scalar field $X^\mu(z, \bar z)$ such that the $\I \pd X^\mu$ and $\I \bar\pd X^\mu$ are of weight $h = (1, 0)$ and $h = (0, 1)$
\begin{equation}
	\I \pd X^\mu
		= \sum_{n} \frac{\alpha^\mu_n}{z^{n+1}},
	\qquad
	\I \bar\pd X^\mu
		= \sum_{n} \frac{\bar\alpha^\mu_n}{\bar z^{n+1}}.
\end{equation}
The commutation relations between the modes are:
\begin{equation}
	\com{\alpha^\mu_m}{\alpha^\nu_n} = m \delta_{m+n,0} \eta^{\mu\nu},
	\qquad
	\com{\bar\alpha^\mu_m}{\bar\alpha^\nu_n} = m \delta_{m+n,0} \eta^{\mu\nu},
	\qquad
	\com{\alpha^\mu_m}{\bar\alpha^\nu_n} = 0.
\end{equation}
The zero-modes of both operators are equal and correspond to the (centre-of-mass) momentum
\begin{equation}
	\alpha^\mu_0
		= \bar\alpha^\mu_0
		= \sqrt{\frac{\alpha'}{2}} \, p^\mu.
\end{equation}
The conjugate of $p^\mu$ is the centre-of-mass position $x^\mu$:
\begin{equation}
	\com{x^\mu}{p^\nu} = \eta^{\mu\nu}.
\end{equation}

Vertex operators are defined by
\begin{equation}
	V_k(z, \bar z)
		= \norder{\e^{\I k \cdot X(z, \bar z)}},
	\qquad
	h = \bar h
		= \frac{\alpha'^2 k^2}{4}.
\end{equation}

The scalar vacuum $\ket{k}$ is annihilated by all positive-frequency oscillators and it is characterized by its eigenvalue for the zero-mode operator
\begin{equation}
	p^\mu \ket{k}
		= k^\mu \ket{k},
	\qquad
	\forall n > 0:
	\quad
	\alpha^\mu_n \ket{k}
		= 0,
	\quad
	\bar\alpha^\mu_n \ket{k}
		= 0.
\end{equation}
The vacuum is associated to the vertex operator $V_k$:
\begin{equation}
	\ket{k} = V_k(0, 0) \ket{0}
		= \e^{\I k \cdot x} \ket{0}.
\end{equation}
The conjugate vacuum is
\begin{equation}
	\bra{k} p^\mu
		= \bra{k} k^\mu,
	\qquad
	\bra{k}
		= \adj{\ket{k}},
	\qquad
	\bra{-k}
		= \ket{k}^t.
\end{equation}

\subsection{Reparametrization ghosts}
\label{app:chap:formulas:cft:bc}

The reparametrization ghosts are described by an anti-commuting first-order system with the parameters (\Cref{cft:chap:systems,cft:tab:ghost-summary}):
\begin{equation}
	\epsilon
		= 1,
	\qquad
	\lambda
		= 2,
	\qquad
	c_{\text{gh}}
		= - 26,
	\qquad
	q_{\text{gh}}
		= - 3,
	\qquad
	a_{\text{gh}}
		= - 1.
\end{equation}
We focus on the holomorphic sector.

The $b$ and $c$ ghosts have weights:
\begin{equation}
	h(b) = 2,
	\qquad
	h(c) = - 1
\end{equation}
such that the mode expansions are:
\begin{subequations}
\begin{gather}
	\label{app:eq:ghost-exp-bc}
	b(z)
		= \sum_{n \in \Z} \frac{b_n}{z^{n + 2}},
	\qquad
	c(z)
		= \sum_{n \in \Z} \frac{c_n}{z^{n - 1}},
	\\
	\label{app:eq:ghost-modes-int}
	b_n
		= \oint \frac{\dd z}{2\pi \I} \, z^{n + 1} b(z),
	\qquad
	c_n
		= \oint \frac{\dd z}{2\pi \I} \, z^{n - 2} c(z).
\end{gather}
\end{subequations}
The anti-commutators between the modes $b_n$ and $c_n$ read:
\begin{equation}
	\label{app:eq:ghost-com-bn-cn}
	\anticom{b_m}{c_n}
		= \delta_{m+n,0},
	\qquad
	\anticom{b_m}{b_n}
		= 0,
	\qquad
	\anticom{c_m}{c_n}
		= 0.
\end{equation}

The energy--momentum tensor and the Virasoro modes are respectively:
\begin{subequations}
\begin{gather}
	T
		= - 2 \, \norder{b \pd c} - \norder{\pd b \, c},
	\\
	\label{app:eq:ghost-Ln}
	L_m
		= \sum_{n} \big( n + m \big) \, \norder{b_{m-n} c_n}
		= \sum_{n} (2 m - n) \, \norder{b_{n} c_{m-n}}.
\end{gather}
\end{subequations}
The expression of the zero-mode is:
\begin{equation}
	\label{app:eq:ghost-L0}
	L_0
		= - \sum_{n} n\, \norder{b_{n} c_{-n}}
		= \sum_{n} n\, \norder{b_{-n} c_{n}}.
\end{equation}
The commutators between the $L_n$ and the ghost modes are:
\begin{equation}
	\label{app:eq:com-Ln-bn-cn}
	\com{L_m}{b_n}
		= \big( m - n \big) b_{m+n},
	\qquad
	\com{L_m}{c_n}
		= - (2 m + n) c_{m+n}.
\end{equation}
In particular, $L_0$ commutes with the zero-modes:
\begin{equation}
	\label{app:eq:com-L0-b0-c0}
	\com{L_0}{b_0}
		= 0,
	\qquad
	\com{L_0}{c_0}
		= 0.
\end{equation}

The anomalous global $\group{U}(1)$ symmetry for the ghost number $N_{\text{gh}}$ is generated by the ghost current:
\begin{equation}
	\label{app:eq:ghost-current}
	j
		= - \norder{b c},
	\qquad
	N_{\text{gh},L}
		= \oint \frac{\dd z}{2\pi \I} \, j(z),
\end{equation}
such that
\begin{equation}
	\label{app:eq:ghost-number-bc-plane}
	N_{\text{gh}}(c) = 1,
	\qquad
	N_{\text{gh}}(b) = - 1.
\end{equation}
Remember that $N_{\text{gh}} = N_{\text{gh},L}$ in the left sector, such that we omit the index $L$.
The modes of the ghost current are
\begin{equation}
	\label{app:eq:ghost-jn}
	j_m
		= - \sum_{n} \norder{b_{m-n} c_{n}}
		= - \sum_{n} \norder{b_{n} c_{m-n}},
	\qquad
	N_{\text{gh},L}
		= j_0
		= - \sum_{n} \norder{b_{-n} c_{n}}.
\end{equation}
The commutator of the current modes with itself and with the Virasoro modes are:
\begin{equation}
	\com{j_m}{j_n}
		= m \, \delta_{m+n,0},
	\qquad
	\com{L_m}{j_n}
		= - n j_{m+n} - \frac{3}{2} \, m (m + 1) \delta_{m+n,0}.
\end{equation}
Finally, the commutators of the ghost number operator are:
\begin{equation}
	\label{app:eq:ghost-com-bc-Ngh}
	\com{N_{\text{gh}}}{b(w)} = - b(w),
	\qquad
	\com{N_{\text{gh}}}{c(w)} = c(w).
\end{equation}

The level operators $N^b$ and $N^c$ and number operators $N^b_n$ and $N^c_n$ are defined as:
\begin{subequations}
\begin{gather}
	\label{app:eq:ghost-N}
	N^b
		= \sum_{n > 0} n \, N^b_n,
	\qquad
	N^c
		= \sum_{n > 0} n \, N^c_n,
	\\
	\label{app:eq:ghost-Nn}
	N^b_n
		= \norder{b_{-n} c_n},
	\qquad
	N^c_n
		= \norder{c_{-n} b_n}.
\end{gather}
\end{subequations}
The commutator of the number operators with the modes are:
\begin{equation}
	\com{N^b_m}{b_{-n}}
		= b_{-n} \delta_{m,n},
	\qquad
	\com{N^c_m}{c_{-n}}
		= c_{-n} \delta_{m,n}.
\end{equation}

The OPE between the ghosts and different currents are:
\begin{subequations}
\label{app:eq:ghost-ope}
\begin{gather}
	c(z) b(w)
		\sim \frac{1}{z - w},
	\qquad
	b(z) c(w)
		\sim \frac{1}{z - w},
	\qquad
	b(z) b(w)
		\sim 0,
	\qquad
	c(z) c(w)
		\sim 0,
	\\
	T(z) b(w)
		\sim \frac{2 b(w)}{(z - w)^2} + \frac{\pd b(w)}{z - w},
	\qquad
	T(z) c(w)
		\sim \frac{- c(w)}{(z - w)^2} + \frac{\pd c(w)}{z - w}.
	\\
	j(z) b(w)
		\sim - \frac{b(w)}{z - w},
	\qquad
	j(z) c(w)
		\sim \frac{c(w)}{z - w}.
	\qquad
	j(z) \mc O(w)
		\sim N_{\text{gh}}(\mc O) \, \frac{\mc O(w)}{z - w},
	\\
	j(z) j(w)
		\sim \frac{1}{(z - w)^2}.
	\\
	\label{app:eq:ghost-ope-T-j}
	T(z) j(w)
		\sim \frac{- 3}{(z - w)^3}
			+ \frac{j(w)}{(z - w)^2}
			+ \frac{\pd j(w)}{z - w}.
\end{gather}
\end{subequations}
any operator $\mc O(z)$ is defined by

The OPE \eqref{app:eq:ghost-ope-T-j} implies that the ghost number is not conserved on a curved space:
\begin{equation}
	N^c - N^b
		= 3 - 3 g,
\end{equation}
and leads to a shift between the ghost numbers on the plane and on the cylinder:
\begin{equation}
	\label{app:eq:ghost-number-cyl}
	N_{\text{gh},L}
		= N_{\text{gh},L}^{\text{cyl}} + \frac{3}{2}.
\end{equation}

The $\group{SL}(2, \C)$ vacuum $\ket{0}$ is defined by
\begin{equation}
	\label{app:eq:ghost-vac-ann}
	\forall n > - 2:
		\quad
		b_n \ket{0} = 0,
	\qquad
	\forall n > 1:
		\quad
		c_n \ket{0} = 0.
\end{equation}
The mode $c_1$ does not annihilate the vacuum and the two degenerate energy vacua are:
\begin{equation}
	\label{app:eq:ghost-vacuum-ud}
	\ket{\downarrow}
		:= c_1 \ket{0},
	\qquad
	\ket{\uparrow}
		:= c_0 c_1 \ket{0}.
\end{equation}
The zero-point energy of these states is:
\begin{equation}
	L_0 \ket{\downarrow}
		= a_{\text{gh}} \ket{\downarrow},
	\qquad
	L_0 \ket{\uparrow}
		= a_{\text{gh}} \ket{\uparrow},
	\qquad
	a_{\text{gh}}
		= - 1.
\end{equation}

The energy for the normal ordering of the different currents is:
\begin{subequations}
\begin{gather}
	\label{app:eq:ghost-Lm-ev}
	L_m
		= \sum_{n} \big(n - (1 - \lambda) m \big) \, \norderv{b_{m-n} c_n}
			+ a_{\text{gh}} \, \delta_{m,0},
	\\
	\label{app:eq:ghost-jnm-ev}
	j_m
		= \sum_{n} \, \norderv{b_{m-n} c_n}
			+ \delta_{m,0}.
\end{gather}
\end{subequations}

The energy--momentum and ghost current zero-modes are explicitly:
\begin{subequations}
\begin{gather}
	\label{app:eq:ghost-L0-ev}
	L_0
		= \sum_{n} n \, \norderv{b_{-n} c_n} + a_{\text{gh}}
		= \what L_0 - 1,
	\\
	\label{app:eq:ghost-jn0-ev}
	N_{\text{gh},L}
		= j_0
		= \sum_{n} \norderv{b_{-n} c_n}
			 + 1
		= \what N_{\text{gh}, L}
			+ \frac{1}{2} \, \big(N^c_0 - N^b_0 \big)
			- \frac{3}{2},
	\\
	\what L_0
		= N^b + N^c,
	\qquad
	\what N_{\text{gh}, L}
		:= \sum_{n > 0} \big(N^c_n - N^b_n \big).
\end{gather}
\end{subequations}

Then, one can straightforwardly compute the ghost number of the vacua:
\begin{equation}
	N_{\text{gh}} \ket{0}
		= 0,
	\qquad
	N_{\text{gh}} \ket{\downarrow}
		= \ket{\downarrow},
	\qquad
	N_{\text{gh}} \ket{\uparrow}
		= 2 \ket{\uparrow}.
\end{equation}
Using \eqref{app:eq:ghost-number-cyl} allows to write the ghost numbers on the cylinder:
\begin{equation}
	N_{\text{gh}}^{\text{cyl}} \ket{\downarrow}
		= - \frac{1}{2} \ket{\downarrow},
	\qquad
	N_{\text{gh}}^{\text{cyl}} \ket{\uparrow}
		= \frac{1}{2} \ket{\uparrow}.
\end{equation}

The $b_n$ and $c_n$ are Hermitian:
\begin{equation}
	\label{app:eq:ghost-adj-bn-cn}
	\adj{b_n}
		= b_{-n},
	\qquad
	\adj{c_n}
		= c_{-n}.
\end{equation}
The BPZ conjugates of the modes are:
\begin{equation}
	b_n^t
		= (\pm 1)^n b_{-n},
	\qquad
	c_n^t
		= - (\pm 1)^n c_{-n},
\end{equation}
using $I^\pm(z)$ with \eqref{cft:eq:modes-bpz}.

The conjugates of the vacuum read:
\begin{equation}
	\eadj{\ket{\downarrow}}
		= \bra{0} c_{-1},
	\qquad
	\eadj{\ket{\uparrow}}
		= \bra{0} c_{-1} c_0.
\end{equation}
The BPZ conjugates of the vacua are:
\begin{equation}
	\bra{\downarrow}
		:= \ket{\downarrow}^t
		= \mp \bra{0} c_{-1},
	\qquad
	\bra{\uparrow}
		:= \ket{\uparrow}^t
		= \pm \bra{0} c_0 c_{-1}.
\end{equation}
We have the following relations:
\begin{equation}
	\label{app:eq:ghost-vac-conj-relations}
	\bra{\downarrow}
		= \mp \eadj{\ket{\downarrow}},
	\qquad
	\bra{\uparrow}
		= \mp \eadj{\ket{\uparrow}}.
\end{equation}

The ghost are normalized with
\begin{equation}
	\bracket{\uparrow}{\downarrow}
		= \bra{\downarrow} c_0 \ket{\downarrow}
		= \bra{0} c_{-1} c_0 c_1 \ket{0}
		= 1,
\end{equation}
which selects the minus sign in the BPZ conjugation.
The conjugate of the ghost vacuum is
\begin{equation}
	\bra{0^c} = \bra{0} c_{-1} c_0 c_1.
\end{equation}

Considering both the holomorphic and anti-holomorphic sectors, we introduce the combinations:
\begin{equation}
	\label{app:eq:bn-cn-pm}
	b_n^\pm = b_n \pm \bar b_n,
	\qquad
	c_n^\pm = \frac{1}{2} \, (c_n \pm \bar c_n).
\end{equation}
The normalization of $b_m^\pm$ is chosen to match the one of $L_m^\pm$ \eqref{app:eq:Ln-pm}, and the one of $c_m^\pm$ such that
\begin{equation}
	\label{app:eq:ghost-com-bc-pm}
	\anticom{b_m^+}{c_n^+}
		= \delta_{m+n},
	\qquad
	\anticom{b_m^-}{c_n^-}
		= \delta_{m+n}.
\end{equation}
We have the following useful identities:
\begin{equation}
	b_n^- b_n^+
		= 2 b_n \bar b_n,
	\qquad
	c_n^- c_n^+
		= \frac{1}{2} \, c_n \bar c_n.
\end{equation}

\section{Bosonic string}

The BPZ conjugates of the scalar and ghost modes are
\begin{equation}
	(\alpha_n)^t = - (\pm 1)^n \, \alpha_{-n},
	\qquad
	(b_n)^t = (\pm 1)^n \, b_{-n},
	\qquad
	(c_n)^t = - (\pm 1)^n \, c_{-n}.
\end{equation}

Combinations of holomorphic and anti-holomorphic modes:
\begin{equation}
	\label{app:eq:Ln-pm}
	L_n^\pm = L_n \pm \bar L_n,
	\qquad
	b_n^\pm = b_n \pm \bar b_n,
	\qquad
	c_n^\pm = \frac{1}{2} \, (c_n \pm \bar c_n).
\end{equation}

The closed string inner product is defined from the BPZ product by an additional insertion of $c_0^-$
\begin{equation}
	\psbr{A}{B} = \bra{A} c_0^- \ket{B},
\end{equation}
while the open string inner product is equal to the BPZ product
\begin{equation}
	\psbr{A}{B} = \bracket{A}{B}.
\end{equation}

The vacuum for the matter and ghosts is
\begin{equation}
	\ket{k, 0}
		:= \ket{k} \otimes \ket{0},
	\qquad
	\ket{k, \downarrow}
		:= \ket{k} \otimes \ket{\downarrow}.
\end{equation}
The vacuum is normalized as
\begin{subequations}
\begin{align}
	\text{open:}&
		\quad
		\bra{k, \downarrow} c_0 \ket{k, \downarrow}
			= \bra{k', 0} c_{-1} c_0 c_1 \ket{k, 0}
			= (2\pi)^D \delta^{(D)}(k + k'),
	\\
	\text{closed:}&
		\quad
		\bra{k, \downarrow\downarrow} c_0 \bar c_0 \ket{k, \downarrow\downarrow}
			= \bra{k', 0} c_{-1} \bar c_{-1} c_0 \bar c_0 c_1 \bar c_1 \ket{k, 0}
			= (2\pi)^D \delta^{(D)}(k + k'),
\end{align}
\end{subequations}

\begin{check}
\subsection{Closed string}
\end{check}

\begin{check}
\subsection{Open string}
\end{check}

\chapter{Quantum field theory}

In this appendix, we gather useful information on quantum field theories.
The first section describes how to compute with path integral with non-trivial measures, generalizing techniques from finite-dimensional integrals.
Then, we summarize the important concepts from the BRST and BV formalisms.

\section{Path integrals}
\label{sec:form:path-integrals}

In this section, we explain how analysis, algebra and differential geometry are generalized to infinite-dimensional vector spaces (fields).

\subsection{Integration measure}

\index{field space!inner-product}%
In order to construct a path integral for the field $\Phi$, one needs to define a notion of distance on the space of fields.
The distance between a field $\Phi$ and a neighbouring field $\Phi + \delta\Phi$ is
\begin{equation}
	\abs{\delta\Phi}^2
		= G(\Phi)(\delta\Phi, \delta\Phi),
\end{equation}
where $G$ is the (field-dependent) metric on the field tangent space (the field dependence will be omitted when no confusion is possible).
\index{field space!norm}%
This induces a metric on the field space itself
\begin{equation}
	\abs{\Phi}^2 = G(\Phi)(\Phi, \Phi),
\end{equation}
from which the integration measure over the field space can be defined as
\begin{equation}
	\label{math:eq:path-int-measure}
	\dd \Phi \sqrt{\det G(\Phi)}.
\end{equation}
Moreover, the field metric also defines an inner-product between two different elements of the tangent space or field space:
\begin{equation}
	\psp{\delta\Phi_1}{\delta\Phi_2}
		= G(\Phi)(\delta\Phi_1, \delta\Phi_2),
	\qquad
	\psp{\Phi_1}{\Phi_2}
		= G(\Phi)(\Phi_1, \Phi_2).
\end{equation}

\begin{remark}[Metric in component form]
	If one has a set of spacetime fields $\Phi_a(x)$, then a local norm is defined by
	\begin{equation}
		\abs{\delta\Phi_a}^2 = \int \dd x \, \rho(x) \gamma_{ab}\big(\Phi(x)\big) \delta\Phi_a(x) \delta\Phi_b(x),
	\end{equation}
	which means that the metric in component form is
	\begin{equation}
		G_{ab}(x, y)(\Phi) = \delta(x - y) \rho(x) \gamma_{ab}\big(\Phi(x)\big).
	\end{equation}
	Locality means that all fields are evaluated at the same point.
	On a curved space, it is natural to write $\gamma$ only in terms of the metric $g$ and to set $\rho(x) = \sqrt{\det g(x)}$, such that the inner-product is diffeomorphism invariant.
\end{remark}

\index{path integral!measure}%
Since a Gaussian integral is proportional to the squareroot of the operator determinant, the integration measure can be determined by considering the Gaussian integral over the tangent space:
\begin{equation}
	\int \dd \delta\Phi \, \e^{- G(\Phi)(\delta\Phi, \delta\Phi)}
		= \frac{1}{\sqrt{\det G(\Phi)}}.
\end{equation}
Note that one needs to work on the tangent space because $G(\Phi)$ can depend on the field, which means that the integral
\begin{equation}
	\int \dd \Phi \, \e^{- G(\Phi)(\Phi, \Phi)}.
\end{equation}
is not Gaussian.

Having constructed the Gaussian measure with respect to the metric $G(\Phi)$, it is now possible to consider the path integral of general functional $F$ of the fields:
\begin{equation}
	\int \dd \Phi \, \sqrt{\det G(\Phi)} \, F(\Phi).
\end{equation}
The (effective) action $S(\Phi)$ provides a natural metric on the field space by defining $\sqrt{\det G} = \e^{- S}$, or
\begin{equation}
	S = - \frac{1}{2} \tr \ln G(\Phi).
\end{equation}
However, it can be simpler to work with a Gaussian measure by considering only the quadratic terms in $S$, and expanding the rest in a power series.
In particular, the partition function is defined from the classical action $S_{\text{cl}}$ by
\begin{equation}
	Z = \int \dd \Phi \, \e^{- S_{cl}(\Phi)}.
\end{equation}

Given an operator $D$, its adjoint $\adj D$ is defined with respect to the metric as
\begin{equation}
	G(\delta\Phi, D \delta\Phi) = G( \adj{D} \delta\Phi, \delta\Phi).
\end{equation}

\index{path integral!measure!free-field}%
The \emph{free-field measure} is such that the metric on the field space is independent from the field itself: $G(X) = G_0$.
In particular, this implies that the metric is flat and its determinant can be absorbed in the measure, setting $\det G_0 = 1$.
In this case, the measure is invariant under shift of the field:
\begin{equation}
	\Phi \to \Phi + \varepsilon
\end{equation}
such that
\begin{equation}
	\label{math:eq:path-int-inv-shift}
	\int \dd \Phi\, \e^{- \frac{1}{2} \abs{\Phi + \varepsilon}^2}
		= \int \dd \Phi\, \e^{- \frac{1}{2} \abs{\Phi}^2}.
\end{equation}
This property allows to complete squares and shift integration variables (for example to generate a perturbative expansion and to derive the propagator).

\begin{computation}[math:eq:path-int-inv-shift]
	\begin{equation}
		\int \dd \Phi\, \e^{- \frac{1}{2} \abs{\Phi + \varepsilon}^2}
			= \int \dd \wtilde \Phi \, \det\frac{\delta \Phi}{\delta \wtilde \Phi}\, \e^{- \frac{1}{2} \abs{\wtilde \Phi}^2} \\
			= \int \dd \wtilde \Phi\, \e^{-\frac{1}{2} \abs{\wtilde \Phi}^2}
	\end{equation}
	The first equality follows by setting $\wtilde \Phi = \Phi + \varepsilon$, and the result \eqref{math:eq:path-int-inv-shift} follows by the redefinition $\wtilde \Phi = \Phi$.
\end{computation}

\subsection{Field redefinitions}

\index{path integral!field redefinition}%
Under a field redefinition $\Phi \to \Phi'$, the norm and the measure are invariant:
\begin{equation}
	\dd \Phi \sqrt{\det G(\Phi)}
		= \dd \wtilde \Phi \sqrt{\det \wtilde G(\wtilde \Phi)},
	\qquad
	G(\Phi)(\delta\Phi, \delta\Phi)
		= \wtilde G(\wtilde \Phi)(\delta \wtilde \Phi, \delta \wtilde \Phi).
\end{equation}
Conversely, one can find the Jacobian $J(\Phi, \wtilde \Phi)$ between two coordinate systems by writing
\begin{equation}
	\dd\Phi
		= J(\Phi, \wtilde \Phi) \dd \wtilde \Phi,
	\qquad
	J(\Phi, \wtilde \Phi)
		= \Abs{\det \frac{\pd \Phi}{\pd \wtilde \Phi}}
		= \sqrt{\frac{\det \wtilde G(\wtilde \Phi)}{\det G(\Phi)}}.
\end{equation}
If the measure of the initial field coordinate is normalized such that $\det G = 1$, or equivalently
\begin{equation}
	\int \dd \delta \Phi \, \e^{- \abs{\delta\Phi}^2} = 1,
\end{equation}
one can determine the Jacobian by performing explicitly the integral
\begin{equation}
	J(\wtilde \Phi)^{-1} = \int \dd \delta \wtilde \Phi \, \e^{- \wtilde G(\delta \wtilde \Phi, \delta \wtilde \Phi)}.
\end{equation}

\begin{remark}[Identity of the Jacobian for $\Phi$ and $\delta\Phi$]
	The Jacobian agrees on the space of fields and on its tangent space.
	This is most simply seen by using a finite-dimensional notation: considering the coordinates $x^\mu$ and a vector $v = v^\mu \pd_\mu$, the Jacobian for changing the coordinates to $\tilde x^\mu$ is equivalently
	\begin{equation}
		J = \det \frac{\pd \tilde x^\mu}{\pd x^\mu}
			= \det \frac{\pd \tilde v^\mu}{\pd v^\mu}
	\end{equation}
	since the vector transforms as
	\begin{equation}
		\tilde v^\mu = v^\nu \, \frac{\pd \tilde x^\mu}{\pd x^\nu}.
	\end{equation}
\end{remark}

\begin{check}

In order to be more explicit, one can consider a change of variables
\begin{equation}
	\wtilde \Phi = F(\Phi),
	\qquad
	\Phi = F(\wtilde \Phi)^{-1}.
\end{equation}
where $F$ is invertible and can contain derivatives and be non-polynomial in $\Phi$.
The variation of both fields are connected by
\begin{equation}
	\delta \wtilde \Phi = F'(\Phi) \delta \Phi,
	\qquad
	\delta \Phi = F'\big(F(\wtilde \Phi)^{-1}\big)^{-1} \delta \wtilde \Phi,
\end{equation}
where $F'$ is the derivative of $F$ with respect to its argument.
It is then possible to compute the Jacobian of the transformation
\begin{equation}
	\label{math:eq:path-int-jacobian-F}
	J(\wtilde \Phi) = \sqrt{\det G'(\wtilde \Phi)}
		= \det F'\big(F(\wtilde \Phi)^{-1}\big)^{-1}
\end{equation}
where one has taken $\det G(\Phi) = 1$ to simplify the analysis.
This can be rewritten as an effective action
\begin{equation}
	J(\wtilde \Phi) = \e^{- S_{\text{eff}}(\wtilde \Phi)},
	\qquad
	S_{\text{eff}}(\wtilde \Phi) = \tr \ln F'\big(F(\wtilde \Phi)^{-1}\big).
\end{equation}
Considering a partition function with a generic action $S_0[\Phi]$, one gets
\begin{equation}
	Z = \int \dd\Phi \, \e^{- S_0[\phi]}
		= \int \dd\wtilde \Phi \, \e^{- S_0\big(F^{-1}(\wtilde \Phi)\big) - S_{\text{eff}}(\wtilde \Phi)}.
\end{equation}
One can imagine using a change of variable to remove derivatives in the action in order to get a purely polynomial action, but the complication would reappear in the form of the determinant.

\begin{computation}[math:eq:path-int-jacobian-F]
	\begin{align*}
		\frac{1}{\sqrt{\det \wtilde G(\wtilde \Phi)}}
			&= \int \dd \delta\wtilde \Phi \, \e^{- \wtilde G(\delta\wtilde \Phi, \delta\wtilde \Phi)}
			= \int \dd \delta\wtilde \Phi \, \e^{- G(\delta\Phi, \delta\Phi)}
			\\
			&= \int \dd \delta\wtilde \Phi \, \exp\left[ - G\Big( F'\big(F(\wtilde \Phi)^{-1}\big)^{-1} \delta\wtilde \Phi, F'\big(F(\wtilde \Phi)^{-1}\big)^{-1} \delta\wtilde \Phi \Big) \right]
			\\
			&= \int \dd \delta\wtilde \Phi \, \exp\left[ - G\Big( \delta\wtilde \Phi, F'\big(F(\wtilde \Phi)^{-1}\big)^{-1\dagger} F'\big(F(\wtilde \Phi)^{-1}\big)^{-1} \delta\wtilde \Phi \Big) \right]
			\\
			&= \frac{1}{\sqrt{\det G(\Phi)}} \left( \det F'\big(F(\wtilde \Phi)^{-1}\big)^{-1\dagger} F'\big(F(\wtilde \Phi)^{-1}\big)^{-1} \right)^{-1/2}.
	\end{align*}
	Assuming that the operator $F'^{-1} \circ F^{-1}$ and its adjoint have the same spectrum, one gets
	\begin{equation}
		\sqrt{\det \wtilde G(\wtilde \Phi)} = \det F'\big(F(\wtilde \Phi)^{-1}\big)^{-1}.
	\end{equation}
\end{computation}

These manipulations look complicated for something as simple as a change of variables, but this helps to ensure that one integrates over the correct variables.

\end{check}

\subsection{Zero-modes}
\label{sec:form:path-integrals:zero-mode}

\index{zero-mode}%
A zero-mode $\Phi_0$ of an operator $D$ is a field such that
\begin{equation}
	D \Phi_0 = 0.
\end{equation}

In the definition of the path integral over the space of fields $\Phi$, the measure is defined over the complete space.
However, this will lead respectively to a divergent or vanishing integral if the field is bosonic or fermionic, because the integration over the zero-modes can be factorized from the rest of the integral.
Writing the field as
\begin{equation}
	\Phi = \Phi_0 + \Phi',
	\qquad
	\psp{\Phi_0}{\Phi'}
		= 0,
\end{equation}
where $\Phi'$ is orthogonal to the zero-mode $\Phi_0$, a Gaussian integral of an operator $D$ reads:
\begin{equation}
	Z[D]
		= \int \dd \Phi \sqrt{\det G} \, \e^{- \frac{1}{2} \psp{\Phi}{D \Phi}}
		= \left( \int \dd \Phi_0 \right)
			\int \dd \Phi' \, \e^{- \frac{1}{2} \psp{\Phi'}{D \Phi'}}
\end{equation}
A first solution could be to simply strip the first factor (for example, by absorbing it in the normalization), but this is not satisfactory.
In particular, the partition function with source
\begin{equation}
	Z[D, J]
		= \int \dd \Phi \sqrt{\det G} \, \e^{- \frac{1}{2} \psp{\Phi}{D \Phi} - \psp{J}{\Phi}}
\end{equation}
will depend on the zero-modes through the sources.
But, since the zero-modes are still singled out, it is interesting to factorize the integration
\begin{equation}
	Z[D, J]
		= \int \dd \Phi_0 \, \e^{- \psp{J}{\Phi_0}}
			\int \dd \Phi' \, \e^{- \frac{1}{2} \psp{\Phi'}{D \Phi'} - \psp{J}{\Phi'}}
\end{equation}
and to understand what makes it finite.
Ensuring that zero-modes are correctly inserted is an important consistency and leads to powerful arguments.
Especially, this can help to guess an expression when it cannot be derived easily from first principles.

To exemplify the problem, consider the cases where there is a single constant zero-mode denoted as $x$ (bosonic) or $\theta$ (fermionic).
The integral over $x$ is infinite:
\begin{equation}
	\int \dd x = \infty.
\end{equation}
Oppositely, the integral of a Grassmann variable $\theta$ vanishes:
\begin{equation}
	\int \dd\theta = 0.
\end{equation}
A Grassmann integral satisfies also
\begin{equation}
	\int \dd\theta \, \theta
		= \int \dd\theta \, \delta(\theta)
		= 1,
\end{equation}
such that an integral over a zero-mode does not vanish if there one zero-mode in the integrand (due to the Grassmann nature of $\theta$, the integrand can be at most linear).
By analogy with the fermionic case, a possibility for getting a finite bosonic integral is to insert a delta function:
\begin{equation}
	\int \dd x \, \delta(x)
		= 1.
\end{equation}
We will see that this is exactly what happens for the ghosts and super-ghosts in (super)string theories.

Since $\ker D$ is generally finite-dimensional, it is interesting to decompose the zero-mode on a basis and to integrate over the coefficients in order to obtain a finite-dimensional integral.
Writing the zero-mode as
\begin{equation}
	\theta_0(x) = \theta_{0i} \psi_i(x),
	\qquad
	\ker D = \Span \{ \psi_i \}
\end{equation}
where the coefficients $\theta_{0i}$ are constant Grassmann numbers, the change of variables $\theta \to (\theta_{0i}, \theta')$ implies:
\begin{equation}
	\label{math:eq:odd-measure-change-zero}
	\dd \theta
		= \frac{1}{\sqrt{\det \psp{\psi_i}{\psi_j}}}
			\, \dd \theta' \prod_{i=1}^n \dd \theta_{0i},
\end{equation}
where $n = \dim \ker D$.

Next, according to the discussion above, one can ask if it is possible to rewrite an integration over $\dd \theta'$ in terms of an integration over $\dd\theta$ together with zero-mode insertions.
This is indeed possible and one finds:
\begin{equation}
	\label{math:eq:odd-measure-add-zero}
	\dd \theta \, \prod_{i=1}^{n} \theta(x_i)
		= \frac{\det \psi_i(x_j)}{\sqrt{\det \psp{\psi_i}{\psi_j}}}
			\, \dd \theta'.
\end{equation}

\begin{computation}[math:eq:odd-measure-change-zero]
	\begin{align*}
		1
			&
			= \int \dd \theta \, \e^{- \abs{\theta}^2}
			= \int \dd \theta' \dd \theta_0 \, \e^{- \abs{\theta}^2 - \abs{\theta_0}^2}
			\\
			&
			= J \int \dd \theta' \prod_i \dd \theta_{0i} \, \e^{- \abs{\theta'}^2 - \abs{\theta_{0i} \psi_i}^2}
			= J \sqrt{\det \psp{\psi_i}{\psi_j}}
	\end{align*}
\end{computation}

\begin{computation}[math:eq:odd-measure-add-zero]
	The simplest approach is to start with the LHS.
	This formula is motivated from the previous discussion: if the integration measure contains $n$ zero-modes, it will vanish unless there are $n$ zero-mode insertions.
	Moreover, one can replace each of them by the complete field since only the zero-mode part can contribute:
	\begin{align*}
		\int \dd \theta_0 \, \prod_{j=1}^{n} \theta(x_j)
			&
			= \int \dd \theta_0 \,
				\prod_{j=1}^{n} \theta_0(x_j)
			= \frac{1}{\sqrt{\det \psp{\psi_i}{\psi_j}}}
				\int \dd^n \theta_{0i} \,
				\prod_{j=1}^{n} \big[ \theta_{0i} \psi_i(x_j) \big]
			\\
			&
			= \frac{\det \psi_i(x_j)}{\sqrt{\det \psp{\psi_i}{\psi_j}}}
				\int \prod_i \dd \theta_{0i} \, \theta_{0i}
			= \frac{\det \psi_i(x_j)}{\sqrt{\det \psp{\psi_i}{\psi_j}}}.
	\end{align*}
	The third equality follows by developing the product and ordering the $\theta_{0i}$: minus signs result from anticommuting the $\theta_{0i}$ such that one gets the determinant of the basis elements.
\end{computation}

\section{BRST quantization}
\label{app:sec:qft:brst}

Consider an action $S_m[\phi^i]$ which depends on some fields $\phi^i$ subject to a gauge symmetry:
\begin{equation}
	\delta \phi^i = \epsilon^a \delta_a \phi^i
		= \epsilon^a R_a^i(\phi),
\end{equation}
where $\epsilon^a$ are the (local) bosonic parameters, such that the action is invariant
\begin{equation}
	\epsilon^a \delta_a S_m = 0.
\end{equation}
The gauge transformations form a Lie algebra with structure coefficients $f_{ab}^c$
\begin{equation}
	\com{\delta_a}{\delta_b} = f_{ab}^c \delta_c.
\end{equation}
It is important 1) that the algebra closes off-shell (without using the equations of motion), 2) that the structure coefficients are field independent and 3) that the gauge symmetry is irreducible (each gauge parameter is independent).

\begin{remark}[Interpretation of the $R_a^i$ matrices]
	If the $\phi^i$ transforms in a representation $\repr{R}$ of the gauge group, then the transformation is linear in the field
	\begin{equation}
		R_a^i(\phi) = \tensor{(T^{\repr{R}}_{a})}{^i_j} \phi^j,
	\end{equation}
	with $T^{\repr{R}}_{a}$ the generators in the representation $\repr{R}$.
	But, in full generality, this is not the case: for example the gauge fields $A^a_\mu$ do not transform in the adjoint representation even if they carry an adjoint index (only the field strength does), and in this case
	\begin{equation}
		R_{a\mu}^b = \delta_a^b \pd_\mu + f_{ab}^c A^b_\mu.
	\end{equation}

	When the fields $\phi^i$ form a non-linear sigma models, the $R_a^i(\phi)$ correspond to Killing vectors of the target manifold.
\end{remark}

In order to fix the gauge symmetry in the path integral
\begin{equation}
	Z = \Omega_{\text{gauge}}^{-1} \int \dd \phi^i \, \e^{- S_m},
\end{equation}
gauge fixing conditions must be imposed:
\begin{equation}
	F^A(\phi^i) = 0.
\end{equation}
Indeed, without gauge fixing, the integration is performed over multiple identical configurations and the result diverges.
The index $A$ is different from the gauge index $a$ because they can refer to different representations, but for the gauge fixing to be possible they should run over as many values.

\index{BRST quantization!Nakanishi--Lautrup auxiliary field}%
Next, ghost fields $c^a$ (fermionic) are introduced for every gauge parameter, anti-ghosts $b_A$ (fermionic) and auxiliary (Nakanishi--Laudrup) fields $B_A$ (bosonic) for every gauge condition.
The gauge-fixing and ghost actions are then defined by
\begin{subequations}
\begin{gather}
	S_{\text{gh}} = b_A c^a \, \delta_a F^A(\phi^i),
	\\
	S_{\text{gf}} = - \I \, B_A F^A(\phi^i)
\end{gather}
\end{subequations}
such that the original partition function is equivalent to
\begin{equation}
	Z = \int \dd\phi^i \, \dd b_A \, \dd c^a \, \dd B_A \, \e^{- S_{\text{tot}}}
\end{equation}
where
\begin{equation}
	S_{\text{tot}} = S_m + S_{\text{gf}} + S_{\text{gh}}.
\end{equation}

\index{BRST quantization!transformations}%
The total action is invariant
\begin{equation}
	\delta_\epsilon S_{\text{tot}} = 0.
\end{equation}
under the (global) BRST transformations
\begin{equation}
	\label{qft:eq:transf-brst}
	\delta_{\epsilon} \phi^i
		= \I \epsilon \, c^a \delta_a \phi^i,
	\qquad
	\delta_{\epsilon} c^a
		= - \frac{\I}{2} \, \epsilon \, f_{bc}^a c^b c^c,
	\qquad
	\delta_{\epsilon} b_A
		= \epsilon \, B_A,
	\qquad
	\delta_{\epsilon} B_A
		= 0,
\end{equation}
where $\epsilon$ is an anti-commuting constant parameter.
Note that the original action $S_m$ is invariant by itself since the transformation acts like a gauge transformation with parameter $\epsilon c^a$.
The transformation of $c^a$ follows because it transforms in the adjoint representation of the gauge group.
Direct computations show that this transformation is nilpotent
\begin{equation}
	\delta_{\epsilon} \delta_{\epsilon'} = 0.
\end{equation}
These transformations are generated by a (fermionic) charge $Q_B$ called the BRST charge
\begin{equation}
	\delta_{\epsilon} \phi^i = \I \, \com{\epsilon Q_B}{\phi^i}
\end{equation}
and similarly for the other fields (stripping the $\epsilon$ outside the commutator turns it to an anticommutator if the field is fermionic).
Taking the ghosts to be Hermitian leads to an Hermitian charge.

An important consequence is that the two additional terms of the action can be rewritten as a BRST exact terms
\begin{equation}
	S_{\text{gf}} + S_{\text{gh}}
		= \anticom{Q_B}{b_A F^A}.
\end{equation}
A small change in the gauge-fixing condition $\delta F$ leads to a variation of the action
\begin{equation}
	\delta S = \anticom{Q_B}{b_A \delta F^A}.
\end{equation}
\index{BRST quantization!change of gauge fixing condition}%
\index{BRST quantization!charge nilpotency}%
The BRST charge should commute with the Hamiltonian in order to be conserved: this should hold in particular when changing the gauge fixing condition
\begin{equation}
	\com{Q_B}{\anticom{Q_B}{b_A \delta F^A}} = 0
	\quad \Longrightarrow \quad
	Q_B^2 = 0.
\end{equation}

Some vocabulary is needed before proceeding further.
A state $\ket{\psi}$ is said to be BRST \emph{closed} if it is annihilated by the BRST charge
\begin{equation}
	\text{$\ket{\psi}$ closed}
	\quad \Longleftrightarrow \quad
	\ket{\psi} \in \ker Q_B
	\quad \Longleftrightarrow \quad
	Q_B \ket{\psi} = 0.
\end{equation}
States which are in the image of $Q_B$ (i.e.\ they can be written as $Q_B$ applied on some other states) are said to be \emph{exact}
\begin{equation}
	\text{$\ket{\psi}$ exact}
	\quad \Longleftrightarrow \quad
	\ket{\psi} \in \Im Q_B
	\quad \Longleftrightarrow \quad
	\exists \ket{\chi}: \;
		\ket{\psi} = Q_B \ket{\chi}.
\end{equation}
\index{BRST quantization!cohomology}%
The \emph{cohomology} $\mc H(Q_B)$ of $Q_B$ is the set of closed states which are not exact
\begin{equation}
	\ket{\psi} \in \mc H(Q_B)
	\quad \Longleftrightarrow \quad
	\ket{\psi} \in \ker Q_B, \quad
		\nexists \ket{\chi}: \;
		\ket{\psi} = Q_B \ket{\chi}.
\end{equation}
Hence the cohomology corresponds to
\begin{equation}
	\mc H(Q_B) = \frac{\ker Q_B}{\Im Q_B}.
\end{equation}
Two elements of the cohomology differing by an exact state are in the same equivalence class
\begin{equation}
	\ket{\psi} \simeq \ket{\psi} + Q_B \ket{\chi}.
\end{equation}

Considering the $S$-matrix $\bracket{\psi_f}{\psi_i}$ between a set of physical initial states $\psi_i$ and final states $\psi_f$, a small change in the gauge-fixing condition leads to
\begin{equation}
	\delta_F \bracket{\psi_f}{\psi_i}
		= \bra{\psi_f} \anticom{Q_B}{b_A \delta F^A} \ket{\psi_i}
\end{equation}
after expanding the exponential to first order.
\index{BRST quantization!change of gauge fixing condition}%
Since the $S$-matrix should not depend on the gauge this implies that a physical state $\psi$ must be BRST closed (i.e.\ invariant)
\begin{equation}
	Q_B \ket{\psi} = 0.
\end{equation}
Conversely, this implies that any state of the form $Q_B \ket{\chi}$ cannot be physical because it is orthogonal to every physical state $\ket{\psi}$
\begin{equation}
	\bra{\psi} Q_B \ket{\chi} = 0.
\end{equation}
This implies in particular that the amplitudes involving $\ket{\psi}$ and $\ket{\psi} + Q_B \ket{\chi}$ are identical, and any amplitude for which an external state is exact vanishes.
\index{BRST quantization!physical states}%
As a conclusion, physical states are in the BRST cohomology
\begin{equation}
	\text{$\ket{\psi}$ physical}
	\quad \Longleftrightarrow \quad
	\ket{\psi} \in \mc H(Q_B).
\end{equation}

If there is a gauge where the ghosts decouple from the matter field, then the invariance of the action and of the S-matrix under changes of the gauge fixing ensures that this statement holds in any gauge (but, one still need to check that the gauge preserves the other symmetries).
If such a gauge does not exist, then one needs to employ other methods to show the desired result.

Note that $B_A$ can be integrated out by using its equations of motion
\begin{equation}
	\frac{\delta F^A}{\delta \phi^i} \, B_A = - \frac{\delta S_m}{\delta \phi^i},
\end{equation}
and this modifies the BRST transformation of the anti-ghost to
\begin{equation}
	\delta_{\epsilon} b_A = - \epsilon \, \left( \frac{\delta F^A}{\delta \phi^i} \right)^{-1} \frac{\delta S_m}{\delta \phi^i}.
\end{equation}
It is also possible to introduce a term
\begin{equation}
	\anticom{Q_B}{b_A B_B M^{AB}} = \I \, B_A M^{AB} B_B
\end{equation}
for any constant matrix $M^{AB}$.
Since this is also a BRST exact term, the amplitudes are not affected.
Integrating over $B_A$ produces a Gaussian average instead of a delta function to fix the gauge.

In the previous discussion, the BRST symmetry was assumed to originate from the Faddeev--Popov gauge fixing.
But, in fact, it is possible to start directly with an action of the form
\begin{equation}
	\label{qft:eq:action-brst}
	S[\phi, b, c, B] = S_0[\phi] + Q_B \Psi[\phi, b, c, B]
\end{equation}
where $\Psi$ has ghost number $-1$.
It can be proven that this is the most general action invariant under the BRST transformations \eqref{qft:eq:transf-brst}.
This can describe gauge fixed action which cannot be described by the Faddeev--Popov procedure: in particular, the latter yields actions which are quadratic in the ghost fields (by definition of the Gaussian integral representation of the determinant), but this does not exhaust all the possibilities.
For example, the background field method applied to Yang--Mills theory requires using an action quartic in the ghosts.

In this section, several hypothesis have been implicit (off-shell closure, irreducibility and constant structure coefficients).
If one of them breaks, then it is necessary to employ the more general BV formalism.

\section{BV formalism}
\label{app:sec:qft:bv}

\index{Batalin--Vilkovisky formalism}%
The Batalin--Vilkovisky (BV, or also field--antifield) formalism is the most general framework to quantize theories with a gauge symmetry.
While the BRST formalism (\Cref{app:sec:qft:brst}) is sufficient to describe simple systems, it breaks down when the structure of the gauge symmetry is more complicated, for example in systems implying gravity.
The BV formalism is required in the three following cases (which can occur simultaneously):
\begin{enumerate}
	\item the gauge algebra is open (on-shell closure);
	\item the structure coefficients depend on the fields;
	\item the gauge symmetry is reducible (not all transformations are independent).
\end{enumerate}
The BV formalism is also useful for standard gauge symmetries to demonstrate renormalizability and to deal with anomalies.

As explained in the previous section, the ghosts and the BRST symmetry are crucial to ensure the consistency of the gauge theory.
The idea of the BV formalism is to put on an equal footing the physical fields and all the required auxiliary and ghost fields (before gauge fixing).
The introduction of antifields -- one for each of the fields -- and the description of the full quantum dynamics in terms of a quantum action (constrained by the quantum master equation) ensure the consistency of the system.
Additional benefits are the presence of a (generalized) BRST symmetry, the existence of a Poisson structure (which allows to bring concepts from the Hamiltonian formalism), the covariance of the formalism and the simple interpretation of counter-terms as corrections to the classical action.

For giving a short intuition, the BV formalism can be interpreted as providing a (anti)ca\-nonical structure in the Lagrangian formalism, the role of the Hamiltonian being played by the action.

\subsection{Properties of gauge algebra}

Before explaining the BV formalism, we review the situations listed above.
The classical action for the physical fields $\phi^i$ is denoted by $S_0[\phi]$ and the associated equations of motion by
\begin{equation}
	\mc F_i(\phi) = \frac{\pd S_0}{\pd \phi^i}.
\end{equation}
Then, a gauge algebra is open and has field-dependent structure coefficients $F_{ab}^c(\phi)$ if:
\begin{equation}
	\com{T_{a}}{T_{b}}
		= F_{ab}^c(\phi) T_{c} + \lambda_{ab}^{i} \mc F_i(\phi).
\end{equation}
On-shell, $\mc F_i = 0$ and the second term is absent, such that the algebra closes.
The fields themselves are constants from the point of view of the gauge algebra, but their presences in the structure coefficients complicate the analysis of the theory.
Moreover, the path integral is off-shell and for this reason one needs to take into account the last term.

Finally, the gauge algebra can be reducible: in brief, it means that there are gauge invariances associated to gauge parameters -- and correspondingly ghosts for ghosts --, and this recursively.
Since there is one independent ghost for each generator, there are too many ghosts if the generators are not all independent, and there is a remnant gauge symmetry for the ghost fields (in the standard Faddeev--Popov formalism, the ghosts are not subject to any gauge invariance).
This originates from relations between the generators $R_a^i$: denoting by $m_0$ the number of level-$0$ gauge transformations, the number of independent generators is $\rank R_a^i$.
Then, the
\begin{equation}
	m_1 = m_0 - \rank R_a^i
\end{equation}
relations between the generators translate into a level-$1$ gauge invariance of the ghosts.
This symmetry can be gauge fixed by performing a second time the Faddeev--Popov procedure, yielding commuting ghosts.
This symmetry can also be reducible, and the procedure can continue without end.
If one finds that the gauge invariance at level $n = \ell$ is irreducible, one says that the gauge invariance is $\ell$-reducible.
If this does not happen, one defines $\ell = \infty$.
The number of generators at level $n$ is denoted by $m_n$.

\begin{example}[$p$-form gauge theory]
	A $p$-form gauge theory is written in terms of a gauge field $A_{p}$ with a a gauge invariance
	\begin{equation}
		\delta A_{p} = \dd \lambda_{p-1}.
	\end{equation}
	But, due to the nilpotency of the derivative, deformations of the gauge parameter satisfying
	\begin{equation}
		\delta \lambda_{p-1} = \dd \lambda_{p-2}
	\end{equation}
	does not translate into a gauge invariance of $A_{p}$.
	Similarly from this should be excluded the transformation
	\begin{equation}
		\delta \lambda_{p-2} = \dd \lambda_{p-3},
	\end{equation}
	and so on until one reaches the case $p = 0$.
	Hence, a $p$-form field has a $p$-reducible gauge invariance.
\end{example}

\subsection{Classical BV}

\index{Batalin--Vilkovisky formalism!fields and antifields}%
Denoting the fields collectively as
\begin{equation}
	\psi^r = \{ \phi^i, B_A, b_A, c^a \},
\end{equation}
the simplest BV action reads
\begin{equation}
	S[\psi^r, \psi_r^*] = S_0[\phi] + Q_B \psi^r \, \psi_r^*
\end{equation}
with the antifields
\begin{equation}
	\psi_r^* = \{ \phi_i^*, B^{A*}, b^{A*}, c_a^* \}.
\end{equation}
The action \eqref{qft:eq:action-brst} is recovered by writing
\begin{equation}
	\psi_r^* = \frac{\pd \Psi}{\pd \psi^r}.
\end{equation}
This indicates that the general BRST formalism could be rephrased in the BV language.
But, in the same way that the BRST formalism generalizes the Faddeev--Popov formalism, it is in turn generalized by the BV formalism.
Indeed, the above action is linear in the antifields: this constraint is not required and one can write more general actions.
In the rest of this section, we explain how this works at the level of the action (classical level) and how the sets of fields and antifields are defined.

Consider a set of physical fields $\phi^i$ with the gauge invariance
\begin{equation}
	\delta \phi^i = \epsilon_0^{a_0} R_{a_0}^i(\phi^i).
\end{equation}
Then, associate a ghost field $c^{a_0}$ to each of the gauge parameters $\epsilon^{a_0}$.
If the gauge symmetry is reducible, a new gauge invariance is associated to the ghosts
\begin{equation}
	\delta c_0^{a_0} = \epsilon_1^{a_1} R_{a_1}^{a_0}(\phi^i, c^{a_0}).
\end{equation}
This structure is recurring and the ghosts of the level-$n$ gauge invariance are denoted by $c^{a_n}$ and they satisfy
\begin{equation}
	\delta c_n^{a_n} = \epsilon_{n+1}^{a_{n+1}} R_{a_{n+1}}^{a_n}(\phi^i, c_0^{a_0}, \ldots, c_n^{a_n}).
\end{equation}
Thus, the set of fields is
\begin{equation}
	\psi^r = \{ c_n^{a_n} \}_{n = -1, \ldots, \ell},
	\qquad
	c_{-1} := \phi.
\end{equation}
A ghost number is introduced
\begin{equation}
	N_{\text{gh}}(\phi^i) = 0,
	\qquad
	N_{\text{gh}}(c_n^{a_n}) = n + 1,
\end{equation}
and the Grassmann parity of the ghosts is defined to be opposite (resp.\ identical) of the parity of the associated gauge parameter for even (resp.\ odd) $n$
\begin{equation}
	\abs{c_n} = \abs{\epsilon_n^{a_n}} + n + 1.
\end{equation}
To each of these fields is associated an antifield $\psi_r^*$ of opposite parity as $\psi^r$ and such that their ghost numbers sum to $-1$
\begin{equation}
	N_{\text{gh}}(\psi_r^*) = - 1 - N_{\text{gh}}(\psi^r),
	\qquad
	\abs{\psi_r^*} = - \abs{\psi^r}.
\end{equation}

The fields and antifields together are taken to define a graded symplectic structure
\begin{equation}
	\omega = \sum_r \dd \psi^r \wedge \dd \psi_r^*
\end{equation}
with respect to which they are conjugated to each other
\begin{equation}
	\psp{\psi^r}{\psi_s^*} = \delta_{rs},
	\qquad
	\psp{\psi^r}{\psi^s} = 0,
	\qquad
	\psp{\psi_r^*}{\psi_s^*} = 0.
\end{equation}
\index{Batalin--Vilkovisky formalism!antibracket}%
The antibracket (graded Poisson bracket) $\psp{\cdot}{\cdot}$ reads
\begin{equation}
	\label{qft:eq:antibracket}
	\psp{A}{B} = \frac{\pd_R A}{\pd \psi^r} \frac{\pd_L B}{\pd \psi_r^*} - \frac{\pd_R A}{\pd \psi_r^*} \frac{\pd_L B}{\pd \psi^r},
\end{equation}
where the $L$ and $R$ indices indicate left and right derivatives.
It is graded symmetric, which means
\begin{equation}
	\psp{A}{B} = - (-1)^{(\abs{A}+1) (\abs{B}+1)} \psp{B}{A}.
\end{equation}
It also satisfies a graded Jacobi identity and the property
\begin{equation}
	N_{\text{gh}}(\psp{A}{B}) = N_{\text{gh}}(A) + N_{\text{gh}}(B) + 1,
	\qquad
	\abs{\psp{A}{B}} = \abs{A} + \abs{B} + 1 \mod 2.
\end{equation}
Moreover, the antibracket acts as a derivative
\begin{equation}
	\psp{A}{B C} = \psp{A}{B} C + (-1)^{\abs{B} {C}} \psp{A}{C} B.
\end{equation}

The dynamics of the theory is described by the (classical) master action $S[\psi^r, \psi_r^*]$ which satisfies
\begin{equation}
	N_{\text{gh}}(S) = 0,
	\qquad
	\abs{S} = 0.
\end{equation}
In order to reproduce correctly the dynamics of the classical system without ghosts, this action is required to satisfy the boundary condition
\begin{equation}
	S[\psi^r, \psi_r^* = 0] = S_0[\phi^i],
	\qquad
	\frac{\pd_L \pd_R S}{\pd c_{n-1,a_{n-1}}^* \pd c_n^{a_n}}\bigg|_{\psi^* = 0} = R^{a_{n-1}}_{a_n}.
\end{equation}
Indeed, if the antifields are set to zero, the ghost fields cannot appear because they all have positive ghost numbers and it is not possible to build terms with vanishing ghost numbers from them.

\index{Batalin--Vilkovisky formalism!BRST transformation}%
In analogy with the Hamiltonian formalism, the master action can be used as the generator of a global fermionic symmetry, and inspection will show that it corresponds to a generalization of the BRST symmetry.
Writing the generalized and classical BRST operator as $s$, the transformations of the fields and antifields read
\begin{subequations}
\begin{align}
	\delta_\theta \psi^r
		&= \theta \, s \psi^r
		= - \theta \, \psp{S}{\psi_r}
		= \theta \, \frac{\pd_R S}{\pd \psi_r^*},
	\\
	\delta_\theta \psi_r^*
		&= \theta \, s \psi_r^*
		= - \theta \, \psp{S}{\psi_r^*}
		= - \theta \, \frac{\pd_R S}{\pd \psi^r},
\end{align}
\end{subequations}
where $\theta$ is a constant Grassmann parameter.
The variation of a generic functional $F[\psi^r, \psi_r^*]$ is
\begin{equation}
	\delta_\theta F
		= \theta \, s F
		= - \theta \, \psp{S}{F}.
\end{equation}
\index{Batalin--Vilkovisky formalism!classical master equation}%
For the BRST transformation to be a symmetry of the action, the action must satisfy the classical master equation
\begin{equation}
	\psp{S}{S} = 0.
\end{equation}
This equation can easily be solved by expanding $S$ in the ghosts: the various terms can be interpreted in terms of properties of the gauge algebra.
Then, the Jacobi identity used with two $S$ and an arbitrary functional gives
\begin{equation}
	\psp{S}{\psp{S}{F}} = 0
\end{equation}
and this implies that the transformation is nilpotent
\begin{equation}
	s^2 = 0.
\end{equation}
A classical observable $\mc O$ satisfies
\begin{equation}
	s \mc O = 0.
\end{equation}

\index{Batalin--Vilkovisky formalism!field redefinition}%
Due to the BRST symmetry, the action is not uniquely defined and the action
\begin{equation}
	S' = S + \psp{S}{\delta F}
\end{equation}
also satisfies the master equation, where $\delta F$ is arbitrary up to the condition $N_{\text{gh}}(\delta F) = - 1$.
This can be interpreted as the action $S$ in a new coordinate system $(\psi'^r, \psi'^*_r)$ with
\begin{equation}
	\psi' = \psi - \frac{\delta F}{\delta \psi^*},
	\qquad
	\psi'^* = \psi^* + \frac{\delta F}{\delta \psi}
\end{equation}
such that
\begin{equation}
	S'[\psi, \psi^*] = S\left[\psi - \frac{\delta F}{\delta \psi^*}, \psi^* + \frac{\delta F}{\delta \psi} \right].
\end{equation}
Indeed, for $F = F[\psi, \psi^*]$, one has
\begin{equation}
	S'[\psi, \psi^*]
		= S[\psi, \psi^*]
			+ \psp{S}{\psi} \frac{\delta F}{\delta \psi}
			+ \psp{S}{\psi^*} \frac{\delta F}{\delta \psi^*}
		= S[\psi, \psi^*]
			- \frac{\pd_R S}{\pd \psi^*} \, \frac{\delta F}{\delta \psi}
			+ \frac{\pd_R S}{\pd \psi} \frac{\delta F}{\delta \psi^*}.
\end{equation}
It can be shown that this transformation preserves the antibracket and the master equation
\begin{equation}
	\psp{\psi'^r}{\psi'^*_s} = \delta_{rs},
	\qquad
	\psp{S'}{S'} = 0.
\end{equation}
More generally, any transformation preserving the antibracket is called an (anti)canonical transformation.
One can also consider generating functions depending on both the old and new coordinates, as is standard in the Hamiltonian formalism.
Under a transformation, any object depending on the coordinates changes as
\begin{equation}
	G' = G + \psp{\delta F}{G}.
\end{equation}
One can consider finite transformation without problems.

\index{Batalin--Vilkovisky formalism!gauge fixing}%
In order to perform the gauge fixing, one needs to eliminate the antifields.
A convenient condition is
\begin{equation}
	S_\Psi[\psi^r] = S\left[ \psi^r , \frac{\pd \Psi}{\pd \psi^r} \right],
	\qquad
	\psi_r^* = \frac{\pd \Psi}{\pd \psi^r},
\end{equation}
where $\Psi[\psi^r]$ is called the gauge fixing fermion and satisfies
\begin{equation}
	N_{\text{gh}}(\Psi) = - 1,
	\qquad
	\abs{\Psi} = 1.
\end{equation}
From the discussion on coordinate transformations this amounts to work in new coordinates where $\psi'^*_r = 0$.
But such a function $\Psi$ cannot be built from the fields because they all have positive ghost numbers.
One needs to introduce \emph{trivial pairs} of fields.

A trivial pair $(B, \bar c)$ is defined by the properties
\begin{subequations}
\begin{gather}
	\abs{B} = - \abs{\bar c},
	\qquad
	N_{\text{gh}}(B) = N_{\text{gh}}(\bar c) + 1,
	\\
	s \bar c = B,
	\qquad
	s B = 0
\end{gather}
\end{subequations}
and the new action reads
\begin{equation}
	\bar S = S[\psi^r, \psi_r^*] - B \bar c^*
\end{equation}
(the position dependence is kept implicit).
In this context $\psi^r$ and $\psi_r^*$ are sometimes called minimal variables.
From this, one learns that
\begin{equation}
	\psp{\bar S}{\bar S} = \psp{S}{S}
		= 0.
\end{equation}

At level-$0$, one introduces the pair
\begin{equation}
	(B_{0 a_0}, \bar c_{0 a_0})
		:= (B^0_{0 a_0}, \bar c^0_{0 a_0})
\end{equation}
and the associated antifields.
The field $\bar c_0 := b$ is the Faddeev--Popov anti-ghost associated to $c_0$ and the trivial pair satisfies
\begin{equation}
	\abs{B_0} = \abs{\epsilon_0},
	\qquad
	\abs{\bar c_0} = - \abs{\epsilon_0},
	\qquad
	N_{\text{gh}}(B_0) = 0,
	\qquad
	N_{\text{gh}}(c_0) = - 1.
\end{equation}
For the level $1$, two additional pairs are introduced:
\begin{equation}
	(B^0_{1 a_1}, \bar c^0_{1 a_1}),
	\qquad
	(\bar B^{1 a_1}_{1}, c^{1 a_1}_{1})
\end{equation}
and the corresponding antifields.
The motivation for adding an additional pair is that the level-$0$ pair only fixes $m_0 - m_1$ of the generators: the additional $m_1$ extra-ghosts $c^{1 a_1}_{1}$ can be fixed by the residual level-$0$ symmetry.
The first level-$1$ pair fixes the level-$1$ symmetry.

Then, the gauge fixed action enjoys a BRST symmetry acting only on the fields
\begin{equation}
	\delta_\theta \psi^r
		= \theta \, s \psi^r
		= \theta \, \frac{\pd_R S}{\pd \psi_r^*}\bigg|_{\psi_r^* = \pd_r \Psi}.
\end{equation}
Note that this BRST operator is generically nilpotent only on-shell
\begin{equation}
	s^2 \propto \text{eom}.
\end{equation}

\subsection{Quantum BV}

At the quantum level,      one considers the path integral
\begin{equation}
	Z = \int \dd \psi^r \dd \psi_r^* \, \e^{- W[\psi^r, \psi_r^*] / \hbar}
\end{equation}
where $W$ is called the quantum master action.
The reason for distinguishing it from the classical master action $S$ is that the measure is not necessarily invariant by itself under the generalized BRST transformation -- this translates into a non-gauge invariance of the measure of the physical fields, i.e.\ a gauge anomaly.

\index{Batalin--Vilkovisky formalism!quantum BRST transformation}%
Quantum BRST transformation are generated by the quantum BRST operator $\sigma$
\begin{equation}
	\delta_\theta F
		= \theta \, \sigma F
		= \psp{W}{F} - \hbar \, \lap F,
\end{equation}
where
\begin{equation}
	\lap = \frac{\pd_R}{\pd \psi_r^*} \frac{\pd_L}{\pd \psi^r}.
\end{equation}
Then, the path integral is invariant if $W$ satisfies the quantum master equation
\index{Batalin--Vilkovisky formalism!quantum master equation}%
\begin{equation}
	\label{qft:eq:master-equation-quantum}
	\psp{W}{W} - 2 \hbar \lap W
		= 0,
\end{equation}
which can also be written as
\begin{equation}
	\lap \e^{- W / \hbar} = 0.
\end{equation}
This can be interpreted as the invariance of $Z$ under changes of coordinates: indeed one finds that
\begin{equation}
	\delta W = \frac{1}{2} \psp{W}{W},
\end{equation}
and the integration measure picks a Jacobian
\begin{equation}
	\sdet J \sim 1 + \lap W.
\end{equation}
In the limit $\hbar \to 0$, one recovers the classical master equation.
More generally, the action can be expanded in powers of $\hbar$
\begin{equation}
	W = S + \sum_{p \ge 1} \hbar^p W_p.
\end{equation}

Observables are given by operators $\mc O[\psi, \psi^*]$ invariant under $\sigma$:
\begin{equation}
	\sigma \mc O = 0,
\end{equation}
which ensures that the expectation value is invariant under changes of $\Psi$
\begin{equation}
	\delta \mean{\mc O} = 0.
\end{equation}
Note that if $\mc O$ depends just on $\psi$ the condition reduces to $s \mc O = 0$, but generically there is no such operators (except constants) satisfying this condition for open algebra.

Consider the gauge fixed integral
\begin{equation}
	Z = \int \dd \psi^r \, \e^{- W_\Psi[\psi^r]},
	\qquad
	W_\Psi[\psi^r] = W\left[\psi^r, \frac{\pd \Psi}{\pd \psi^r}\right].
\end{equation}
Varying the gauge fixing fermion by $\delta \Psi$ gives
\begin{equation}
	Z = \int \dd \psi^r \, \e^{- W_\Psi[\psi^r]} \left( \frac{\pd_R S}{\pd \psi_r^*} \right)_{\psi^* = \pd_\psi \Psi} \frac{\pd(\delta \Psi)}{\pd \psi^r}.
\end{equation}
Integrating by part gives the quantum master equation.

\refchapter

\begin{itemize}
	\item Manipulations of functional integral are given in~\cites[sec.~15.1, 22.1]{Hatfield:1998:QuantumFieldTheory}[chap.~14]{Nakahara:2003:GeometryTopologyPhysics}{Polchinski:1986:EvaluationOneLoop}{DHoker:1988:GeometryStringPerturbation}.

	\item Zero-modes are discussed in~\cite{Blau:1989:DeterminantsDiracOperators}.

	\item A general summary of path integrals for bosonic and fermionic fields can be found in~\cite[app.~A]{Polchinski:2005:StringTheory-1}.

	\item BRST formalism: most QFT books contain an introduction, more complete references are~\cites[chap.~15]{Weinberg:2005:QuantumTheoryFields-2}{vanHolten:2004:AspectsBRSTQuantization}{Henneaux:1994:QuantizationGaugeSystems};

	\item BV formalism~\cites[chap.~15]{Weinberg:2005:QuantumTheoryFields-2}{vanHolten:2004:AspectsBRSTQuantization, Fuster:2005:BRSTantifieldQuantizationShort, Gomis:1995:AntibracketAntifieldsGaugeTheory}{Henneaux:1994:QuantizationGaugeSystems}{DeWitt:2014:GlobalApproachQuantum} (several explicit examples are given in~\cite[sec.~3]{Gomis:1995:AntibracketAntifieldsGaugeTheory}, see~\cite{Batalin:1983:QuantizationGaugeTheories, Witten:1990:NoteAntibracketFormalism, Sen:1994:NoteGaugeTransformations} for more specific details).
\end{itemize}

\printbibliography[heading=bibintoc]

\phantomsection
\printindex

\end{document}